\documentclass[letterpaper,headsepline,DIV=15,BCOR=20mm,font=small,labelfont=bf]{scrbook}

\let\mypdfximage\pdfximage
\protected\def\pdfximage{\immediate\mypdfximage}

\usepackage[automark]{scrpage2}
\lehead{CBETA Design Report  }

\usepackage[USenglish]{babel}
\usepackage{graphicx,pstricks}
\usepackage{graphics}
\usepackage{epsfig}
\usepackage{subfig}
\usepackage{verbatim}
\usepackage{multirow}
\usepackage{enumerate}
\usepackage[pdftex=true,hyperindex=true,colorlinks=true,breaklinks=true, urlcolor=blue, linkcolor=black,citecolor=black,bookmarksnumbered=true,pdfpagelabels=true,plainpages=false]{hyperref}

\usepackage{booktabs} 
\ifpdf \usepackage{pdflscape} \else \usepackage{lscape} \fi 

\usepackage{savesym}
\usepackage[intlimits]{amsmath}
\savesymbol{iint}
\restoresymbol{TXF}{iint}
\usepackage{amssymb}
\numberwithin{equation}{section}
\numberwithin{figure}{section}
\numberwithin{table}{section}

\renewcommand{\bibname}{References}

\usepackage[sectionbib]{chapterbib} 

\usepackage[sectionbib,square,sort&compress]{natbib} 
\setcitestyle{numbers}
\setcitestyle{citesep={,}}  

\renewcommand\bibname{References} 

\usepackage{ifthen}

\usepackage[nottoc]{tocbibind}

\newboolean{FullDocument}
\setboolean{FullDocument}{true}
\newcommand{\buildingFullDocument}[1]{\def\@buildingFullDocument{#1}}
\buildingFullDocument{}

\newcommand{\FiguresDirectory}{figures}
\newcommand{\FullDocumentRoot}{.}

\newcommand{\unit}[1]{\,\ensuremath{\mathrm{#1}}}

\newcommand{\Eq}[1]{Eq.~\eqref{#1}}

\newcommand{\Fig}[1]{Fig.~\ref{#1}}
\newcommand{\Figure}[1]{Figure~\ref{#1}}
\newcommand{\Tab}[1]{Tab.~\ref{#1}}
\newcommand{\Table}[1]{Table~\ref{#1}}

\newcommand{\Section}[1]{\S\ref{#1}} 

\newcommand{\Ref}[1]{\citep{#1}} 

\newcommand{\Leader}[1]{}

\begin{document}

\bibliographystyle{\FullDocumentRoot/references/pddr}
\renewcommand{\bibname}{References}

\title{CBETA Design Report}

\subtitle{Cornell-BNL ERL Test Accelerator}

\author{
\begin{small}
\begin{minipage}{\textwidth}
\emph{Principle Investigators:} G.H.~Hoffstaetter, D.~Trbojevic\\
\\
\emph{Editor:} C.~Mayes\\
\\
\emph{Contributors:}  N.~Banerjee, J.~Barley, I.~Bazarov, A.~Bartnik, J.~S.~Berg, S.~Brooks, D.~Burke, J.~Crittenden, L.~Cultrera, J.~Dobbins, D.~Douglas, B.~Dunham, R.~Eichhorn, S.~Full, F.~Furuta, C.~Franck, R.~Gallagher, M.~Ge, C.~Gulliford, B.~Heltsley, D.~Jusic, R.~Kaplan, V.~Kostroun, Y.~Li, M.~Liepe, C.~Liu, W.~Lou, G.~Mahler, F.~M\'eot, R.~Michnoff, M.~Minty, R.~Patterson, S.~Peggs, V.~Ptitsyn, P.~Quigley, T.~Roser, D.~Sabol, D.~Sagan, J.~Sears, C.~Shore, E.~Smith, K.~Smolenski, P.~Thieberger, S.~Trabocchi, J.~Tuozzolo, N.~Tsoupas, V.~Veshcherevich, D.~Widger, G.~Wang, F.~Willeke, W.~Xu \\ \\
\begin{center}
\includegraphics[width=\textwidth]{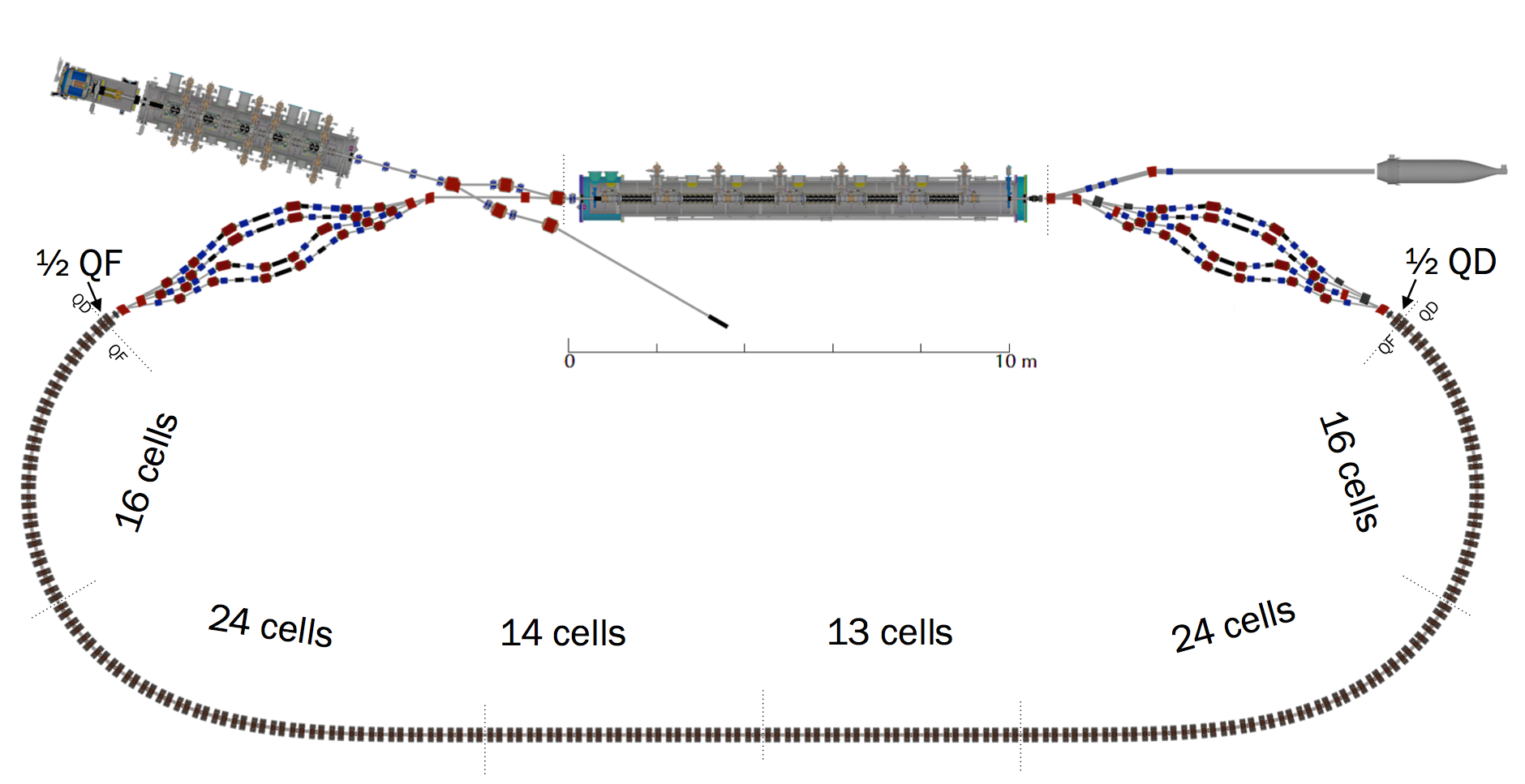}
\end{center}
\end{minipage}
\end{small}
}
\date{June 8, 2017}

%




\maketitle

\section*{Preface}
This Design Report describes the baseline design of the Cornell-BNL ERL Test Accelerator, as it exists on the date of its publication in June 2017.


The Design Report will not change frequently in the future.  In contrast, the parameter sheets that summarize the CBETA design will respond as quickly and as thoroughly as necessary to maintain configuration control.
\clearpage

\tableofcontents

%


\ifdefined \buildingFullDocument

\renewcommand{\FiguresDirectory}{introduction/figures}

\else
\newcommand{\FullDocumentRoot}{..}
\newcommand{\FiguresDirectory}{figures}

\begin{document}
\fi

\chapter{Introduction}\label{chapter:introduction}

\section{Executive Summary}

The Cornell-BNL ERL Test Accelerator (CBETA) will be a unique resource to carry out accelerator science and enable exciting research in nuclear physics, materials science, and industrial applications. Initially it will prototype components and evaluate concepts that are essential for Department of Energy (DOE) plans for an Electron-Ion Collider (EIC).

CBETA is an Energy-Recovery Linac (ERL) that is being constructed at Cornell University. It will be the first ever multi-turn ERL with superconducting RF (SRF) acceleration, and the first ERL based on Fixed Field Alternating Gradient (FFAG) optics. Its ERL technology recovers the energy of the accelerated beam and reuses it for the acceleration of new beam, using accelerator components already constructed and tested at Cornell University \Ref{Cornell-ERL-PDDR}. Additionally, energy is saved because the FFAG optics is built of permanent magnets instead of electro magnets.

The Nuclear Physics (NP) division of DOE has been planning for an EIC for more than a decade. Research and development on this project is mostly performed at Brookhaven National Laboratory (BNL) and at the Thomas Jefferson National Accelerator Facility (TJNAF). BNL on Long Island, NY is planning to transform the Relativistic Heavy Ion Collider (RHIC) into eRHIC \Ref{aschenauer14} while TJNAF is planning to collide an existing electron beam with a new, electron-cooled ion beam in the Jefferson Lab Electron-Ion Collider (JLEIC).

Both EIC projects need an ERL as an electron cooler for low-emittance ion beams. For eRHIC, a new electron accelerator would be installed in the existing RHIC tunnel at BNL, colliding polarized electrons with polarized protons and $^3$He ions, or with unpolarized ions from deuterons to Uranium. There are two concepts, one where electron beam is stored in a ring for a ring-ring collider and another where it is provided by an ERL for a linac-ring collider. Because experiments have to be performed for all combinations of helicity, bunches with alternating polarization have to be provided for the collisions. An electron ring can provide these conditions only when it is regularly filled by a linac. Both eRHIC concepts therefore have a recirculating linac with return loops around the RHIC tunnel. 

Significant simplification and cost reduction is possible by configuring eRHIC with non-scaling FFAG (NS-FFAG) optics in combination with a recirculating linac where several beams with different energy pass through the same FFAG lattice.  Two NS-FFAG beamline arcs placed on top of each other allow multiple passes through a single superconducting linac. For a large accelerator like eRHIC, where each separate return loop is many kilometers long, an FFAG produces a significantly more cost-optimized accelerator.

CBETA will establish the operation of a multi-turn ERL. The return arc, made of an FFAG lattice with large energy acceptance, will be commissioned, establishing this cost-reducing solution for eRHIC. Many effects that are critical for designing the EIC will be measured, including the Beam-Breakup (BBU) instability, halo-development and collimation, growth in energy spread from Coherent Synchrotron Radiation (CSR), and CSR micro bunching. In particular, CBETA will use an NS-FFAG lattice that is very compact, enabling multiple passes of the electron beam in a single recirculation beamline, using the SRF linac four times.

Because the prime accelerator-science motivations for CBETA are essential for an EIC, and address items that are perceived as the main risks of eRHIC, its construction is an important milestone for the NP division of DOE and for BNL. 

The scientific merits of CBETA are even broader, because it produces significantly improved, cost-effective, compact continuous wave (CW) high-brightness electron beams that will enable exciting and important physics experiments, including dark matter and dark energy searches \Ref{milner13}, Q-weak tests at lower energies \Ref{androi13}, proton charge radius measurements, and an array of polarized-electron-enabled nuclear physics experiments. High brightness, narrow line-width gamma rays can be generated by Compton scattering \Ref{albert11} using the ERL beam, to be used for nuclear resonance fluorescence, the detection of special nuclear materials, and an array of astrophysical measurements.  The energy and current range of CBETA will also be ideal for studying high power free electron laser (FEL) physics for materials research and for industrial applications.

CBETA brings together the resources and expertise of a large DOE National Laboratory, BNL, and a leading research university, Cornell.  CBETA will be built in an existing building at Cornell, using many components that have been developed at Cornell under previous R\&D programs that were supported by the National Science Foundation (NSF), New York State, and Cornell University. These components are a fully commissioned world-leading photoemission electron source, a high-power injector, and an ERL accelerator module, both based on SRF systems, and a high-power beam stop.  The only elements that require design and construction from scratch are the permanent-magnet FFAG transport lattices of the return arc.

The collaborative effort between Brookhaven National Laboratory and Cornell University to build a particle accelerator will be a model for future projects between universities and national laboratories, taking advantage of the expertise and resources of both to investigate new topics in a timely and cost-effective manner.

The CBETA project and several associated topics have been presented at IPAC 2017 \Ref{IPAC2017:TUOCB3}. 

\section{The L0E experimental hall at Cornell}

\begin{figure}[htbp]
\centering
\includegraphics[width=0.95\textwidth]{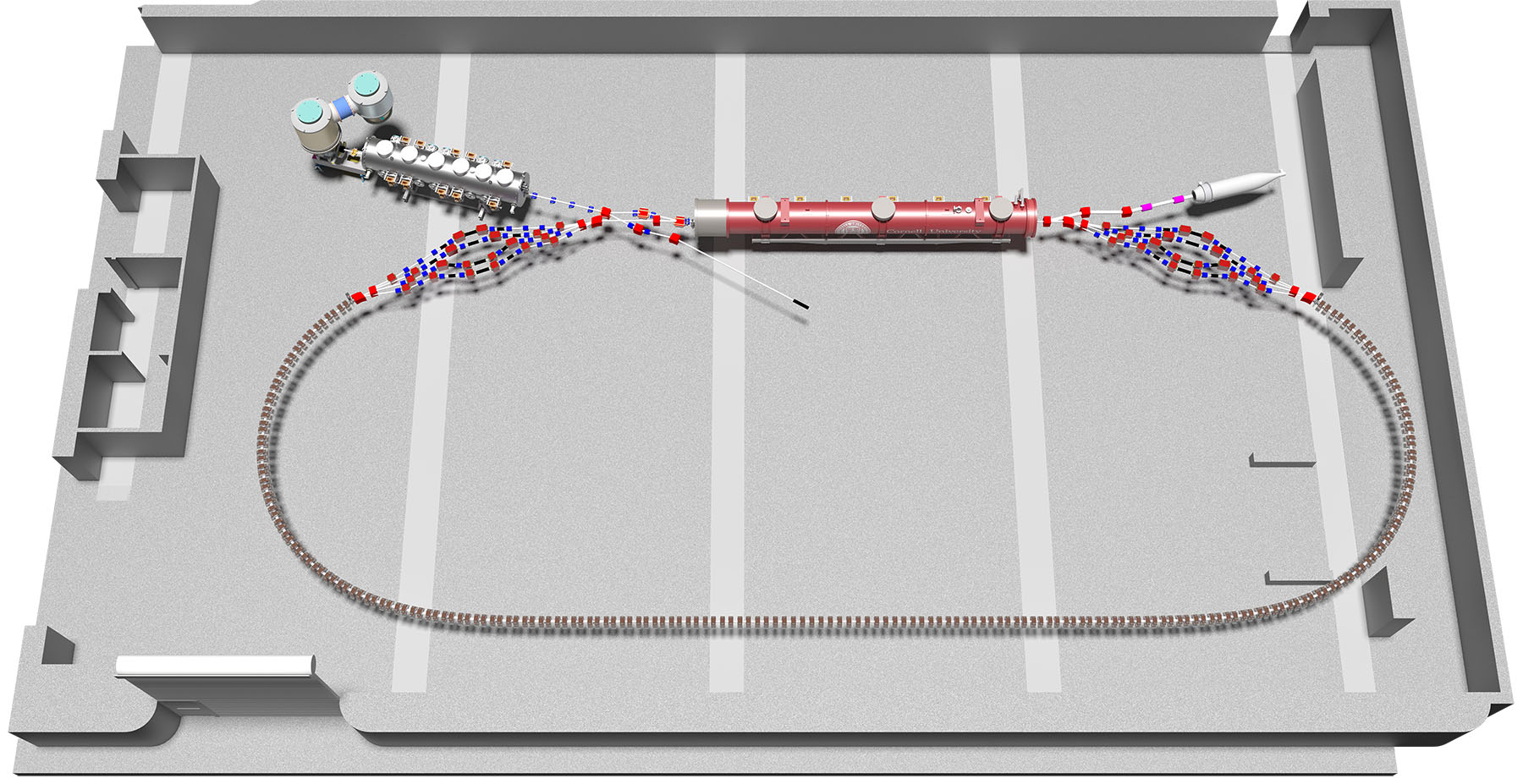}
\caption[]{Floor plan of the Cornell-BNL ERL Test Accelerator in the L0E experimental hall at Cornell's Wilson laboratory.}
\label{fig:L0Ehall}
\end{figure}

Cornell's Wilson laboratory  has an experimental hall that has already largely been freed up for the installation of CBETA. It was originally constructed as the experimental hall for extracted-beam experiments with Cornell's 12~GeV Synchrotron. It is equipped with a high ceiling and an 80 ton crane, with easy access and a suitable environment, mostly below ground level. The dimensions of CBETA fit well into this hall, as shown in \Fig{fig:L0Ehall} with the parameters of \Tab{tab:cbeta_parameters}. 

The DC photo-emitter electron source, the injector linac, the ERL merger, the high-current ERL linac module, and the ERL beam stop are already installed in this hall and are connected to their cryogenic systems and to other necessary infrastructure.

\begin{table}[htbp]
\caption[]{Primary parameters of the Cornell-BNL ERL Test Accelerator.}
\begin{tabular*}{\columnwidth}{@{\extracolsep{\fill}}lll}
\toprule
Parameter & Value & Unit \\
\midrule
Largest energy & 150 & MeV \\
Injection energy & 6 & MeV \\
Linac energy gain & 36 & MeV \\ 
Injector current (max) & 40 & mA \\
Linac passes & 8 (4 accel. + 4 decel.) & \\
Energy sequence in the arc & $42\rightarrow78\rightarrow114\rightarrow150\rightarrow114\rightarrow78\rightarrow42$ & MeV \\
RF frequency & 1300. & MHz \\
Bunch frequency (high-current mode) & 325. & MHz \\
Circumference harmonic & 343 & \\
Circumference length & 79.0997 & m \\
Circumference time (pass 1) & 0.263848164 & \unit{\mu s} \\
Circumference time (pass 2) & 0.263845098 & \unit{\mu s} \\
Circumference time (pass 3) & 0.263844646 & \unit{\mu s} \\
Circumference time (pass 4) & 0.265003298 & \unit{\mu s} \\
Normalized transverse rms emittances & 1 & \unit{\mu m} \\
Bunch length & 4 & ps \\
Typical arc beta functions & 0.4 & m \\
Typical splitter beta functions & 50 & m \\
Transverse rms bunch size (max) & 1800 & \unit{\mu m} \\
Transverse rms bunch size (min) & 52 & \unit{\mu m} \\
Bunch charge (min) & 1 & pC \\
Bunch charge (max) & 123 & pC \\
\bottomrule
\end{tabular*}
\label{tab:cbeta_parameters}
\end{table}

CBETA is not only an excellent accelerator for prototyping components and for developing concepts for the EIC, and in particular for eRHIC, in its experimental hall at Cornell. It is also an important part of future plans at Cornell for accelerator research, nuclear physics research, materials studies, and for ongoing ERL studies.

\section{Existing Components at Cornell}

\paragraph{DC photo-emitter electron source:}
High voltage DC photoemission electron guns offer a robust option for photoelectron sources, with applications such as ERLs. A DC gun for a high brightness, high intensity photoinjector requires a high voltage power supply (HVPS) supplying hundreds of kV to the high voltage (HV) surfaces of the gun. At Cornell, the gun HV power supply for 750~kV at 100~mA is based on proprietary insulating core transformer technology. This technology is schematically  shown in \Fig{fig:diagram_Cornell_gun_intro} for Cornell's DC photoemitter gun. This gun holds the world record  in sustained current of up to 75mA.

\begin{figure}[tb]
\centering
\includegraphics[width=0.7\textwidth]{\FiguresDirectory/newgun_cutout.png}
\caption[A schematic view of the DC gun.]{A cutaway view of the DC photoemission gun.  Photocathodes are prepared in a load lock system mounted on the large flange at the left, and transported through the cathode cylinder to the operating position in the Pierce electrode shape on the right.  The beam exits through the small flange to the right.}
\label{fig:diagram_Cornell_gun_intro}
\end{figure}

\begin{figure}[tb]
\centering
\subfloat[Injector cryomodule]{\includegraphics[width=0.45\textwidth]{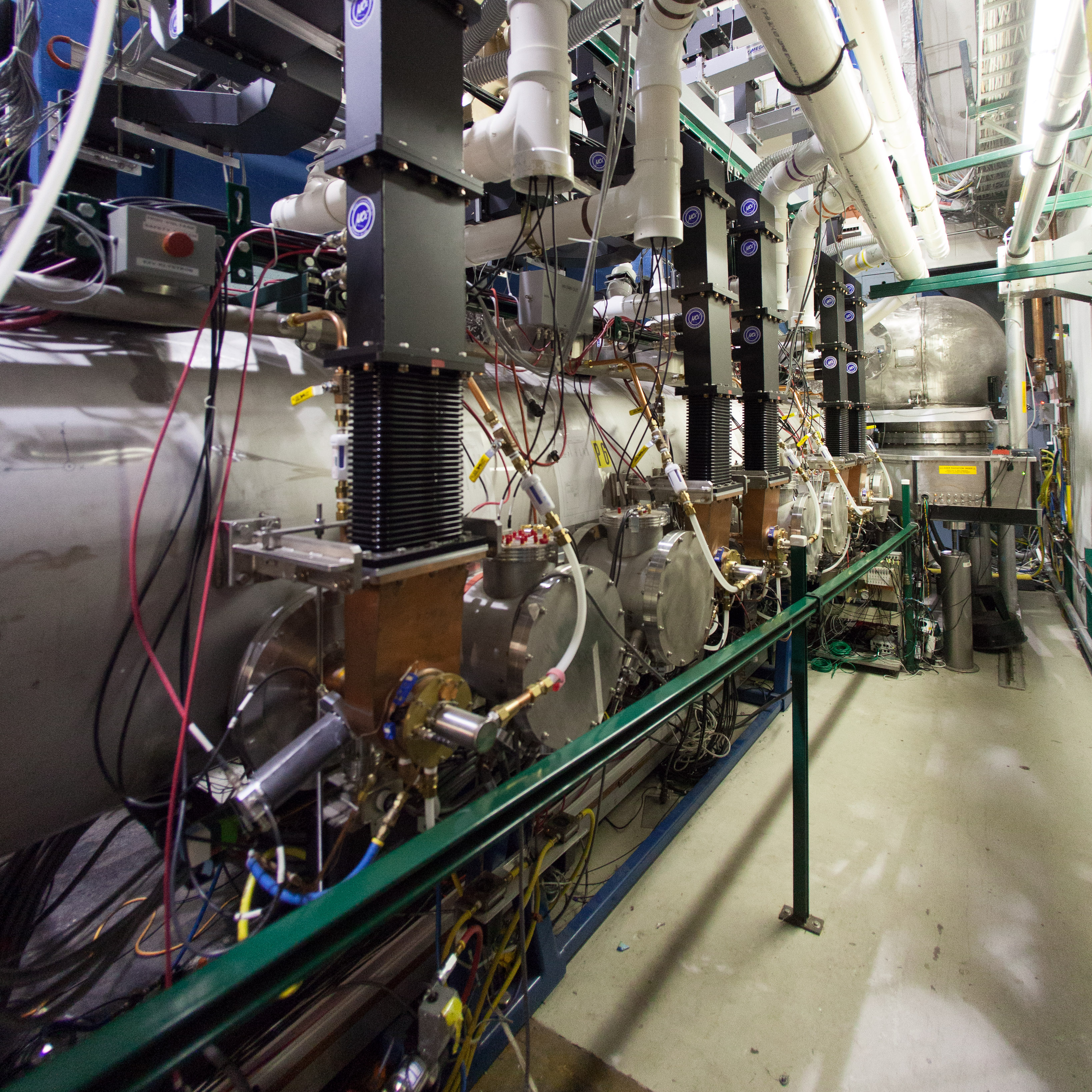}\label{fig:photo_cornell_injector_intro}}
\hspace{0.05\textwidth}
\subfloat[DC gun]{\includegraphics[width=0.45\textwidth]{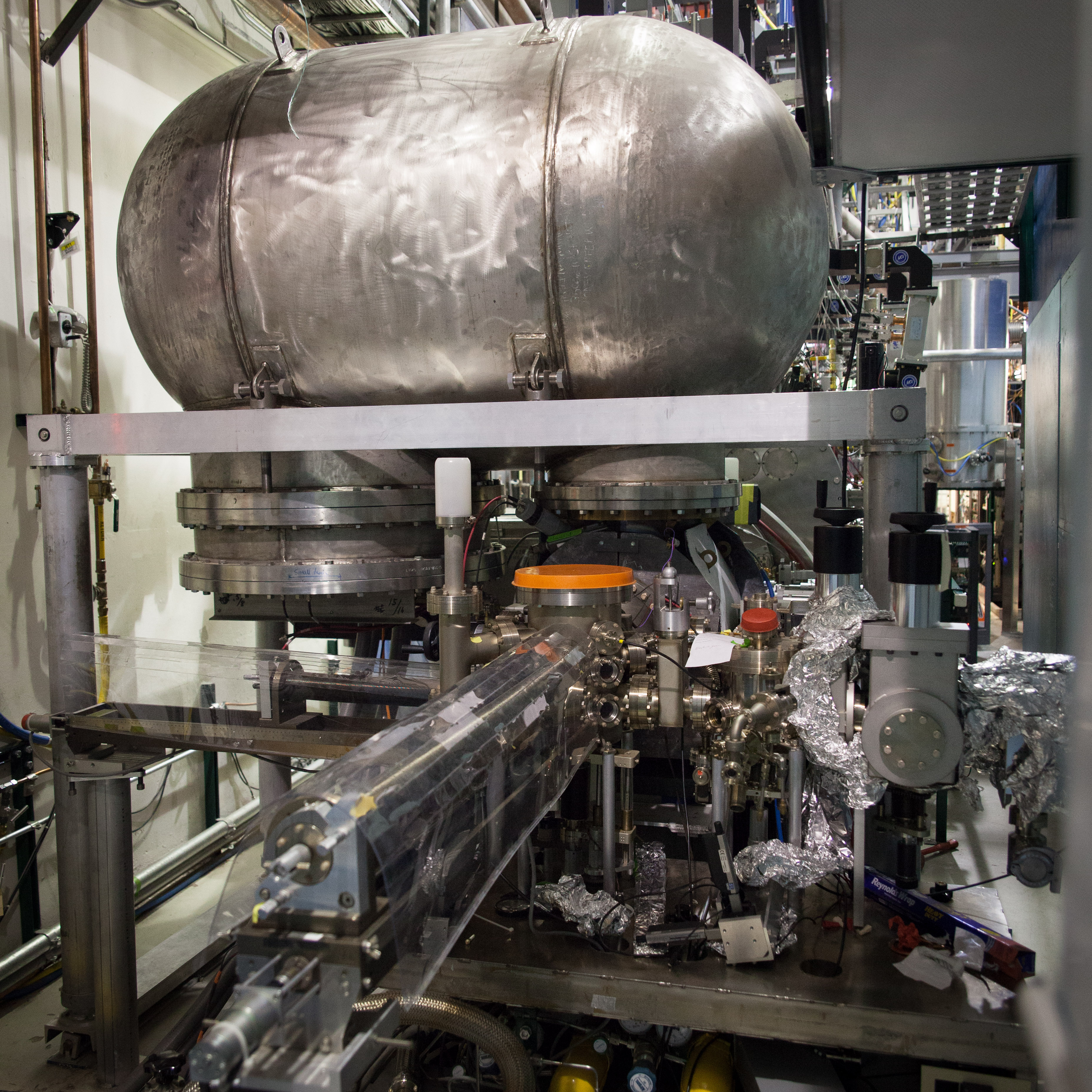}\label{fig:photo_cornell_dc_gun_intro}}
\caption[]{ The photographs show (from right to left) the high voltage DC gun, an emittance compensation section, the RF buncher, and the cryomodule.  Accelerated beam is then directed into a beamline or into the beam stop.}
\label{fig:photos_cornell_injector_intro}
\end{figure}

\paragraph{High-Power CW SRF injector linac:}
The photoemission electron injector shown in \Fig{fig:photos_cornell_injector_intro} is fully operational, and requires no further development.   It has achieved the world-record current of 75~mA \Ref{cultrera13, dunham13, cultrera11}, and record low beam emittances for any CW photoinjector \Ref{gulliford13}, with normalized brightness that outperforms other sources by a substantial factor.  Cornell has established a world-leading effort in photoinjector source development, in the underlying beam theory and simulations, with expertise in guns, photocathodes, and lasers. The injector delivers up to 500 kW of RF power to the beam at 1300~MHz.  The buncher cavity uses a 16~kW IOT tube, which has adequate overhead for all modes of operation.  The injector cryomodule is powered through ten 50 kW input couplers, using five 130 kW CW klystrons.  The power from each klystron is split to feed two input couplers attached to one individual 2-cell SRF cavity.  An additional klystron is available as a backup, or to power a deflection cavity for bunch length measurements.  

\begin{figure}[tb]
\centering
\includegraphics[width=\textwidth]{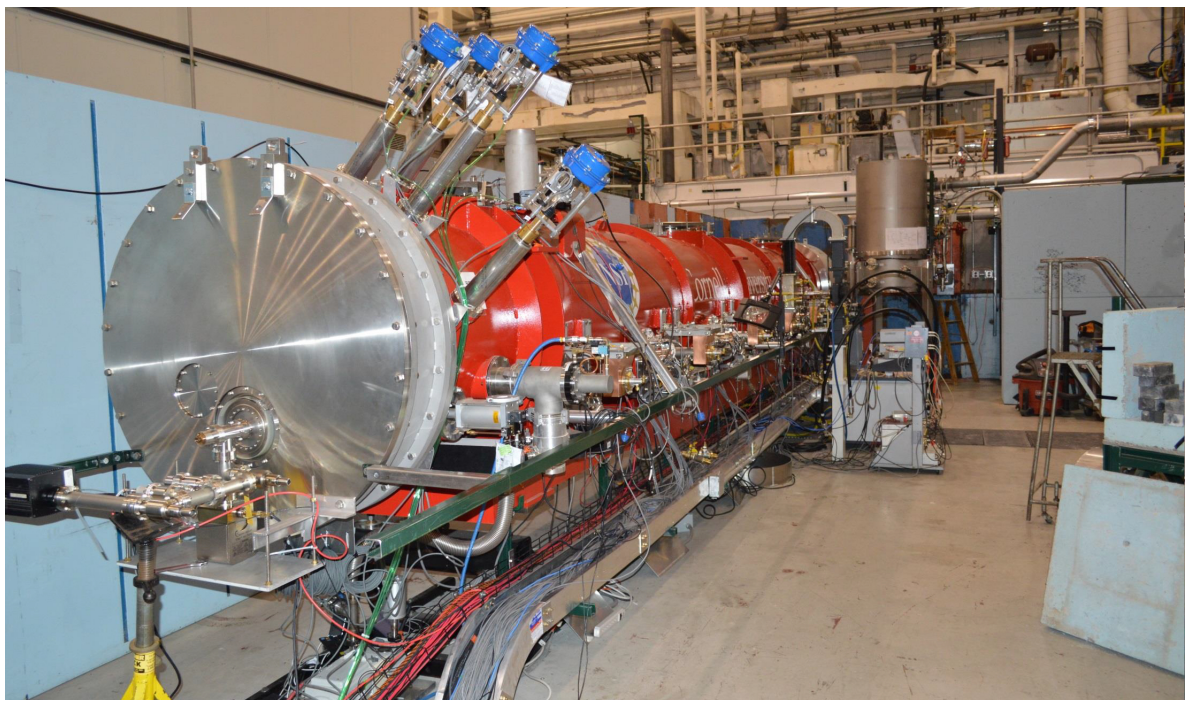}
\caption[MLC_completion]{The Cornell Main Linac Cryomodule (MLC) installed for RF testing in the experimental hall L0E.}
\label{fig:MLC_completion_intro}
\end{figure}

\paragraph{High-current ERL cryomodule:}
For CBETA, the main accelerator module will be the Main Linac Cryomodule (MLC), which was built as a prototype for the NSF-funded Cornell hard-X-ray ERL project. This cryomodule houses six 1.3GHz SRF cavities, powered via individual CW RF solid state amplifiers. Higher order mode (HOM) beamline absorbers are placed in-between the SRF cavities to ensure strong suppression of HOMs, and thus enable high current ERL operation.  The module, shown in \Fig{fig:MLC_completion_intro} was finished by the Cornell group in November 2014 and successfully cooled-down and operated starting in September 2015. The MLC will be powered by 6 individual solid-state RF amplifiers with 5 kW average power per amplifier.  Each cavity has one input coupler.  One amplifier is currently available for testing purposes, so an additional 5 amplifiers are needed for this project.

\paragraph{ERL merger and ERL beam stop:}
In \Fig{fig:L0Ehall}, three merger magnets are shown between the Injector Cryomodule (ICM) and the MLC. These merger magnets steer the injected beam with 6~MeV from the ICM into the MLC, bypassing the recirculated beams of higher energy. This merger has already been tested after the ICM, and it was shown that its influence on the beam emittances can be minimized. The beam stop in the top left of that picture also already exits, and with a power limit 600kW it can absorb all beams that are specified for CBETA.

\section{Components to be development for CBETA}
\begin{figure}[htb]
\centering
\includegraphics[width=0.95\textwidth]{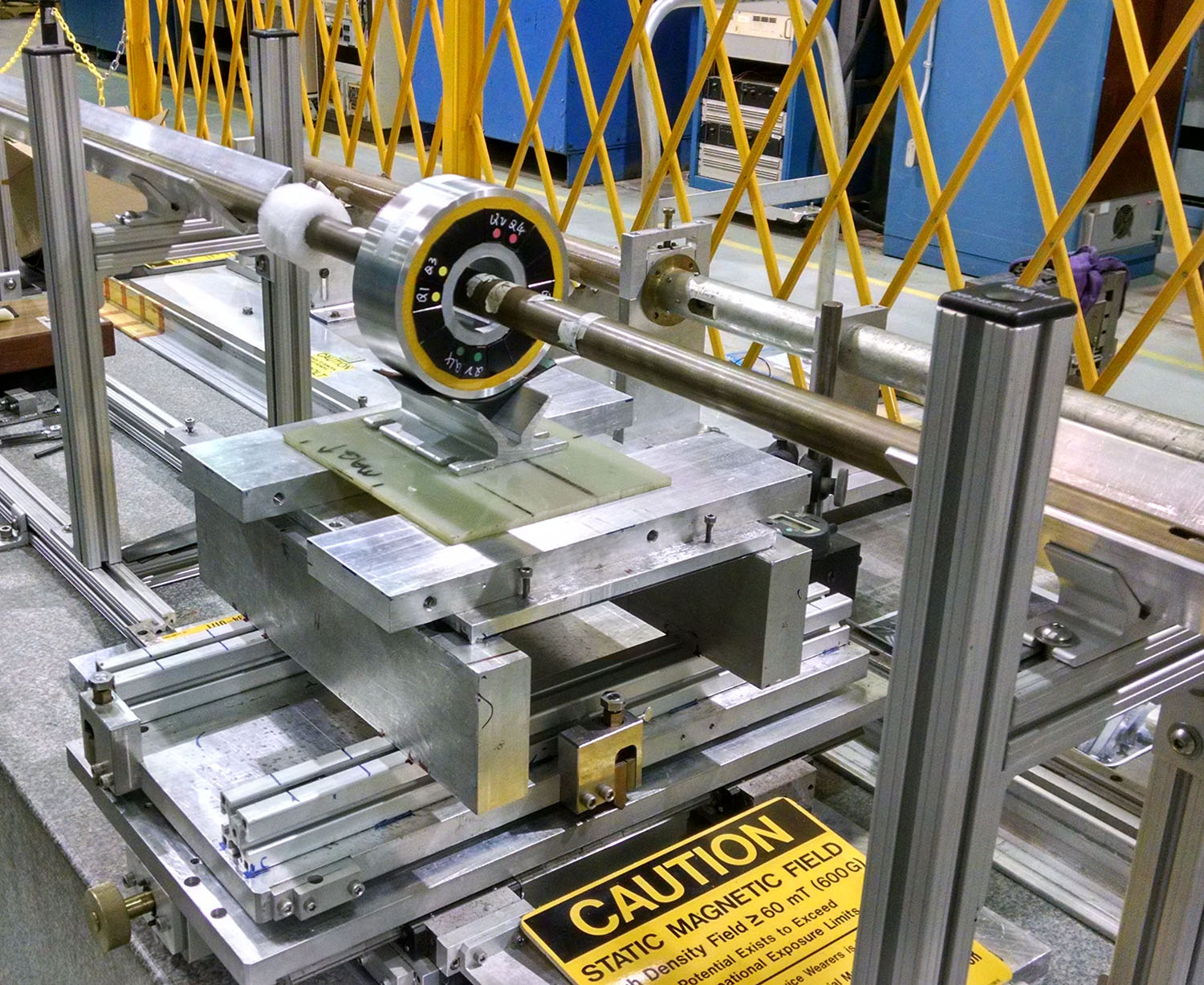}
\caption[]{The Halbach FFAG magnet QF on a rotating coil test bench.}
\label{fig:RotatingCoilMeasurementQf_intro}
\end{figure}

While the splitter sections to the right and the left of the MLC in \Fig{fig:L0Ehall} are equipped with conventional electro magnets, the magnets of the FFAG arc are made of permanent magnets. The field of these quadrupoles is shaped only by permanent magnet pieces, not by iron poles. This has the advantage of being very compact. A prototype of this design is shown in \Fig{fig:RotatingCoilMeasurementQf_intro} on a field-measuring bench at BNL.

This report describes the following other larger systems that will be developed for CBETA:
\begin{itemize}
\item The vacuum system
\item Girders for magnets and beamlines
\item Beam-Position Monitor (BPM) system
\item The control system and other beam instrumentation 
\end{itemize}

This Design Report describes the baseline design of the Cornell-BNL ERL Test Accelerator, as it exists on the date of its publication in late January 2017.

Some details of the accelerator design will continue to evolve, for example the collimation system and its shielding will evolve as more knowledge of beam-loss mechanisms is gained. The baseline lattice shown in this report was established on November 17, 2016 and is optimization for the use of hybrid magnets in the FFAG return arcs.  Since then, the design has changed to Halbach magnets in the return arcs, which are more compact, have less field cross talk between neighboring magnets, and can be somewhat stronger. The baseline lattice will therefore change in minor ways while it is fine-tuned to make optimal use of all CBETA components.

The detailed evolutions of design components for CBETA can be viewed at \\\url{https://www.classe.cornell.edu/CBETA_PM}. Major changes beyond the content of this report are not envisioned and such changes are under close control of a baseline control board.

\section{Construction of the prototype FFAG girder with Halbach magnets}

\begin{figure}[htb]
\centering
\includegraphics[width=0.75\textwidth]{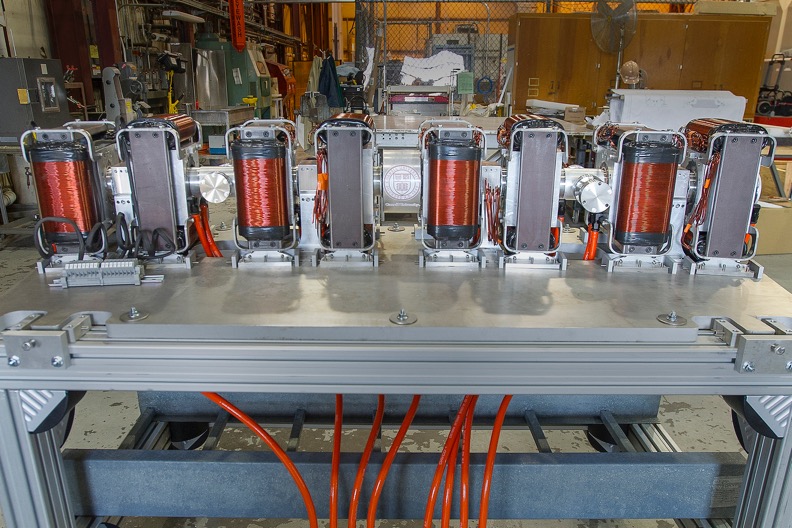}
\caption[]{The prototype FFAG girder with 8 Halbach magnets, it's vacuum chamber, cooling and beam diagnostics ports.}
\label{fig:first_girder}
\end{figure}

On April 30th 2017, the CBETA project reached the major funding milestone, ``Prototype FFAG Girder.'' For this setup, 8 Halbach magnets with appropriate strength for CBETA were assembled on the first girder around the associated vacuum system. Figure \ref{fig:first_girder} also shows the horizontal and vertical correctcor magnets that are constructed around every alternate Halbach magnet.

\section{Operation of the complete accelerating system with beam}

\begin{figure}[htb]
\centering
\includegraphics[width=0.75\textwidth]{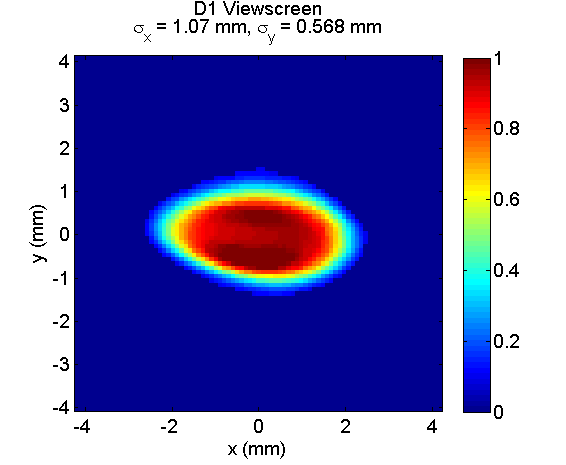}
\caption[]{The first beam accelerated by the MLC, captured on a viewscreen directly after the MLC.}
\label{fig:MLC_first_beam}
\end{figure}

On May 15th 2017, the CBETA project reached the major funding milestone, ``Beam through the MLC.'' For this test, the team had to successfully accelerate the electron beam to 6 MeV in the Injector Cryomodule (ICM), and then to a final energy of 12 MeV in the Main Linac Cryomodule (MLC). The MLC contains six superconducting accelerating cavities; for this initial test only a single cavity was powered. Figure \ref{fig:MLC_first_beam} is a viewscreen image of the first 12 MeV beam accelerated by the MLC.

Though many tests of the MLC have been performed prior to this, this milestone is both the first test of the MLC with an electron beam, and also the first test of the LLRF system’s ability to regulate and stabilize the cavity field, made difficult by the exteremely high Q of the cavity and the presence of microphonics from the helium vessel. It is also the first time all accelerating systems from the elecron source to the Main Linac Cryomodule have been operated together and with beam. Despite these challenges, and the aggressive schedule of the CBETA project, this test has been completed more than 3 months ahead of schedule.


\ifdefined \buildingFullDocument

\bibliographystyle{references/pddr}
\bibliography{references/pddr}

\else

\bibliographystyle{\FullDocumentRoot/references/pddr}
\renewcommand{\bibname}{References}
\bibliography{\FullDocumentRoot/references/pddr}
\renewcommand{\bibname}{References}

\end{document}

\fi





\ifdefined \buildingFullDocument

\renewcommand{\FiguresDirectory}{accelerator_physics/figures}

\else
\newcommand{\FullDocumentRoot}{..}
\newcommand{\FiguresDirectory}{figures}

\begin{document}
\fi

\chapter{Accelerator Physics}\label{chapter:accelerator_physics}

\section{Accelerator Layout \Leader{Chris}}

CBETA is a four-pass energy recovery linac. It incorporates the existing Cornell ERL high-power injector, MLC, and beam stop, to demonstrate four passes up in energy and four passes down in energy through a single arc section consisting of FFAG magnets with a common vacuum chamber. In order to properly inject into and extract from this FFAG arc, splitter sections are inserted between the MLC and the FFAG arc.

\begin{landscape}
\begin{figure}[htbp]
\centering
\includegraphics[width=\linewidth]{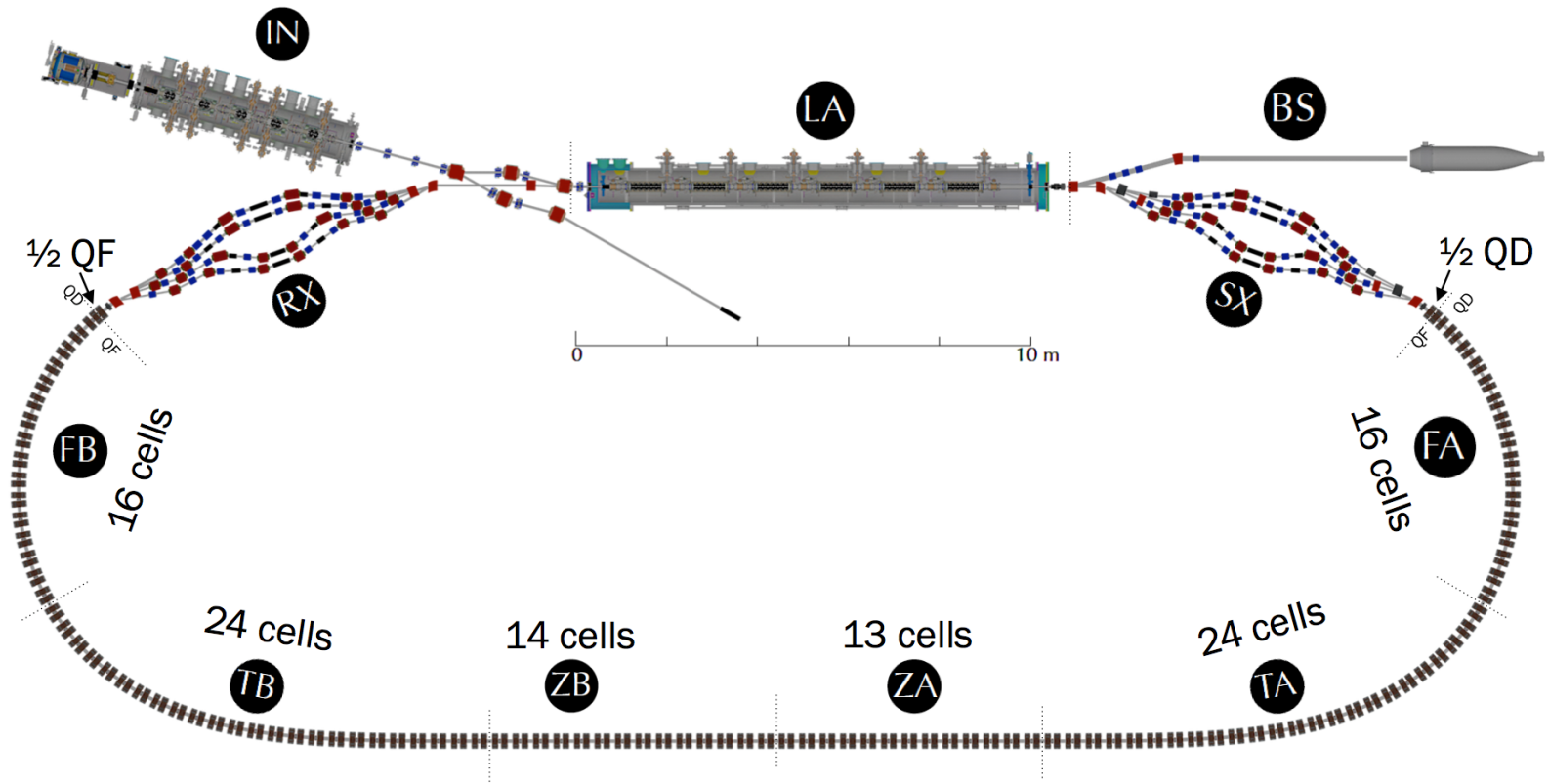}
\caption[]{CBETA layout with all major sections.}
\label{fig:cbeta_layout}
\end{figure}
\end{landscape}

The layout is broken into nine major sections:
\begin{description}
\item[IN] Injector: DC gun, front-end, injector cryomodule, and merger.
\item[LA] Linac, containing the MLC.
\item[SX] Splitter sections S1, S2, S3, and S4.
\item[FA] FFAG arc 
\item[TA] Transition from arc-to-straight
\item[ZA] Straight FFAG section.
\item[ZB] Straight FFAG section. This is a mirror of ZA.
\item[FB] FFAG straigth-to-arc, arc. This is a mirror of FA.
\item[TB] Transition from straight-to-arc
\item[RX] Splitter sections R1, R2, R3, R4. This is similar to a mirror of SX sections.
\item[BS] Beam stop, including demerging. 
\end{description}
These are shown in \Fig{fig:cbeta_layout}. Tables~\ref{tab:conventional_magnet_count} and \ref{tab:ffag_magnet_count} show the magnet counts for all sections.

\begin{table}[htbp]
\caption[]{Component count for splitter sections. A design study for these conventional electromagnets is described in \Section{sec:splitter_magnets}. }
\begin{tabular*}{\columnwidth}{@{\extracolsep{\fill}}lcccccc}
\toprule
Section & Dipole & Common & Septum & Quad  & BPM & Corrector (V) \\
\midrule
S1      &	6       &       2	&       0       &	8	&	8	&	4	\\
S2  	&	8       &       0	&       2       &	8	&	8	&	4	\\
S3	&	4       &       0	&       0       &	8	&	8	&	4	\\
S4  	&	2       &       0	&       2       &	8	&	8	&	4	\\
R1	&	6       &       2	&       0       &	8	&	8	&	4	\\
R2  	&	8       &       0	&       2 	&       8	&	8	&	4	\\
R3	&	4       &       0	&       0       & 	8	&	8	&	4	\\
R4  	&	2       &       0       &       2	& 	8	&	8	&	4	\\
\midrule
Total 	&	40      &       4      &       8	& 	64	&	64 	&	32 \\
\bottomrule
\end{tabular*}
\label{tab:conventional_magnet_count}
\end{table}

\begin{table}[htbp]
\caption[]{Component count for FFAG sections. Corrector counts are for separated horizontal (H) and vertical (V) correctors. }
\begin{tabular*}{\columnwidth}{@{\extracolsep{\fill}}llllll}
\toprule
Section & Focusing (F) Quad  & Defocusing (D) Quad   & BPM & Corrector (H) & Corrector (V) \\
\midrule
FA		&	16	&	17	&	17	&	16 	&	16	\\
TA		&	24	&	24	&	24	&	24	&	24	\\
ZA		&	13	&	13	&	13	&	13	&	13	\\
ZB		&	14	&	14	&	14	&	14	&	14	\\
TB		&	24	&	24	&	24	&	24 	&	24	\\
FB		&	17	&	16	&	17	&	16	&	16\\
\midrule
Total	&	108	&	108	&	109	&	107	& 107\\
\bottomrule
\end{tabular*}
\label{tab:ffag_magnet_count}
\end{table}

\clearpage
\section{Optics overview \Leader{Dejan} }
The last two decades have seen a remarkable revival of interest in Scaling Fixed Field Alternating Gradient (S-FFAG) accelerators. Originally developed in the 1950s~\Ref{Symon,Okhawa,Kolomenski}, S-FFAGs have very large momentum 
acceptances, with a magnetic field that varies across the aperture 
according to
\begin{equation}
B 	\;\sim\;	 B_{o}\left(\frac{r}{r_o}\right)^{k}
\end{equation}
where the scaling exponent $k \sim 150$ is as large as possible. The revival began in Japan, with a proof-of-principle 
proton accelerator at KEK followed by a 150 MeV proton accelerator at Osaka University (now at Kyushu University) and many smaller electron S-FFAGs built for a variety of applications such as food irradiation. 

Although S-FFAGs have the advantages of fixed magnetic fields, zero chromaticities, and fixed tunes,
synchrotrons mostly dominate, despite their need for magnet cycling, because synchrotrons have much smaller apertures --- a few centimeters compared to of order one meter for S-FFAGs. The international muon collider collaboration proposed that S-FFAGs accelerate short lifetime muons, avoiding the need for very rapid cycling synchrotrons~\Ref{Ankenbrandt}.  Similarly, ERLs cannot use synchrotron-like arcs, because it is not possible to rapidly change the magnetic field during electron acceleration. Except for CBETA, all proposed and operational multipass ERLs (and recirculating linacs) use multiple beamlines --- one beamline for every electron energy. 

Non-Scaling FFAG (NS-FFAG) optics~\Ref{trbojevic,Johnstone} reduce the number of beamlines required in a multipass ERL to one, while preserving centimeter-scale apertures.  Aperture reduction is enabled by ensuring very small values of horizontal dispersion $D_x$, because
\begin{equation}
\Delta x	\;=\;	D_x \, \frac{\Delta p}{p}
\end{equation}
where $\Delta x$ is the orbit offset, and $p$ is the electron momentum.  For example, if the dispersion $D_x \approx 40$~mm, then the orbit offsets are only $\Delta x \approx \pm20$~mm for a momentum range $\Delta p/p \approx \pm50$\% that corresponds to an energy range of 3 for relativistic particles.  The magnetic field in NS-FFAG optics is purely linear, with
\begin{eqnarray}
B_y 		\;= &	 B_0 \;\;+	&G(s) \, x, \\ \nonumber
B_x 		\;= &			&G(s) \, y.
\end{eqnarray}
This in stark contrast to the nonlinear field variation in S-FFAG magnets.   All NS-FFAG magnets are linear combined function magnets, often just transversely displaced quadrupoles.  However, abandoning nonlinear scaling has a price --- the tunes and the chromaticities now vary with energy, and the time-of-flight is a parabolic function of energy. 

\begin{figure}[tb]
\centering
\includegraphics[width=0.7\textwidth]{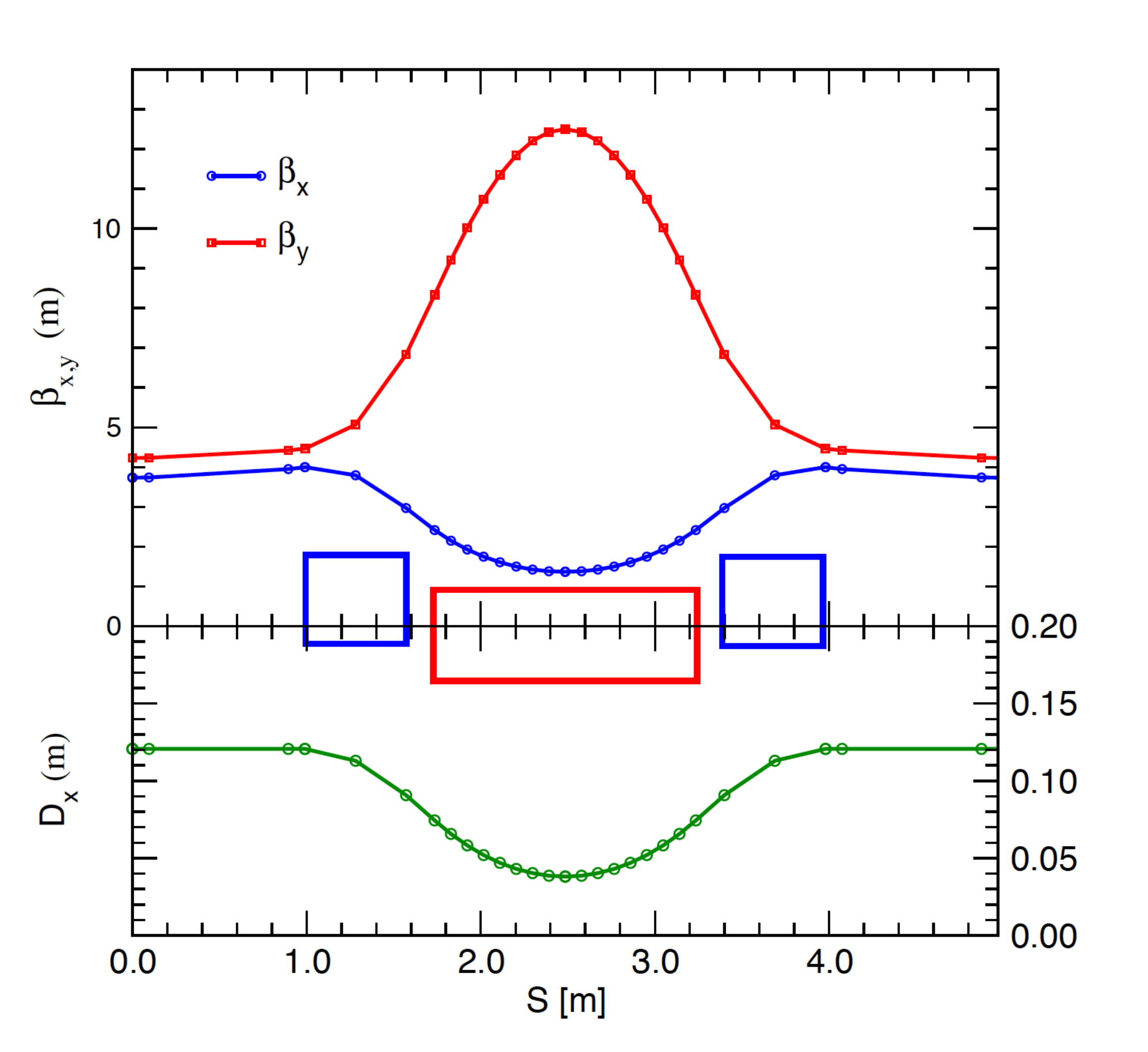}
\caption{Betatron and dispersion functions at the central energy in a triplet cell with a central combined function dipole.}
\label{fig:Low-Emittance-Lattice}
\end{figure}

\begin{figure}[tb]
\centering
\includegraphics[width=\textwidth]{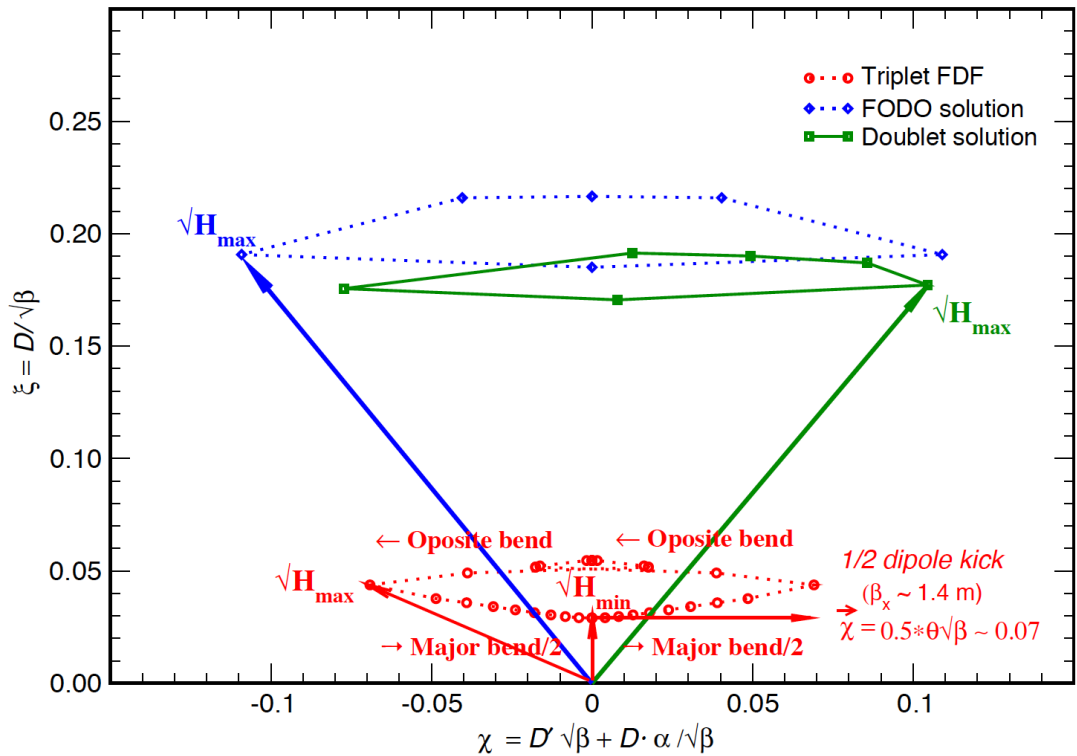}
\caption{Normalized dispersion function for three basic cell configurations --- triplet, FODO, and doublet -- using the same magnets.}
\label{fig:TogetherWithFODO-Doublet}
\end{figure}

Tuning the NS-FFAG optics to minimize the orbit offsets is similar to minimizing the natural emittance of a synchrotron light source, because both refer to the dispersion action function
\begin{equation}
H	 	\;=\;	\left(\frac{D}{\sqrt{\beta}}\right)^{2} 	\;+\;	 \left(D' \sqrt{\beta} + \frac{\alpha D}{\sqrt{\beta}} \right)^2
\end{equation}
where $D'$ is the slope of the dispersion function, and $\beta$ and $\alpha$ are Twiss functions~\Ref{trbojevicPRSTAB1,Machida}.  The minimum emittance is achieved in a synchrotron light source by minimizing the average dispersion action
\begin{equation}
\langle H \rangle 	\;=\;	 \frac{1}{L} \int_{0}^{L} H(s) ds
\end{equation}
For a given lattice cell geometry, such as the triplet configuration shown in Figure~\ref{fig:Low-Emittance-Lattice}, this is accomplished by adjusting the parameters so that
\begin{equation}
\frac{\partial\langle H \rangle}{\partial D_{0}}	\;=\;		\frac{\partial \langle H \rangle}{\partial\beta_{0}}		\;=\; 0
\end{equation}
where  $D_0$ and $\beta_0$ are the periodic values at the cell ends~\Ref{trbojevicPRSTAB1}.  The performance of three such configurations --- triplet, FODO, and doublet --- is shown in 
Figure~\ref{fig:TogetherWithFODO-Doublet}, by following the evolution of the normalized dispersion vector $\sqrt{H}$ in $(\chi,\xi)$ space, where
\begin{eqnarray}
\chi	&\;=\;&	D'\sqrt{\beta}	\;+\;	\frac{\alpha D}{\sqrt{\beta}} \\ \nonumber
\xi	&\;= \;&	\frac{D}{\sqrt{\beta}}
\end{eqnarray}

Superficially the triplet configuration is the most advantageous, with smaller maximum orbit offsets and smaller path length differences.  However, it is difficult to implement a triplet configuration when space is limited, as it was for the first NS-FFAG accelerator, the Electron Model for Many Application (EMMA)~\Ref{Machida2}.  Space is also very constrained for CBETA, making it difficult to include the two small drifts between the two focusing magnets and the central defocusing combined function magnet.

The outer transverse size of the magnets depends directly on the maximum orbit offsets within the magnets --- smaller orbit offsets allow a smaller pole tip radius. The optimum solution for the CBETA cell is to use two kinds of combined function magnet: one with a larger focusing gradient and a small dipole field, and the other with a smaller defocusing gradient providing most of the bending.  These combined function fields are achieved by displacing two different kinds of quadrupole.  The smallest possible gradient values are achieved by displacing focusing and defocusing quadrupoles in opposite directions, maximizing the good field region when the orbit displacement is at its maximum.  The focusing quadrupole is displaced away from the center of the circular arc, and the defocusing quadrupole is displaced inwards.

\begin{figure}[tb]
\centering
\includegraphics[width=1\textwidth]{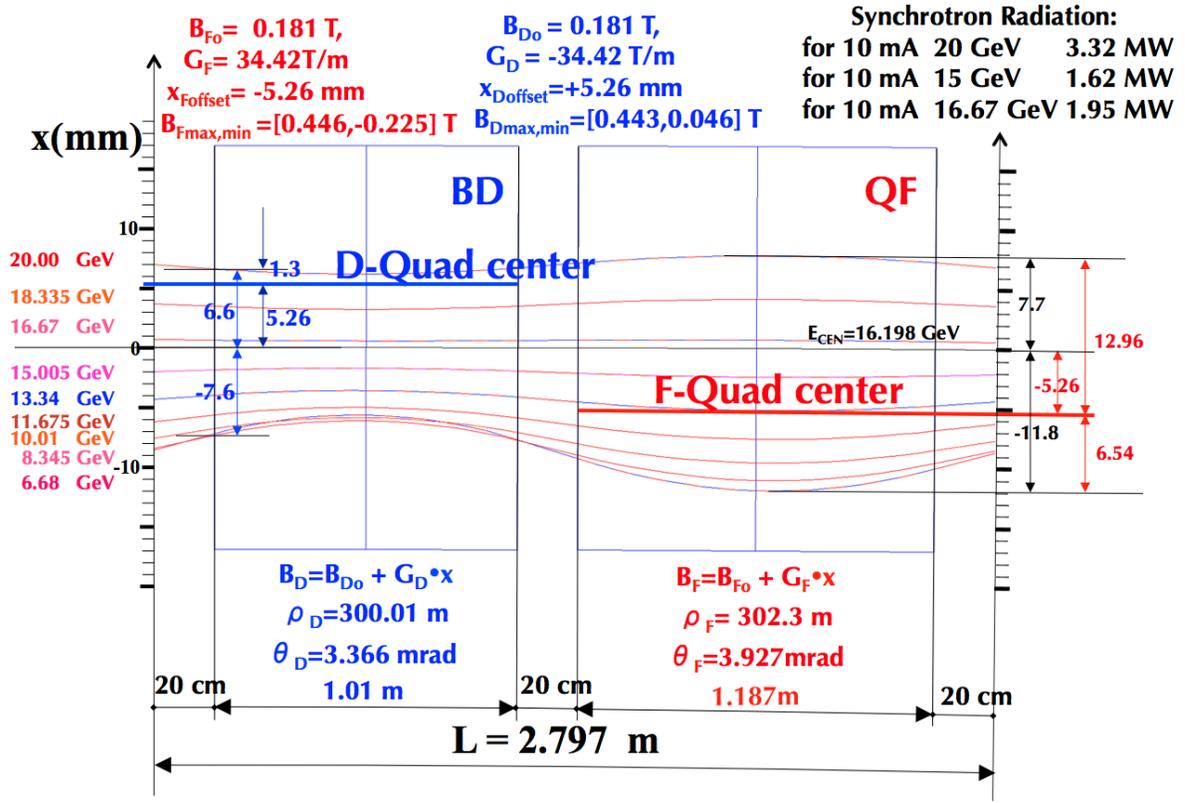}
\caption{Orbits and dimensions of the eRHIC NS-FFAG basic arc cell.}
\label{fig:eRHIC-HighEnergy-Cell}
\end{figure}

The NS-FFAG lattice for the electron-Ion collider eRHIC has far milder magnet size and bending radii requirements.   Synchrotron radiation is a serious challenge at the highest energy (20 GeV) in eRHIC, even though the average radius of the RHIC tunnel is $\sim$ 330 meters. To simplify the eRHIC magnet design, the gradients and the bending fields of the focusing and defocusing combined function magnets are made equal. Combined function magnets are obtained by radially displacing the quadrupoles by only $\pm$~5.26~mm in opposite directions, as shown in Figure~\ref{fig:eRHIC-HighEnergy-Cell} . 

The most important difference between eRHIC magnets and CBETA magnets is in the orbit offsets. The maximum orbit offsets at the highest energy in eRHIC (with respect to the magnetic axis of the quadrupole) are around 13~mm and 14~mm for focusing and defocusing magnets, respectively. The offsets in CBETA are significantly larger because the arc radius of the curvature is only $\sim$~5~m. This is discussed in detail in the later chapters.  

\clearpage
\section{Injector (IN) \Leader{Colwyn}}
The design, layout, and performance of the Cornell high brightness, high current photoinjector has been well documented \Ref{Gulliford13_01,Dunham13_01,Gulliford15_01}.  The injector dynamics are strongly space charge dominated.  Consequently, the 3D space charge simulation code General Particle Tracer (GPT) is used as the base model of the injector dynamics~\Ref{Greer07-01}.  All of the beamline elements relevant for the space charge simulations in this work have been modeled using realistic field maps. Poisson Superfish \Ref{bib:psfish} was used to generate 2D cylindrically symmetric fields specifying the electric fields $E_r(r,z)$ and $E_z(r,z)$ as well as the magnetic fields $B_r(r,z)$ and $B_z(r,z)$, for the high-voltage DC and emittance compensation solenoids, respectively.  Detailed plots of these fields can be found in \Ref{Gulliford13_01}.

The dipole and quadrupole fields are described using the following off axis expansion for the fields \Ref{Wei00_01}:
\begin{eqnarray}
B_x^D \sim \mathcal{O}(4),\hspace{1cm}B_y^D = B_0(y) - \frac{y^2}{2}\frac{d^2B_0}{dz^2} + \mathcal{O}(4),\hspace{1cm}B_z^D = y\frac{dB_0}{dz} +  \mathcal{O}(4)
\end{eqnarray}

\begin{eqnarray}
B_x^Q &=& y\left[G(z) - \frac{1}{2}\left(3x^2+y^2\right)\frac{d^2G}{dz^2}\right] + \mathcal{O}(5),
\nonumber
\\
B_y^Q &=& x\left[G(z) - \frac{1}{2}\left(3y^2+x^2\right)\frac{d^2G}{dz^2}\right] + \mathcal{O}(5),
\nonumber
\\
B_z^Q &=& xy\frac{dG(z)}{dz} + \mathcal{O}(4).
\end{eqnarray}
In this expression $B_0=B_y^D(r=0,z)$ is the on-axis field in the dipole and $G(z)=\partial B_y^Q(r=0,z)/\partial x$ is the on-axis quad field gradient.  Both of these quantities are extracted from full 3D maps generated using Opera3D \Ref{bib:opera3D}.

All RF cavity fields were generated using the eigenmode 3D field solver in CST Microwave Studio \Ref{bib:MWS}. For the SRF cavities, two solutions for the fields are generated and then combined using the method described in \Ref{Gulliford11_01} in order to correctly include the asymmetric quadrupole focusing of the input power couplers.  Currently, these combined field maps have been constructed with powers in their fully retracted position, typically used for low current.

In order to facilitate using both GPT and ASTRA~\Ref{Flottman00-01}, a standalone input particle distribution generator was written in C++, and was used for all space charge simulations discussed in this section.  This code uses standard sub-random sequences to generate particle phase space coordinates and allows for nearly arbitrary six-dimensional phase space distributions specified either from a file or using the combination of several continuously variable basic distribution shapes.  For example, it allows one to load the measured transverse laser profile on the cathode along with a simulated longitudinal laser profile using a model of the longitudinal shaping crystals. See \Ref{Gulliford15_01} for examples of generating realistic particle distributions from measured laser profiles.

The determination of the appropriate optics in the injector is set by the requirements for the FFAG ring.  In particular it is important to inject with a beam that is suitably matched into the first splitter section S1.  Because the dynamics in the injector and largely through the first pass through the linac are dominated by space charge forces, multi-objective genetic algorithm  (MOGA) optimizations of the injector model have been performed (see \Ref{Gulliford13_01,Gulliford15_01} for a detailed description of the optimization procedure) for beams traveling through the first pass of the linac.

\begin{figure}[tb]
\centering
\includegraphics[width=0.65\textwidth]{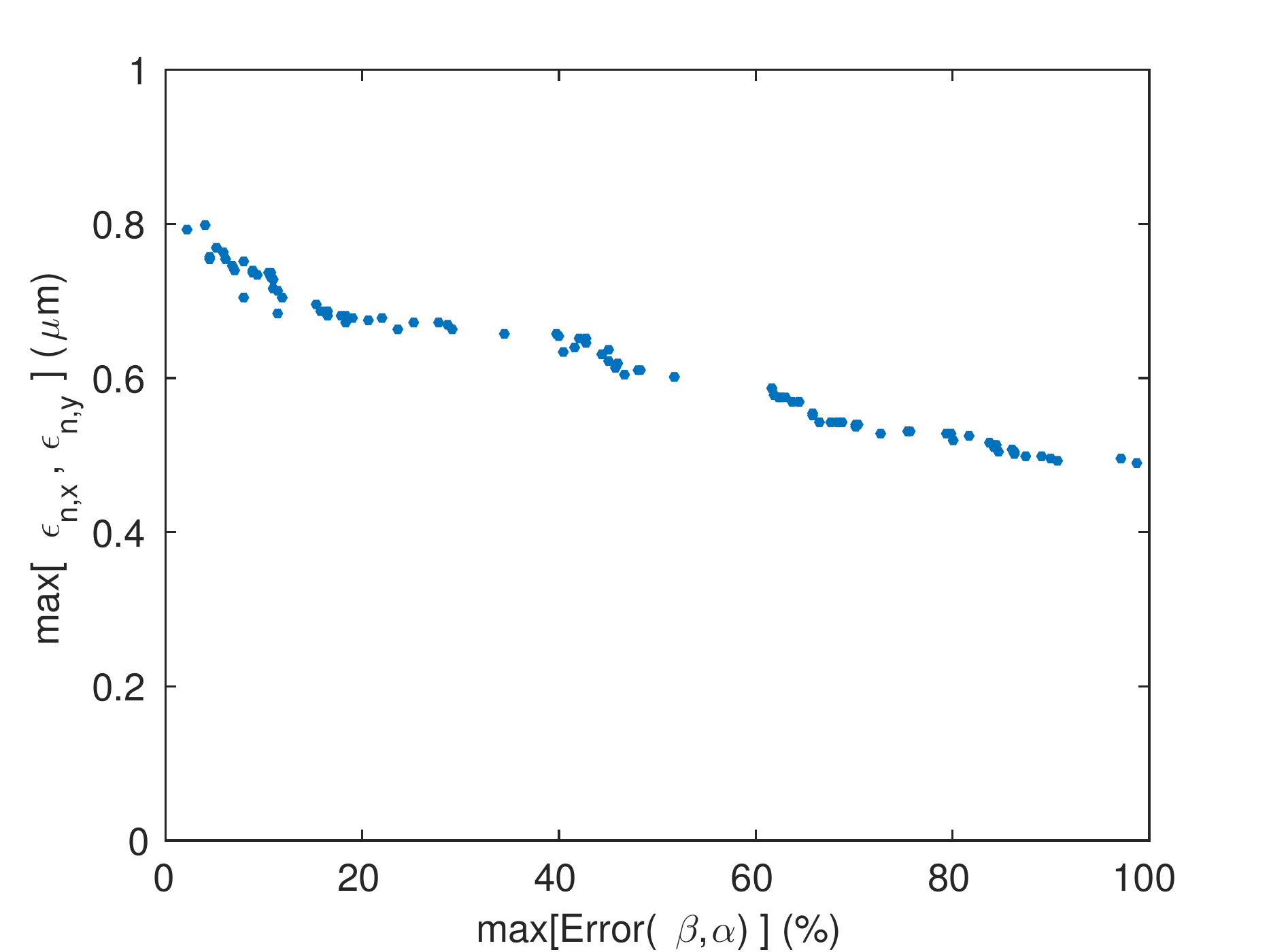}
\caption[]{Trade-off between emittance and matching into the lowest energy FFAG passes, extra quadrupole added in front of MLC.}
\label{fig:twiss_vs_emitt}
\end{figure}

\begin{figure}[tb]
\centering
\subfloat[beta functions]{\label{fig:betaxy}\includegraphics[width=0.49\textwidth]{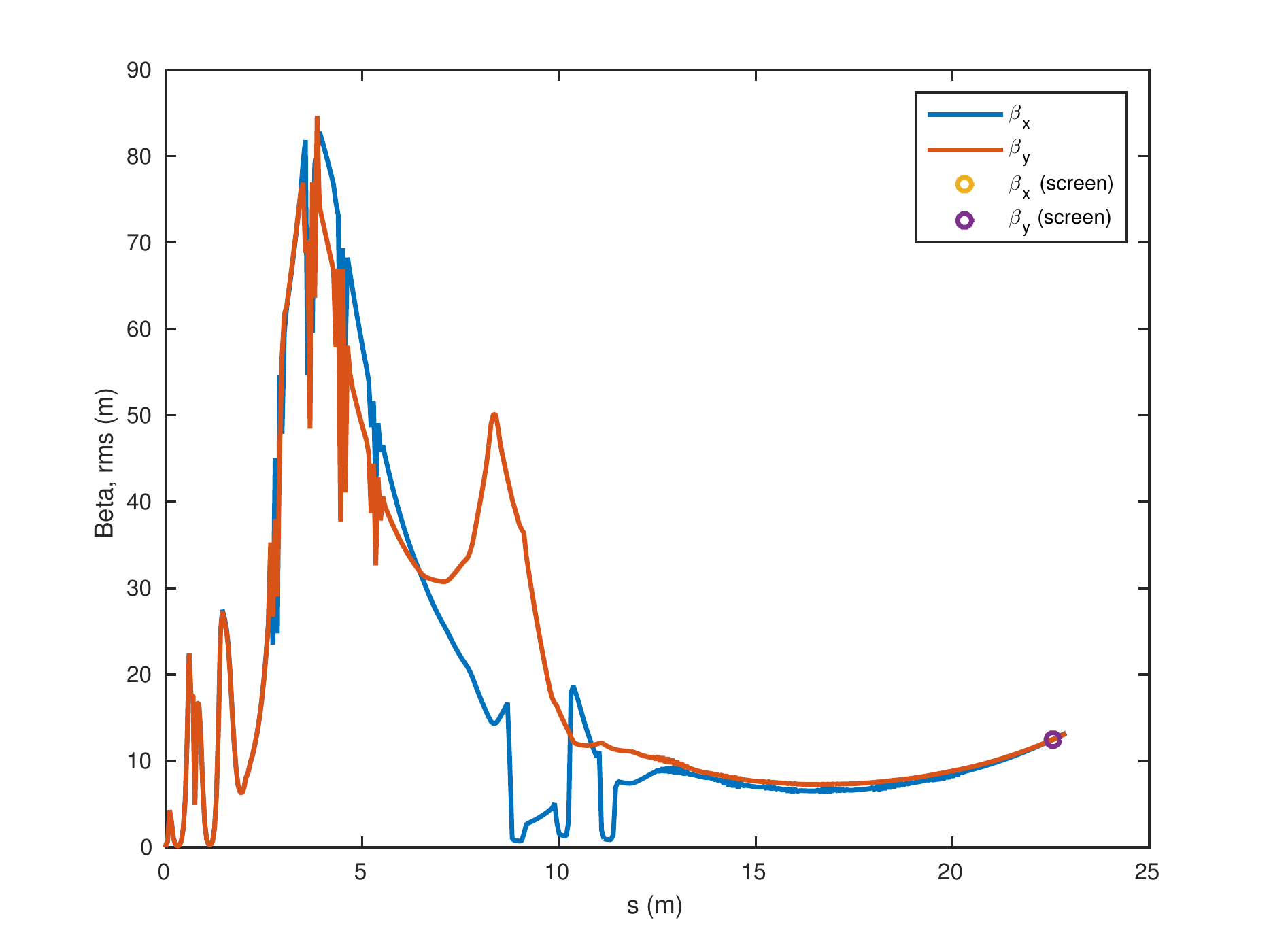}}
\subfloat[alpha functions]{\label{fig:alphaxy}\includegraphics[width=0.49\textwidth]{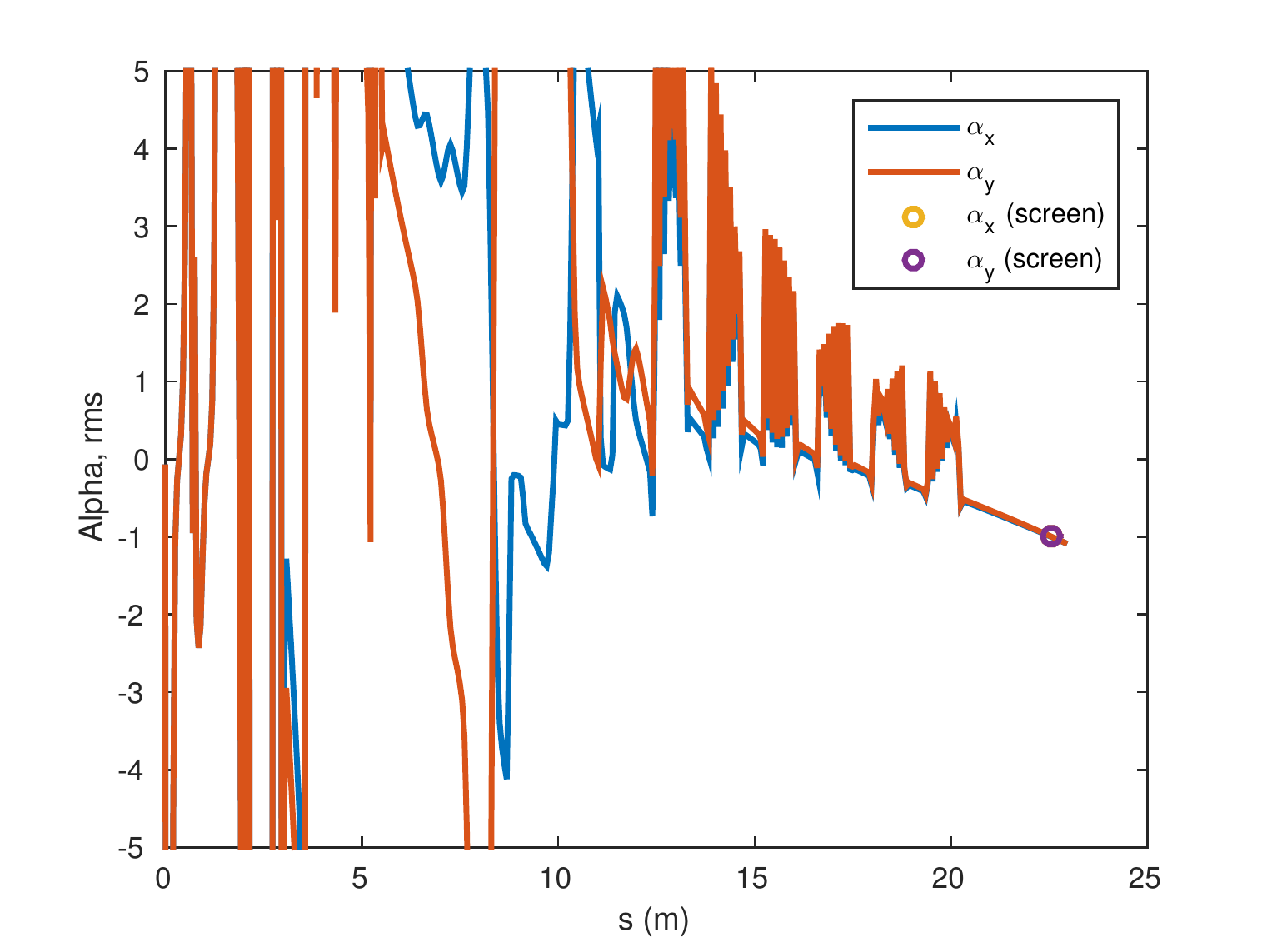}}
\caption[]{Beta and alpha functions through injector and main linac for 100 pC bunches. $\beta_{x, y}=12.5$ m and $\alpha_{x, y}=-1$ at the screen position marked by the circles.}

\end{figure}

\begin{figure}[tb]
\centering
\includegraphics[width=0.8\textwidth]{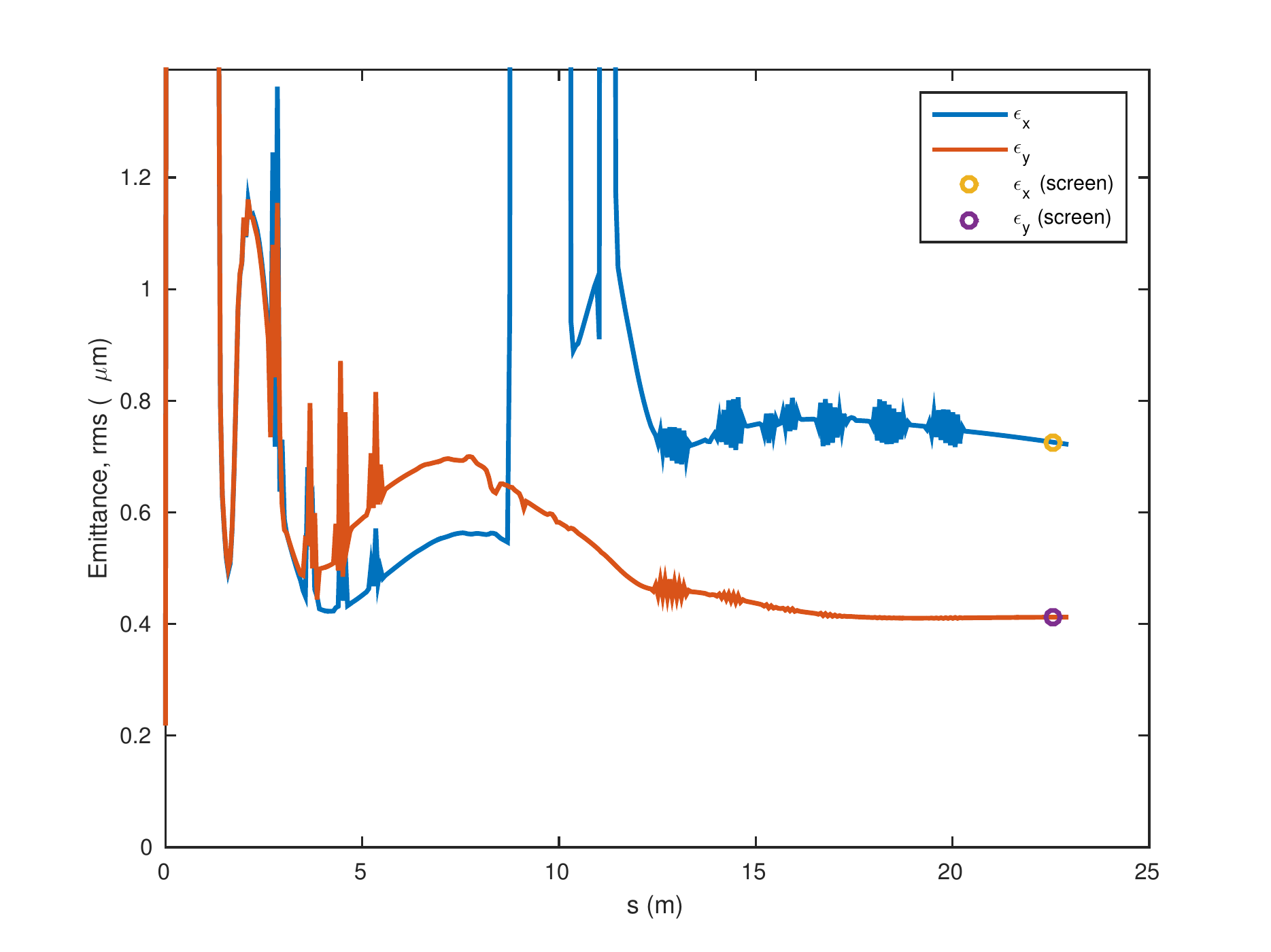}
\caption[]{Transverse emittances through injector and main linac for 100 pC bunches.}
\label{fig:emitxy}
\end{figure}

MOGA optimizations for 100 pC/bunch, 6 MeV injector energy, and linac energy gain of roughly 48.5 MeV have been performed in order to investigate the trade-off between the transverse normalized emittances and how well the Twiss parameters after the linac can be matched.  Initial results showed a strong trade-off between these parameters.  To remove this, an additional quadrupole was placed right in front of the main linac.  \Figure{fig:twiss_vs_emitt} shows the resulting trade-off.
From this front, the solution with the smallest maximum Twiss error was further refined by using a standard root finding algorithm to more closely match the desired Twiss parameters.  \Figure{fig:betaxy}, \Fig{fig:alphaxy}, and \Fig{fig:emitxy} show the beta functions, alpha functions, and emittances along the injector and main linac beamlines.
The particle distribution resulting from this simulation has been converted into a Bmad form and simulated through the machine.

\clearpage
\section{Linac (LA) \Leader{Chris}}
\begin{figure}[htbp]
\centering
\includegraphics[width=0.85\textwidth]{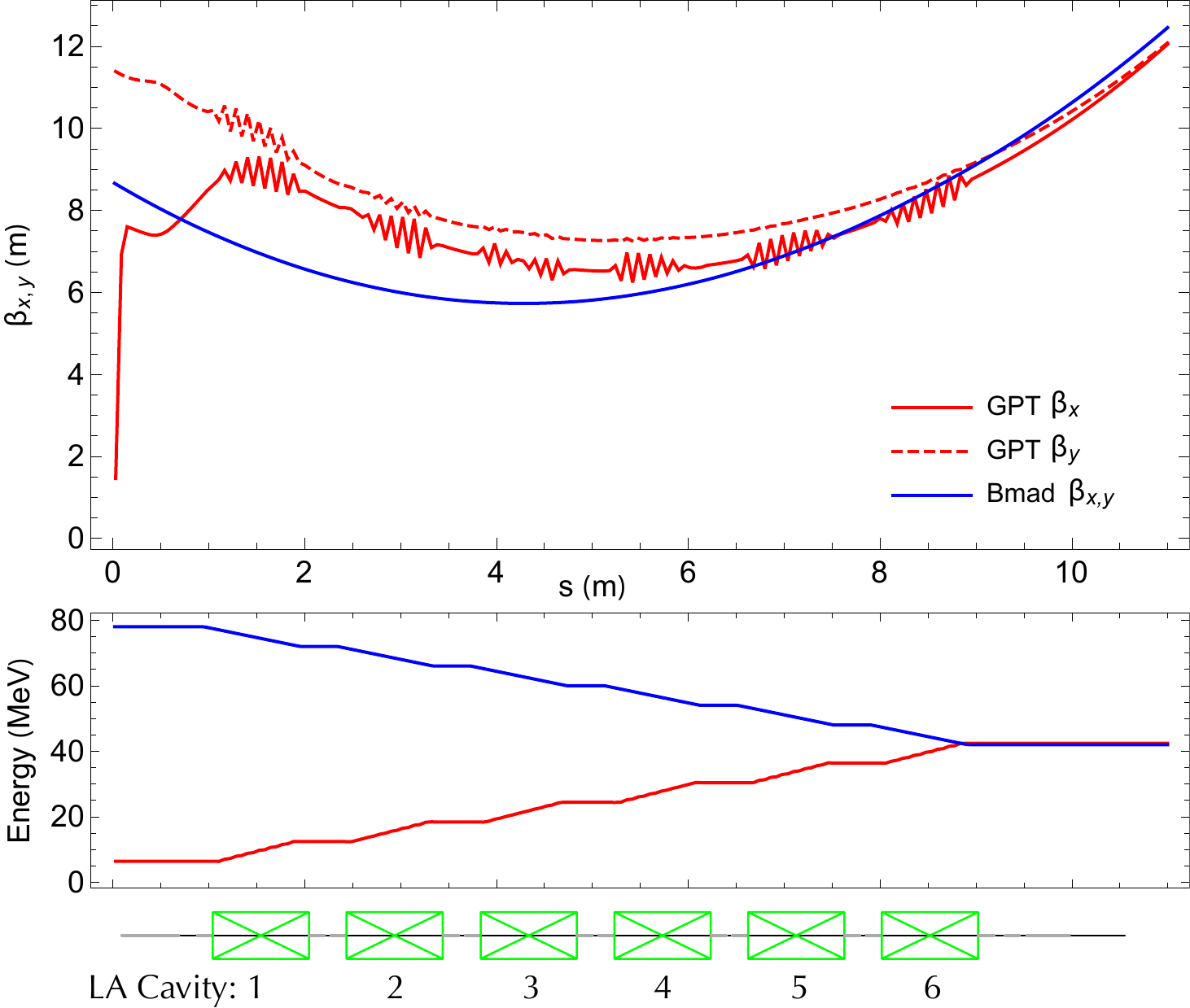}
\caption[]{Beam optics for the first pass accelerating beam, calculated with space charge using GPT, and the second to last pass decelerating beam using Bmad without space charge. Both beams continue to go through the same S1 splitter section.  Optics for subsequent passes are nearly identical to the Bmad curve. }
\label{fig:la_focusing}
\end{figure}

The linac section (LA), consisting mainly of the MLC, has no independent adjustments for beam focusing. The beam's behavior is thus governed almost entirely by its incoming properties. Roughly speaking, all beams are focused through LA with minimum average horizontal and vertical beta functions. For simplicity we specify that $\beta_{x,y} = 12.5\unit{m}$ and $\alpha_{x, y} = -1$ at the end of this section for the first seven passes of the beam, with the eighth pass adjusted to match into the beam stop (BS). This imposes a matching criteria for all eight beams propagating through the linac. 

The first and last passes start and end, respectively, at low energy. In this regime, cavity focusing and space charge can be significant, so space charge is used in the calculation. \Figure{fig:la_focusing} shows how the first pass beam, starting at 6~MeV and ending at 42~MeV, is matched to the same optics second-to-last pass beam starting at 78~MeV and decelerating to 42~MeV. Both beams then propagate through the splitter section S1.

\clearpage
\section{Splitters (S1-4, R1-4) \Leader{Chris} }\label{sec:splitters}

The splitter design matches four on-axis beams of differing energy with relatively large beam sizes into the FFAG arc with off-axis orbits. The first three lines (S1, S2, S3) accommodate two beams each, one that had just been accelerated and one that had just been decelerated. These sections S1--S4 must:
\begin{itemize}
\item Have path path lengths so that passes 1--3 have the same harmonic.
\item Match beam sizes and dispersion into FA arc (6 parameters).
\item Match orbits into FA.
\item Provide $r_{56}$ adjustment.
\item Provide $\pm 10^\circ $ of RF phase path length adjustment
\end{itemize}
The last line (S4) only receives a beam that has been accelerated and is only traversed once. It has a path length that is an integer plus 1/2 of the linac RF wavelength, thus setting up the beam for deceleration for its fifth pass through the linac. The second splitter section RX, consisting of lines R1--R4, are nearly symmetric and identical to lines S1--S4.

\begin{figure}[tb]
\centering
\includegraphics[width=\linewidth]{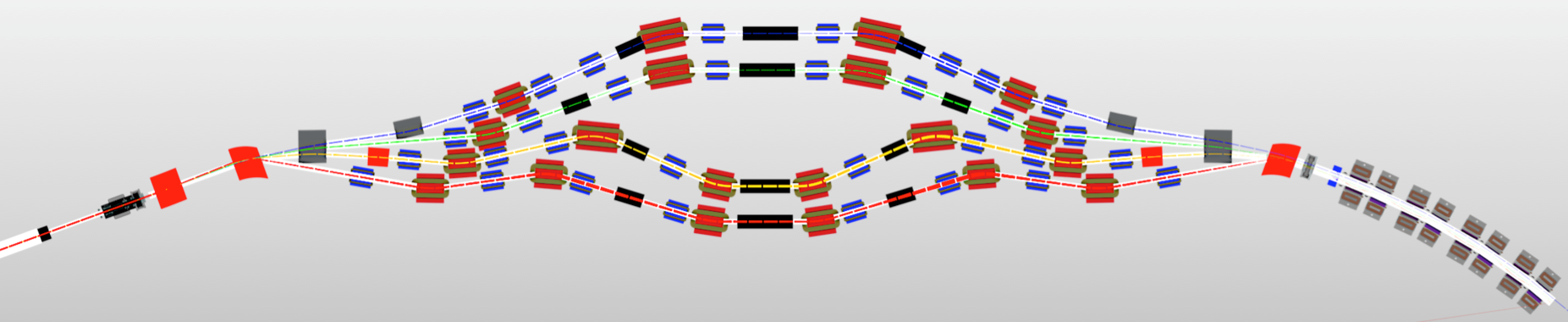}
\caption[]{SX layout}
\label{fig:cbeta_layout_sx}
\end{figure}
\Figure{fig:cbeta_layout_sx} shows the layout of these lines. For path length adjustment, the inner four dipole magnets and the pipes connecting them are designed to move along sliding joints, without changing the bending angles. A maximum change in linear length of $\pm8.5\unit{cm}$ for all sliding joints can provide $\pm 10^\circ $ of RF phase path length adjustment for each of S1--S4. Combined with an identical path length adjustment scheme in RX, each pass for CBETA can be adjusted by $\pm 20^\circ $ of an RF wavelength (about 13~mm path length) for linac phasing control.

\begin{landscape}
\begin{figure}[tb]
\centering
\includegraphics[width=\linewidth]{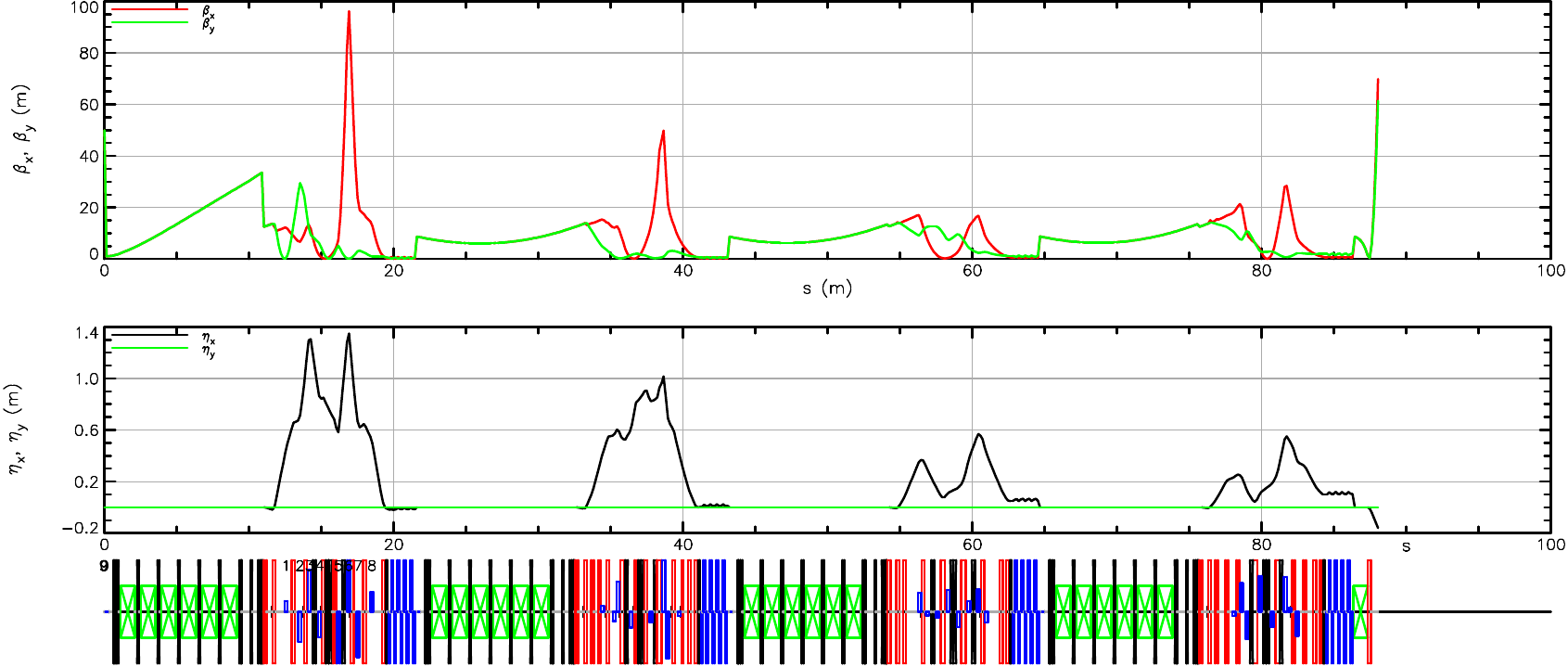}
\caption[]{Optics in SX and LA for the first four accelerating passes, including a short piece of the FA beamline following each SX line. Remaining FFAG beamline and RX are omitted.}
\label{fig:optics_sx_summary}
\end{figure}
\end{landscape}

In order to achieve the optics requirements, each line would need at least seven independent quadrupole magnets to satisfy the six optics parameters and single $r_{56}$ parameter. For additional flexibility and symmetry, we use eight quadrupole magnets per line. \Figure{fig:optics_sx_summary} summarizes the optics for each splitter, and shows how each matches into the appropriate FFAG optics. 

Unfortunately it is very difficult to make the $r_{56}$ of each pass zero. However, the $r_{56}$ of the four splitters can be adjusted so that the map after a symmetric four passes has $r_{56}$ close to zero, so that the machine as a whole is roughly isochronous. This fact is evident from start-to-end beam tracking as shown in \Fig{fig:optics_4pass}, where the bunch length is nearly the same as injection in the center of the fourth pass. 

The machine will be commissioned in a staged approach. The total path lengths of this section will be reconfigured for 1, 2, 3, and 4-pass operation by lengthening or shortening the final pass by $1/4$ of an RF wavelength for both SX and RX lines to provide energy recovery. In the future, an eRHIC style 650~MHz cavity will be tested in this layout, which will require further lengthening of each line. 

\subsection{Splitter Magnet Feasibility Study} \Leader{Jim Crittenden}

The close proximity of the splitter beamlines on the girders, each of surface area approximately $4\times2\unit{m^2}$, imposes strict limits on the transverse size and fringe fields of the 18 dipole and 32 quadrupole magnets in each of the SX and RX sections. A design feasibility study~\Ref{crittenden16} has been carried out to determine the field quality which can be obtained with conventional electromagnets satisfying the stringent space constraints. An H-magnet design was adopted for the dipoles in order to minimize horizontal fringe fields. The dipole gap (36~mm) and quadrupole bore (45~mm) values were chosen on the basis of a vacuum chamber with outer dimensions $42 \times 30$~mm with a wall thickness of 3~mm. Sufficient field quality in the dipoles was achieved with a 7-cm-wide pole. The transverse space constraints limited the width of the backleg and thus the region of linearity such that the 20~cm design became nonlinear at the percent level for central field values greater than 6~kG. For this reason, the 30-cm-long design was used where greater field integrals were required. The quadrupole design is a scaled-down version of the CESR storage ring quadrupoles. The vertical correctors are designed as C-magnets rotated by 90 degrees in order to provide sufficient field integral for 10~cm length in the (S3, S4) and (R3, R4) lines. In the (S1, S2) and (R1, R2) lines, a corrector length of 5~cm suffices. 

\begin{figure}[tb]
\centering
   \includegraphics[width=0.45\textwidth]{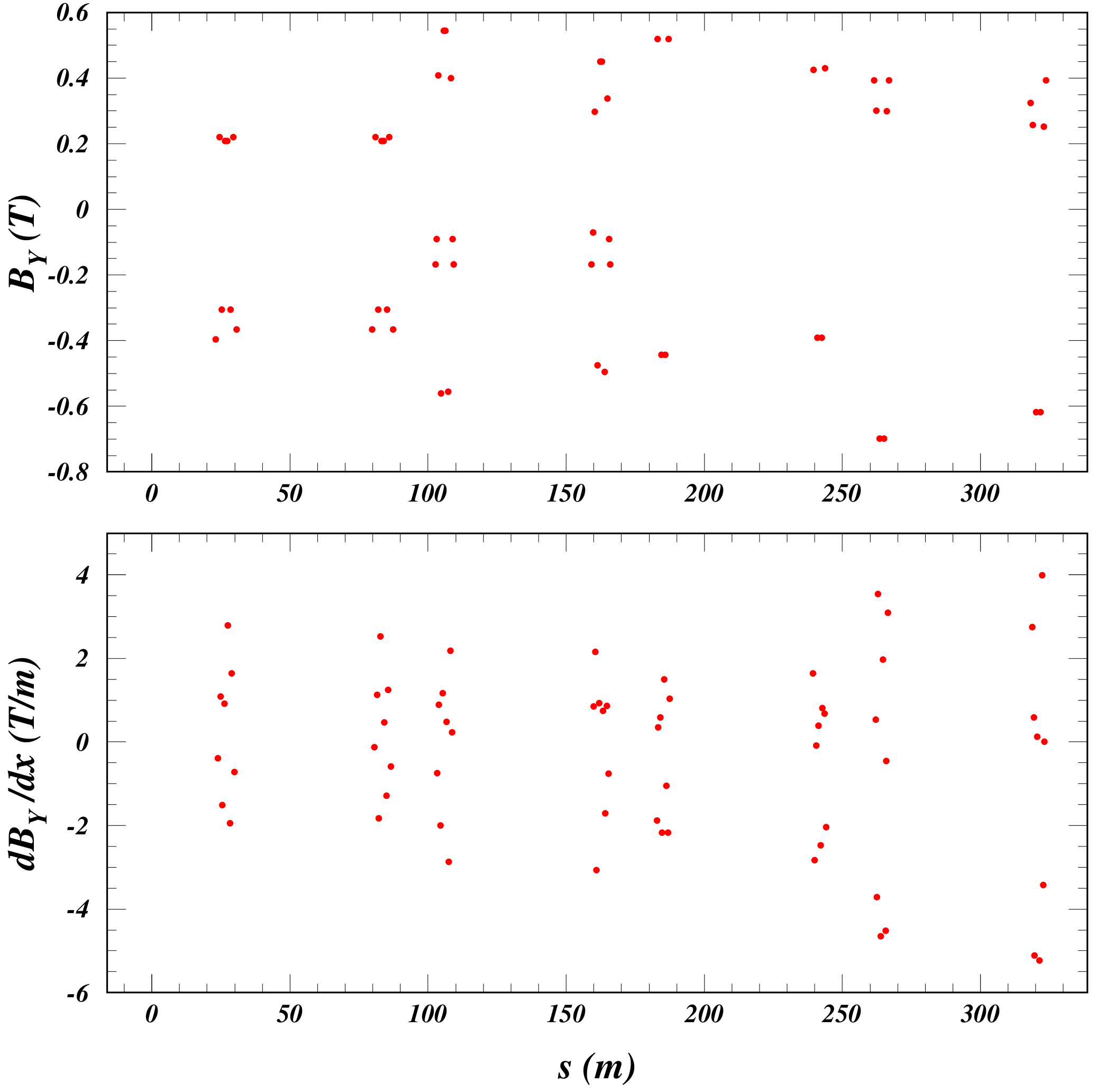}
   \caption{Distributions of required splitter dipole and quadrupole magnet field strengths versus the longitudinal orbit coordinate $s$. The magnets are thus ordered from left to right S1-R1-S2-R2-S3-R3-S4-R4.}
   \label{fig:splitter_magnets_20161117}
\end{figure}

The 30-cm H-dipole is intended for use in the cases where the present baseline lattice specifies field integrals greater than $20~\text{cm}~\times~5.4$~kG (see Fig.~\ref{fig:splitter_magnets_20161117}). 
In order to limit additional fabrication cost, the transverse dimensions of the two dipole designs are identical. The non-linearity in the field/current relationship between 1.7~kG and 5.4~kG is less than 0.1\%.

\begin{figure}[tb]
\centering
   \includegraphics[width=0.25\textwidth]{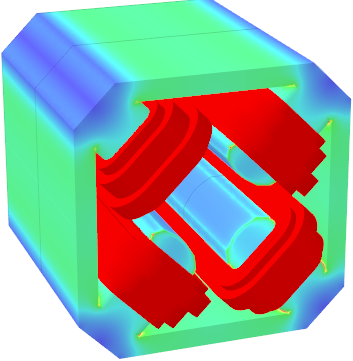}
   \caption{
            Color contours of the modeled field magnitude on the steel surface of the quadrupole magnet at 4.9~T/m excitation. The field ranges up to 13.0~kG (red) on the end corners of the pole. 
}
   \label{fig:quadB}
\end{figure}

\begin{figure}[tb]
\centering
   \includegraphics[width=0.45\textwidth]{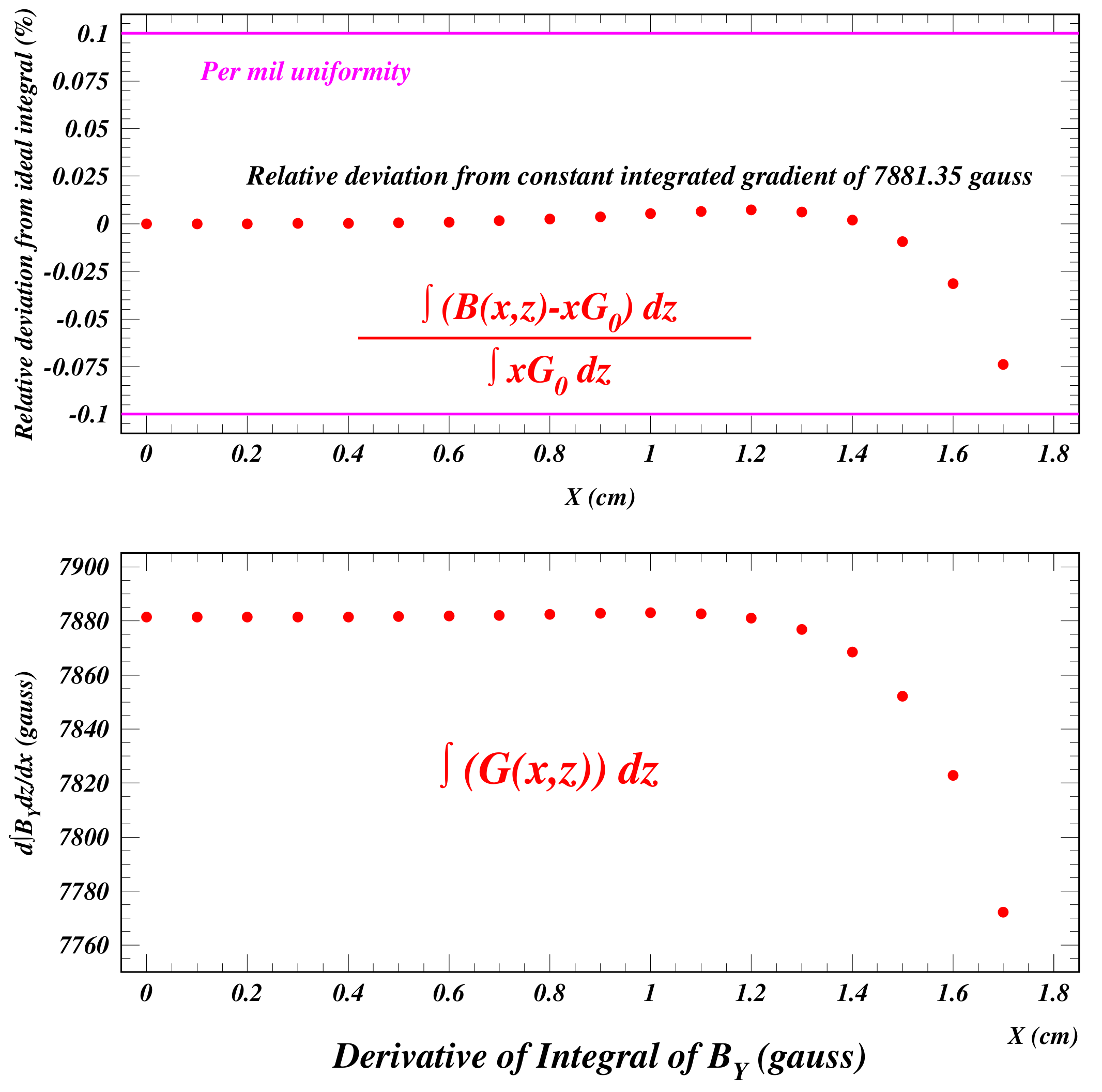}
   \caption{The upper plot shows the relative deviation of the field gradient integral from that of the perfectly uniform case. The lower plot shows the variation of the field gradient integral. The $10^{-4}$ specification is satisfied over 30~mm of the 34-mm width of the vacuum chamber.}
   \label{fig:quaduniformity}
\end{figure}

The chosen quadrupole poleface and yoke design was adapted from a large-aperture design developed at the Cornell Electron Storage Ring (CESR) in 2004~\cite{Palmer05}. Figure~\ref{fig:quadB} shows surface color contours of the field magnitude on the steel for the 4.9~T/m excitation. 
A model approximately scaled to a 45-mm bore diameter, resulting in a 15-cm square outer cross section and a 3.36~cm pole width, was found to exhibit excellent linearity up to an excitation of over 12~T/m with the central field and field integral nonuniformities shown in Fig.~\ref{fig:quaduniformity}, as modeled with the Vector Fields Opera 18R2 software~\cite{bib:opera3D}.
The maximum value of the field gradient required in the baseline lattice of 4.9~T/m is obtained with a current density of 2.8~A/mm$^2$. Of the 64 quadrupole magnets in the splitters, 47 of them run at current densities less than 1.5~A/mm$^2$.

\begin{figure}[tb]
\centering
   \includegraphics[width=0.4\textwidth]{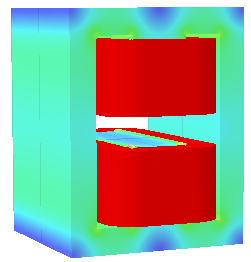}
   \includegraphics[width=0.4\textwidth]{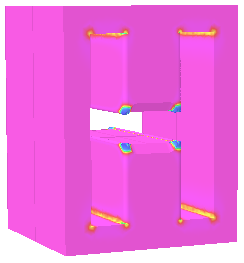}
   \caption{
            Color contours of the modeled field magnitude on the steel surface of the 20-cm-long dipole magnet models (left) and of the
magnetic permeability (right). The field values ranges up to 18~kG (red) in the return yoke for the central field value of 3.8~kG. The magnetic permeability value ranges from 2.4 (blue) to $10^{4}$ (pink). 
}
   \label{fig:dipoleB}
\vskip -5mm
\end{figure}

\begin{figure}[tb]
\centering
   \includegraphics[width=0.45\textwidth]{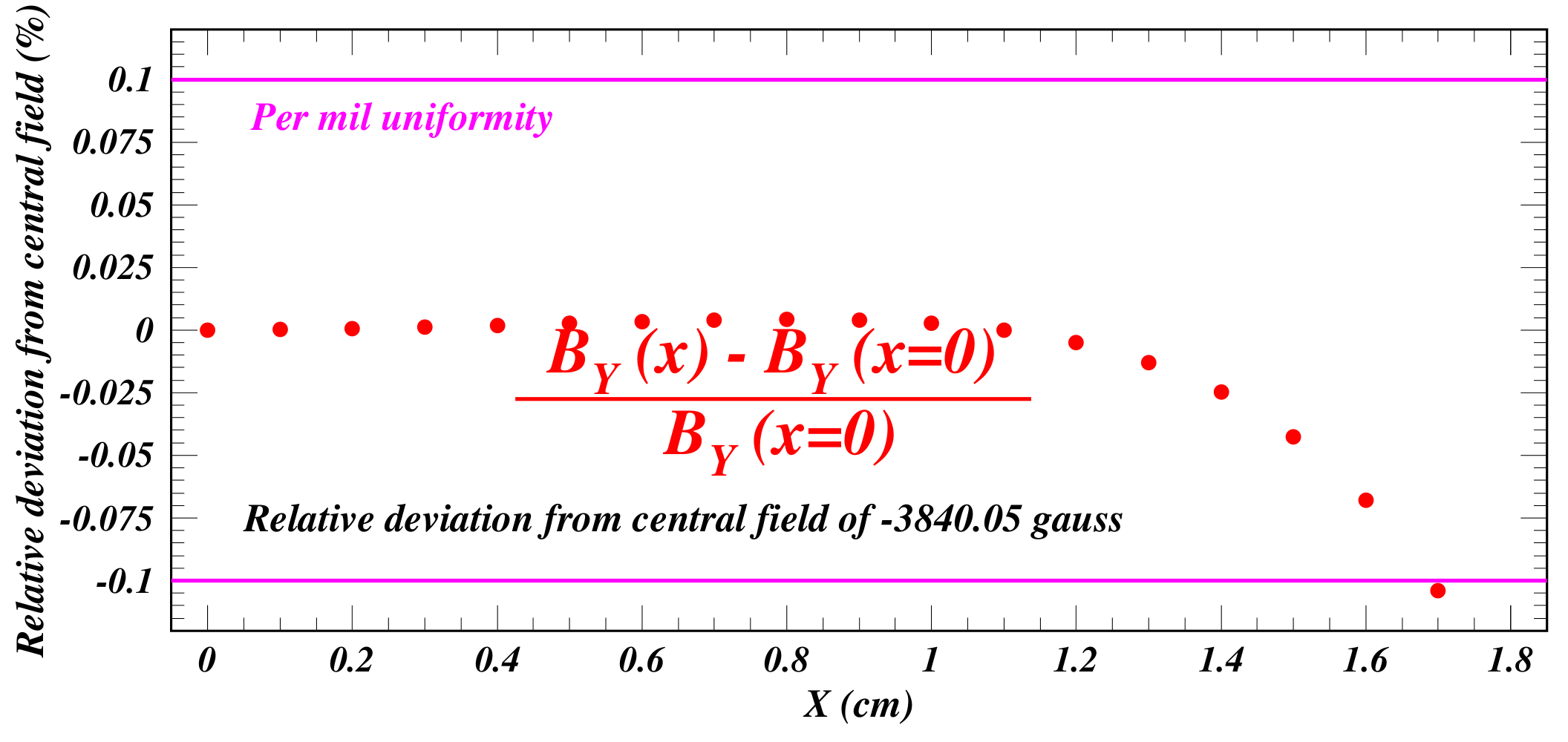}
   \includegraphics[width=0.45\textwidth]{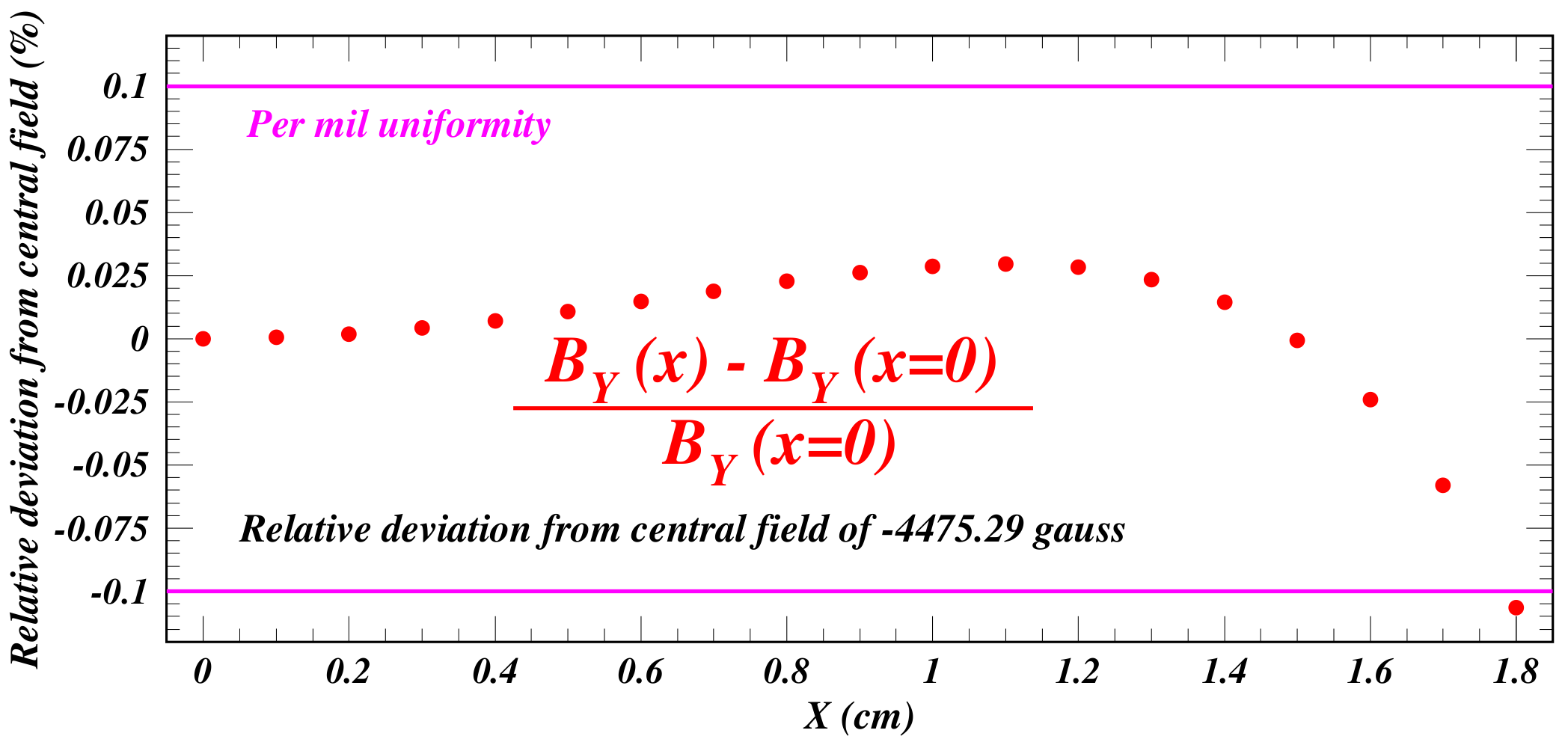}\\
   \includegraphics[width=0.45\textwidth]{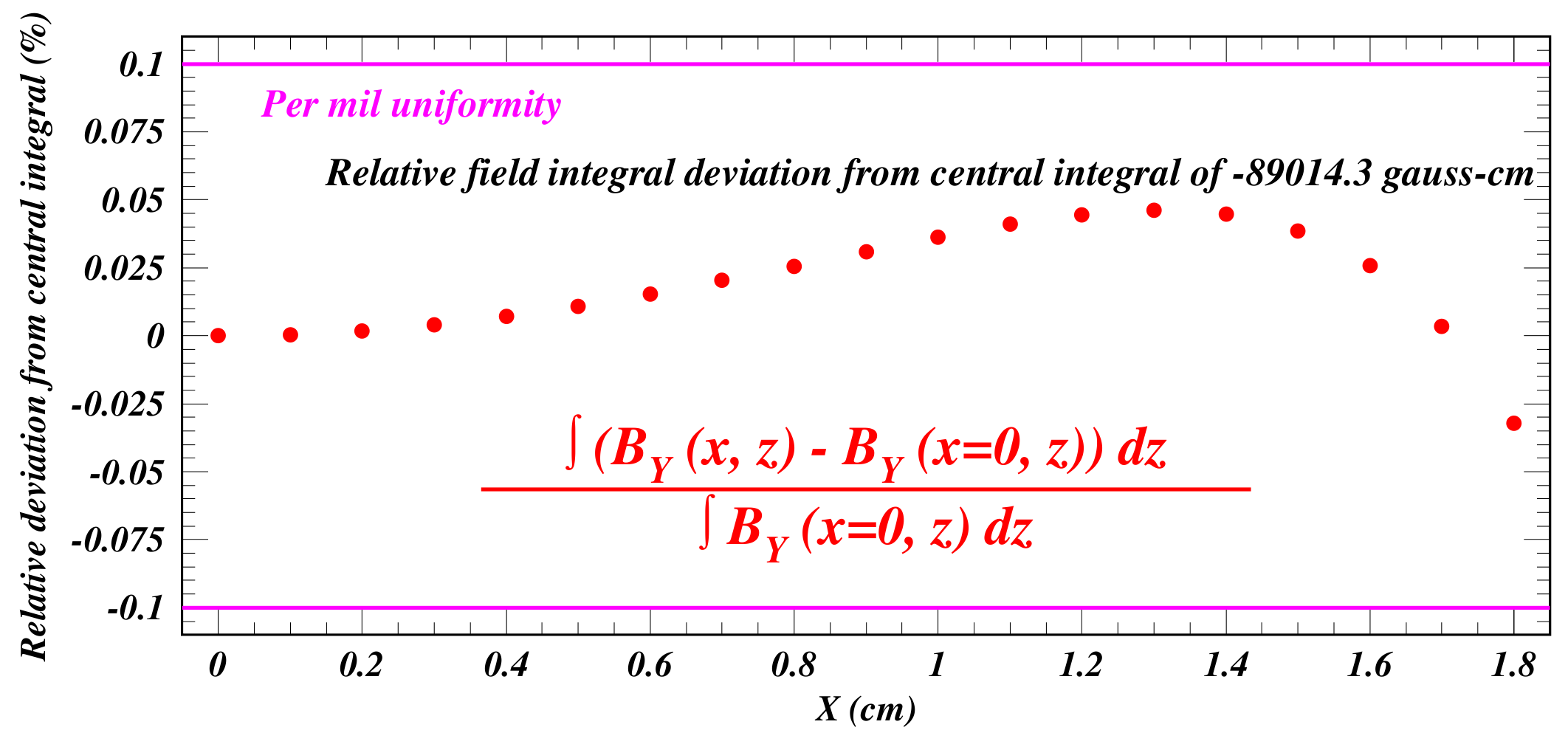}
   \includegraphics[width=0.45\textwidth]{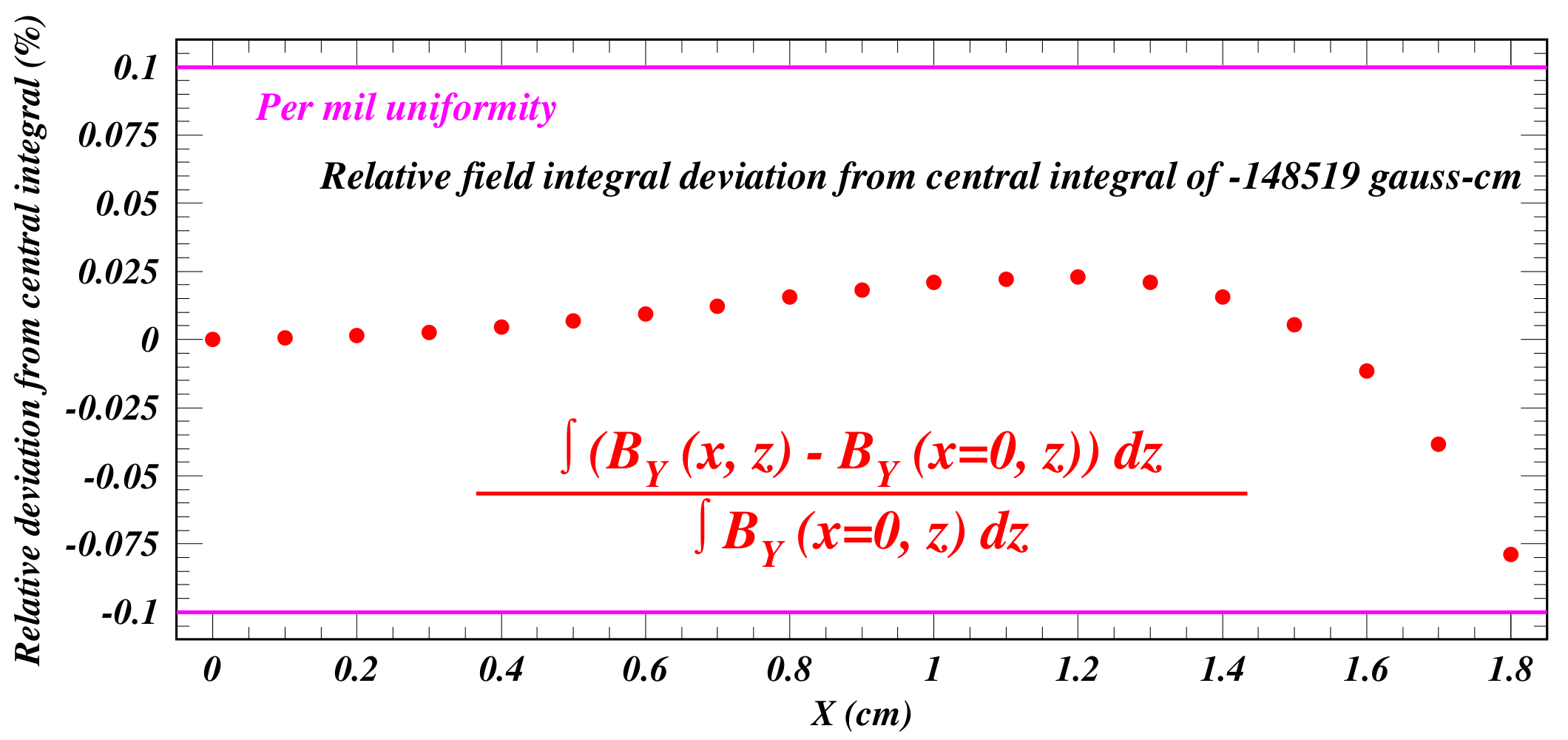}
   \caption{Modeled field uniformity for the H-dipole magnet designs. The upper plots show the relative deviation of the field from the central field value in the middle of the magnet. The lower plots show the relative deviation of the field integral from the ideal case. Left: 20-cm-long model at 3.8~kG central field strength. Right: 30-cm-long model at 4.5~kG central field strength.}
   \label{fig:20cmdipoleuniformity}
\end{figure}

The maximum field central field of the H-dipole is restricted to 5.4~kG in order to limit flux leakage out of the central pole steel. This 
limitation arises from the height of the magnet, bounded below by the need for space for the coil and the number of conductor turns required to allow use of a 100-A power supply. Surface color contours of the field magnitude on the surface of the steel for the 20- and 30-cm-long models are shown in Fig.~\ref{fig:dipoleB}. 
Figure~\ref{fig:20cmdipoleuniformity} shows the resulting field and field integral uniformity.
An alternative model employing bedstead-shaped coils was found to provide a field integral similar to that of the 30-cm-long model with racetrack coils with good linearity and a 22-cm steel length. In addition, it was 5-shorter when accounting for the coils and the uniformity was within specifications. However, when scaled to the length of the 20-cm racetrack model, saturation in the poleface resulted in insufficient field integral uniformity over the required excitation range. The choice of bedstead or racetrack coils will be determined during discussions with
the chosen magnet vendors.
 
\begin{figure}[tb]
\centering
   \includegraphics[width=0.35\textwidth]{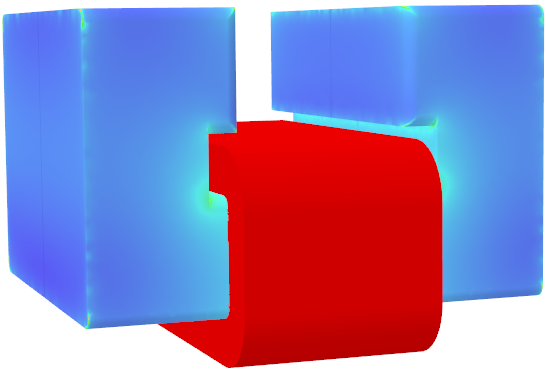}
   \includegraphics[width=0.32\textwidth]{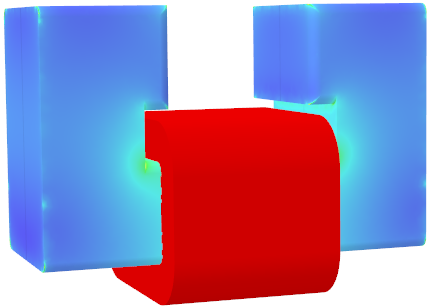}
   \caption{
            Color contours of the modeled field magnitude on the steel surface of the vertical corrector magnet models for a central field value of 160~G. The field ranges from 12~G (blue) to 400~G (green) in the visible region. Under the coil the field reaches 700~G.
Left: 10-cm-long model with field integral 0.022~T-m. Right: 5-cm-long model with field integral 0.015~T-m.
}
   \label{fig:vcorB}
\end{figure}

The design of the vertical correctors was driven by the requirement of a 5~mrad kick for the 150~MeV beam, corresponding to a field integral of 0.025 T-m, together with a maximum 10-cm steel length. After a window-frame design proved not to produce enough field integral given the
constraints of the vacuum chamber size, the rotated C-magnet design shown in Fig.~\ref{fig:vcorB} was chosen. 
 In the interest of saving space and providing positioning flexibility, the steel length is reduce from 10 cm to 5 cm for the 16 correctors in the S1-2 and R1-2 lines.

\begin{figure}[tb]
\centering
   \includegraphics[width=\textwidth]{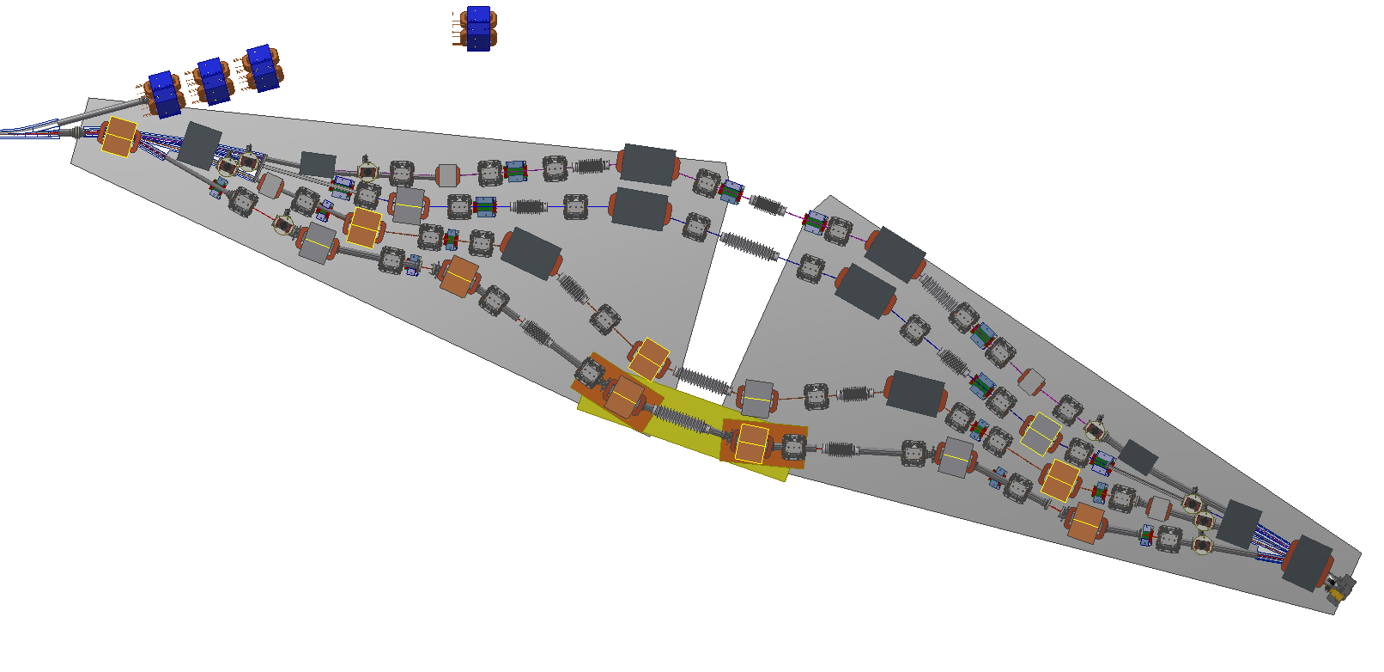}
   \caption{
            Placement of the H-dipole (dark gray) and quadrupole (light gray) magnets in the SX region. Four vertical corrector magnets are shown (blue) 
to exemplify how the space constraints can be satisfied. The magnet shown in purple on the upper left is one of the four septum magnets
under design. The bellows in the S1 line shows how the path length adjustment for energy recovery is achieved.
}
   \label{fig:sx_magnets}
\end{figure}

Figure~\ref{fig:sx_magnets} shows an engineering schematic of the SX region exhibiting the placement and clearances of the magnets
described above. This feasibility study concluded that magnets of sufficient field quality and satisfying the space constraints
can be be obtained.

\subsection{Splitter Magnet Design Development} \Leader{Jim Crittenden}

Responsibility for design development, engineering and manufacture of the splitter dipole, quadrupole and vertical corrector magnets has been assumed by Elytt Energy of Madrid, Spain. 
It has been determined that the requirements of the main dipoles are satisfied by yoke lengths of 16~cm and 31~cm. Coil designs have been re-optimized. The quadrupole magnet coil designs
have also been re-optimized and 44 of the 64 magnets can be operated without water cooling, taking into account a 20\% operational adjustment margin.
Table~\ref{tab:splitter_magnet_types} shows an overview of geometrical, field quality, and electrical parameters of the present status of these  magnet designs. 
\begin{landscape}
\begin{table}[tb]
\centering
\caption{Operational parameters for the splitter dipole, quadrupole and vertical corrector magnets.
The values include the required adjustment margins on the maximum field values of 10\% for the dipoles and 20\% for the quadrupoles.}
\tabcolsep 2mm
\begin{tabular}{lcccccc}
\toprule
\textbf{\small Parameter} & \textbf{\small H-Dipole} & \textbf{\small H-Dipole} & \textbf{\small Quad-Air}  & \textbf{\small Quad-Water}  & \textbf{\small V Corr 1}& \textbf{\small V Corr 2}\\
& \textbf{\small 21x31x16} & \textbf{\small 21x31x31} & \textbf{\small 15x15x15}  & \textbf{\small 15x15x15}  & \textbf{\small 12x7.5x10}& \textbf{\small 12x7.5x5}\\

Number of magnets & 24 & 12 & 44 & 20 & 16 & 16 \\

Gap or Bore (cm) & 3.6 & 3.6 & 4.5 & 4.5 & 4.4 & 4.4\\

Steel height (cm) & 30.5 & 30.5 & 15.0 & 15.0 & 8.6 & 8.6\\

Steel width (cm) & 21.0 & 21.0 & 15.0 & 15.0 & 12.0 & 12.0\\

Steel length (cm) & 16.0 & 31.0 & 15.0 & 15.0 & 10.0 & 5.0\\

Width including coil (cm) & 21.0 & 21.0 & 15.0 & 15.0 & 12.0 & 12.0 \\

Length including coil (cm) & 24.1 & 37.4 & 15.0 & 15.0 & 15.0 & 8.5\\

Pole width (cm) & 7.66 & 7.66 & 3.57 & 3.57 & 5.0 & 5.0\\

Field (T)/Gradient (T/m) & 0.035-0.611 & 0.412-0.649 & 0.01-2.63 & 2.96-6.28 & 0-0.016 & 0-0.016\\

\parbox{150pt}{Field/Gradient Integral\\ (T-m/T) at X=0/1 cm} & 0.008-0.120 & 0.142-0.224 & 0.002-0.394 & 0.445-0.941 & 0-0.00224 & 0-0.00150 \\

Good Field Region (mm) & $\pm$~15 & $\pm$~15 & $\pm$~5 & $\pm$~5 & $\pm$~18 & $\pm$~18 \\

Central Field Unif (\%) & $\pm$~0.03 & $\pm$~0.03 & $\pm$~0.05& $\pm$~0.05 & $\pm$~3.0 & $\pm$~3.0 \\

Field Integral Unif (\%) & $\pm$~0.03 & $\pm$~0.03 & $\pm$~0.05& $\pm$~0.05 & $\pm$~3.0 & $\pm$~3.0 \\

Bend Angle Unif (\%) & $\pm$~0.1 & $\pm$~0.1 & -- & -- & $\pm$~3.0 & $\pm$~3.0 \\

NI per coil (Amp-turns) & 583-9131 & 6073-9588 & 2-535 & 603-1277 & 0-570 & 0-570\\

Turns per coil & 4 x 13 & 4 x 13 & 82 & 11 & 57 x 10 & 57 x 10 \\

Coil cross section (cm x cm) & 2.44 x 8.58 & 2.44 x 8.58 & 0.9 x 7.5 & 2.5 x 2.1 & 9.2 x 1.5 & 9.2 x 1.5\\

Cond. cross sect (cm x cm) &  \multicolumn{2}{c}{(0.56x0.56)/0.36-diam hole} & 0.12 x 0.4 & (0.56x0.56)/0.36 & \multicolumn{2}{c}{AWG 17/0.115-diam}\\

Cond. straight length (cm) & 16.4 & 28.6 & 10.0 & 10.0 & 10.0 & 5.0 \\

Cond. length/turn, avg (cm) & 59.78 & 84.2 & 32.2 & 32.7 & 28.4 & 18.4\\

R$_{\text{coil}}$~ ($\Omega$) & 0.0229 & 0.0322 & 0.107 & 0.0030 & 2.8 & 1.8\\

L (mH) & 2~x~10.0~=~20.0  & 2~x~17.0~=~34.0 &  4~x~0.11~=~0.44 &  4~x~0.11~=~0.44 & 22.4 & 15.8\\

Power supply current (A) & 11.2-175.6 & 97.8-184.4 & 0.0-6.5 & 54.8-116.1 & 0-1.0 & 0-1.0\\

Current density (A/mm$^2$) & 0.5-7.6 & 4.24-8.0 & 0.0-1.4 & 2.6-5.5 & 0-0.4 & 0-0.4\\

Voltage drop/magnet (V) & 0.5-8.0 & 6.3-11.9 & 0-2.8 & 0.7-1.4 & 0-2.8 & 0-1.8\\

Power/magnet (W) & 3-705 & 621-2188 & 0-18 & 36-162 & 0-2.8 & 0-1.8\\
\bottomrule
\end{tabular}
\label{tab:splitter_magnet_types}
\end{table}
\end{landscape}

\section{FFAG arcs (FA, FB, TA, TB, ZA, ZB)  \Leader{Scott}}

\begin{table}
  \caption{Basic parameters for the FFAG return line.}
  \label{tab:ffag:param}
  \begin{tabular*}{1.0\linewidth}{l@{\extracolsep{0pt plus1fil}}r}
    \hline
    Total energy, pass 1 (MeV)&42\\
    Total energy, pass 2 (MeV)&78\\
    Total energy, pass 3 (MeV)&114\\
    Total energy, pass 4 (MeV)&150\\
    Focusing quadrupole length (mm)&133\\
    Defocusing magnet length (mm)&122\\
    Minimum short drift length (mm)&66\\
    Minimum long drift length (mm)&123\\
    Arc radius of curvature, approximate (m)&5.1\\
    Arc cell bend angle (deg.)&5\\
    Cells per arc&16\\
    Cells per transition section&24\\
    \hline
  \end{tabular*}
\end{table}
The FFAG beamline consists of a sequence of doublet cells with one defocusing and one defocusing magnet. There are three types of sections: the arcs (FA at the beginning, FB at the end), which have identical cells bending the beam along a circular path; the straight (ZA/ZB), containing identical cells transporting the beam parallel to the linac in the opposite direction; and the transitions (TA after FA, TB before FB) where every cell is different, adiabatically changing from cells like the arc cells to cells like the straight cells. The parameters that apply to the entire FFAG beamline are given in \Tab{tab:ffag:param}.

Each arc has 16 cells, giving 80 degrees of bend. The transition will be designed with a symmetry such that the average bend per cell is half the arc cell bend angle. Thus each transition section supplies 60 degrees of bend. Thus each spreader/combiner supplies the remaining 40 degrees of bend for half the machine.

Every focusing quadrupole will have a horizontal corrector (vertical dipole field), while every defocusing magnet will have a vertical corrector.

Engineering requirements create the basic constraints for the design:
\begin{itemize}
\item The long drift will be at least 11~cm of usable space to accommodate a variety of devices.
\item The short drift will be at least 5~cm long to accommodate a BPM.
\item There will be at least 12~mm clearance from the closed orbits to the inside of the beam pipe, and the beam pipe could be up to 3~mm thick.
\item The size of the room dictates a maximum arc radius of around 5~m.
\end{itemize}

\subsection{Arc Cell Design (FA, FB Sections)}
\begin{figure}[!tb]
  \includegraphics[width=\linewidth]{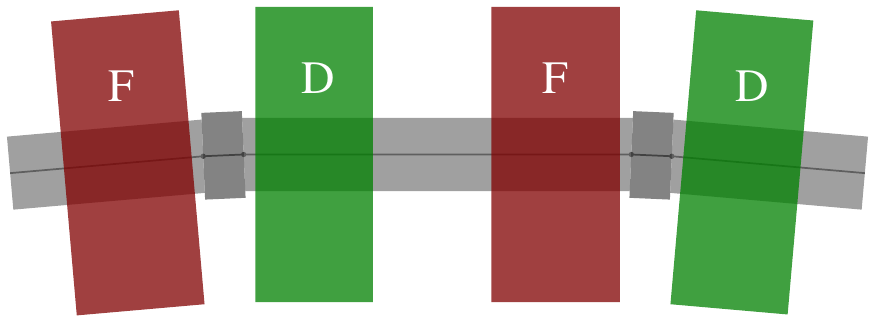}
  \caption{Illustration of FFAG arc cell geometry, showing two full cells. Lines show the reference geometry, with dots delimiting the ends of the segments. Magnet offsets are the distance of the magnet ends from the nearest dot. Segments bend by half the cell bend angle at each dot.}
  \label{fig:ffag:geom}
\end{figure}

The arc cell is the basic building block for the FFAG beam line. An illustration is given in \Fig{fig:ffag:geom}. The basic cell is a doublet, consisting of a focusing quadrupole and a combined function magnet with a dipole and defocusing quadrupole component. The geometry is defined to relate to the vacuum chamber design, which consists of 42~mm BPM blocks connected by straight beam pipes. It is thus defined by a sequence of straight lines, which bend by half the cell angle where they join. The parameters that define the geometry are given in \Tab{tab:ffag:arc}. The BPM blocks are centered in the short drift between the magnets. The precise value for the pipe length was chosen to help get the correct value of the time of flight for the entire machine.

\begin{table}
  \caption{Parameters for the arc cell.}
  \label{tab:ffag:arc}
  \begin{tabular*}{1.0\linewidth}{l@{\extracolsep{0pt plus1fil}}r}
    \hline
    Bend angle (deg.)&5\\
    BPM block length (mm)&42\\
    Pipe length (mm)&402\\
    Magnet offset from BPM block (mm)&12\\
    Focusing quadrupole length (mm)&133\\
    Defocusing magnet length (mm)&122\\
    Single cell horizontal tune, 42~MeV&0.368\\
    Single cell vertical tune, 150~MeV&0.042\\
    Integrated focusing magnet strength (T)&$-$1.528\\
    Integrated defocusing magnet strength (T)&+1.351\\
    Integrated field on axis, defocusing (T\,m)&$-$0.03736\\
    \hline
  \end{tabular*}
\end{table}
Once the longitudinal lengths are fixed, there are three free parameters: two magnet gradients, and the dipole field in the defocusing magnet. The parameters are chosen so that the maximum horizontal closed orbit excursion at 150~MeV and the minimum horizontal closed orbit excursion at 42~MeV, relative to the line defining the coordinate system, are of equal magnitude and opposite sign. The remaining two degrees of freedom are used to set the tunes at the working energies. 

\begin{figure}[!tb]
  \includegraphics[width=\linewidth]{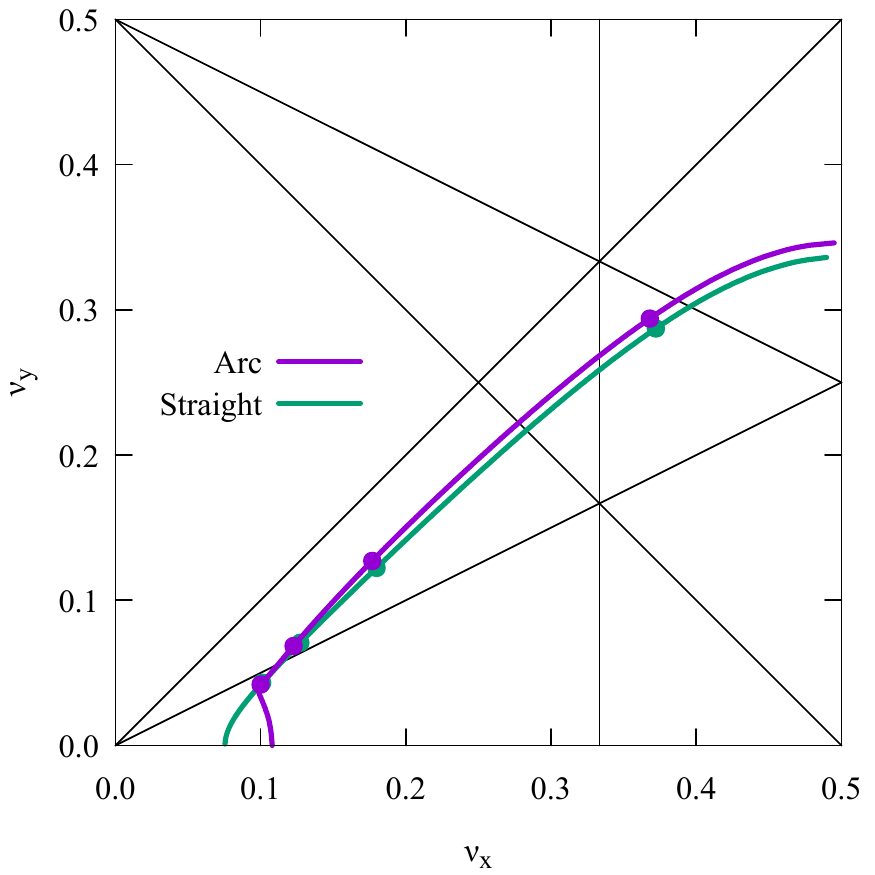}
  \caption{Tune per cell for the arc and straight cells, treated as periodic. Design energies are shown with dots. Computations are made with field maps for Halbach magnet designs.}
  \label{fig:ffag:tuneplane}
\end{figure}
\begin{figure}[!tb]
  \includegraphics[width=\linewidth]{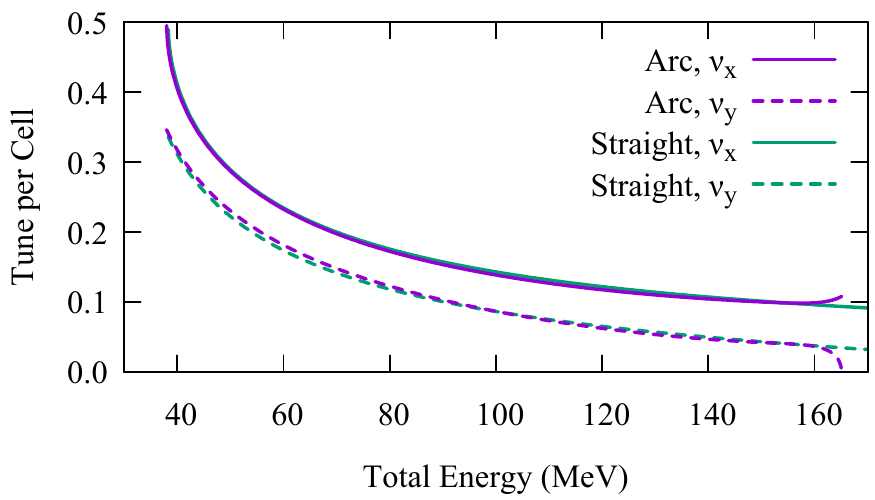}
  \caption{Tune per cell for the arc and straight cells, treated as periodic, as a function of energy. Computations are made with field maps for Halbach magnet designs.}
  \label{fig:ffag:tunevsE}
\end{figure}
\begin{figure*}
  \includegraphics[width=\linewidth]{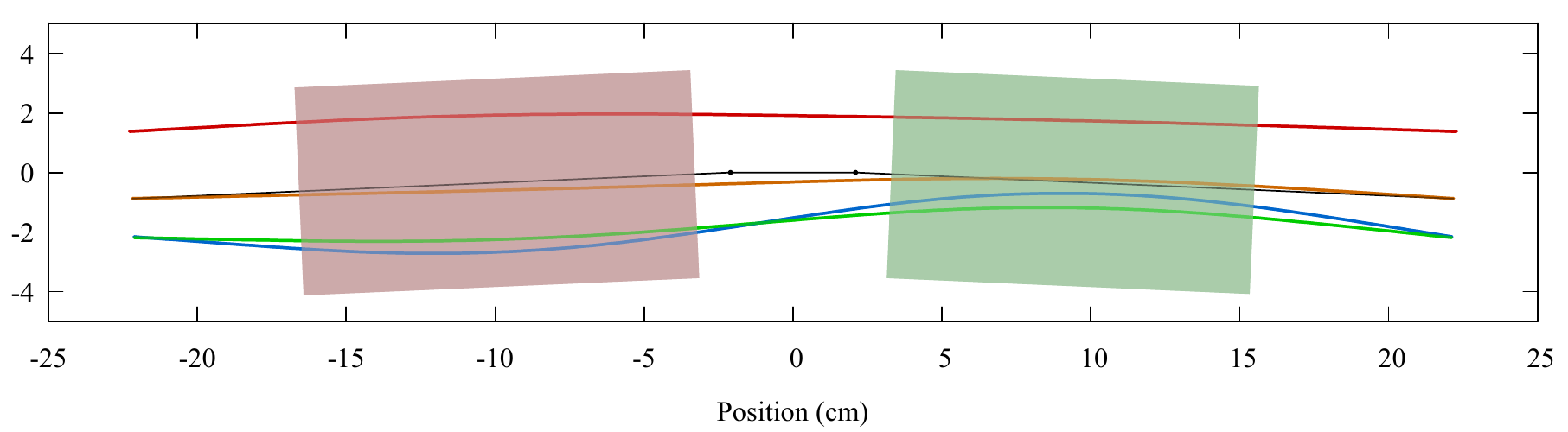}
  \caption{Periodic orbits in the arc cell. Also shown are the coordinate reference segments and the nominal magnet positions. The width of the magnets shown is equal to the pipe aperture in the mid-plane.}
  \label{fig:ffag:orbits}
\end{figure*}
The computation of the parameters is performed using field maps generated by the finite element software OPERA. Field maps for an initial estimate for the magnet designs are created, and these field maps are scaled and shifted to achieve the desired orbit centering and tune working point. Magnet designs are then modified to have the resulting integrated gradient and central field, field maps are computed from those designs, and the results are checked (and were found to be in good agreement). \Fig{fig:ffag:tuneplane} and \Fig{fig:ffag:tunevsE} show the tune per cell for the arc cell, and \Fig{fig:ffag:orbits} shows the periodic orbits in the arc cell.
\clearpage
\subsection{ZA, ZB sections}
The transition will adiabatically distort the lattice cell from the arc cell to a straight cell. We should thus first decide the parameters of the straight cell. To keep the transition smooth, all magnets of a given type (focusing/defocusing) will have the same integrated gradient and length. In addition, we will use the same focusing quadrupole everywhere. We will, however, use different types of defocusing magnets, differing in the integrated field on their axis. In particular, the defocusing magnet for the straight section will have zero field on its axis.

\begin{table}
  \caption{Parameters for the straight cell.}
  \label{tab:ffag:str}
  \begin{tabular*}{1.0\linewidth}{l@{\extracolsep{0pt plus1fil}}r}
    \hline
    BPM block length (mm)&42\\
    Pipe length (mm)&413\\
    Magnet offset from BPM block (mm)&17.5\\
    Focusing quadrupole length (mm)&133\\
    Defocusing magnet length (mm)&122\\
    Straight cell count&27\\
    \hline
  \end{tabular*}
\end{table}
If the longitudinal lengths in the straight cell are identical to those of the arc cell, the tunes and Courant-Snyder betatron functions would differ between the arc and the straight cells due to additional focusing occurring due to the curved paths the particles take through the arc magnets. Our goal is to make the tunes of the straight cell as close as possible to those of the arc cell. The only parameters available to do this are the drift lengths. The criterion used to determine the best fit is
\begin{equation}
  \sum_p\left[\sum_i T_{p,\text{str}}(E_i)-T_{p,\text{arc}}(E_i)\right]^2
\end{equation}
where $T_{p,\text{str}}(E)$ is the trace of the transfer matrix at energy $E$ for plane $p$ (\textit{i.e.}, twice the cosine of the phase advance) for the straight cell, and similarly $T_{p,\text{arc}}(E)$ for the arc cell. The chosen parameters are given in Table~\ref{tab:ffag:str}. The corresponding tunes are shown in Figs.~\ref{fig:ffag:tuneplane} and \ref{fig:ffag:tunevsE}.

\subsection{TA, TB sections}
The goal of the transition is to bring the orbits in the arc at and near the design energies onto the axis in the straight. It accomplishes this by adiabatically varying the cell parameters from those in the arc to those in the straight. The adiabatic variation allows the entire energy range to end up very close to the axis in the straight. At that point, to get the correction exactly right at the design energies, the correctors can be used, and the strengths required will be very small if the transition works well.

To measure the effectiveness of the transition, we begin with the periodic orbit in the arc cell, transport it through the transition, and determine the normalized action in the straight cell when the straight cell is treated as periodic. The normalized action is
\begin{equation}
  J_{\text{str}}(E) = \dfrac{1}{2m_ec}\left(\gamma_xpx^2+2\alpha_xxp_x+\dfrac{\beta_x}{p}p_x^2\right)
\end{equation}
where $\beta_x$, $\alpha_x$, and $\gamma_x$ are the Courant-Snyder functions for the straight cell, $p$ is the total momentum for the orbit, $x$ is the horizontal position and $p_x$ is the horizontal momentum. The values of $J_{\text{str}}$ give an approximation to the emittance growth, and should therefore be compared to the normalized emittance of the beam, which is 1~$\mu$m.

\begin{figure}[!tb]
  \includegraphics[width=\linewidth]{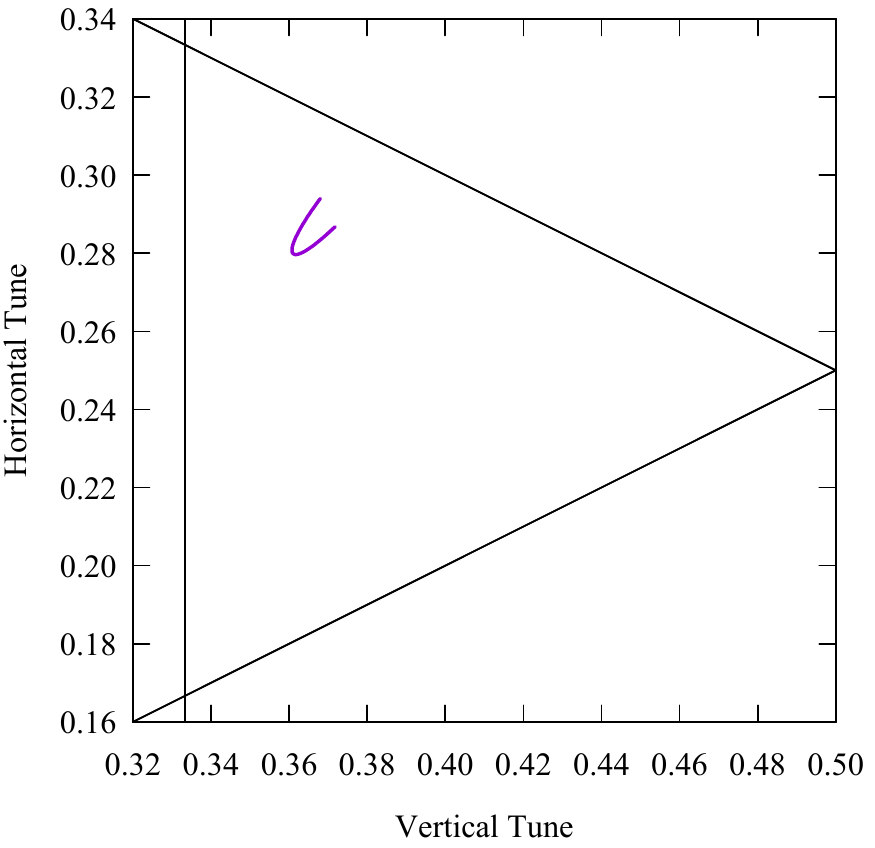}
  \caption{Tunes for a periodic cell, with angles, lengths, magnet displacements, and gradients for a hard edge  model varied linearly as described in the text. Parameters vary from arc parameters to straight parameters. Computation is done using the hard-edge model described in the text.}
  \label{fig:tunetaper}
\end{figure}
Each parameter $p$ being varied has a value $p_i$ at cell $i$ given by
\begin{equation}
  p_i = \left[1-f_T\left(\dfrac{i}{n_T+1}\right)\right]p_{\text{arc}}+
  f_T\left(\dfrac{i}{n_T+1}\right)p_{\text{str}}\label{eq:pi}
\end{equation}
where cell 1 is adjacent to the straight and cell $n_T=24$ is adjacent to the arc. The parameters varied are the lengths of the drifts, and bend angle at the BPM block, and the distance of the axis where the integrated field of the defocusing magnet is zero from the coordinate axis for the cell. The start/end of the cell is such that the distance from the end of the BPM block to the corresponding end of the cell is the same on either side of the cell.

The transition function $f_T$ is of the form
\begin{equation}
  f_T(x)=\dfrac{1}{2}+\left(x-\dfrac{1}{2}\right)\sum_{k=0}a_k\binom{2k}{k}x^k(1-x)^k
\end{equation}
where we will determine the coefficients $a_k$ that given the best behavior. We don't choose parameters that give the absolute minimum for the maximum $J_{\text{str}}$ over the energy range for a couple reasons. First, we prefer to ensure adiabatic reduction in $J_{\text{str}}$ at lower energies rather than adjusting parameters for the absolute minimum at higher energies; this allows lower energies to in a sense take care of themselves without being dependent on the precise choice for the $a_i$ and fine-tuning by correctors. Second, because the doublet is not reflection symmetric in the longitudinal direction, the two transitions behave somewhat differently, and thus the optimal coefficients are somewhat different for the two transitions. However, they are close enough that it is reasonable to choose the same coefficients for both transitions, and the penalty for doing so is small.

\begin{figure}[!tb]
  \includegraphics[width=\linewidth]{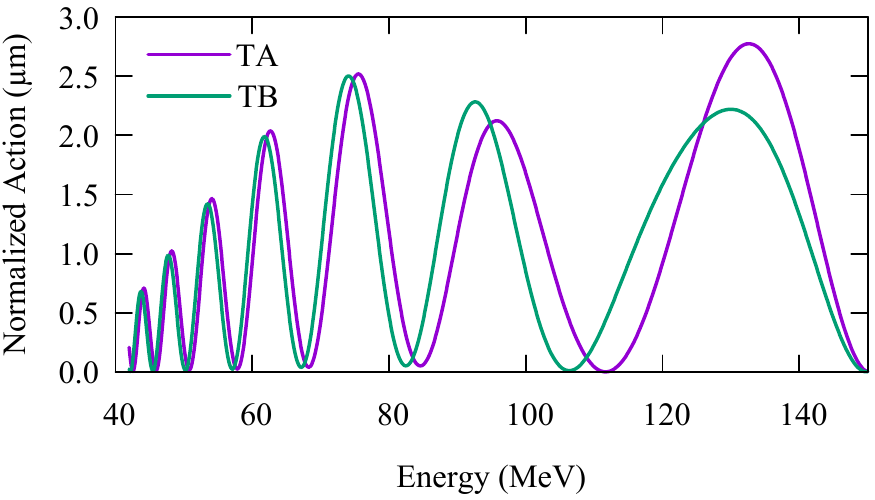}
  \caption{$J_{\text{str}}(E)$ for the transition using the taper parameters, using a hard edge model.}
  \label{fig:ffag:taperJ}
\end{figure}
\begin{figure}[!tb]
  \includegraphics[width=\linewidth]{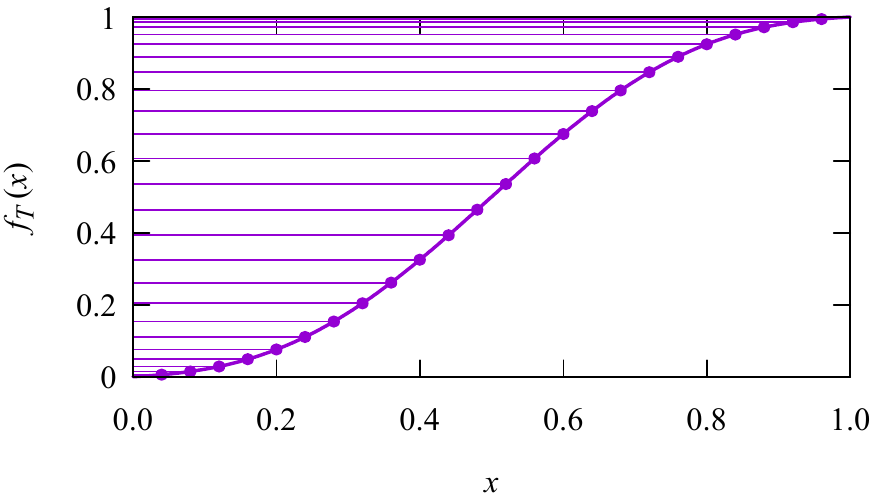}
  \caption{$f_T(x)$ used for the transition. Dots show the values used for the individual transition cells.}
  \label{fig:ffag:fT}
\end{figure}
\begin{table}
  \caption{$a_k$ in $f_T$ used for the transitions.}
  \label{tab:ffag:ak}
  \begin{tabular*}{1.0\linewidth}{r@{\extracolsep{0pt} }l@{\extracolsep{0pt plus 1 fil}}r@{\extracolsep{0pt} }l@{\extracolsep{0pt plus 1 fil}}r@{\extracolsep{0pt} }l@{\extracolsep{0pt plus 1 fil}}r@{\extracolsep{0pt} }l}
    \hline
    $a_0$:&1.000&
    $a_1$:&0.894&
    $a_2$:&0.659&
    $a_3$:&0.329\\
    \hline
  \end{tabular*}
\end{table}
\begin{table}
  \caption{Magnet types used in the FFAG return line, and horizontal positions, relative to the physical magnet center, of where their integrated fields are zero.}
  \label{tab:mags}
  \begin{tabular*}{1.0\linewidth}{r@{\extracolsep{0pt} }r@{\extracolsep{0pt plus 1 fil}}r@{\extracolsep{0pt} }r}
    \hline
    BD:&27.642 mm&
    BDT2:&24.080 mm\\
    QD:&0.000 mm&
    BDT1:&9.629 mm\\
    \hline
  \end{tabular*}
\end{table}
\begin{figure}[!tb]
  \includegraphics[width=\linewidth]{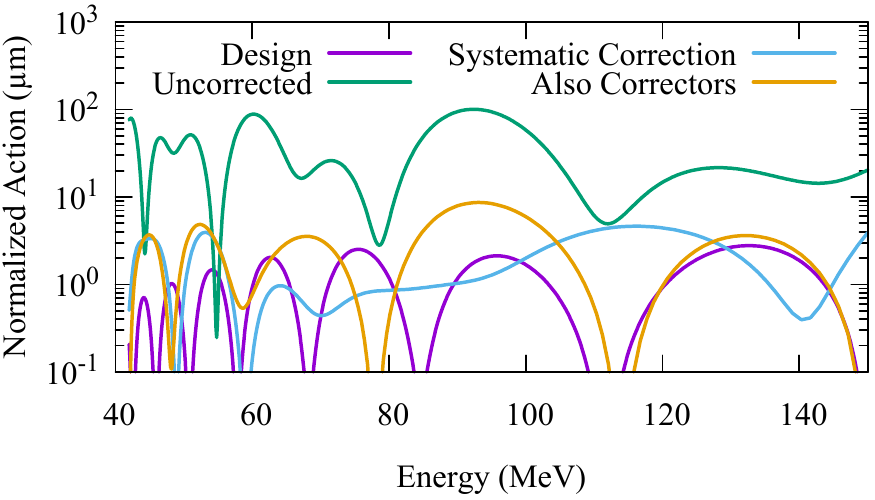}
  \caption{$J_{\text{str}}(E)$ for TA. ``Design'' is for the hard edge model, and is the same as Fig.~\ref{fig:taperJ}. ``Uncorrected'' is with field maps, when $f_T$ alone is used to position the magnets. ``Systematic Correction'' applies an additional systematic correction to each magnet type as described in the text. ``Also Correctors'' additional applies correctors on the QF magnets in the transition to make $J_{\text{str}}(E)$ be zero at the design energies.}
  \label{fig:ffag:TAmaps}
\end{figure}
The coefficients were optimized using a hard-edge approximation to the lattice that attempts to give a good approximation to the low and high energy tunes and orbits. The tunes and the orbit positions at the center of the long pipe are matched at the low and high energy by adjusting quadrupole and dipole fields of the hard edge model, as well as adding thin quadrupoles to the magnet ends, offset so they have the same zero field axis as the magnet they correspond to. The drift lengths are adjusted as described above, and the modeled quadrupole gradients and the offset of the zero field axis are adjusted using $f_T$ as well (note the gradients of the real magnets do not change). The resulting $J_{\text{str}}(E)$ is shown in \Fig{fig:ffag:taperJ}, with the $f_T$ used shown in \Fig{fig:ffag:fT}. The corresponding $a_k$ are shown in \Tab{tab:ffag:ak}.

The FFAG beamline uses the same focusing quadrupole throughout, but four distinct types of defocusing magnets. While all the defocusing magnets have the same integrated gradient, they have different integrated fields on-axis, or equivalently, a different horizontal position where the integrated field is zero. The horizontal positions where the integrated fields are zero for the different magnet types are shown in \Tab{tab:mags}. BD is used in the arc, QD in the straight, and BDT1 and BDT2 are used in the transition.

The horizontal positions of BDT1 and BDT2 are changed depending on which cell the magnets are in, so as to have the position of the zero axis vary as described in \Eq{eq:pi}. We use BDT2 when $f_T>0.71$ and BDT1 for $f_T<0.71$ so that, for each magnet the positive and negative shifts are approximately equal. The resulting $J_{\text{str}}(E)$ is shown in \Fig{fig:ffag:TAmaps}. As can be seen, the performance of the transition is significantly worse with the maps. The underlying reason is that two different magnet types, placed with their zero-field axes in the same location, do not behave precisely the same.

\begin{table}
  \caption{Additional offsets of magnets in the transition, given at the endpoints of the transition section for each defocusing magnet type. These values are for TA.}
  \label{tab:ffag:BDTshifts}
  \begin{tabular*}{1.0\linewidth}{r@{\extracolsep{0pt plus1fil}}rr}
    \hline
    $f_T$&$\Delta x$ ($\mu$m)&$\Delta x$ ($\mu$m)\\
    \hline
    0.00&BDT1\hspace{2em}\hfill+210&QF\hspace{2em}\hfill+220\\
    0.71&BDT1\hspace{2em}\hfill+110&QF\hspace{2em}\hfill+100\\
    0.71&BDT2\hspace{2em}\hfill+50&QF\hspace{2em}\hfill+150\\
    1.00&BDT2\hspace{2em}\hfill$-$40&QF\hspace{2em}\hfill+40\\
    \hline
  \end{tabular*}
\end{table}
\begin{table}
  \caption{Offset of magnets in each TA transition cell.}
  \label{tab:ffag:BDTall}
  \begin{tabular*}{1.0\linewidth}{l@{\extracolsep{0pt plus1fil}}rrr}
    \hline
    D type&$f_T$&Offset, QF (mm)&Offset, D (mm)\\
    \hline
    BDT1&0.0056&0.219&$-$9.266\\
    BDT1&0.0146&0.218&$-$9.017\\
    BDT1&0.0286&0.215&$-$8.634\\
    BDT1&0.0486&0.212&$-$8.083\\
    BDT1&0.0757&0.207&$-$7.337\\
    BDT1&0.1106&0.201&$-$6.377\\
    BDT1&0.1535&0.194&$-$5.198\\
    BDT1&0.2041&0.186&$-$3.807\\
    BDT1&0.2617&0.176&$-$2.223\\
    BDT1&0.3252&0.165&$-$0.475\\
    BDT1&0.3933&0.154&+1.396\\
    BDT1&0.4641&0.142&+3.344\\
    BDT1&0.5359&0.129&+5.319\\
    BDT1&0.6067&0.117&+7.267\\
    BDT1&0.6748&0.106&+9.138\\
    BDT2&0.7383&0.139&$-$3.630\\
    BDT2&0.7959&0.117&$-$2.056\\
    BDT2&0.8465&0.098&$-$0.673\\
    BDT2&0.8894&0.082&+0.498\\
    BDT2&0.9243&0.069&+1.452\\
    BDT2&0.9514&0.058&+2.194\\
    BDT2&0.9714&0.051&+2.741\\
    BDT2&0.9854&0.046&+3.122\\
    BDT2&0.9944&0.042&+3.370\\
    \hline
  \end{tabular*}
\end{table}
To attempt to correct for this, we add a systematic offset to BDT1 and BDT2 as well as the QF magnets in the corresponding sections. This function will be linear in $f_T$ for the corresponding section:
\begin{equation}
  \Delta x(f_T) = \Delta x(f_0)\dfrac{f_1-f_T}{f_1-f_0}+
  \Delta x(f_1)\dfrac{f_T-f_0}{f_1-f_0}
\end{equation}
For the end point in the middle, we use 0.71. The values of $\Delta x$ for the focusing and defocusing magnets at the end points for each transition section with a given defocusing magnet type (8 values in all) are adjusted to minimize the maximum $J_{\text{str}}(E)$ over the energy range. The resulting offsets at the endpoints are given in \Tab{tab:ffag:BDTshifts}, and the corresponding $J_{\text{str}}(E)$ are shown in \Fig{fig:ffag:TAmaps}.

\subsubsection{Applying Dipole Correctors}
Dipole correctors can be applied to get the design energies precisely correct. The goal of the taper is to bring $J_{\text{str}}(E)$ as close as possible to zero over the full energy range, to reduce the required corrector strengths required to zero $J_{\text{str}}(E)$ at the design energies, and to make the design robust against systematic errors. The correctors are then applied on top of this, and the required strengths should be small.

To compute the corrector strengths, I used an iterative algorithm where a matrix computing the response of $x$ and $p_x$ at the straight for the design energies to changes in dipole corrector strengths is computed. A linear computation is made to determine approximately the changes in corrector strengths that would zero $J_{\text{str}}(E)$ at the design energies, while minimizing the sum of the squares of the changes in the corrector strengths. Starting with the corrector strengths at zero, this algorithm is repeated until the $J_{\text{str}}(E)$ are zero at the design energies; in fact, one step of the algorithm gives a more than adequate estimate. The resulting $J_{\text{str}}(E)$ is shown in \Fig{fig:ffag:TAmaps}. The maximum required corrector strength is 16~$\mu$T\,m.

\begin{figure}[htbp]
\centering
\includegraphics[width=\linewidth]{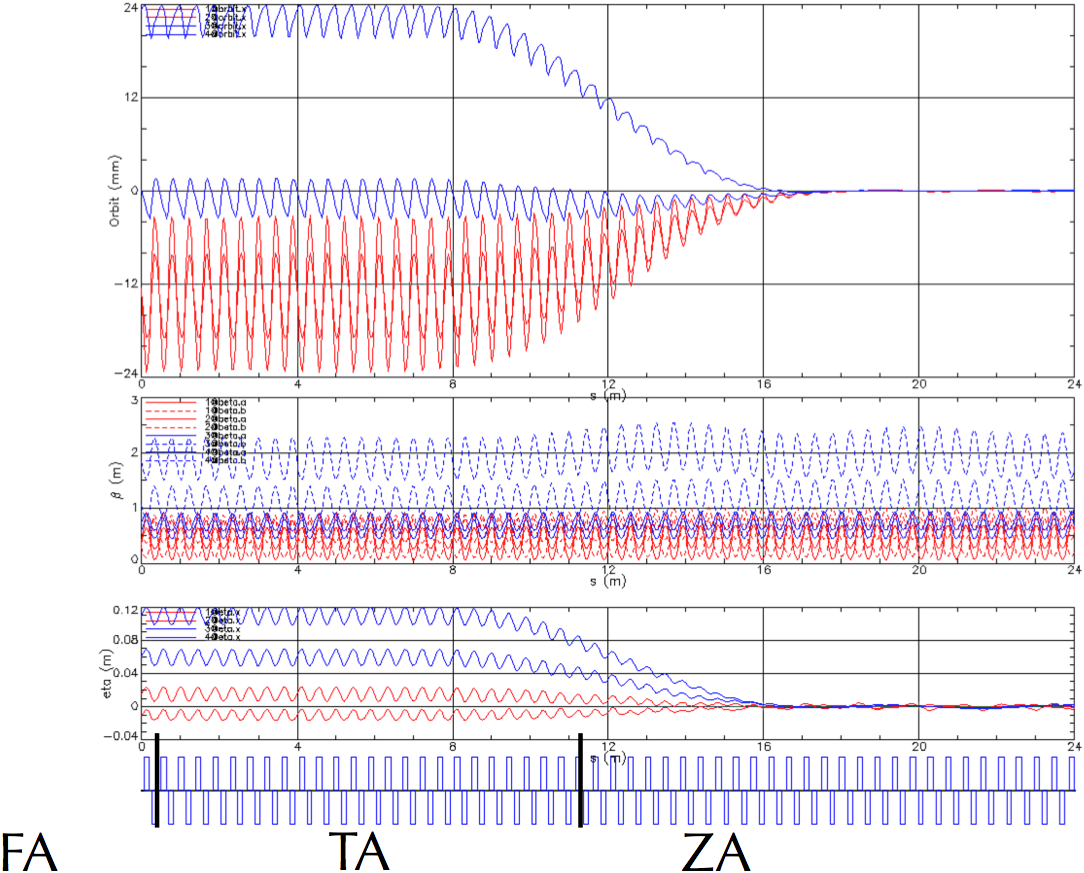}
\caption[]{Complete optics for the FA, TA, and ZA sections. }
\label{fig:optics_ffag_symmetric}
\end{figure}

Putting this all together results in the optics shown in \Fig{fig:optics_ffag_symmetric}.

\clearpage
\section{High-Energy Loop for Users \Leader{Dejan}}
In the early operational stages of CBETA energy recovery begins after four acceleration passes through the linac.  A $\pi/2$ change in RF phase is achieved by adjusting the path length in both the S4 section after the fourth acceleration pass, and in the R4 section before the first deceleration pass. 

\begin{figure}[ht]
	\begin{center}
		\includegraphics[width=0.9\textwidth]{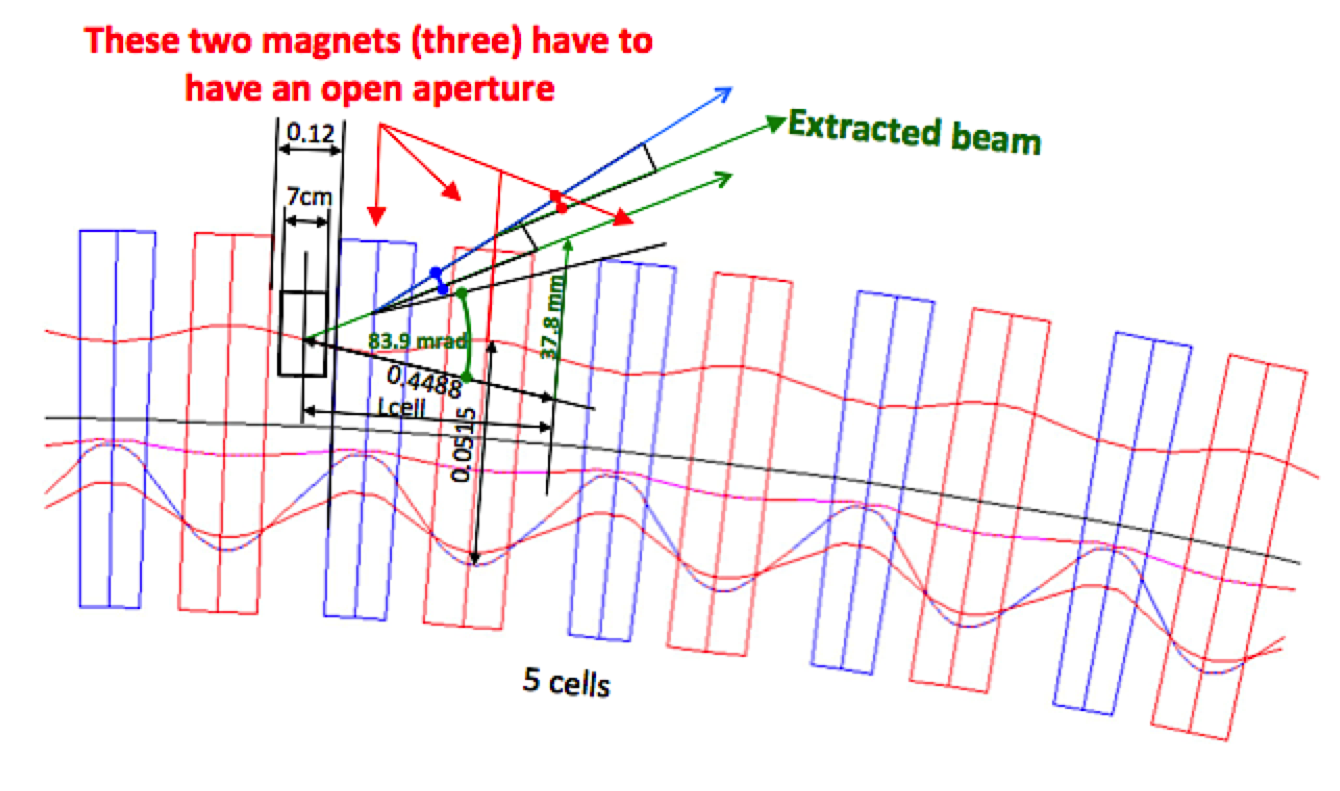}
	\end{center}
	\caption{Orbit of the extracted beam through couple of FFAG arc magnets.}
	\label{fig:ExtractedOrbit}
\end{figure}

\begin{figure}[h]
	\begin{center}
	 	\includegraphics[width=0.5\textwidth]{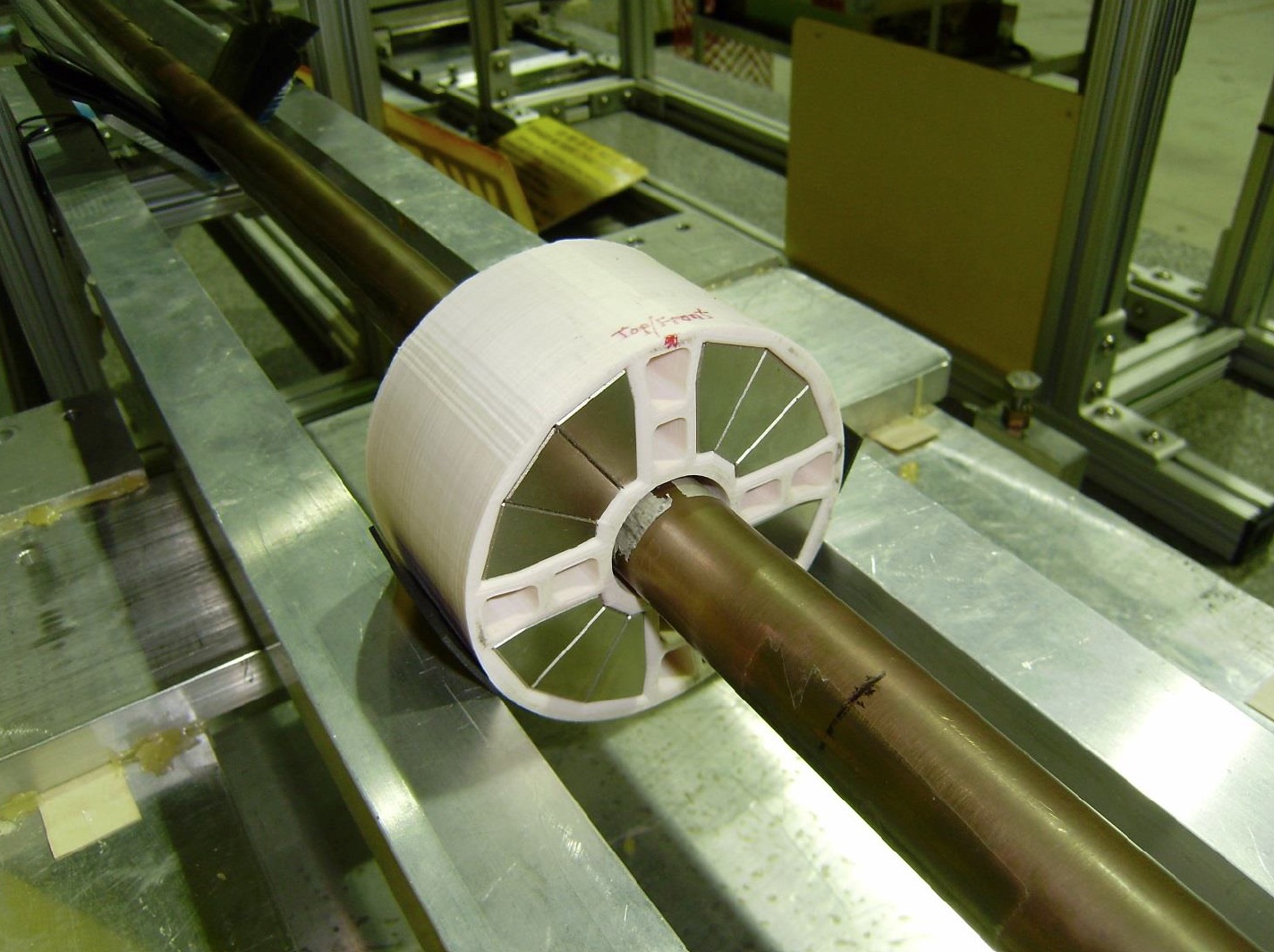}
	\end{center}
	\caption{A prototype eRHIC Halbach style quadrupole, with an open mid-plane that allows synchrotron radiation to escape. }
	\label{fig:HalbachOpenMagnet}
\end{figure}

 In the final operational stage, a straight CBETA beamline will be made available for experimental users, delivering highest energy electrons.  Extraction to the experimental line from the arc is made possible by including a couple of special open mid-plane magnets, as indicated in \Fig{fig:ExtractedOrbit}.  These can be achieved by using Halbach-style magnets.  Prototype Halbach quadrupoles intended for eRHIC have already been built and successfully tested at BNL, as shown in \Figure{fig:HalbachOpenMagnet}.

\begin{figure}[h]
	\begin{center}
		\includegraphics[width=0.7\textwidth]{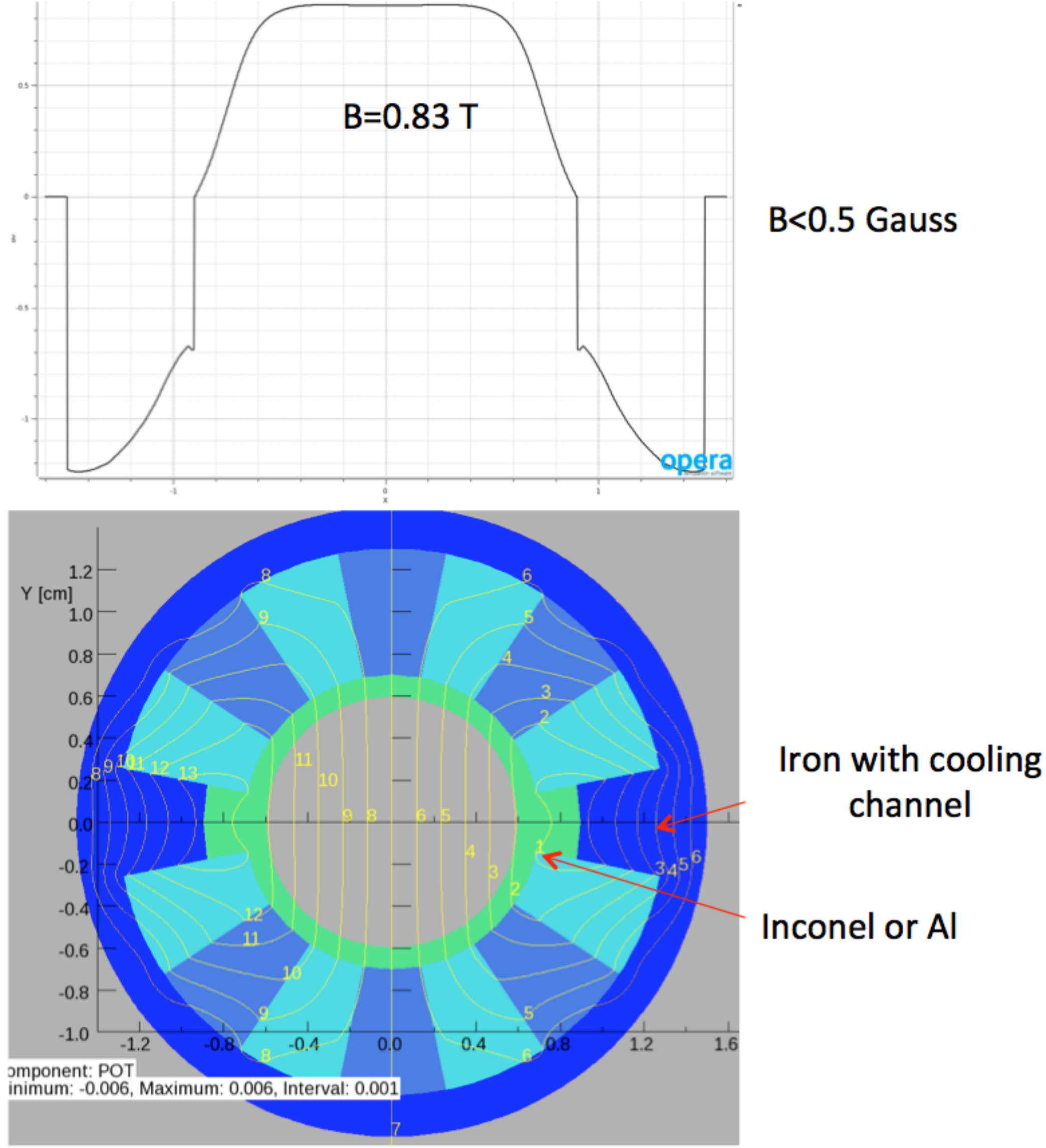}
	\end{center}
		\caption{Preliminary design of the Halbach extraction dipole.  (Courtesy of N. Tsoupas).  } 
		\label{fig:HalbachExtractionMagnet}
\end{figure}
The highest energy beam is naturally radially displaced by more than 20~mm.  It passes through the field of a 70~mm long Halbach dipole inside the beam pipe, placed in the 120~mm drift between two adjacent quadrupoles. Plated permanent magnet blocks within the pipe are ultra-high vacuum compatible.  A preliminary design of the extraction dipole is shown in \Fig{fig:HalbachExtractionMagnet}.

\clearpage 
\section{Bunch patterns\Leader{Stephen}\label{sect:bunch-patterns}}

\begin{table}[tb]
\begin{center}
\caption[]{Operational Modes}
\begin{tabular*}{0.8\columnwidth}{@{\extracolsep{\fill}}lcccc}
\toprule
 & Commissioning & eRHIC & High-current & \\
\midrule
Injection Rate & 0.95 & 41.9 & 325 & MHz\\
Max Bunch Charge & 125 & 125 & 125 & pC \\
Max Current & 0.12 & 5.2 & 40 & mA\\
Probe Bunch Rate & N/A & $\le$0.95 & $\le$0.43 & MHz \\
\bottomrule
\end{tabular*}
\label{tab:bunchpatterns}
\end{center}
\end{table}

CBETA will support multiple operating modes, single pass and multi-pass/multi-energy, pulsed and CW, with and without energy recovery. Many of these modes are only intended for commissioning and machine studies. These modes must cover a wide range of average current, suitable for the wide range of necessary diagnostics --- nanoamp for view screens, microamp for BPMs, and milliamp for full current operation. But, within each of these ranges, the full range of bunch charge may need to be explored, in order to better isolate possible limiting effects. In general, the beam modes must be well matched the goals of the commissioning, the precise manner that they will be achieved, and the diagnostics that will be used. 

The RF cavities in the CBETA linac operate at 1300 MHz. The injector must supply bunches at a sub-harmonic of this frequency. The multiple passes of these bunches through CBETA produce inter-bunch timing patterns which depend on the injection frequency and the revolution period. Additionally the decelerating bunches must have a timing which is an integer + one-half RF cycles offset from the accelerating bunches. Additional path length in the highest-energy splitter lines delays the highest energy turn by 1.5 RF periods.

\begin{figure}[htbp]
\centering
\includegraphics[width=0.95\textwidth]{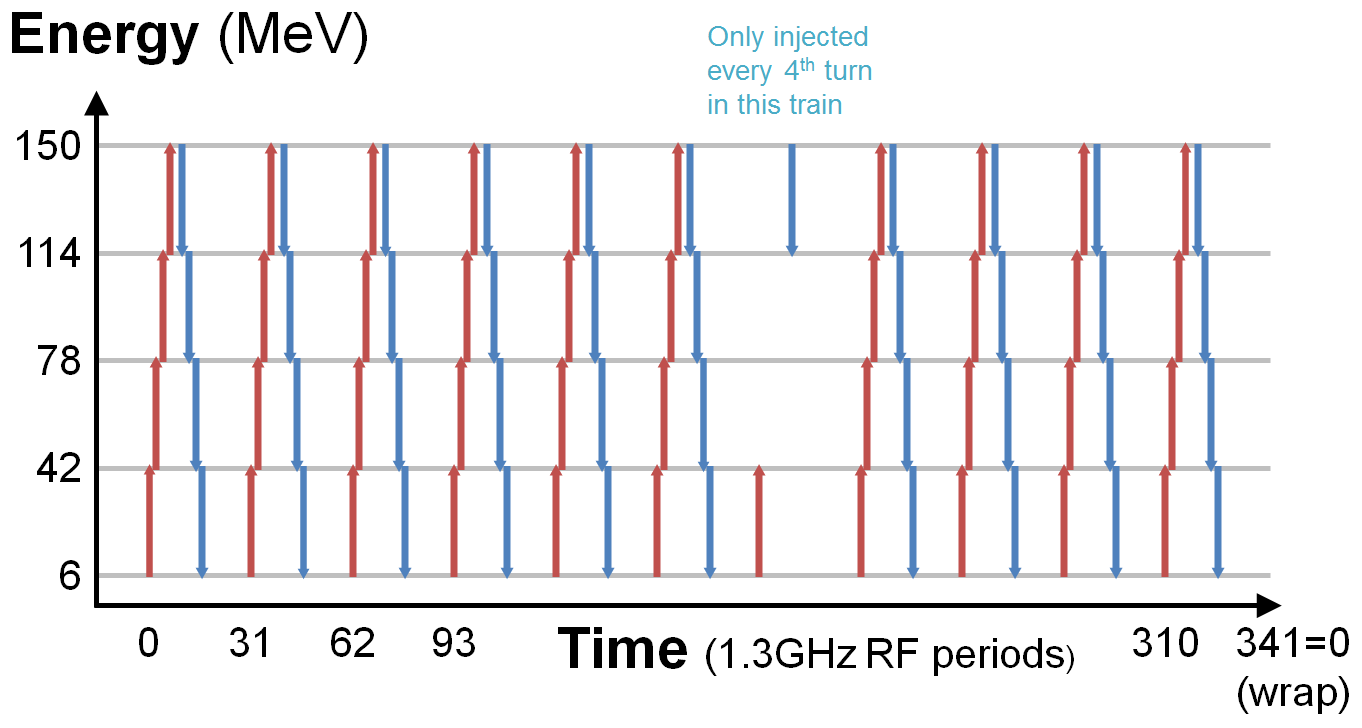}
\caption[]{Bunch pattern produced in the eRHIC-like mode that has a circumference of $h=343$ RF wavelengths but a time periodicity of 341 RF periods.}
\label{fig:bunchpattern343}
\end{figure}

\begin{figure}[htbp]
\centering
\includegraphics[width=0.95\textwidth]{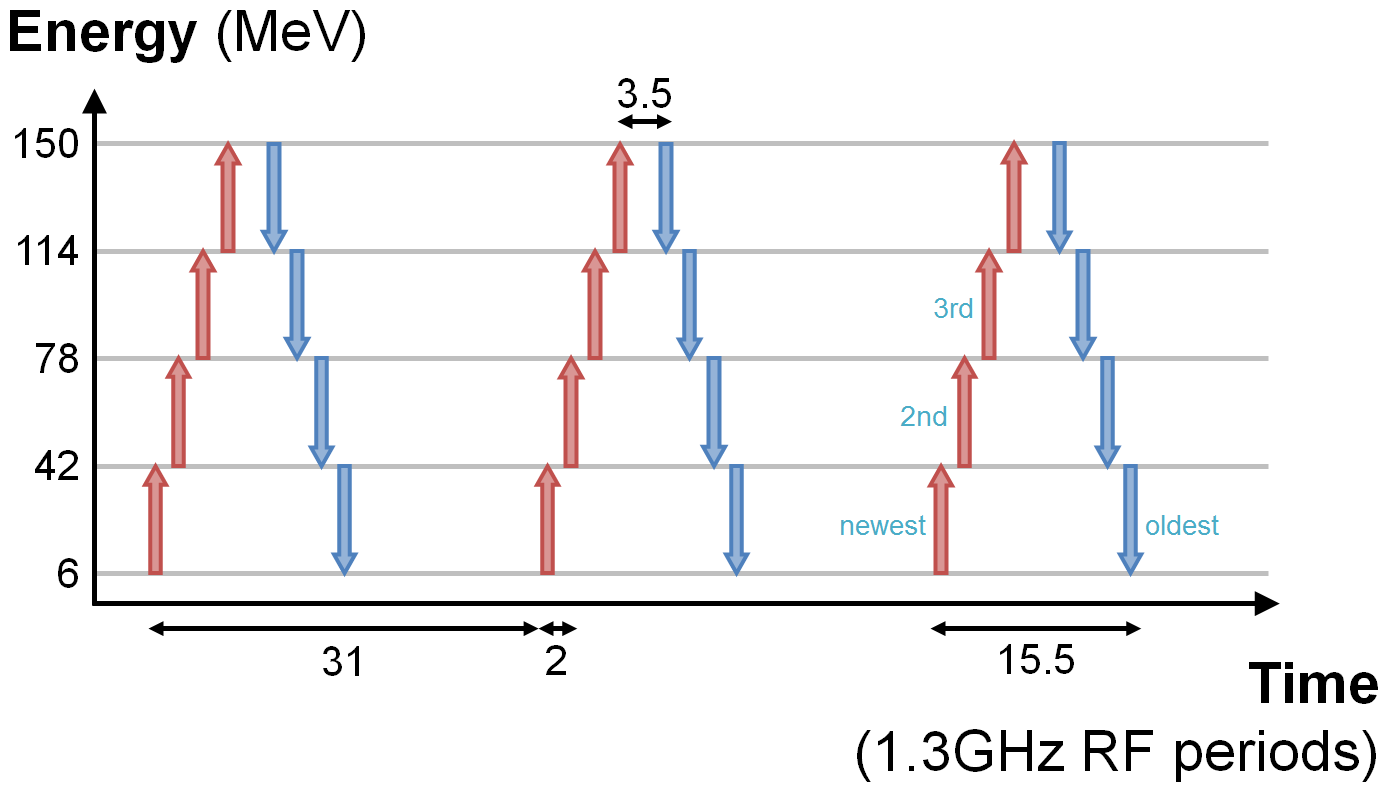}
\caption[]{Bunch train details in the eRHIC-like mode.  Note the top-energy bunches are well-separated (by 3.5 RF wavelengths).}
\label{fig:bunchpattern343zoom}
\end{figure}

The baseline scheme with three operational modes is shown below.  There will also be a ``single shot'' mode during early commissioning.

\begin{figure}[htbp]
\centering
\includegraphics[width=0.95\textwidth]{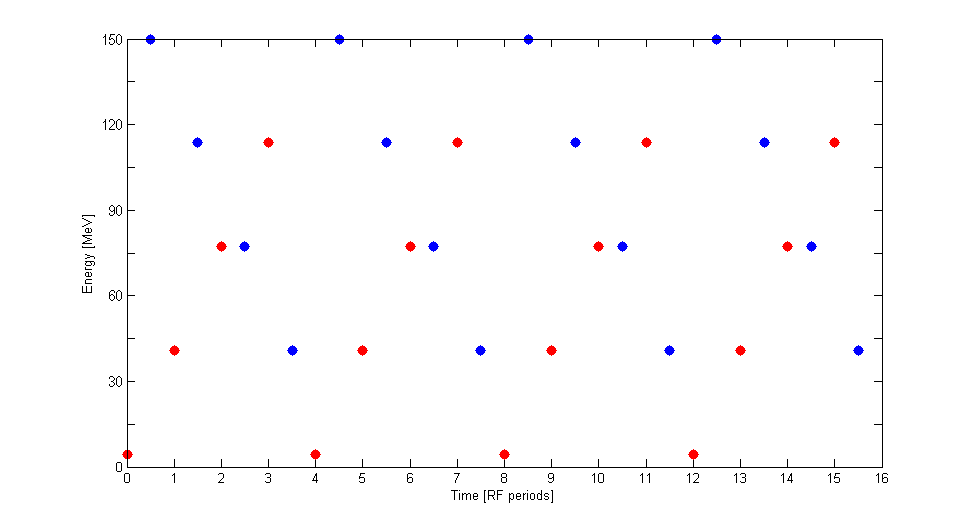}
\caption[]{Bunches passing a reference point at the start of the linac in the high-current mode, with one accelerating bunch and one decelerating bunch in every RF period.  Red and blue dots represent accelerating and decelerating bunches, respectively.}
\label{fig:bunchpattern_highcurrent}
\end{figure}

\begin{itemize}
  \item 343 RF period circumference.
  \item \textbf{Commissioning mode}:
Probe bunch injected every 341*4 periods ($\sim$4 turns).
There would be two passing through the linac at any time, one accelerating, one decelerating, 
separated by 9.5 periods.  This 9.5 period gap is designed for conventional 
diagnostics to be able to resolve the two bunches. In this manner, complete knowledge of the bunches at all energies will be available from the BPM system. Full current from the gun in this mode, assuming a typical bunch charge of 125 pC, is around $119\unit{\mu A}$.

  \item \textbf{eRHIC-like mode}:
Bunches injected at a 1.3GHz/31 frequency
(341/11 = 31).  This produces 11 bunch trains simultaneously in the 
ring.  Within each train the bunch-to-bunch separation is 2 wavelengths (``650MHz''), with 1.5 wavelengths added at the top energy, giving 8 bunches spread over 15.5 wavelengths.  One of the trains is replaced by the above pair of probe bunches by selectively suppressing laser pulses for it on 3 out of every 4 turns.  This is shown in \Fig{fig:bunchpattern343}.  This allows the BPM electronics to continue operating in an identical fashion to the commissioning mode, and complete knowledge of the probe bunches is still maintained in this operating mode. Full current from the gun in this mode, assuming a typical bunch charge of 125 pC, is around 5.2 mA.

\item \textbf{High-current mode}:
Inject at 1.3GHz/4=325MHz.  The circumference 343 is equivalent to -1 (modulo 4), so each successive turn's CW bunch train would slip 1 RF wavelength and fill every RF peak.  Then, during deceleration, after the 1.5 wavelength offset, all
the decelerating troughs would also be filled.  So in total all the available RF peaks and troughs
would be used (at 2.6GHz total rate). This is shown in \Fig{fig:bunchpattern_highcurrent}.  Pilot bunches may still be injected, but would have to be injected at a lower rate than in the previous two modes. To do so, a gap in the bunches needs to be introduced once per turn, for at least 7 turns on either side of the pilot bunch, allowing it to be seen in isolation from the rest of the bunches. Full current from the gun in this mode, assuming a typical bunch charge of 125 pC, is around 40 mA.
\end{itemize}

Note that this initial 1.3GHz configuration of the machine has an odd harmonic number, because this is required for the high-current operating mode.  When the 650MHz eRHIC prototype cavity is installed, the circumference will be shortened to 342 1.3GHz wavelengths, or h=171 for the new 650MHz frequency, by shifting the splitter lines.  Since 171=170+1, any small factor of 170 (F=1,2,5,10), may be chosen for the number of trains and injection would happen at 650*F/170 MHz (38.2MHz for F=10).  In both of these schemes, the circumference is larger than the injection periodicity, so the bunches in the ring `fall behind' as the newly injected ones arrive.

Both of the lower-current modes include the `probe bunches', of which two would be in the ring at any one time (one accelerating and one decelerating), as they are injected every fourth turn.  These would replace one of the 11 eight-bunch trains in the case of the eRHIC-like scheme (the seventh train in \Fig{fig:bunchpattern343}).  The two probe bunches are separated by 9.5 RF periods, or 7.3~ns, which allows the BPMs to distinguish their signals.  The probe bunches are also separated from the bunches of adjacent bunch trains by at least this time interval.  To be explicit, the bunch trains are 15.5 RF periods long and recur every 31 periods, thus the gap between them is 15.5 RF periods.  Each BPM measures the probe bunch as it passes with successively increasing then decreasing energy on each turn, giving orbits for all seven FFAG passes.

The same time domain BPM electronics that is used in the low frequency commissioning mode serves for the eRHIC-like mode also, with the denser bunch `trains' remaining unmeasured (except perhaps on an average basis).  The important principle to be demonstrated is that measurements on well-separated probe bunches can provide enough information to operate the machine with the high-average-current trains in it.

Injection using the eRHIC-like mode will allow for for a per energy beam current of 1 mA at around 24 pC bunch charge and therefore enables achievement of all the Key Performance Parameters. The Ultimate Performance Parameters can only be achieved using the high-current mode.

In the high-current mode, the pilot bunch pattern will need to be changed. In this case, seven gaps needs to be introduced before, during, and after a pilot bunch, separated in time by the circumference of the ring. In that manner, the gaps will overlap with each other as they pass around the ring. Thus, the maximum rate of the pilot bunches is once per every 9 round trips. In this mode, there is a single pilot bunch, which will gain energy 4 times before it gives the energy back, so the maximum rate of pilot bunches may be additionally limited by that transient load on the RF.

\clearpage
\section{CSR\Leader{Chris}}

When a charged particle is transversely accelerated in a bending magnet, it produces radiation according to the well-known synchrotron radiation spectrum. When $N$ such particles are bunched on a scale of length $\sigma$, the power spectrum per particle at frequencies smaller than $c/\sigma$ in this spectrum is enhanced by roughly a factor $N$. This results in increased radiation, and hence increased energy losses from the individual particles. This coherent synchrotron radiation was first calculated in a seminal paper by Schwinger \Ref{Schwinger45}.

The CSR wake $W_{\text{CSR}}(z)$ is the energy change per unit length of a particle with longitudinal position $z$ in a bunch, and it can be shown that for ultra-relativistic particles this $W_{\text{CSR}}(z)$ scales with the factor
\begin{equation}
W_0 = N r_c m c^2 \frac{\kappa^{2/3}}{\sigma^{4/3}}\end{equation}
where $m$ is the mass of a single particle, $r_c$ is its classical electromagnetic radius, and $\kappa$ is the trajectory curvature (e.g.~see \Ref{PhysRevSTAB.12.024401}). For a Gaussian bunch moving on a continuous circle, called `steady-state CSR', the average energy loss per unit length is approximately $0.35\times W_0$, and the maximum energy loss per unit length is approximately $0.6\times W_0$, near the center of the bunch.

Bmad simulates the effect of CSR using the one-dimensional model described in \Ref{PhysRevSTAB.12.040703}. The formalism accounts for arbitrary geometries, and also includes the effect of the beam chamber via an image charge method. The code has recently been modified to include well off-axis orbits, and uses the actual orbit history to compute the CSR force.

The simulation results on CSR have been presented at IPAC 2017 \Ref{IPAC2017:THPAB076}. 
\begin{figure}[tb]
\centering
\includegraphics[width=0.6\textwidth]{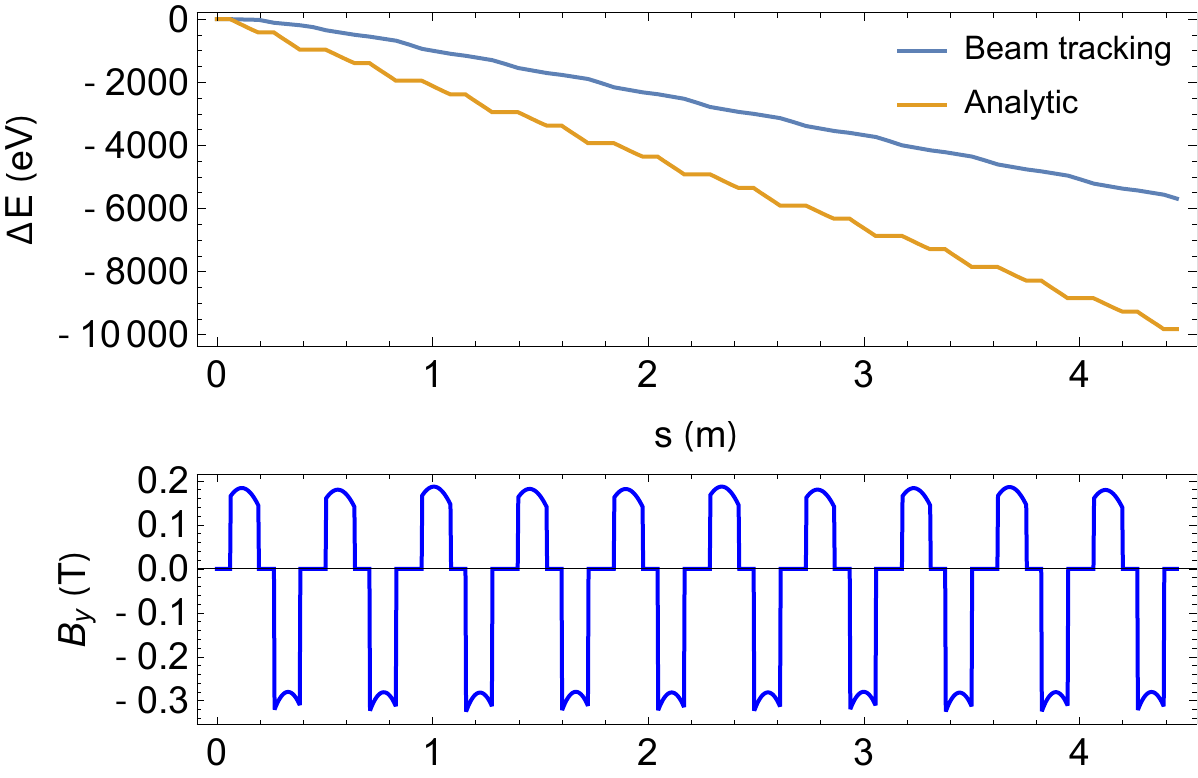}
\caption[]{CSR average energy losses through 10 FFAG cells for a bunch at energy 42~MeV, 77~pC charge, and 3~ps duration by tracking in Bmad. The analytic calculation is simply accumulating $0.35\times W_0$ with curvature based on the local magnetic field $B_y$, and represents a lower limit on the average loss.}
\label{fig:csr_loss}
\end{figure}

\Figure{fig:csr_loss} shows tracking results with CSR in Bmad. The difference in the curves implies that the bunch is in a partially steady state regime due to the finite lengths of the magnets and accounting for CSR propagation between magnets. The slope implies a loss of about 600~eV per cell. For $280^\circ$ of FFAG arc in CBETA, this implies an average relative energy loss of $0.8\times 10^{-3}$, and a maximum induced energy spread of about twice that. Beam tracking through the real CBETA FFAG section results in an average relative energy loss of $1.0\times 10^{-3}$

Detailed CSR studies are quite involved and require extensive testing. These will be performed over the coming months.

\section{Space Charge\Leader{Colwyn}}

\begin{figure}[htbp]
\centering
\includegraphics[width=0.55\textwidth]{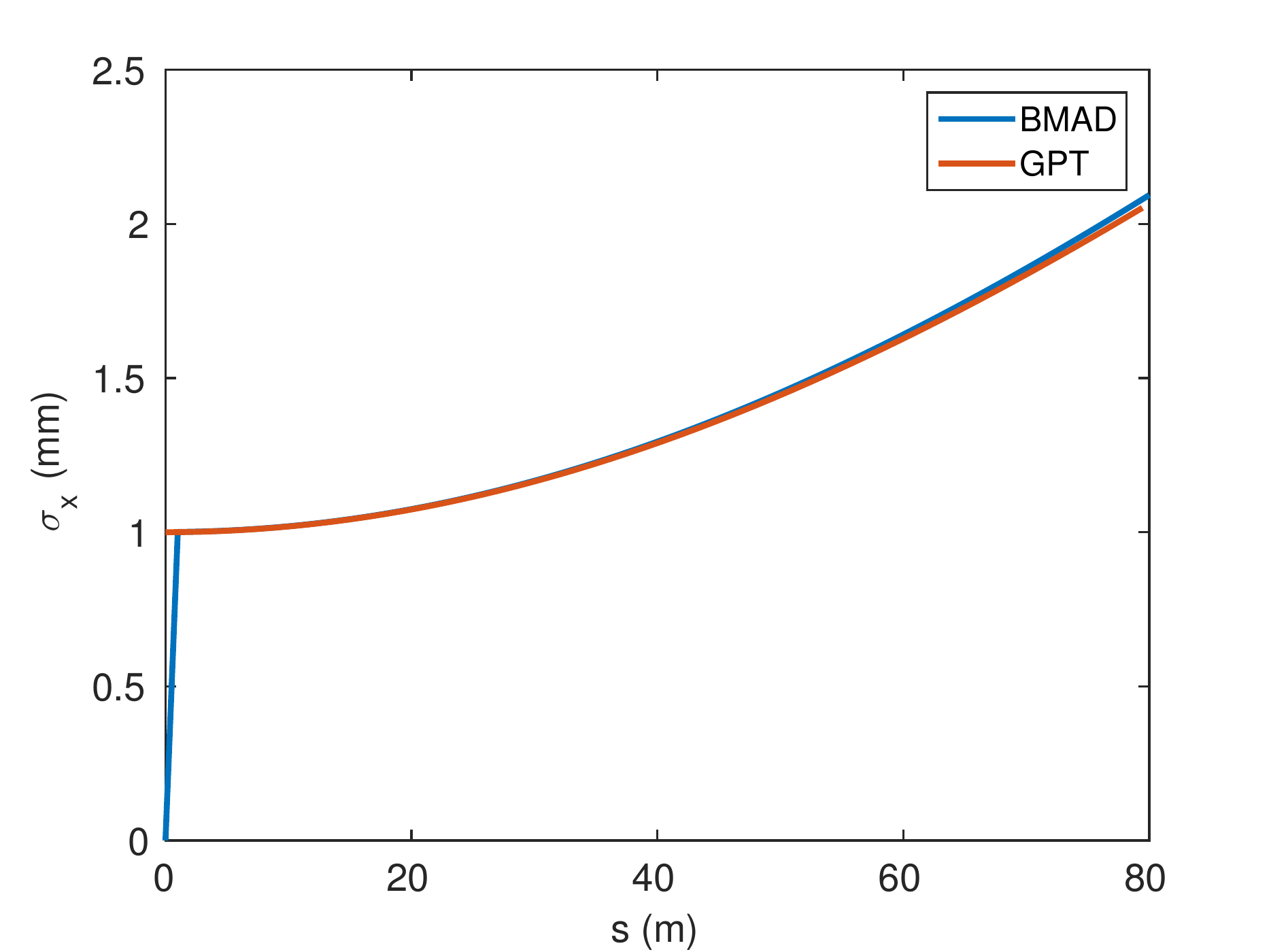}
\caption[]{Comparison of Bmad and GPT space charge models for a drifting 42 MeV Gaussian beam.}
\label{fig:compstdx}
\end{figure}

\begin{figure}[htbp]
\centering
\includegraphics[width=0.55\textwidth]{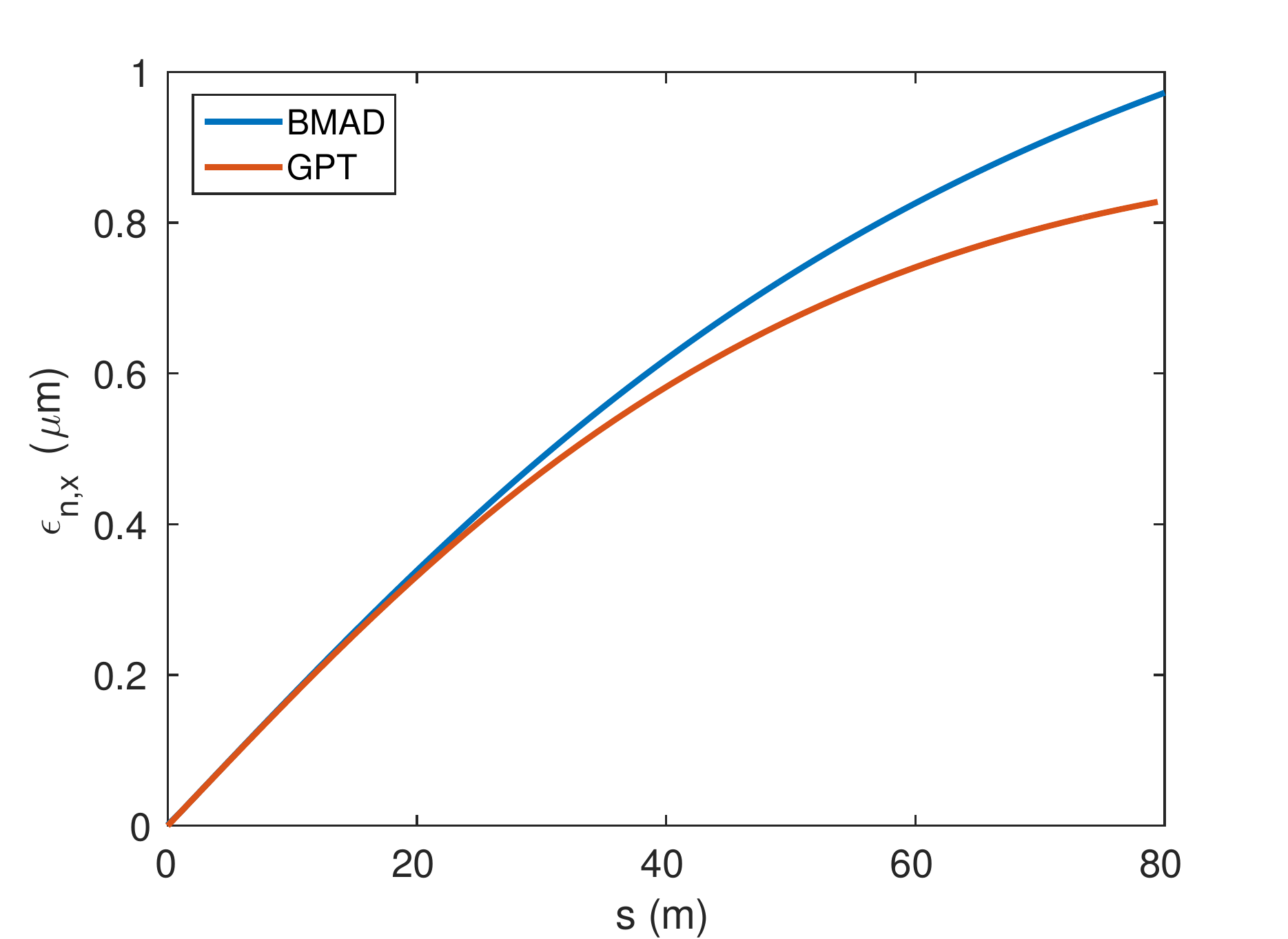}
\caption[]{Comparison of Bmad and GPT space charge models for a drifting 42 MeV Gaussian beam.}
\label{fig:compenx}
\end{figure}

\begin{figure}[htbp]
\centering
\includegraphics[width=0.55\textwidth]{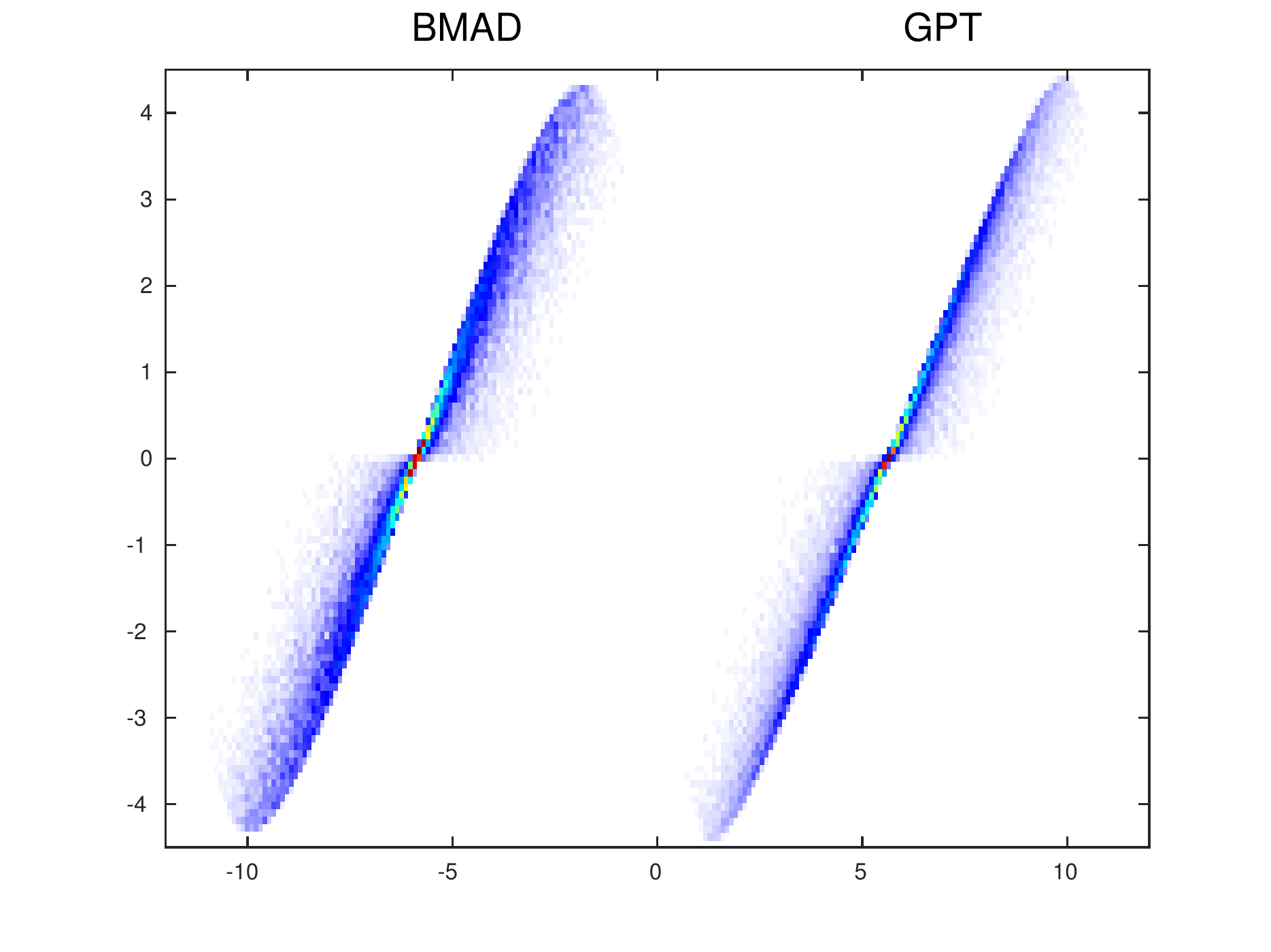}
\caption[]{Comparison of Bmad and GPT space charge models for a drifting 42 MeV Gaussian beam.}
\label{fig:compxpx}
\end{figure}

\begin{figure}[htbp]
\centering
\includegraphics[width=0.85\textwidth]{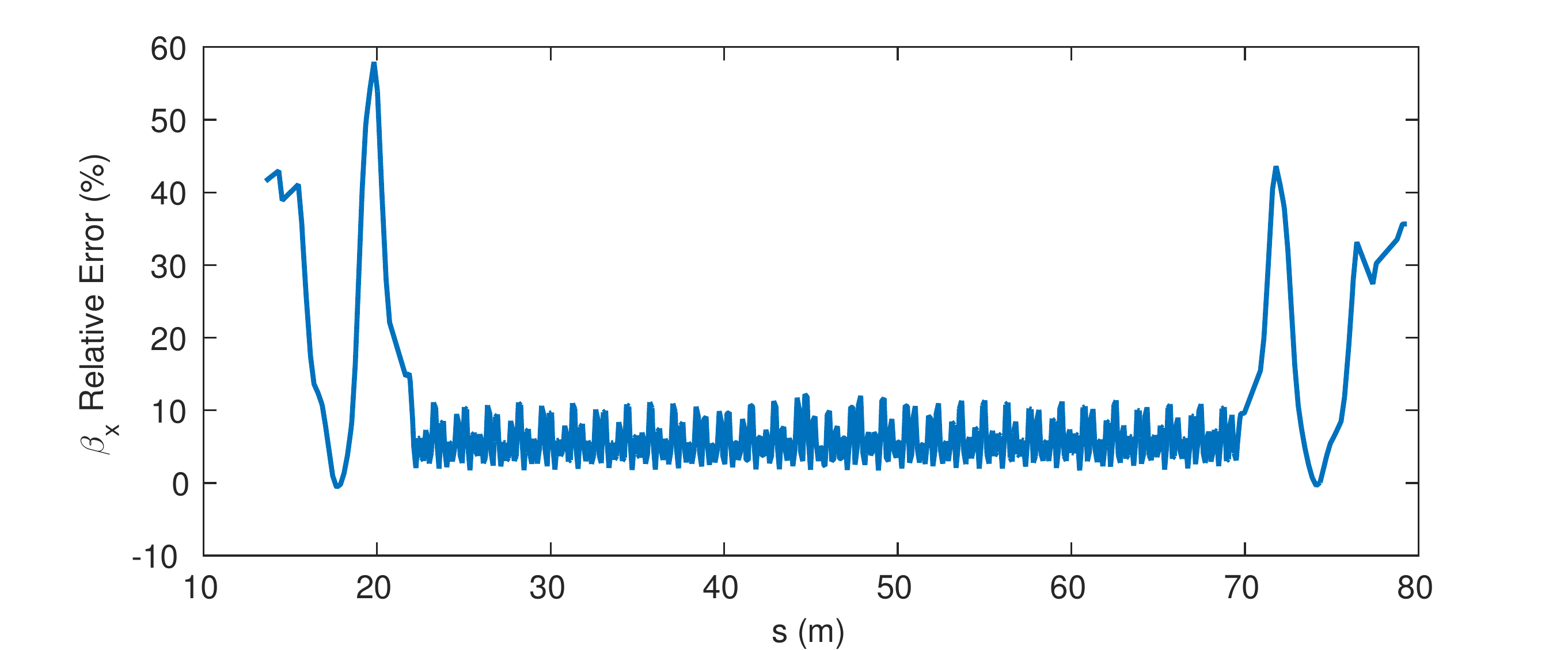}
\caption[]{Relative error in the horizontal beta function computed with and without the Bmad space charge model.}
\label{fig:compSC}
\end{figure}

The relatively low energy of the first pass through CBETA (42~MeV) requires estimation of the effects of both transverse and longitudinal space charge.  To compute space charge fields Bmad uses an approximate relativistic for the fields from longitudinal slices of the beam which is Gaussian transversely (see \Ref{PhysRevSTAB.12.040703} for details).  To test the validity of this model, simulations of a zero emittance Gaussian beam drifting for 80 meters at 42 MeV was simulated in Bmad as well as GPT, a standard 3D space charge code.  The bunch charge/length for this comparison was 100 pC and 4 ps, respectively. \Figure{fig:compstdx} and \Fig{fig:compenx} show the horizontal rms beam size and normalized emittance, respectively. As the comparison shows, at this energy the beam size growth is well modeled in Bmad, however there is some discrepancy with the emittance growth.  This can be seen in the final transverse phase space, shown in \Fig{fig:compxpx}.
Note that for this examples the Bmad model overestimates the emittance growth.  The effects of longitudinal space charge at this energy, bunch charge, and bunch length were negligible.  For a final comparison, the example bunch shown in the linac optics section was sent through one pass of the machine with space charge on and off.  \Figure{fig:compSC} shows the relative error in the horizontal beta function through one pass computed with space charge on and off.  The relative error through the FFAG section is roughly 10\%.
\begin{figure}[htbp]
\centering
\includegraphics[width=0.85\textwidth]{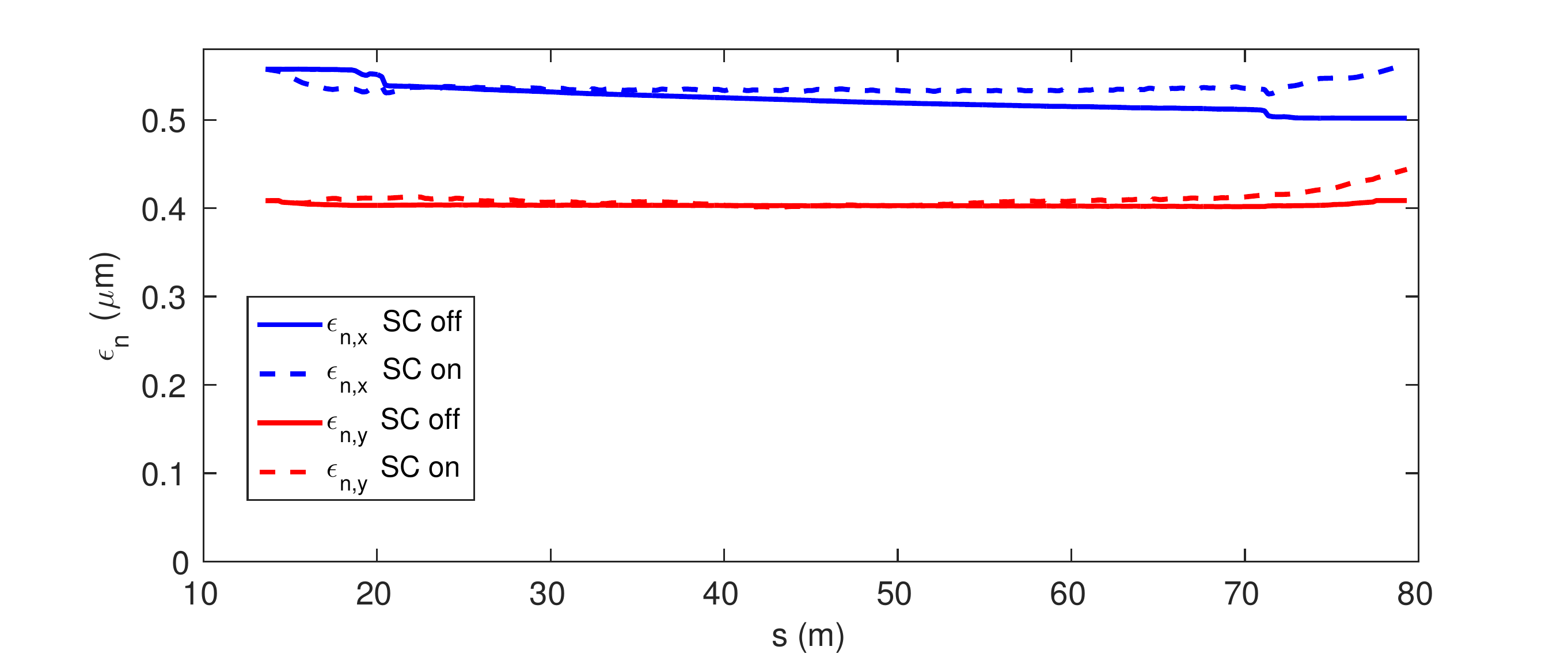}
\caption[]{Comparison of horizontal emittances with (dashed lines) and without (solid lines) the Bmad space charge model.}
\label{fig:compSCem}
\end{figure}
\Figure{fig:compSCem} shows the corresponding resulting emittances through the same pass.  From both these plots, it appears that space charge is not a major effect at this bunch charge (100 pC) and energy (42 MeV).  Initial simulations with CSR show that CSR will have a larger effect on the dynamics than space charge.

\section{Wakefields, \Leader{Blaskiewicz} }
Assuming the beam remains stable, the primary difficulty due to wakefields will be increased energy spread. 
The resistive wall and surface roughness contributions are usually dominant but individual devices will need 
to be looked at to make sure there are no problems. For resistive wall we used the low frequency
approximation for the longitudinal wake potential
\begin{equation}
W(s) = {d\over\strut\displaystyle ds} H(s) {cL\over\strut\displaystyle 2\pi b} \sqrt{Z_0 \rho_e \over\strut\displaystyle \pi s},
\label{eqmmb0}
\end{equation}
where $Z_0 = 377\Omega$, $s\ge 0$ is the lag distance, $c$ is the speed of light, $L$ is the length of the resistive section,
$b$ is the pipe radius, and $\rho_e$ is the electrical resistivity. When applying equation (\ref{eqmmb0}) and in formulas below
we use integration by parts to obtain actual voltages. The numerics are very straightforward and will not be discussed.

For the wake potential due to surface roughness we  used Stupakov's formula \Ref{stupakov2000}.
Define 
$$W(s) = {d\over\strut\displaystyle ds} H(s) Re(\Phi(s)).$$ 
 In MKS units
\begin{equation}
\Phi(s) = \int\limits_0^\infty dk_z \int\limits_{-\infty}^\infty dk_x
|k_z|^{3/2} { \langle|\hat{s}(k_x,k_z)|^2\rangle\over\strut\displaystyle \epsilon_0 b^2 \sqrt{\pi}} { 1-i \over \sqrt{s} }\exp \left( i {k_x^2 + k_z^2 \over\strut\displaystyle 2 k_z} \right)
\label{eqmmb1}
\end{equation}
where the angular brackets $\langle\rangle$ denote statistical averages and 
\begin{equation}
\hat{s}(k_x,k_z) = \int\limits_0^L dz \int\limits_0^{2\pi b} dx { h(z,x)\over\strut\displaystyle (2\pi)^2} \exp( i k_z z + i k_x x ),
\label{eqmmb2}
\end{equation}
with surface roughness $h(x,z)$ where $z$ is measured along the beam direction. For $h = h_0 \cos k z$ one has
$$ \langle|\hat{s}(k_x,k_z)|^2\rangle = { h_0^2 L b \over\strut\displaystyle 8\pi}\delta(k_x)\delta(k_z-k).$$
and
\begin{equation}
W_0(s) = {d\over\strut\displaystyle ds} H(s) L h_0^2 k^{3/2} {\cos(ks/2) + \sin(ks/2) \over\strut\displaystyle 8 \epsilon_0 \pi^{3/2} b}.
\label{eqmmb3}
\end{equation}

\begin{figure}[htbp]
\centering
\includegraphics[width=0.6\textwidth]{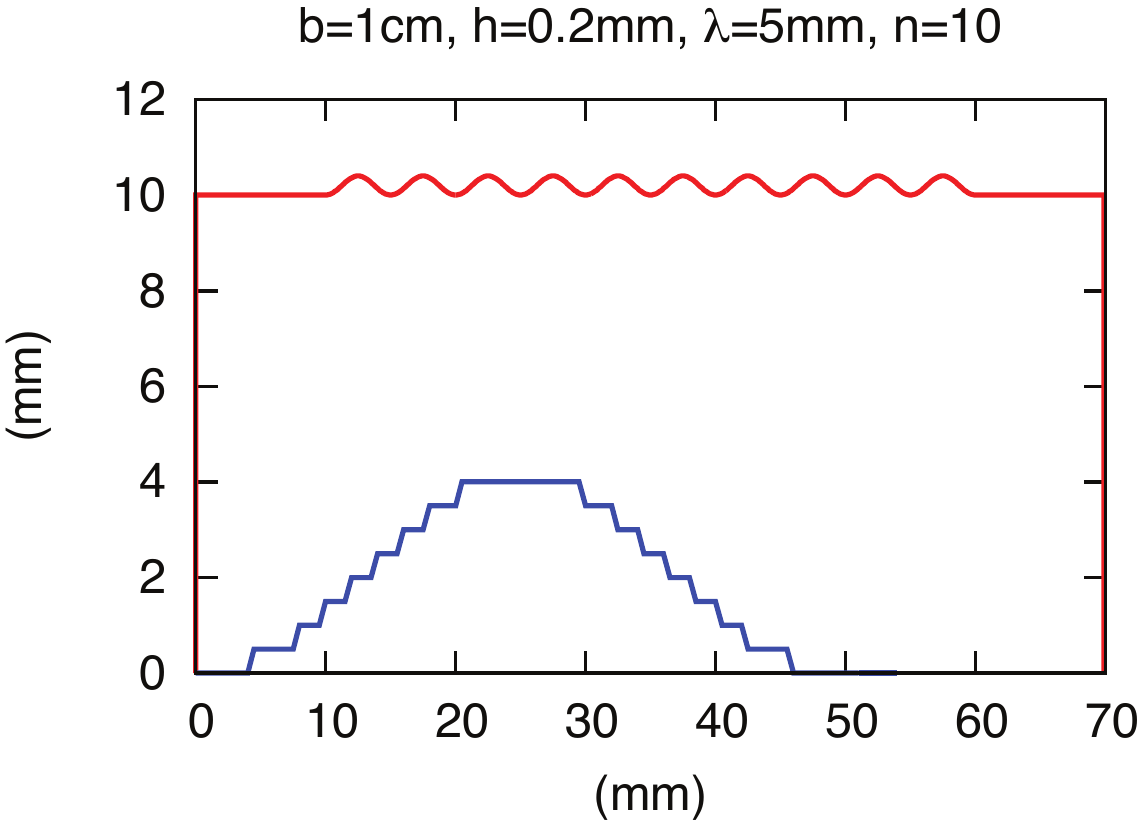}
\caption[]{geometry to compare Stupakov's formula with ABCI. The red curve is the ideal geometry.
 The blue staircase is a 10 fold zoom of an actual convolution used in the simulation.}
\label{fig:plot_roughness_wakefield_geometry}
\end{figure}

\begin{figure}[htbp]
\centering
\includegraphics[width=0.6\textwidth]{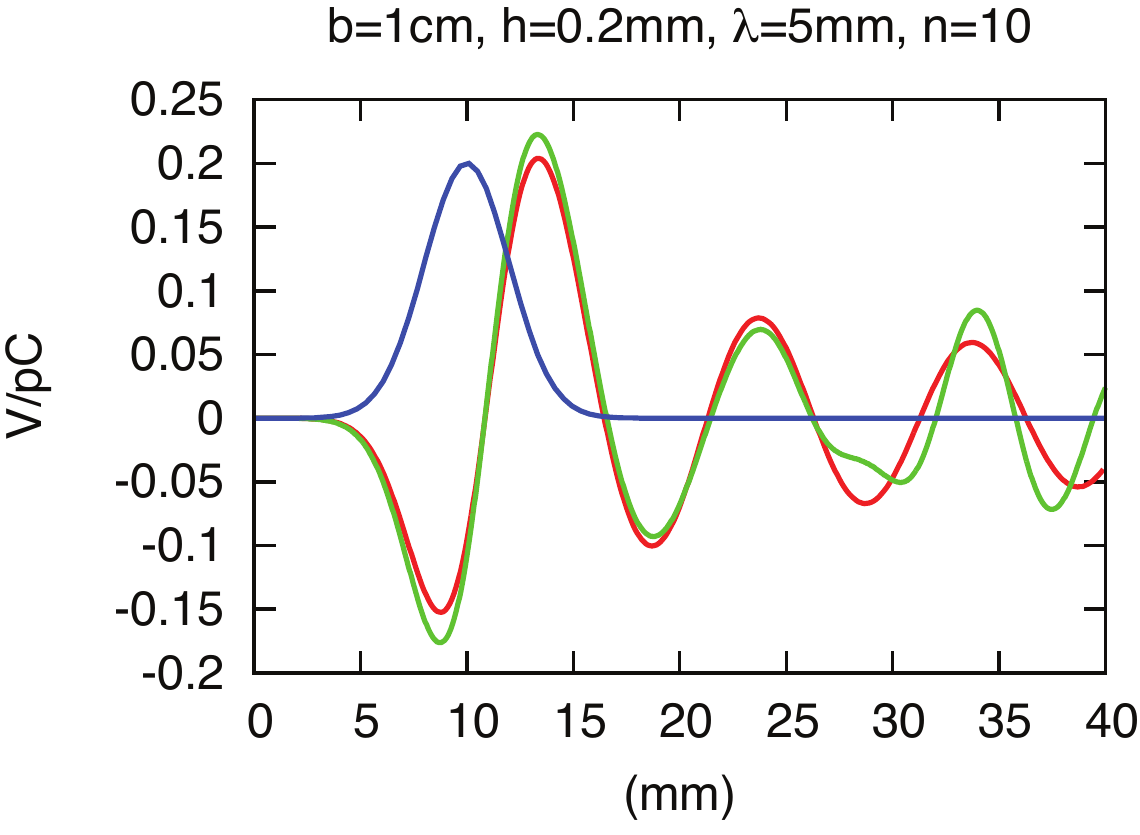}
\caption[]{wakefields from Stupakov's formula and ABCI.}
\label{fig:plot_roughness_wake}
\end{figure}
\Figure{fig:plot_roughness_wakefield_geometry} and \Fig{fig:plot_roughness_wake} show the input and results of an ABCI \Ref{ABCI} simulation and equation (\ref{eqmmb3}). 
For these parameters the agreement is excellent. Other parameters have been checked and the amplitude
of the wake is always good within a factor of 2. 

To get the impedance due to wall roughness requires a statistical model. For simplicity we take a stationary random process
and a correlation function given by
\begin{eqnarray}
\langle h(x_1,z_1)h(x_2,z_2)\rangle&& = C(x_1-x_2,z_1-z_2)\\
&&=h_0^2 \exp\left( - {(x_1-x_2)^2 \over \strut\displaystyle 2\sigma_x^2  } - {(z_1-z_2)^2 \over \strut\displaystyle 2\sigma_z^2  } \right),
\label{eq1mmb}
\end{eqnarray}
where $h_0$ is the rms distortion, $\sigma_z$ is the correlation length along the axis of the pipe and $\sigma_x$ is along the circumference.
Using the Wiener-Khinchin theorem 
\begin{eqnarray}
 \langle|\hat{s}(k_x,k_z)|^2\rangle&& = {b L\over\strut\displaystyle (2\pi)^3} \int\limits_{-\infty}^\infty dxdz\ C(x,z)\exp(ik_z z + i k_x x)\\
&&                        = {b L h_0^2 \sigma_x\sigma_z \over \strut\displaystyle (2\pi)^2} \exp(-k_x^2\sigma_x^2/2 - k_z^2\sigma_z^2/2).
\label{eqmmb4}
\end{eqnarray}
Inserting (\ref{eqmmb4}) in (\ref{eqmmb1}) and doing the $k_x$ integration yields.
\begin{equation}
\Phi(s) = {\sqrt{2}(1-i) h_0^2 \sigma_x\sigma_z \over\strut\displaystyle (2\pi)^2 \epsilon_0 b \sqrt{s}} \int\limits_0^\infty dk_z 
{k_z^2 \exp(-k_z^2\sigma_z^2/2 + i k_zs/2) \over\strut\displaystyle \sqrt{\sigma_x^2 k_x -is}}.
\label{eqmmb5}
\end{equation}
In (\ref{eqmmb5}) the square root has a positive real part and a negative imaginary part for for $s>0$. The integral is done numerically.

\begin{figure}[htbp]
\centering
\includegraphics[width=0.6\textwidth]{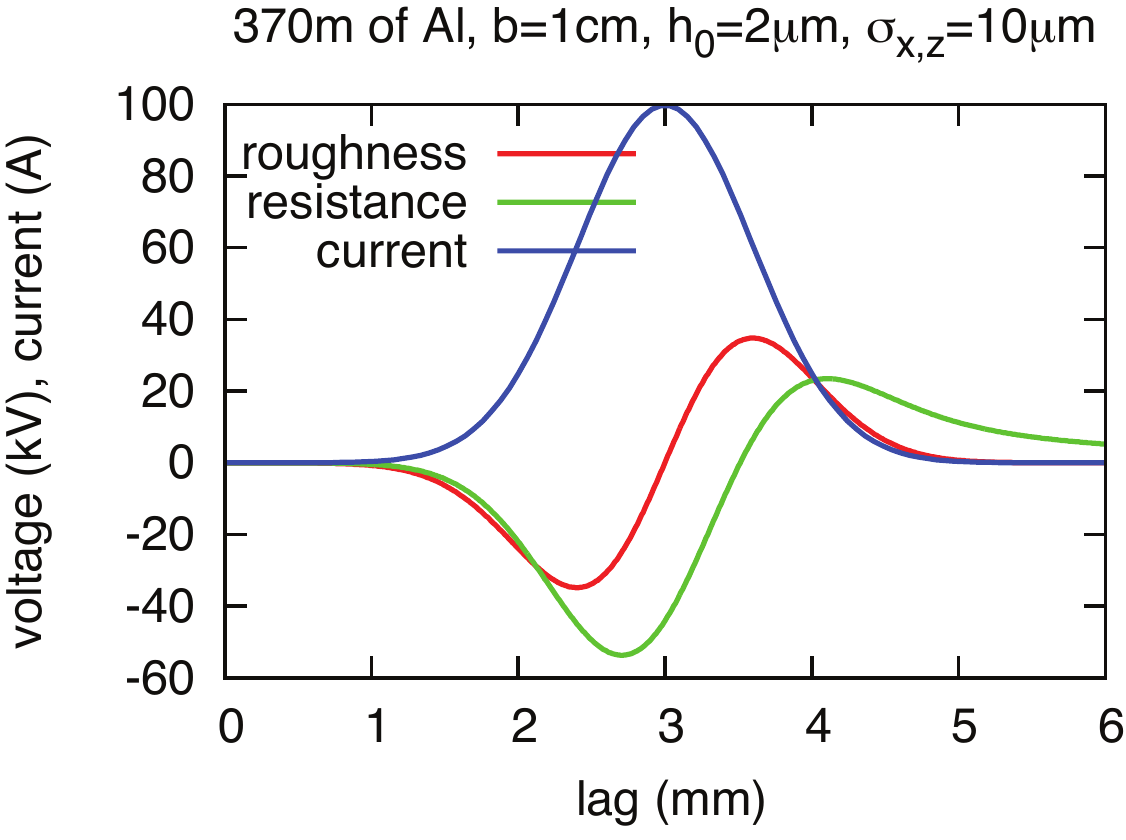}
\caption[]{bunch current and induced voltage}
\label{fig:plot_wakefield_current}
\end{figure}
\Figure{fig:plot_wakefield_current} shows the net voltage for a pipe of radius 1~cm and length equal to 4~passes up and 4~passes down. 
The roughness of $h_0 = 2\unit{\mu m}$ with $10 \unit{\mu m}$ correlation length will require some care but is well within
the state of the art. The extraction energy is 6~MeV so the $\pm 2\%$ energy spread should be easy to accommodate. 
 The chamber profile is a flat oval chamber of 24~mm full height. Generalizing equation (\ref{eqmmb1}) to general apertures requires knowledge of all transverse electric and transverse magnetic microwave modes as well as a tractable approximation for their cutoff frequencies at high energy, a formidable task. On the other hand
we note that the surface roughness acts much like a surface impedance. 
Figure 8 in \Ref{yokoya1992} shows the low frequency, longitudinal resistive wall wake for elliptical pipes.
For all values the impedance is within 10\% of the wake for a round pipe with the smaller aperture. Because a flat chamber can be taken as the limit
of one semi-principle axis going to infinity, the flat oval chamber should be very close to the round chamber results.

These estimates therefore indicate that the energy spread from resistive wakes and from roughness wakes is acceptably small. It does not prohibit the clean transport of the decelerated beam to the beam stop.


\section{Beam Loss due to Gas Scattering \Leader{Gang Wang}}

Electrons in the beam can interact with residue gas molecules left in the vacuum chamber, leading to beam losses and formation of the beam halo. In addition, the lost high energy electrons may further induce desorption of the vacuum chamber and quenches the superconducting components. 

\begin{figure}[htbp]
\centering
\includegraphics[width=0.49\textwidth]{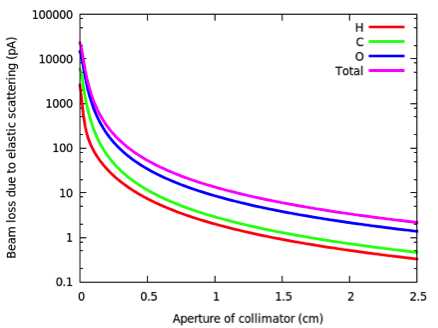}
\includegraphics[width=0.49\textwidth]{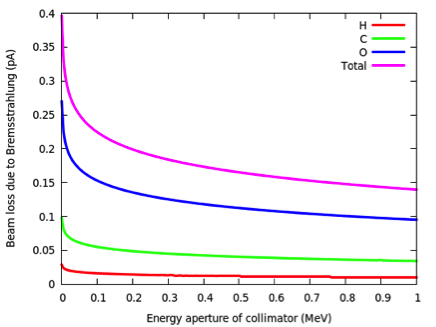}
\includegraphics[width=0.49\textwidth]{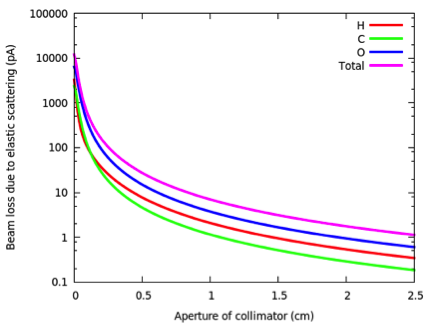}
\includegraphics[width=0.49\textwidth]{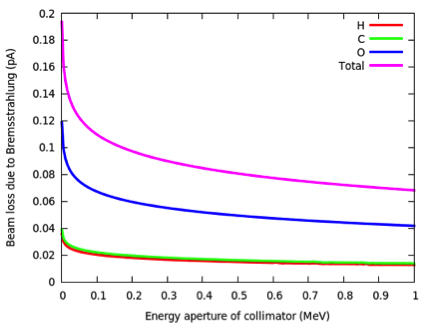}
\caption[]{Analytical estimation of electron beam losses due to scattering off residue gases in CBETA. Electrons pass through the FFAG arc for 7 times and the linac for 8 times. (Left) beam losses due to elastic scattering as a function of the aperture of a collimator located at the last pass of the linac; (Right) beam losses due to the energy aperture of a collimator located at the last pass of linac. Parameters from \Tab{tab:gasscattering} are used. The top graphs were computed for gas species of the initial operation. For the bottom two graphs, gas species of the later, stable operation are used.}
\label{fig:gasScattering1}
\end{figure}

Beam losses due to two types of beam-gas scattering have been analytically estimated for the CBETA ring: elastic scattering and Bremsstrahlung. The elastic scattering of the electrons in the beam off the residue gas molecules can change the trajectory of the electrons and excite betatron oscillations. If the scattering angle is larger than the deflection angle aperture set by the collimator, the electrons will get lost at the location of the collimator \Ref{Tenenbaum01_01, Temnykh08_03}. In the process of Bremsstrahlung, an electron in the beam scatters off the gas nucleus and emits a photon, which results in an abrupt energy change of the electron. If the energy change is beyond the energy deviation aperture, the electron will also be lost \Ref{Tenenbaum01_01}. Using the parameters listed in \Tab{tab:gasscattering}, the beam losses due to gas scattering in the CBETA ring are analytically estimated and shown in \Fig{fig:gasScattering1} for the both the initial and the stable operation modes. Assuming that the limiting transverse aperture locates at the last linac pass, the analytical estimate shows that in the initial operation stage, the beam loss due to elastic scattering ranges from 2.16 pA (2.5~cm aperture) to 13.4~pA (1~cm aperture) and the beam loss due to Bremsstrahlung ranges from 0.22~pA (0.1~MeV energy aperture) to 0.14~pA (1~MeV energy aperture). At the stable operation stage, the beam loss due to both processes reduces by a factor of 2.

More accurate estimates can be achieved through element-by-element simulation with the detailed lattice design and environment parameters. 

\begin{table}[htbp]
\centering
\caption[]{Parameters used in the estimates of beam losses due to beam-gas scattering.}
\begin{tabular}{|c|c|c|}
\hline
&Arcs&Linac\\
Electron bunch charge & 123 pC       & \\
Repetition frequency  & 325 MHz        & \\
Number of FFAG passes  & 7              & \\
Energy gain per pass  & 36 MeV        & \\
Avg. beta function    & 0.5 m          & 50 m \\
Temperature           & 300 K          & 2 K \\
Gas Pressure          & 1 nTorr        & 10-3 nTorr \\
Length                & 54.34m         & 10m \\
Initial operations    & H2 (50\%), CO (30\%), H2O (20\%) & \\
Stable operations	    & H2 (78\%), CO (12\%), H2O (10\%) & \\
\hline
\end{tabular}
\label{tab:gasscattering}
\end{table}

\clearpage
\section{Orbit \& Optics correction\Leader{Chris}}\label{sec:orbit_and_optics_correction}

This machine has the unique requirement that beams with four different energies must propagate through the same FFAG section, and any correction applied will affect all beams simultaneously. At first glance it may seem impossible to correct all beams perfectly at every BPM, and this is true. Fortunately this correction only needs to be approximate at every BPM, with some locations more important than others (e.g. the ends of the FFAG section, and the straight section), and correction with this understanding is possible.  

Practically this correction is achieved by using the response matrix from all correctors to all BPMs, calculating its pseudoinverse by singular value decomposition (SVD), and applying this to measured offsets from ideal BPM readings. As long as the computer model of the machine is not wildly different from the actual machine, this response matrix can be calculated from the computer model and not the `true' corrector-to-BPM response in the live machine.

\begin{figure}[tb]
\centering
\subfloat{\includegraphics[width=\linewidth]{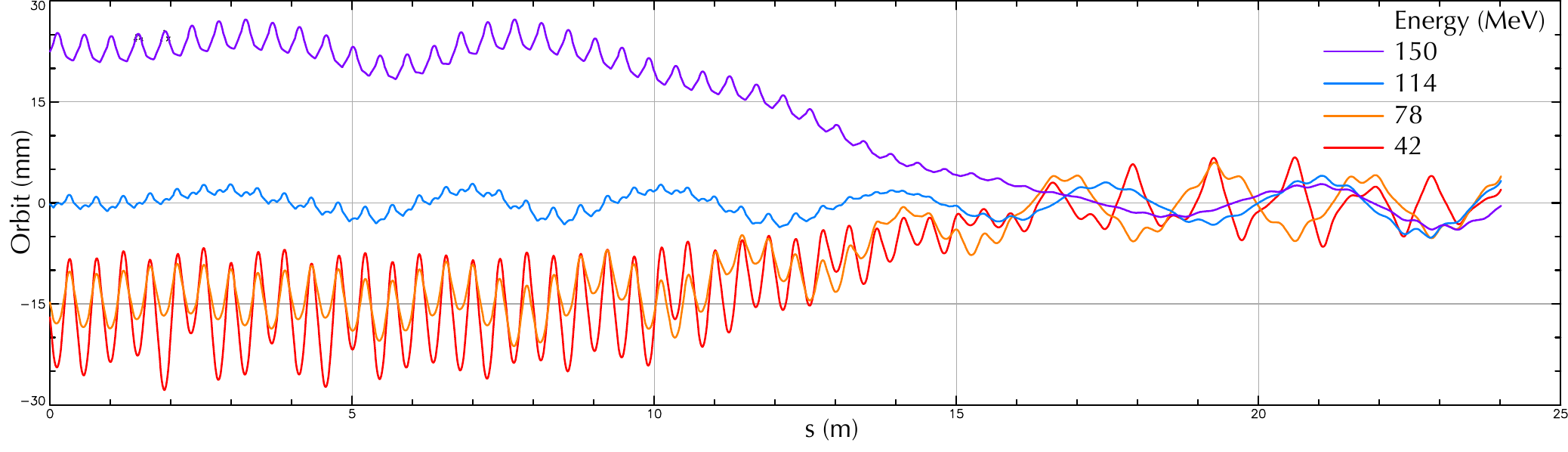}}
\hspace{0.05\textwidth}
\subfloat{\includegraphics[width=\linewidth]{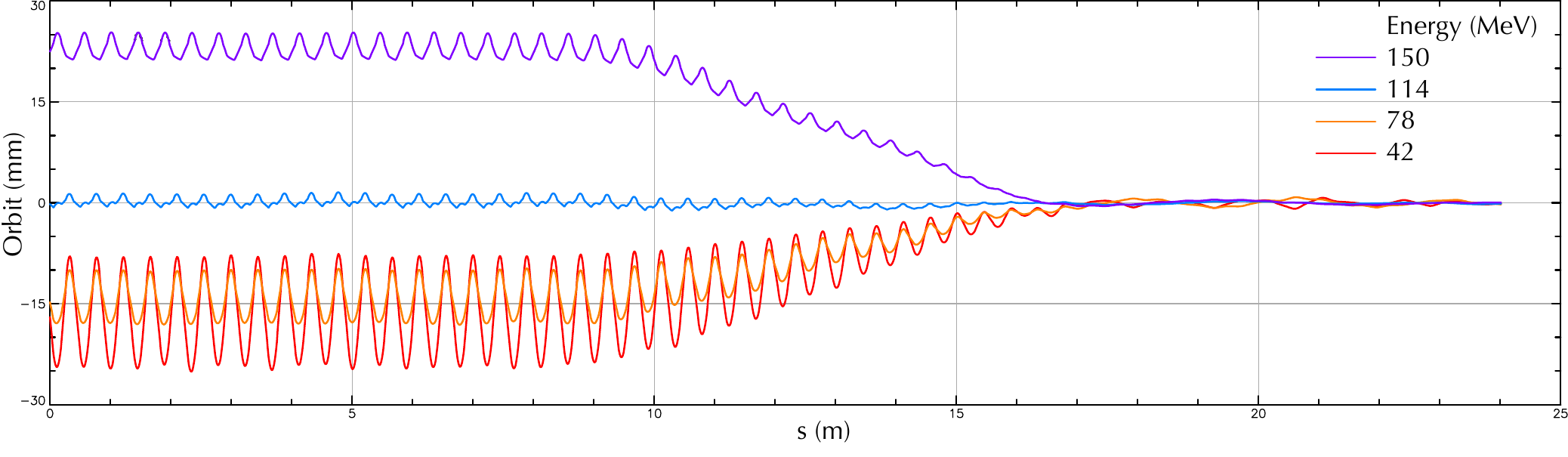}}
\caption[]{Example of simultaneous orbit correction for four beams with uniform horizontal and vertical offset errors in all FFAG magnets.
Offsets are uniformly distributed within $\pm 200\unit{\mu m}$. Every F quadrupole has a horizontal corrector, and every D quadrupole has a vertical corrector. BPMs are placed in the short drift between these magnets in each cell.}
\label{fig:ffag_errors_200um_offsets}
\end{figure}
\Figure{fig:ffag_errors_200um_offsets} shows how this correction works with offset errors in all FFAG magnets. It assumes that all beams enter the FFAG perfectly, and that the BPMs can read each beam position independently. Even though the computer could calculate an exact corrector-to-BPM response matrix in the perturbed system, we use the method described above where the design optics are used to calculate this matrix once and for all, in order to simulate how the actual machine will be operated.

\begin{table}[tb]
\caption[]{Orbit correction analysis procedure. Typically this procedure is iterated for $N=100$ times.}
\begin{tabular*}{\columnwidth}{@{\extracolsep{\fill}}rl}
\toprule
Step	&	Procedure \\
\midrule
  1	&	Initialize design lattice \\
  2 &	Calculate orbit and dispersion response matrices\\
  3 &	Perturb the lattice with random set of errors \\
  4 &	Apply the SVD orbit correction algorithm\\
  5 &	Save this perturbed lattice\\
  6 &	Track particles through, and save statistics\\
  7 &	Reset the lattice\\
  8 	& 	Repeat steps 3-7 $N$ times\\
\bottomrule
\end{tabular*}
\label{tab:orbit_correction_analysis_procedure}
\end{table}

Orbit correction studies are important to estimate what errors can be corrected, and what corrector strengths are required for this correction. In order to get meaningful statistical information, simulations must be preformed repeatedly. \Table{tab:orbit_correction_analysis_procedure} outlines the procedures for such studies. 

\begin{figure}[htbp]
\centering
\subfloat{\includegraphics[width=0.45\textwidth]{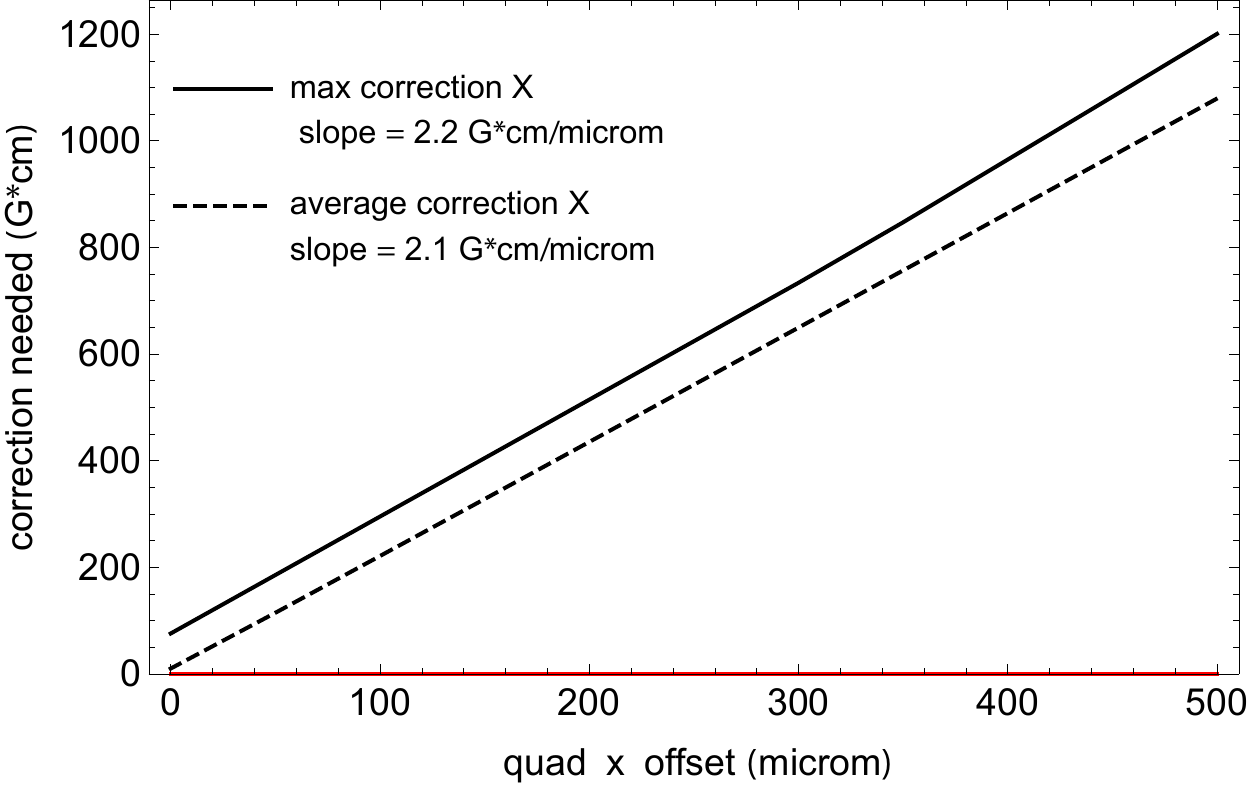}}
\hspace{0.05\textwidth}
\subfloat{\includegraphics[width=0.45\textwidth]{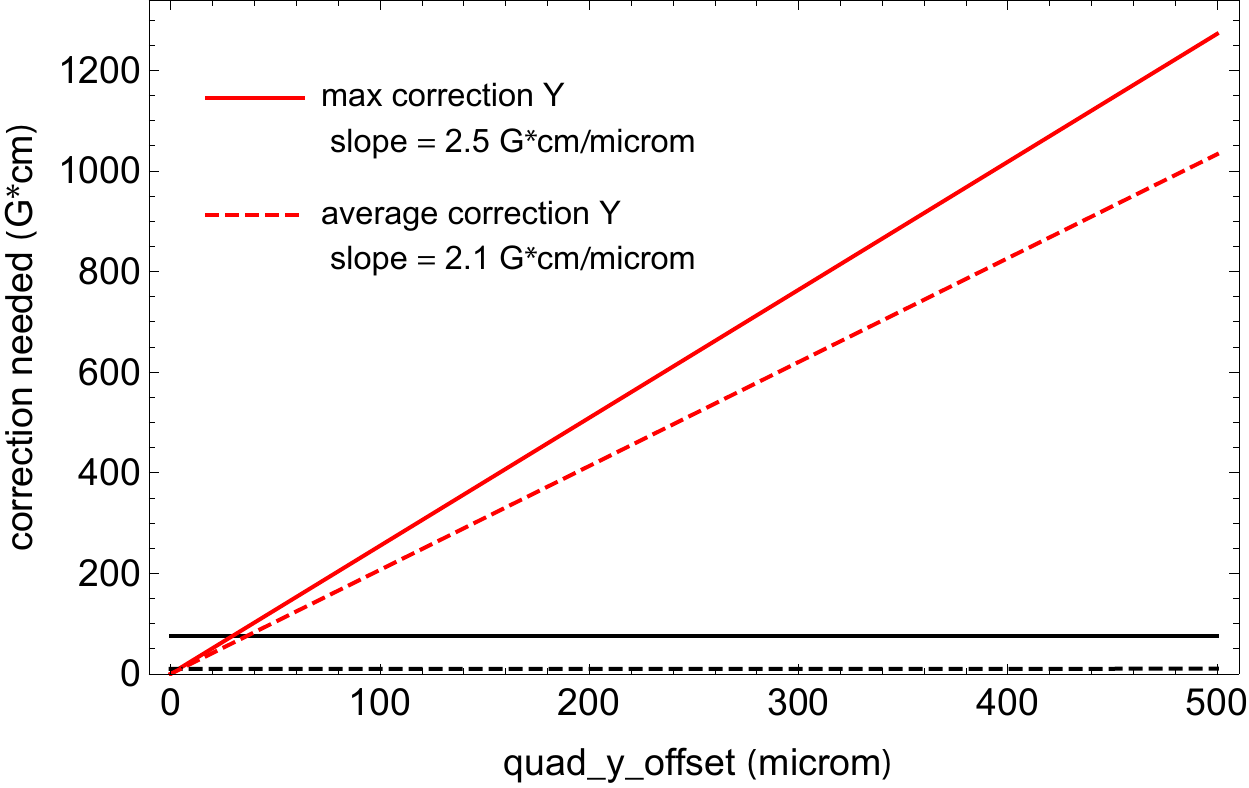}}
\caption[]{Summary of orbit correction of horizontal and vertical FFAG magnet offset errors.  The layout, corrector layout, and definition of the errors are the same as in \Fig{fig:ffag_errors_200um_offsets}. For example, if one expects horizontal offset errors of all quadrupole magnets distributed uniformly in the range $\pm 200\unit{\mu m}$, then the figure implies that the average horizontal corrector strength needed will be approximately 420~G-cm, and anticipates that the maximum strength will need to be approximately 500~G-cm. }
\label{fig:cor_scaling_ffag_quad_offset}
\end{figure}

\Figure{fig:cor_scaling_ffag_quad_offset} summarizes some results from such a study.

\section{Tolerances\Leader{Dave}}

\subsection{Analysis Process}
ERLs --- as they are intended to generate and precisely control phase/energy correlations within the beam --- are, architecturally, time-of-flight spectrometers. As such, they may require very stringent tolerances on timing, energy, and transport component excitation, location/alignment, and field quality. A preliminary validation of a design must therefore include estimates of magnet alignment and excitation tolerances, response to RF phase and amplitude errors, and validate the impact of magnetic field inhomogeneity. The latter effect can result in significant degradation of beam quality during energy recovery via a coupling of field-error-driven betatron oscillations to energy by way of RF phase errors \Ref{Douglas02_01, Douglas10_01}.
	Estimates of performance response to errors then are used to inform simulation studies of machine behavior in the presence of realistic errors. When system sensitivity to individual errors is thus evaluated (and tolerances at which error response become nonlinear thereby known), an error budget can be developed by imposing multiple error classes in an ensemble of test cases. Potential interactions amongst errors can then be studied, and operationally realistic correction (local) and compensation (global) schemes tested. The impact of residual errors (after correction/compensation) is thus characterized, ``large'' individual error terms can be assessed to determine signatures associated with out-of-tolerance hardware (blunders), and the ability of control algorithms to deal with accumulated ``subliminal'' errors --- those below the resolution of local sets of diagnostics, but large enough to degrade performance --- assessed. 

\subsection{Estimates}

	Standard methods can be used to evaluate the impact of perturbations on accelerator performance \Ref{Douglas99_01, Douglas99_02, Douglas96_01, Powers07_01}. These involve evaluating the linear response of a particular parameter (or parameters) to single perturbations, and superposing `the effect of sequential random errors of the same class. For accelerators, typical perturbations include RF phase and amplitude errors, magnet misalignments, excitation errors, and field inhomogeneity. We now provide estimates for examples of each of these in the context of CBETA.

\subsection{Alignment Sensitivity}\label{sec:alignment_sensitivity}

Misalignment of a quadrupole by dx from its nominal location will result in a deflection $\delta x' \sim \delta x/f$ (where $1/f = B'L/(B\rho)$ is the quadrupole inverse focal length) of a beam entering the quad on its reference orbit, resulting in a betatron oscillation downstream. If independent offsets are encountered at $N$ quadrupoles along a mono-energetic beamline, an average of the mean square betatron displacement over an ensemble of misalignments and betatron phase along the line give the following result for the rms orbit offset at the end of the line \Ref{Douglas99_01, Douglas99_02}.
\begin{equation}
\left<x\right>\sim \sqrt{\frac{N}{2}} \frac{\bar{\beta}}{\bar{f}} \left<\delta x\right>
\end{equation}
Here, $\bar{\beta}$ is the average lattice Twiss envelope in the line, $\bar{f}$ the average quad focal length, and $\left<\delta x\right>$ the rms misalignment. 

If --- as in CBETA --- multiple passes through a single line occur but the passes are separated by notionally arbitrary (or random) phase advance, betatron phase averaging simply replaces $N$ by $2N$. If the effects of the perturbation are to be observed at a different energy, a factor of $\sqrt{p_{\text{perturbation}} / p_{\text{observation}}}$ is applied to account for adiabatic damping or anti-damping. 

For CBETA, $1/f=B'L/(B\rho) \sim (100\unit{kG/m} \times 0.1\unit{m})/(33.3564\unit{kG.m/(GeV/c)} p_{\text{arc}}$. The average $\beta$ is $\sim0.5\unit{m}$, and $N\sim200$ for each pass. For the first arc only, with $p\sim0.04\unit{GeV/c}$, this yields $\left<x\right>\sim 37.5\unit{m}\left<\delta x\right>$. An rms alignment tolerance of $100\unit{\mu m}$ would thus yield $\sim 4\unit{mm}$ rms orbit error at the end of the first turn. This is notionally operationally manageable. 
\begin{equation}
\left<x\right> \sim \sqrt{N} \frac{\bar{\beta}}{f_0}\left[  
  \frac{1}{\sqrt{p_1 p_{\text{dump}}}}   
+ \frac{1}{\sqrt{p_2 p_{\text{dump}}}} 
+ \frac{1}{\sqrt{p_3 p_{\text{dump}}}} 
+ \frac{1}{\sqrt{p_4 p_{\text{dump}}}} \right] \left<\delta x\right> 
\end{equation}

Over multiple turns, the uncorrected orbit will wander significantly further afield, especially when decelerated. Denoting by $1/f_0$ the ``generic'' focusing strength $B'L/(B\rho_0)$ of $100\unit{kG/m}\times 0.1\unit{m}/33.3564\unit{kG.m/(GeV/c)} \sim (0.3\unit{GeV/c)/m}$, a roll-up of the contributions described above gives the following result for the rms orbit excursion at the dump (where $p_n$ is (for $n$=1, 2, 3, and 4) the momentum for each of the four nominal beam energy levels). 
The  $\sqrt{2}$ associated with the fourth energy accounts for the single passage through the FFAG system at the highest energy. For injection/extraction at 10~MeV/c and full energy of 150~MeV, $p_1=45\unit{MeV/c}$, $p_2=80\unit{MeV/c}$, $p_3=115\unit{MeV/c}$, and $p_4=150\unit{MeV/c}$, yielding $\left<x\right> \sim 280\unit{m} \left<\delta x\right> $. A $100\unit{\mu m}$ alignment tolerance thus --- over the full system --- results in a significant potential offset. Commissioning and operational practices must therefore make provision for local --- or at least pass-to-pass --- orbit correction. As the system transports multiple beams in a common structure, orbit optimization will thus likely be iterative and may involve degrading lower energy passes so as to bring higher-energy passes within operating tolerances \Ref{Bodenstein07_01}.

\subsection{Impact of Excitation Errors}
	Excitation errors in beamline components can arise due to fabrication or powering errors, or the variation in magnetic properties of materials used in magnets. An error in gradient will result in deviations from design focusing, with beam envelope and/or lattice function mismatch evolving as a consequence. The scale of various effects along a mono-energetic section of a beamline is set by the number of perturbed elements N, the average lattice functions at perturbations, and the deviation of focusing from nominal. In notation consistent with that used above, the envelope, divergence, phase, and dispersion errors at the end of the beamline segment under consideration are --- in terms of the rms focal length deviation  $\left<\delta(1/f)\right>$, as follows \Ref{Douglas96_01}.

\begin{equation}
\left<\frac{\Delta\beta}{\beta}\right>= \left<\alpha\right> = \sqrt{\frac{N}{2}} \bar{\beta} \left<\delta\left(\frac{1}{f}\right)\right>
\end{equation}

\begin{equation}
\left<\Delta\psi\right> =\frac{1}{4\pi} \sqrt{\frac{3N}{2}} \bar{\beta} \left<\delta\left(\frac{1}{f}\right)\right>
\end{equation}

\begin{equation}
\left<\Delta\eta\right> =\sqrt{\frac{N}{2}} \bar{\beta}\bar{\eta} \left<\delta\left(\frac{1}{f}\right)\right>
\end{equation}

The impact of the lattice and beam parameters to focusing errors may then be evaluated in much the same manner as in the case of misalignments. For the lowest energy (45 MeV) pass in CBETA, $N\sim{200}$, $\bar{\beta}\sim0.5\unit{m}$, $\bar{\eta}\sim0.03\unit{m}$, and $\left<\delta(1/f)\right> \sim T(1/\bar{f})$, with $T$ an error tolerance and $1/\bar{f}$) the average inverse focal length of $\unit{(100 kG/m \times 0.1 m/(33.3564 kG.m/(GeV/c) \times 0.045 GeV/c) \sim 6.7/m}$. The scaling then gives
\begin{align}
\left<\frac{\Delta\beta}{\beta}\right> = \left<\Delta\alpha\right> \sim& 33.5\unit{T} \\
\left<\Delta\psi\right>	 \sim& 4.6\unit{T} \\
\left<\Delta\eta\right>	\sim& 1.0\unit{T} 
\end{align}
For absolute excitation errors at the $1\%$ level (either due to deviations in excitation, errors in energy, or to being offset  --- as the low and high energy beams will be --- in an inhomogeneous region of the quad field), dispersion and envelope errors are at the $33\%$ level, with phase advance errors of a few degrees of betatron phase. The envelope and dispersion effects are significant, and may be expected to accumulate through the machine, with associated deviation of the transport system response to errors --- and the beam envelopes/sizes --- from design. Operational algorithms --- such as those employed in CEBAF \Ref{Lebedev97_01}
 --- may be required to negotiate corrections amongst the various passes of beams through the system so as to avoid lattice error hypersensitivity, beam size blowup, or problems with halo and beam transmission.

\subsection{RF Phase/Amplitude Response}
	As noted, ERLs are time of flight spectrometers; consequently, lattice perturbations that couple to path length may have an effect on time of flight (RF phase), resulting in variations in parameters that influence RF performance. Details of RF power/beam transient effects are described in Reference \Ref{Powers07_01}, which characterizes the RF power required to control cavity fields under various scenarios for pass-to-pass phasing (including the impact of phase transients) and energy recovery. 

Here, we estimate the impact of two path-length-related effects: the impact of alignment errors on path length, and the magnitude of pass-to-pass phase errors resulting from field inhomogeneities. In the first case, misalignment of quadrupoles from design position will --- in addition to the positional errors evaluated in \Section{sec:alignment_sensitivity} --- lead to path length errors via the coupling of path length to angular deflection via $R_{52}$: $\delta l = R_{52} \delta x'$.
If the deflection occurs at a non-dispersed location, the $R_{52}$ from the deflection to the next pass through the linac will be zero (the evolved path length is second order in the deflection); when the deflection is at a dispersed location, $R_{52}=\eta$, yielding the following scaling of path length with misalignment.
In this case, $\delta l = R_{52} \delta x' = \eta \left<\delta x\right>/f$. 
Summing the rms path length over $N$ quadrupoles at dispersed locations with average focal length with average focal length gives the following expression.
\begin{equation}
\delta l = \sqrt{N} \bar{\eta} \left(\frac{1}{f}\right)\left<\delta x\right>
\end{equation}
Inserting first arc values as outlined above (with $N\sim100$ rather than 200, as the backleg --- about half the quads --- is nondispersed) gives $\left<\delta l\right> \sim  10 \times 0.03 \times [100\times0.1/(33.3564\times0.045)] \left<\delta x\right>\sim 2\unit{m}\left<\delta x\right>$.  An alignment tolerance of $100\unit{\mu m} \sim 0.2\unit{mm}$ path length error --- about 0.3 RF degrees. This estimate can be used to guide the design of path length correction algorithms, and to set the scale for turn-to-turn path length discrepancies.

Magnetic field inhomogeneities are a significant challenge for ERL performance. To date, successful microtrons, recirculating linacs, and ERLs have had field homogeneity at the $0.01\%$ level. Variations in field quality couple to path length, thence RF phase, and from there downstream beam behavior. Local discrepancies in field will generate angular deflections of portions of the beam phase spaces (or, in an FFAG, of one beam relative to another) leading to phase --- and eventually energy --- errors \Ref{Douglas02_01, Douglas10_01}. In the first/last pass of CBETA, a relative field error $\Delta B/B$ in a magnet of length $L$ would lead to an angular deflection $\Delta BL/(B\rho)$, which, integrated over the entire beamline ($360^\circ$ of bending) would lead to a path length error as follows:
\begin{equation}
\delta l = R_{52}\delta x' = \frac{\eta \Delta B L}{B \rho} = \eta \left(\frac{\Delta B}{B}\right) \theta
\end{equation}
For $\eta\sim 0.03\unit{m}$, $\theta \sim 2\pi\unit{radians}$ (full arc), this suggests that there will be a sensitivity of order $\delta l\sim0.2 (\Delta B/B)\unit{m}$.
Thus, a $1\%$ field error would result in 2~mm path length error of one pass relative to another --- or an order of $3^\circ$ RF at 1.3~GHz.

An accurate assessment of these effects will require simulation of an ensemble of machines based on the final lattice design while modeling errors at levels similar to those observed in prototype hardware. Results of such simulations can be used to set specifications on RF power requirements and/or define phase/path length stabilization methods and requirements.

\subsection{FFAG Tolerances}\Leader{Chris}

To estimate the FFAG magnet multipole tolerances, we add multipole moments randomly to the FFAG magnets in the simulation model, and apply the correction scheme described in \Section{sec:orbit_and_optics_correction}. 

Magnetic fields due to normal $b_n$ and skew $a_n$ multipole coefficients of order $n$ are added to the model so that the horizontal and vertical fields are:
\begin{equation}
B_x + i B_y = \frac{b_n + i a_n}{L}\left(x + i y \right)^n,
\end{equation}
where $L$ is the length of the element. The fraction of the field $f$ that a normal multipole coefficient adds to a quadrupole magnet with gradient $G$ at a reference radius $r_0$ is 
\begin{equation}
f = b_n \frac{r_0^{n-1}}{G L}.
\end{equation}
For convenience we define a `unit' of multipole at a radius $r_0$ as $u_0 \equiv 10^4 f$, so that the multipole coefficients scale as
\begin{equation}
b_n = \left[10^{-4} \frac{GL}{r_0^{n-1}}\right]u_0.
\end{equation}
Conversion of units at another radius $r_1$ is then $u_1 = (r_1/r_0)^{n-1}u_0$. Here we use a standard radius of 25~mm, which is near the extent of the highest and lowest energy beams. 

We determine limits on the allowable multipoles by adding pure multipoles, order $n$ of type normal or skew, to the complete FFAG part of the lattice with all four beams. These are added randomly to each FFAG magnet, and distributed as a uniform distribution. The magnitude of the multipole, in terms of units is quoted as the maximum of this distribution: 100~u normal sextupole errors means that each FFAG magnet was given a normal sextupole moment between [-100~u, 100~u]. 

For a given magnitude, many (on the order of 100) sets of errors are introduced and then corrected in the simulation, and statistics are gathered about the effect on a tracked beam. The lowest energy beam is typically the most sensitive. We define a limit on this magnitude as when the error on average will add a 10\% increase to either the beam size or emittance. This magnitude is then called the individual limit. 

The results from these simulations are summarized in \Tab{tab:ffag_multipole_tolerances_normal} and \Tab{tab:ffag_multipole_tolerances_skew}. As a byproduct of these simulations, we can determine how strong the dipole correction will need to be at a given level of error. This is the dipole correction scaling. For example, if normal sextupoles are distributed uniformly as [-10u, 10u], then the dipole correctors will need about 40~G-cm of field in order to correct the orbits. 

Finally, once the individual limits are determined, we run error simulations that add all multipoles up to 20-pole with magnitudes as a fraction of their individual limit. A fraction of about 30\% leads to a 10\% increase to either the beam size or emittance, and this becomes our recommended limit. 

\begin{table}[]
\centering
\caption{Normal multipole error limits. Units u are at a 25~mm radius. }
\label{tab:ffag_multipole_tolerances_normal}
\begin{tabular}{lllll}
\toprule
type       & $n$ & dipole correction scaling (G-cm/u) & individual limit (u) & recommended limit (u) \\
\midrule
sextupole  & 2 & 4                                  & 60                   & 18                    \\
octupole   & 3 & 5                                  & 40                   & 12                    \\
decapole   & 4 & 6                                  & 25                   & 7.5                   \\
dodecapole & 5 & 7                                  & 19                   & 5.7                   \\
14-pole    & 6 & 9                                  & 11                   & 3.3                   \\
16-pole    & 7 & 11                                 & 10                   & 3                     \\
18-pole    & 8 & 15                                 & 6                    & 1.8                   \\
20-pole    & 9 & 20                                 & 5                    & 1.5                  \\
\bottomrule
\end{tabular}
\end{table}

\begin{table}[]
\centering
\caption{Skew multipole error limits. Units u are at a 25~mm radius. }
\label{tab:ffag_multipole_tolerances_skew}
\begin{tabular}{lllll}
\toprule
type       & $n$ & dipole correction scaling (G-cm/u) & individual limit (u) & recommended limit (u) \\
\midrule
sextupole  & 2 & 8                                  & 150                  & 45                    \\
octupole   & 3 & 8                                  & 100                  & 30                    \\
decapole   & 4 & 11                                 & 70                   & 21                    \\
dodecapole & 5 & 12                                 & 50                   & 15                    \\
14-pole    & 6 & 17                                 & 35                   & 10.5                  \\
16-pole    & 7 & 19                                 & 26                   & 7.8                   \\
18-pole    & 8 & 26                                 & 20                   & 6                     \\
20-pole    & 9 & 30                                 & 15                   & 4.5                  \\
\bottomrule
\end{tabular}
\end{table}

\subsection{Summary}
	The strong focusing inherent in FFAG transport reduces sensitivity to typical beamline errors and accelerator component imperfections. Estimates of typical errors suggest that appropriately conservative operational algorithms can compensate and/or correct the effect of imperfections expected in CBETA; of particular ongoing interest will be errors that can lead to irreproducible and/or time-transient path length errors, inasmuch as these will constrain the RF drive power required for stable operation.

\clearpage

\section{Start-to-End Simulation\Leader{Chris}}

\begin{figure}
  \begin{center}
\subfloat{\includegraphics[width=0.3\textwidth]{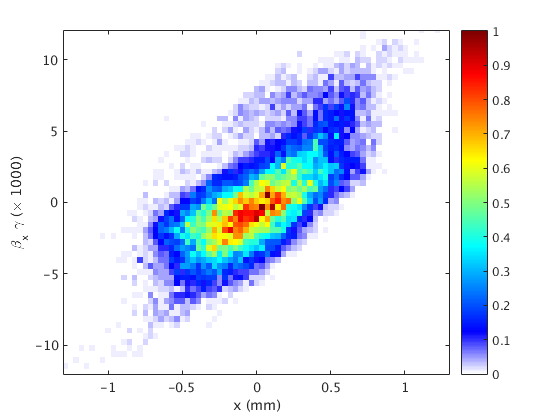}}
\hspace{0.04\textwidth}
\subfloat{\includegraphics[width=0.3\textwidth]{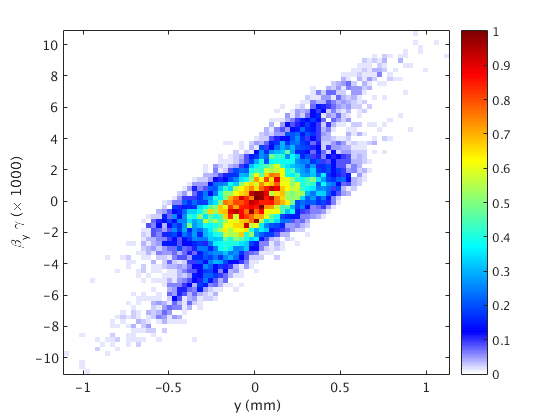}}
\hspace{0.04\textwidth}
\subfloat{\includegraphics[width=0.3\textwidth]{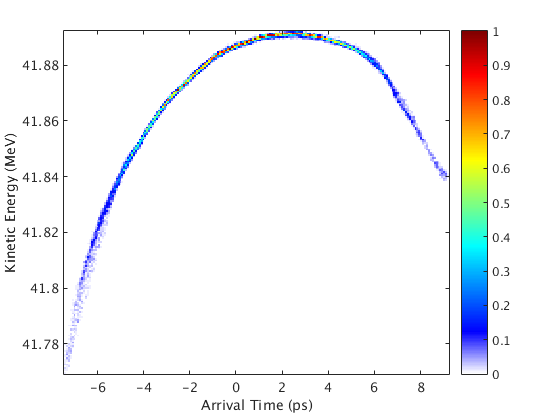}}
  \end{center}
  \caption{Bunch phase space sections at the end of the first pass through LA using GPT with space charge. The bunch charge is 100~pC. This bunch is used for start-to-end tracking using Bmad.}
  \label{fig:LA_end_pass1_phase_space}
\end{figure}

\begin{landscape}
\begin{figure}[htbp]
\centering
\includegraphics[width=\linewidth]{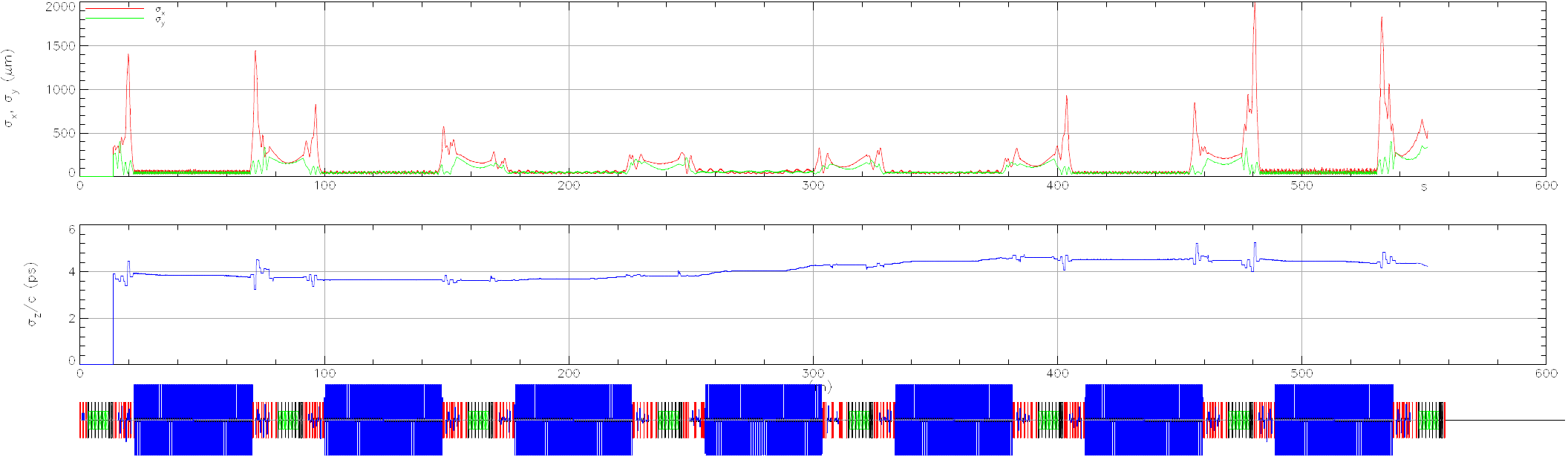}\\
\hspace{10pt}
\includegraphics[width=\linewidth]{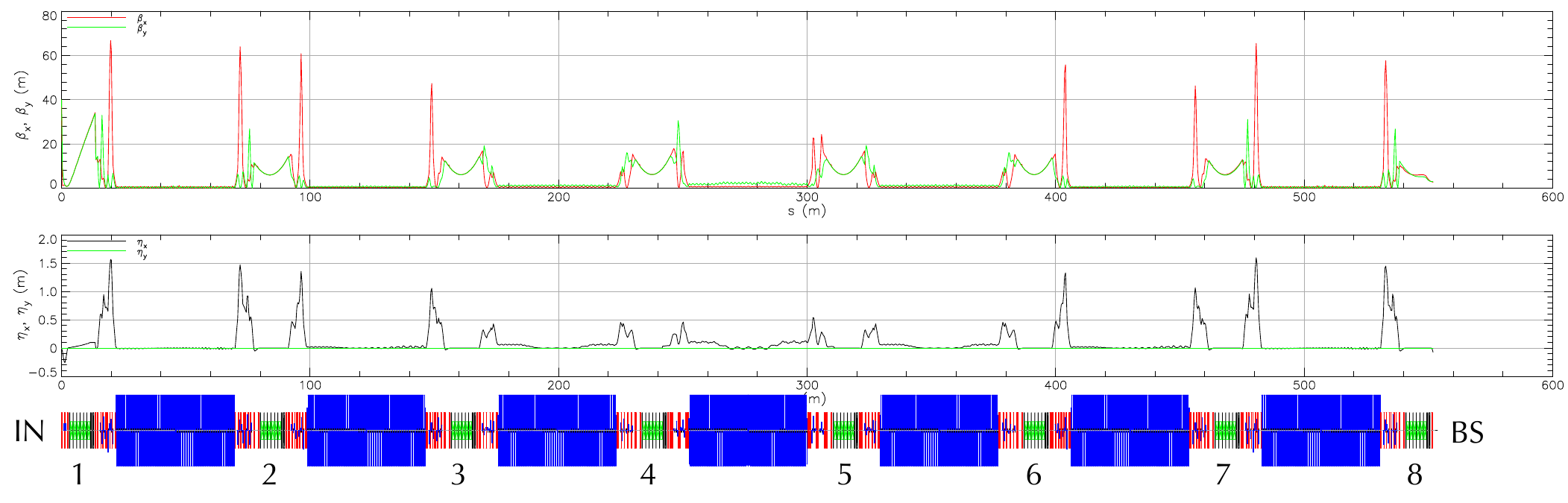}
\caption[]{Optics and beam sizes summary for all passes from injection to dump. Numbers below the linac indicate the pass number through the linac.}
\label{fig:optics_4pass}
\end{figure}
\end{landscape}

\begin{landscape}
\begin{figure}[htbp]
\centering
\includegraphics[width=\linewidth]{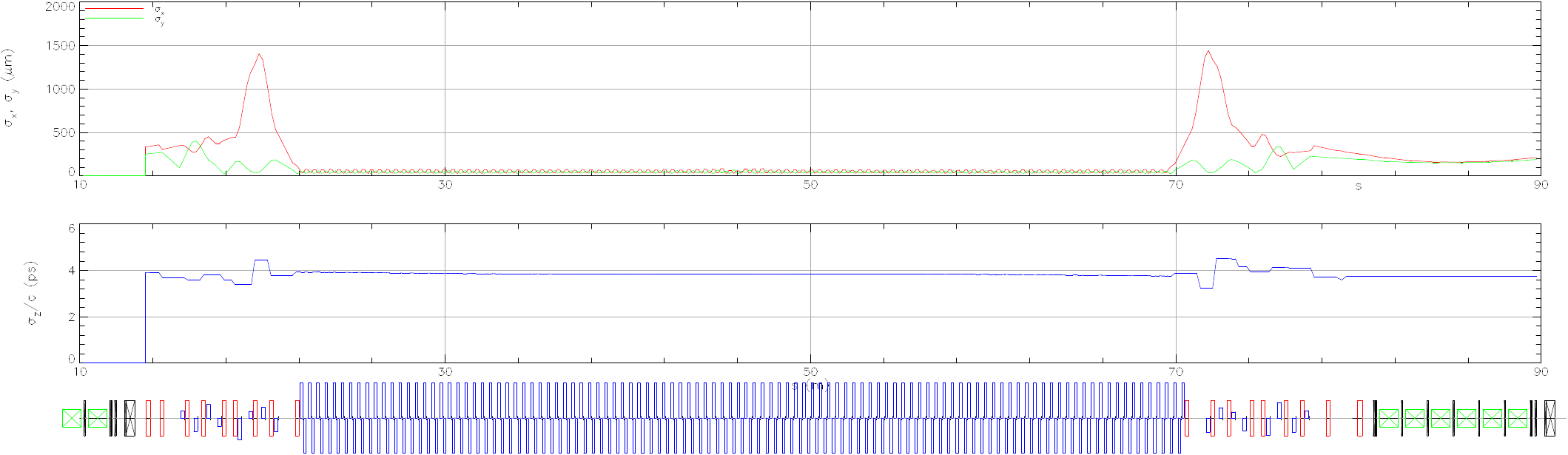}\\
\hspace{10pt}
\includegraphics[width=\linewidth]{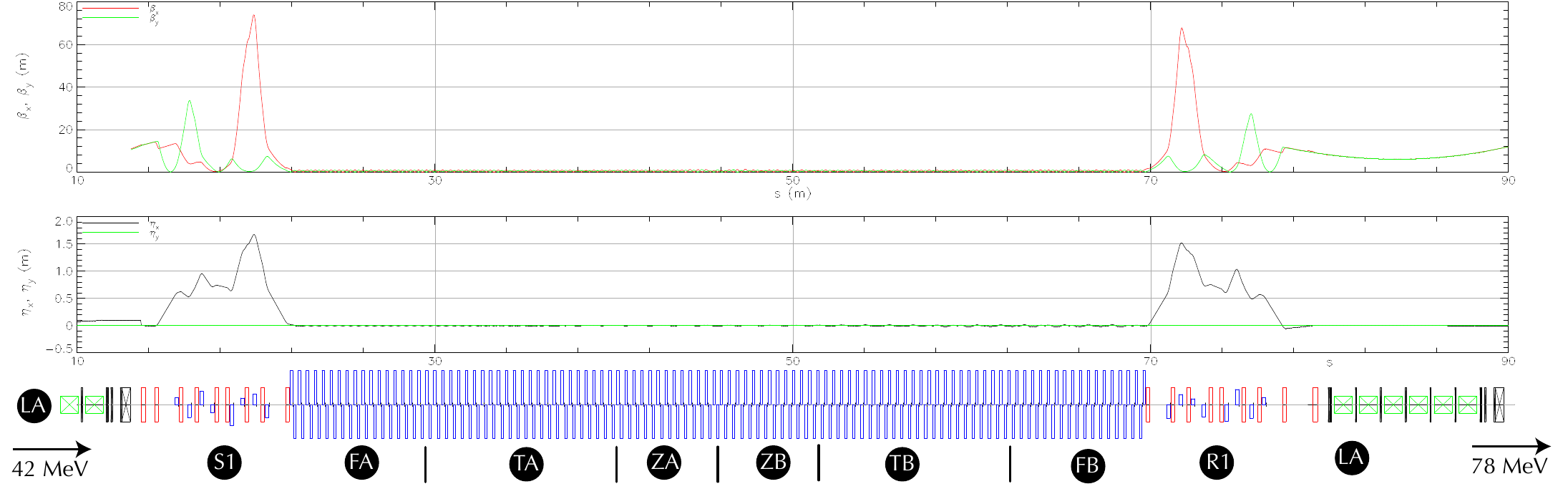}
\caption[]{Optics and beam sizes for pass 1, 42 MeV}
\label{fig:optics_4pass_1}
\end{figure}
\end{landscape}

\begin{landscape}
\begin{figure}[htbp]
\centering
\includegraphics[width=\linewidth]{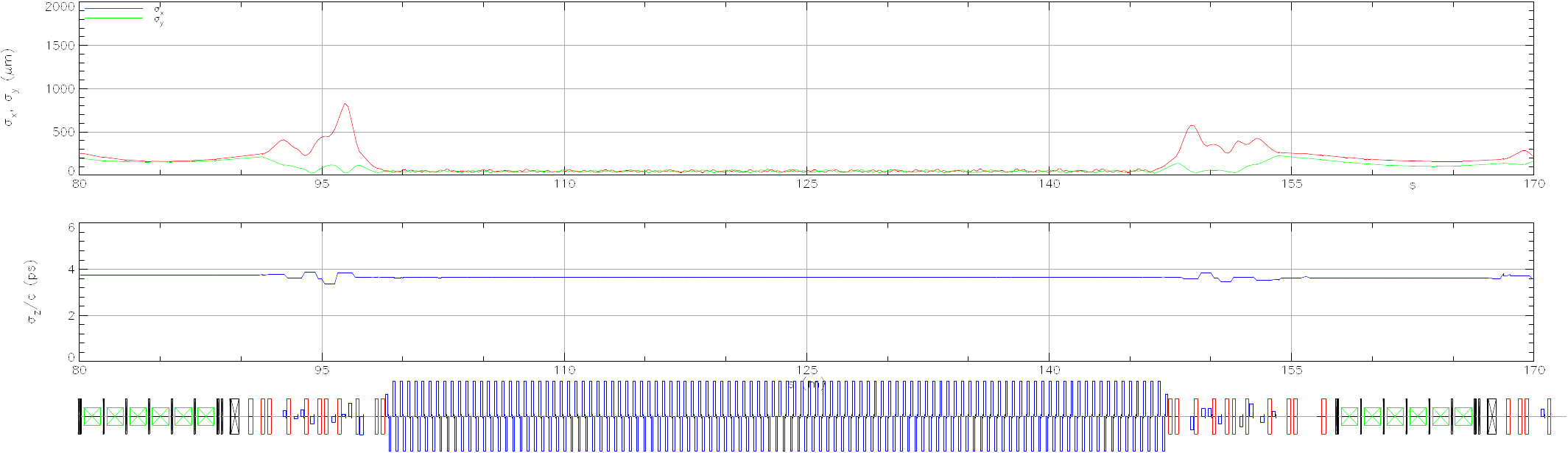}\\
\hspace{10pt}
\includegraphics[width=\linewidth]{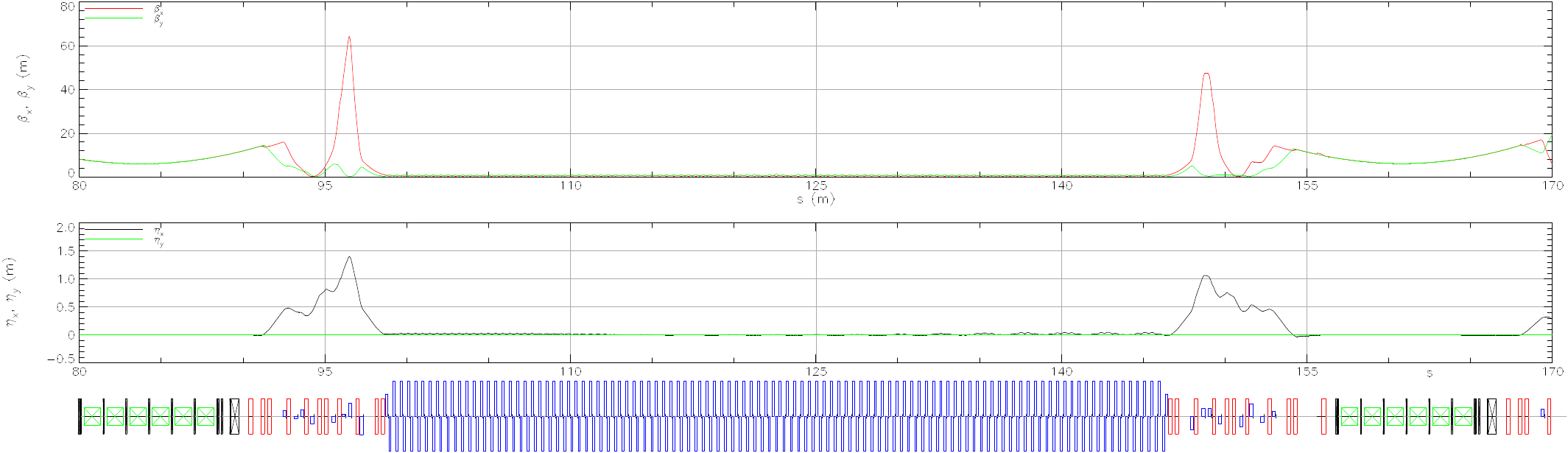}
\caption[]{Optics and beam sizes for pass 2, 78 MeV}
\label{fig:optics_4pass_2}
\end{figure}
\end{landscape}

\begin{landscape}
\begin{figure}[htbp]
\centering
\includegraphics[width=\linewidth]{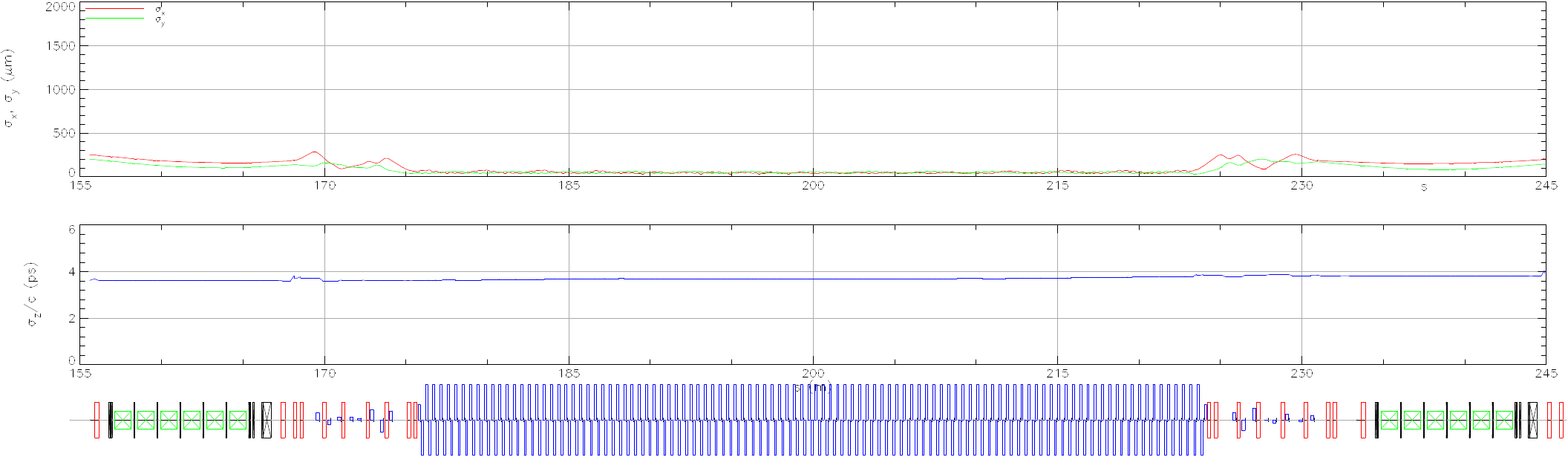}\\
\hspace{10pt}
\includegraphics[width=\linewidth]{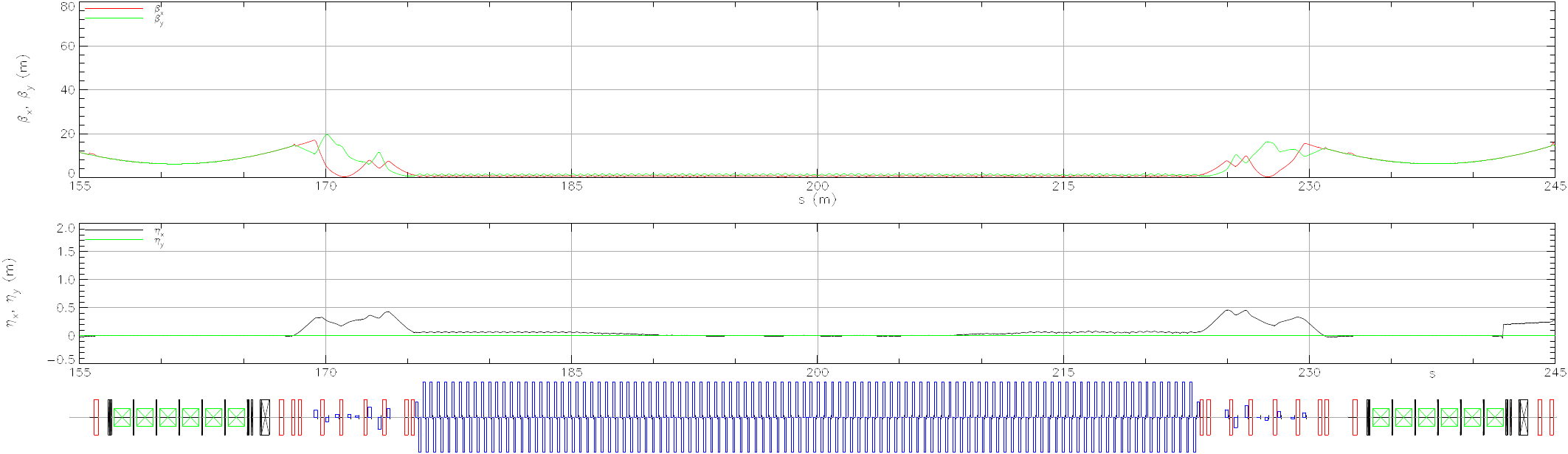}
\caption[]{Optics and beam sizes for pass 3, 114 MeV}
\label{fig:optics_4pass_3}
\end{figure}
\end{landscape}

\begin{landscape}
\begin{figure}[htbp]
\centering
\includegraphics[width=\linewidth]{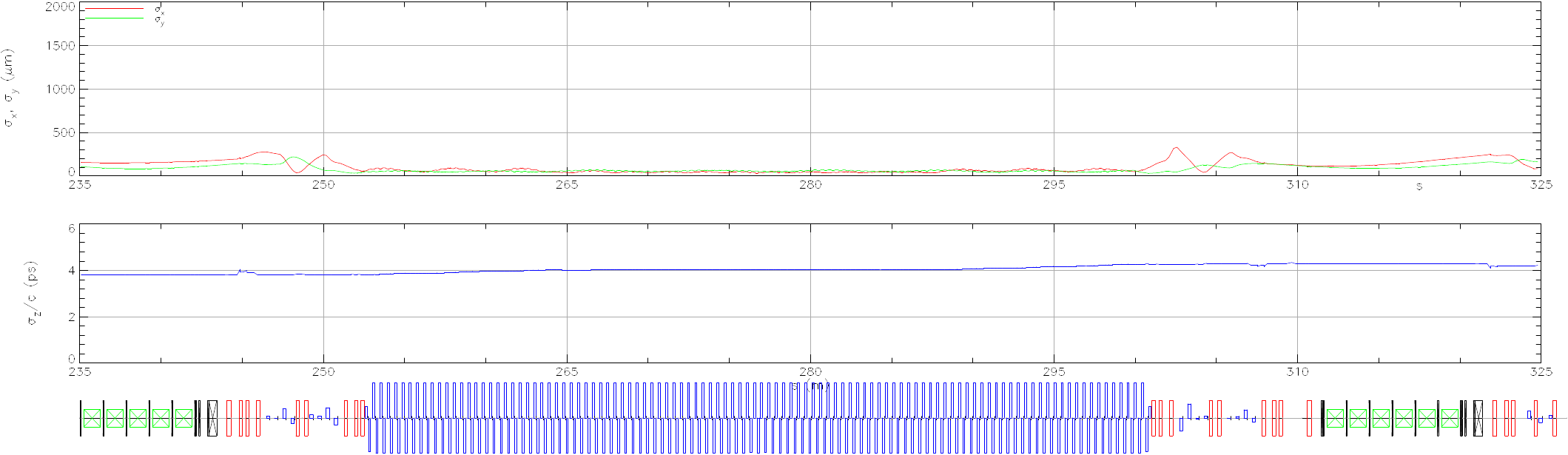}\\
\hspace{10pt}
\includegraphics[width=\linewidth]{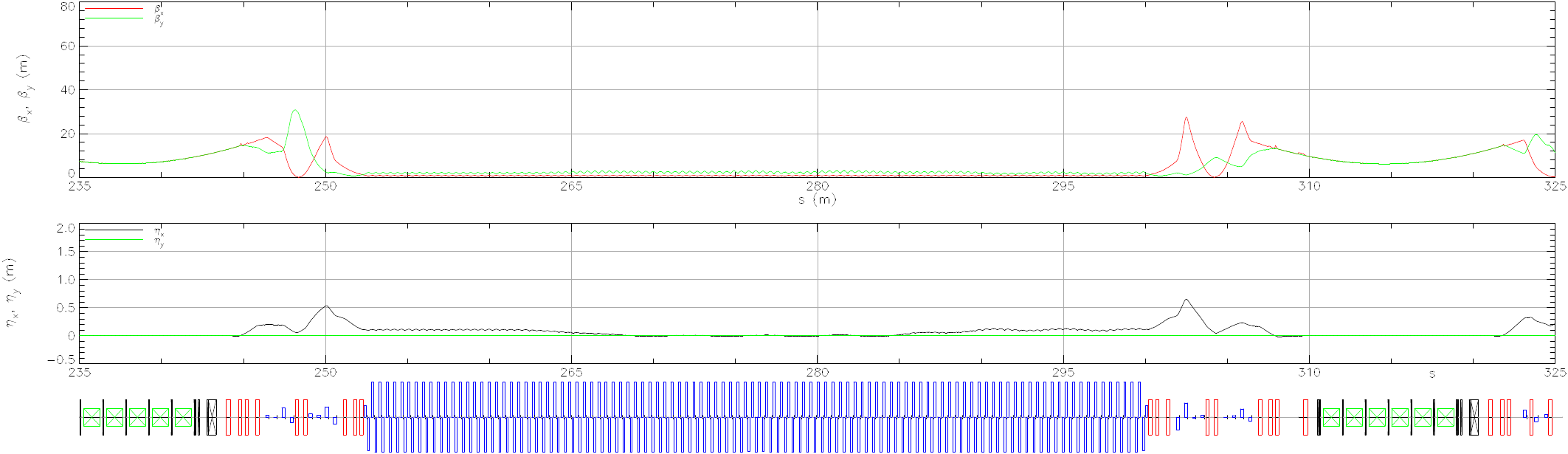}
\caption[]{Optics and beam sizes for pass 4, 150 MeV}
\label{fig:optics_4pass_4}
\end{figure}
\end{landscape}

\begin{landscape}
\begin{figure}[htbp]
\centering
\includegraphics[width=\linewidth]{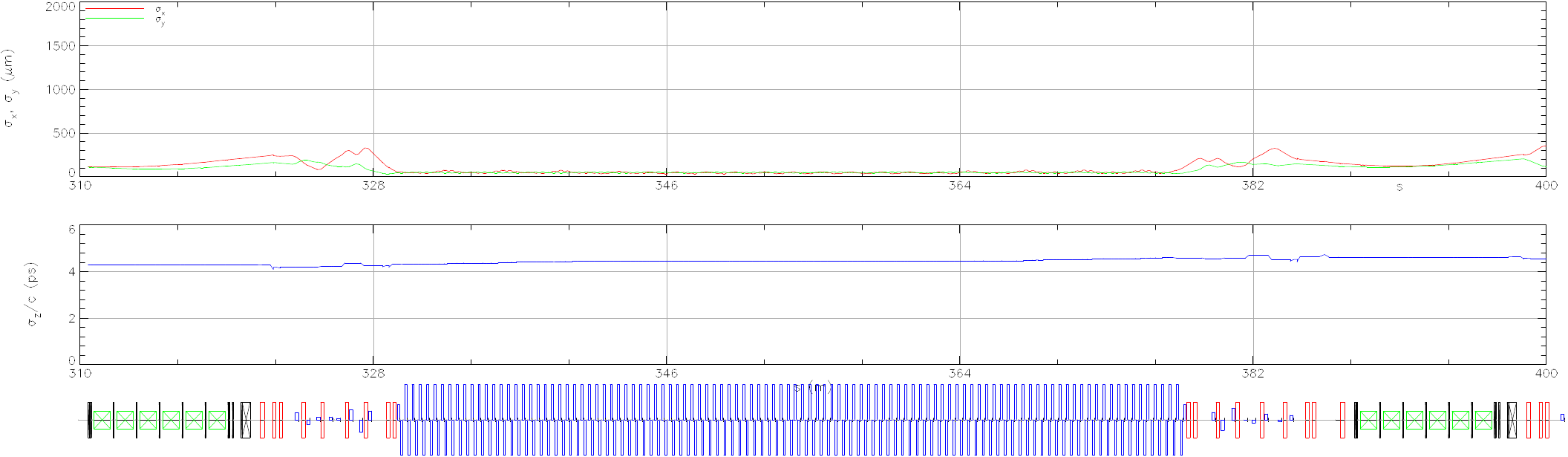}\\
\hspace{10pt}
\includegraphics[width=\linewidth]{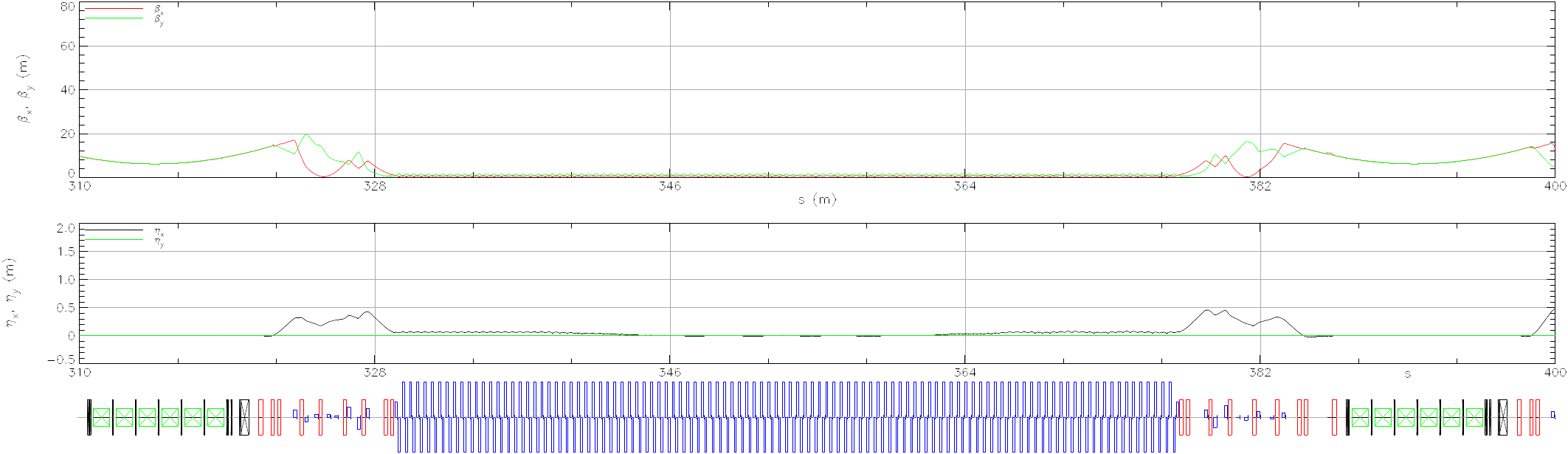}
\caption[]{Optics and beam sizes for pass 5, 114 MeV}
\label{fig:optics_4pass_5}
\end{figure}
\end{landscape}

\begin{landscape}
\begin{figure}[htbp]
\centering
\includegraphics[width=\linewidth]{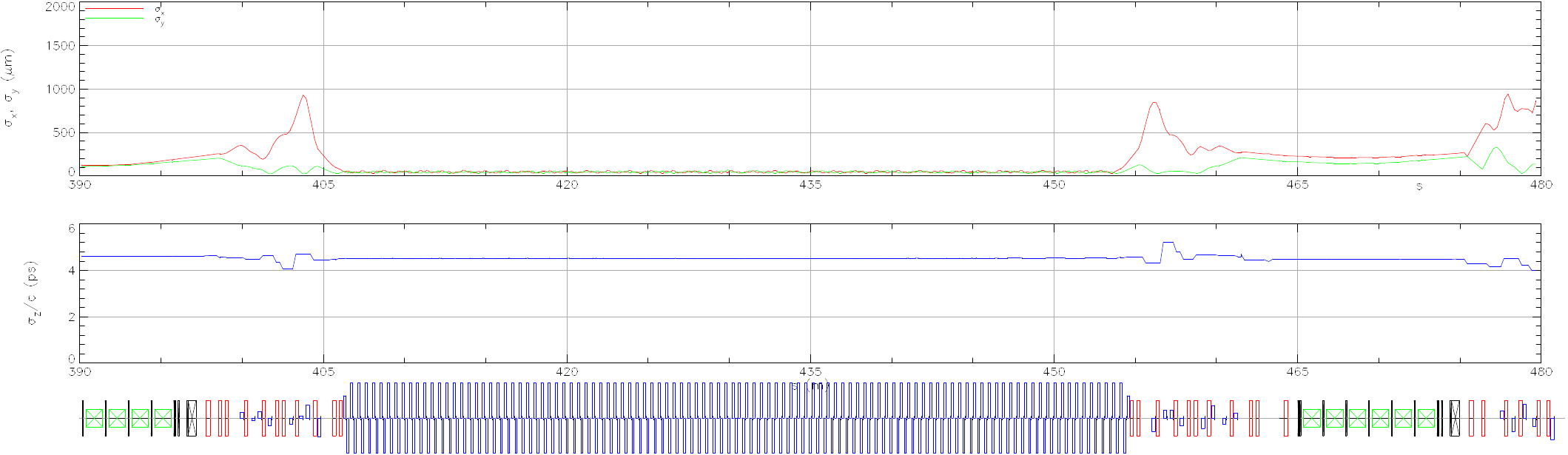}\\
\hspace{10pt}
\includegraphics[width=\linewidth]{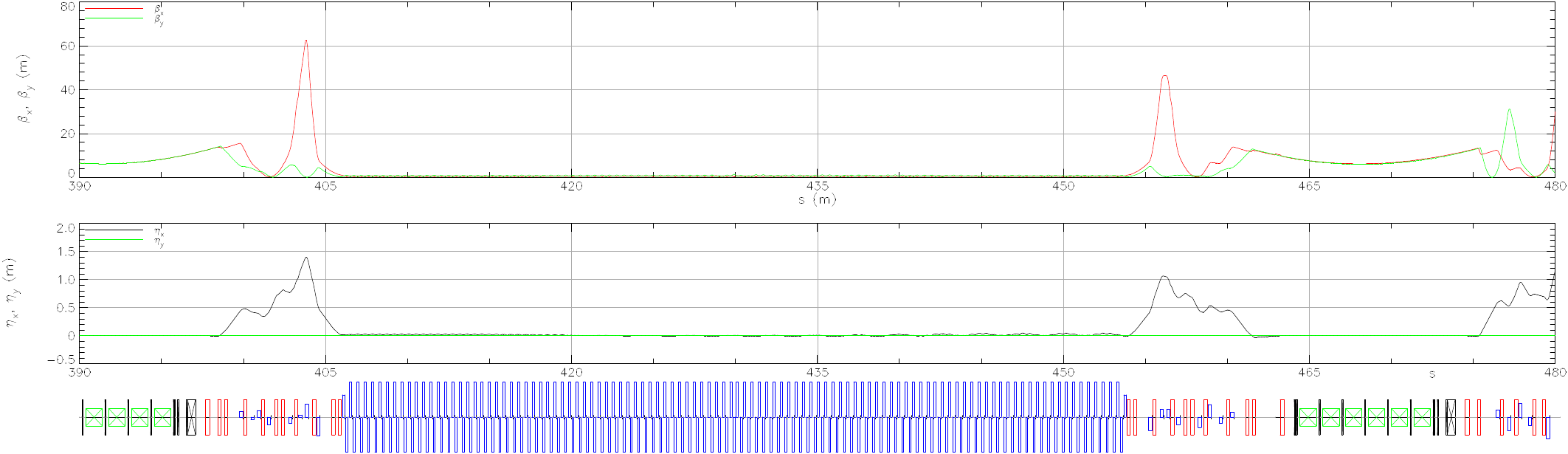}
\caption[]{Optics and beam sizes for pass 6, 78 MeV}
\label{fig:optics_4pass_6}
\end{figure}
\end{landscape}

\begin{landscape}
\begin{figure}[htbp]
\centering
\includegraphics[width=\linewidth]{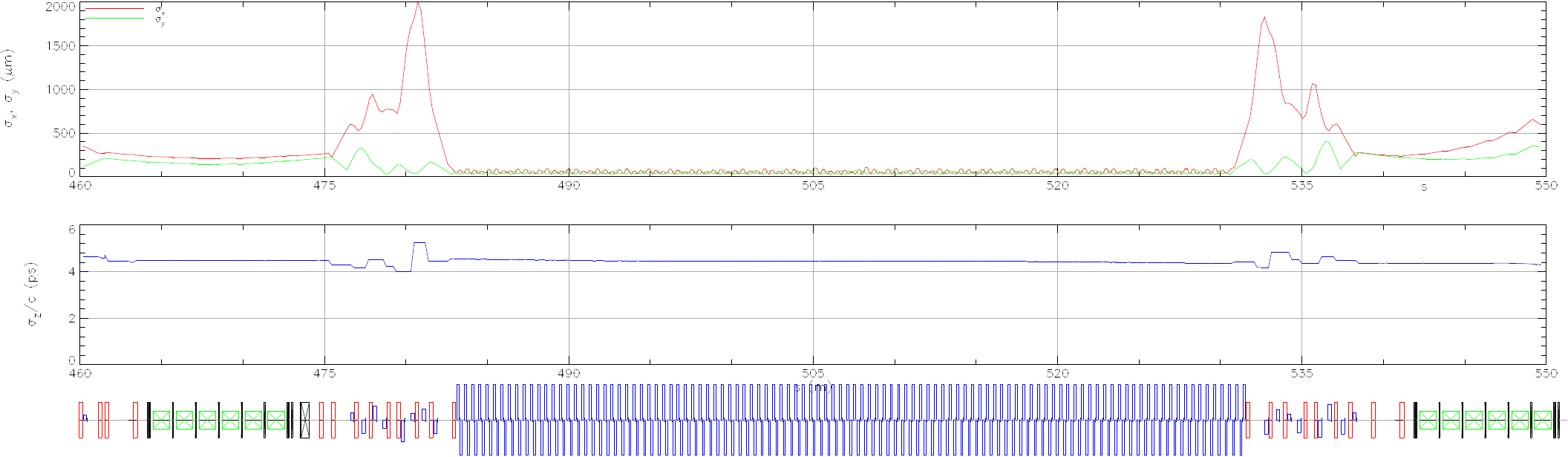}\\
\hspace{10pt}
\includegraphics[width=\linewidth]{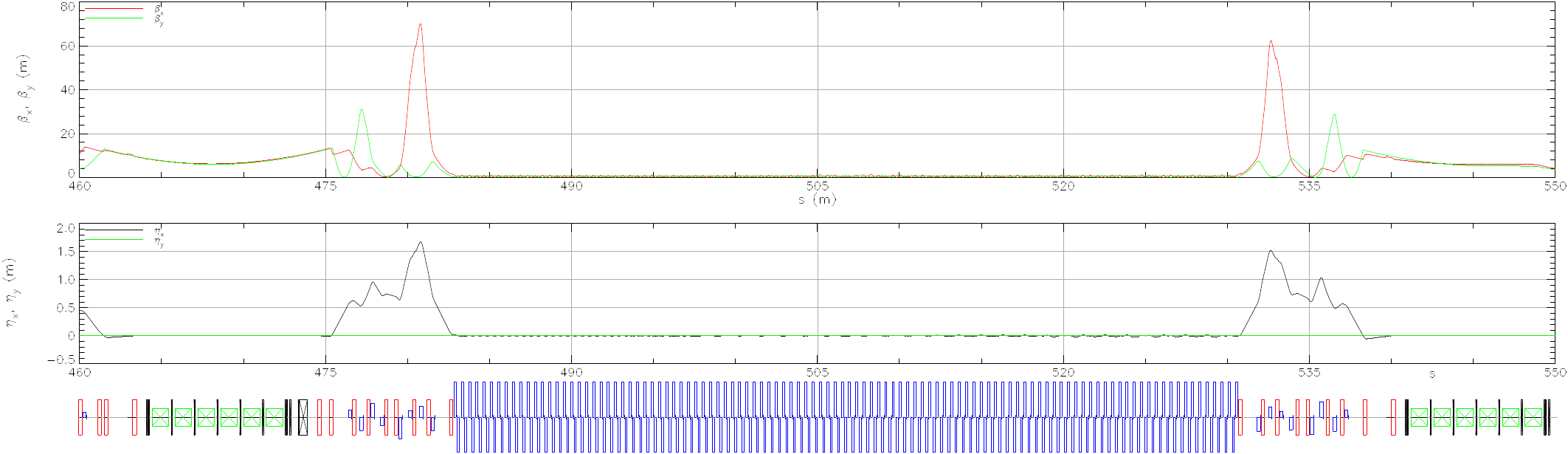}
\caption[]{Optics and beam sizes for pass 7, 42 MeV}
\label{fig:optics_4pass_7}
\end{figure}
\end{landscape}

Our initial start-to-end simulations start by optimizing and tracking a bunch from creation through the LA section using GPT with space charge. \Figure{fig:LA_end_pass1_phase_space} shows such a bunch. This bunch is then tracked through all passes of the machine using Bmad. \Figure{fig:optics_4pass} shows optics and beam sizes for this tracking. \Figure{fig:optics_4pass_1} to \Fig{fig:optics_4pass_7} show the details of each pass. 

A few extra simulation results on the CBETA lattice have been presented at IPAC 2017 \Ref{IPAC2017:MOPIK123, IPAC2017:THPAB058}. 

In the future, these simulations will include space charge, micro-bunching, CSR, resistive wall wakefields, and other wakefields into the calculation.  Furthermore, with the orbit correction scheme in place, we will track bunches though lattices with corrected errors. We can also study emittance growth by tracking bunches with large energy spread, as well as the dynamic aperture.

\section{Beam instabilities: BBU\Leader{William}}

\subsection{Introduction}
   Beam breakup (BBU) occurs in recirculating accelerators when a recirculated beam interacts with HOMs of the accelerating cavities. The most dominant HOMs are the dipole HOMs which give transverse kick to the beam bunches. The off-orbit bunches return to the same cavity and excite more dipole HOMs which, if in phase with the existing dipole HOMs, can kick the bunches more in the same direction. The effect can build up and eventually result in beam loss. Therefore, BBU is a primary limiting factor of the beam current, and the maximum achievable current is called the threshold current $I_\text{th}$. With more recirculation passes, bunches interact with cavities for more times, and $I_\text{th}$ can significantly decrease \Ref{bbu_Georg_Ivan}. The target current of CBETA is 100mA for the 1-pass machine, and 40mA for the 4-pass machine. Simulations are required to check whether $I_\text{th}$ is above this limit.  

The latest studies on BBU has been presented at IPAC 2017 \Ref{IPAC2017:THPVA143}. 

\subsection{Bmad Simulation Overview}

Cornell University has developed a simulation software called Bmad to model relativistic beam dynamics in customized accelerator lattices. Subroutines have been written to simulate BBU effect and find $I_\text{th}$ for a specific design. A complete lattice provided to the program must include at least one multi-pass cavity with HOM(s) assigned to it. It is possible to assign HOMs of different orders to a single cavity, and also a different set of HOMs to other cavities. Parameters such as bunch frequency and numerical tolerances can also be specified to the program. 

For each simulation, the program starts with a test current and records the voltage of all assigned HOMs over time. As the beam pass by the cavities, the momentum exchange between the bunches and wake fields are calculated, as well as the new HOM voltages. If all HOM voltages are stable over time, the test current is considered stable, and a new greater current will be tested. In contrast, if at least one HOM voltage is unstable, the test current is regarded unstable, and a smaller current will be tested. Usually $I_\text{th}$ can be pinned down within 50 test currents.

In BBU simulation, only cavities with HOM(s) assigned are essential, so other lattice structures can be hybridized. Hybridization is a process of merging certain lattice components into an equivalent Taylor map (up to linear order). A single BBU simulation on a CBETA 1-pass hybridized lattice takes up to 20 minutes, better than hours without hybridization. To efficiently find $I_\text{th}$ for various HOM assignments or design change, hybridization is necessary.

\subsection{Bmad Simulation Result}                                                                                 
                                                                                                                    
Dipole HOMs of a single CBETA SRF cavity have been simulated by Nick Valles \Ref{bbu_NickValles}. Random errors were introduced to each ellipse parameter of the cavity shape, resulting in a spectrum of dipole HOMs, and their characteristics ( shunt impedance $(R/Q)$, quality factor $Q$, and frequency $f$) were recorded. Each random error comes from a uniform distribution, with 4 different error cases: $\pm$ 125, 250, 500, and 1000$\unit{\mu m}$. For simplicity, we use $\epsilon$ to denote the error case: "$\epsilon$ = 125$\unit{\mu m}$" means the errors introduced come from a $\pm$ 125$\unit{\mu m}$ uniform distribution.
A cavity with smaller $\epsilon$ has better manufacture precision. For each error case, 400 unique cavities were provided, and the top 10 "worst" dipole HOMs (ones with greater HOM figure of merit $\xi = (R/Q) \sqrt{Q}/f $) were recorded for each cavity. 

Practically the 6 CBETA cavities are not identical, but manufactured with similar precision. Thus, for simulation each cavity is assigned with a different (randomly chosen) set of 10 dipole HOMs, and all 6 sets have the same $\epsilon$. Hundreds of simulations with different HOM assignments were run, and to statistical distributions of $I_\text{th}$ were obtained for each specific design and choice of $\epsilon$. Three distributions will be presented as histograms in this section:

1)	CBETA 1-pass with $\epsilon = 125\unit{\mu m}$

2)  CBETA 4-pass with $\epsilon = 125\unit{\mu m}$

3)	CBETA 4-pass with $\epsilon = 250\unit{\mu m}$

Since modern cavities are built with manufacture precision below 250$\unit{\mu m}$, the $\epsilon=500\unit{\mu m}$ and $\epsilon=1000\unit{\mu m}$ cases will not be investigated.

\subsubsection{(1) CBETA 1-pass with $\epsilon = 125\unit{\mu m}$}
\begin{figure}[h]
\centering
\includegraphics[width=0.6\textwidth]{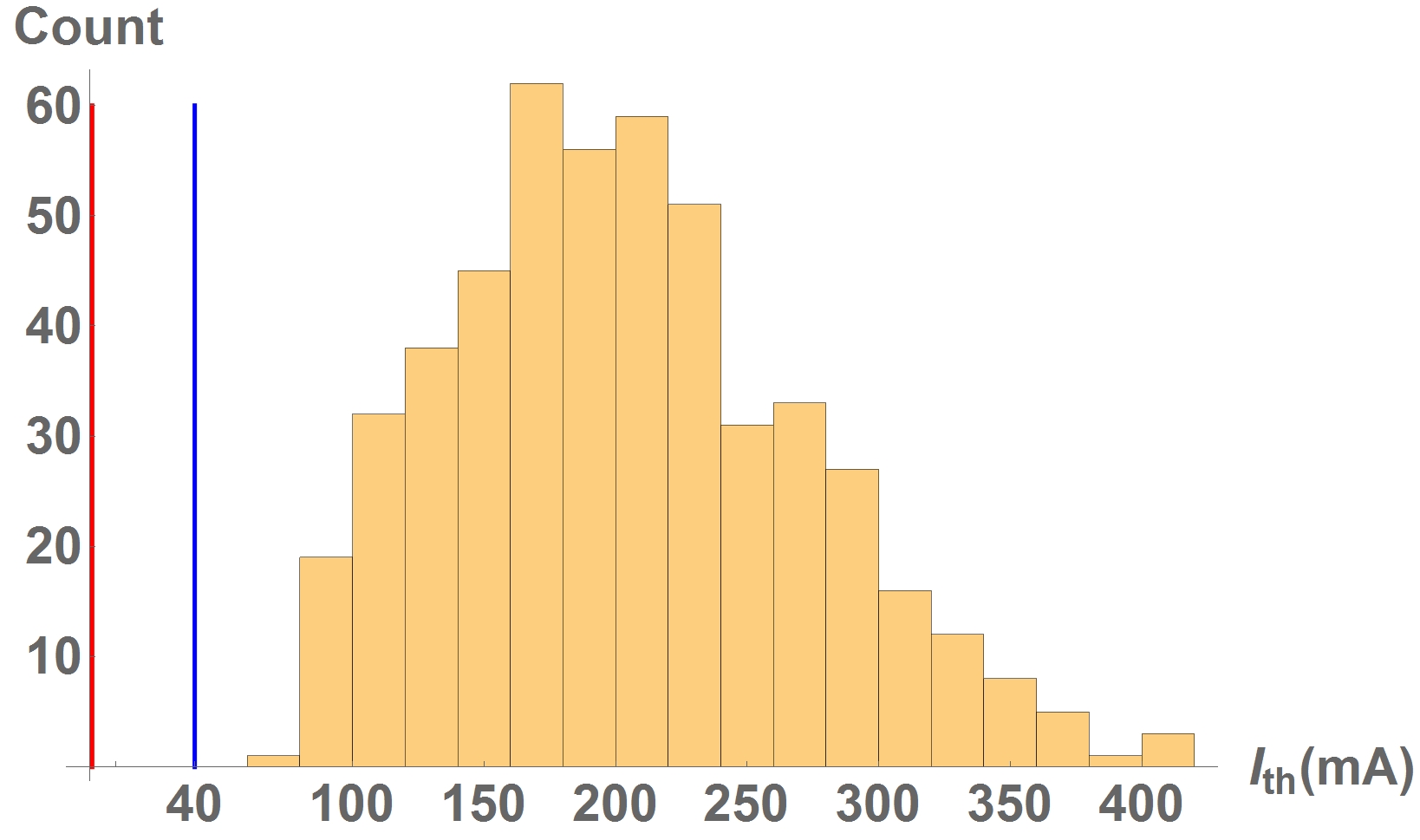}
\caption{ 500 BBU simulation results of $I_\text{th}$ for the CBETA 1-pass lattice. Each cavity is assigned with a random set of 10 dipole HOMs ($\epsilon=125\unit{\mu m}$). The red line indicates the KPP of 1mA, while the blue line indicates the UPP 40mA for 1-pass machine. }
\label{bbu_1pass_125um}
\end{figure}

\Figure{bbu_1pass_125um} shows that all 500 simulations results exceed the UPP of 40mA for the 1-pass machine. The lowest simulated $I_\text{th}$ is 70mA, well above the KPP of 1mA. The result is quite promising.

\subsubsection{(2) CBETA 4-pass with $\epsilon = 125\unit{\mu m}$}
\begin{figure}[h]
\centering
\includegraphics[width=0.6\textwidth]{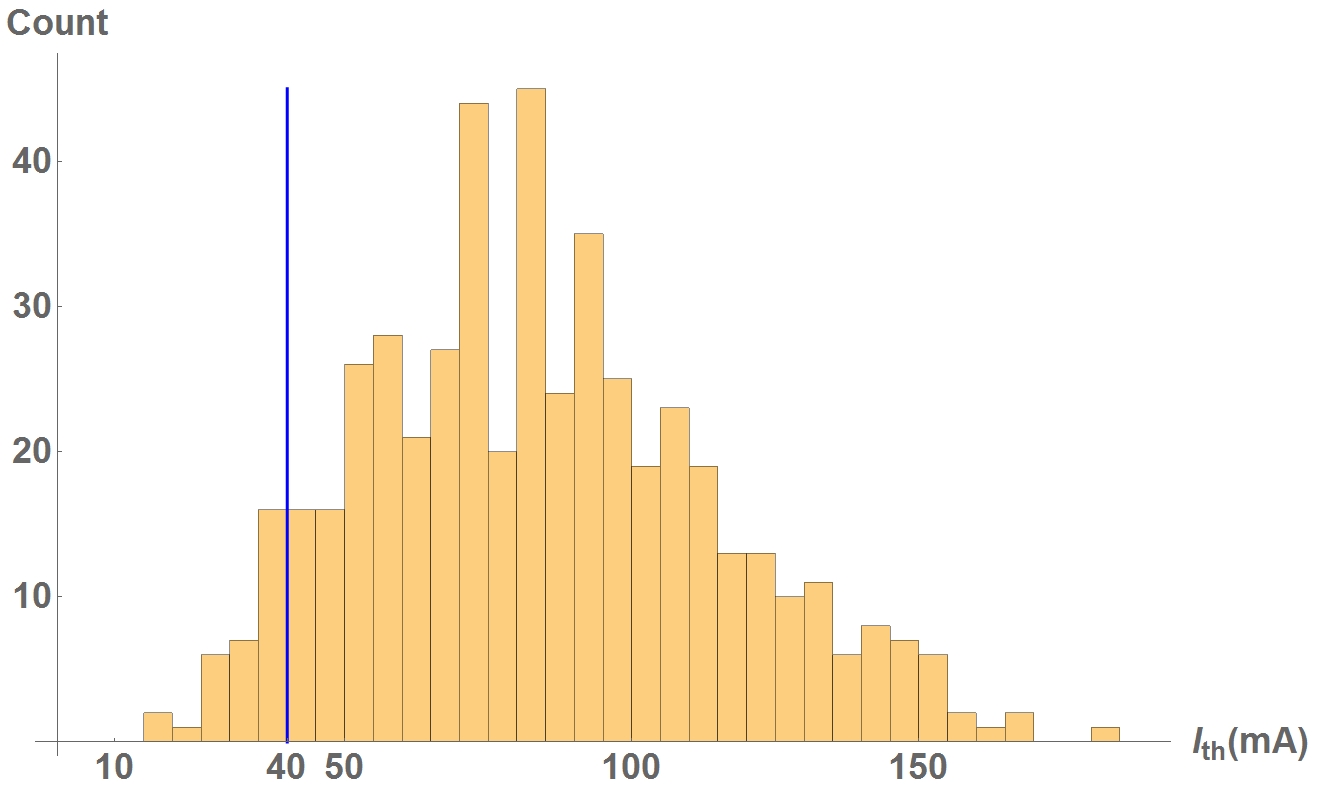}
\caption{500 BBU simulation results of $I_\text{th}$ for the CBETA 4-pass lattice.
Each cavity is assigned with a random set of 10 dipole HOMs ($\epsilon=125\unit{\mu m}$). The blue line indicates the UPP of 40mA for 4-pass machine. }

\label{bbu_4pass_125um}
\end{figure}

\Figure{bbu_4pass_125um} shows that out of 500 simulations, 468 of them exceed the 40mA goal for CBETA 4-pass machine, and the remaining 32 are above 10mA. This implies that with certain undesirable combinations of HOMs in the cavities, $I_\text{th}$ can be limited.

\subsubsection{(3) CBETA 4-pass with $\epsilon = 250\unit{\mu m}$}

\begin{figure}[h]
\centering
\includegraphics[width=0.6\textwidth]{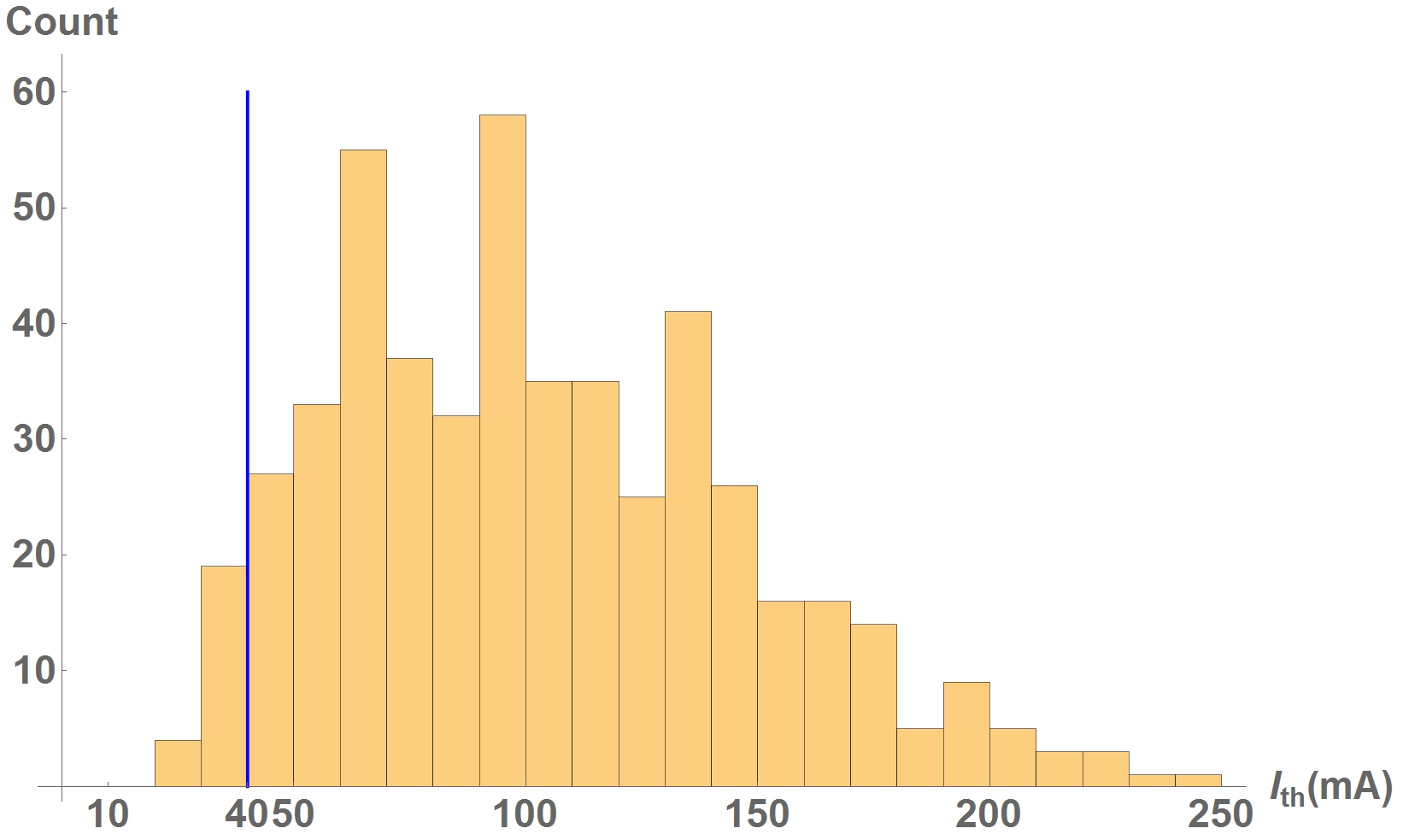}
\caption{500 BBU simulation results of $I_\text{th}$ for the CBETA 4-pass lattice.
Each cavity is assigned with a random set of 10 dipole HOMs ($\epsilon=250\unit{\mu m}$). The blue line indicates the UPP of 40mA for the 4-pass machine. }
\label{bbu_4pass_250um}
\end{figure}

It is interesting to see how $I_\text{th}$ behaves differently with a different $\epsilon$ for the 4-pass lattice. For $\epsilon=250\unit{\mu m}$, 477 out of 500 simulations are above the UPPs of 40mA and all are well above 10mA (See \Fig{bbu_4pass_250um}). This is slightly better than the $\epsilon=125\unit{\mu m}$ case, with similar implications. Some might wonder if a greater $\epsilon$ could result in higher threshold current. Indeed, the more the cavity shapes deviate, the HOM frequency spread becomes greater. A greater spread means the HOMs across cavities act less coherently when kicking the beam, thus statistically increases the $I_\text{th}$. However, a greater deviation also tends to undesirably increase the $Q$ (and possibly $R/Q$) of the HOMs, which usually lowers $I_\text{th}$ \Ref{bbu_NickValles}. A compensation between the frequency spread and HOM damping means a greater manufacture error in cavity shapes can not reliably improve $I_\text{th}$.

There are several ways to improve the accuracy of the simulation results. Perhaps the most important one is to assign the actual HOM spectra measured directly from the built SRF cavities. However, measurement takes time, and complication may arise in identifying the HOM mode orders and choosing which ones to be included in simulation. Besides improving the simulation accuracy, a more important issue is to achieve a greater $I_\text{th}$, as discussed in the following section.

\subsection{Aim for higher $I_\text{th}$}

To achieve a higher $I_\text{th}$, three ways have been proposed, and their effects can be simulated. The first way is to change the bunch frequency $f_{b}$ (repetition rate) by an integer multiple. Simulations on a CBETA 1-pass lattice show a change of $I_\text{th}$ fewer than $5\%$ over various choices of $f_{b}$. Such result may imply that $I_\text{th}$ could not be systematically improved by changing $f_{b}$. On the other hand, rigorous calculation \Ref{bbu_Georg_Ivan} has shown that $I_\text{th}$ depends on $f_{b}$ in a non-linear way for a multi-pass ERL, and it will be interesting to experiment this effect on the realistic CBETA. 
The other two ways involve varying the phase advances and introducing x-y coupling between the cavities. The simulation results foe these two methods are presented in the following sections.

\subsection{Effect on $I_\text{th}$ by varying phase advance}
\label{bbu_decoupled}

$I_\text{th}$ can potentially be improved by changing the phase advances (in both x and y) between the multi-pass cavities. This method equivalently changes the $T_{12}$ (and $T_{34}$) element of the transfer matrices, and smaller $T_{12}$ values physically correspond to a greater $I_\text{th}$ in 1-pass ERLs \Ref{bbu_Georg_Ivan}. To vary the phase advances in Bmad simulations, a zero-length matrix element is introduced right after the first pass of the linac (See section \Section{sec:bbu_tech} for details). In reality the phase advances are changed by adjusting the quad strengths around the accelerator structure. In simulation the introduction of the matrix may seem arbitrary, but this gives us insight on how high $I_\text{th}$ can reach as phase advances vary. 

For each simulation, each cavity is assigned with three $ ``\epsilon = 125\unit{\mu m}$" dipole HOMs in x, and three identical HOMs in y (polarization angle = $\pi /2$). The $I_\text{th}$ is obtained for a choice of ($\phi_x, \phi_y$), each from 0 to 2$\pi$. Several simulations were run for both the 1-pass and 4-pass CBETA lattice, and the results are presented below.

\subsubsection{(1) 1-pass results}

\begin{figure}[h!]
\centering
\includegraphics[width=0.6\textwidth]{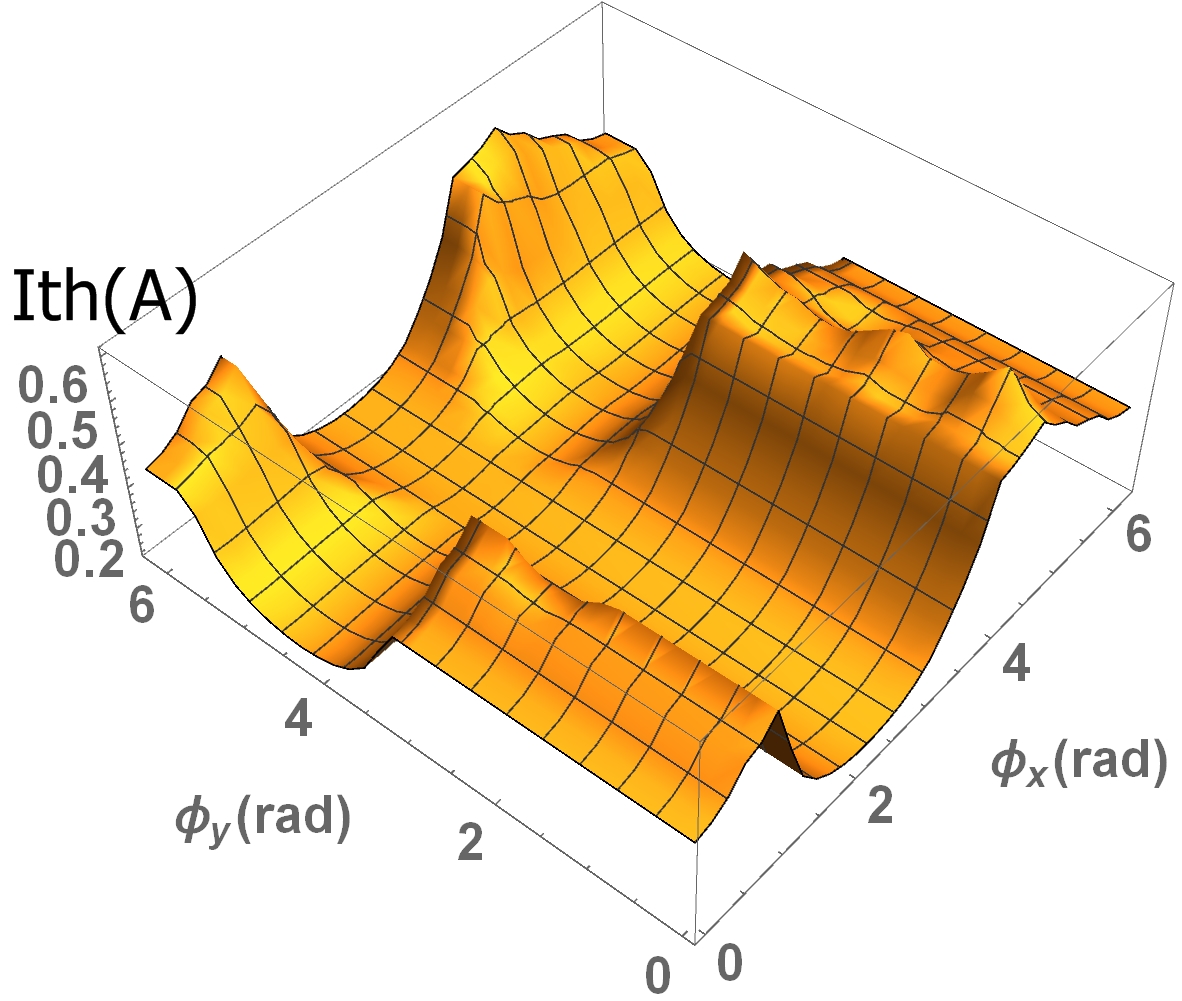}
\caption{ A scan of BBU $I_\text{th}$ over the two phase advances for the CBETA 1-pass lattice. Each cavity is assigned with a random set of 3 dipole HOMs in both x and y polarization. ($\epsilon=125\unit{\mu m}$). For this particular HOM assignment, $I_\text{th}$ ranges from 140mA to 610mA.}
\label{bbu_1pass_125um_decoup}
\end{figure}

\Figure{bbu_1pass_125um_decoup} shows a typical way  $I_\text{th}$ varies with the two phase advances. Depending on the HOM assignment, the $I_\text{th}$ can reach up to 650mA with an optimal choice of ($\phi_x, \phi_y$). This implies that changing phase advances does give us advantages in improving $I_\text{th}$ for the 1-pass CBETA lattice (the improvement can range from +200mA to +400mA depending on the HOMs assigned).  Note that $\phi_x$ and $\phi_y$ affect $I_\text{th}$ rather independently. That is, at certain $\phi_x$ which results in a low $I_\text{th}$ (the ``valley"), any choice of $\phi_y$ does not help increase $I_\text{th}$, and vise versa. It is also observed that $I_\text{th}$ is more sensitive to $\phi_x$, and the effect of $\phi_y$ becomes obvious mostly at the ``crest" in $\phi_x$. Physically this is expected since many lattice elements have a unit transfer matrix in y, and the effect of $T_{12}$ is more significant than $T_{34}$. In other words, HOMs in x polarization is more often excited. As we will see this is no longer true when x-y coupling is introduced. 

It is also observed that the location of the ``valley" remains almost fixed when HOM assignements are similar. Physically the valley occurs when the combination of phase-advances results in a great $T_{12}$ which excites the most dominant HOM. Therefore, the vally location depends on which cavity is assigned with the most dominant HOM, and is consistent with the simulation results. We will also see that the valley location becomes unpredictable when x-y coupling is introduced.  

\subsubsection{(2) 4-pass results}

\begin{figure}[h!]
\centering
\includegraphics[width=0.6\textwidth]{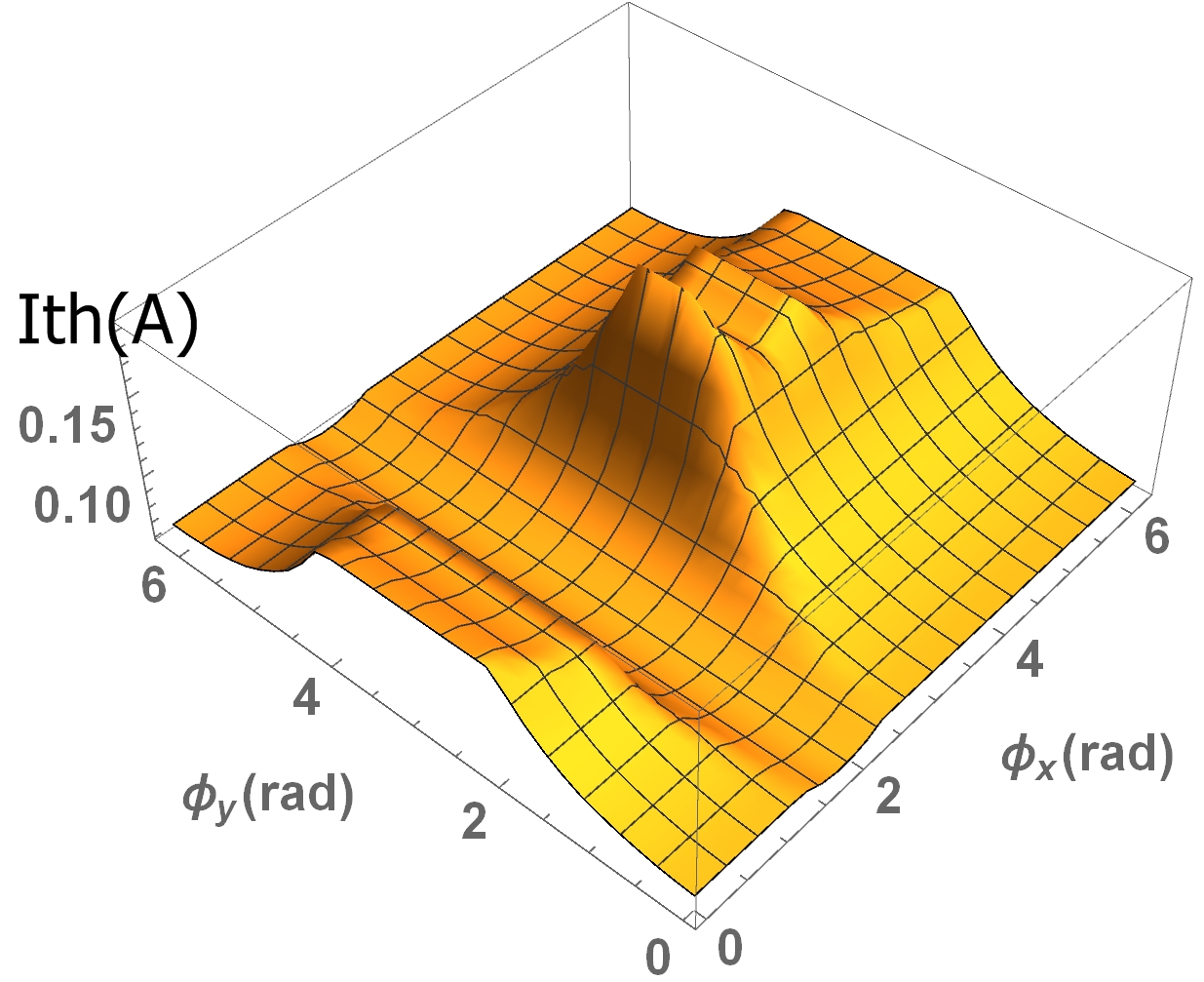}
\caption{ A scan of BBU $I_\text{th}$ over the two phase advances for the CBETA 4-pass lattice. Each cavity is assigned with a random set of 3 dipole HOMs in both x and y polarization. ($\epsilon=125\unit{\mu m}$). For this particular HOM assignment, $I_\text{th}$ ranges from 61mA to 193mA.}
\label{bbu_4pass_125um_decoup}
\end{figure}

\Figure{bbu_4pass_125um_decoup} shows a typical result for the 4-pass lattice. Depending on the HOM assignment, the $I_\text{th}$ can reach up to 200mA with an optimal choice of ($\phi_x, \phi_y$). Therefore the method of varying phase advances can still improve $I_\text{th}$. However, the improvement is more significant for the 1-pass lattice (+300mA to +400mA) than the 4-pass lattice (about +150mA).

\subsection{Effect on $I_\text{th}$ with x-y coupling}
\label{bbu_coupled}
The third way involves x/y coupling in the transverse optics, so that horizontal HOMs excite vertical oscillations and vise versa. This method has been shown very effective for 1-pass ERLs \Ref{bbu_Georg_Ivan_coupled}. To simulate the coupling effect in Bmad simulation, a different non-zero length is again introduced right after the first pass of the linac (See \Section{sec:bbu_tech} for details). The matrix couples the transverse optics with two free phases ($\phi_1,\phi_2$) to be chosen. These two phases are not the conventional phase advances, but can also range from 0 to 2$\pi$. The HOM assignment is the same as in the second method. Again, several simulations were run for both 1-pass and the 4-pass CBETA lattice, with results presented below.

\subsubsection{(1) 1-pass results}

\begin{figure}[h!]
\centering
\includegraphics[width=0.6\textwidth]{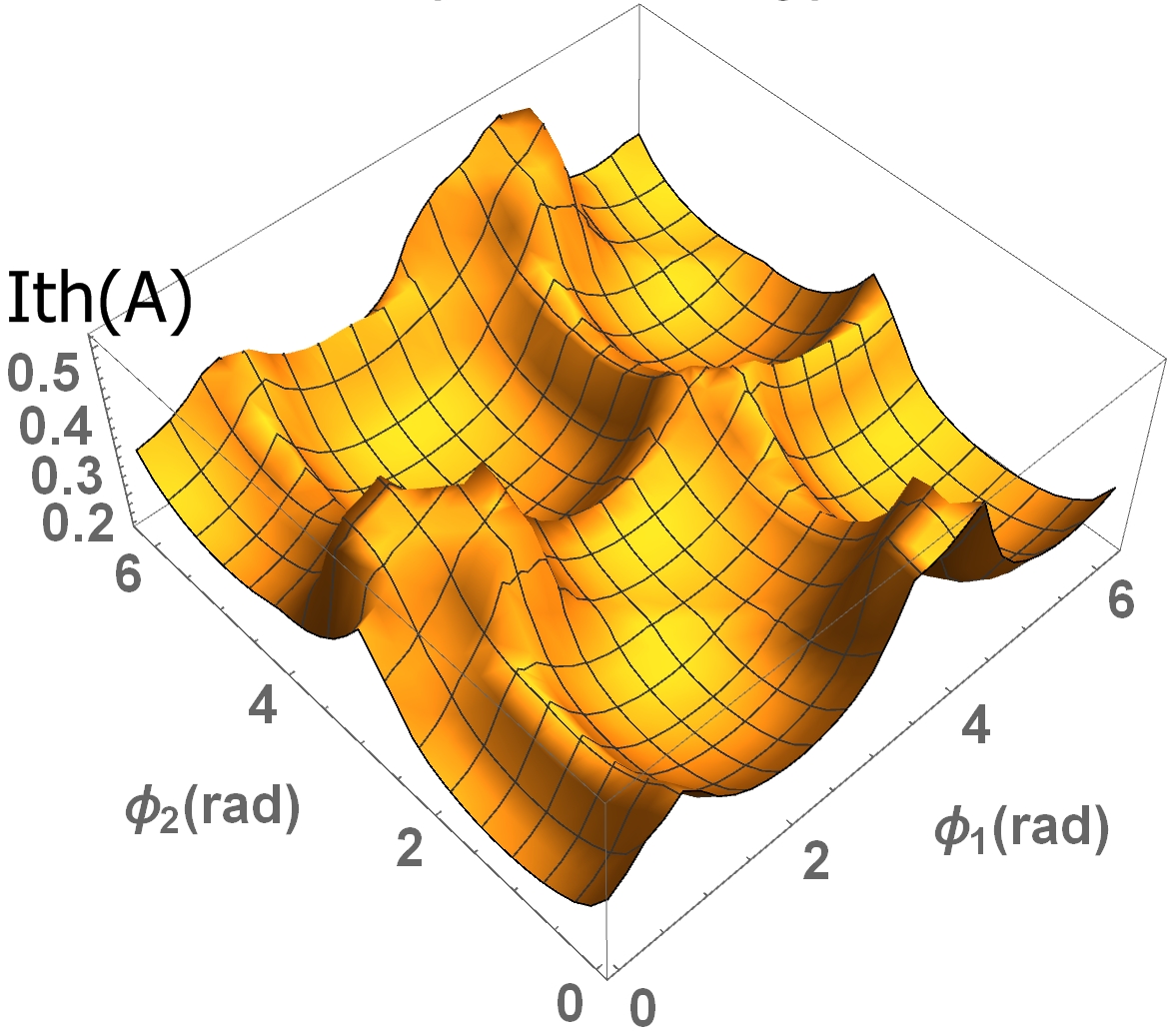}
\caption{ A scan of BBU $I_\text{th}$ over the two free phases for the CBETA 1-pass lattice with x-y coupling. Each cavity is assigned with a random set of 3 dipole HOMs in both x and y polarization. ($\epsilon=125\unit{\mu m}$). For this particular HOM assignment, $I_\text{th}$ ranges from 140mA to 520mA.}
\label{bbu_1pass_125um_coup}
\end{figure}

\Figure{bbu_1pass_125um_decoup} shows a typical way  $I_\text{th}$ varies with the two free phases.  Depending on the HOM assignment, the $I_\text{th}$ can reach up to 680mA with an optimal choice of ($\phi_1, \phi_2$). Because the transverse optics are coupled, the two phases no longer affect $I_\text{th}$ in an independent manner. That is, there is no specific $\phi_1$ which would always result in a relatively high or low $I_\text{th}$. Both phases need to be varied to reach a relatively high $I_\text{th}$. In contrast to varying phase advances, the potential improvement in $I_\text{th}$ for 1-pass depends heavily on the HOMs assigned. The improvement observed ranges from +200mA to +400mA with x-y coupling.

\subsubsection{(2) 4-pass results}

\begin{figure}[h!]
\centering
\includegraphics[width=0.6\textwidth]{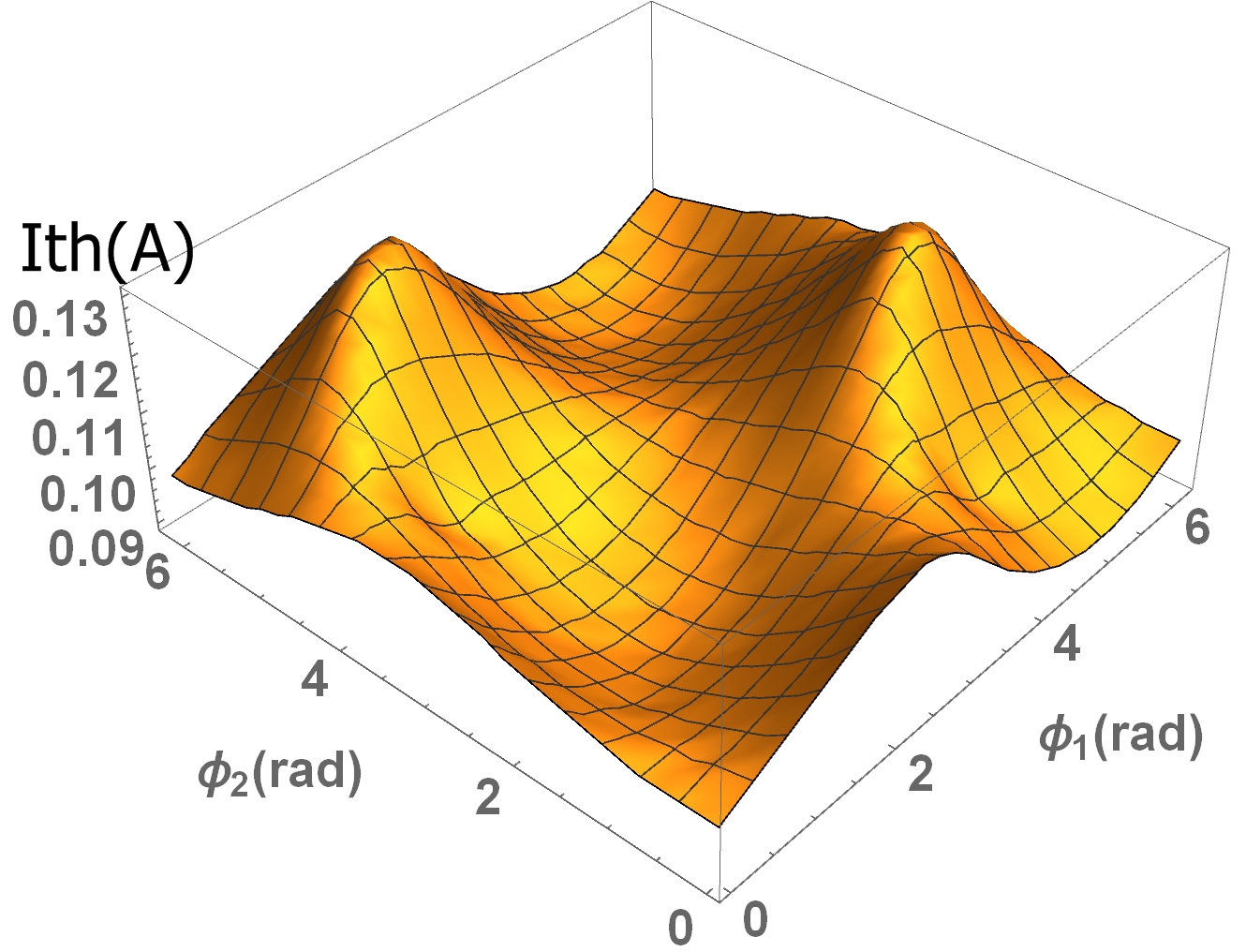}
\caption{ A scan of BBU $I_\text{th}$ over the two free phases for the CBETA 4-pass lattice with x-y coupling. Each cavity is assigned with a random set of 3 dipole HOMs in both x and y polarization. ($\epsilon=125\unit{\mu m}$). For this particular HOM assignment, $I_\text{th}$ ranges from 89mA to 131mA.}
\label{bbu_4pass_125um_coup}
\end{figure}

\Figure{bbu_4pass_125um_decoup} shows a typical result for the 4-pass lattice. Depending on the HOM assignment, the $I_\text{th}$ can reach up to 120mA with an optimal choice of ($\phi_1, \phi_2$). Therefore introducing x-y coupling can still improve $I_\text{th}$ for the 4-pass lattice (about +60mA), but not as significantly as varying phase advances. In contrast to 1-pass with x-y coupling, the range of improvement also becomes less sensitive to the HOMs assigned.

\subsection{BBU conclusion}
Bmad simulation has shown that with the current design lattice, both the 1-pass and 4-pass machine can always reach the KPP design current (1mA), and can surpass the UPP (40mA) over 93\% of time depending on the actual HOMs.

To potentially increase the threshold current, we can either vary the injector bunch frequncy, or change the lattice optics (by introducing additional phase advances or x-y coupling). While the former is shown ineffective by simulation, the later provides room for improvement. For the 1-pass lattice, both optics methods allow great improvement in $I_\text{th}$ (about +200mA to +400mA), and the range of improvement with x-y coupling depends more on the HOMs assigned. For the 4-pass lattice, the method of varying phase advances allow more improvement (about +150mA) than x-y coupling (about +60mA). 

In short, varying phase advances is the most promising and cost-effective method to increase the threshold current of CBETA.

\subsection{Technical details to introduce phase advances and x-y coupling}
\label{sec:bbu_tech}

In terms of Twiss parameters, the 2x2 transfer matrix $M(\phi)$ from location 0 to 1 is:
\[
M_{1 \leftarrow 0}(\phi) =
\begin{pmatrix}
  \sqrt{\frac{\beta_{1}}{\beta_{0}}} (\cos\phi+\alpha_{0}\sin\phi) & \sqrt{\beta_{1} \beta_{0}}\sin\phi \\ 
  \frac{1}{\sqrt{\beta_{1}\beta_{0}}}[(\alpha_{0}-\alpha_{1})\cos\phi-(1+\alpha_{0}\alpha_{1})\sin\phi] & \sqrt{\frac{\beta_{1}}{\beta_{0}}} (\cos\phi-\alpha_{1}\sin\phi) 
\end{pmatrix}
\]

To vary x and y phase advances, the 4x4 $T_{decoupled}$ matrix is used. The Twiss parameters are specified at the end of linac first pass.
\[
T_{decoupled}(\phi_{x},\phi_{y}) =
\begin{pmatrix}
   M_{x\leftarrow x} (\phi_{x}) & \boldsymbol{0}    \\
  \boldsymbol{0}  &  M_{y\leftarrow y} (\phi_{y}) 
\end{pmatrix}
\]

To introduce x-y coupling, the 4x4 $T_{coupled}$ matrix is used.
\[
T_{coupled}(\phi_{1},\phi_{2}) =
\begin{pmatrix}
  \boldsymbol{0}   & M_{x\leftarrow y} (\phi_{1})   \\
   M_{y\leftarrow x} (\phi_{2})   & \boldsymbol{0}
\end{pmatrix}
\]
Both 4x4 matrices are symplectic.

\clearpage
\section{Halo\Leader{Dave}}

``Halo'' comprises the notionally diffuse components of the beam that can potentially sample large amplitudes. It is typically multicomponent, non-Gaussian, and mismatched to the beam core. As a result, it often is at very low density and may not be observable without use of extremely large dynamic range (LDR) diagnostics \Ref{Evtushenko12_01}. Halo emittance can readily exceed ERL transport system acceptance, leading to beam loss at aperture restrictions and/or locations with large lattice beam envelope function. During high-power CW operation, halo possesses  significant levels of beam power, and thus is a serious machine protection concern - it has, in fact, posed one of the most severe operational challenges for all ``true'' CW SRF ERLs (those with full beam power exceeding installed linac RF drive). 

There are many sources of halo. In the electron source, scattered light generated in the drive laser transport can illuminate active areas of the cathode outside the nominal source spot. Reflections off the cathode and gun structure can generate photoelectrons at essentially arbitrary phase and location. Masking of the cathode mitigates, but does not eliminate, the resulting large amplitude beam components. Depending on the choice of cathode material, cathode relaxation may lead to temporal tails at bunch formation. When operating with subharmonic bunch repetition rates, finite extinction ratio in drive laser pulse train gating can lead to so-called ``ghost pulses'', which will (due to their very low charge) be mismatched to the injector focusing (which is optimized for high charge) and will evolve as halo. The beam dynamics of of bunch formation and capture also generate halo \Ref{Evtushenko11_01}. Finally, field emission in the gun structure or in early cells of the first injector SRF cavities can be captured and accelerated, forming well defined beam components \Ref{EvtushenkoPC_01}.

Various beam dynamical processes contribute to halo formation at higher energy. Intrabeam scattering, especially Touschek scattering, can potentially lead to significant intensity at large amplitude. Beam/gas scattering similarly contributes. Very bright and/or intense beams can, in principle, evolve halo in the form of long beam tails generated via nonlinear collective effects, CSR, LSC, microbunching, wakefields, and beam interactions with environmental impedances. 
	
Halo management is critical to successful ERL operation. As noted, this has been a primary operational challenge in all high power SRF ERLs operated to date (see, e.g. \Ref{Alacron13_01}); beam loss must be avoided so as to avoid hardware damage and/or activation. This will be particularly critical in CBETA because of the widespread use of permanent magnet materials. At high full energy beam powers (MW levels), typical losses - which must be restricted to $\sim$ W/m levels - thus call for halo control and confinement at the few part per million. Machine design and operations must allow for beam components with emittances significantly exceeding that of the beam core, and which are not characterized by the same Twiss parameters. In particular, apertures cannot be established by use of ``rms beam sizes'' as these are not directly connected to the halo (inasmuch as it has different emittance, phase space distributions, and envelopes  - which themselves may be mismatched to the transport lattice focusing structure that has been optimized for the core \Ref{Douglas10_01}.

ERL designs thus generally provide a ``working aperture allowance'' to accommodate largely unknown and machine-dependent effects. LDR (with as much as six orders of magnitude range) instrumentation is needed, and control algorithms that optimize halo transmission as well as core behavior are required to achieve very high powers. Collimation can prove palliative - but as in ring injection systems, it is not in general curative - and has not yet been successfully demonstrated in a high-power ERL environment, despite testing on the JLab IR Demo \Ref{NeilPC_01}. Sensitivity of system designs to halo can be evaluated by evaluating halo ``maps'' \Ref{Johnson02_01, Evtushenko11_02} - which are basically the envelope of all possible orbits in the machine. This can be generated by displacing the beam in position and/or angle at regular, closely spaced locations along the beamline and tracking its motion throughout the rest of the system. Transverse halo maps determine potential aperture restrictions and possible points of significant beam loss (and thus serve to guide the location of beam loss monitors for machine protection and/or collimators \Ref{Johnson02_01}; longitudinal maps define a momentum aperture that can be compared to the deviations anticipated from Touschek scattering. Similar interpretation of intrabeam scattering and beam-gas scattering results will provide guidance on potential aperture requirements and restrictions.









\ifdefined \buildingFullDocument

\renewcommand{\FiguresDirectory}{ffag_magnets/figures}

\else
\newcommand{\FullDocumentRoot}{..}
\newcommand{\FiguresDirectory}{figures}

\begin{document}
\fi

\chapter{FFAG Magnets \Leader{Dejan}}\label{chapter:ffag_magnets}

The Halbach magnet type was selected after multiple confirmations were obtained from the 
building and measuring of 12~prototype CBETA magnets for the lattice with a 
maximum energy of $E_\text{max}=250$~MeV. This decision was due to the following:
\begin{enumerate}
\item Extremely magnetic field quality was obtained after wire shimming correction.
\item Magnetic field measurements of the two Halbach magnets, placed with a spacer between 
them, showed no cross talk between the magnets. After the second magnet was radially rotated 
by $90^\circ$ the obtained results were the same without measurable effect from the other magnet. 
\item The obtained errors in the integral gradients had a relative RMS of 0.29\%. 
\item Measurements of the window frame correctors placed around the the Halbach magnet showed 
very little effect on the quality of the Halbach-produced magnetic field. The very small difference 
occurred due to a value of $\mu_r=1.025$ instead of exactly $\mu_r=1.0$. The superposition 
theory of the magnetic field is shown very clearly in measurements.
\item Very good confidence was obtained in assembly procedure and production steps.
\item The size of the Halbach magnets is within a diameter of less than 15.1 cm, while with the aluminum 
block surrounding it, it fits inside a frame of 17.8 cm diameter.
\item The system of a window frame containing the quadrupole, horizontal, and vertical dipoles 
is a standard, well-known way of making correctors, and is used everywhere in beam lines, synchrotron light sources, etc.
\item Electron orbits are placed in the middle of both focusing and combined function magnets, where
there is the best magnetic field quality. The Halbach magnet design is made such that the electron beam orbit 
variations are in the middle of the focusing quadrupole and the defocusing combined function magnet. 
\item It is easy to fit the Halbach magnets around the round vacuum pipe. This can be very advantageous 
in using the four-button Beam Position Monitors (BPMs) and can provide additional significant 
savings in the BPM electronics.
\end{enumerate}

The CBETA lattice design requires a strong bending element in the middle 
of the basic cell with a defocusing gradient. This is a defocusing combined function magnet 
and is surrounded by focusing quadrupole magnets. The NS-FFAG principle is based on a very strong 
focussing, linear radial magnetic field, and with the minimum of the dispersion function to allow large 
momentum range. It has been shown \Ref{trbojevicPRSTAB1,Machida} that the optimum solution 
is a triplet solution, with the defocusing combined function magnet in the middle,
very similar to the light sources low emittance lattice requirements. The Halbach magnets 
represent the most efficient way of using the permanent magnet material in creating strong magnetic field but 
using the smallest possible volume for obtaining the combined function or quadrupole magnets.
The magnetic field dependence on the radial axis is supposed to be linear function $B_{x} = B_{0} + G \cdot x$. 
Unfortunately in the CBETA project the available space is very limited and bending radius of the FFAG 
cells is defined as $R=5$~m. This requires very short magnets with the doublet, not triplet, structure
for the most effective filling factor. 

Due to the short magnet lengths relative to their apertures in CBETA, the magnetic field is far from being 
the hard edge type with a long flat part. As it will be shown later, the magnetic field profile in the
longitudinal direction varies very much, and the end fields from one magnet overlap into the next. Previous measurements confirmed the superposition theory. The dynamical aperture, or acceptance, of the NS-FFAG with
linear magnetic field dependence is extremely large. The magnetic field distribution obtained by the 
OPERA 3D field calculations showed that the flat dependence of the field does not exist and the field is 
varying from the maximum value at the middle of the magnet with a bell shape curve dropping down to a small value 
in the drift region. The field is very nonlinear. To obtain design values for orbit offsets and horizontal 
and vertical betatron tunes, the Halbach magnet design needs adjustment of the gradient strengths until the 
designed values are obtained. It becomes very clear that the infinite acceptance of the NS-FFAG will drop 
significantly due to non-linear magnetic fields, but it remains large enough for safe operation. In addition, a 
robust correction system allows very good orbit and tune control.

In this chapter, we present the magnetic design of the Halbach magnets and
the magnetic field accuracy obtained by prototypes.  Separate sections are dedicated to the window-frame 
corrector design and the assembly and disassembly procedure.

\clearpage 
    \section{Halbach-type Design\Leader{Stephen}}
An iron-free Halbach permanent magnet design is proposed for CBETA, which allows 150~MeV top energy (36~MeV linac energy gain).  Halbach magnets are smaller in size than iron-yoked magnets and exhibit less cross-talk.  This design is described in the subsections below.

\subsection{Comparison of Features vs. Iron Poled Magnets}
A summary is provided in \Tab{tab:halbachcomparison} to illustrate the technology differences between choosing between a Halbach magnet design and an iron-dominated permanent magnet design in an accelerator.  Permanent magnet materials all have a temperature dependence and this can be compensated in the magnet in various ways.  The iron quadrupole uses a technique from the Fermilab recycler where the permanent magnet blocks sandwiched in the iron yoke are mixed with NiFe alloy whose magnetization contribution varies in the opposite way as magnetization of the blocks, to provide a temperature range of 20$^\circ$C or more with virtually no field strength variation.  In the Halbach magnets, the field and magnetization directions are not parallel, so this method does not work because the NiFe alloy would not provide compensating magnetization in the correct direction.  Instead, the dipole and quadrupole correctors, which would be present in the design anyway, are used to compensate the field variation, which manifests as an overall reduction factor in field strength and is therefore linear as a function of position like the magnets themselves.

\begin{table}[htb]
\caption[]{Comparison of iron and Halbach-type magnets.}
\begin{tabular*}{\columnwidth}{@{\extracolsep{\fill}}p{3cm}p{6cm}p{6cm}}
\toprule
&	Iron Poles&	Halbach \\
\midrule
Field quality + tuning	&Determined by iron pole shape.  Adjustment would be via conventional pole shimming.&	Determined by block magnetization vectors.  Adjustment via floating shims/iron wires just inside aperture.\\\\
Field strength + tuning&	Iron shunts to partially short-circuit flux applied to outside.  Also block pre-measurement and sorting.  EM quad corrector coils around poles. &	Determined by block magnetization vectors.  Tune with EM normal quad and dipole online correctors (see ‘correctors' below).\\\\
Temperature sensitivity + compensation&	0.1\%/K for NdFeB but can (at $\sim$20\% strength penalty) incorporate NiFe material to passively compensate.	&0.1\%/K for NdFeB, cancelled by using EM normal quad and dipole online correctors.\\\\
Cross-talk in doublet + compensation	&Few percent cross-talk, can be corrected with shunts.&	Negligible cross talk, mu$\sim$1 linear field superposition.\\\\
Correctors (online/EM)&	Normal quadrupole can be coils would around each pole.  Others require special coils put within the bore.	&Window-frame outside Halbach magnet using field superposition, because Halbach is magnetically transparent.\\
\bottomrule
\end{tabular*}
\label{tab:halbachcomparison}
\end{table}

To compensate temperature changes, the correctors are set using data from the orbit position feedback (although a local magnetic field monitor could in principle be used).  It is possible to circulate channels of water around the edge of the holder for the Halbach magnet blocks, to stabilize their temperature to so that heat from the surrounding window-frame corrector's coils do not significantly affect the temperature of the permanent magnets.

\subsection{Halbach Magnet Design}
\begin{figure}[tb]
\centering
\subfloat[]{\includegraphics[width=0.4\textwidth]{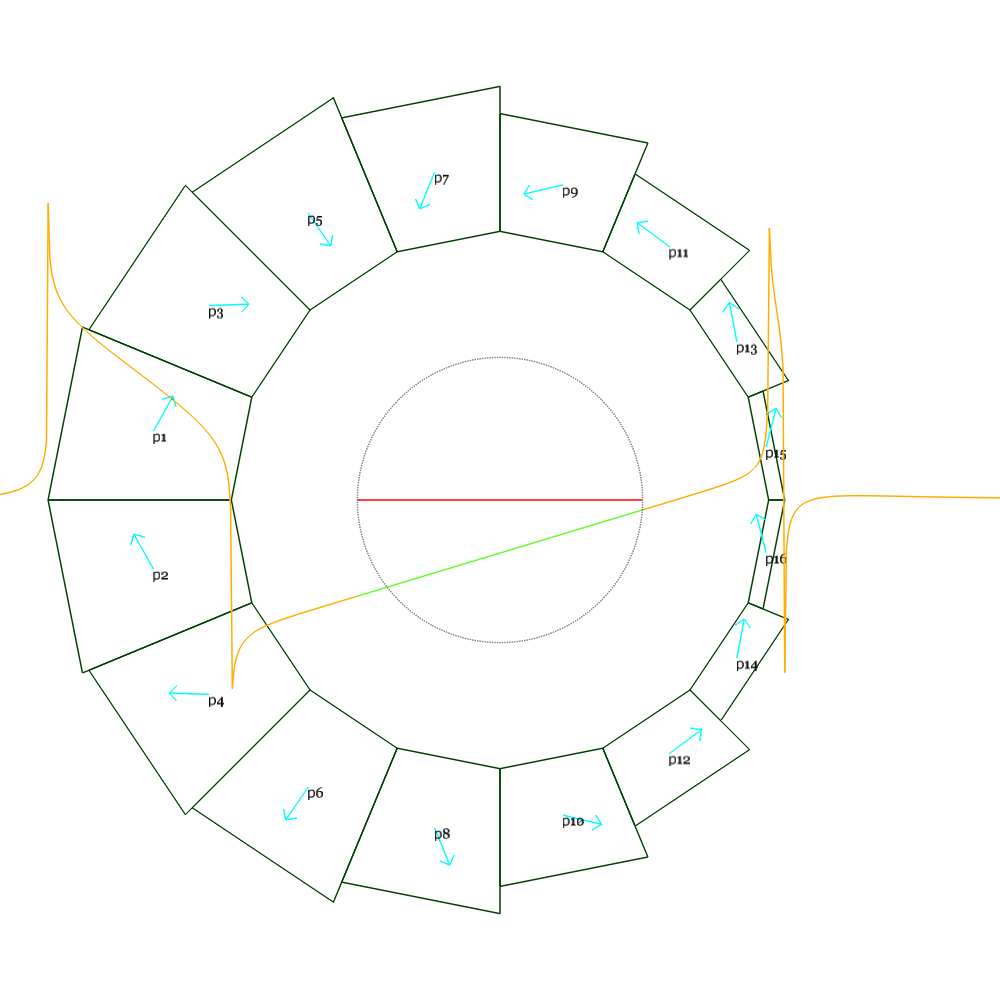}\label{fig:left}}
\hspace{0.05\textwidth}
\subfloat[]{\includegraphics[width=0.4\textwidth]{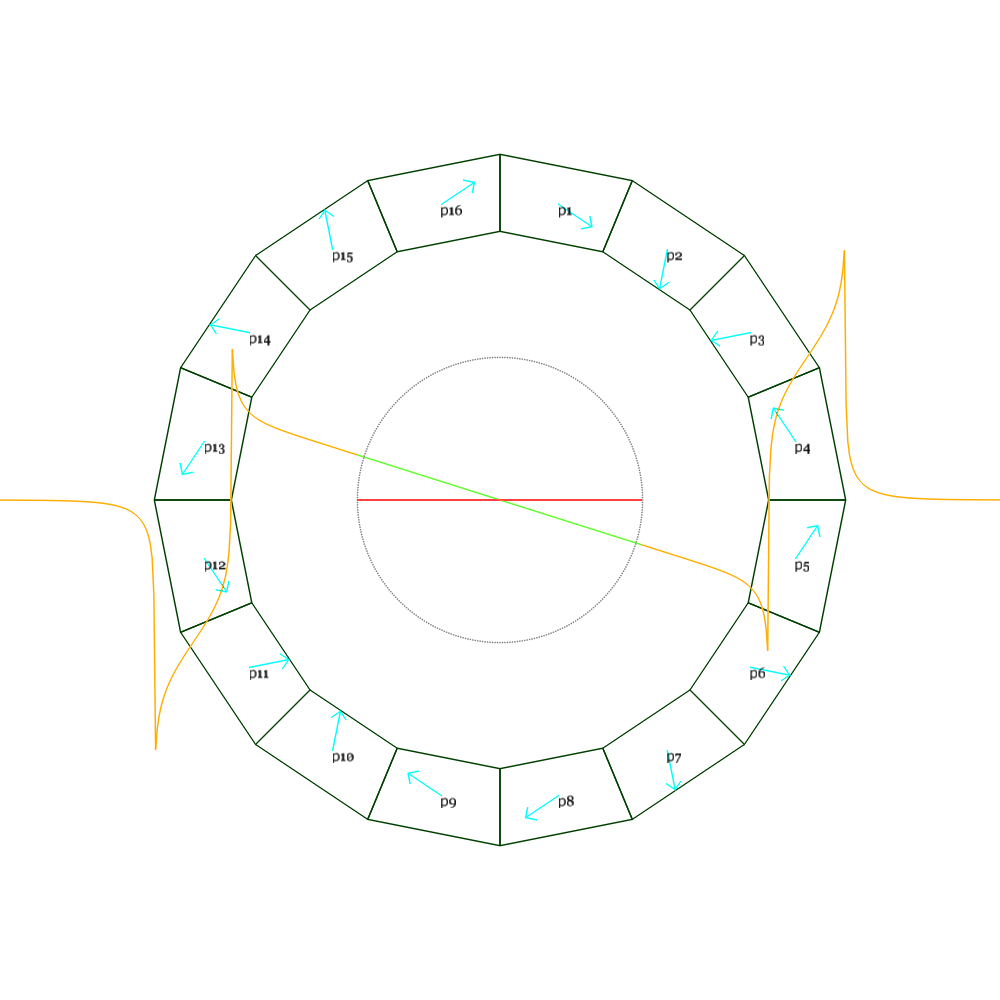}\label{fig:right}}
\caption[]{(Left) BD magnet.  (Right) QF magnet.  Orange graph is vertical field component $B_y$ on the y=0 axis, with varying x position.  The green segment is the field within the region required by the beams ($\pm$23mm), which is meant to be linear.  Magnetization axis is shown by blue arrows in each block.}
\label{fig:halbach-magnets-firstgirder}
\end{figure}

The optimized FFAG cell requires the QF magnet to be very close to a symmetrical quadrupole, i.e. with zero field at the center.  To simplify matters, the bore location is adjusted slightly so that QF really was exactly symmetrical, so that its design is that of a conventional Halbach quadrupole.  Cross-sections of the two magnets are shown in \Fig{fig:halbach-magnets-firstgirder}. The latest studies on the FFAG magnets have been presented at IPAC 2017 \Ref{IPAC2017:THPVA151, IPAC2017:THPIK007, IPAC2017:THPVA094, IPAC2017:TUPIK130}. 

The BD magnet, on the other hand, contains a significant dipole component.  In fact, all of the beams go through the negative $B_y$ field region, which bends electrons clockwise in the L0E hall.  The design of the BD magnet is not a conventional Halbach arrangement: it requires a combination of dipole and quad, whereas conventional annular arrangements can only do one pure multipole at a time.  It was considered to nest conventional dipole and quadrupole Halbach magnets, but the outer magnet has to be quite large in that case.  It was also noticed that on one side of the nested magnet the magnetizations were mostly cancelling anyway, so optimization was run on a design with only a single layer of permanent magnet wedges, but with variable thickness and different magnetization directions.  This achieved a very accurate ($<10^{-5}$ in the linear model) combined function integrated field as required, a result that was replicated to high accuracy ($\sim 10^{-4}$) by OPERA-3D simulations.  It also uses much less material than a nested design.

\subsubsection{Magnet Simulation and Codes}
Two codes were used in the design and simulation of these Halbach magnets, which have shown good agreement as shown in this section.  The simpler of the two is PM2D written by Stephen Brooks, which is a current sheet approximation of the fields from permanent magnet polygons in 2D.  This provides an accurate model of the ``average'' field (integrated field divided by permanent magnet piece length) through the magnet, provided two conditions hold:
\begin{itemize}
\item The materials stay in the linear part of their B-H curve.  In fact, if this is violated, the magnets will experience permanent demagnetization, so any valid design ought to satisfy this condition.  PM2D can also evaluate the demagnetizing flux from the other blocks at any point to ensure it does not go beyond the coercive force ($H_{cj}$) of the material.
\item $\mu_r=1$ for all materials.  This is almost true of NdFeB, which has a $\mu_r$ of about 1.025.
\end{itemize}

PM2D was used for the initial optimization of the wedge sizes in the BD magnet, which tried to reduce the error multipoles to zero by changing their thickness and magnetization direction independently keeping the required symmetry in the $y=0$ mid-plane.  This requires many iterations of the design to be simulated, so a faster code is preferred during this design stage, before coordinates of the wedge corners are generated as input for the 3D magnet simulation.

The second code used, by Nick Tsoupas for 3D simulations, is OPERA-3D, which is the industry standard.  Very good agreement was attained between the two codes (on integrated field multipoles) when the materials were not in the demagnetizing regime.  Once the design was set, OPERA-3D was always used to do the final simulation and 3D field map generation.

\begin{figure}[tb]
\centering
\includegraphics[width=0.95\textwidth]{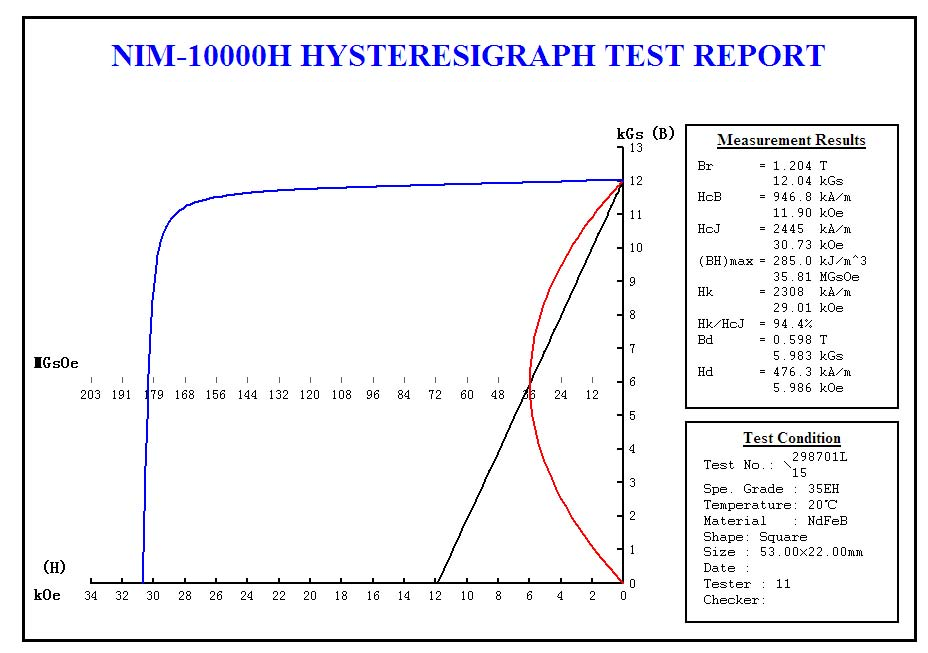}
\caption[]{B-H curve of the AllStar Magnetics N35EH NdFeB permanent magnet material.}
\label{fig:halbach-BHcurve}
\end{figure}
 
Running in OPERA-3D required that a specific material grade and B-H curve was chosen for the permanent magnet blocks.  These grades and curves vary by manufacturer, so a grade from AllStar Magnetics was selected, which is the manufacturer for blocks for the CBETA first girder prototype magnets currently under order and shipping in March~2017.  The grade N35EH was selected, which combines a medium strength of 35 MGauss.Oe (the maximum available being $\sim$53 MGauss.Oe) with a good resistance to external demagnetizing fields.  This is what the ``EH'' suffix means: a strong resistance to heat, which stems from its high $H_{cj}$ demagnetizing field value ($\ge$3.0 T or 30kOe) at room temperature.  Its B-H curve is shown in \Fig{fig:halbach-BHcurve}.  The strength translates into a residual field $B_r$ of 1.207~T.

After OPERA-3D models were run, a best fit with the magnetization ``$B_r$'' value used in PM2D, which assumes $\mu_r=1$, was found (1.1939~T), which represents the average magnetization from the material including the small reduction from regions of reverse flux with $\mu_r$ being slightly larger than 1.  This lies between $B_r$ and $H_{cb}$ of the material as expected.  With this value, the PM2D designs could be loaded directly into OPERA-3D (with the N35SH material) and the strength would be correct, with no further design modifications required.

\subsection{Tracking and Compatibility with FFAG Lattice}
Once OPERA-3D field maps have been generated, they can be loaded back in to the Muon1 \cite{Muon1} tracking code, which is the same code used for the original lattice optimization done with field models rather than field maps.
 
\begin{figure}[tb]
\centering
\includegraphics[width=0.95\textwidth]{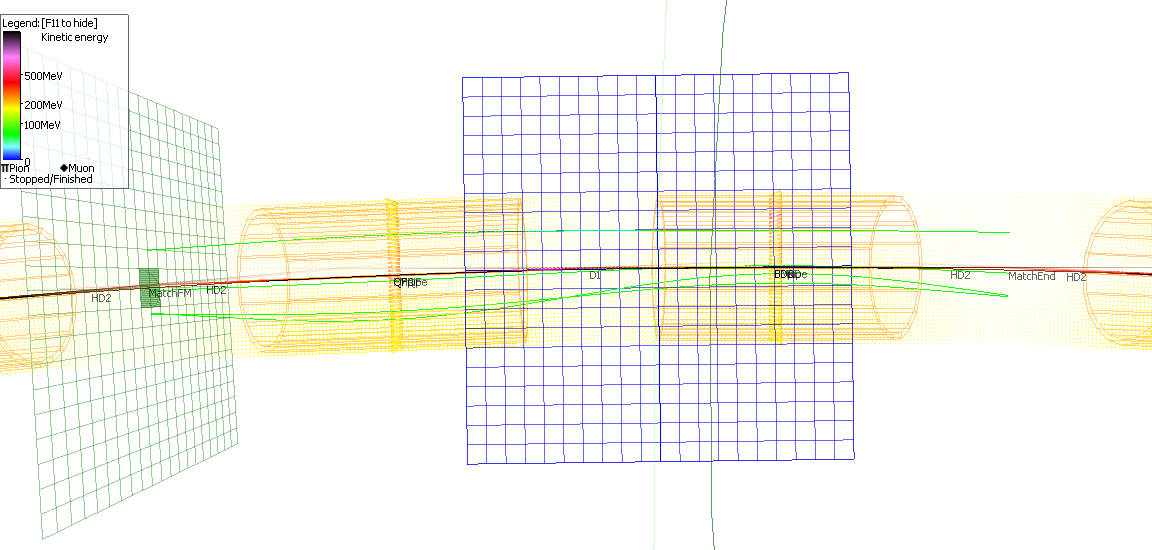}
\caption[]{Matched orbits (green) for the four CBETA energies through an FFAG arc cell made with OPERA-3D field maps generated from Halbach magnets.  The orange cylinders represent the approximate apertures of the vacuum pipe and the grids are 1~cm per square.}
\label{fig:halbach-tracking}
\end{figure}

\begin{figure}[tb]
\centering
\includegraphics[width=0.7\textwidth]{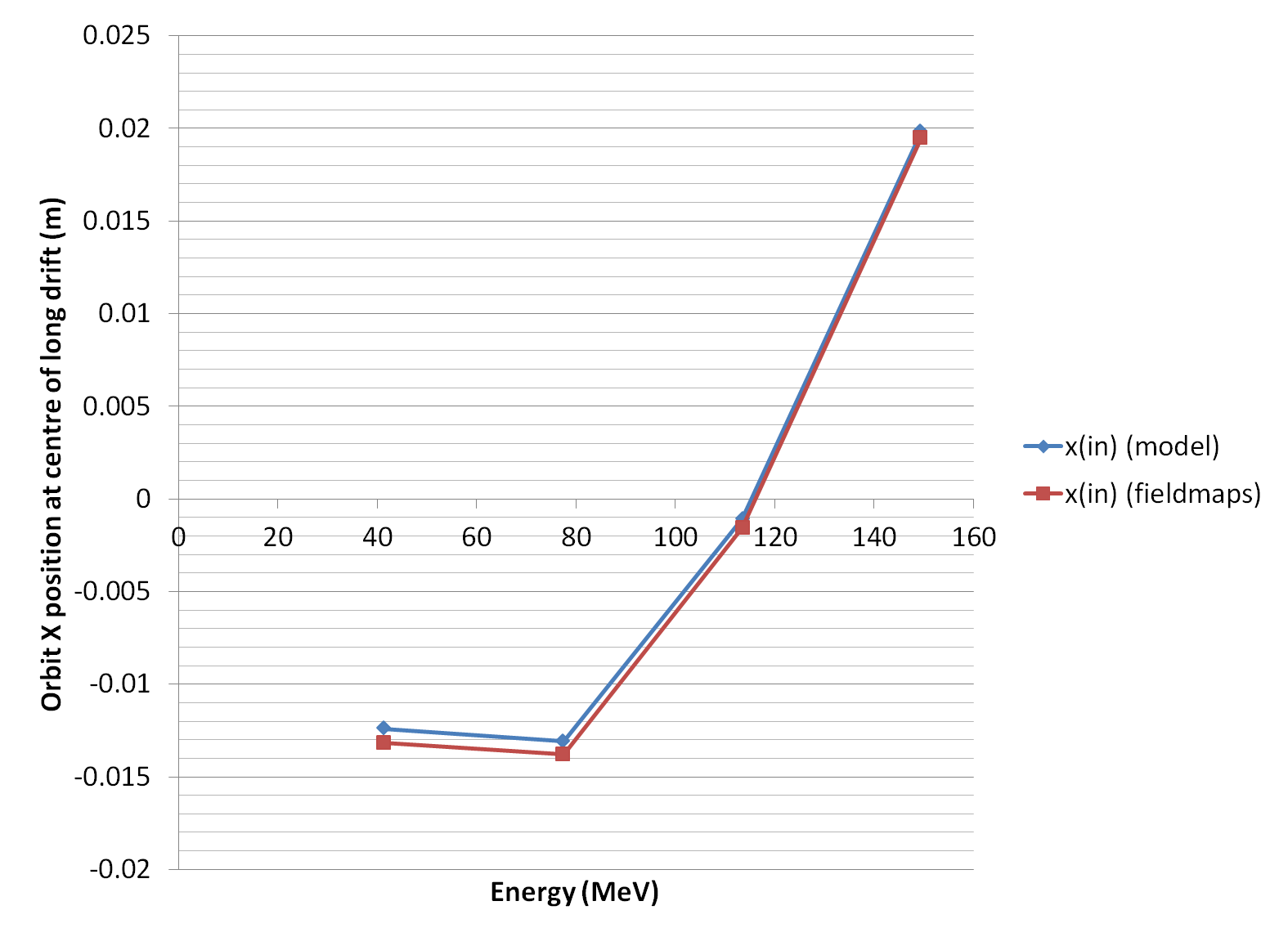}
\caption[]{Transverse position X of the four closed orbits, as a function of energy, at the midpoint of the long drift in the CBETA FFAG arc cell.  Blue dots are from a Muon1 simulation using field models and red dots from a Muon1 simulation using OPERA-3D field maps.}
\label{fig:halbach-orbitcomp}
\end{figure}

\Figure{fig:halbach-tracking} shows such a simulation, where Muon1 has found ``closed orbits'' for each energy, which exit the cell at the same position and angle that they enter.  The closed orbits found through field maps will be slightly different than those found for the original field model in the lattice-design optimization, but as shown in the \Fig{fig:halbach-orbitcomp} and \Tab{tab:fieldmappositions}, the discrepancy is not very large ($<$1mm).

\begin{table}[tb]
\caption[]{Transverse position $x$ of the four closed orbits, calculated with Muon1 soft-edged field models or OPERA-3D field maps, at the midpoint of the long drift in the CBETA FFAG arc cell.}
\begin{tabular*}{\columnwidth}{@{\extracolsep{\fill}}rrr}
\toprule
Energy (MeV)&	$x$ (m) (model)&	$x$ (m) (field maps)\\
\midrule
150&0.019811&0.019505\\
114&	-0.001057&-0.001544\\
78&-0.013086&-0.013790\\
42&-0.012414&-0.013162\\
\bottomrule
\end{tabular*}
\label{tab:fieldmappositions}
\end{table}
 
This good agreement is partly due to a fortunate choice of fringe field length in Muon1's soft-edged Maxwellian field model.  Muon1 models the fall-off of multipole components near the entrance of exit of a magnet as proportional to $\frac12+ \frac12 \tanh(z/f)$ where $z$ is the longitudinal position relative to the magnet end and f is a ``fringe length'' parameter ($f$).  It was chosen to be 2.5~cm here, roughly the same order of size as the magnet apertures.  Detailed studies suggested the best agreement with these field maps is obtained with $f=2.7$~cm.  For these short magnets in CBETA, the fringe field makes up a large part of the field so it is important to include it consistently (some hard-edged models do not have good agreement with the optics).

\begin{figure}[htbp]
\centering
\includegraphics[width=0.7\textwidth]{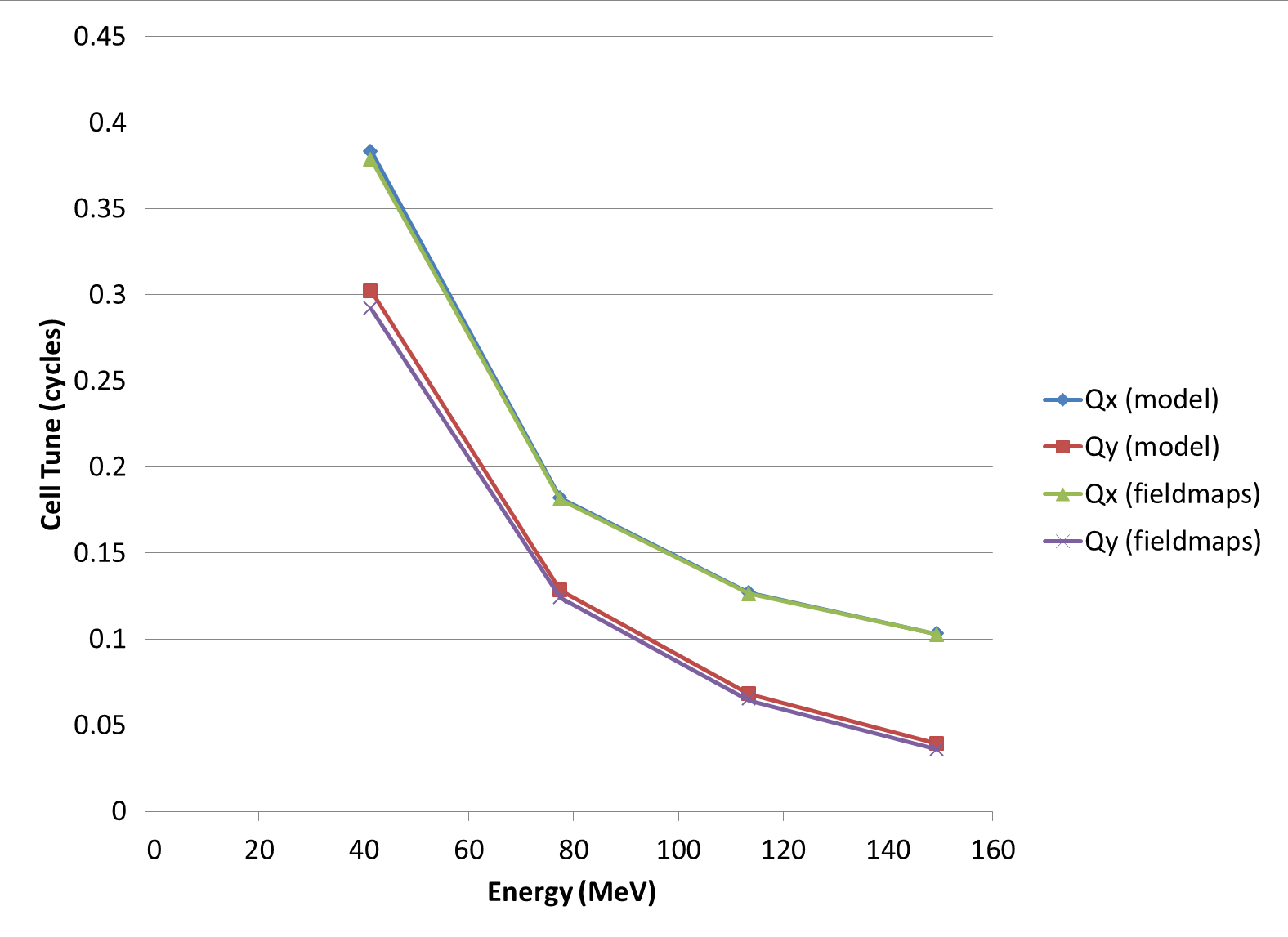}
\caption[]{Comparison of the calculated X and Y tunes of the FFAG cell using Muon1's model field and OPERA-3D field maps as a function of energy.}
\label{fig:halbach-tunecomp}
\end{figure}

The closed orbit matching process also determines the shape of the beam (optical alpha and beta functions) that will be preserved on traversing once through the cell.  This also allows the single-cell tunes in the $x$ and $y$ planes to be calculated.  A similar comparison of tunes from the field map versus the original optimizer's field model is shown in \Fig{fig:halbach-tunecomp} and \Tab{tab:fieldmaptunes}.
 
\begin{table}[htb]
\caption[]{Comparison of the calculated X and Y tunes of the FFAG cell using Muon1's model field and OPERA-3D field maps.}
\begin{tabular*}{\columnwidth}{@{\extracolsep{\fill}}lcccc}
\toprule
Energy (MeV)&	$Q_x$ (model)&	$Q_y$ (model)&	$Q_x$ (field maps)&	$Q_y$ (field maps)\\
\midrule
150&0.102835&0.039079&0.102565&0.035741\\
114&0.126853&0.068072&0.126273&0.064546\\
78&0.182031&0.128300&0.180967&0.124042\\
42&0.383192&0.301829&0.378423&0.291816\\
\bottomrule
\end{tabular*}
\label{tab:fieldmaptunes}
\end{table}

The cell tunes are important because they determine the limits on the stability of the beam (0 and 0.5 being the unstable limits) and its response to errors, the tune determining the frequency of error oscillations.  The largest discrepancy between field map and model field is found in the low-energy (42MeV) beam vertical tune ($Q_y$), where the model predicts 0.3018 and the field maps predict 0.2918, a difference of 0.0100 cycles per cell.  This is not a large enough difference to put the beam into a resonance or drastically affect the optical behavior of the machine.

\clearpage
\subsection{Halbach Prototype Magnet Series}
\begin{figure}[htbp]
\centering
\includegraphics[width=0.6\textwidth]{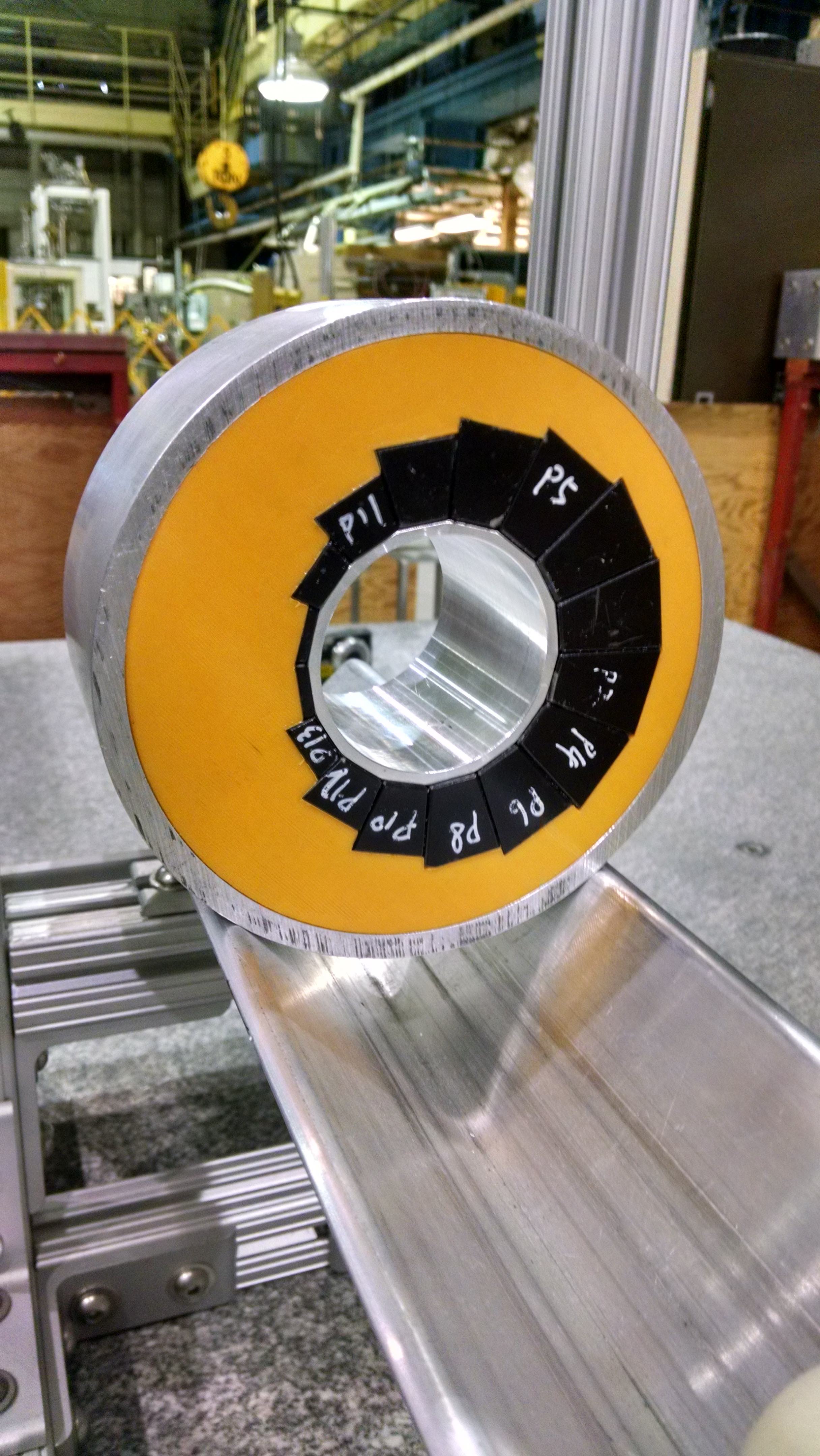}
\caption[]{An unshimmed prototype BD magnet awaiting measurement at BNL.}
\label{fig:halbach-Cbeta_proto_BD}
\end{figure}

\begin{figure}[htbp]
\centering
\includegraphics[width=0.6\textwidth]{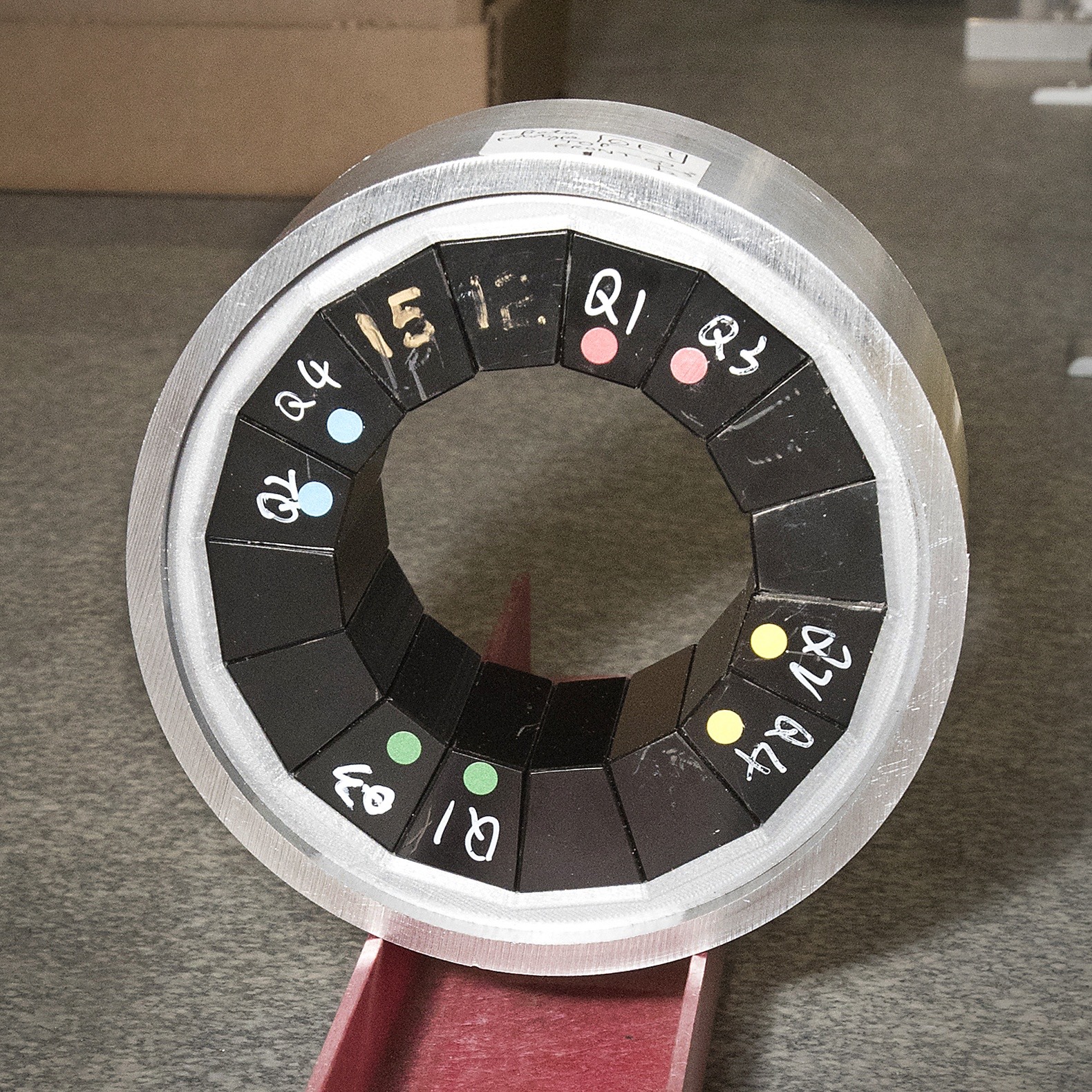}
\caption[]{An unshimmed prototype QF (pure quadrupole) Halbach magnet.}
\label{fig:halbach-Cbeta_proto_QF}
\end{figure}

A first round of prototype magnets for CBETA have been built, including the ``lopsided Halbach'' magnet BD (\Fig{fig:halbach-Cbeta_proto_BD} and \Fig{fig:halbach-Cbeta_proto_QF}).  Due to the 2--3 month magnet lead times, these are from an old lattice design Cell\_Brooks\_2015-12-11 rather than the most recent cell presented in this report, but they have many similarities.  In fact, the prototypes have about twice the field required because their lattice was designed for a 250~MeV top energy rather than the current 150~MeV.  A comparison of the magnets in the two versions is given in \Tab{tab:halbachprototypecomparison}.  Six magnets of each type (QF and BD) have been built and most of these have also been measured after one iteration of shimming discussed below.

\begin{table}[b]
\caption[]{Comparison of current lattice Halbach magnets to those of the prototypes ordered.}
\begin{tabular*}{\columnwidth}{@{\extracolsep{\fill}}lcccc}
\toprule
Parameter&	QF current&	QF prototype&	BD current&	BD prototype\\
\midrule
Length&	133.3mm&	114.9mm&	121.7mm&	123.7mm\\
Gradient&	-11.50 T/m&	-23.62 T/m&	11.00 T/m&	19.12 T/m\\
Dipole at center&	0&	0&	-0.3110 T&	-0.3768 T\\
Max good field radius&	23.0mm&	20.2mm&	23.0mm&	13.7mm\\
Block inner radius&	42.5mm&	37.2mm&	42.5mm&	30.7mm\\
Block outer radius&	55.8mm&	62.4mm&	72.9mm&	59.4mm\\
Max field in good field region&	0.26 T&	0.48 T&	0.56 T&	0.64 T\\
Max field at ``pole tip''&	0.49 T&	0.88 T&	0.78 T&	0.96 T\\
\bottomrule
\end{tabular*}
\label{tab:halbachprototypecomparison}
\end{table}

The NdFeB wedges were ordered from AllStar Magnetics rather than Shin-Etsu due to cost reasons, which means a larger magnetization angle error in the blocks of $\pm$5 degrees as specified by their factory.  However, the observed variation of magnetization angle within the batch received, based on Hall probe point field measurements, was much better than this, with an RMS below 1 degree.  The factory later confirmed that they could specify $\pm$3 degrees with 80\% of the blocks within $\pm$2 degrees on all but the smallest blocks of the BD magnet.

\subsubsection{Shimming Method with Iron Wires}

\begin{figure}[htbp]
\centering
\includegraphics[width=0.95\textwidth]{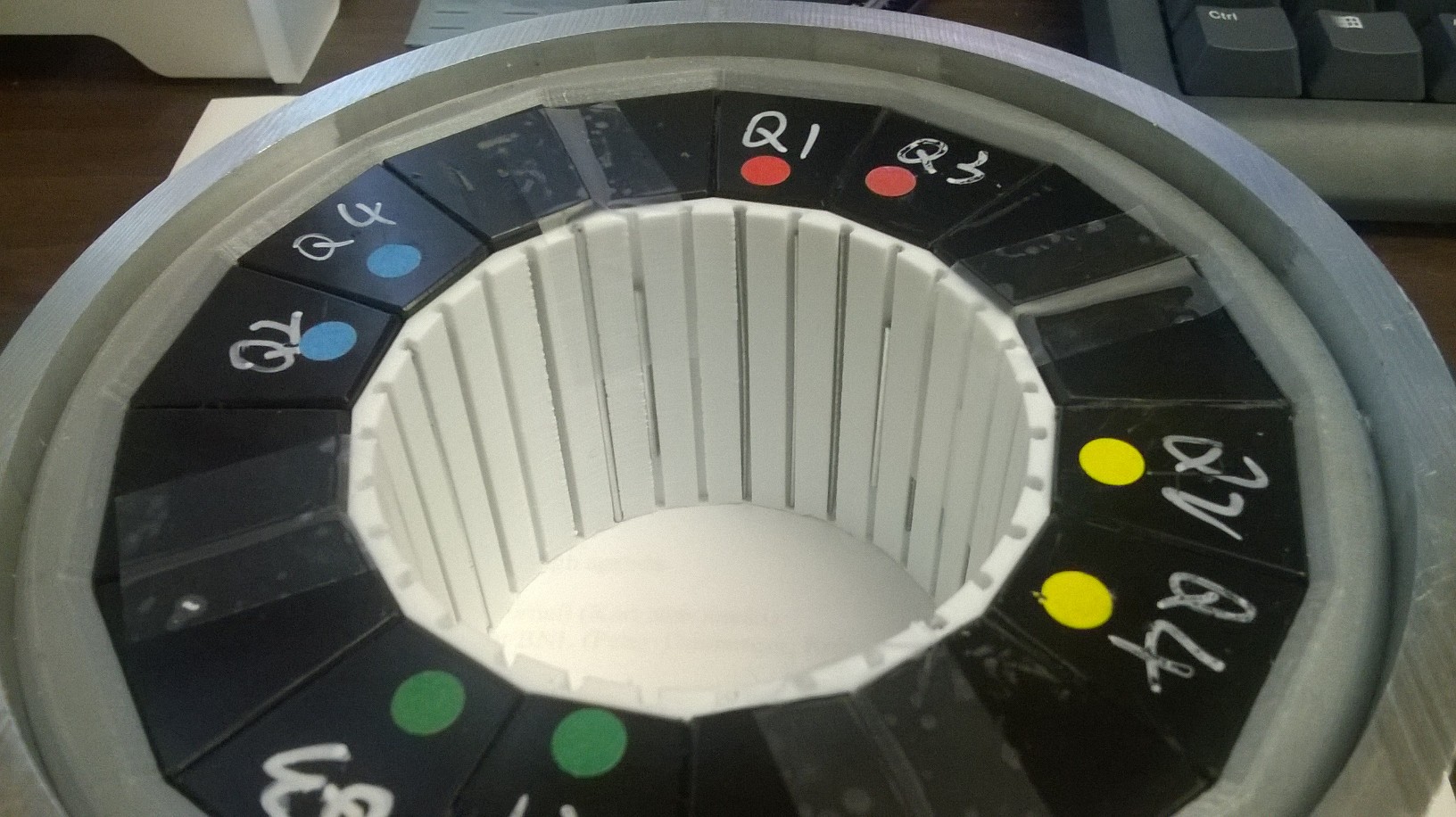}
\caption[]{A shimmed prototype QF Halbach magnet, showing the 3D printed plastic shim holder (white) containing iron wires.}
\label{fig:halbach-shimmed_proto_QF}
\end{figure} 

In order to get from the $\sim$1\% level magnetization errors present in the NdFeB blocks from the factory to a magnet with better than $10^{-3}$ field accuracy, a plastic shim holder containing iron wires of various sizes is inserted into the magnet bore, shown in \Fig{fig:halbach-shimmed_proto_QF}.  There are 32 wires evenly-spaced around the bore of the magnet, two per permanent magnet wedge, although in some cases holders for 48 or 64 wires were printed to increase the shimming strength.

Each wire is magnetized by the surrounding field of the magnet and produces an external dipole field, as would be produced outside a $\cos(\theta)$ cylindrical current winding.  The strength of these dipoles is proportional to the average cross-sectional area of each wire.  It was found that, given a rotating coil measurement of the unshimmed magnet, an optimizer can find arrangements of wire sizes that can cancel any multipole combination provided it is not too large.  This is now implemented as a program that takes the rotating coil output file and produces lists of wire lengths.  In the case of the prototype magnets, several wire thicknesses (.014, .018, .035, .041 and .063 inch diameter) were also allowed to give better resolution for small corrections.  It may be possible to simplify this and use a single thickness in the production magnets.

\subsubsection{Rotating Coil Results}

\begin{figure}[htbp]
\centering
\includegraphics[width=0.95\textwidth]{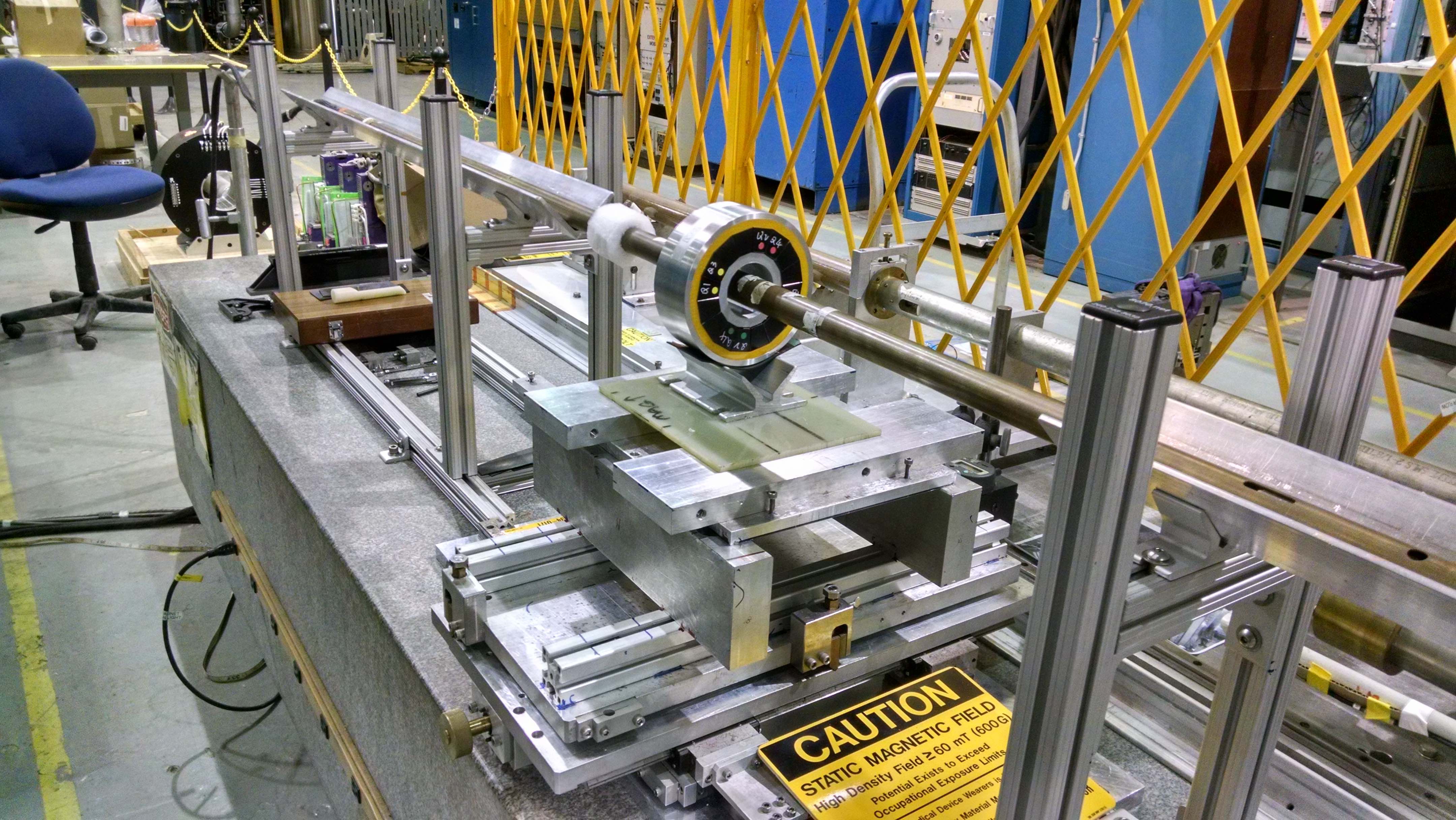}
\caption[]{A prototype QF Halbach magnet being measured on BNL's rotating coil.}
\label{fig:halbach-rotcoil}
\end{figure} 

\begin{figure}[htbp]
\centering
\includegraphics[width=0.8\textwidth]{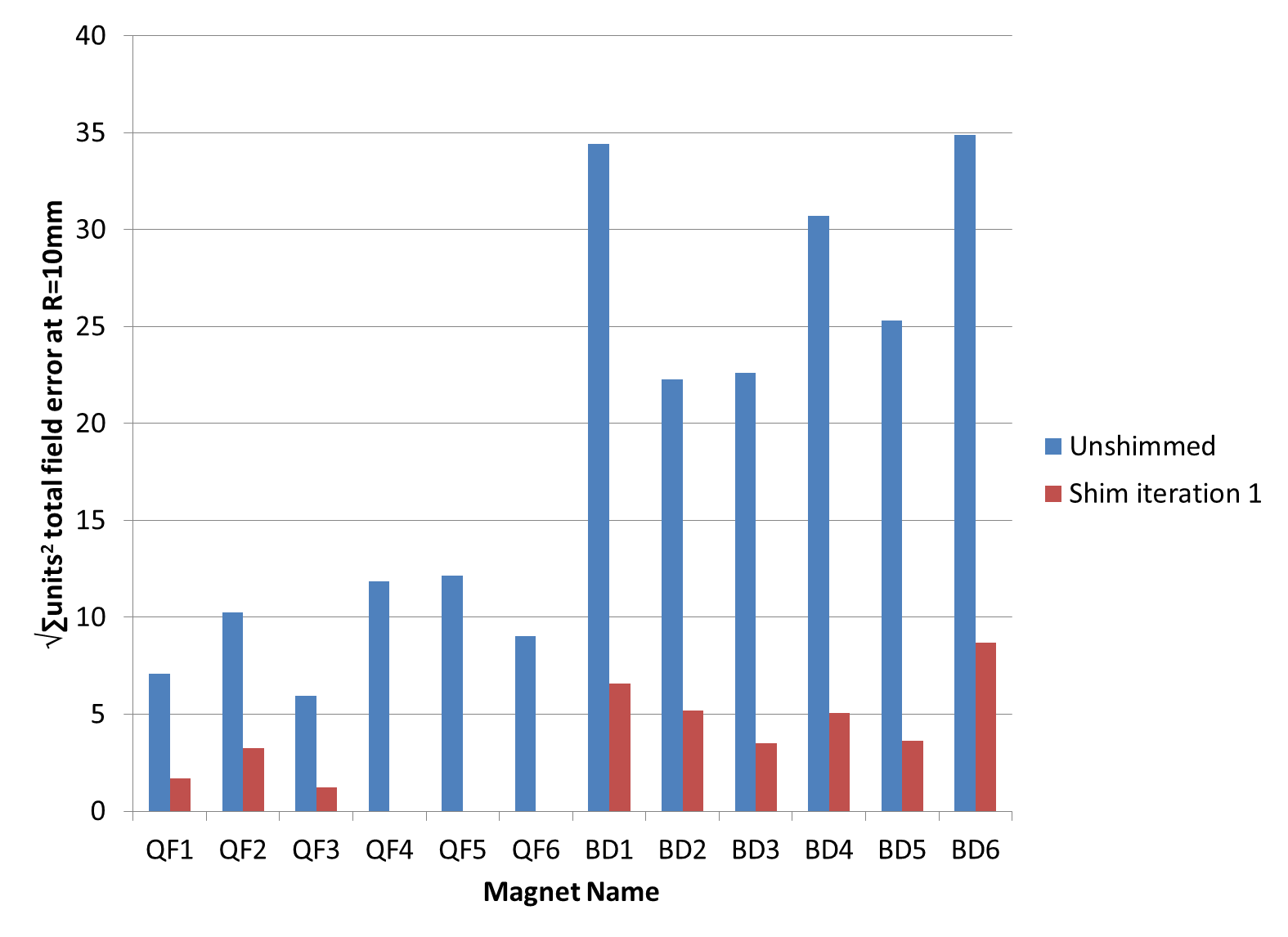}
\caption[]{The quadrature sum of higher harmonic units $\sqrt{\sum_{n=3}^{15} a_n^2+b_n^2}$ in the Halbach prototypes, converted to R=10mm for comparison (because only the smaller coil size would fit through the BD magnet aperture).}
\label{fig:halbach-proto-fom}
\end{figure}

The rotating coils at BNL (\Fig{fig:halbach-rotcoil}) were used to measure the multipole components of the prototype magnets before and after shimming.  The results for the BD magnets were measured with a coil of radius $\sim$12mm (normalized to R=10mm) because its aperture is slightly smaller than the QF magnets, which were measured with coil of radius $\sim$27mm (normalized to R=25~mm).  For comparison purposes, both can be normalized to the smaller radius and an RMS-like measure of the multipoles can be calculated in all cases, as shown in \Fig{fig:halbach-proto-fom}.  Normalizing the smaller coils to the larger radius is not recommended due to the rapid amplification of coil electrical noise in the higher harmonics by this conversion.  Note that both measurements are normalized to the quadrupole being 10000 units, even though the dipole is a larger field magnitude in the case of BD.

\begin{figure}[htbp]
\centering
\includegraphics[width=\textwidth]{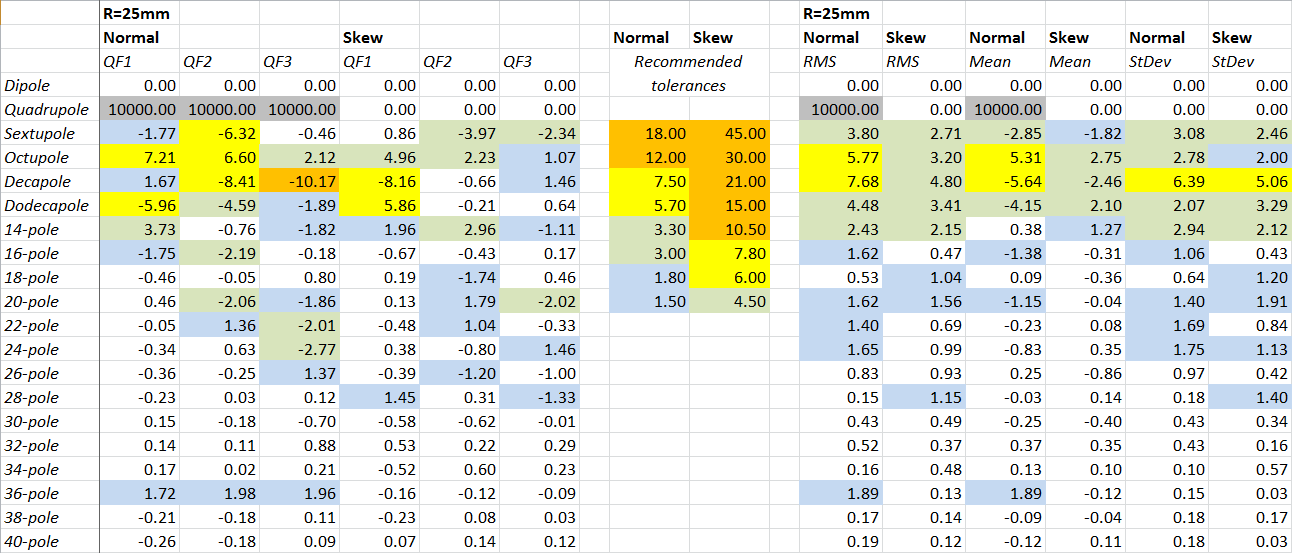}
\caption[]{Results at R=25mm from the three shimmed QF magnets measured so far.  (left) Harmonics in units; (centre) recommended tolerances for the FFAG magnets; (right) summary statistics, including overall RMS and separate average (systematic) and estimated population standard deviation.  Color code is for entries whose magnitude exceeds 1, 2, 5, 10 units.}
\label{fig:halbach-proto-qf-multipoles}
\end{figure} 

The results show that a single iteration of shimming gives approximately a fourfold reduction in the multipole content.  In the case of the QF magnet, this already brings the multipoles for the magnet below the tolerances required of the CBETA FFAG, as shown in \Fig{fig:halbach-proto-qf-multipoles}.  The BD prototype magnets have a larger multipole content at a given radius than the QF magnets but they also have a smaller aperture, which tends to make comparisons difficult.  It is expected that with larger apertures, BD magnets will also give good results.  There is also the option of doing a second shimming iteration, which has worked on a previous R\&D magnet, where the multipole content reduced from 29.62 $\rightarrow$ 6.69 $\rightarrow$ 1.94 units over two iterations.

\begin{figure}[htbp]
\centering
\includegraphics[width=0.8\textwidth]{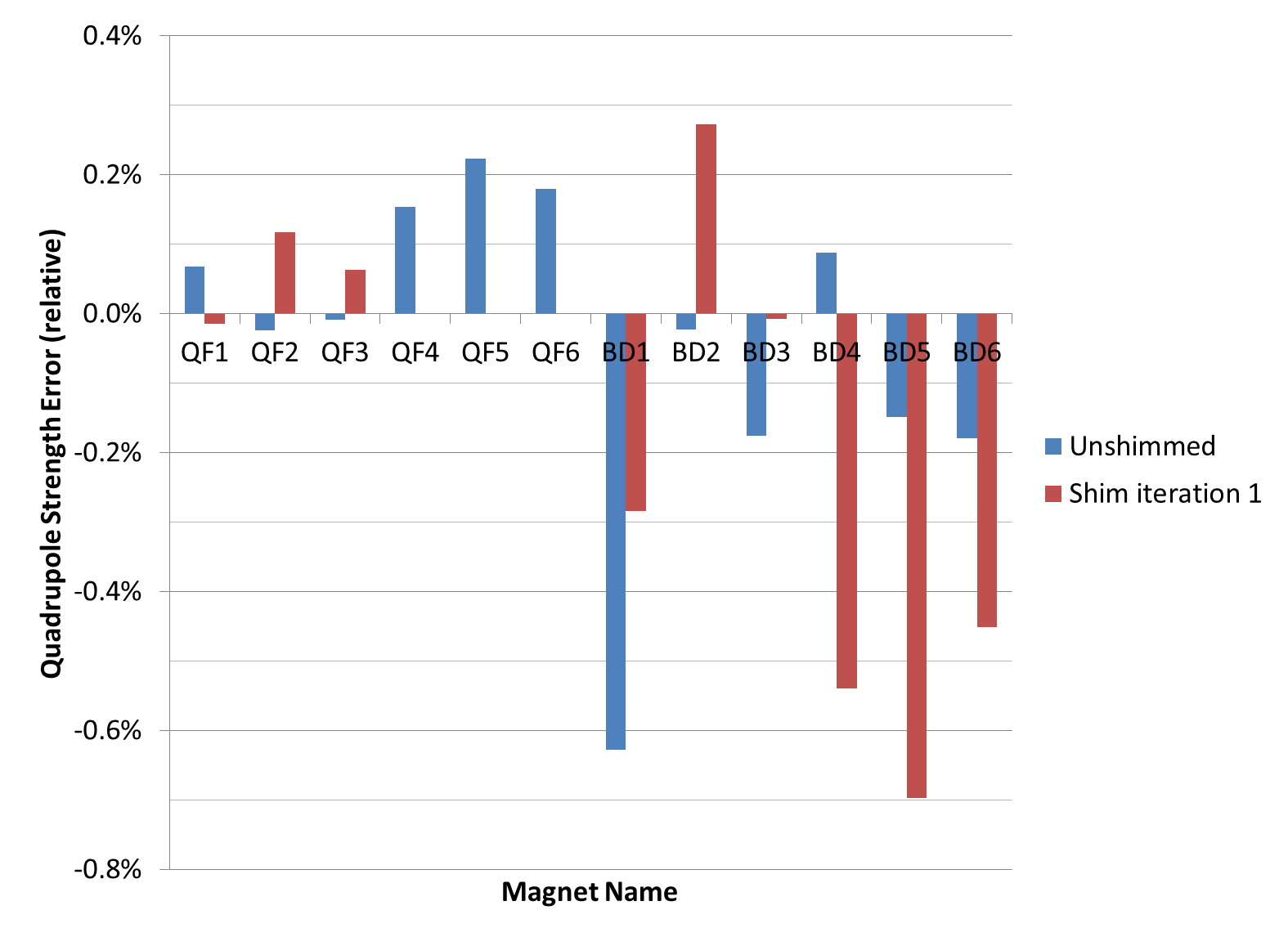}
\caption[]{Error in the quadrupole strength of each magnet.  This is potentially affected by temperature variation via the NdFeB temperature coefficient of $-1.1\times10^{-3}$/K.}
\label{fig:halbach-proto-quad}
\end{figure} 

The rotating coil also measured the absolute strength of the magnets, shown in \Fig{fig:halbach-proto-quad}.  As no attempt was made to temperature control the prototype magnets during the measurement and shimming process, this strength will pick up any differences in temperature between the two measurement times (which were separated by up to 6 months in some cases).  The wire correction method can target a particular gradient strength in the magnet (over a range of not more than 1\%), however the target gradient will be achieved only at the same temperature the first rotating coil measurement was done at.  The largest gradient error would correspond to a 6.3~K temperature change.

When the first girder magnets for CBETA are built, this shimming process will be done with a temperature-controlled water circuit running in the magnets and/or a reference magnet also placed in the coil so that the subsequent magnets are shimmed to the same strength regardless of thermal conditions.

\clearpage 
\subsection{Manufacturing Pipeline and Vendors}
Discussions are starting with magnet manufacturing companies about what they can build for CBETA.  The pipeline of magnet manufacture and assembly onto the machine breaks down into the four stages below.

\subsubsection{Permanent Magnet Wedges}
These will be purchased, directly or indirectly, from a company.  As mentioned previously, Shin-Etsu Corporation is a large manufacturer of the permanent magnet blocks with reasonably high quality.  AllStar Magnetics has also provided BNL permanent magnets block in the past (for instance the radiation damage experiment), although they specify larger tolerances on their magnetization angles.  Electron Energy Corporation (EEC) has recently succeeded in an SBIR proposal for CBETA and eRHIC magnet development. 
EEC manufactures both the blocks and magnet assemblies on-site in their machine shop.  Finally, VacuumSchmelze GmbH has been contacted by Holger Witte for magnet blocks for the iron-poled quadrupole.  Other companies not contacted yet include the undulator manufacturer KYMA.

Of these companies, AllStar generally provides the lowest cost but the least accurate magnetization vector guarantee ($\pm$3 degrees).  Shin-Etsu provides $\pm$1 degree tolerance with some additional cost for tooling.  EEC say even $\pm$0.5 degrees is possible but there is an associated cost because additional steps of demagnetizing the block, re-grinding it to an accurate shape and re-magnetizing it have to occur.

\subsubsection{Magnet Assembly}
Although in theory this could be done on the BNL or Cornell sites, it seems that several companies are willing to bid for this work and are capable of doing it.  EEC could be used as an end-to-end vendor for these first two steps.  RadiaBeam LLC will make assemblies and girders but have to get the PM blocks from another company.  They previously gave a cost estimate for assembling the CBETA magnets and girders and are the only ones to have significant accelerator field experience (in fact they also make Halbach magnets for electron microscopes).  Their absolute tolerances on positioning magnets on the girders were 0.1~mm.

\subsubsection{Shimming and Rotating Coil Measurements}
Discussions so far with magnet manufacturers are indicating that the rotating coil is a specialized piece of measurement equipment for accelerator applications.  None of the companies contacted so far have functioning rotating coils, although Radiabeam and EEC have Hall probes for field mapping.  The shimming method works best using a rotating coil, so this stage is likely to be done in the BNL magnet division, where they have done it before.

\subsubsection{Alignment and Girder}
Because the magnets will need to be removed to do separate rotating coil and shimming steps, a fully integrated manufacture (measurement while on girder) does not look possible.  Instead, survey fittings will be included in the non-magnetic body during the magnet assembly step and these will be used in the hall at Cornell to fit in with their on-site survey system.  The survey references may also be used in the rotating coil stage to ensure alignment between the magnetic field and the magnet holder.

\clearpage
\section{Window-Frame Correctors \Leader{Nick}}
\begin{figure}[tb]
\centering
\includegraphics[width=0.6\textwidth]{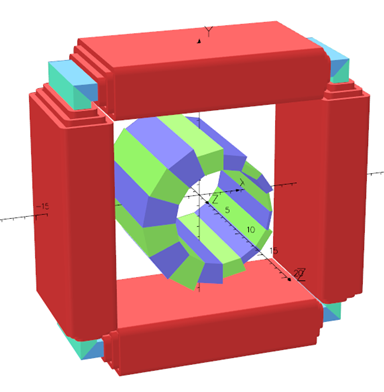}
\caption[]{A window frame magnet with four coils (one coil in each side) can generates a normal quadrupole field, by connecting to a power supply in series the two outer layers of each of the four coils, and a normal dipole field by connecting in series to a power supply the  innermost layer of the left and right coil, and a skew dipole  by connecting in series to a power supply the  innermost layer of the top and bottom coil.}
\label{fig:halbach-windowframe}
\end{figure}

\begin{figure}[tb]
\centering
\includegraphics[width=0.8\textwidth]{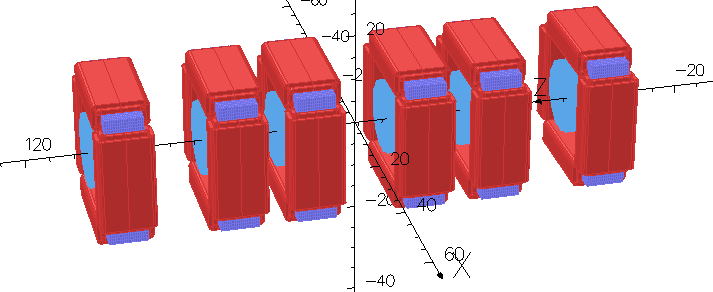}
\caption[]{Isometric view of six of the permanent magnets of the CBETA arc.}
\label{fig:halbach-sixcorrectorsiso}
\end{figure}

Each FFAG Halbach magnet will have a normal quadrupole corrector and either a normal dipole or a skew dipole corrector in a way that no two consecutive FFAG Halbach magnets have the same type of dipole multipole.  \Figure{fig:halbach-windowframe} shows an isometric view of a window frame magnet around an FFAG Halbach magnet. Each of the four coils is shown with three layers. The right and left inner layers or the top and bottom inner layers of the coils are powered with a single power supply to generate an integrated normal or skew dipole field of 1200~Gauss-cm respectively. The outer two layers of each of the four coils are connected to a power supply in a way to produce an integrated quadrupole field of 275~Gauss. The current density in the wires of the coils is 1 A/mm$^{2}$. 
\Fig{fig:halbach-sixcorrectorsiso} is an isometric view of six window frame magnets around the FFAG magnets.

Because the permanent magnets are fully saturated magnetically it allows the magnetic field of the corrector to penetrate the permanent magnet material with no distortion of the field therefore the field of the corrector will superpose on the magnetic field of the permanent magnet. This superposition has been experimentally proven (Jain Animesh).
What follows is a list of statement which are based on results from the 3D OPERA calculations on the window frame magnet and other observation:

\begin{enumerate}
\item The window frame magnets, in spite their large aperture and short length, do not excite significant transverse magnetic multipoles except the ones are designed to produce.
\item An excited window frame magnet placed around a Halbach-type permanent magnet as in \Fig{fig:halbach-windowframe} does not alter significantly the multipoles of the Halbach-type magnet (experimental measurements by Jain Animesh) and there is an almost perfect superposition of the field of the widow frame magnet corrector to that of the FFAG Halbach magnet. 
\item	Four Halbach-type magnets were placed next to each other along their symmetry axis with the magnets touching each other and the integrated multipoles of all four magnets was measured to be equal to the sum of the of the integrated multipoles of each magnet measured separately. This measurement provides an almost perfect proof of field superposition. Because this statement is based on measurements we will not present calculations which prove this statement.
\item The Halbach-type magnets lend themselves easily to window-frame corrector magnets and do not interfere with possible access to the beam instrumentation which is placed in the short drift spaces between the magnets.
\end{enumerate}

\subsection{Multipole Strengths and Power Dissipation}
\begin{figure}[tb]
\centering
\includegraphics[width=0.8\textwidth]{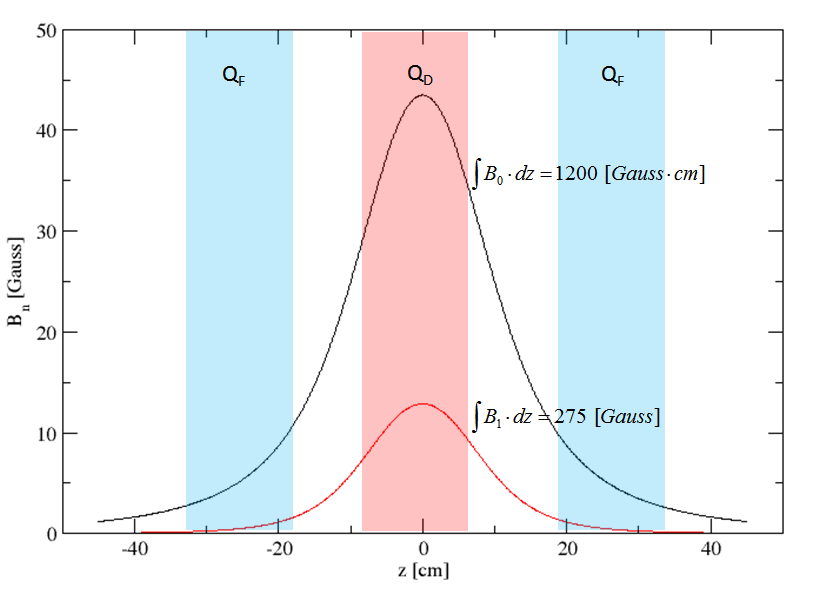}
\caption[]{The black and red lines are the plots of the strengths of the dipole and quadrupole multipoles of the window frame magnet which surrounds the FFAG magnet (pink renctangle). The values of integrated strengths of these multipoles are 1200 [Gauss.cm] and 275 [Gauss] }
\label{fig:halbach-B-multiple-vs-z}
\end{figure}

The window frame magnet as shown in \Fig{fig:halbach-windowframe} has an 8~cm long iron frame and the 12~cm coil around each side does not extend beyond the drift space defined by the FFAG magnets. \Figure{fig:halbach-B-multiple-vs-z} plots the strength of the dipole and quadrupole multipoles generated by the window frame magnet at radius R=1~cm as a function of distance $z$ along the beam direction, when the current of the power supplies of the correctors are set at maximum. These multipoles were calculated from the 3D fields as calculated by modeling the corrector magnet using the OPERA computer code. 

The maximum power dissipation in the coils which generates the dipole and quadrupole fields is shown in columns 5 and 6 respectively of \Tab{tab:windowframecoils} 

\begin{table}[tb]
\caption[]{Power dissipation in the rcoils to generate the required dipole or quadrupole field.}
\begin{tabular*}{\columnwidth}{@{\extracolsep{\fill}}lccccr}
\toprule
Type	&Integrated&	\# Racetrack coils&	$J$&	MaxPower\\
&Strength& &(A/mm$^2$)&(W)\\
\midrule
Dipole&	$\pm$1200 [Gauss.cm]&	2 ( L, R) or (T, B) &	1&	5.4\\
Quadrupole&	$\pm$275 [Gauss]&	4 (R, L, T, B)&	1&	31.5\\
\bottomrule
\end{tabular*}
\label{tab:windowframecoils}
\end{table}

\subsection{Temperature Stabilisation}
\begin{figure}[tb]
\centering
\includegraphics[width=0.9\textwidth]{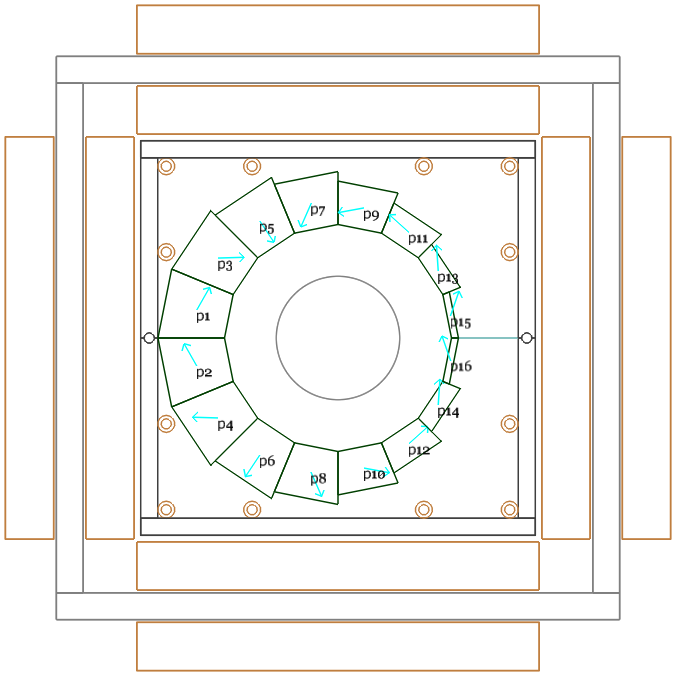}
\caption[]{Cross-section of the Halbach magnet and window-frame corrector assembly.  Cooling water flows through channels in the magnet holder to act as a barrier for heat from the window frame corrector coils.}
\label{fig:halbach-cooldrawing}
\end{figure}

As water cooling will be used for the window-frame corrector coils, it is inexpensive to add an additional layer of water in the magnet holder to stabilize the temperature of the permanent magnet blocks.  This ideally will be the first place the cool water flows, before it gets heated up in the hollow copper conductors.  A schematic of this scheme is shown in \Fig{fig:halbach-cooldrawing}.
 
\clearpage
\section{Magnet Assembly Method}
\begin{figure}[tb]
\centering
\includegraphics[width=0.95\textwidth]{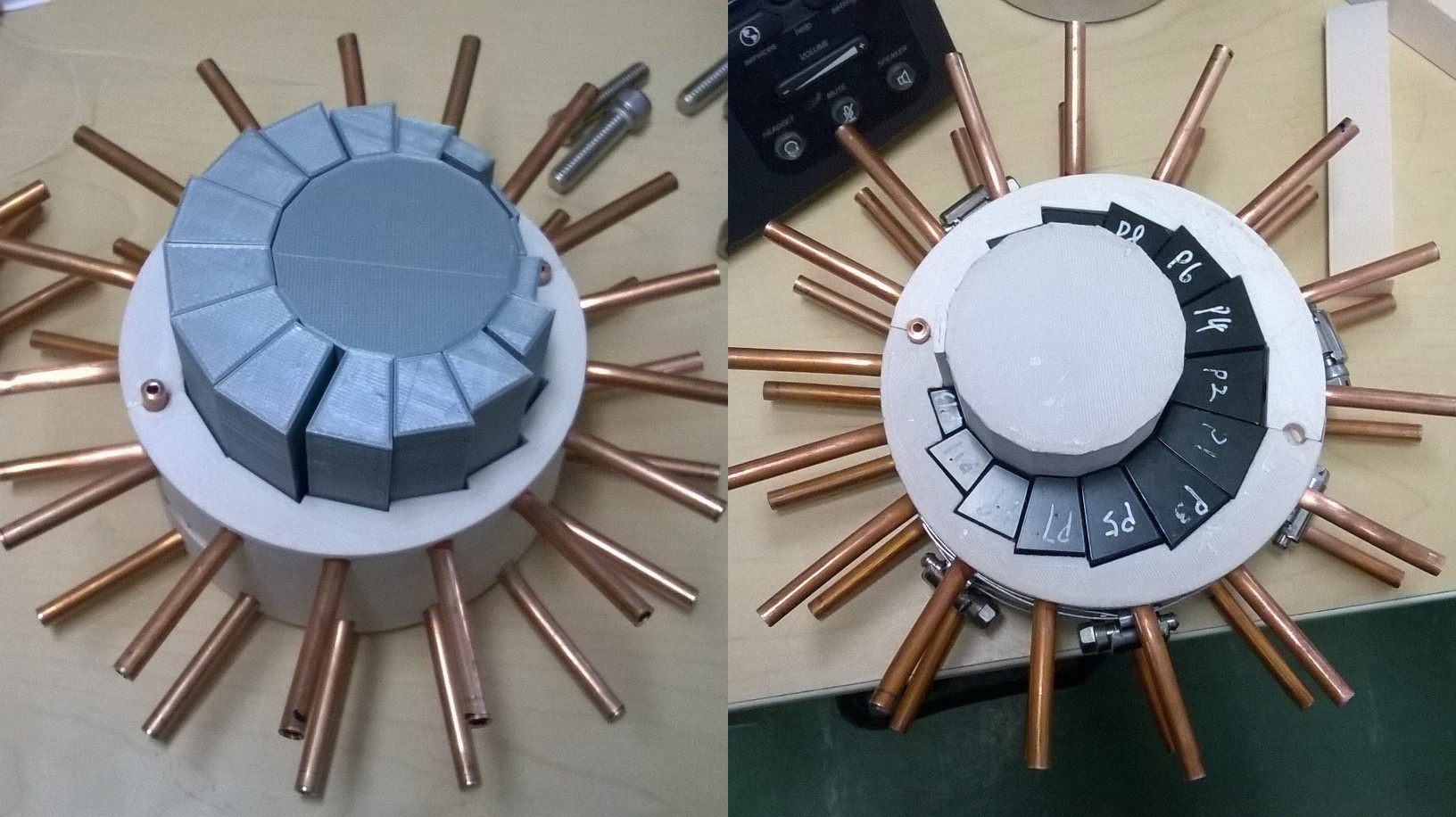}
\caption[]{Previous assembly of the permanent magnet block. One by one they replaced the 3D printer plastic molds.}
\label{fig:PermanentBlocksAndMolds}
\end{figure}
\begin{figure}[htbp]
\centering
\includegraphics[width=0.8\textwidth]{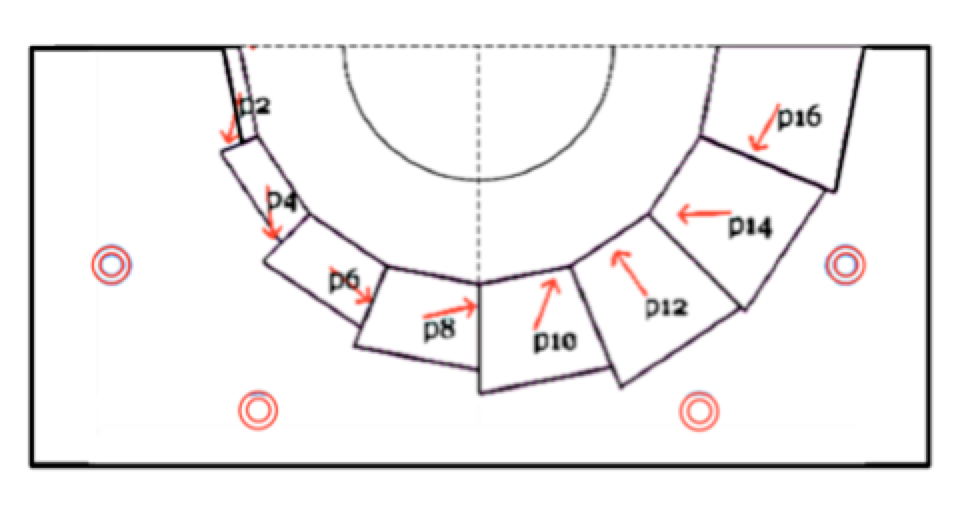}
\caption[]{Extruded aluminum surrounding the permanent magnet blocks. The blocks get glued onto aluminum.}
\label{fig:ExtrudedAluminum}
\end{figure}

 
There are couple of advantages by using the aluminum blocks instead of combinations between the aluminum frames and plastic molds made by the 3D printers: the water cooling is very effective for the permanent magnet blocks as well, as it would not be the case when the plastic mold from the 3D printers as it is an insulator. The only disadvantage might be if the extrusion cannot make accurately enough slots for the magnet blocks placement. Additional tooling might add the cost to the project.
The ANSYS analyses of the effect of the forces to the aluminum frame ends in the middle of the frame have shown possible horizontal deformations in the radial directions. To completely remove this effect it is possible to use the extruded aluminum support instead of a combination between the aluminum and plastic mold form made by the 3D printers, as shown in \Fig{fig:ExtrudedAluminum}.
Production starts with building the plastic molds by the 3D printers to be able to connected the permanent magnetic blocks.  The second part in production is connection of the of extruded aluminum upper and lower part with already build molds and central plastic 3D printer pipe a mold as shown in the previous assembly (\Fig{fig:PermanentBlocksAndMolds}).

\begin{figure}[htbp]
\centering
\includegraphics[width=0.95\textwidth]{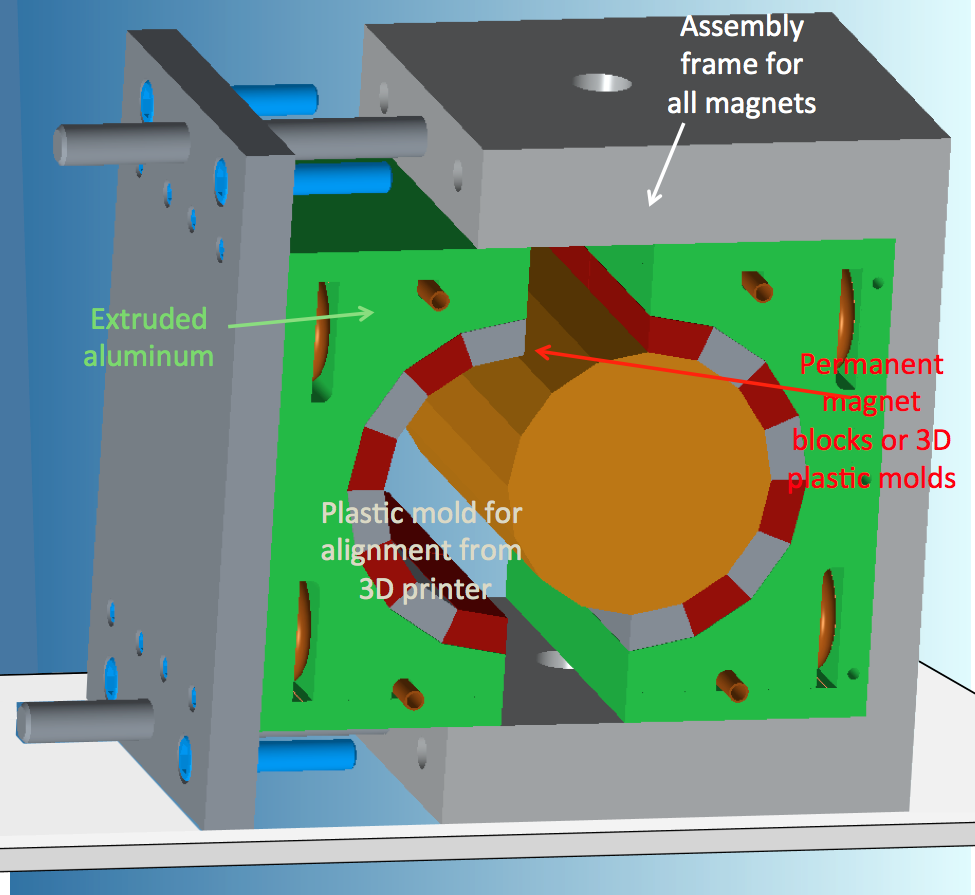}
\caption[]{Assembly procedure starts with the 3D mold built with plastic but using an aluminum future made for all magnets.}
\label{fig:AssemblyAlignmentPins}
\end{figure}

\begin{figure}[htbp]
\centering
\includegraphics[width=0.95\textwidth]{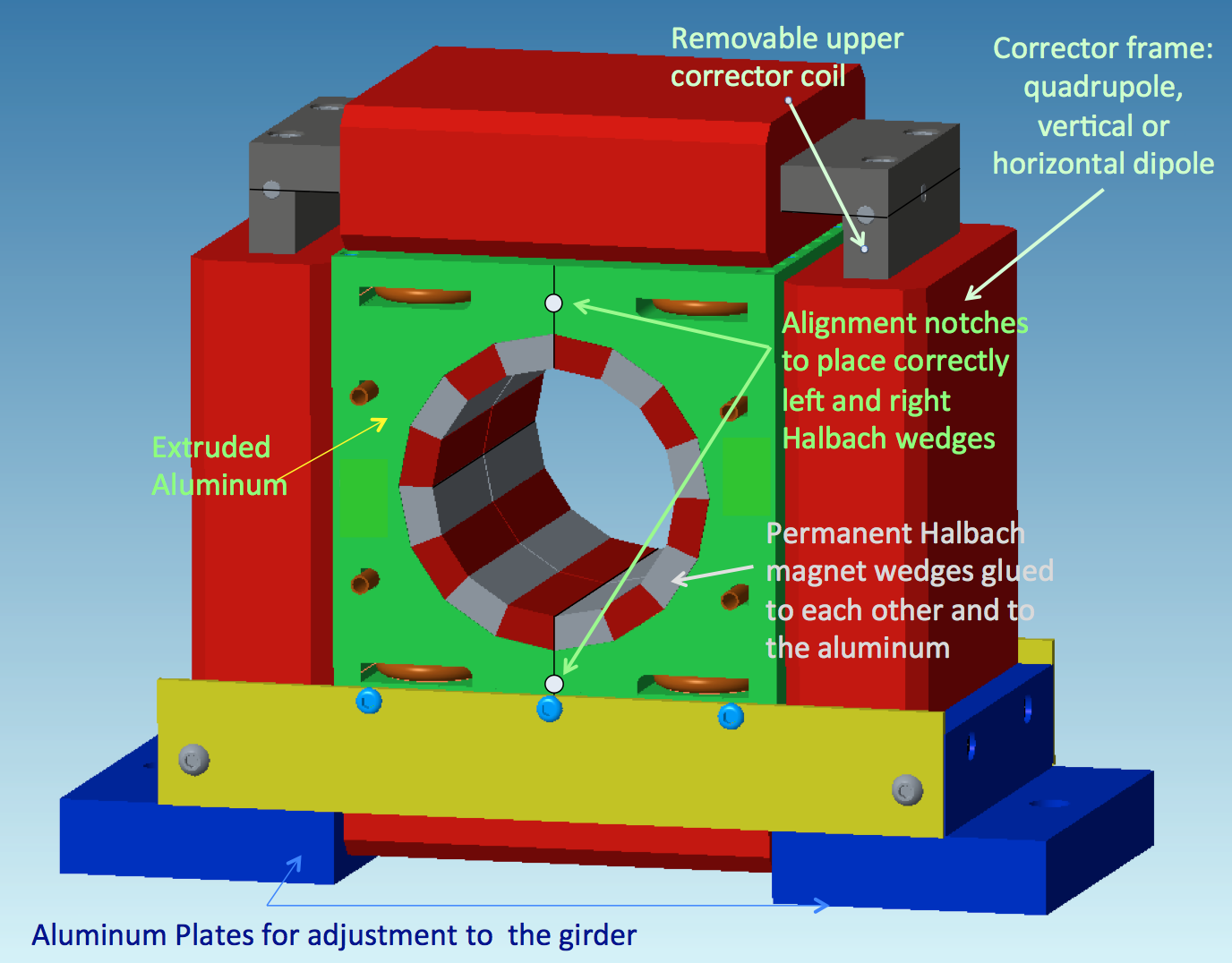}
\caption[]{The window frame with the Halbach defocusing magnet assembled.}
\label{fig:AssembledMagnet}
\end{figure}

The magnet blocks are glued into the aluminum plates with the 3M 5200 marine adhesive sealant (used to extremely hard bond fiberglass deck to hull, wood to fiberglass, motors on fiberglass transom, stern joints etc.). The sealant has already been tested with gluing the permanent magnetic blocks on the plastic mold made by the 3D printer. Additional molds are necessary in assembly procedure, as it was found out during the previous prototype production: each pair (in the case of the combined function magnet) of permanent magnetic blocks has copies made of plastic molds produced by the 3D printers, as shown in \Fig{fig:ExtrudedAluminum}.  A necessary mold in the middle of the magnet is also made of plastic and two halves of the magnet (lower and upper) are first made of plastic molds. The next step is a creation of the alignment holes (\Fig{fig:AssemblyAlignmentPins}).
A necessary mold in the middle of the magnet is also made of plastic and two halves of the magnet (lower and upper) are first made of plastic molds. The next step is a creation of the alignment holes.
The whole magnet assembly is shown in \Fig{fig:AssembledMagnet}.

\begin{figure}[htbp]
\centering
\includegraphics[width=0.95\textwidth]{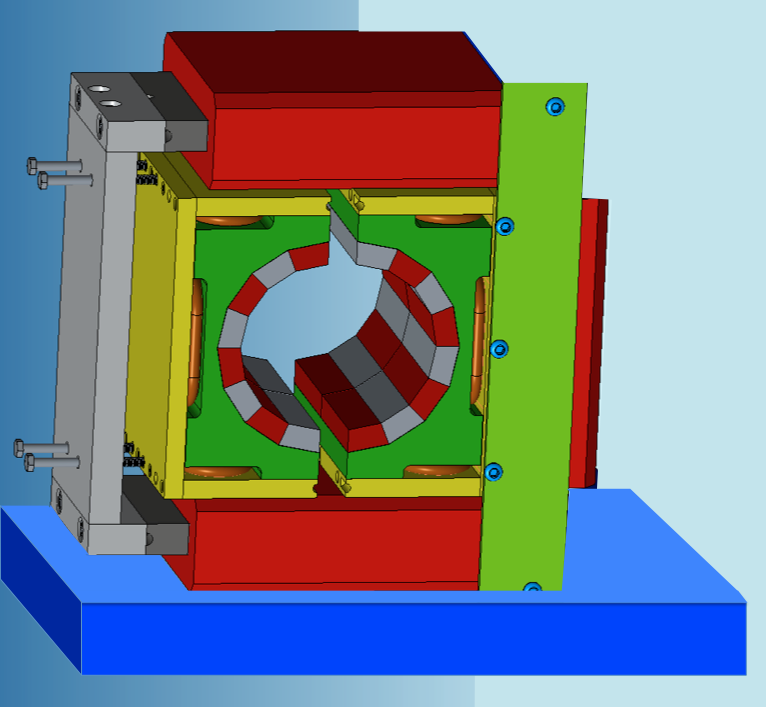}
\caption[]{An additional aluminum upper frame is built for the assembly and disassembly procedure.}
\label{fig:Disassembly}
\end{figure}

Disassembly procedure requires removal of the upper corrector coil as shown in \Fig{fig:Disassembly}.

\clearpage 
\section{Magnet Girder Support System \Leader{George Mahler}}

\begin{figure}[htbp]
\centering
\includegraphics[width=0.95\textwidth]{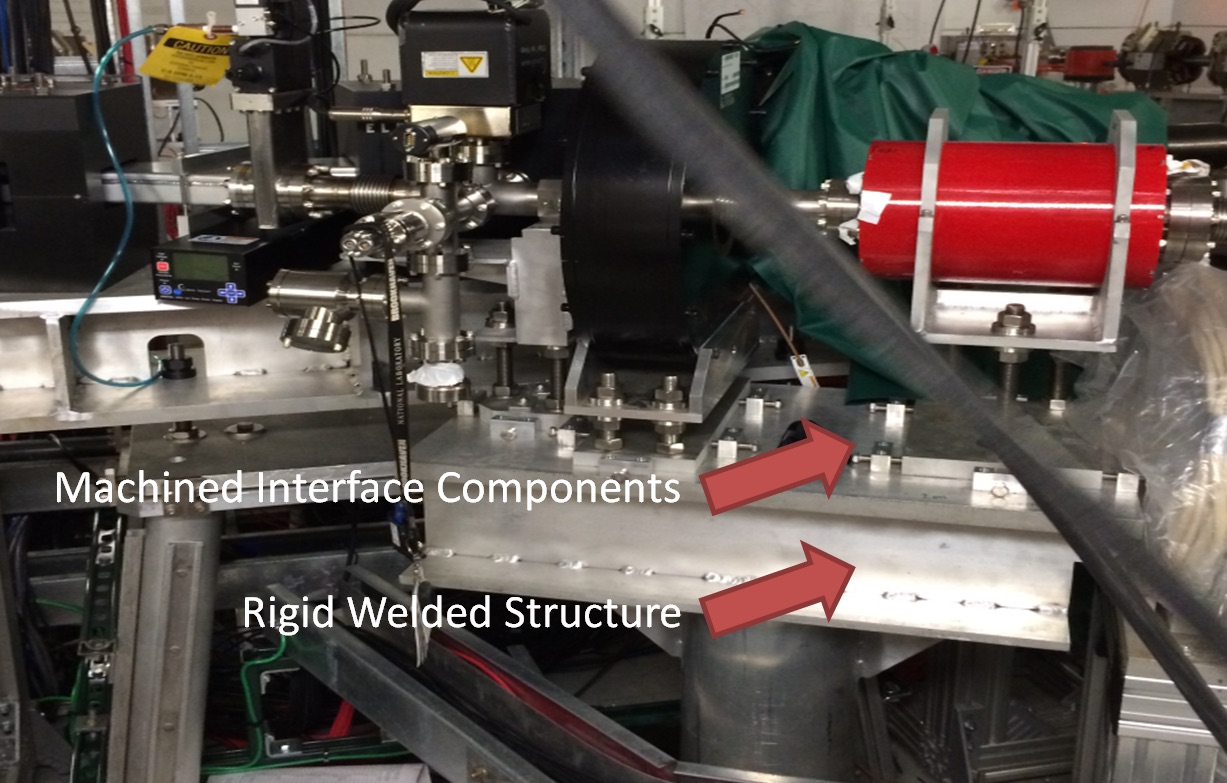}
\caption[]{Welded structure.}
\label{fig:girder-welded_structure}
\end{figure} 

\begin{figure}[htbp]
\centering
\includegraphics[width=0.85\textwidth]{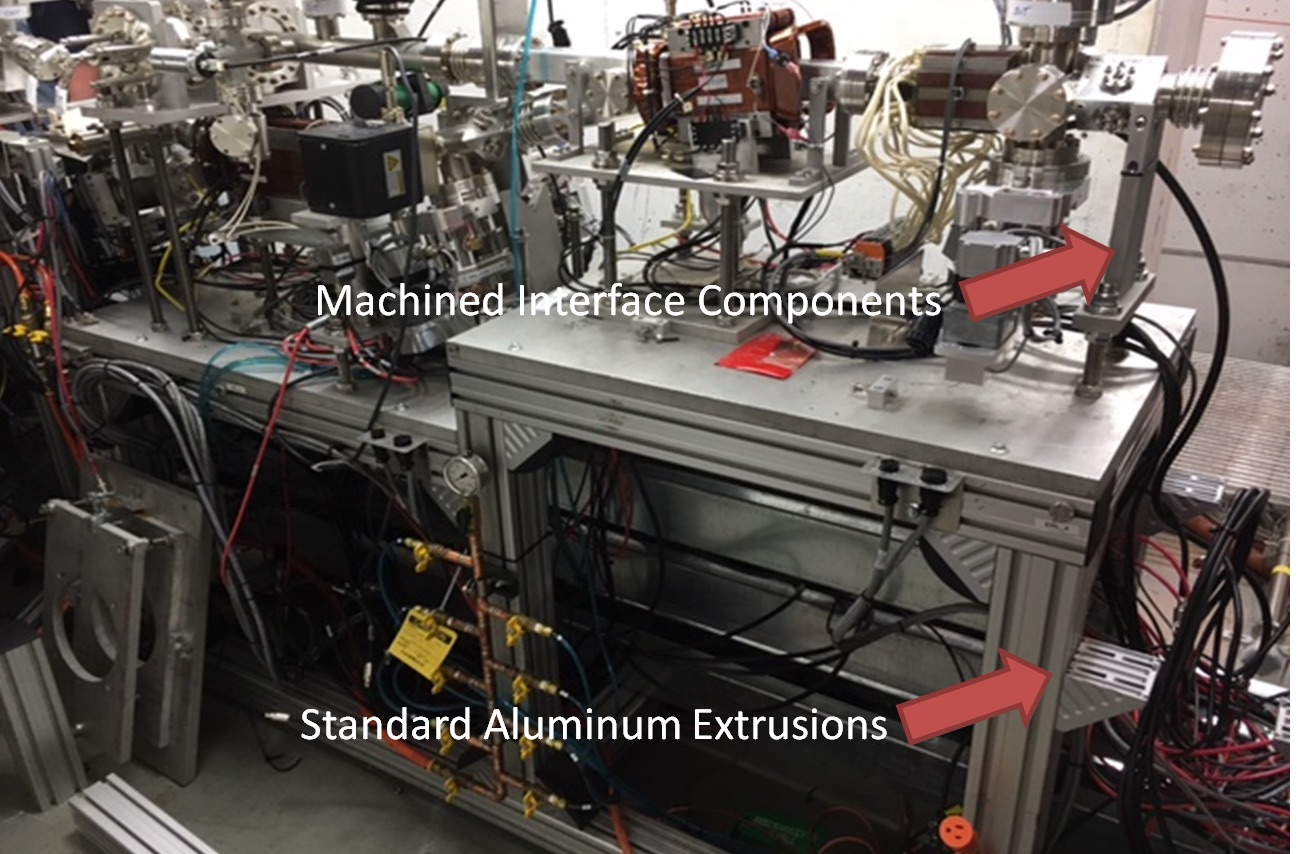}
\caption[]{Modular structure.}
\label{fig:girder-modular_structure}
\end{figure} 

\begin{figure}[htbp]
\centering
\includegraphics[width=0.85\textwidth]{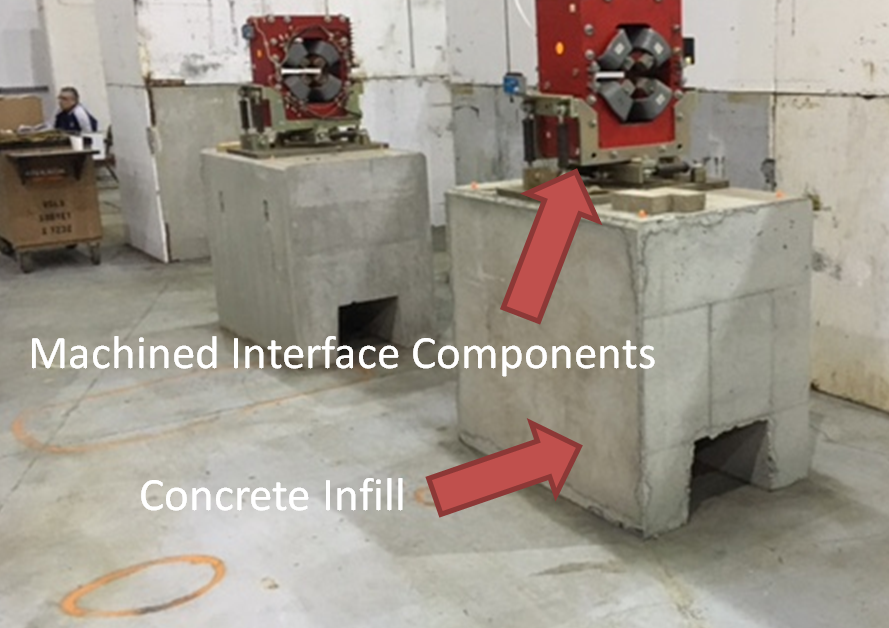}
\caption[]{Concrete infill.}
\label{fig:girder-concrete_infill}
\end{figure} 

A conceptual design has been proposed based upon the designs of existing girder support systems in a variety of other machines. Three particular designs had been proposed, those being:

\begin{enumerate}
\item A rigid welded structure with post process stress relief and subsequent machining combined with machined components that interface and adjust position of accelerator constituents (\Fig{fig:girder-welded_structure});
\item Modular structure using standardized aluminum extrusions and machined components that interface and adjust position of accelerator constituents (\Fig{fig:girder-modular_structure});
\item Concrete infill and machined components that interface and adjust position of accelerator constituents (\Fig{fig:girder-concrete_infill}).
\end{enumerate}

Typically the favored design is the rigid welded structure (\Fig{fig:girder-welded_structure}) but the cost of such must also be considered, as it is the most expensive to produce.  We will pursue the favored design until budget constraints dictate otherwise. In general, the magnet girder support system must provide a stable platform for the accelerator constituents during assembly and after installation to tolerances defined by the required magnet stability.  It must provide adjustment for the initial component alignment as well as subsequent alignment due to long term facility movement. We will also determine the resonant frequencies of the girder system and adjust structure to avoid building and machine frequencies.

\clearpage
\section{Power supplies \Leader{John Barley}}
    
\subsection{Splitter Transport Line Converters}

The Splitter Transport Lines are used to transfer beam from the MLC Linac to the FFAG ring. The beam makes a full turn around the FFAG ring and returns beam back to the MLC through another splitter. Both of these source and return lines are each made up of four transport lines tuned to very specific energies. Magnets on these transport lines are categorized: Dipole (Bend), Quadrupole, and Vertical Corrector.  They are all categorized and listed in \Tab{tab:splitter_magnet_power_supply_parameters}. In addition, there are four unipolar power supplies for each of: The common splitter dipole, the 3-beam septum magnets, and the 2-beam septum magnets. \Table{tab:splitter_power_supply_counts} lists counts for these items. 
    
\begin{table}[tb]
\centering
\caption{Splitter Magnet Power Supply Parameters}
\label{tab:splitter_magnet_power_supply_parameters}
\begin{tabular}{lrrrrr}
\toprule
                 & H-Dipole 20 cm & H-Dipole      &  Quadrupole             & V Corr 1       & V Corr 2        \\
\midrule                
Quantity         & 28             & 12            & 64            & 16      & 16      \\
Magnet R (\unit{\Omega})     & 0.0914         & 0.1202         & 0.00876       & 0.434   & 0.267   \\
PS Curr (A)      & 27.1--86.0      & 38.2--86.0     & 5.0--140.7     & 2.98    & 2.98    \\
PS Volts (V)     & 2.5--7.9        & 4.6--10.3      & 0.0--1.2       & 1.3     & 0.8     \\
PS Watts (W)     & 67--676         & 175--889       & 0.2--173       & 3.9     & 2.8     \\
Stability/ripple (ppm) &                &               &               & 500  & 500\\
Resolution(bits) &                &               &               & 16      & 16      \\
Operation        & unipolar       & unipolar      & unipolar      & bipolar & bipolar \\
\bottomrule
\end{tabular}
\end{table}

\begin{table}[tb]
\centering
\caption{Splitter magnet Power supply Types (QDC=Quad, Dipole, or Common)}
\label{tab:splitter_power_supply_counts}
\begin{tabular}{llrrrr}
\toprule
Type          & Quantity & Spare & PS Curr (A) & PS Volts (V) & PS Power (W) \\
\midrule
QDC 1         & 59       & 6     & 50          & 10           & 500          \\
QDC 2         & 44       & 5     & 100         & 15           & 1500         \\
QDC 3         & 5        & 1     & 150         & 10           & 1500         \\
3 Beam Septum & 4        & 1     &             &              &              \\
2 Beam Septum & 4        & 1     &             &              &              \\
V Corrector   & 32       & 6     & 3           & 8            & 24          \\
\bottomrule
\end{tabular}
\end{table}

\subsection{FFAG Corrector Magnet Power Supplies}

The FFAG ring guides beam from the MLC Linac and returns the beam to the input of the MLC Linac. This ring carries the beam through four acceleration cycles before finally returning the beam to the MLC for deceleration. There are a great number of corrector magnets involved.  They have been categorized and listed in \Tab{tab:ffag_corrector_power_supplies}.    
\begin{table}[]
\centering
\caption{FFAG Corrector Magnet Power Supply Parameters:}
\label{tab:ffag_corrector_power_supplies}
\begin{tabular}{lrrrrr}
\toprule
                & (F) Quadrupole               & (D) Quadrupole             & (H) Corrector              & (V) Corrector      \\
\midrule
Quantity          &             107 &            106   & 107 & 107 \\
Magnet R (\unit{\Omega})       &             7.6 &               7.06 &              3.8   &              3.8      \\
PS Curr (A)       &           2.08  &              2.08  &          2.08      &             2.08         \\
PS Volts (V)      &           15.8  &               15.8 &           7.9      &             7.9  \\
PS Watts (W)      &           33    &                33  &             17     &              17          \\
Stability/ ripple (ppm) & \textless 500        & \textless 500          & \textless 500           & \textless 500           \\
Resolution(bits)  & 16                           & 16                         & 16                         & 16                     \\
Operation         & bipolar                      & bipolar                    & bipolar                    & bipolar              \\
\bottomrule
\end{tabular}
\end{table}

\begin{table}[]
\centering
\caption{FFAG Corrector Magnet Power Supply counts}
\label{my-label}
\begin{tabular}{lrrrrr}
\toprule Quantity (channels) & Spare (chassis) & PS Curr (A) & PS Volts (V) & PS Power (W) \\
\midrule
   426               & 10              & 3           & 15           & 45          \\
\bottomrule
\end{tabular}
\end{table}

The Basic Requested specifications are:
\begin{itemize}
\item Bipolar operation constant current mode
\item Output 3A, 8V
\item Current Ripple  \textless0.01\% FS
\item Stability \textless500 ppm
\item Temperature coefficient \textless 100ppm
\item Zero crossing distortion \textless .01\% FS
\item Minimum 16 bit programming and read backs
\item Ethernet communication
\item 19 in  rack mountable
\item 426 channels plus spares
\end{itemize}

    \subsubsection{BiRa MCOR 12/30}
    
\begin{figure}[htbp]
\centering
\includegraphics[width=0.7\textwidth]{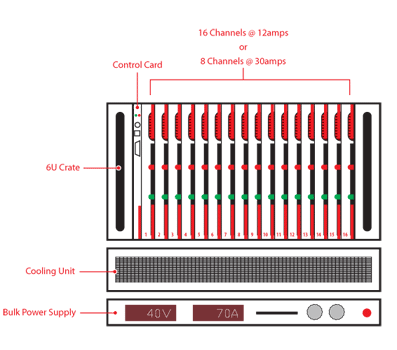}
\caption[Bira Module]{A schematic of the BiRa crate}
\label{fig:bira_module}
\end{figure}

BiRa produces blade power supplies in a 6U crate, with 2U for cooling fans and 1U for the bulk power supply (see \Fig{fig:bira_module}).  Each crate has 1 ethernet control card, 16 $\pm$12~A modules (or 8 $\pm$30~A modules).  The 12~A modules can be configured to 2~A, 6~A, 9~A or 12~A maximum current.

%

\subsubsection{CAEN 20-10 Easy Driver}
    
\begin{figure}[tb]
\centering
\includegraphics[width=0.6\textwidth]{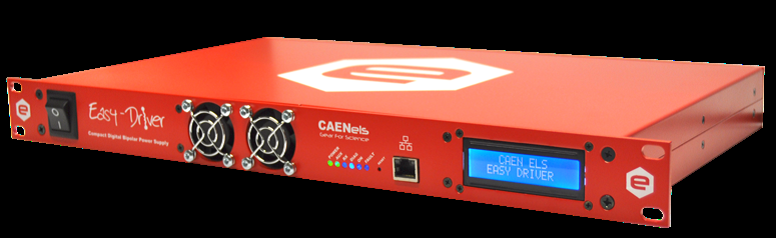}
\caption[Caen driver]{The Caen magnet power supply unit.}
\label{fig:caen_driver}
\end{figure}

Caen has 1U standalone units with $\pm$5~A and $\pm$10~A version, with an ethernet interface (see \Fig{fig:caen_driver}.  They have low-noise output with digital regulation loops, internal protections and auxiliary readbacks.  An 8 foot rack can hold 54 channels.

The specifications are:
\begin{itemize}
\item{$\pm$20~V, $\pm$10~A},
\item{160 $\mu$A current setting resolution},
\item{20 bit output resolution for voltage and current},
\item{output current ripple 50~ppm/FS},
\item{output current stability 50~ppm/FS},
\item{switching frequency 100~kHz},
\item{closed loop bandwidth 1~kHz}.
\end{itemize}









\ifdefined \buildingFullDocument

\renewcommand{\FiguresDirectory}{injector/figures}

\else
\newcommand{\FullDocumentRoot}{..}
\newcommand{\FiguresDirectory}{figures}

\begin{document}
\fi


\chapter{Injector \Leader{Colwyn Gulliford}}\label{chapter:injector}

\section{Overview and Operational Status}

\begin{figure}[htbp]
\centering
\includegraphics[width=0.9\textwidth]{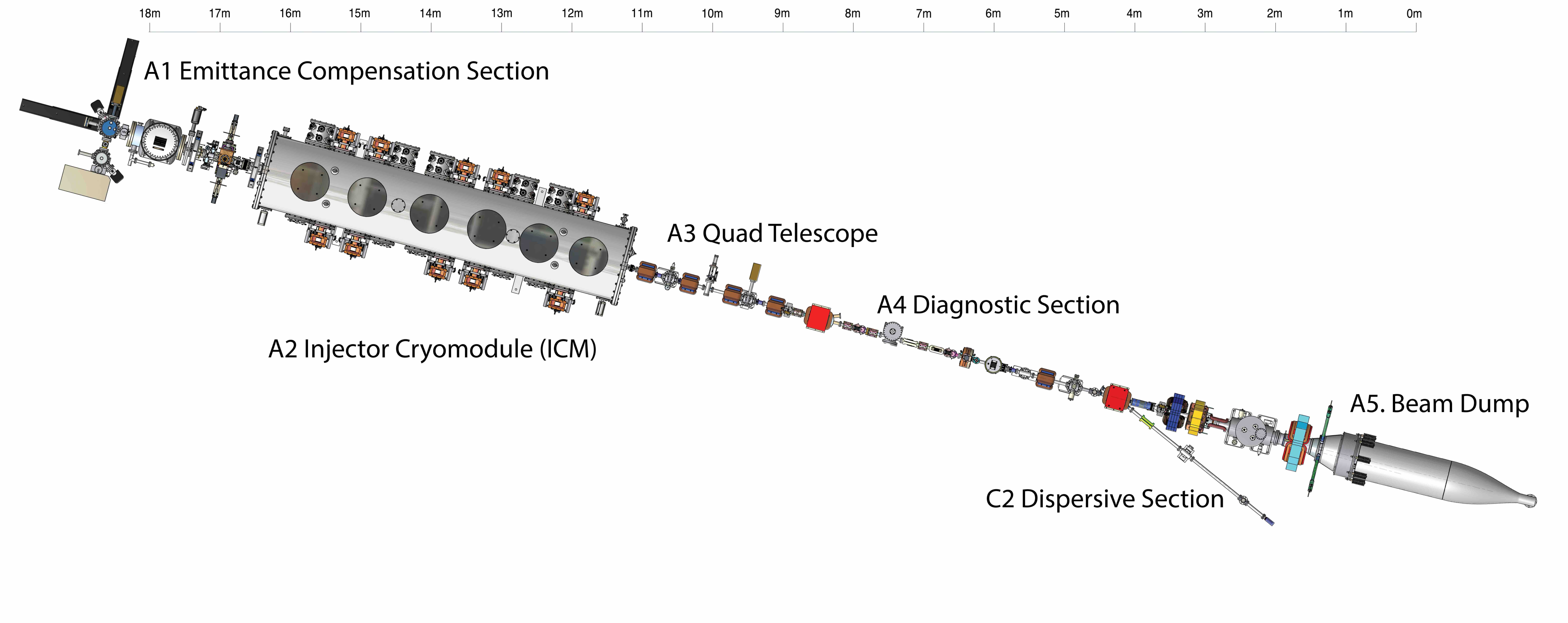}
\caption[A schematic of the Cornell photoinjector]{A schematic of the Cornell photoinjector after installation and recommissioning for use in CBETA.}
\label{fig:diagram_Cornell_injector}
\end{figure}

The Cornell accelerator group now operates the world'��s highest average current, high brightness photoinjector, a prerequisite for emerging machines in nuclear physics, high energy physics, FEL facilities, and industrial applications. Among the achievements of this machine are: the world-record average current of 75 mA from a photocathode injector with many-day cathode lifetimes at currents above 60 mA [12,13], and beam emittances approaching the thermal (intrinsic) photocathode emittance at bunch charges up to 300 pC in the space-charge dominated regime [14]. \Fig{fig:diagram_Cornell_injector} shows the layout (as of 1/17/2017) of the Cornell photoinjector with all critical subsystems installed in their final location for the CBETA experiment.  In this machine the beam is produced in a laser-driven photoemission gun (A1), accelerated to as much as 10~MeV in superconducting RF cavities (A2) housed in the injector cryomodule (ICM), after which beam parameters are measured (A3, A4,C2), and finally absorbed in a beam dump (A5).    



\begin{figure}[htbp]
\centering
\includegraphics[width=0.9\textwidth]{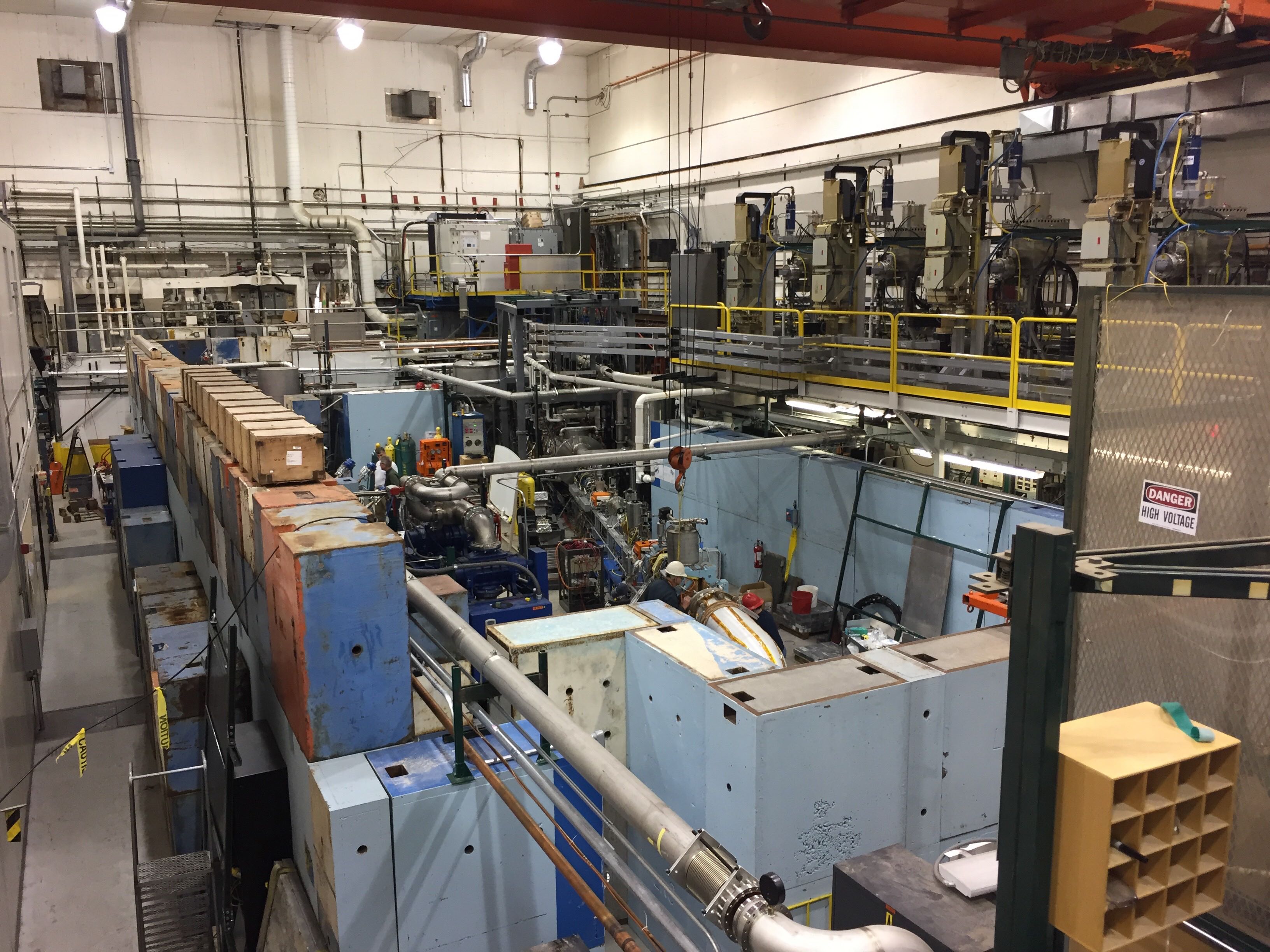}
\caption[Photo of the CBETA experimental hall]{A photo of the CBETA experimental hall housing the recommissioned Cornell photoinjector.}
\label{fig:photo_Cornell_injector}
\end{figure}

\begin{table}[htbp]
\caption[]{Status of Critical Injector Subsystems}
\begin{tabular*}{\columnwidth}{@{\extracolsep{\fill}}lccc}
\toprule
Parameter & Target & Current Status & Comment\\
\hline
Avg. Laser Power* &	2 Watt & 10 Watt & Sufficient for KPP\\
Laser Pulse Length* & 9 ps & 9 ps & Sufficient for KPP\\
Laser Power stability* & Better than 2\% & 1\% & Sufficient for KPP \\
Laser Position stability*	&	10 $\mu$m rms  & 5 $\mu$m & Sufficient for KPP \\
Cathode QE & 1-10\%  & $>$1\% & Sufficient for KPP \\
DC Gun Voltage   & 300-400 kV     & 350 kV   & Sufficient for KPP \\
Buncher Voltage   & 30-70 kV       & 55  kV   & Sufficient for KPP \\
SRF Cavity 1 Voltage & 1125 kV   & 1000 kV  &  Needs further processing \\
SRF Cavity 2 Voltage & 1790 kV   & 1000 kV  &  Needs further processing \\
SRF Cavity 3 Voltage & 1580 kV   & 1000 kV  &  Needs further processing \\
SRF Cavity 4 Voltage & 835  kV   & 1000 kV  &  Sufficient for KPP \\
SRF Cavity 5 Voltage & 465  kV   & 1000 kV  &  Sufficient for KPP \\
\bottomrule
*at cathode \\
\end{tabular*}
\label{tab:system_status}
\end{table}

Originally located in a separate experimental hall, the Cornell photoinjector has been disassembled and installed in its final location for the CBETA experiment.  All of the principle beamline elements and supporting infrastructure, including but not limited to: the DC gun, buncher cavity, ICM, magnets, beam diagnostics, cooling water, cavity crygogenics, cavity power supplies (IOT and klystrons), and magnet power supplies, have been installed and tested.  Recommissioning of the photoinjector began immediately following its reassembly.  For this purpose,an existing 50 MHz laser system was used (note only minor adjustments to this system are required to produce the 42 MHz required for CBETA). The recommissioning of the injector included high voltage processing of the DC gun, buncher cavity, and SRF cavities.  To date, the DC gun is now capable of operation at 350 kV, slightly lower than anticipated 400 kV, but sufficient for CBETA.  The buncher cavity has been processed up to roughly 55 kV, which meets the CBETA requirements.  The SRF cavities have been processed, with the input couplers set for low current, up to roughly 1 MV on each cavity.  \Tab{tab:system_status} summarizes the status of the injector subsystem parameters most relevant to production of a beam suitable for CBETA. Note that all cavity voltage targets were determined by detailed space charge simulations through injector. After both the DC and RF processing, beam was generated and transported to the main dump.  Up to 4.2 mA of average current was successfully produced and transported. The current is believed to be currently limited by the SRF coupler positions, which will be adjusted in the future when higher currents are required.

\begin{figure}[htbp]
\centering
\includegraphics[width=0.9\textwidth]{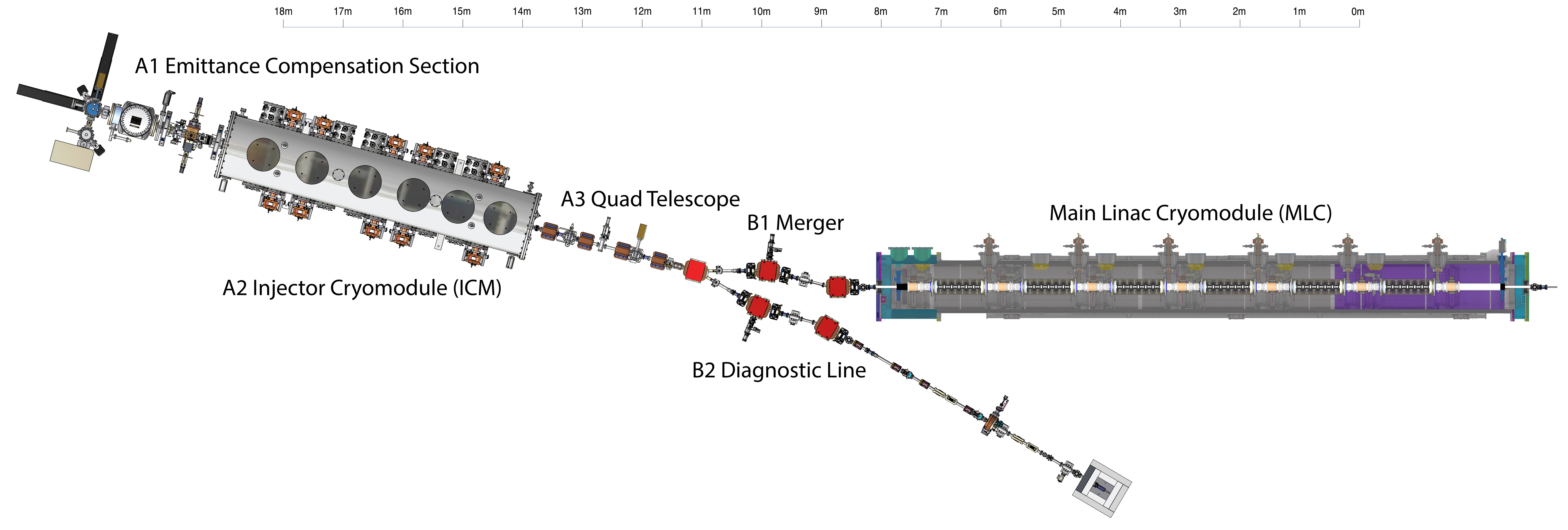}
\caption[Photo of the CBETA experimental hall]{A photo of the CBETA experimental hall housing the recommissioned Cornell photoinjector.}
\label{fig:final_Cornell_injector}
\end{figure}

\Fig{fig:final_Cornell_injector} shows the final layout of the injector with both the merger (B1) to the main linac cryomodule (MLC) as well as the diagnostic beam (B2).  In this layout, the position of the DC gun, ICM, and quad telescope remain unchanged, with the elements currently installed in place in the experimental hall. Work is underway to move the MLC into its final position as shown, as well as installing the merger and diagnostic sections.  The majority of the parts for these beam lines are already built and tested, and with the exception of a handful of magnets (small quads), ready for installation up completion of the MLC move.  Thus as a whole, the photoinjector through to the MLC is near completed.

\section{Technical Data for the Main Injector Subsystems}

\subsection{The DC Photoemission Gun}

\begin{figure}[htbp]
\centering
\includegraphics[width=0.7\textwidth]{\FiguresDirectory/newgun_cutout.png}
\caption[A schematic view of the DC gun.]{A cutaway view of the DC photoemission gun.  Photocathodes are prepared in a load lock system mounted on the large flange at the left, and transported through the cathode cylinder to the operating position in the Pierce electrode shape on the right.  The beam exits through the small flange to the right.}
\label{fig:diagram_Cornell_gun}
\end{figure}

\begin{figure}[htbp]
\centering
\includegraphics[width=0.3\textwidth]{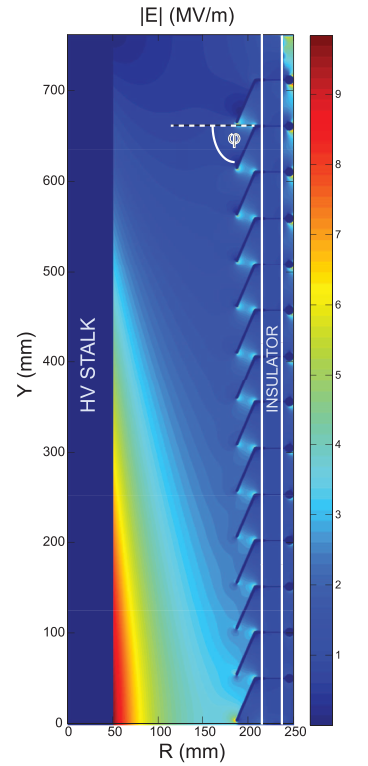}
\caption[HV simulations]{ Simplified HV model made in the software Opera 2D. Shown is the HV stalk, insulator, and guard rings for the gun operating at 750~kV. Color corresponds to the magnitude of the electric field in MV/m. For simplicity, this HV model is monolithic, rather than made of two insulator assemblies (as built), and similarly does not show the triple point protection rings.}
\label{fig:gun_simulation}
\end{figure}
\begin{figure}[htbp]
\centering
\includegraphics[width=0.7\textwidth]{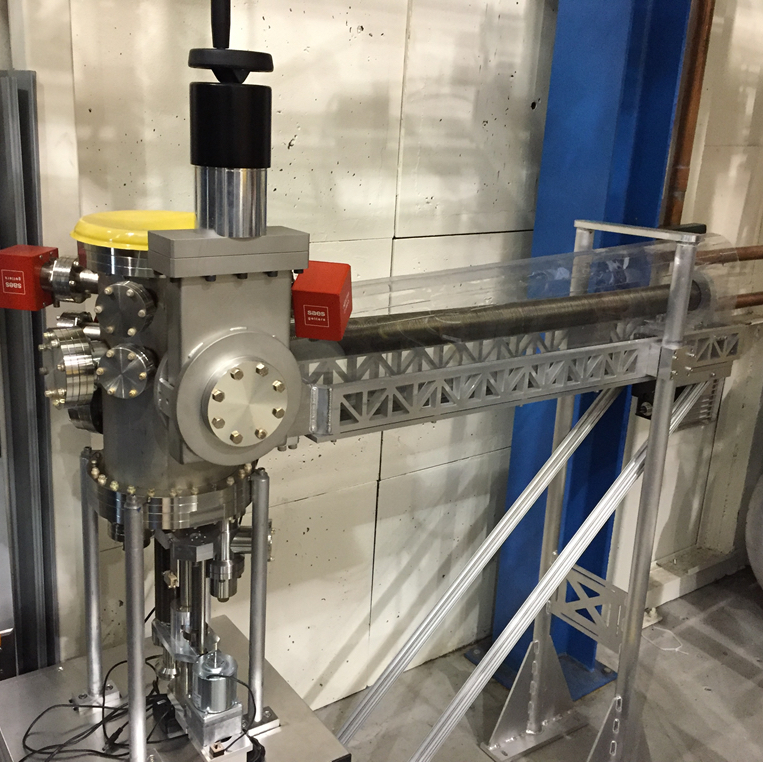}
\caption[The cathode transfer system.]{Cathodes are transported in a vacuum suitcase (not shown) and attached to the manual gate valve. Cathodes are moved from the suitcase into the central vacuum chamber, which can hold three cathode pucks.  The bellows translation mechanism  moves the pucks from the chamber into the gun. }
\label{fig:photo_cornell_gun_loadlock}
\end{figure}

\begin{figure}[htbp]
\centering
\includegraphics[width=0.35\textwidth]{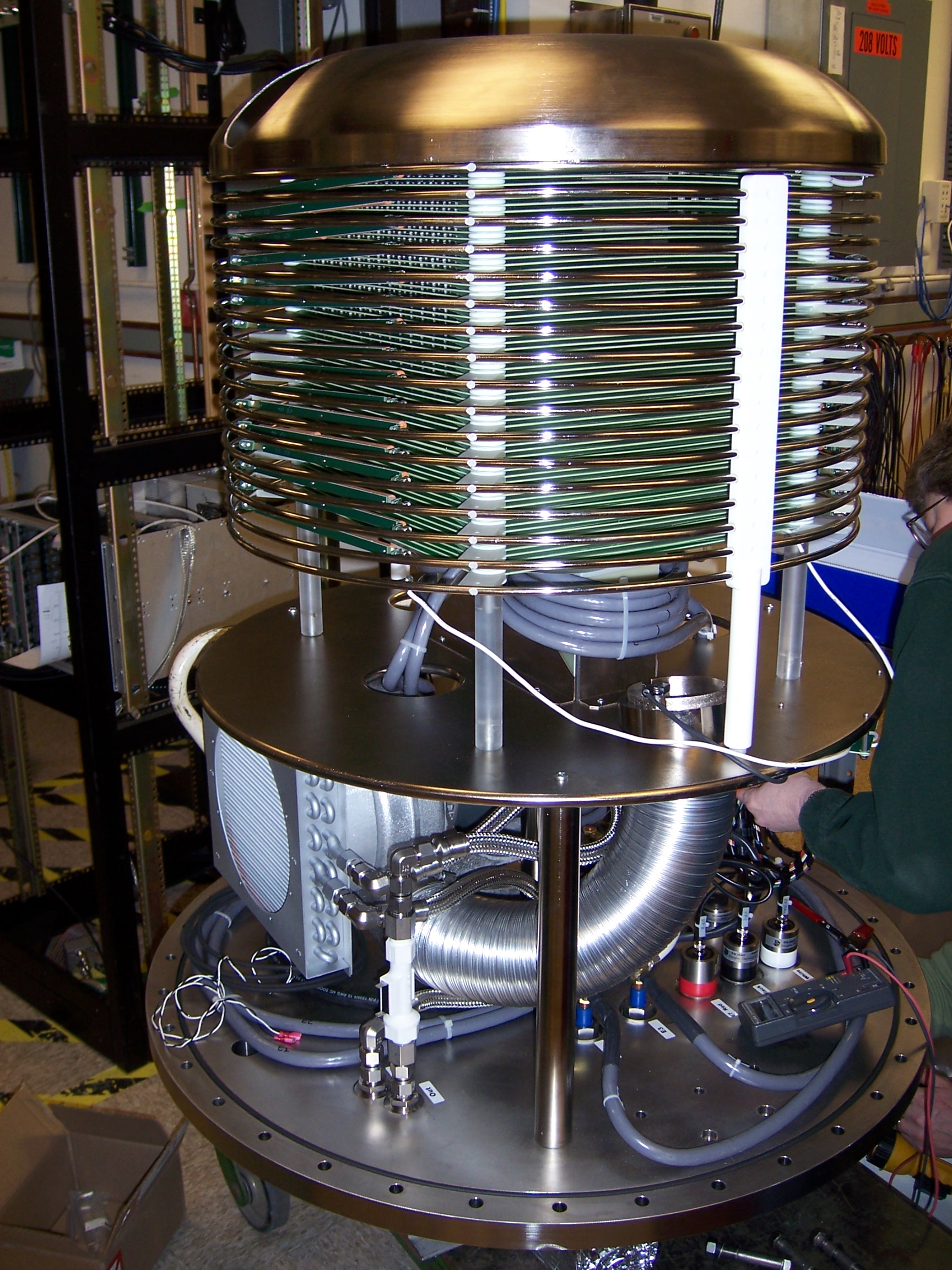}
\caption[The HV section of the 750kV power supply.]{The HV section of the 750kV power supply.  Individual circuit boards are visible inside the potential grading rings.  The primary winding of the insulating core transformer is the large diameter gray-insulated wire below the circuit boards.  The HV section is mounted in the pressurized SF$_6$ tank with its axis parallel to the axis of the gun ceramic. }
\label{fig:photo_hv_section_of_power_supply}
\end{figure}

An overview of the gun and its major components is
shown in \Fig{fig:diagram_Cornell_gun}. The power supply (A) and outer gun structures at HV are enclosed by a chamber (B) pressurized with sulfur hexafluoride (SF$_6$) as a dielectric buffer gas to suppress arcing in
the HVPS. The HV surfaces of the gun chamber are held off
from the grounded chamber surfaces via an insulator structure (E). Inside the chamber, the cathode electrode (G) is suspended
in the chamber center via a support stalk (F), which also connects
the HV to the cathode electrode. The anode electrode (H) is constructed from titanium, and is isolated from ground using small alumina washers.  This allows the anode to be biased in order to reject ions that backstream into the anode-cathode gap region from further down the beamline.  The anode can hold off up to 5~kV. At the rear of the gun  a vacuum load-lock system allows cathodes to be inserted and removed without disturbing the gun vacuum. The cathode is illuminated from the front side of the gun using mirrors that are mounted in a light box between the first solenoid and the buncher.

One of the most important design differences between this gun and the original Cornell DC gun (17) is the
use of a segmented insulator structure (\Fig{fig:diagram_Cornell_gun}(E)). The entire insulating structure is composed of two smaller insulator
assemblies. Both insulator assemblies were manufactured by
Friatec AG. Each insulator assembly has 7 segments, or 14
in total installed on the gun. Each segment is a ring of Al$_2$O$_3$
with an inner diameter of 435~mm, 50~mm tall, and 20~mm
thick. The dielectric strength of the Al$_2$O$_3$ is quoted by Friatec to be beyond 30~kV/mm, with a resistivity of $10^{15} ~\Omega  \cdot cm$
at room temperature. The top segment and bottom segment of
each assembly is brazed into a 22.125 inch 316LN wire seal flange.

A kovar ring is brazed between each of the insulator segments. In vacuum, the kovar ring allows the attachment of
the insulator guard rings (described below). These rings also
extend outside the insulator body into the SF$_6$ environment.
In the SF$_6$, a resistor chain from HV (at the top) to ground (at
the bottom) connecting to each kovar ring directly defines the
voltage on all segment interfaces and inner guard rings. Kovar
was chosen as the interface ring material for its similar 
coefficient of thermal expansion to that of the Al$_2$O$_3$, so that
the braze joint would be minimally stressed during thermal cycling. The kovar rings at the top and bottom of the insulator are shaped to fit into 22.125 inch 316LN wire seal flanges and welded.

The insulator guard rings between each segment block field emitted electrons from landing on the insulator surface, reducing the possibility of insulator damage, key concern in previous gun designs.  The rings were made of copper due to its ease of
machining and high thermal conductivity, thereby minimizing
the heating of the ring and nearby braze joints from any stray
field emission. A simplified high voltage model of the guard
rings, insulator, and HV stalk are shown in \Fig{fig:gun_simulation}. The angle
of each ring with respect to the horizontal,  $67.5^{\circ}$ , was
chosen such that no field emitted electrons could reach the surface of an insulator segment, based on particle tracking. The angle of the lowermost ring was increased to
$72.0^{\circ}$ as the electric field lines changed near the ground plane, and some electrons could have reached the insulator with the $67.5^{\circ}$ angle.

One of the most vulnerable locations with respect to field
emission on any vacuum insulator is the so-called triple point
junction, which is the interface between the stainless steel, ceramic, and vacuum. These junctions are shielded by additional
triple point protection (TPP) rings which attach directly to the
interior of the flanges, and are also made of stainless steel.
The HV stalk (\Fig{fig:diagram_Cornell_gun}(F)) is a 1.25~m long hollow cylinder
with 125~mm outer diameter. The stalk center axis is co-linear
with the insulator's center axis (in \Fig{fig:gun_simulation}, the line at R=0),
and is attached to a support plate (\Fig{fig:diagram_Cornell_gun}(D)) which rests on
top of the uppermost TPP ring. As the stalk supports the cathode electrode, its height, angle with respect to the y-axis, and
rotation angle about the y-axis are of direct importance for
the symmetry of fields in the photoemission region. Thus, adjustment screws on the stalk plate permit the adjustment of
height, both angular offsets from the y-axis, and the rotation
of the stalk.

The insulators and stalk sit on the gun vacuum chamber,
which is held at ground potential, along with the SF$_6$ chamber walls. The gun chamber is a 600~mm diameter cylinder
shell, 5~mm thick, with its symmetry axis along the direction of the beam. Pumping is provided by a three non-evaporable getter modules (SAES Capacitorr D-2000) and two 40~l/s
ion pumps. Each ion pump is placed
behind a NEG module in a single assembly, so that any gas
load produced by the ion pumps would first be pumped by the
NEG modules before entering the gun chamber. Each of these
combined pumping assemblies is attached to a 8 inch ConFlat\textregistered\ 
flange on either side of gun chamber. A leak valve manifold
 with both a turbomolecular pump and
a source of ultra-pure noble gas is attached to a UHV right-angle valve on the gun chamber for noble gas processing.
The cathode electrode (\Fig{fig:diagram_Cornell_gun}(G)) design is identical to
that of the original Cornell gun (17), as this design was shown
to be an effective balance between providing optimal focusing
and high photocathode field strength while having minimal
electric fields outside of the photoemission region (18). Furthermore, this cathode design was shown to give excellent emittance for bunch charges up to 80 pC (16) for a gun operating
at 350 kV. The cathode design is Pierce-type, with a focusing
angle of 25$^\circ$. The focusing introduced by this electrode angle
serves to counteract the initial space charge expansion of an
intense photoemitted beam.

The cathode electrode is made of vacuum remelt 316LN
stainless steel. It features a leaf spring assembly inside the
cathode interior to hold and register a removable photocathode puck (\Fig{fig:diagram_Cornell_gun}(1)) in the center of the Pierce electrode.
The back of the cathode electrode is terminated with a half-torus to keep electric fields low, with the torus hole permitting the transfer of photocathodes in and out of the interior
holder.

In general, we follow the procedures developed for cleaning SRF cavities whenever
possible for surfaces supporting high electric fields. Although field emission is a poorly understood process, it is
well known that the condition of the HV surface in terms of
both roughness and contaminants strongly affect the fields at
which field emission or pre-breakdown activity begins. Both
particulate contaminants and scratches or roughness can cause
field enhancements which precipitate field emission or vacuum breakdown \Ref{Fremerey99_01,Temnykh01}. Furthermore, both surface contaminants
and dielectric inclusions in the metal can alter the work function of the material.

First, all metallic HV surfaces of the gun, including stalk,
cathode electrode, copper rings, TPP rings, and anode, were
mechanically hand-polished using silicon carbide. For the
stainless steel electrodes, an additional polishing with diamond paste is performed.  Then, a chemical polishing step
was applied to all mechanically polished parts. For stainless
steel parts (stalk, cathode electrode TPP rings), standard electropolishing was performed, removing 10~$\mu$m of material. For
the copper rings, a weak citric acid etch was performed, as
this was shown via interferometric microscopy to produce a
surface with smaller root-mean-square (RMS) roughness than
a more powerful copper etchant (such as nitric and sulfuric
acid). After electropolishing, all
stainless steel vacuum components (including chamber, stalk,
and electrodes) are baked in air at 400~C for 100~hours to reduce hydrogen outgassing. 

After surface treatment, all vacuum surfaces (including
the chamber itself, but excluding the vacuum pumps) in the
gun were high pressure rinsed with DI
water in clean room conditions equivalent to ISO~5 or better, to remove particulate contamination. The insulator itself
was rinsed with 300~psi water, whereas all metallic surfaces
were rinsed with pressures 600 psi (minimum), for approximately 3~h
per part. Both copper and TPP rings were rinsed on a separate
rinse stand, apart from the insulator.

After a full air dry of all parts in the clean room, the insulators were populated with rings manually. The cathode electrode was assembled on a test stand, and then suspended in the chamber via
a temporary support from the back of the gun. The insulators were installed on to the
chamber and cathode assembly via a clean room crane. The
stalk without top plate was similarly lowered through the insulators via the crane, and was attached to the cathode and
temporary cathode support. Finally, the top plate was installed
and attached to the stalk, allowing the cathode electrode support to be removed. Using surveying mounts installed in the
photocathode holding structure, the height and angle of the
electrodes were adjusted to be concentric with the axis defined by the anode and load-lock chamber flanges. The gun
was then sealed, evacuated and checked for leaks, then transported to its final location. A vacuum
bake was performed at 150~C for approximately two days,
followed by NEG activation at 500~C for 1 h. The final vacuum prior to processing was $2 \times 10^{-11}$ torr.


The gun HV power supply, 750~kV at 100~mA, is based on proprietary insulating core transformer technology and was manufactured by Excelitas Technologies (formerly Kaiser Systems). It is comprised of a stack of circuit boards insulated from each other. Each board contains two ferrite cores which couple a high frequency magnetic field from one board to the next. Each board can deliver up to 12.5~kV at 100~mA, and is only 5~mm thick, leading to a very compact supply. 62 boards are used in the full stack, which is shown in \Fig{fig:photo_hv_section_of_power_supply}. The primary of the ferrite insulating core transformer is powered by an external high frequency driver.

\begin{figure}[htbp]
\centering
\includegraphics[width=0.95\textwidth]{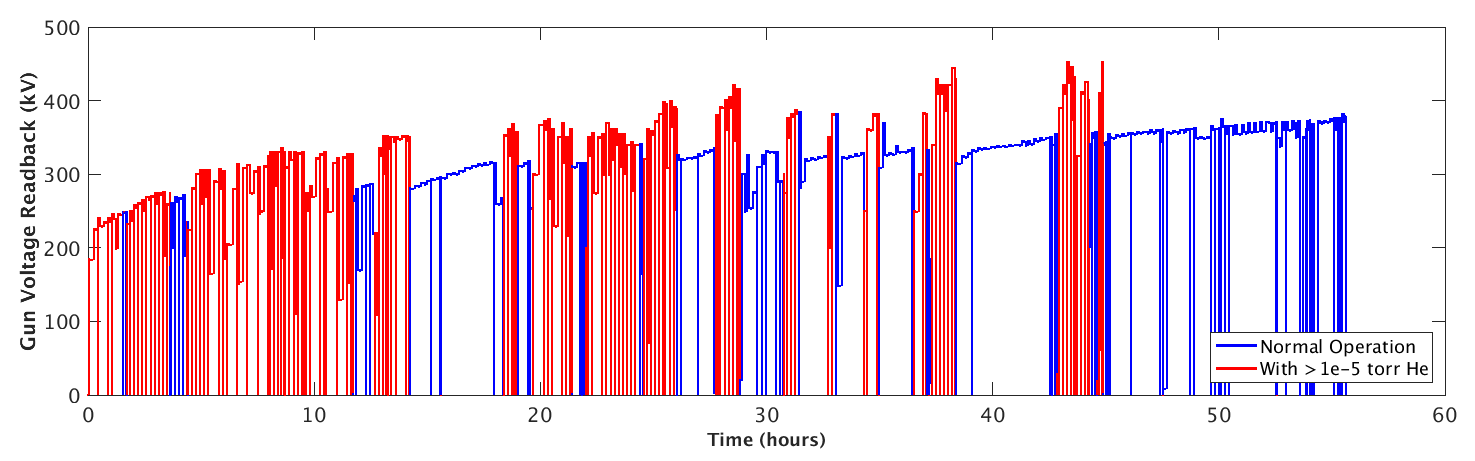}
\caption[Graph of HV processing.]{The voltage applied to the gun during conditioning. Data points are colored for UHV (blue dot) and helium (red) conditioning. }
\label{fig:gun_processing}
\end{figure}

The DC gun was initially processed for 150~h and used in previous experiments. Installation of the gun in the CEBTA experimental hall required re-baking of the gun as and required further high voltage processing.  The gun was installed with a 600~kV DC power
supply from Excelitas Technologies (we have a 750~kV and a 600~kV supply), in an SF$_6$ environment at 3~bar.
The gun is connected to the power supply via three 300~M$\Omega$
 film resistors (Nichrome Electronics, model 1000.300), each 1 m long, connected in parallel for redundancy. This processing resistor
limits the current applied to the gun in case of a short circuit through
an arc in the gun. Each segment of the insulator has two parallel sets of 1~G$\Omega$ resistors (Caddock resistor model MG815), giving the insulator a total resistance
of 7~G$\Omega$, as verified by an electrometer. Thus, a given voltage V
applied by the power supply corresponds to a voltage across
the gun of ($1 − 0.014)\times V$, as 1.4\% of the supply voltage is
dropped across the processing resistor.


The conditioning was performed with an anode-cathode gap of 50~mm. The voltage across the gun was slowly
increased to a state of pre-breakdown, most often to the
point of tripping off due to exceeding the current limit setting, with
subsequent attempts permitting higher and higher voltages before tripping. 
We found that alternating cycles of processing with/without (shown in red/blue in \Fig{fig:gun_processing}) helium to be effective in suppressing and removing field emitters. For gas
processing, ~$5 \times 10^{-5}$~torr of   gas was introduced into
the gun via leak valve, with all ion pumps off and an external turbo pump connected.  Noble gases are not well 
pumped by the NEG's, so do not saturate them. The use of gas routinely allowed a voltage
setting above 450~kV. Conditioning with alternating rounds of UHV and gas conditions allowed us to apply conditioning voltages around 400~kV in UHV. At this time, the gun is stable enough for operations at 350~kV, but further processing is needed to be stable at 400~kV.

\section{Photocathodes}

\begin{figure}[htbp]
\centering
\includegraphics[width=0.8\textwidth]{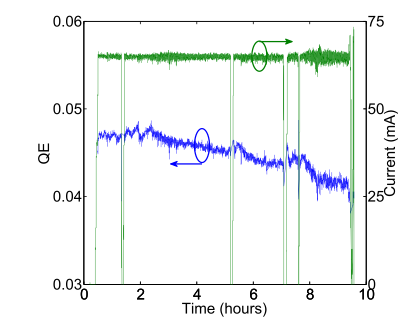}
\caption[Cathode lifetime]{Photocurrent and QE of a NaKSb photocathode operated in a high
average current run of the ERL injector.
65 mA of average current was been delivered for about 9 h. QE degradation
with a 1/e time constant  of  66~h was  observed}
\label{fig:lifetime}
\end{figure}

Alkali photocathodes are known to be robust from previous RF gun use \Ref{diag_Akre2008} and use as detectors in photomultiplier tubes.  Following on the work of others, we developed recipes to grow various alkali cathodes and measured their properties.  Numerous papers by our team give details on thermal emittance measurements and high current operation.  Overall, alkali-type cathodes have roughly the same thermal emittance as GaAs cathode (at 520 nm), but are much less sensitive to vacuum disturbances and thus have reasonable lifetime.  For operations, we typically use NaKSb type photocathodes, which are known to survive even at elevated temperatures.  We performed high average current lifetime measurements using a cathode of this type.  \Fig{fig:lifetime} shows a plot of average current and quantum efficiency as a function of time, demonstrating that, with adequate laser power, one can operate for approximately 1 week using a single cathode.

\subsection{The Laser System}
The key design parameters of  the Cornell photoinjector laser system are given in \Tab{tab:injector_laser_parameters}. A brief justification for these parameter choices follows. The wavelength chosen for the laser is 520~nm -- corresponding to the frequency of doubled Ytterbium fiber laser. This wavelength is a reasonable match to the desirable properties of alkali cathodes, and it is relatively easy to generate significant average optical powers. Yb lasers allow the generation of a high frequency comb of pulses with a range of optical pulse widths. It is also relatively easy to shape these visible optical pulses transversely and longitudinally, and to control the light reaching the photocathode with fast electro-optic devices.	

\begin{table}[tb]
\caption[]{Key design parameters for the ERL photoinjector laser system}
\begin{tabular*}{\columnwidth}{@{\extracolsep{\fill}}lr}
\toprule
Wavelength							&	520 nm \\
Average power at the cathode			 &	2 Watt \\
Repetition rate						&	$1300/31 = 41.935$ MHz \\
Synchronization to external an RF signal		&	 Better than 1 ps rms \\
Pulse duration (rms)				 &	9 ps \\
Pulse temporal shape					&	flat top, $< 2$ ps rise and fall \\
Transverse shape						&	Truncated Gaussian \\
Power stability						 &	Better than 2\% \\
Position stability					&	10 microns rms  \\
\bottomrule
\end{tabular*}
\label{tab:injector_laser_parameters}
\end{table}

Assuming a 1\% quantum efficiency photocathode, 2 watts of laser power is sufficient to generate 200 pC of bunch charge at 41.9 MHz, which corresponds to 8 mA of beam current, well above our KPP goal of 1 mA. Between the exit of the laser and the photocathode, a large number of optical and electro-optical devices are necessary, to transversely and longitudinally shape the optical pulses, transport them from the laser exit to the photocathode, focus them on the photocathode, provide a suitable means to start up beam delivery for both tuning and full power operation, and finally to rapidly terminate beam delivery in the case of a fault. The large number of optical elements means that even with antireflection coatings on all surfaces, there will be a very significant optical power loss, from both reflection and absorption, between the laser and the photocathode. A factor of two loss is not exceptional, and indeed, requires care to achieve. Accordingly, we require that the laser system provide at least 4~W of average optical power at its exit, which has already been achieved. An even larger value may be required to provide additional headroom for optical losses, laser beam shaping, and feedback overhead.

The synchronization of the laser output pulses with the RF signal from the Master Oscillator affects the timing jitter of the beam bunches with the accelerator RF. This timing jitter is compressed during the bunching that takes place in the injector by a factor of 10--20. The present laser system has already achieved $< 1$~ps rms jitter.

The 9~ps optical pulse duration requirement is based on simulations showing that this pulse width gives the smallest final beam emittance from the injector (though the value is not very critical due to the presence of the tunable RF buncher). This pulse width has been obtained by external optical pulse shaping, using birefringent crystals to stack 16 laser pulses longitudinally. 

A power stability of 1\% is typically the best such a high power laser can produce without feedback. The sources of instability are thermal drift in mechanical components, vibrations in gain fibers or crystals, and noise in the pump lasers. For Yb-fiber lasers, the inversion time is on the order of milliseconds, producing noise at kHz rates, but the pulse-to-pulse stability at 42~MHz will be very good as the time between pulses is much shorter than this. Pump lasers can generate noise at many frequencies, from typical line frequencies to $100+$~kHz for those using switching supplies. The electron beam current stability will need to be better than 1\%, thus requiring a series of slow and fast feedback systems between the beam and the laser. 

Poor pointing stability leads to a smearing out of the electron beam size (and shape), leading to emittance growth. Beam simulations show that a 10~$\mu$m rms position jitter is acceptable from the point of beam centroid jitter, which generally responds differently to the accelerator optics than the beam envelope in the space charge dominated regime of the photoinjector. Based on our experience, the laser can achieve 10~$\mu$m rms jitter directly after second harmonic generation (SHG) crystal, at the position of a beam waist. Using a series of 1:1 imaging telescopes to transport the beam to the photocathode, the low jitter after the laser can be maintained. 



\begin{figure}[htbp]
\centering
\includegraphics[width=0.7\textwidth]{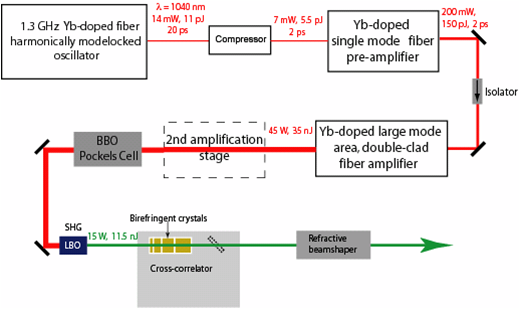}
\caption[Drive laser system schematic.]{Drive laser system schematic. Current 1.3 GHz oscillator will be replaced with a 41.9 MHz oscillator for CBETA.}
\label{fig:diagram_drive_laser}
\end{figure}

The laser system is shown schematically in \Fig{fig:diagram_drive_laser}. The oscillator is a commercially built- harmonically mode-locked fiber laser \Ref{Pritel}. It provides a 1.3~GHz train of 20~ps-long pulses synchronized to an external clock. For this project, a new oscillator operating at 41.9 MHz will be needed. We have an existing alternate home-built oscillator operating at 50.0 MHz, and that will require only a small modification to the fiber length to operate at 41.9 MHz, and should be a simple modification. The  pulses are fed to a single mode fiber pre-amplifier where the pulse energy is boosted to 150~pJ (200~mW average power). This pulse energy is small enough to avoid nonlinear effects in the fiber. The required pulse energy of 100~nJ is achieved through amplification in a double-clad large-mode-area fiber amplifier built to work in a nearly single mode regime. The pulses are compressed to 2 ps after the amplifier using a pair of gratings.  The amplified IR pulses are frequency doubled in a LBO crystal to produce pulses centered at 520~nm. Currently, with one high power amplification stage we have achieved 110 W average IR power and 60W average green power with good stability. This is the highest average power achieved with a fiber laser at this frequency. 


As is well known, generating low emittance beams from a photocathode gun depends strongly on the laser shape incident on the cathode. We have developed a technique to shape the pulses longitudinally by stacking $2^n$ short pulses in $n$ birefringent crystals \Ref{Bazarov08_01}. This technique produces a nearly flat-top laser pulse, has low optical losses, is easy to implement.

In the transverse plane, a truncated Gaussian profile is desirable for minimize emittance. We have tried a number of commercial devices with only moderate success. The old-fashioned method of expanding the laser before it passes through a pin-hole, then imaging the pinhole to the cathode is still the most reliable method. However, it wastes considerable beam power, leading to a rather high requirement for power at the laser exit. In our experience, the commercial devices are similarly inefficient in practical use, and will not be used here.

\section{Buncher System}

To reduce emittance dilution due to space charge effects in the beamline between the gun and the first superconducting cavity, the electron bunches are created at the photocathode with the rms duration of 10-30~ps or 5-14$^\circ$ at 1.3~GHz. On the other hand, to minimize a nonlinear energy spread due to RF waveform in the main superconducting linac, a much shorter bunch duration of about 2~ps rms is desirable. Hence, the bunch length has to be compressed after the gun. The first stage of the bunch compression  happens in the beamline between the gun and the injector superconducting linac. As the beam is still non-relativistic at this point, the simplest method of bunch compression is the velocity bunching, a well-known technique used, for example, in klystrons. Rather moderate requirements for the buncher cavity voltage (up to 200~kV) make it possible to use a normal conducting structure. \Table{tab:buncher_rf_specifications} summarizes buncher cavity and RF system specifications.

\begin{table}[tb]
\caption[]{Buncher RF system specifications}
\begin{tabular*}{\columnwidth}{@{\extracolsep{\fill}}lr}
\toprule
Operating frequency	&	1.3~GHz \\Cavity shunt impedance, Rsh = $V^{2}_{acc}/ 2P$	&	 1.7 M$\Omega$ \\Cavity quality factor	 &	20,000 \\Nominal accelerating voltage		&	 120 kV \\Cavity detuning by beam current at nominal voltage	&	46.0 kHz \\Cavity wall dissipation power at nominal voltage		&	4.24 kW \\Maximum accelerating voltage		 &	200 kV \\Cavity detuning by beam current at maximum voltage	&	27.6 kHz \\Cavity wall dissipation power at maximum voltage		 &	11.8 kW \\Maximum IOT output power		 &	16 kW \\Amplitude stability	&	$8\times10^{-3}$ rms \\Phase stability	 &	 $0.1^\circ$ rms \\
\bottomrule
\end{tabular*}
\label{tab:buncher_rf_specifications}
\end{table}

In order to maximize the energy variation along the bunch at a given cavity accelerating voltage $V_{\text{acc}}$, the beam passes the buncher cavity $-90^\circ$ off crest, i.e. at its zero-crossing. The RF power required to maintain a constant field in the cavity is then given by
\begin{equation}
P_{\text{forw}} = \frac{V_{\text{acc}}^2}{R/Q\cdot Q_{\text{ext}}} \frac{ \left(1+\beta\right)^2}{4 \beta^2} \left[1+\frac{Q_0^2}{\left(1+\beta\right)^2}\left(2\frac{\Delta\omega}{\omega} - \frac{I_b\, R/Q}{V_{\text{acc}}}\right)^2 \right]
\end{equation}
where $\beta = Q_0 / Q_{\text{ext}}$ is the coupling factor of the input coupler, $\omega_c$ is the cavity resonant frequency, $\Delta\omega = \omega_c -\omega$, and $\omega$ is the RF frequency. It is desirable to minimize the required RF power with and without beam passing through the cavity. The minimum power of 5.8~kW is required at nominal accelerating voltage, if the cavity frequency is tuned to 1300.000 + 0.023~MHz and if the coupling factor is $\beta = 1.7$. Amplitude fluctuations of the buncher cavity voltage will affect the resulting bunch length. If the bunch length fluctuation should not be more than 0.1~ps rms, the amplitude stability requirement is only $8\times10^{-3}$ rms. The phase stability is derived from the required energy error and is $0.1^\circ$ rms.

\begin{figure}[htbp]
\centering
\includegraphics[trim = 2mm 2mm 2mm 2mm, clip, width=0.5\textwidth]{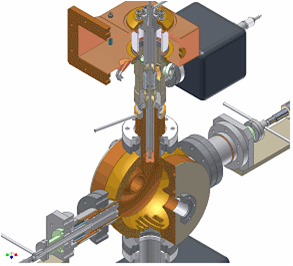}
\caption[3D view of the buncher cavity.]{3D view of the buncher cavity showing input coupler, plunger-type frequency tuner, pumping slots.}
\label{fig:diagram_buncher_cavity_3D}
\end{figure}

The buncher cavity \Ref{Veshcherevich03_01} is a copper single-cell cavity that has an optimized spherical reentrant shape. A 3D view is shown in \Fig{fig:diagram_buncher_cavity_3D}.
The cavity input coupler is of a water-cooled coaxial loop type. Its coaxial part is short and ends with a coax-to-waveguide transition, which incorporates a ceramic window. The coupling can be adjusted during installation by rotation of the coupling loop. The coupling loop, inner conductor and part of the outer conductor of coaxial line are water cooled. The cavity has two tuners with water-cooled 40~mm pistons. The pistons are moved by linear motion actuators with stepper motors. Two tuners provide a better field symmetry on the beam axis. Only one tuner is used for routine operation, the other one is used for preliminary frequency adjustment. During operation, the tuner has to compensate thermal effects (roughly 400~kHz from cold cavity to maximum voltage) and beam detuning. That corresponds to plunger travel of 2~mm. The full 15~mm stroke of one tuner gives a tuning range of 2.5~MHz.

\begin{figure}[htbp]
\centering
\includegraphics[width=0.8\textwidth]{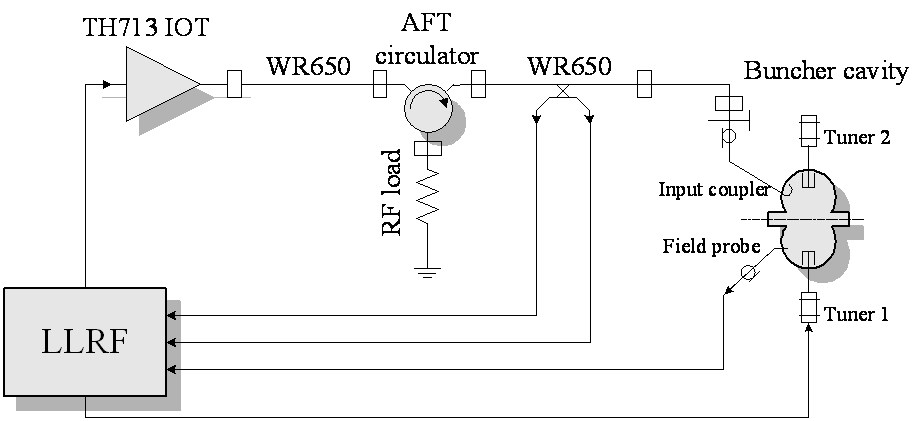}
\caption[Buncher cavity RF system]{Block diagram of the buncher cavity RF system.}
\label{fig:diagram_buncher_cavity_RF_system}
\end{figure}

\begin{figure}[htbp]
\centering
\includegraphics[width=0.6\textwidth]{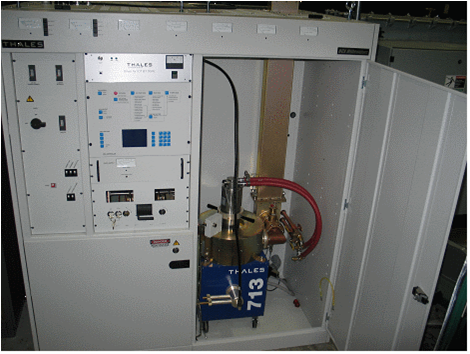}
\caption[IOT]{IOT inside the transmitter.} 
\label{fig:photo_TH_713_IOT}
\end{figure}

The buncher RF power station in the  injector \Ref{Belomestnykh08_01} comprises low level RF (LLRF) electronics, a high power amplifier (HPA), and waveguide transmission line components connecting the HPA to the cavity.    The block diagram of the buncher RF is shown in \Fig{fig:diagram_buncher_cavity_RF_system}. The HPA incorporates a 16~kW IOT tube in a commercial broadcast unit a photo of which is shown in \Fig{fig:photo_TH_713_IOT}.  The HPA efficiency is 60\% with a gain of 21~dB at maximum power output.  The amplitude and phase ripple noise without the LLRF feedback are 0.13\% and $0.5^\circ$ respectively.

\section{Injector Linac}\label{sec:injector:linacdesign}

\subsection{Introduction}
The ERL injector linac contains 5 2-cell superconducting RF cavities, each providing an energy gain of up to 1.2~MeV at maximum 100~mA beam current. RF power is transferred to each cavity via two input couplers, `twin couplers', up to 120~kW per cavity. Efficient absorption of the HOM power is achieved by placing broadband HOM absorbers in the beam tube sections between the cavities. The cryomodule design is based on the TTF-III technology with modifications for CW operation.

A five-cavity  injector cryomodule was designed and fabricated as part of the Cornell ERL Phase~1A effort with the goal of de In the following sections, the designs of the 2-cell SRF cavity, input coupler, HOM absorbers, LLRF system, and cryomodule are discussed in detail. The original injector was designed to support 100~mA beam currents, so this text reflects designs for 100~mA, even though we do not plan to operate at such a high level for the multi-pass ERL machine.

\subsection{Injector cavities}

\begin{figure}[htbp]
\centering
\includegraphics[width=0.8\textwidth]{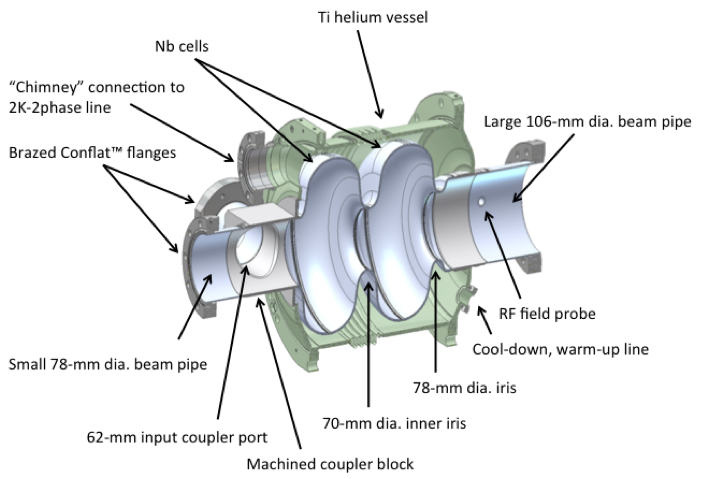}
\caption[]{ERL injector cavity.}
\label{fig:diagram_injector_cavity}
\end{figure}

Two-cell, 1.3~GHz superconducting cavities were developed for the ERL injector . The cavity design (\Fig{fig:diagram_injector_cavity}) was optimized for handling high-current, low-emittance CW beams \Ref{Shemelin03_01}. The cavity parameters are listed in \Tab{tab:injector_cavity_parameters}. Efficient damping of the HOMs is essential to reduce resonant heating due to monopole HOMs and to avoid beam breakup instabilities due to dipole HOMs. Since the TTF-III technology was chosen as the baseline for the injector design, the inner iris diameter (70 mm) and the beam pipe diameter (78~mm) are identical to those of the TESLA cavity \Ref{PhysRevSTAB.3.092001}. However, in this geometry the lowest dipole HOM (TE11-like) is trapped. To facilitate propagation of this mode toward a beamline HOM absorber, the diameter of one of the cavity beam pipes was increased to 106~mm. A 78~mm diameter iris at this end of the cavity keeps the electromagnetic fields of fundamental mode from leaking out of the cell. The cell shapes were optimized for a maximum value of $G\cdot R/Q$ to minimize the cryogenic load while ensuring that the frequency of the lowest TE11-like mode stays at least 10~MHz above the large beam pipe cut-off frequency.

\begin{table}[tb]
\caption[]{Parameters of the injector cavity}
\begin{tabular*}{\columnwidth}{@{\extracolsep{\fill}}lr}
\toprule
Resonant frequency ($\pi$ mode)			&	1.3~GHz \\
Accelerating voltage						&	1.2 MV \\
Accelerating gradient, $E_{\text{acc}}$		&	5.5 MV/m \\
Cells per cavity							&	2 \\
$R/Q$									&	222 $\Omega$ \\
Geometry factor, $G$ 					&	261 $\Omega$ \\
Cavity quality factor, $Q_0$				& 	$>1\times10^{10}$ \\
Nominal external quality factor, $Q_{\text{ext}}$ &	$5.4\times10^4$ \\
Cell-to-cell coupling					&	0.7\% \\
$E_{\text{pk}} / E_{\text{acc}}$ 				&	1.94 \\
$H_{\text{pk}} / E_{\text{acc}}$ 				&	42.8 Oe/(MV/m) \\
Small beam pipe diameter					&	78 mm \\
Large beam pipe diameter					&	106 mm \\
Inner iris diameter						&	70 mm \\
Active cavity length						&	0.218 m \\
Cavity length flange to flange			&	0.536 m \\
\bottomrule
\end{tabular*}
\label{tab:injector_cavity_parameters}
\end{table}

\begin{figure}[htbp]
\centering
\includegraphics[width=0.7\textwidth]{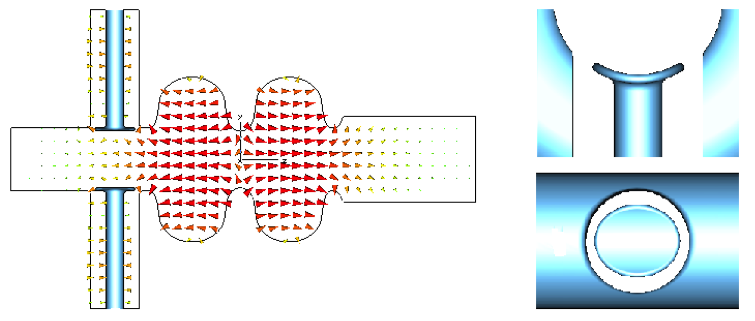}
\caption[]{Geometry of the cavity and details of the coupler antenna with the electric field lines of the fundamental mode indicated}
\label{fig:diagram_injector_cavity_geometry}
\end{figure}

To support a 100~mA CW beam, the input coupler has to be strongly coupled to the cavity and this induces a strong, non-symmetric local perturbation of the otherwise axially symmetric cavity fields. This produces a transverse kick to the beam even if it traverses the cavity on axis. To compensate for this kick, the injector cavity uses two identical symmetrically placed antenna type couplers (twin couplers) that are described below. An additional benefit of using twin couplers is a 50\% reduction in the RF power per coupler. Optimization of the coupler antenna tip was part of the cavity design process. The result is a bent elliptic disc, which conforms to the radius of the beam pipe \Ref{2005pac..conf.4290L} and is shown in \Fig{fig:diagram_injector_cavity_geometry}. Bending of the disc increased the coupling by 20\%. Since one of the goals for the ERL injector  was to explore a range of beam energies from 5 to 15~MeV, the input coupler was designed to be adjustable. 

\begin{figure}[htbp]
\centering
\includegraphics[width=0.7\textwidth]{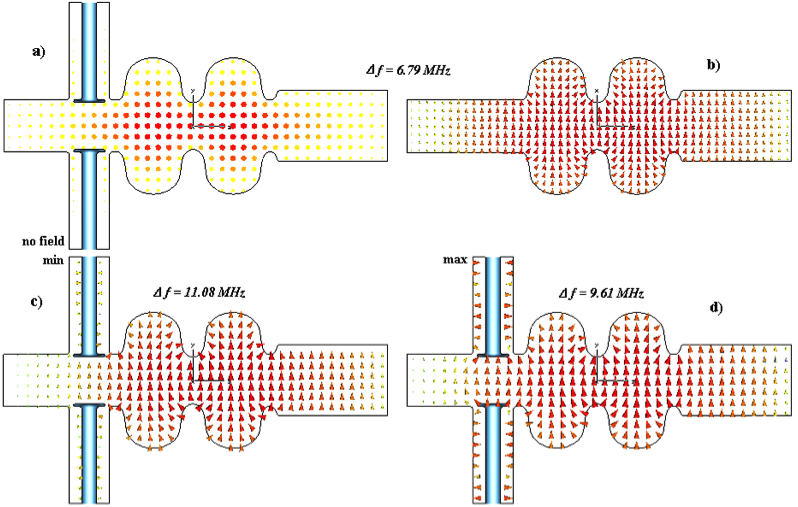}
\caption[]{Electric field of the two dipole modes: a) and b) the `perpendicular' mode; c) and d) the `in-plane' mode with electric and magnetic walls at the ends of coaxial lines, respectively}
\label{fig:diagram_injector_cavity_fields}
\end{figure}

The TE11-like mode can have two polarizations resulting in two degenerate modes with identical resonant frequencies. The geometric perturbation introduced by the input couplers resolves the degeneracy and splits the modes into an `in-plane' mode and a `perpendicular' mode with respect to input couplers as shown in \Fig{fig:diagram_injector_cavity_fields}. The frequencies of the modes are different from the original one but stay high enough above the cut-off frequency. The in-plane mode is strongly coupled not only to the beam pipe but also to the input couplers resulting in an external $Q$ of 250, compared to the $Q_{\text{ext}}$ of 1000 for the perpendicular mode.

Six cavities, one  and five production cavities, were fabricated for the ERL injector . The inner surface of each completed cavity was etched to remove $120\unit{\mu m}$ with BCP 1:1:2 at a temperature below $15^\circ\unit{C}$ maintained by water-cooling the exterior of the cavity. Because of the vertical orientation during etching, the cavity needed to be flipped to eliminate asymmetric removal across the cells. Brazed joints and knife edges at the ConFlat\textregistered\ flanges were protected with Teflon plugs to shield them from being attacked by the acid. After chemical etching, the cavity was rinsed with a closed-loop DI water system overnight followed by a four-hour session of high-pressure water rinsing in a clean room. All cavities reached the performance goal during vertical RF tests \Ref{Geng07_01}.

\begin{figure}[htbp]
\centering
\includegraphics[width=0.7\textwidth]{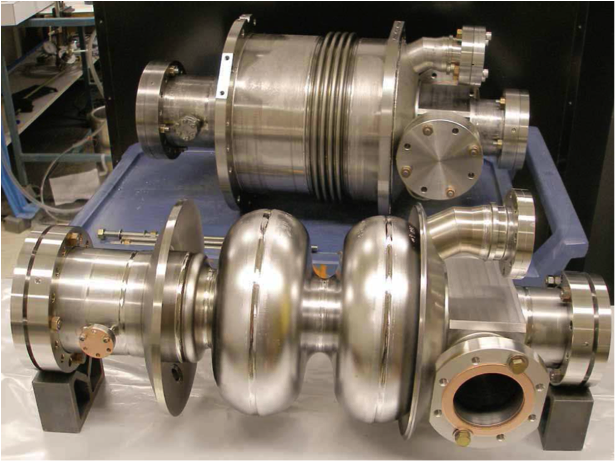}
\caption[]{2-cell cavity before and after welding the helium vessel}
\label{fig:photo_injector_cavity}
\end{figure}

%
%
%
%
%

\subsection{Injector input coupler}\label{sec:injector:coupler}

The input coupler is one of the key components of the injector linac due to strict requirements such as a high CW power transferred to the beam (up to 120~kW per cavity), strong coupling, wide range of coupling adjustment, and small distortion of transverse beam motion. Each injector cavity is equipped with two identical antenna type couplers symmetrically attached to a beam pipe of the cavity. This is a remedy to reduce RF power per single coupler, coupling to the cavity, and the transverse kick to the beam.

 The design of the ERL injector couplers is based on the design of TTF III input coupler \Ref{Dwersteg01_01}, consisting of a cold section mounted on the cavity in the clean-room and sealed by a `cold' ceramic window, and a warm section incorporating a transition from the evacuated coaxial line to the air-filled waveguide. The warm coaxial line is sealed by a `warm' ceramic window. Both windows are made of alumina ceramics and have anti-multipacting titanium nitride coating. Bellows in the inner and outer conductors of the coaxial line of the coupler allow a few mm of motion between the cryomodule cold mass and the vacuum vessel when the cavities are cooled from room temperature to 2~K. A low thermal conductivity is achieved by using stainless steel pipes and bellows with a 10--30~$\mu$m copper plating at the radio frequency conducting surfaces. Also, the bellows allow 16~mm of center conductor movement for coupling adjustment.

\begin{figure}[htbp]
\centering
\includegraphics[width=0.8\textwidth]{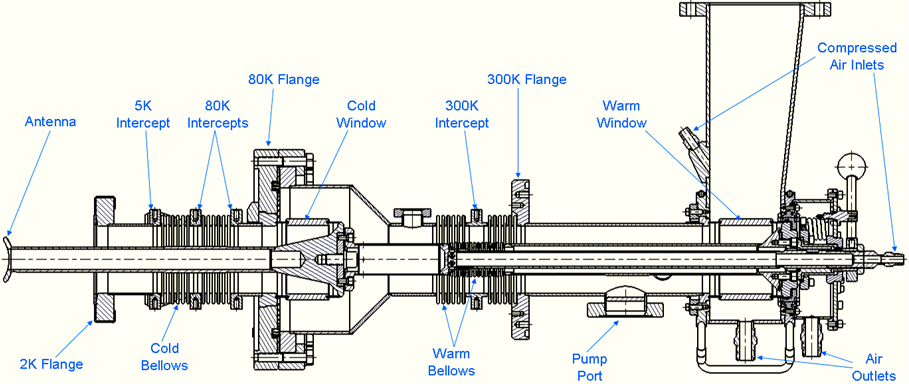}
\caption[Injector input coupler]{2D section view of the injector input coupler.}
\label{fig:diagram_2D_input_coupler}
\end{figure}

The ERL injector coupler design has, however, significant modifications necessary to handle much higher average RF power \Ref{Veshcherevich05_02}:
\begin{itemize}
  \item The cold part was completely redesigned using a 62~mm, 60~$\Omega$ coaxial line (instead of a 40~mm, 70~$\Omega$) for stronger coupling, better power handling, and alleviating multipacting.
  \item The antenna tip was enlarged and shaped for stronger coupling.
  \item The `cold' window was enlarged to the size of the `warm' window.
  \item The outer conductor bellows design (both in warm and cold coaxial lines) was improved for better cooling (heat intercepts were added).
  \item Forced air cooling of the warm inner conductor bellows and warm ceramic window was added.
\end{itemize}
The parameters of couplers for the injector cavities are summarized in \Tab{tab:injector_input_coupler_parameters}. The general design of the coupler is shown in \Fig{fig:diagram_2D_input_coupler}.

\begin{table}[tb]
\caption[]{Parameters if the injector input power couplers}
\begin{tabular*}{\columnwidth}{@{\extracolsep{\fill}}lr}
\toprule
Central frequency	&	1.3~GHz \\Bandwidth	&	$\pm 10$ MHz \\Maximum RF power transferred to matched load		&	 60 kW \\Number of ceramic windows	&	2 \\Qext range	&	$9.2\times10^4$ to $8.2\times10^5$ \\Cold coaxial line impedance	&	 60~$\Omega$ \\Warm coaxial line impedance	&	46~$\Omega$ \\Coaxial line OD	&	62 mm \\Antenna stroke	&	 16 mm \\Heat leak to 2 K		&	$< 0.2$ W \\Heat leak to 5 K		 &	 $< 3$ W \\Heat leak to 80 K	&	$< 75$ W \\
\bottomrule
\end{tabular*}
\label{tab:injector_input_coupler_parameters}
\end{table}

Installed in a cryomodule, high power input couplers require conditioning at high RF power, especially if they were not pre-conditioned before installation. However, \emph{in situ} conditioning is not as flexible as that in a dedicated set up: it is limited to only standing wave (full reflection) mode of operation. All input couplers were processed in pulsed mode up to 25~kW per coupler (50~kW klystron power) at full reflection. All couplers conditioned well, reaching these power levels within 25 to 75~hours (RF on time) of processing multipacting. If the conventional RF processing of multipacting is a limiting factor, two additional built-in measures of alleviate this phenomenon can be employed. First, the warm couplers can be baked \emph{in situ} using special heating elements install on them. Second, a special capacitor assembly can be installed, isolating the center conductor from ground and allowing use of DC bias for multipactor suppression.

\subsection{Wakefield and HOM calculations}

When the 100~mA beam current passes though the  beamline in the injector cryomodule, the electron bunches will leave behind significant electromagnetic fields. The power transferred to these wakefields needs to be intercepted in the HOM absorbers located in the beam pipe sections between the individual cavities. In addition, these HOM absorbers need to damp monopole and dipole modes sufficiently to avoid excessive HOM power in case of resonant excitation of a monopole mode and to guarantee beam stability.

\begin{figure}[htbp]
\centering
\includegraphics[width=\textwidth]{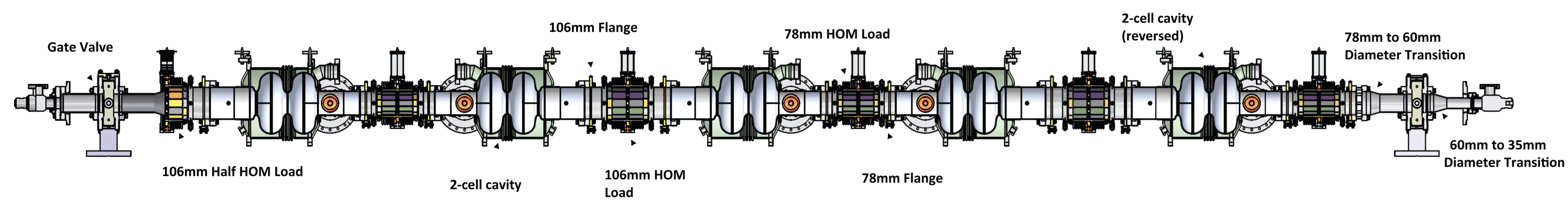}
\caption[]{CAD model view of the Cornell ERL  injector cryomodule beamline with five 2-cell cavities. Note that the full ERL injector will have 12 2-cell cavities. Beamline components from left (beam entrance) to right (beam exit): gate valve; 106 mm half HOM load; first SRF cavity; 78~mm HOM load; second SRF cavity; 106 mm HOM load; third SRF cavity; 78~mm HOM load; fourth SRF cavity; 106~mm HOM load; fifth SRF cavity; 78~mm HOM load; 78~mm to 60~mm diameter transition; gate valve; 60~mm to 35~mm diameter transition.}
\label{fig:diagram_injector_cryomodule_CAD_model}
\end{figure}

\begin{figure}[htbp]
\centering
\includegraphics[width=0.6\textwidth]{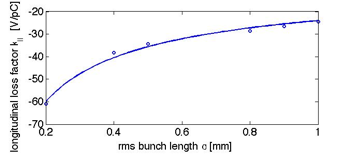}
\caption[]{Total loss factor in the  injector cryomodule beamline with 5 2-cell cavities as a function of bunch length.}
\label{fig:plot_injector_loss_factor_vs_bunch_length}
\end{figure}

The longitudinal loss factor $k_{||}$ of a beamline section can be used to estimate the average power transferred from the beam to electromagnetic fields excited by the beam:	
\begin{equation}
P_{\text{average}} = k_{||}\cdot q \cdot I \, ,
\end{equation}
where $q$ is the bunch charge and $I$ is the average beam current. The total longitudinal loss factor of the beamline section with five 2-cell injector cavities as shown in \Fig{fig:diagram_injector_cryomodule_CAD_model} was calculated \Ref{Liepe09_01}. The result is a longitudinal loss factor of 6.4~V/pC per one-cavity section (32~V/pC for 5 cavities) at the design bunch length of $\sigma = 0.6\unit{mm}$, see \Fig{fig:plot_injector_loss_factor_vs_bunch_length}. Accordingly, the average monopole mode HOM power excited by the 100~mA, 77~pC beam is found to be $\approx 50\unit{W}$ per cavity section, i.e. per HOM absorber.

\begin{figure}[htbp]
\centering
\includegraphics[width=0.75\textwidth]{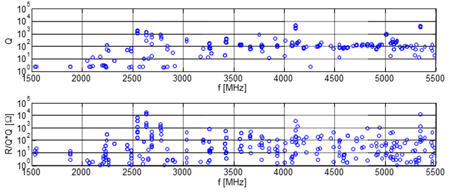}
\caption[]{Simulated monopole mode damping in the full ERL injector (CLANS results). Top: Quality factor of all monopole modes between 1.5~GHz and 5.5~GHz. Bottom: $R/Q\cdot Q$ of these modes. Realistic complex dielectric properties where used in these simulations for the RF absorbing materials in the HOM dampers.}
\label{fig:plot_injector_Q_monopole_vs_frequency}
\end{figure}

To verify the effectiveness of the HOM damping scheme with HOM beam pipe absorbers located between the cavities as shown in \Fig{fig:diagram_injector_cryomodule_CAD_model}, the resulting HOM damping was studied both numerically and experimentally. \Figure{fig:plot_injector_Q_monopole_vs_frequency} shows simulation results for the quality factors of monopole modes between 1.5~GHz and 5.5~GHz, as well as the product of $(R/Q)Q$, which is the figure of merit in case of resonant excitation of an HOM by the beam. The quality factors of the modes are reduced strongly to very low values of typically 100 to a few 1000. Only the modes of the accelerating TM010 passband at 1.3~GHz remain unaffected by the HOM dampers because their frequencies are below the cut-off frequency of the beam pipes at the cavity ends.  Even in the unlikely event of resonant mode excitation, the power transferred to any of these strongly damped modes would be modest and well below the maximum power handling specifications of the HOM dampers. HOM measurements at the Cornell ERL injector  cryomodule have confirmed these simulation results \Ref{Liepe09_02}.

\subsection{Injector HOM dampers}\label{sec:injector:HOM}

The requirements on the beam pipe HOM absorbers in the ERL injector are similar to the HOM damping requirements in the ERL main Linac. The only differences are (1) a  factor of $\approx 4$ smaller average power to be intercepted per load and (2) slightly different beam pipe radii (39~mm and 53~mm instead of 55~mm in the main Linac). Therefore, the HOM dampers in the ERL injectors will be a  modified version of the beam pipe HOM dampers developed for the ERL main Linac. 

\begin{figure}[htbp]
\centering
\includegraphics[width=0.8\textwidth]{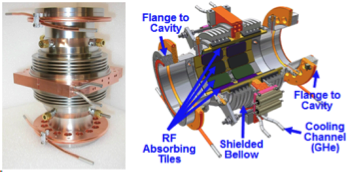}
\caption[]{Cornell ERL injector  HOM load. Left: Finished load. Right: Cut-open CAD model of the  load showing the RF absorber tiles.}
\label{fig:diagram_HOM_load_}
\end{figure}

\begin{figure}[htbp]
\centering
\includegraphics[width=\textwidth]{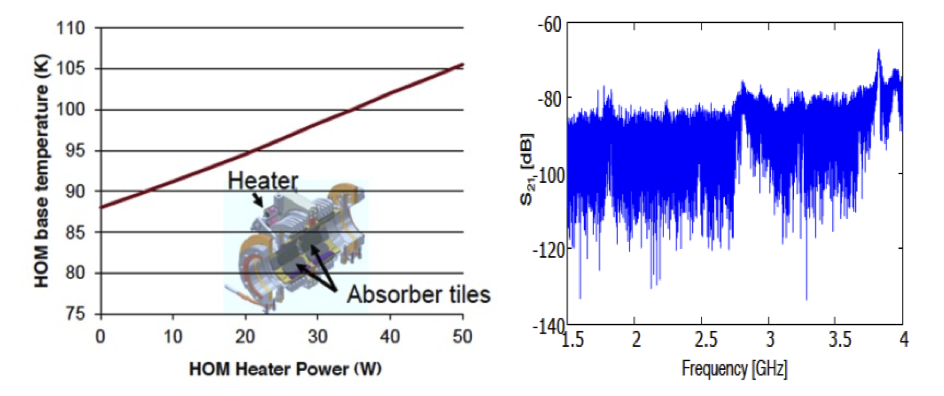}
\caption[]{Left: Temperature of the HOM load temperature as function of power intercepted. The test was done at a relatively low cooling gas flow speed. Right: Vector network analyzer scan for HOMs between 1.5~GHz to 4~GHz. Shown is the transmission amplitude vs. scan frequency. Pick-up antennas on the cavities and HOM loads were used to couple to the HOMs.}
\label{fig:plot_HOM_load_temperature_vs_power}
\end{figure}

Cryogenic HOM beam pipe absorbers have been tested successfully in the Cornell ERL injector. \Figure{fig:diagram_HOM_load_} shows one of the  HOM loads prior to installation in the ERL  injector beamline. The damping of HOMs in the injector cavities by these beamline absorbers was investigated using a vector network analyzer to excite modes via pick-up antennas located at the cavity beam tubes and at the HOM loads (see \Fig{fig:plot_HOM_load_temperature_vs_power}). Preliminary results confirm very strong suppression of monopole and dipole modes with typical quality factors of only a few 1000 as predicted by simulations. Heater elements on the HOM absorber load bodies were used to verify the effective heat exchange to the high pressure cooling gas up to the maximum design heat load of 200~W; see \Fig{fig:plot_HOM_load_temperature_vs_power}. The measured temperature increase of the HOM load body was found to be in good agreement with simulation results.

The   injector  HOM designs suffered from several problems.  The RF tile soldering was not robust, and several tiles detached and fell, generating dust and particles.  In addition, two of the three tile types became insulating enough at 80~K that any charge accumulated on their surfaces would not bleed off.  This charge could be from  electrons scattered during beam tuneup, or from x-rays and UV light generated during cavity processing.  The electrostatic fields generated from the charge buildup severely distorted the beam passing through the cryomodule, making the beam unusable.  The tiles facing the beam were removed, and the solder joints improved on the others to eliminate these problems.

\subsection{Injector RF stability requirements and LLRF}

The intra-bunch energy spread after the injector is about $\sigma_{\text{inj}} = 15\unit{keV}$. It is desirable for the bunch-to-bunch energy fluctuation (bunch centroid energy) at the end of the injector to be below the intra-beam energy spread so that the total energy spread of the beam is dominated by the intra-bunch energy spread. The gun laser timing jitter, the buncher cavity as well as the five superconducting injector cavities each contribute to a bunch-to-bunch energy variation in the injector. We have to distinguish between uncorrelated and correlated (from cavity to cavity) errors. For the ERL injector cavities, small fluctuations in the 100~mA beam loading will be the dominating source of field perturbation, which will cause correlated field errors. Accordingly, we shall assume here correlated field errors in the injector cavities. We will require that the bunch-to-bunch energy fluctuation caused by the injector SRF cavities  increases the total energy spread at the end of injector by no more than 20\%, i.e. to a total of 18~keV rms. Accordingly, the maximum allowable bunch-to-bunch centroid energy gain fluctuation is 10~keV, assuming no correlation between the intra-bunch energy spread and the bunch-to-bunch gain fluctuation. We will allow for 5~keV energy spread contribution from each phase errors and amplitude errors in the 12 injector cavities. This simple estimate results in a requirement for the relative amplitude stability of $\sigma_A/A=5\unit{keV}/15\unit{MeV}=3.3\times10^{-4}$. Assuming acceleration with a phase within 5 deg of on-crest then gives a requirement for the phase stability of $\sigma_p =0.2^\circ$.

A digital LLRF control system will be used to stabilize the RF fields in the injector cavities in amplitude and phase to these stability levels. A combination of feedforward and feedback control will be used to stabilize the cavity fields in the presence of strong beam loading and other perturbations of the RF fields. Sensors will be used to monitor all relevant signals, including the cavity fields, the incident and reflected RF power, and the beam current. Any disturbances due to klystron noise and ripple can be handled using feedforward.   Extremely reliable hardware, a high degree of automation, and sophisticated built-in diagnostics will ensure a high degree of operability, availability and maintainability of the LLRF system.

The  LLRF control system  has been tested extensively, showing excellent performance.
This LLRF system is an improved generation of the LLRF system previously developed for CESR \Ref{Liepe03-01}, with lower loop latency ($<1\mu$s), reduced noise, and increased sample rates and ADC resolution (16 bits). The integral and proportional gains of the fast feedback loop used to stabilize the RF fields in the cavities were optimized. At optimal gains, exceptional field stabilities of $\sigma_A/A < 2\times10^{-5}$ in relative amplitude and $\sigma_p < 0.01^\circ$ in phase (in-loop measurements) have been achieved, far exceeding the ERL injector and ERL main linac requirements. In addition to the fast feedback loop, the system employs feedforward control to compensate beam loading and fluctuations in the high voltage of the klystrons, a state machine for automatic start-up and trip recovery, trip detection, and cavity frequency control.


\subsection{RF Power System for the Injector Linac}

\begin{table}[tb]
\caption[]{Main parameters of the injector cryomodule RF system and power source}
\begin{tabular*}{\columnwidth}{@{\extracolsep{\fill}}lr}
\toprule
Number of RF channels	&	5 \\RF power per cavity	&	120 kW \\Maximum useful klystron output power with incremental gain of 0.5 dB/dB	&	$\geq 120$ kW \\Klystron efficiency at maximum useful power	&	$>50$\%  \\Tube bandwidth at $-1$ dB	&	$\pm 2$ MHz \\Tube bandwidth at $-3$ dB	&	$\pm 3$ MHz \\Klystron gain at nominal operating conditions	&	$>45$ dB \\Klystron beam high voltage	&	45 kV \\Typical klystron current		&	5.87 A \\Maximum klystron CW output power		&	135 kW \\Klystron saturated output power (pulsed)		&	165 kW \\Tube efficiency at saturated power	&	 $>60$\% \\Cavity field amplitude stability		&	$9.5\times10^{-4}$ rms \\Cavity field phase stability		&	$0.1^\circ$ rms \\
\bottomrule
\end{tabular*}
\label{tab:injector_cryomodule_parameters}
\end{table}

\begin{figure}[htbp]
\centering
\includegraphics[width=0.9\textwidth]{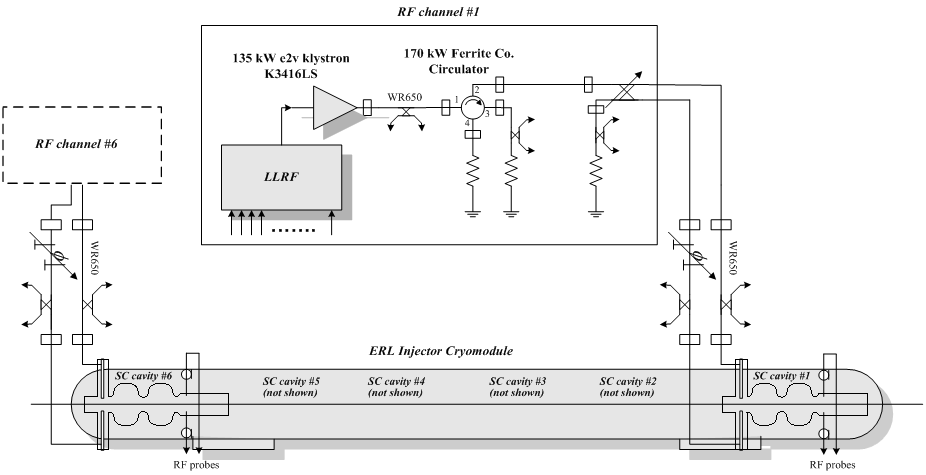}
\caption[ERL injector RF system]{Block diagram of the ERL injector RF system (one cryomodule is shown).}
\label{fig:diagram_ERL_injector_RF_system}
\end{figure}

The injector cryomodules house five 2-cell SC cavities, each delivering up to 120~kW to the beam.  Because the cavities operate independently, the system consists of five identical channels.  Each channel includes a set of LLRF electronics and RF interlocks, a klystron based HPA, and a waveguide distribution network.  RF power is delivered to the cavities via twin input couplers \Ref{Veshcherevich05_02} each carrying 60~kW.  The main parameters of the system are given in \Tab{tab:injector_cryomodule_parameters} and a block diagram is presented in \Fig{fig:diagram_ERL_injector_RF_system}.  A motorized, adjustable short-slot hybrid power splitter and a two stub phase shifter in one of the waveguide arms are used to are used to tune relative amplitude and phase between the two couplers \Ref{Belomestnykh04_03}.  A 170~kW ferrite circulator is used for klystron protection.

The klystrons were designed and manufactured by e2V cavity klystrons and have   165~kW saturated power output.  Similar tubes are anticipated for the 12 cavity injector complement.  To provide stable regulation  of the cavity field the klystron must have finite gain and thus cannot run in saturation.  The maximum power output for the  tube was defined as 0.5~dB/dB of drive and specified to be no less than 120~kW.  At this level the efficiency of the  tubes is at least 50\% and the bandwidth not less than $\pm2.5\unit{MHz}$ at -1~dB level and not less than $\pm3\unit{MHz}$ at -3~dB level.

\subsection{Injector cryomodule}
The ERL injector cryomodule design is based on TTF III technology with modifications for CW operation. This builds upon the considerable development work performed for this linac technology over the past 15 years. TTF III technology is at the forefront of SRF linac performance in regard to cavity gradient, $Q$, power coupled to the beam, cavity tuning, minimal cryogenic heat load, industrial fabrication, and operational reliability. 

The modifications to TTF III technology for CW operation of an injector cryomodule are structurally subtle, but have significant operational differences. Among the modifications to the TTF III cryomodule are the following:
\begin{itemize}
  \item Use 2 coax RF input couplers per cavity, where one 120~kW CW klystron feeds a cavity coupler pair, each coupler rated at 60~kW CW.
  \item The coax RF input couplers have outer conductors with 62 mm diameter and increased cooling for high average power.
  \item The SRF cavities have only 2 cells per cavity with a 0.2 m active length, operated at a nominal gradient of 6~MV/m (1.2~MeV) to deliver the 120~kW klystron power to the beam.
  \item 5 SRF cavities in the injector cryomodule.
  \item One side of the SRF cavity has a larger beam tube diameter, 106~mm, to allow better propagation and damping of HOMs.
  \item Implement beamline HOM Loads for strong broadband damping of HOMs generated by the high current and short bunches.
  \item Cooling of thermal intercepts is provided by small `jumper' tubes with flowing He gas, such as to the HOM loads and the RF couplers, as opposed to copper straps.
  \item Use the INFN blade tuner with the addition of piezos for fast tuning.
  \item Locate access ports in the vacuum vessel to allow the tuner stepper motor to be accessible for replacement while the string is in cryomodule.
  \item Use precision fixed surfaces between the beamline components and the Gas Return Pipe(GRP) for easy `self' alignment of the beamline.
  \item Use rails mounted on the inside of the vacuum vessel and rollers on the composite support posts to insert the cold mass into the vacuum vessel, as opposed to Big Bertha.
  \item Increase the magnetic shielding so that the cavity $Q$ is limited only by the BCS resistance.
  \item Do not include a 5~K shield.
  \item Increase the diameter of the cavity helium vessel port to 10~cm for the high CW heat load.
  \item Increase the diameter of the 2-phase 2~K He pipe to 10~cm for the high CW gas load.
  \item Use a module end-cap and cryogenic feed-cap with reduced length.
\end{itemize}

The ERL injector cryomodule  is based on the TTF III module structure. All of the cavity helium vessels are pumped to 1.8~K (12 Torr) through a common 25 cm inside diameter GRP which also serves as the mechanical support from which the beamline components are suspended. To minimize the heat load to the refrigeration plant, all of the 1.8~K cryomodule components are surrounded by 5~K intercepts to minimize the heat leak to 1.8~K, and the 5~K intercepts are likewise surrounded by 100~K intercepts, which absorb the heat load from the 293~K vacuum vessel. The GRP is suspended from composite support posts that are constructed from low-thermal conductivity G-10 fiberglass. The composite posts have integral metal stiffening disks and rings that also serve as thermal intercepts at  5~K and 100~K between the 1.8~K face that attaches to the GRP and the 293~K face that attaches to the vacuum vessel bosses that support the cold mass. There are stainless steel manifolds of smaller diameter than the GRP running the length of the module that transport the supply of liquid helium and the supply and return of 5~K and 100~K helium gas for the thermal intercepts. Jumper tubes with 5~mm inner diameter are connected between the 5~K and 100~K supply and return manifolds to the various thermal intercepts within a module. A shell of 6~mm thick, grade 1100 aluminum sheet surrounds the beamline and the GRP and is linked to the 100~K manifold to serve as a thermal radiation shield between the 293~K vacuum vessel and the cold mass. The aluminum 100~K shield has apertures through which the RF couplers pass and also has panels with instrumentation feedthroughs. The 100~K shield is mechanically suspended from one of the integral metal stiffeners in the composite support posts. Multi-layer insulation is wrapped around the exterior of the 100~K shield as well as all of the 1.8~K and 5~K cold mass components.

The magnetic shielding in the cryomodule must keep the field in the region of the SRF cavity to $< 2$~mG to have negligible residual SRF wall loss and provide a good safety margin for the goal of cavity $Q_0 = 2\times10^{10}$. Such a low field is accomplished by degaussing the carbon-steel vacuum vessel, lining it with Co-NETIC \textregistered\ mu-metal shielding that will be at 293~K, and then wrapping each cavity's 1.8~K helium vessel with a magnetic shield that is formulated to have maximal shielding at the low temperatures around 4~K \Ref{Amuneal01}.


The injector cryomodule delivers high average power to the injected beam. Even with a modest cavity gradient of 6~MV/m and only 2 cells per cavity, the input RF power of 120~kW CW per cavity to the 100~mA beam is pushing the limits of input couplers, as described in  \Section{sec:injector:coupler}. Having two RF couplers per cavity requires the vacuum vessel RF ports to be symmetrically located on each side of the cryomodule, as opposed to one coupler per cavity with ports along only one side of the module. Having only two cells per cavity makes the cavity much shorter than 7-cell or 9-cell cavities, and the cryomodule structure in the vicinity of the cavities more congested. The blade tuner is then slightly longer than the cavity helium vessel and the helium pumping port must be located on the end cone rather than on the OD of the helium vessel.

\begin{figure}[htbp]
\centering
\includegraphics[width=0.8\textwidth]{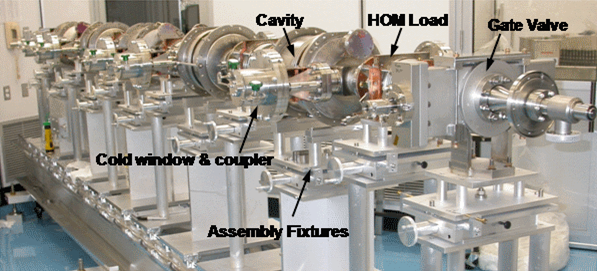}
\caption[Injector  beamline string]{Assembled ERL injector  beamline string in the clean room.}
\label{fig:photo_beamline_string}
\end{figure}

The beamline consisting of the cavities, HOM loads, cold couplers, tapers, and gate valves 
is assembled in a class 100 or better clean room. All components are flushed with filtered water or alcohol and individually receive a mild vacuum bake at $120^\circ$~C for 24 hours. The components are mounted on an assembly fixture one by one in the clean room. Each added component is aligned to the other components with the only critical alignment being the azimuthal position about the beam axis. The azimuthal alignment is needed so that the flat precision mounting surface at their tops will mate to the planar precision surfaces on the GRP. This alignment can be accomplished with a simple accurate spirit level. Any longitudinal spacing or planar shift errors of the mounting surfaces are accommodated by the flex in the HOM load bellows. The component mating vacuum flanges are then bolted together. A photograph of the assembled ERL injector  beamline string in the clean room is shown in \Fig{fig:photo_beamline_string}. After all components are assembled, the beamline string is vacuum leak tested while still in the clean room so that only filtered particulate-free air will pass through any potential leak. The pumping and purging during the leak test is performed at a slow rate of 1--2 Torr/minute through the viscous flow range of 760~Torr--1~Torr to minimize propagation of any particulate contamination throughout the beamline.

 As a parallel operation to the beamline string assembly in a clean room, the cold mass assembly fixture can be set up in a high-bay area with overhead crane access. The composite support posts are attached to the GRP and the GRP is hung from the assembly fixture by the composite posts. The 2-phase pipe is then mounted to one side of the GRP and its exhaust is welded into the GRP.



\begin{figure}[htbp]
\centering
\includegraphics[width=0.9\textwidth]{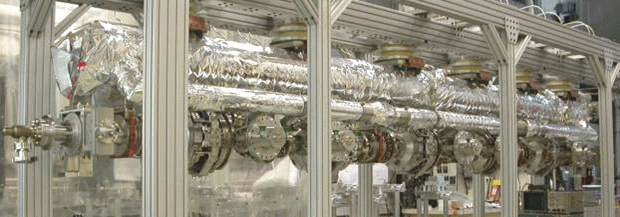}
\caption[Beamline string hung from the GRP]{Beamline string hung from the GRP for the ERL injector .}
\label{fig:photo_hung_beamline_string}
\end{figure}

After the beamline string passes the vacuum leak test, it is removed from the clean room and positioned underneath the cold mass assembly fixture. The string is raised and the precision mounting surfaces on the string and the GRP are brought together with integral alignment pins being engaged. The mating surfaces are then bolted together.  A photograph of the injector  beamline hung from the GRP is shown in \Fig{fig:photo_hung_beamline_string}. String attachment to the GRP in this manner proved to be quick and easy for the ERL injector, the entire procedure taking about 1~hour.

After the beamline is hung from the GRP, magnetic shielding layer I is attached to the helium tanks of the cavities. This shielding will reside at 1.8~K. Traditional Co-NETIC \textregistered\ `mu-metal' shielding derates at cryogenic temperatures to about 15\% of its 300~K shielding capacity, so the magnetic shield I material is formulated to have maximal shielding at low temperatures \Ref{Amuneal01}.

The cavity blade tuners are attached after the magnetic shielding. The stepping motors of the tuners have to be wrapped in a copper sleeve that is tied to 5~K to prevent the motor heat from propagating to the helium vessel. The stepping motors are also wrapped with low temperature magnetic shielding since they can have stray fields of a few hundred mG, which would otherwise be present in close proximity to apertures in magnetic shield I.


Several cryogen manifolds run the length of the cryomodule. These manifolds include a 1.8~K liquid helium supply to the `fill' ports located at the bottom of each of the cavity helium vessels, the supply and return of 5~K helium gas, and the supply and return of 100~K helium gas.


The liquid helium and 5~K gas manifolds are mounted close to the GRP using G-10 standoffs, thus keeping similar temperatures in close proximity to each other with low thermal conductivity connections between them. These manifolds are the next components mounted on the cold mass. Jumper tubes with 6~mm ID are then routed from the liquid helium manifold to the helium vessel fill ports. Jumper tubes from the 5~K gas manifolds are then connected to thermal intercepts on the HOM loads and RF couplers. This final joining of the stainless steel jumper tubes from the manifold to the thermal intercepts is performed by orbital welding. In standard TTF technology, the connections between the manifolds and the thermal intercepts are accomplished by copper straps. For the ERL injector with CW operation, both the 5~K and 100~K heat loads are large enough to require gas flow from the manifolds to the intercepts through jumper tubes.


The 100~K manifolds are mounted outboard of the 5~K manifolds and are attached to the 100~K thermal radiation shield. One of the 100~K supply lines cools the shield, and the return lines are hung from low thermal conductivity hangars. The material of the 100~K shield is grade 1100 aluminum, chosen for its high thermal conductivity and light weight. The shield is fabricated from standard flat panels that are cut and formed to shape. The top portion of the shield is attached to the 100~K ring of the composite support post and is 6~mm thick to support the weight of the cryogen manifolds and the lower portion of the shield.
\begin{figure}[htbp]
\centering
\includegraphics[width=0.9\textwidth]{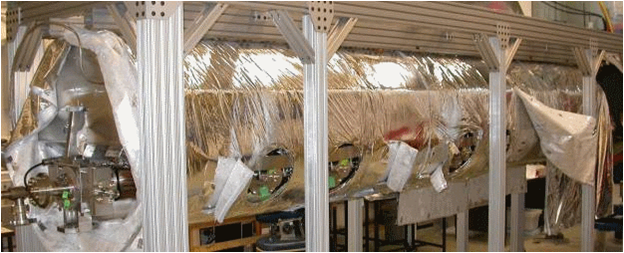}
\caption[Completed ERL injector  100~K shield.]{Photograph of the completed ERL injector  100~K shield being wrapped with MLI.}
\label{fig:photo_completede_ERL_injector_}
\end{figure}

After the cryogen manifolds and intercept jumpers are attached to the cold mass, low thermal conductivity coax cable is routed from the cavity RF field probes with thermal anchoring to the 5~K manifold, along with cabling from temperature sensors, helium level sticks, and other instrumentation. The lower half of the 100~K shield is attached and the instrumentation cabling is thermally anchored at this point to a 100~K instrumentation feed-through panel on the shield. The 100~K shield is then wrapped with 30 layers of Multi Layer Insulation (MLI) and the cold mass is ready for insertion into the vacuum vessel. A photograph of the completed   100~K shield being wrapped with MLI is shown in \Fig{fig:photo_completede_ERL_injector_}.

The cold mass that is wrapped with MLI is pushed into the vacuum vessel and then leveled and aligned inside of the vacuum vessel using jack screws connected to the composite support posts at the top ports. The warm portions of the RF couplers are attached to the cold portions through side ports on the vacuum vessel while under small portable clean rooms. The vessel end plates are attached to the vacuum vessel and it is pumped out and vacuum leak tested.



\section{Beam Stop}\label{sec:beam_dump}

The primary beam stop (dump) must intercept the full beam current at the end of the energy
recovery process, and safely dissipate the beam power as waste heat. The dump was originally designed to handle $600\unit{ kW}$ of average power at a maximum energy of $15\unit{ MeV}$. For this project, the maximum current is $40\unit{ mA}$ with an recovered energy of $6\unit{ MeV}$, so the existing hardware will be more than adequate. 

The range of $15\unit{ MeV}$ electrons is less than $8 \unit{g/cm^2}$ in practical beam dump materials, and thus the beam power is deposited over a very small depth.
The natural beam spot size is quite small, even after energy recovery. The effective area of the beam then needs to be expanded to more than $1\unit{m^2}$ where it intercepts the surface of the dump, to reduce the power density in the dump material to a level that can be safely handled. This expansion can be accomplished by several techniques, such as strongly defocusing the beam, rastering the beam over a larger area, or intercepting the dump surface at a shallow angle. All of these methods will be employed for the primary dump. Clearly the dump material must have a reasonably high-thermal conductivity, to limit the maximum
temperature at the uncooled entrance face of the dump. As there is no significant shower
multiplication from $15\unit{MeV}$ electrons, the surface of the dump, which is furthest from the cooling water, will have the highest temperature.

The only practical choice for the primary dump material is aluminum. Aluminum offers the
very significant advantages of a high photoneutron threshold ($13.3\unit{MeV}$) and relatively low-residual radioactivity comprised primarily of short-lived isotopes. The relatively low-residual radioactivity of aluminum is a significant consideration for the ultimate disposal of a decommissioned beam dump. The aluminum used will be an alloy, and the various alloying elements have lower photo-neutron thresholds. These elements will be responsible for a fraction of the residual radioactivity of a $15\unit{MeV}$ aluminum dump. Copper has a significantly lower photo-neutron threshold, and much higher residual radioactivity of longer-lived isotopes. Beryllium would be exceptionally expensive, and has a very low photoneutron threshold. Carbon, as pyrolytic graphite, is mechanically difficult, and has an extremely anisotropic thermal conductivity.

The dump must remain fully functional during several decades of operation at very high
average power. With an aluminum dump, it is especially critical to control the water chemistry to avoid corrosion. Therefore, heat will be removed from the primary beam dump with a closed circuit de-ionized (DI) water circulation system, which will be continuously powered. The only acceptable metals in this system are aluminum and stainless steel. The water chemistry will be carefully monitored at all times to assure proper pH, resistivity, and the absence of harmful ions.

It is very desirable to minimize the deposition of beam power directly in the cooling
water, to minimize hydrogen production through radiolysis \Ref{Walz67_01}. At the same time, it is desirable to locate the cooling water as close as practical to the interior surface of the dump to minimize thermal effects. These realities led directly to the use of a dump shaped like an ogive (pointed arch) of revolution, similar to a high-power klystron collector. Even with an optimum thickness dump wall, there will be enough radiolysis in the cooling water to require monitoring the hydrogen level in the closed cooling circuit. It is anticipated that the modest quantities of hydrogen generated can be vented to the atmosphere, with no need for hydrogen recombination systems.
Were hydrogen recombination to prove necessary, reliable hydrogen recombination systems
were developed for the high-power beam dumps at SLAC, and were duplicated, with improved
instrumentation, for the high-power dumps at Jefferson Lab \Ref{Wiseman97_01, Walz67_01}. The $15\unit{MeV}$  beam energy is far too low to produce either tritium or $^7$Be through spallation of oxygen, so there will be no direct long-lived radioactivity in the DI water circuit. Heat will be removed from the closed DI water circuit with a water-to-water heat exchanger. The pumps, deionization and filtration equipment,
surge tank, hydrogen-venting scheme, and water-to-water heat exchanger will be located
remote from the dump itself, to allow servicing and to eliminate any potential for radiation damage. All plumbing and piping in the closed-circuit system will be of either aluminum or stainless steel.

The primary dump will be a powerful source of prompt, low-energy gamma radiation as
well as a modest flux of low-energy neutrons. The primary radiation shielding for the dump will consist of steel and concrete blocks that completely surround it. Detailed calculations of the total radiation from the dump have been made with the code MCNPX. 

If the dump were to be operated in normal air, significant quantities of nitric acid could be produced by radiolysis of nitrogen, leading to the production of nitric oxide, which oxidizes to form nitrogen dioxide, which, with water, forms nitric acid. As a consequence, dump area may need to be purge with an inert gas to eliminate the possibility of nitric acid formation. This solution has proven very effective with the two high-average power ($1\unit{MW}$) beam dumps routinely operated at Jefferson Laboratory.

\begin{figure}[htbp]
\centering
\includegraphics[width=\textwidth]{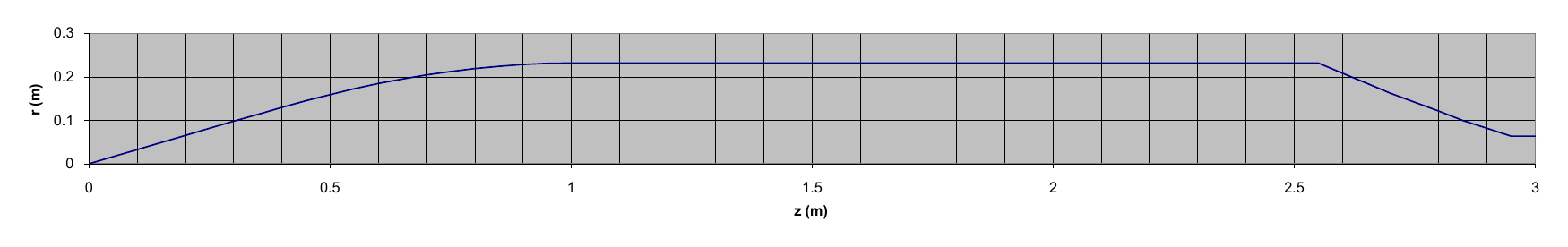}
\caption[]{The inner surface profile of the beam dump}
\label{fig:profile_phase1a_beam_stop}
\end{figure}

Although it is very desirable to isolate the dump from the accelerator vacuum system, this is simply not possible. For example, even in a beryllium window, the power deposition from the $dE/dx$ losses of a $100\unit{mA}$ average current beam is $30\unit{kW}$ per $\unit{mm}$ of window thickness (the window thickness is irrelevant for cooling considerations). It is certainly not practical, and likely not possible, to remove such a large amount of heat from a thin window in vacuum. Thus, the beam dump will of necessity be within the accelerator vacuum system. A differential vacuum pumping system will be used to isolate the high-gas load from the dump when operating at high
average beam power from the much lower pressure in the beamline from the accelerator.  Finally, a reasonably fast-acting, RF shielded gate valve will be located well upstream of the beam dump, to provide protection to the accelerator in the event of a dump failure. This is very important as the superconducting Linac is relatively close to the primary beam dump.

\begin{figure}[htbp]
\centering
\includegraphics[width=0.6\textwidth]{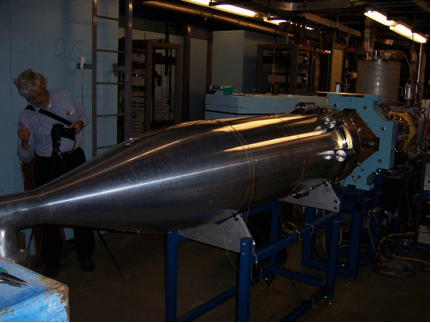}
\caption[]{The completed  beam dump before installation of the shielding blocks.}
\label{fig:photo_phase1a_beam_stop}
\end{figure}

\begin{figure}[htbp]
\centering
\includegraphics[width=\textwidth]{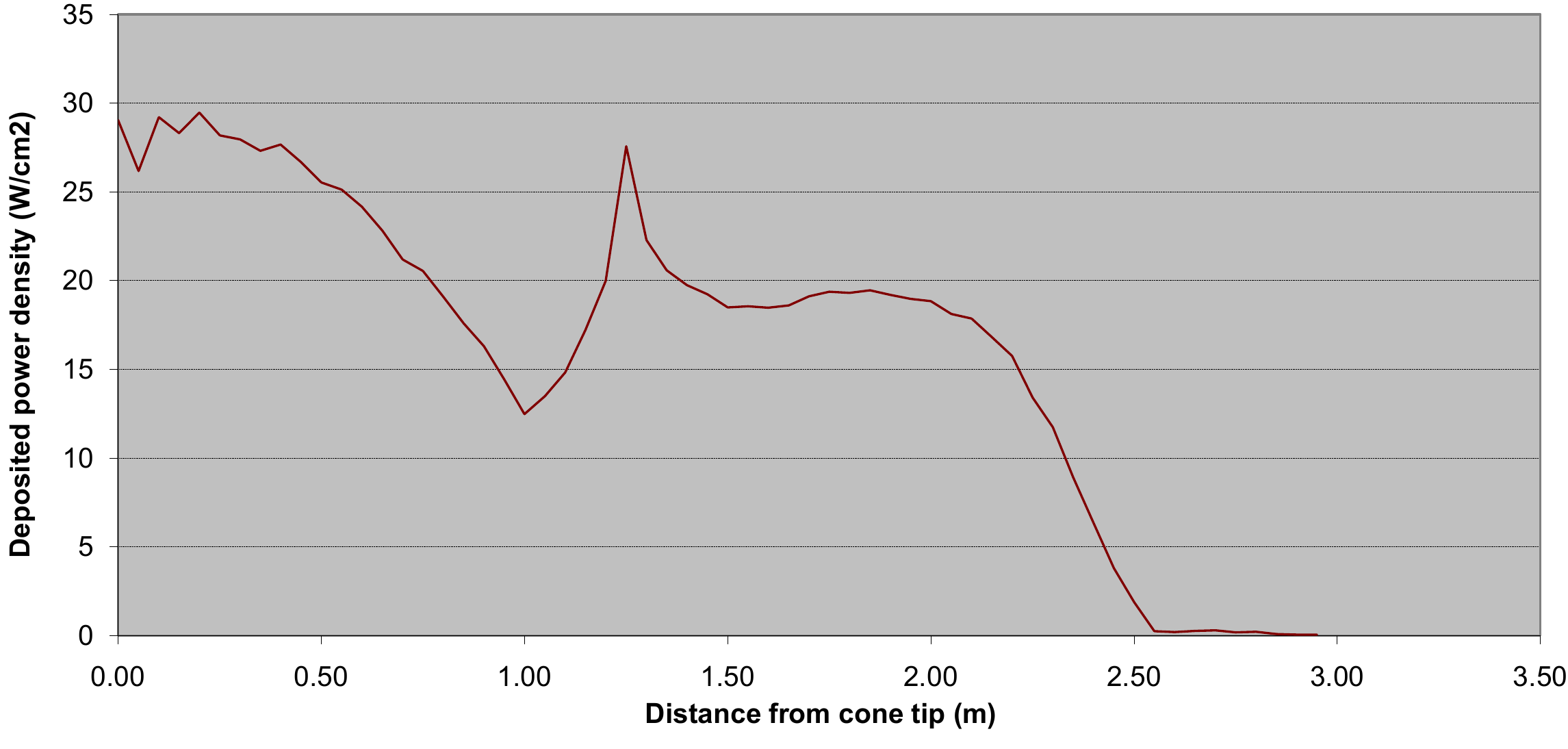}
\caption[]{Energy deposition for 600 kW beam power at optimized condition}
\label{fig:plot_beam_stop_energy_deposition}
\end{figure}

The profile of the inner surface   is shown in \Fig{fig:profile_phase1a_beam_stop}. The 3-meter-long dump was assembled from three shorter segments by electron beam welding. A photograph of the completed dump is shown in \Fig{fig:photo_phase1a_beam_stop}. Water cooling channels are machined in the outer surface of the dump body, which is mounted inside an aluminum jacket. To reduce thermal stresses, the dump body is free to move longitudinally within the jacket. GEANT was used to calculate the power deposition in the dump body, and ANSYS calculations then determined the temperatures throughout the dump, the thermal stresses, etc.
The results of some of these calculations are given in \Fig{fig:plot_beam_stop_energy_deposition}.
Beam on-off cycles are sudden, and result in rapid temperature changes, which in turn may lead to eventual fatigue failure. The water flow was chosen to limit the maximum temperature differentials in the dump, leading to a very large number of temperature cycles before the onset of fatigue failure. For the design of a $60\unit{gpm}$ water flow, the flow velocity is only $1.71 \unit{m/sec}$. Erosion of water channels
will therefore not be a problem.

Two active devices are used to enlarge the beam area at the dump surface --
a quadrupole that strongly over-focus the beam, and a sextupole with the coils powered such that one can make an arbitrary ellipse shape.  The final result of the quadrupole field and sextupole kick is a  raster pattern that move the beam spot in a circular path at $60\unit{Hz}$. If either of these devices failed, the dump would rapidly overheat, quite possibly to the point of damaging, or even melting the dump surface, particularly if there were a transition from nucleate to film boiling at the water-metal interface. Redundant hardwired interlocks will assure that each of the beam focusing and rastering magnets is properly powered. On any interlock failure, the beam will be aborted. Similar interlocks will be provided on the cooling water flow, pressure differential, and temperature. Field strengths, cooling requirements, and sweep amplitudes of the said system for this design are based on previous operational experience. 

It is important that the beam is not only properly enlarged, but that it is also correctly positioned in the dump. A quadrant detector at the entrance to the dump will assure the correct beam size and position at the dump entrance, while upstream BPMs will assure the correct entrance angle. Each element of the quadrant detector will cover close to $90$ degrees of azimuthal angle, and will intercept a very small fraction of the beam. The elements must be water-cooled, protected from RF heating, and the ceramics providing electrical isolation shielded from the possibility of charging from stray scattered electrons. Basically, each element is a low-efficiency Faraday cup, and thus must be thick enough to assure beam electrons are dumped. Interlocks on the amplitude of the DC and $60\unit{Hz}$ left-right and up-down difference signals assure that the quadrupole over-focusing and raster amplitude are correctly set, and that the beam centroid is properly centered on the dump.

The existing beam dump has been tested up to power levels of $350\unit{ kW}$, so no problems are expected with the maximum beam parameters of $6\unit{ MeV}$ and $40\unit{ mA}$ for the CBETA project.



\ifdefined \buildingFullDocument

\renewcommand{\FiguresDirectory}{linac/figures}

\else
\newcommand{\FullDocumentRoot}{..}
\newcommand{\FiguresDirectory}{figures}

\begin{document}
\fi

\chapter{Linac and RF systems\Leader{Matthias} }\label{chapter:linac}

\normalsize
\section{Introduction}
\begin{figure}[htbp]
\centering
\includegraphics[width=\textwidth]{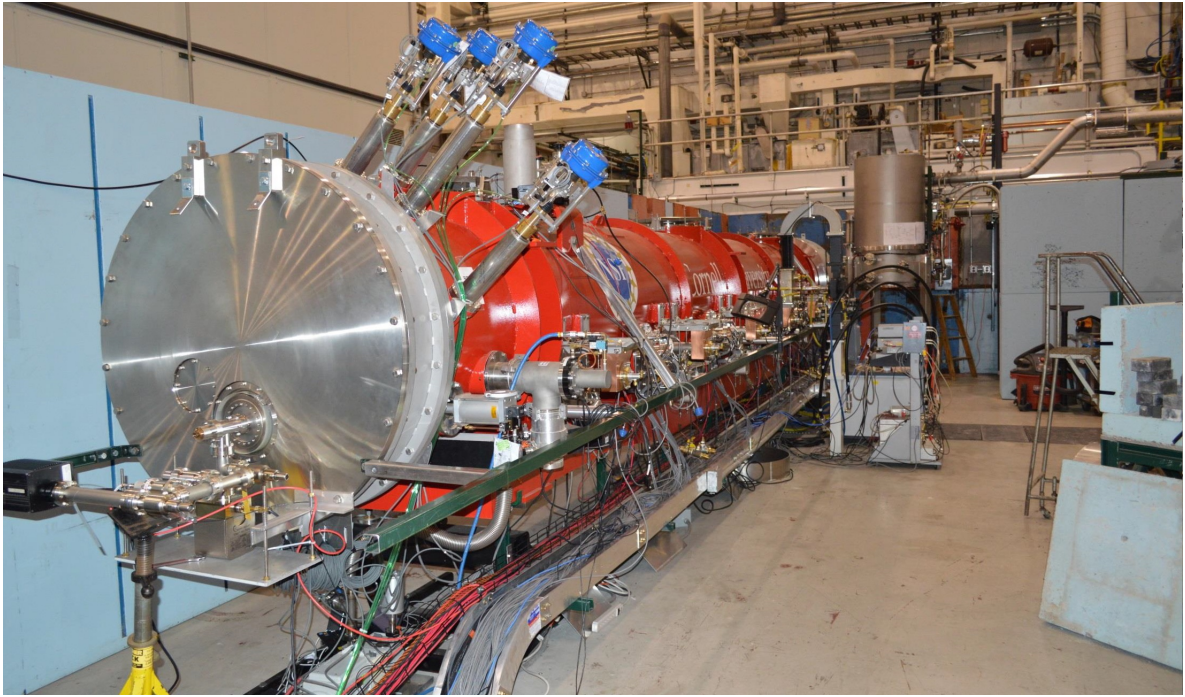}
\caption[MLC_completion]{The Cornell Main Linac Cryomodule (MLC) installed for RF testing at Cornell.}
\label{fig:MLC_completion}
\end{figure}

For CBETA, the accelerator module in the ERL loop will be the MLC, which has been built as a prototype for the Cornell ERL project. This cryomodule houses six SRF 1.3 GHz, 7-cell cavities, powered via individual 5~kW CW RF solid state amplifiers, providing a total single-pass energy gain of up to 75~MeV. HOM beam line absorbers are placed in-between the SRF cavities to ensure strong suppression of HOMs, and thus enable high current ERL operation. The module, shown in \Fig{fig:MLC_completion}, was finished by the Cornell group in November 2014 and was successfully cooled-down and operated starting in September 2015. Detailed design considerations and parameters for this cryomodule can be found in Cornell ERL PDDR~\Ref{Cornell-ERL-PDDR}. In the following sections we will summarize the main features of the MLC as relevant for CBETA and describe the performance that has been measured so far.
%
%
\section{MLC Overview}
\begin{figure}[htbp]
\centering
\includegraphics[width=\textwidth]{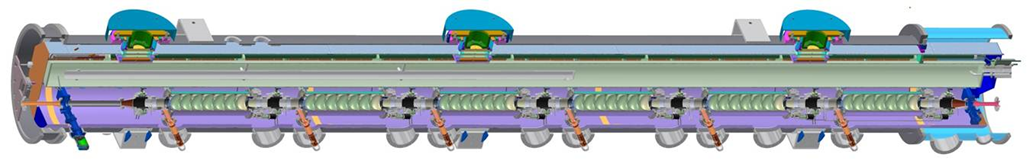}
\caption[MLC_concept]{3-D CAD model of the MLC prototype. It houses 6 superconducting 7-cell cavities with HOM loads located between them. The vacant space on the left could house a magnet/ BPM section. The overall length will be 9.8~m.}
\label{fig:MLC_CAD_model}
\end{figure}

The layout of the linac cryomodule is shown in \Fig{fig:MLC_CAD_model}.  The 10~m long module houses six superconducting cavities, operated in CW mode at 1.8~K. These 7-cell, 1.3 GHz cavities with a design $Q_0$ of $2 \times 10^{10}$ at 1.8~K can provide an average accelerating field of up to 16~MV/m (corresponding to 12.8 MeV energy gain per cavity). Each cavity is driven by a separate 5~kW solid state RF amplifier to ensure maximal flexibility and excellent RF field stability in high loaded Q operation. The RF power is coupled into the cavities via RF input couplers with fixed coupling ($Q_{ext} \approx 6 \times 10^7$). The shape of the cavities has been optimized to achieve a high BBU limit in ERL operation.  Due to the high beam current combined with the short bunch operation, a careful control and efficient damping of the HOMs is essential, leading to the installation of beam line RF absorbers in-between the cavities for strong suppression of HOMs. \\
The cryomodule design has been guided by the ILC cryomodule \Ref{Horlitz1995_01, Pagani1999_01, Peterson08_01}, while  modifications have been made to enable CW operation (instead of pulsed ILC operation with $\approx 1$\% duty cycle) with correspondingly higher dynamic cryogenic loads. All beam line components within the cryomodule are suspended from the Helium Gas Return Pipe (HGRP), a 280~mm titanium pipe. This large diameter  pipe returns the gaseous helium boiled off by the cavities to the liquefier and also acts as a central support girder. All beam line components inside the module are aligned within $\pm 1$~mm via fixed reference surfaces on the HGRP.
\section{Beam Line Components}
This section gives an overview of the key beam line components of the MLC:  its accelerating SRF cavities, the RF power input couplers used to couple RF power into the cavities, the cavity frequency tuners used to adjust the RF frequency of the fundamental (accelerating) mode, and the HOM beam line absorbers used for HOM suppression and HOM power extraction. 
%
\subsection{SRF Cavities}
\begin{figure}[htbp]
\centering
\includegraphics[width=0.95\textwidth]{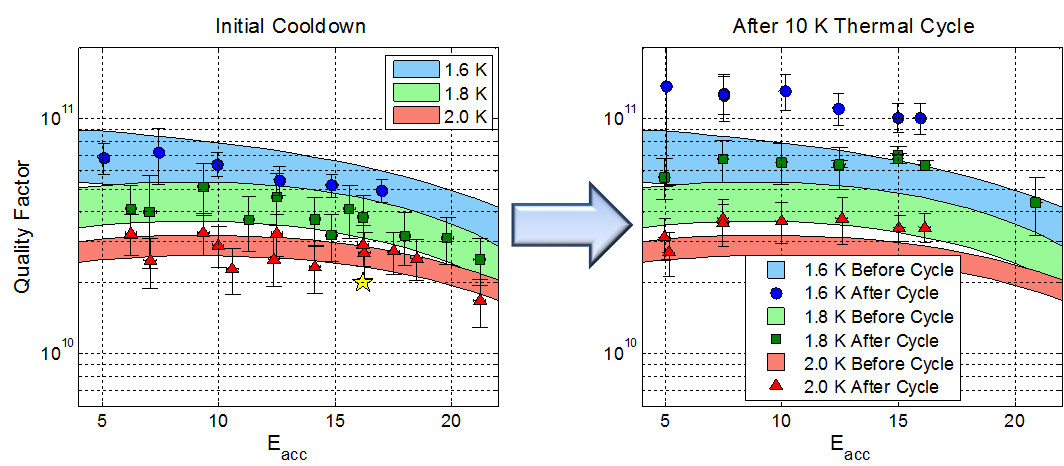}
\caption[]{Performance of the 7-cell, superconducting cavity. The quality factor (which is inversely proportional to the cryogenic losses) was measured for different temperatures as a function of the accelerating field. The initial findings are shown on the left, the performance reported on the right diagram were yielded after a 10~K thermal cycle which increased the quality factor further. At 16~MV/m accelerating field and 1.8~K, a Q of 6$\times 10^{10}$ was achieved, surpassing the design goal by a factor of 3.}
\label{fig:HTC_3_results}
\end{figure}

\begin{figure}[htbp]
\centering
\subfloat[]{\includegraphics[width=0.43\textwidth]{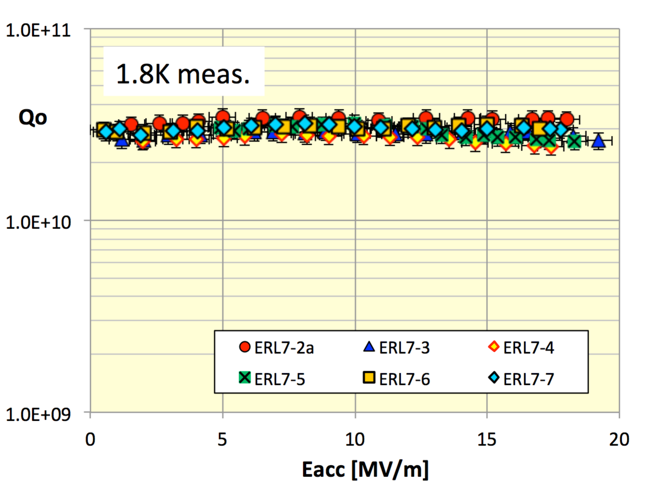}\label{fig:vertical_test_results1}}
\subfloat[]{\includegraphics[width=0.47\textwidth]{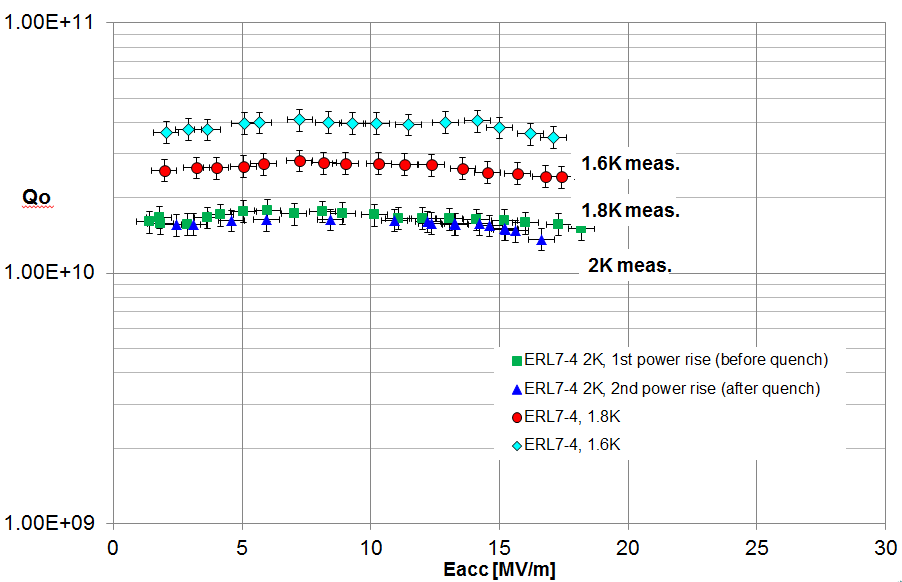}\label{fig:vertical_test_results2}}
\caption[]{Left: Vertical test results for all 6 MLC production cavities. All cavities exceeded the design specifications for the ERL ($Q_0=2 \times 10^{10}$ at 1.8 K). The right figure shows the Q vs E curves at different temperatures of a single cavity.  The reproducibility of the results, gained without any reprocessing of a cavity, is very good.}
\label{fig:vertical_test_results}
\end{figure}

The design and surface preparation of the 1.3 GHz, 7-cell SRF cavities for the MLC have been optimized for (1) supporting high beam currents in ERL operation, and (2) efficient cavity operation in CW mode. \Table{tab:linac_ERL_cavity_design_parameters} lists key parameters of the optimized cavity. An operation temperature of 1.8 K has been chosen based on an optimization process aiming to minimize AC cooling power demands. Design specifications were set at an accelerating gradient of 16 MV/m and a quality factor of $2\times10^{10}$ at 1.8~K, which, at the time of the decision, was an ambitious goal. An extensive R\&D program was started ten years ago to ensure that these parameters can be achieved reliably. In addition, to verify strong HOM suppression and thus a high BBU-limit, HOM damping and high-current beam operation were studied in detail.   \\
A prototype 1.3~GHz, 7-cell main-linac cavity was fabricated first and tested extensively. The cavity received a simple, high $Q_0$ surface preparation (bulk buffer-chemical polish (BCP) of 150~$\mu$m,  outgassing at 650~C for 12 hours, tuning to 1297.425~MHz,  final 10~$\mu$m BCP, 120~C heat treatment for 48 hours). After meeting quality factor and gradient specifications in the vertical test, the cavity was removed from the vertical test stand, and a helium jacket was welded to the cavity. The prototype cavity was then tested in a short single-cavity horizontal cryomodule (Horizontal Test Cryostat - HTC) in a three tier approach: With HTC-1, the cavity was only equipped with a low power probe coupler; the HTC-2 rebuilt added the RF power coupler to the cavity; and the final HTC-3 step added the HOM absorbers to the cavity package. The prototype cavity performance of the fully assembled prototype cavity in the test cryomodule  (HTC-3) is shown in \Fig{fig:HTC_3_results}. The cavity performance again exceeded specifications,  qualifying our cavity preparation recipe.\\
\begin{table}[tb]
\caption[]{MLC cavity design parameters. Note that $R/Q$ is always in the circuit definition.}
\begin{tabular*}{\columnwidth}{@{\extracolsep{\fill}}ll}
\toprule
Parameter&Value\\
\midrule
Type of accelerating structure&Standing wave\\
Accelerating mode&TM$_{0,1,0}$ $\pi$  \\
Fundamental frequency&1.3 GHz\\
Design gradient&16 MV/m\\
Intrinsic quality factor&$2\times10^{10}$\\
Loaded quality factor&$6 \times 10^7$\\
Cavity half bandwidth at $Q_{\text{L}}= 6\times 10^7$&11 Hz\\
Operating temperature&1.8~K\\
Number of cells&7\\
Active length&0.81 m\\
Cell-to-cell coupling (fundamental mode)&2.2\%\\
Iris diameter center cell / end cells&36 mm / 36 mm\\
Beam tube diameter&110 mm\\
Geometry factor (fundamental mode)&$270.7\,\Omega$\\
$R/Q$ (fundamental mode)&$387\,\Omega$\\
$E_{\text{peak}}/E_{\text{acc}}$ (fundamental mode)&2.06\\
$H_{\text{peak}}/E_{\text{acc}}$ (fundamental mode)&41.96 Oe/(MV/m)\\
$\Delta f / \Delta L$&350 Hz/ $\mu$m\\
Lorentz-force detuning constant&1 Hz / (MeV/m)$^2$\\
Cavity total longitudinal loss factor for $\sigma=0.6\unit{mm}$&14.7 V/pc\\
Cavity longitudinal loss factor for $\sigma=0.6\unit{mm}$,\\
non-fundamental&13.1 V/pC\\
Cavity transverse loss factor for $\sigma=0.6\unit{mm}$&13.7 V/pC/m\\
\bottomrule
\end{tabular*}
\label{tab:linac_ERL_cavity_design_parameters}
\end{table}
%
All six production cavities for the MLC have been produced in-house starting from flat metal 3~mm, RRR 300 niobium sheets. To investigate microphonics, we decided to build 3 unstiffened cavities as well as 3 cavities with stiffening rings between the cells.  The process began with half cells formed by a deep drawing process in which sheet metal of  RRR niobium is radially drawn into a forming die by a first press at 3~tons, then a second forming press (100~tons). The dies for the center cells were carefully designed to deal with the spring back effect. The equators of each cup have an additional straight length on them (1.5~mm). The purpose of this extra length is to allow for trimming later on to meet the target frequency and length. Those dumbbells are built in an intermediate step by welding two cups together on their irises. Ultimately six dumbbells were be welded together by electron beam welding to form the center-cells of the seven cell cavity and end-cells with end-groups are added. During the cavity production, we improved the mechanical tolerances in the cavity forming and welding, leading to a mean length deviation of the last 3 cavities by only 0.2 mm \Ref{Eichhorn13_04}.\\
For the preparation of the production cavities, a simple recipe based on BCP similar to the preparation of the prototype cavity was chosen. Starting after fabrication, the damage layer was removed by bulk BCP (140 $\mu m$). A hydrogen degassing was then done at 650 C for 4 days, followed by final 10~$\mu$m BCP, a 120~C heat treatment for 48 hours to reduce BCS surface resistance, and a HF rinse removing and regrowing the oxide layer to reduce residual surface resistance. All cavities were performance tested vertically. The summary of these test is given in \Figure{fig:vertical_test_results}. All six cavities exceeded the design quality factor, averaging to $2.9 \times 10^{10}$ at 1.8K. The reproducibility of the Q versus E curves for all cavities was very good. None of the cavities needed additional processing - giving a 100\% yield.
\subsection{Fundamental Power Couplers}
\begin{figure}[htbp]
\centering
\includegraphics[width=0.60\textwidth]{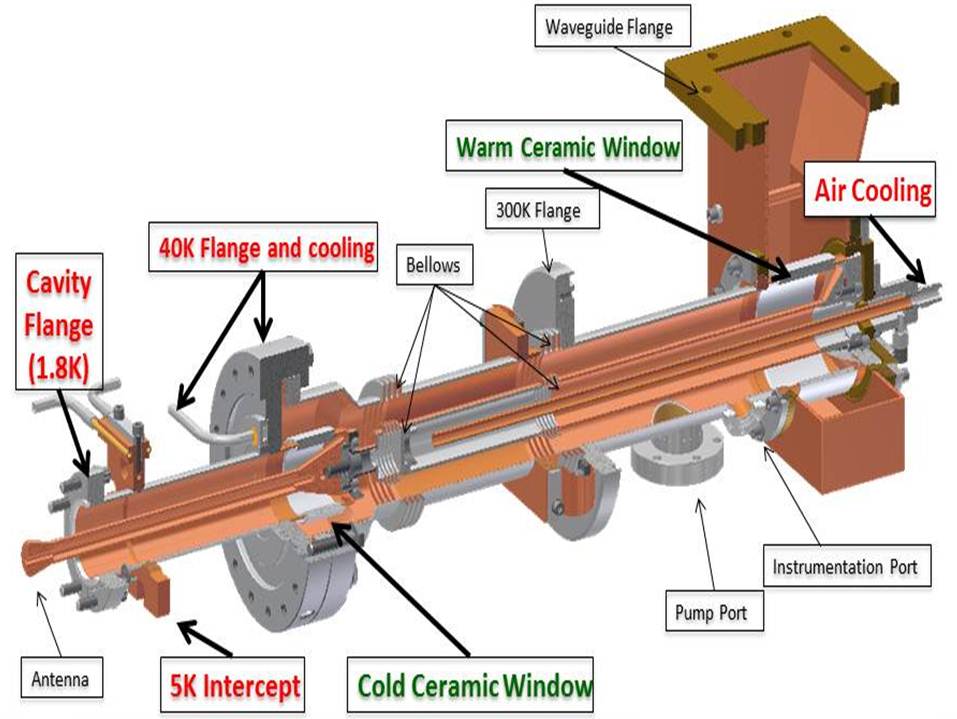}
\caption[]{RF Power coupler. The coaxial transmission line has two bellows which allow for lateral movement during cool-down being up to 10 mm.}
\label{fig:RF_power_coupler}
\end{figure}

\begin{figure}[htbp]
\centering
\includegraphics[width=0.80\textwidth]{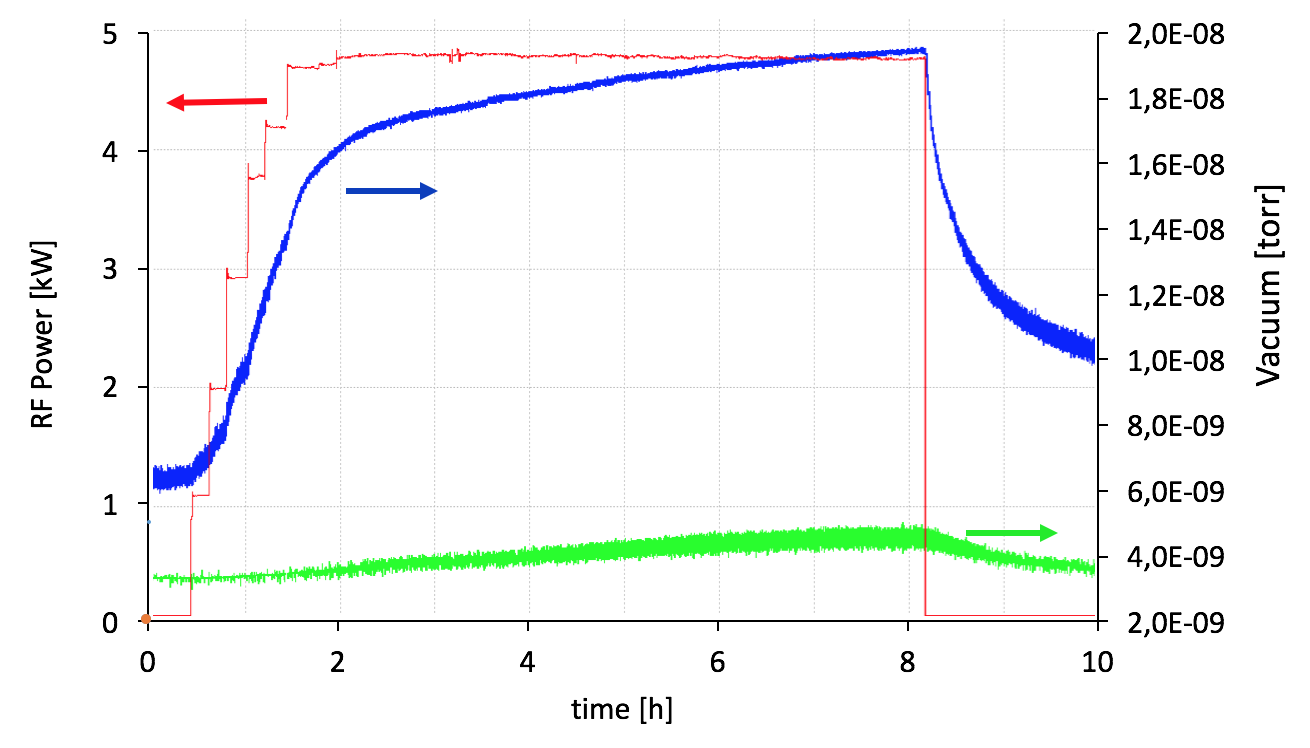}
\caption[]{Typical testing results of the couplers, measured on a test-stand before the coupler was mounted to the string. Top curve shows the RF power as being ramped up, the curve below is the coupler vacuum, rising to $2\times10^{-8}$ Torr while the cavity vacuum (bottom curve) is only slightly affected.}
\label{fig:RF_coupler_tests}
\end{figure}

Even though the ERL main linac input couplers only have to deliver up to 5~kW~CW RF power to the cavities, the design is rather sophisticated: The design approach chosen for the whole module requires compensating the lateral movement between the straight mounting situation and the dog-leg geometry gained after cool-down (with an off-set of up to 1~cm). In addition, the mounting procedure is such that the cold part is fully mounted inside the cleanroom, hermitically closing the cavity. This approach has been found to be very successful as we have seen no cavity Q degradation after the coupler mounting procedure. The warm portion of the coupler is mounted outside the cleanroom with no special precaution on extreme cleanliness.   \Figure{fig:RF_power_coupler}  shows a CAD model of the coupler, and \Tab{tab:linac_coupler_parameters} lists key parameters. \\
\begin{table}[tb]
\caption[]{MLC cavity RF power input coupler parameters. }
\begin{tabular*}{\columnwidth}{@{\extracolsep{\fill}}ll}
\toprule
Parameter&Value\\
\midrule
Operating frequency&1300 MHz\\
Maximum power (CW)&5 kW\\
$Q_{ext}$ (fixed)& $6 \times 10^7$ \\
Static heat load to 2K& 0.05 W\\
Dynamic heat load to 2K& 0.06 W\\
Static heat load to 5K&0.64 W\\
Dynamic heat load to 5K&0.32 W\\
Static heat load to 40K&3.78 W\\
Dynamic heat load to 40K&5.94 W\\
\bottomrule
\end{tabular*}
\label{tab:linac_coupler_parameters}
\end{table}
%
The external Q of the coupler is $6 \times 10^7$ (with the option to adjust it using a three-stub waveguide tuner) in order to minimize the RF power requirements taking into account the microphonics inside the module. A first prototype of the coupler was successfully tested to 5~kW~CW RF power under full reflection.  This was done without any conditioning required to reach this power level using a commercial 5~kW solid state RF amplifier. All six production couplers have been procured at CPI and were tested upon receiving on a test stand, applying 5~kW~CW RF power under full refection without seeing any vacuum action. Essentially, no conditioning was required to reach this power level; see \Fig{fig:RF_coupler_tests}. 
%
\subsection{Cavity Frequency Tuners}
\begin{figure}[htbp]
\centering
\includegraphics[width=0.80\textwidth]{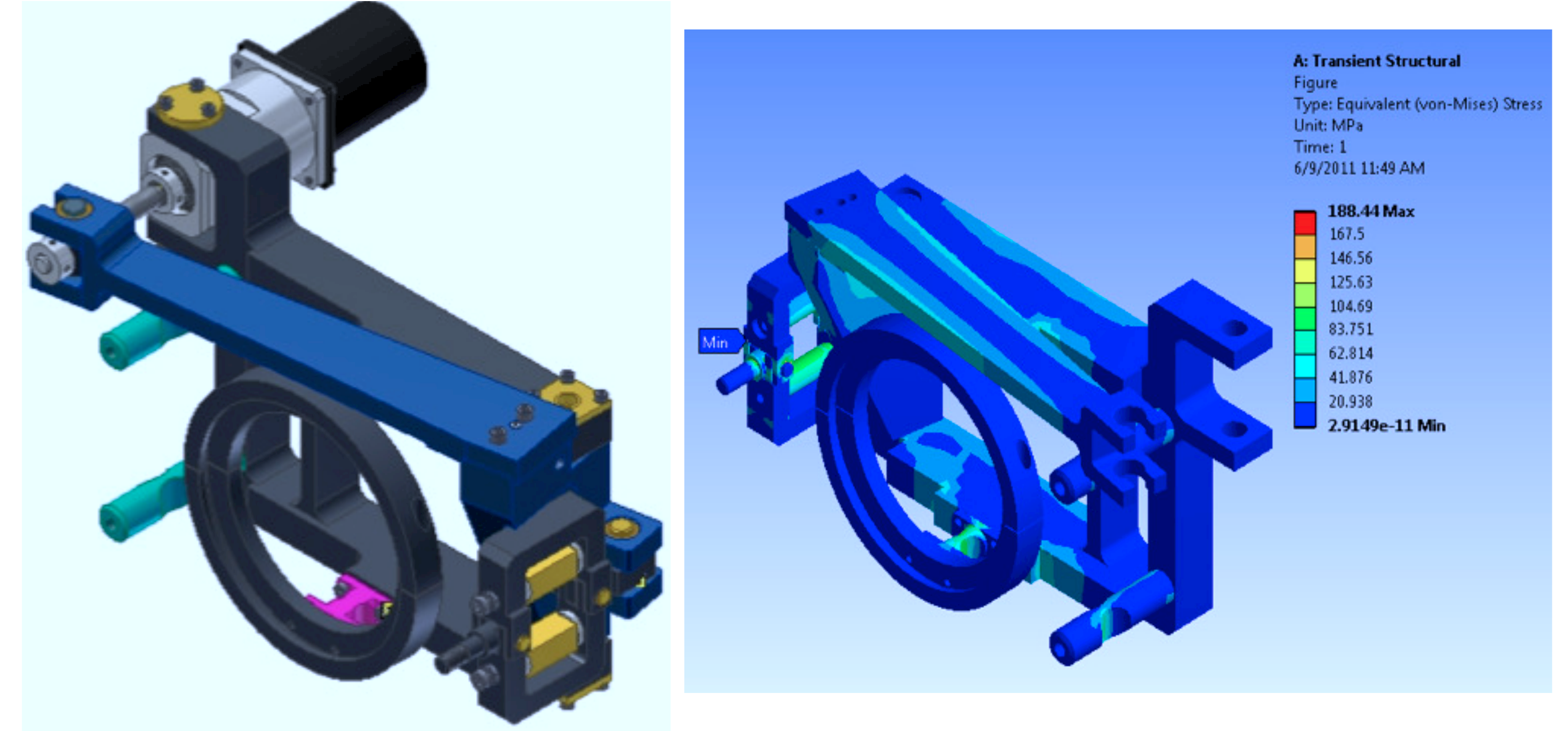}
\caption[]{MLC cavity frequency tuner with fast piezoelectric actuators and increased stiffness for the main linac cavities. Left: CAD model. Right: ANSYS simulation of the von Mises stress for a tuning force of 26 kN applied to the cavity. The maximum stress is well below the yield strength of stainless steel at cryogenic temperatures.}
\label{fig:tuner}
\end{figure}

The MLC cavity frequency tuner is based on the Saclay I tuner with a stepping motor drive for slow frequency adjustment ($>500$ kHz tuning range) and  a piezo-electric actuators for fast frequency control ($>1$ kHz tuning range) for Lorentz-force compensation and microphonics compensation.  The tuner frame and the piezo stacks were modified for increase stiffness to support high tuner forces of up to 26 kN for the main linac cavity version with stiffening rings. An illustration of the modified tuner is shown in \Fig{fig:tuner}.
\subsection{HOM Absorbers}
\begin{figure}[htbp]
\centering
\includegraphics[width=0.95\textwidth]{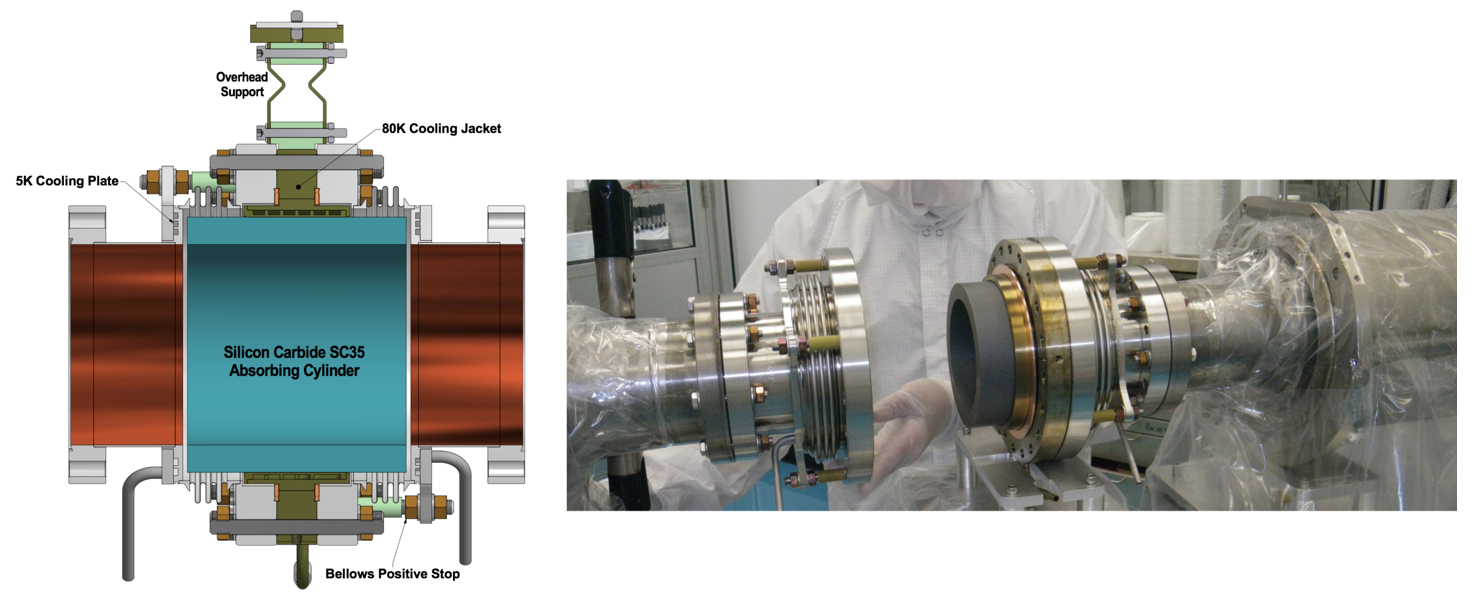}
\caption[]{Cornell's HOM absorber used in the MLC. The absorbing material is SC\_35 from CoorsTek, shrink-fitted into a Ti cylinder.}
\label{fig:HOM_absorber}
\end{figure}

\begin{figure}[htbp]
\centering
\includegraphics[width=0.6\textwidth]{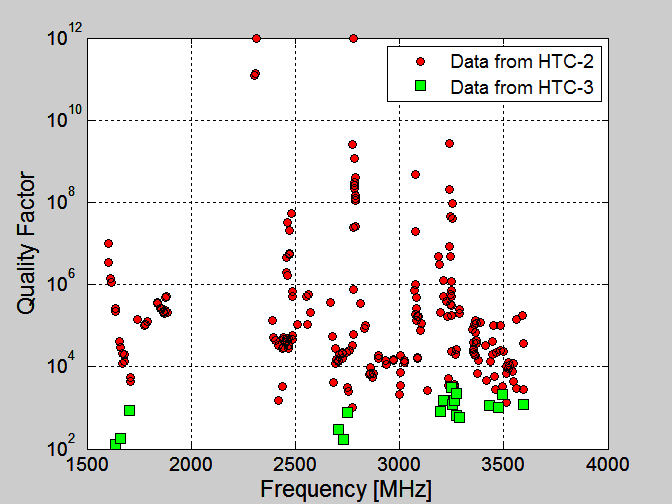}
\caption[]{External Quality factor of the Cornell 7-cell ERL cavity without (red) and with (green) HOM absorbers installed.}
\label{fig:External_quality_factors_HOMs}
\end{figure}

\begin{figure}[htbp]
\centering
\includegraphics[width=0.9\textwidth]{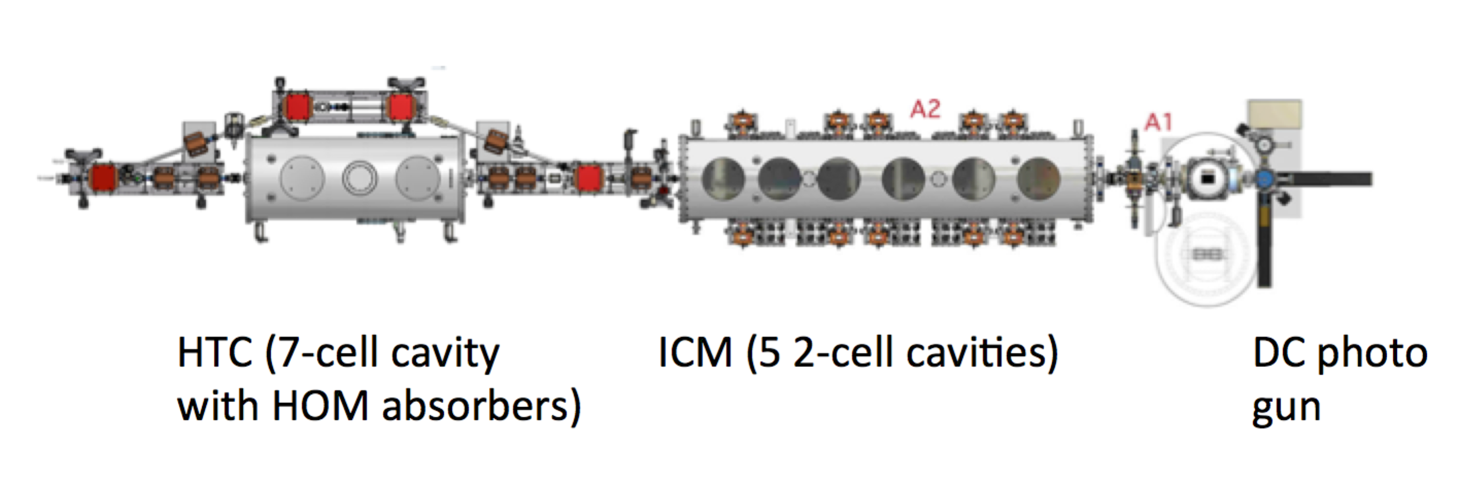}
\caption[]{The horizontal test cryostat (HTC) placed behind the photo injector for testing of the 7-cell cavity equipped with HOM absorbers with beam. A beam current of up to 40 mA was run through the set-up.}
\label{fig:HTC_diagram}
\end{figure}

The guiding concept for the beam line HOM absorbers in the MLC is to have a broad band absorbing material covering the whole range from 1 to $>$40 GHz in the shape of a cylinder, located in the beam tube sections between the cavities, with the beam passing through the center. After several iterations, a  final design was developed and tested, as depicted in \Fig{fig:HOM_absorber}.
The center assembly consists of a SiC cylinder from CoorsTek (SC\_35) which is shrink fit into a titanium cooling jacket and flange. The cooling jacket and flange locate, support, and provide cooling at 80~K to the absorbing cylinder using a cooling channel inside the titanium to ensure 400~W of HOM power can be extracted. The end pieces of the assemblies contain a 3 convolution bellows, a 5~K cooling plate, and taper seal flange to mate with the cavities. The bellows allows for small length variations in the string, small angular misalignments of cavity flanges, as well as adds a long thermal path from 80~K to 5~K. There are positive stops that prevent the bellows from compressing and closing the gap between the 5~K cooling jacket and the absorbing cylinder to less than 1~mm. This prevents any rubbing of metal to ceramic that could create particles. The beam tubes have a copper plating about 10~micron thick to prevent beam induced heating. More details on the design can be found in \Ref{Eichhorn13_02, Eichhorn14_01}.\\
Two prototype HOM absorbers where build with a slightly mechanically different design: the absorbing ceramic was identically, but it but brazed into a tungsten cylinder. As part of designing and verifying parameters for the MLC, these prototypes were tested in a HTC together with the prototype 7-cell cavity. Excellent higher order mode damping was found with no observed dipole mode having an external Q higher than $10^4$ while the Q of the fundamental mode was unaffected ($6\times10^{10}$ at 1.8 K). \Figure{fig:External_quality_factors_HOMs} shows the quality factors of the cavity with and without the absorber. More details are published in \Ref{Valles13_01}.\\
To quantify the HOM power extracted and measure the heating effects, tests with beam were conducted on the prototype absorbers. For that, the HTC was moved to the Cornell ERL photo-injector and located directly behind the injector cryomodule. The layout of the beam line is show in \Fig{fig:HTC_diagram}.  For the beam test, we ran different beams with currents up to 40~mA through the HTC and measured the heating in various locations. As expected, the heating scaled with beam current, shorter bunches lead to higher heating. The total HOM power absorbed  was about 6~W at 40~mA. One must consider however that a portion of the HOM spectrum is not bound to the region of the cavity and its adjacent HOM loads due to it being above the cut-off frequency of the beam pipe. A high-frequency HOM wakefield can therefore propagate beyond the HOM load without losing all its energy. Simulations show that this underestimation is no greater than a factor of 2, and that therefore no more than 12~W of HOM power were excited by the 40~mA beam. This then results in an estimate of the longitudinal loss factor of the cavity of $\approx 10$ V/pC, slightly lower than the simulated value. In addition to the high current operation, the beam was used to excite and probe individual HOMs in the cavity, with the purpose of measuring their loaded quality factor $Q_L$.  This beam-based HOM search revealed the presence of only 8 modes with a $Q_L > 10^5$, all of which are highly likely quadrupole or sextupole modes, and thus due not to pose any danger of causing BBU in the ERL.
\section{MLC Design}
\subsection{Mechanical Design}
All components within the cryomodule are suspended from the HGRP. This large diameter titanium pipe has a combined function: it returns the gaseous helium boiled off the cavity vessels to the liquefier and acts as a central support girder. The HGRP is Grade~2 Titanium, 280~mm in diameter with a wall thickness of 9.5~mm which is  supported by 3 support post, the middle one being fixed while the outer two slide by 7 and 9~mm during cool-down, respectively.  With a 1 ton weight force of the beam line string, the maximum vertical displacement of the HGRP is 0.1~mm and the natural frequency is 88 Hz.  Simulations  show that a 3-posts support system is well suited to ensure an acceptable vertical displacement and vibration characteristics. A series of ANSYS modal finite element analyses were performed to evaluate the cryomodule design mechanical stability in more detail, including individual piping lines inside the module. The location and stiffness of piping supports were considered in the design to make sure the natural frequencies are higher than 60~Hz. A 2-Phase 2~K pipe feeds helium to the helium vessels of the individual cavities  through 6 chimneys. The diameter of this pipe is strongly increased (compared to the ILC cryomodule design) to accommodate CW operation.
The latest results on the ERL cryomodule testing and beam capabilities have been presented at IPAC 2017 \Ref{IPAC2017:WEPIK036}. 
\subsection{Cryogenic Cooling Scheme}
\begin{figure}[htbp]
\centering
\includegraphics[width=0.9\textwidth]{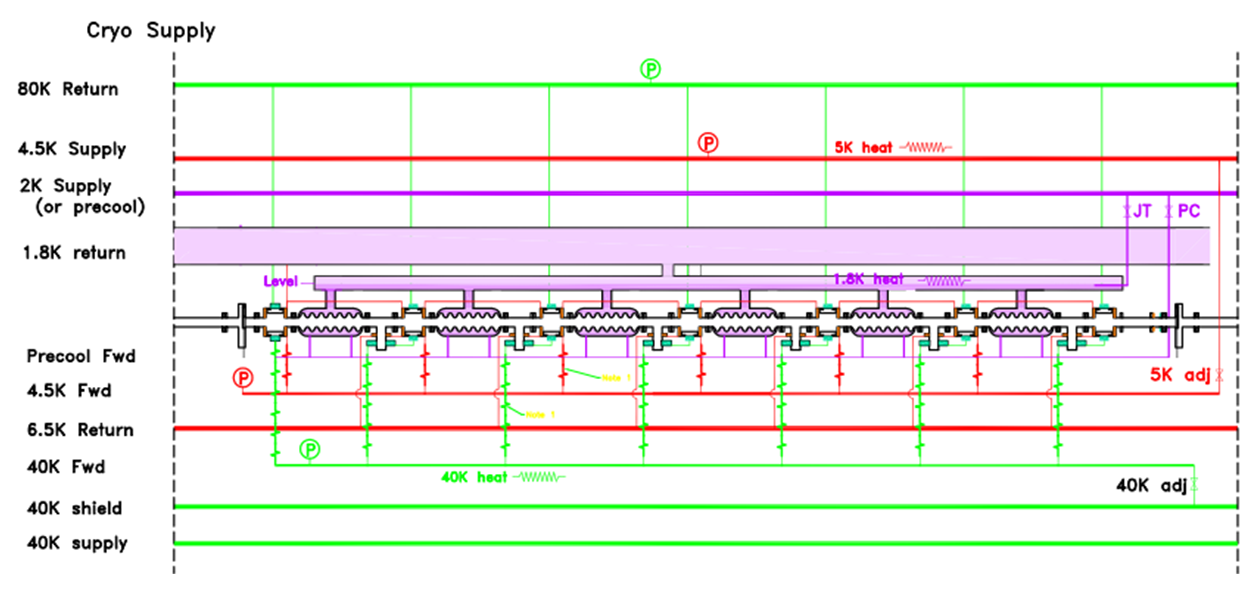}
\caption[]{Diagram of the different cooling loops within the MLC: The 1.8 K loop feeds the cavities, the 5 K loop intercepts the cavity, coupler and HOM flanges and the 80 K loop cools the radiation shield, the HOM Absorbers and intercepts the power coupler.}
\label{fig:cooling_loops}
\end{figure}

\begin{figure}[htbp]
\centering
\includegraphics[width=0.75\textwidth]{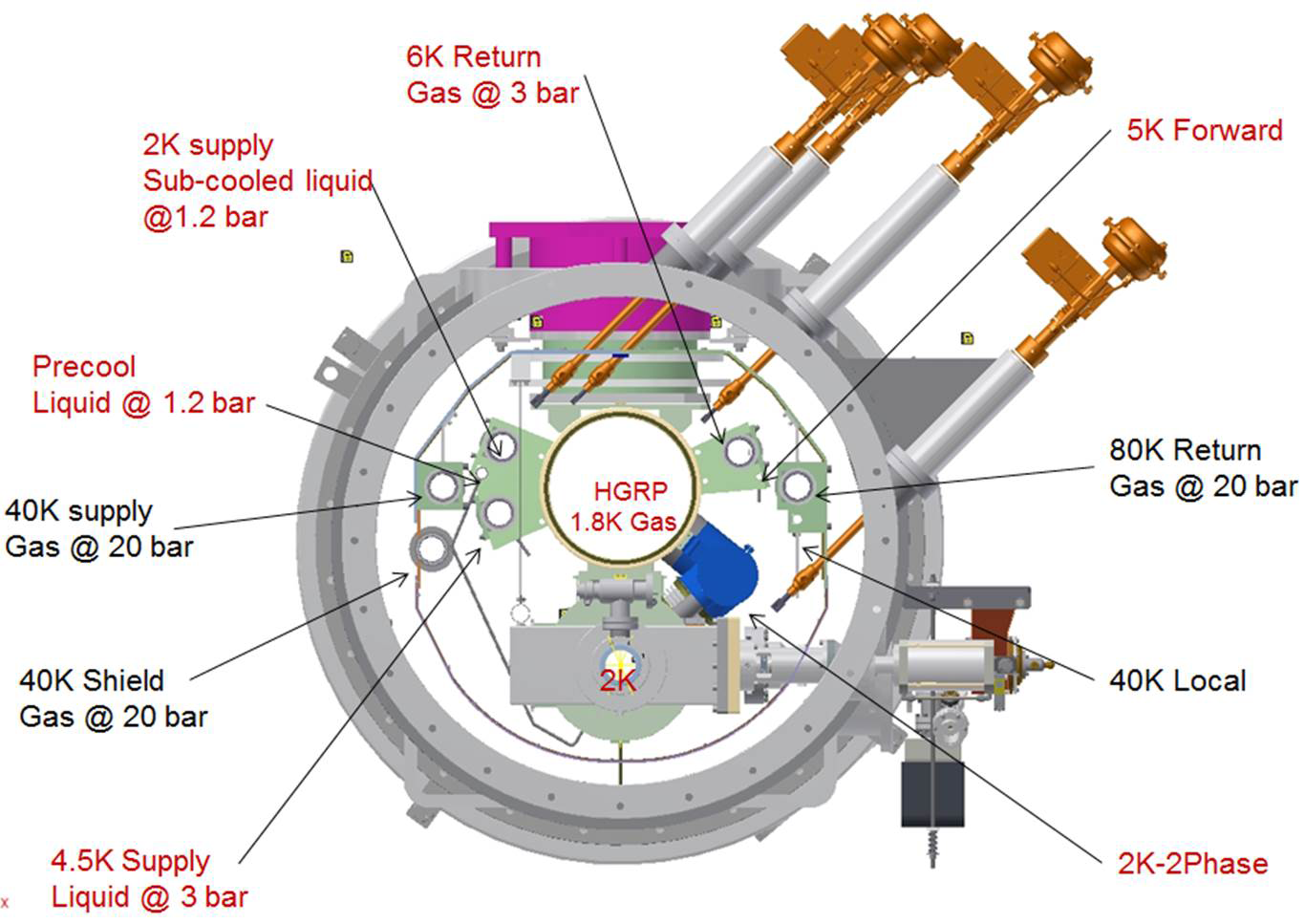}
\caption[]{Cross-section of the MLC, giving more details on the piping and positioning of the components.}
\label{fig:MLC_cross_section}
\end{figure}

The cryogenic cooling of the module consists of 3 different temperature loops. The cavities are cooled by liquid helium obtained by a JT-Valve located at the module entrance. Sub-cooled to 1.8~K by pumping the He-atmosphere down to 20~mbar ensures an optimum operation regime for the superconducting cavities. The second loop consisting of supercritical 5~K helium is used to cool the intercept all transitions to warmer temperatures in order to assure a minimal heat transfer to the 1.8~K system. Finally, a 80~K loop provides cooling for the coupler intercepts, cools the thermal radiation shield of the module and removes the heat generated in the HOM absorbers. As shown in \Fig{fig:cooling_loops}, the loads are partially in parallel, partially in series. As the expected heat load especially at the HOM absorbers  are expected to vary individually on a scale of 0 to 400 W, a careful investigation of the stability of parallel flows was performed \Ref{Eichhorn13_02}.  \Figure{fig:MLC_cross_section} gives the cross-section of the module showing the spatial arrangement of the cooling loops.
\subsection{80~K Thermal Shield}
\begin{figure}[htbp]
\centering
\includegraphics[width=0.75\textwidth]{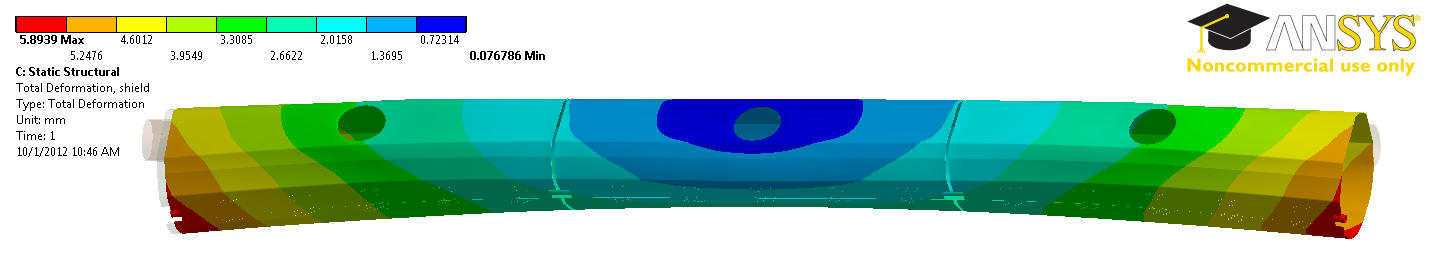}
\includegraphics[width=0.75\textwidth]{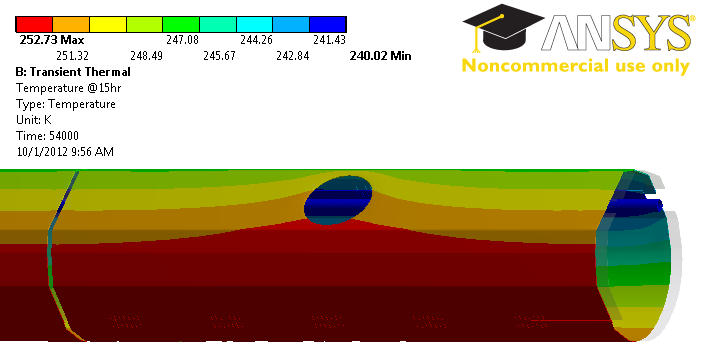}
\caption[]{Cool-down analysis of the thermal shield (Aluminum). As it is cooled only on one side, the cool-down will not be uniform, resulting in an up to 13K temperature difference (a) and a lateral bending of up to 6 mm (b).  }
\label{fig:cooldown_analysis}
\end{figure}

The thermal shield (Al1100-14) of the module is cooled by the 80~K delivery line which is connected to the shield on one side. As a result, the cool-down process is asymmetric. To avoid damage to the shield due to thermal stress during cool-down, thermal and thermo-structural finite element analyses (FEA) were performed to study the temperature gradient, deformation and thermo-mechanical stresses of the 80~K thermal shield with a helium gas cool-down rate of 4~K per hour - being an adequate assumption. The simulated results indicated that with a cooling rate of 4K/hour, the temperature gradient reaches a maximum of 13~K on the entire shield, occurring 15 hours after the start of the cool down. This situation is shown in \Fig{fig:cooldown_analysis}. Once fully cooled down, the steady state maximum temperature gradient is only 2.2~K.
It should be mentioned that the thermal shield will be bent during cool-down estimated to be 6 mm at max.
\subsection{Magnetic Shielding}
High $Q_0$ cavity operation requires excellent shielding of the Earth's magnetic field, so that the residual ambient magnetic field at the cavity locations is reduced to $<2$~mG. A 80~K  Mu-metal magnetic shield is attached to the outside of the 80~K thermal shield. In addition,  every cavity is enclosed by a second, individual magnetic shield, made from cryoperm. 
\subsection{Vacuum Vessel}
\begin{figure}[htbp]
\centering
\includegraphics[width=0.95\textwidth]{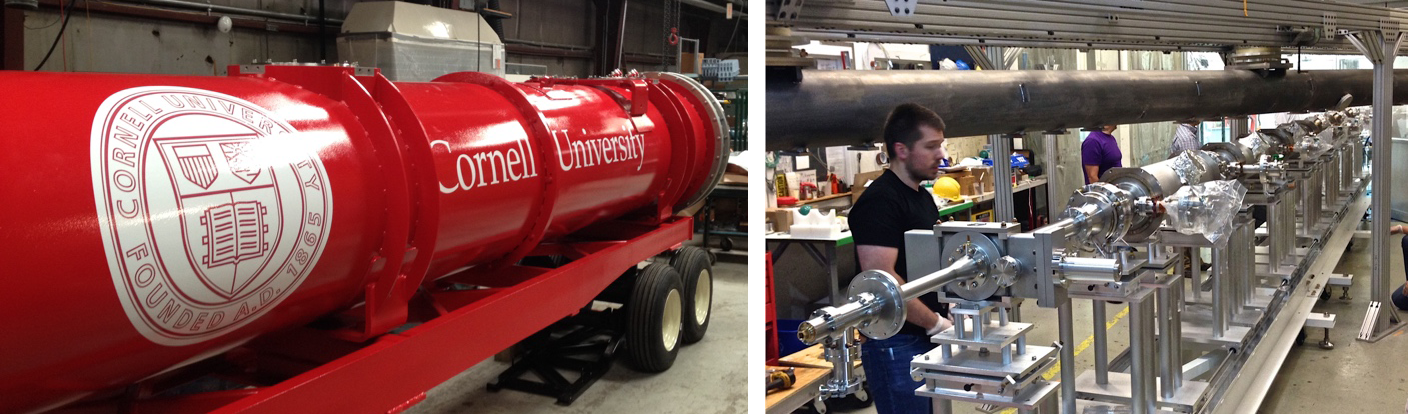}
\caption[]{Left: Vacuum Vessel as received by Cornell University. Right: Marriage step: the cavity string coming from the clean room assembly is rolled under the HGRP providing precision alignment.}
\label{fig:Vacuum_vessel}
\end{figure}

The cylinder of the vacuum vessel is 965.2 mm in diameter, rolled longitudinally welded carbon steel (A516 GR70) pipe. All precision required surfaces were machined in a single spool piece setup. \Figure{fig:Vacuum_vessel} shows a photograph of the vacuum vessel.
\subsection{Alignment}
\begin{figure}[htbp]
\centering
\includegraphics[width=0.7\textwidth]{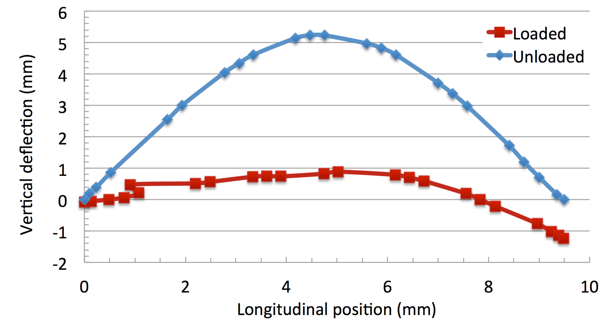}
\caption[]{Vertical position of the reference surfaces on the helium gas return pipe defining the positions of all beam line components. In a module string, the right end will receive more support resulting in a better vertical positioning.}
\label{fig:HGRP_vertical_deflection}
\end{figure}

The alignment concept of the cryomodule relies on the helium gas return pipe, acting as a strong-back of the module. Suspended by 3 composite post assemblies from the outer vessel, it provides precision surfaces for all cavity mounts and beam line components. The position of these surfaces where surveyed upon receiving the pipe, initially displaying larger than the specified $\pm1$ mm accuracy. However, when preloaded with the approximate weight of the cold-mass, alignment was within specifications. This data is given in \Fig{fig:HGRP_vertical_deflection}. 
\section{Assembly Process, Cool-down, and Performance Testing}
%
\subsection{MLC Assembly}
While production of components started earlier, the actual assembly began in March 2014, when the first cavities where connected to a string inside the cleanroom. For space reasons, two half strings, consisting of three cavities with attached cold section of the coupler and 4 HOM absorbers as well as a gate valve were assembled. Each substring was leak checked and connected to the other later on. In May, cold mass assembly continued outside the cleanroom. Mating the pre-aligned cavity string with the precision surfaces on the HGRP strong-back turned out to be not an issue at all. All bellows located at the HOM absorber package were able to compensate for deviations and only the longitudinal position of one HOM absorber had to be adjusted.\\
Installation of the cavity magnetic shield, the tuner, the thermal and magnetic shield, all cryogenic piping and jumpers, instrumentation and cabling as well as a wire position monitor to track alignment during cool-down took 3 months. The cold-mass was moved it into the vacuum vessel in September 2015. As final assembly steps, the warm portion of the couplers, feed-throughs and cryogenic valves were installed. All circuits were leak checked and pressure tested.
\subsection{Preparation for Testing}
\begin{figure}[htbp]
\centering
\includegraphics[width=0.95\textwidth]{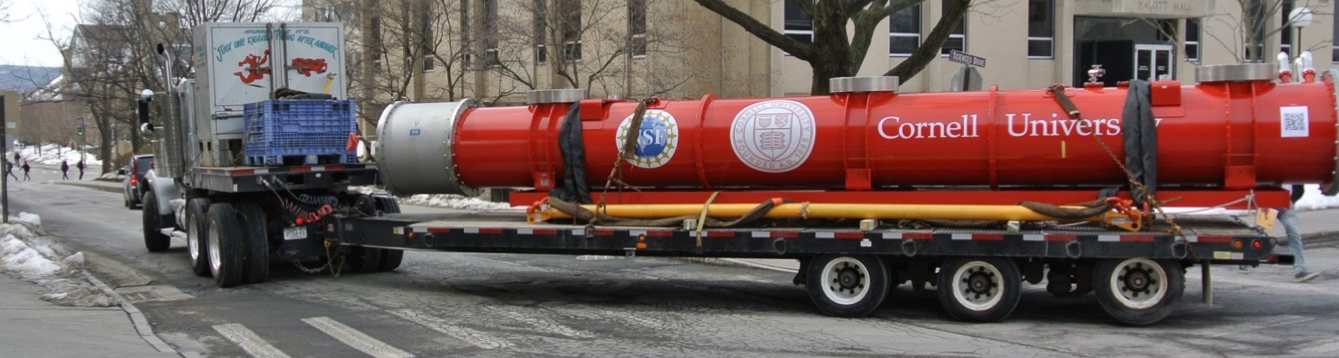}
\includegraphics[width=0.95\textwidth]{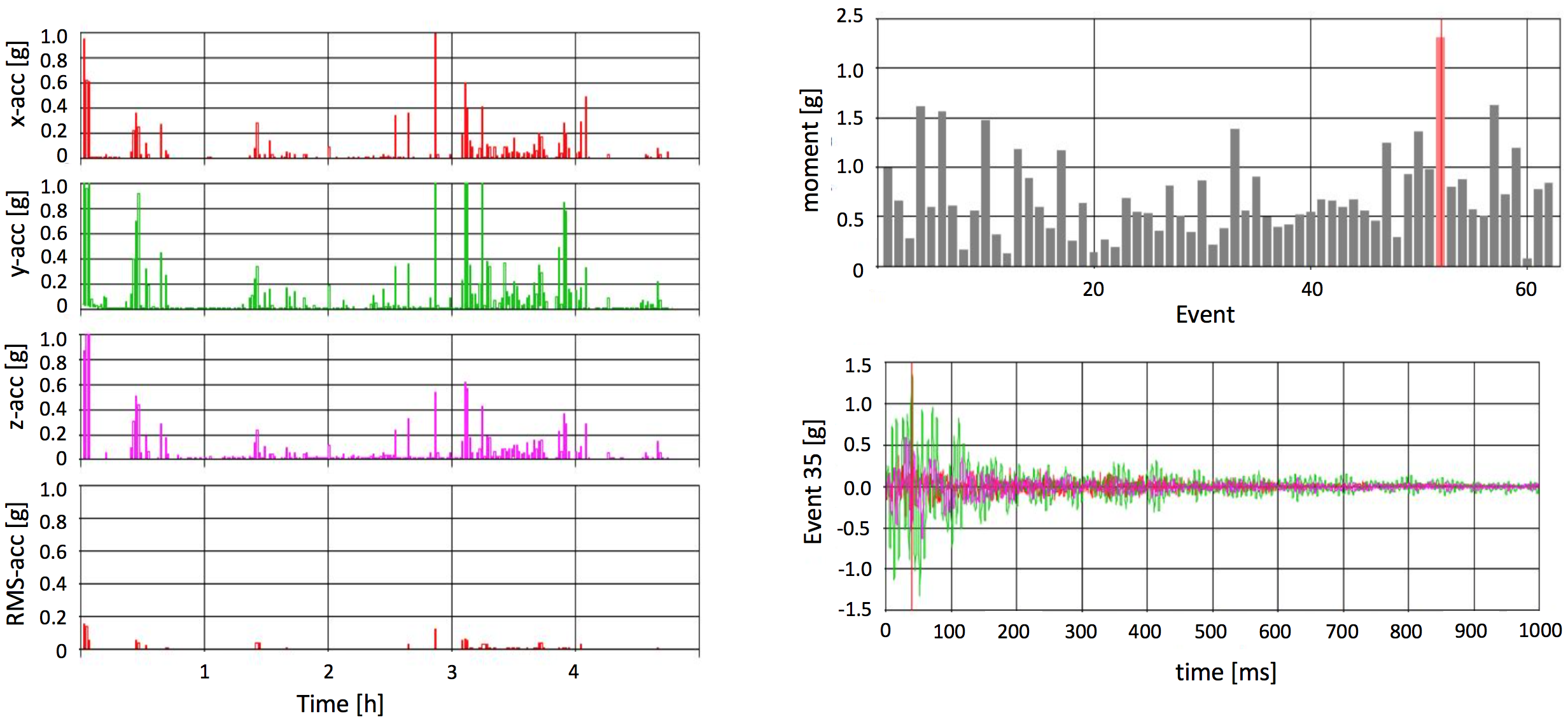}
\caption[]{Accelerometer data taken during the transportation of the cryomodule across the campus.}
\label{fig:cryomodule_moving}
\end{figure}

In preparation for the testing of the MLC, the module was transported across the Cornell campus. No special damping frame was used. However, all movements were done with extreme care and transportation speed was set to 5 km/h max. In addition, we measured mechanical shocks using accelerometers. The data, given in \Fig{fig:cryomodule_moving}, revealed a maximum g force of 2.3 (lasting less than 10~ms), which occurred while pulling the module (siting on its own wheels) in place after lifting it from the truck. During the road trip, max g-factors were below 1.5.
\subsection{Initial Cool-down}
\begin{figure}[htbp]
\centering
\subfloat[]{\includegraphics[width=0.5\textwidth]{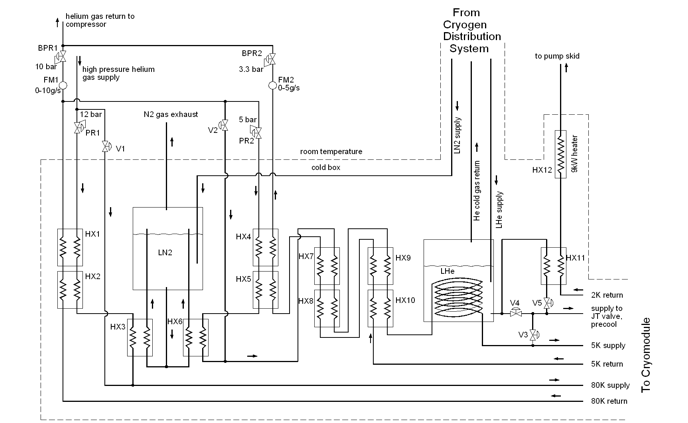}\label{fig:Heat_exchanger1}}
\subfloat[]{\includegraphics[width=0.5\textwidth]{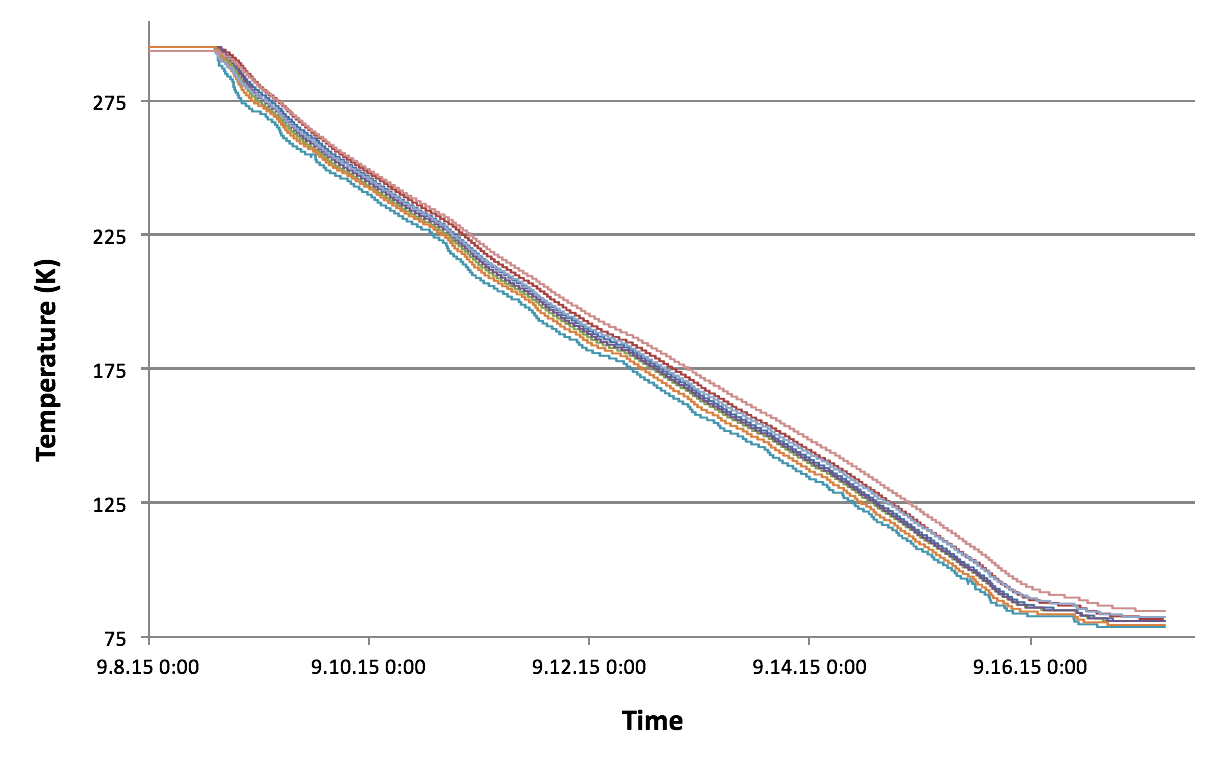}\label{fig:Thermal_shield_temperatures}}
\caption[]{Left: Heat exchanger can build to cool-down and operate the main linac cryomodule. Special emphasis was given towards achieving a smooth and controlled cool-down to 77 K.  
Right: Temperatures on the thermal shield during the cool-down. Due to the design of the thermal shield the temperature spread across the shield had to stay below 20~K, leading to a cool-down rate of approximately 1.25~K/h.}
\label{fig:Heat_exchanger}
\end{figure}

\begin{figure}[htbp]
\centering
\includegraphics[width=0.75\textwidth]{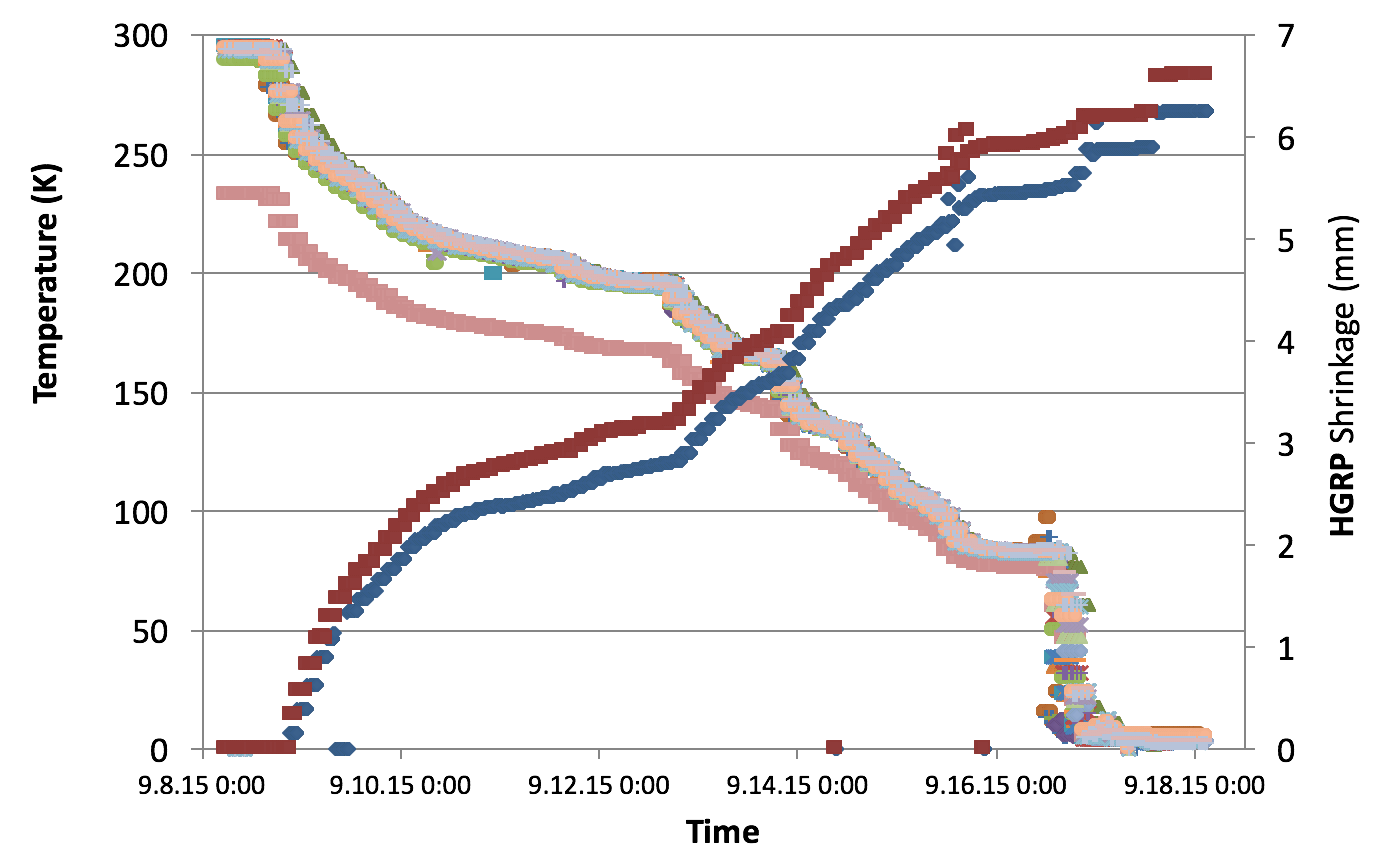}
\caption[]{Movement of the outer (sliding) support posts: as the temperature of the HGRP goes down the posts move inwards. The movement was as expected.}
\label{fig:Support_post_movement}
\end{figure}

In order to facilitate a smooth and controlled cool-down, a new heat exchanger can was built -- the piping diagram is given in \Fig{fig:Heat_exchanger}. It allows to add a warm stream of gas forwarded to the cold-mass, resulting in a very controlled cool-down, as shown in \Fig{fig:Heat_exchanger}. This was mandatory as the thermal shield is cooled by conduction only with an extruded pipe running just along one side. As a result, the cool-down of the shield is asymmetric and we calculated stress limits on the aluminum transitions which required us to keep the temperature spread across the shield below 20~K. In the initial cool-down we maintained 10~K, becoming 15~K below 200~K with an average cool-down rate of 1.25~K/h. \\
During cool-down, we also monitored the movement of the two outer, sliding support posts.  The cavities in the MLC are  aligned via the helium gas return pipe, made out of titanium, being suspended from the vacuum vessel by three support posts. As a result of the cold-mass shrinking as a whole with the central post fixed, the two outer posts are sliding. \Figure{fig:Support_post_movement} gives the movement of these posts as the temperatures on the cool-mass go down. The movement was as expected. 
\subsection{Cool-down Cycles}
\begin{figure}[htbp]
\centering
\subfloat[]{\includegraphics[width=0.5\textwidth]{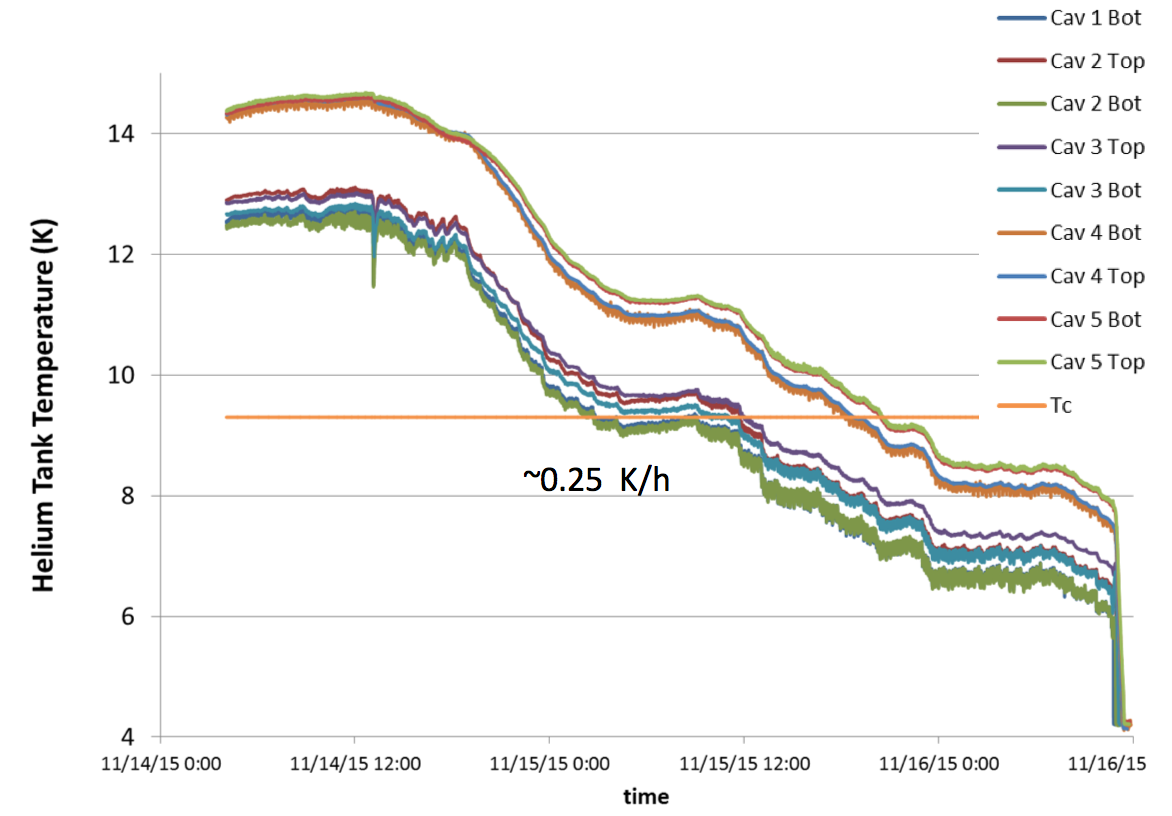}\label{fig:slow_fast_cooldown1}}
\subfloat[]{\includegraphics[width=0.5\textwidth]{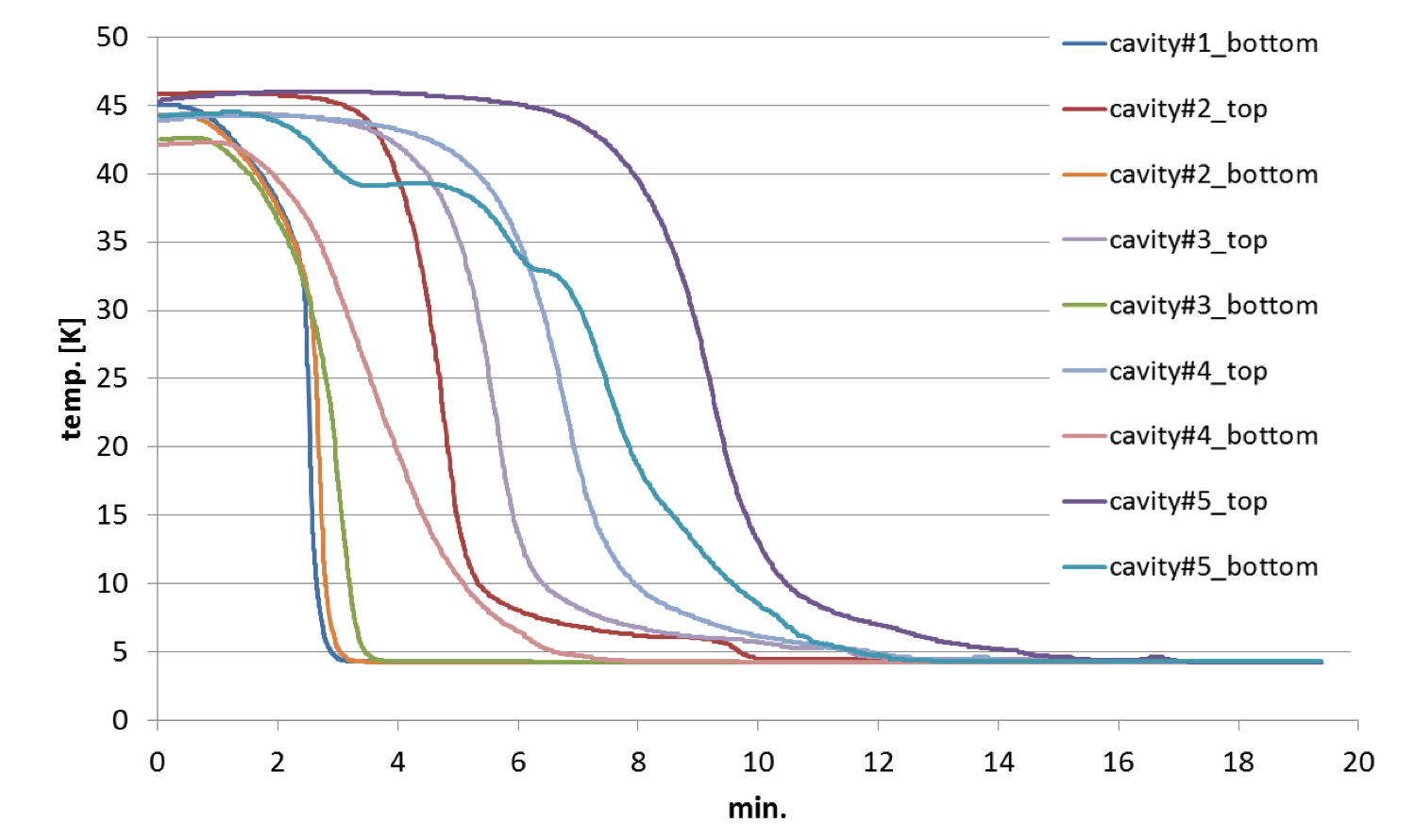}\label{fig:slow_fast_cooldown2}}
\caption[]{Slow (left) and fast (right) cool-down cycle performed in order to measure cycle dependent impacts on the cavity performance. For the slow cool-down we were able to get 0.25 K/h while the fast cool-down resulted in 0.5 K/min -- 2 K/min, depending on how close the cavity was to the JT valve.}
\label{fig:slow_fast_cooldown}
\end{figure}

Recent findings have indicated that the performance of an SRF system also depends on details of the cool-down process. Findings at Cornell indicate that for conventionally treated cavities a slow cool-down leads to a higher quality factor of the cavity.  We were able to explain this finding by describing the role of thermo-currents that are excited at the material transitions between the niobium (cavity) and the titanium (enclosing helium vessel), driven by temperature gradients \Ref{Eichhorn16_01, Eichhorn15_03}. So-called nitrogen-doped cavities, however, seem to require a fast cool-down and it was found that this helps expelling residual magnetic field more efficiently than a slow cool-down. It should be noted that N-doped cavities are stronger effected by trapped magnetic flux compared to conventional treated cavities. To study the impact of slow and fast cool-down speeds on the MLC cavities' performances as well as to verify the reproducibility of these cycles, a total of 5 thermal cycles were completed on the MLC, trying very slow and extremely fast cool-downs. Results are given in \Fig{fig:slow_fast_cooldown}. We found that for a slow cool-down some cavities went through transition several times with some unwanted warm-up in between. We also learned that on the fast procedure the final cool-down speed depended on the cavity position, especially how close the cavity was in relation to the Joule-Thompson valve.
\subsection{RF Test Results}
\begin{figure}[htbp]
\centering
\includegraphics[width=0.95\textwidth]{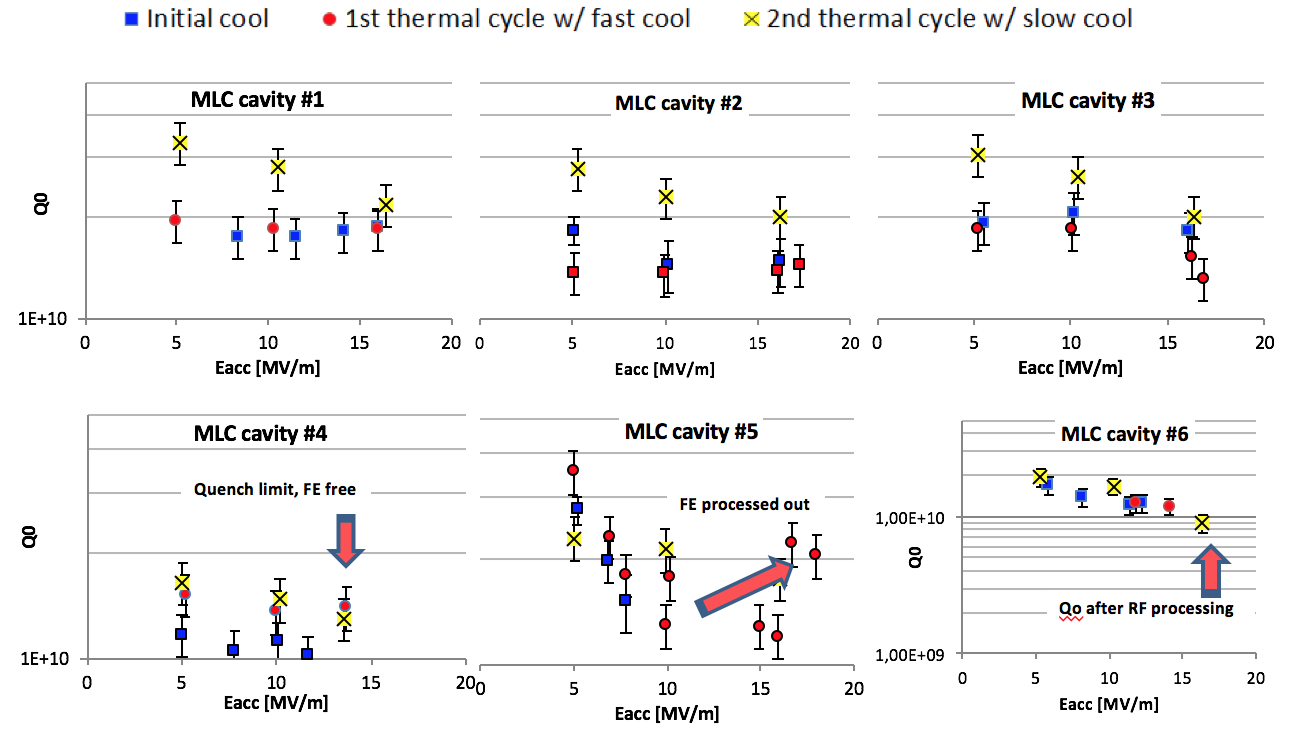}
\caption[]{Summary of the MLC cavity performance. All data were taken at 1.8~K. Two cavities had to be RF processed to increase the field and to remove field emission. Cavity \#4 is still limited by an early quench, but administrative procedures (for now) prevented us from processing this cavity further.}
\label{fig:cavity_performance_summary}
\end{figure}

Test results from all 6 cavities are summarized in \Fig{fig:cavity_performance_summary} and \Tab{tab:cavity_performance}. After some initial processing 5 of the 6 cavities perform close to their design specifications. One cavity is currently limited by a premature quench which we hope to overcome by a thermal cycle and pulse processing. Even so the quality factors are slightly lower than the design, the cavities (except \#4) easily reach the design gradient. From \Fig{fig:cavity_performance_summary} one could also conclude that in our case the cool-down speed did not strongly affect the cavity performance. The data shows slightly higher Qs for the slow cool-down. This might indicate that we have a slightly higher residual magnetic field inside the MLC (as compared to our short test module HTC). 
\begin{table}[tb]
\centering
\caption{Cavity performance inside the MLC.}
\begin{tabular*}{0.8\columnwidth}{@{\extracolsep{\fill}}llccc}
\toprule
           &          & $Q_0$/$10^{10}$ & $E_{acc}$ {[}MV/m{]} & $Q_{ext}$/$10^7$ \\ \midrule
Cavity \#1 & ERL 7-3  & 1.88          & 16.0               & 5.13           \\
Cavity \#2 & ERL 7-5  & 1.98          & 16.2               & 5.38           \\
Cavity \#3 & ERL 7-4  & 2.01          & 17.2               & 6.90           \\
Cavity \#4 & ERL 7-7  & 1.45          & 13.7               & 5.67           \\
Cavity \#5 & ERL 7-2a & 1.78          & 16.0               & 5.38           \\
Cavity \#6 & ERL 7-6  & 1.91          & 16.0               & 6.14           \\ \midrule
Design     &          & 2.0           & 16.2               & 6.5            \\ \bottomrule
\end{tabular*}
\label{tab:cavity_performance}
\end{table}
\subsection{HOM Damping Studies}
\begin{figure}[htbp]
\centering
\subfloat[]{\includegraphics[width=0.47\textwidth]{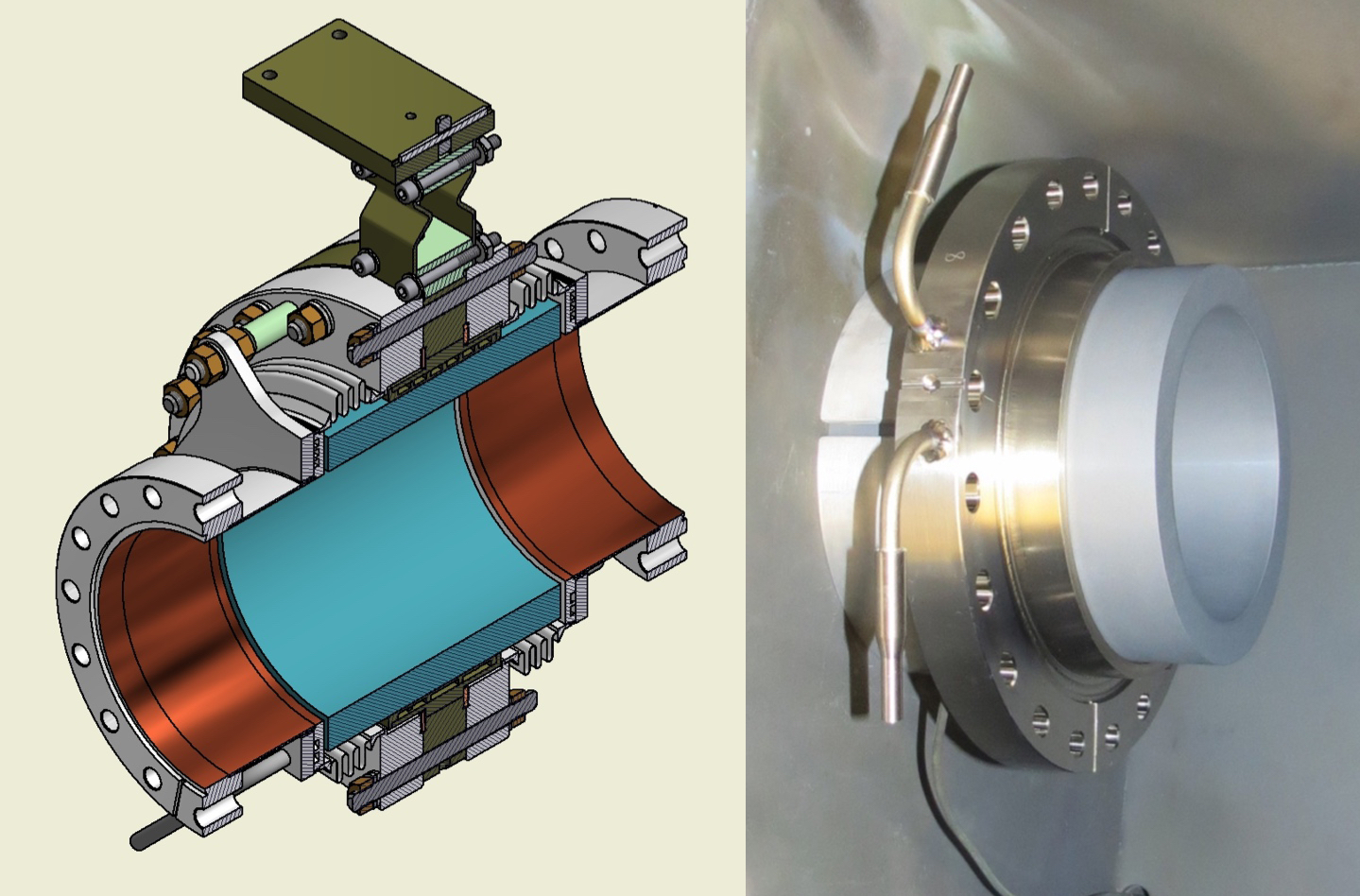}\label{fig:HOM_absorber_performance1}}
\subfloat[]{\includegraphics[width=0.53\textwidth]{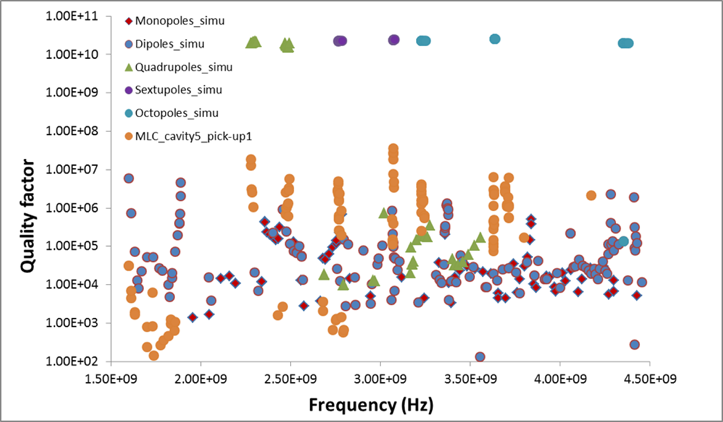}\label{fig:HOM_absorber_performance2}}
\caption[]{Performance of the higher order mode absorber. On the left a schematic view of the absorber is given, where is absorbing material (a SiC composite) is shown in blue. The absorbing cylinder is pictured in the center. On the right, the measured quality factors of some higher order modes, compared to undamped resonances are given showing that the absorber operates as designed.}
\label{fig:HOM_absorber_performance}
\end{figure}

Higher order mode scans on multiple cavities in the MLC were completed, showing no unexpected results, i.e. no high Q dipole modes. \Figure{fig:HOM_absorber_performance} shows a comparison of the MLC HOM loaded quality factor ($Q_L$) measurements and simulation. This comparison indicates that (1) the measured HOM frequencies agree well with simulation results, and (2) that the $Q_L$ of dipole HOMs of the MLC cavities are strongly damped below the target value of $\approx 10^4$ for high $R/Q$ dipole modes. The MLC results also agreed well with results from a previous HOM study on the prototype 7-cell cavity in the HTC.  The measured $Q_L$  values $>10^7$ shown in \Fig{fig:HOM_absorber_performance}  belong to quadrupole and sextupole modes, which do not contribute to the BBU limit. Currently, HOM scans on the remaining cavities continue in order to confirm the results.
\subsection{Tuning and Microphonics}
\begin{figure}[htbp]
\centering
\subfloat[]{\includegraphics[width=0.5\textwidth]{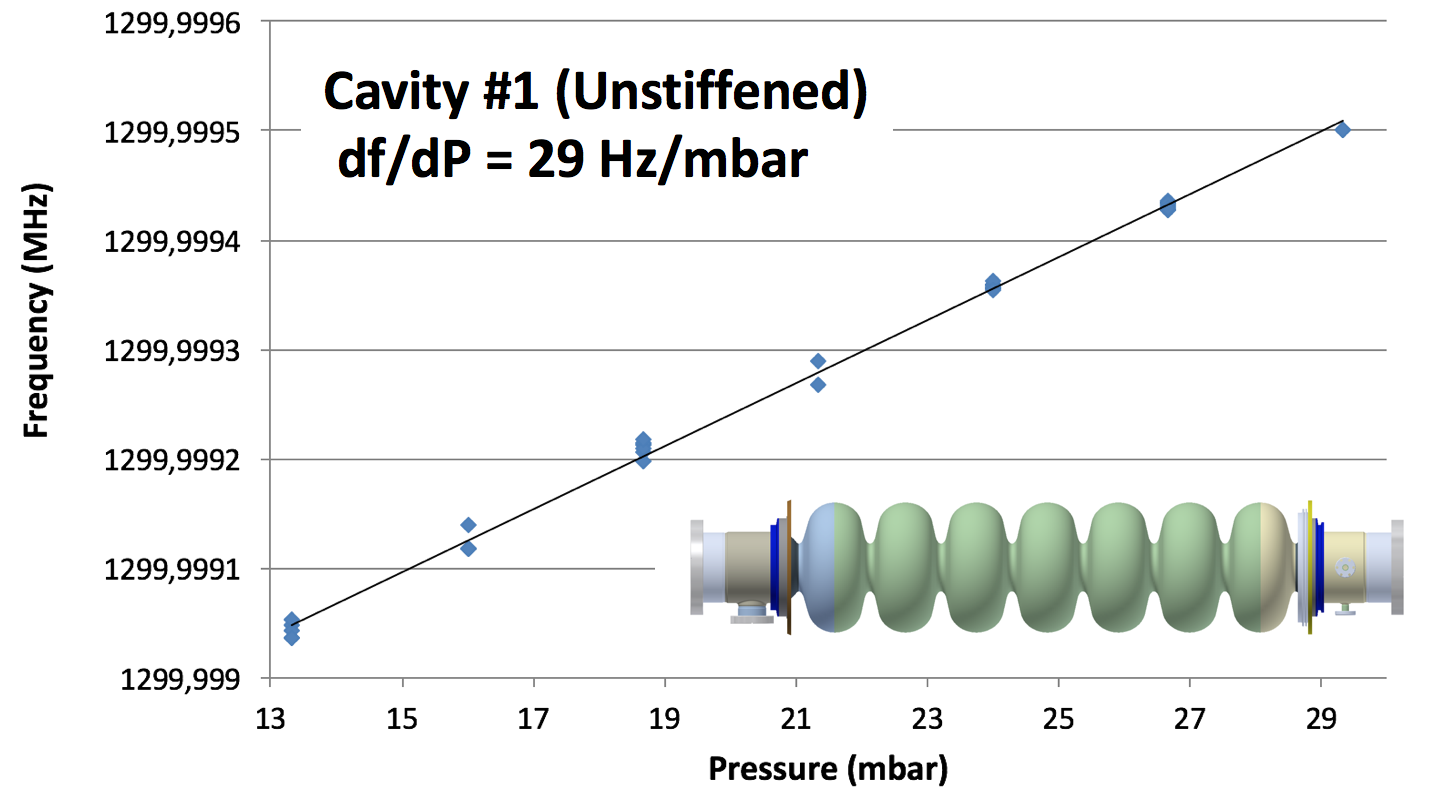}\label{fig:tuning_sensitivities1}}
\subfloat[]{\includegraphics[width=0.5\textwidth]{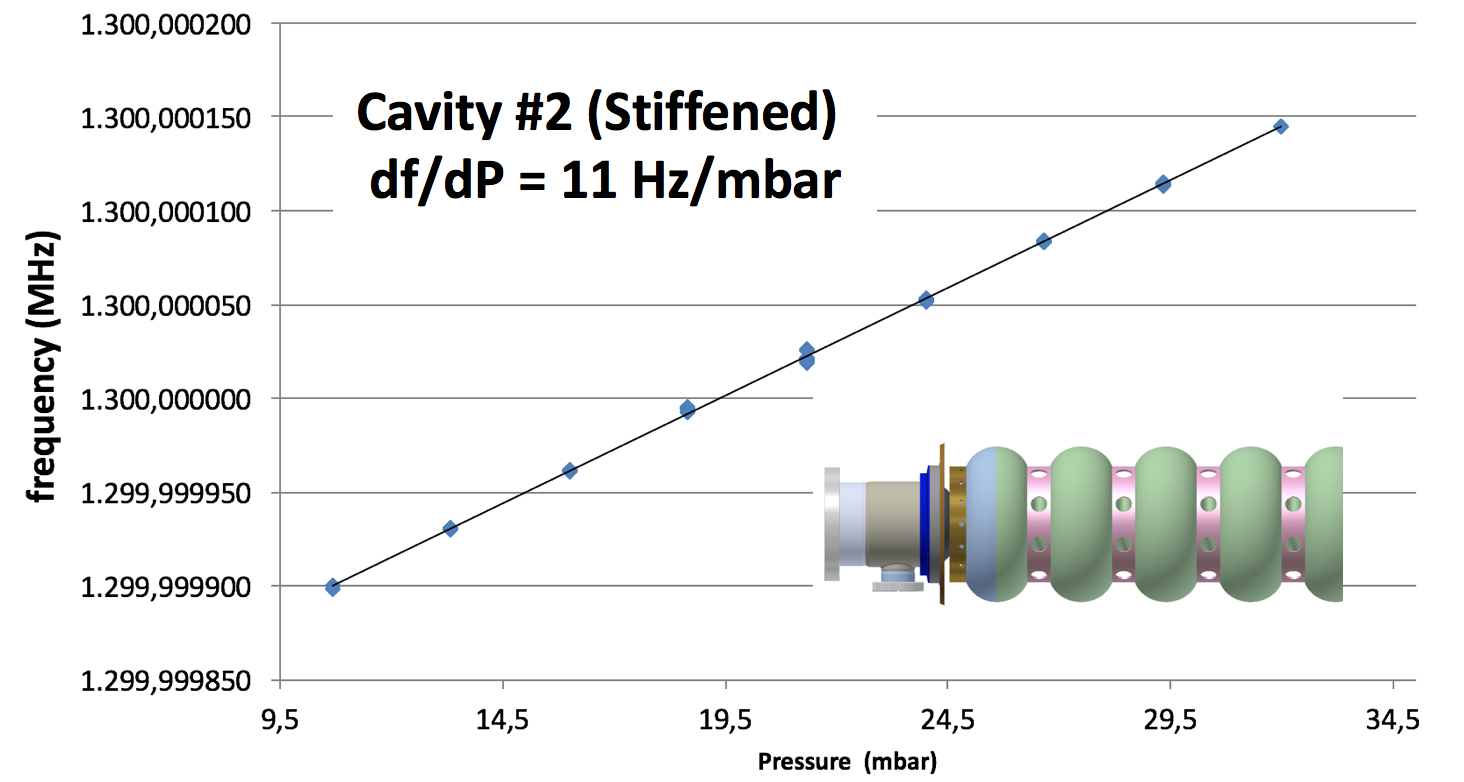}\label{fig:tuning_sensitivities2}}
\caption[]{Tuning sensitivities of the unstiffened and the stiffened cavities.}
\label{fig:tuning_sensitivities}
\end{figure}

To better understand the trade-offs in the mechanical behavior of the cavities, the MLC was built with two types of cavities: three had stiffening rings added in order to minimize the pressure responds $(df/dp)$ and to minimize microphonics, while the remaining three cavities were built without stiffening. One concern was that the stiffened cavities might overload the tuner and can not be tuned to frequency. In addition, our initial testing indicated that and unstiffened cavity would meet our requirements and in addition would be easier to fabricate.\\
\Figure{fig:tuning_sensitivities} shows the pressure response curve for a stiffened and an unstiffened cavity. As expected, the pressure sensitivity of the stiffened cavity is smaller by more than a factor of two. However, with the pressure stability in our system being 0.1 mbar both values are well acceptable. \Table{tab:cavity_tuning} gives measured values for all cavities, while also reporting on the tuning that was necessary after cool-down. As two cavities are close to their tuning range limits, we will slightly lower our operating frequency to 1299.9 MHz.

\begin{figure}[htbp]
\centering
\includegraphics[width=0.9\textwidth]{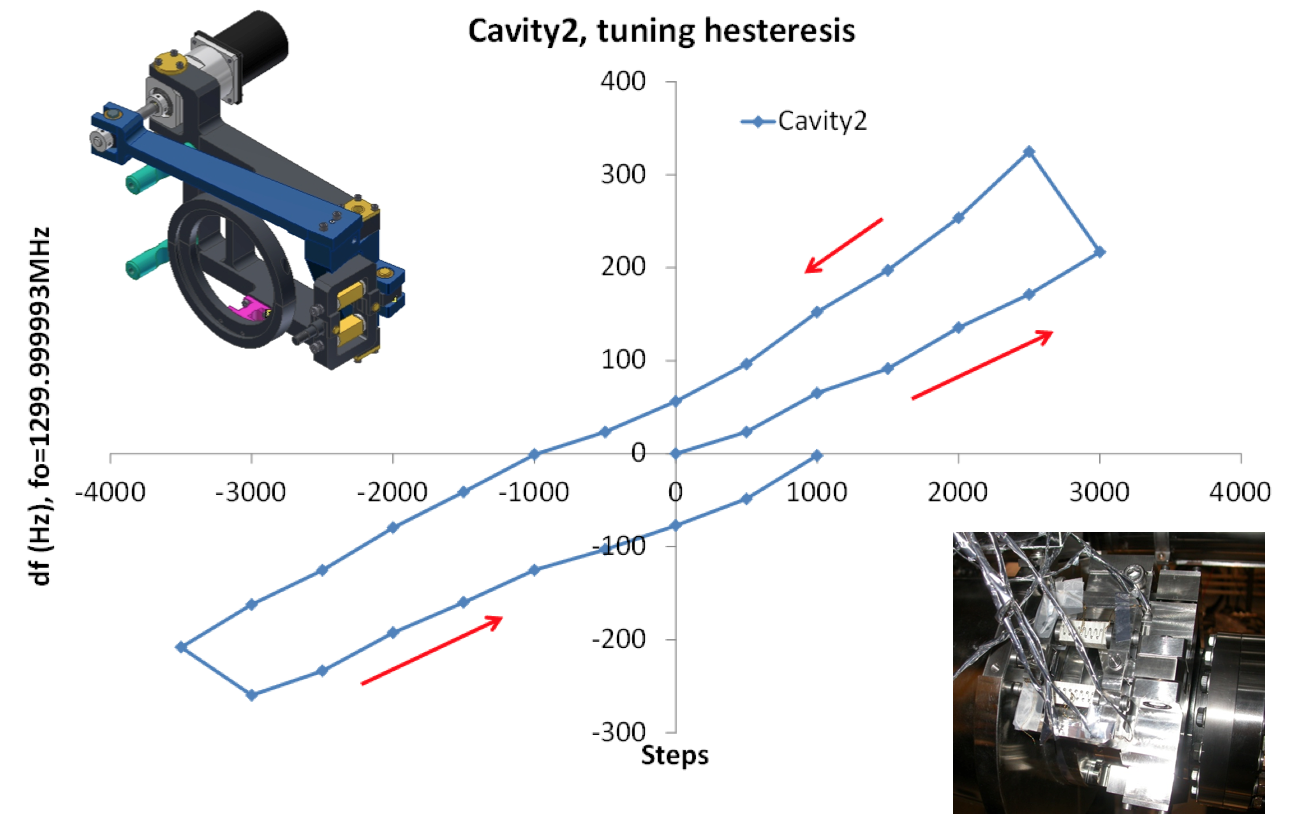}
\caption[]{Result of testing of the course tuner driven by a stepper motor. The observed hysteresis of 150 Hz is as expected and will be compensated by the action of the fine tuner driven by piezo actuators.}
\label{fig:tuning_hysteresis}
\end{figure}

\begin{table}[tb]
\centering
\caption{Tuning and pressure sensitivity of the MLC cavities. unstiffened (U) and stiffened (S) cavities are denoted. Cavity \#4 and \#6 were close to the tuning range.}
\begin{tabular*}{\columnwidth}{@{\extracolsep{\fill}}llcccc}
\toprule
           &              & Before tuning & Post tuning & Tuner range & Pressure sensitivity \\
           &              & {[}MHz{]}     & {[}MHz{]}   & {[}kHz{]}   & {[}Hz/mbar{]}        \\ \midrule
Cavity \#1 & ERL 7-3 (U)  & 1299.525      & 1300.000    & +470        & 29                   \\
 Cavity \#2 & ERL 7-5 (S)  & 1299.724      & 1300.00     & +270        & 11                   \\
Cavity \#3 & ERL 7-4 (U)  & 1299.650      & 1300.00     & +340        & 35                   \\
Cavity \#4 & ERL 7-7 (S)  & 1299.615      & 1299.996    & +381        & 13                   \\
Cavity \#5 & ERL 7-2a (U) & 1299.677      & 1300.00     & +323        & 25                   \\
Cavity \#6 & ERL 7-6 (S)  & 1299.554      & 1299.939    & +385        & 13                   \\ \midrule
Design     &              & 1299.700      & 1300.000    & +400        &        \\ \bottomrule
\end{tabular*}
\label{tab:cavity_tuning}
\end{table}
A typical curve that we measured on the coarse tuner is given in \Fig{fig:tuning_hysteresis}, displaying a 150 Hz hysteresis which will be counteracted by the piezo actuators being used for fine tuning. First measurements of the cavity microphonics levels indicated a rather high level of mechanical motions, likely driven by vibration sources outside of the MLC.  Especially the unstiffened cavities seem to have very strong mechanical vibrations. We are currently in the process of taking accelerometer data as well as fast He-pressure sensor data to understand the source of the microphonics and the path under which the couple into the module. First results indicate that vibrations by the MLC insulation vacuum pump strongly were coupled into the MLC, which can be resolved easily by adding vibration isolation between the pump and the MLC. 

The latest studies on microphonics has been presented at IPAC 2017 \Ref{IPAC2017:MOPVA122}. 



\ifdefined \buildingFullDocument

\renewcommand{\FiguresDirectory}{cryogenics/figures}

\else
\newcommand{\FullDocumentRoot}{..}
\newcommand{\FiguresDirectory}{figures}

\begin{document}
\fi

\chapter{Cryogenics\Leader{Eric Smith}}\label{chapter:cryogenics}

\section{Overview}
\subsection{Thermal loads expected }
There are two places in the CBETA experiment where cooling to cryogenic temperatures is required: the ICM, and the MLC.   The ICM contains five 2-cell niobium superconducting cavities which will be operated at 2.0K to accelerate an electron beam from about 350keV as it emerges from the gun to about 6MeV.  The six 7-cell superconducting cavities in the MLC will be operated at 1.8K to achieve a maximal energy of 150MeV of the electron beam after four passes through this assembly.  There are additional significant thermal loads from input couplers and higher-order-mode absorbers which will be taken out at temperatures near 80K and 5K, where it is more thermodynamically efficient to cool these loads.  The primary source for the cooling near 80K is supplied from a central liquid nitrogen facility in the laboratory, which is a 30,000 liter tank refilled at regular intervals upon demand by an external vendor.  The primary cooling source for both the 5K and 2K cooling is provided by a 4.2K helium liquefaction/refrigeration system whose main function is cooling for the main Wilson Lab synchrotron ring, but which has additional capacity that will be used for the CBETA project.  The approximate heat loads that must be extracted from each of these cryomodules at the operating temperature points are shown below in \Tab{tab:MLC_ICM_heat_loads}.
\begin{table}[tb]
\small
\centering
\caption{An indication of the expected heat loads from the two cryomodules under various operating conditions.  Many of these values have been directly measured, some involve extrapolations (the MLC has not been tested with a beam current passing through it). The acceleation gradient for this table has been chosen larger (sufficient for 200MeV beam) in order to add a safety margine for cavities that do not reach the full $Q_0$-facotr.}

\begin{tabular*}{\columnwidth}{@{\extracolsep{\fill}}lllllll}
 & &          & &          &  &          \\ 
\textbf{MLC Thermal Loads} &  &          &  &          &  &          \\ 
\toprule
Coolant temperature                       & 1.8 K &          & 5 K  &          & 80 K &          \\ 
                                          & Watt  & g/sec He & Watt & g/sec He & Watt & g/sec He \\ \midrule
Static Heat Load, No RF, No Beam          & 16    & 0.8      & 40   & 1.6      & 150  & 0.75     \\
Heat Load with 48 MeV, 1 mA ($\times$ 4)  & 43    & 2.15     & 45   & 1.5      & 200  & 1        \\
Heat Load with 48 MeV, 40 mA ($\times$ 4) & 50    & 2.5      & 50   & 1.67     & 900  & 4.5      \\ \bottomrule
\end{tabular*}

\begin{tabular*}{\columnwidth}{@{\extracolsep{\fill}}lllllll}
  & &          & &          &  &          \\ 
\textbf{ICM Thermal Loads} &  &          &  &          &  &          \\ 
\toprule
Coolant temperature                       & 1.8 K &          & 5 K  &          & 80 K &          \\ 
                                          & Watt  & g/sec He & Watt & g/sec He & Watt & g/sec He \\ \midrule
Static Heat Load, No RF, No Beam          & 15    & 0.75     & 40   & 1.6      & 150  & 0.75     \\
Heat Load with 48 MeV, 1 mA ($\times$ 4)  & 40    & 2        & 50   & 1.67     & 200  & 1        \\
Heat Load with 48 MeV, 40 mA ($\times$ 4) & 45    & 2.25     & 90   & 3.00     & 1200 & 6        \\ \bottomrule
\end{tabular*}

\label{tab:MLC_ICM_heat_loads}
\end{table}

Because of very substantial heat loads expected at intermediate temperatures in the cryomodules, it was considered necessary to provide cooling from cryogen streams in close proximity to the heat loads (too much temperature drop between coolant and heat source if conductive cooling through copper braid were attempted), the decision was made to provide cooling to the localized heat sources by a series of parallel flow channels in each cryomodule, and to assure uniform cooling power between the different channels, it was important to have single phase fluid rather than two-phase flow for the coolant.  So pressurized helium gas is the most desirable coolant for the 5K and 80K thermal anchor points.  The coolant flows for each cryomodule were provided in separate heat exchanger cans (HXC's) which cooled high pressure helium gas flow from compressors in the central helium liquefaction facility by heat exchange with the primary LN$_2$ and LHe coolant sources.  A separate HXC was used for each cryomodule, partly because the cryomodules were built and tested at different times and the HXC's already existed, partly because the ideal characteristics for the coolant streams are somewhat different for the two cryomodules.  The temperature of the pumped helium for the MLC has been chosen as 1.8K to achieve a very high intrinsic $Q$ for the superconducting cavities not simply to reduce the heat load into the liquid helium but to reduce the power requirements for the RF amplifiers being used to accelerate the beam.  Since almost all of the RF power absorbed by the beam being accelerated is compensated by the returning beam being decelerated, it is primarily the losses in the cavities which load the amplifiers.  It would be even better from the point of view of the RF amplifiers to run at the even lower temperature of 1.6K, but the increases in RF efficiency at the lower temperature come at a large additional cryogenic cost to provide cooling at the lower temperature.  Operation at 2K is a better compromise for the ICM, since the energy going to the beam is not recovered in any case, and the demands on the pumping system are much lower when operating at twice the helium vapor pressure.  The cavities in the ICM have lower intrinsic $Q$ in any case (for reasons that are not completely understood), and so there is in any case not much gain in operating at reduced temperature.  Extensive testing of the ICM at both 1.8K and 2K was carried out over a period of some years, while developing higher current low emittance operation, and the 2K operation was found to be much more economical.  A temperature below the helium superfluid transition (2.17K) is however highly desirable, since the thermal conductivity of the liquid helium is much higher in this condition, and the uniformity of the cooling on the cavity walls is greatly improved.To achieve the low pressures and moderately high mass throughput of helium required for the 1.8K/2K heat loads in these cryomodules, we have used a parallel combination of multiple pump skids, each consisting of a Tuthill MB2000 Roots blower backed by a Tuthill KT-500 Rotary Piston pump.  The ICM testing in the past has used two of these skids in parallel, as has the MLC testing.  At the present time we have a total of three of these pump skids, two of which are intended to be used for the ICM (which is intended to continue development operations during much of the time period for the installation of the new beamline and magnet systems), and one of which will be used to finalize testing of the MLC over the next several months, on a basis of testing one cavity operation at a time.  The purpose of this initial testing is to better understand and improve microphonics in the RF cavities, and to experiment with stabilization of cavity tuning using piezoelectric feedback, which was not yet completed in the MLC testing before reorganizing the LOE site to accommodate the new test ring.  We will need an additional 3 pump skids (as soon as money becomes available for their purchase) to provide the full cooling power required for simultaneous operation at full field of the six SRF cavities in the MLC.

\subsection{Use of existing Wilson Lab cryoplant}

The liquid helium refrigeration plant in Wilson Lab consists of 3 independent liquefier/refrigerator systems each capable of producing 600W of refrigeration at 4.2K, each with its own compressor system.  In its primary role, this system provides cooling for the CESR RF system and several superconducting wigglers that are used in the operation of the CHESS facility.  During the testing phases of the cryomodules to be used in the CBETA project, it has also provided the refrigeration for these systems.  Normally, the available refrigeration power has been in a range which can be handled by two of the three systems in operation, while the third system has been available as a near-immediate backup in case of performance degradation or failure of one of the operating refrigerators.  Simultaneous operation of the CESR/CHESS operations along with the CBETA project will bring the total refrigeration demand to a level where it will in general be necessary to use all three of the systems simultaneously, with only modest additional capacity available.  This should not present excessive risk to the operation of CESR/CHESS under the assumption that if maintenance or repairs to one of the systems should become necessary between scheduled down periods, the CBETA operations would be put into a standby mode for several days while the unscheduled maintenance was performed.  Prior system performance for many months of continuous operation would indicate that very few, if any, shutdowns of CBETA would occur because of main cryoplant failures during the test period, and that they would typically represent periods on the order of a few days if a significant repair were in fact necessary.The compressors used for the helium refrigeration have additional capacity beyond what is minimally needed to run the 3 expansion-engine refrigeration systems, and this extra capacity is what has been used in the past to provide the compressed helium gas flow for operating the 80K and 5K cooling systems.  The available capacity appears to be adequate to meet the needs for both the 1mA benchmark operation and the eventual 40mA conditions, but the figures for the load that must be absorbed from higher-order modes generated in the MLC involve extrapolation from the single 7-cell cavity that has already been tested with beam in it, and at a lower average beam current.  As there is expected to be some variation from cavity to cavity, and as the HOM power goes as the beam current squared, there is some level of uncertainty as to whether there will be adequate compressor power at the higher current operation.  If not, there might be a need to add an additional smaller compressor to make up for the shortage in capacity.  Should there be a problem, it should be possible to determine at intermediate beam currents between 1mA and 40mA what level of shortage there might be, and how much additional compressor capacity might be needed (or whether it would alternatively be acceptable to deal with slightly greater temperature rise in the HOM loads).

\subsection{Subsystem Components}
The majority of the changes to the building infrastructure needed to deliver cryogenic fluids to the cryomodules in the LOE location have already been completed.  This has included a re-routing of the LN$_2$/LHe transfer line and a local valve box from the earlier Phase 1a test area in LO to the new location in LOE, moving the ICM from its former test location to the final location for the CBETA testing, connection of the HEX can for the ICM into its new location (after rebuilding it for more convenient/efficient operation), re-installing the control wiring for the ICM in the new location, moving the pumping skids for the ICM into a temporary location for testing of the ICM during the installation of the new beam tube and the magnet ring, installation of the temporary 6" pumping line to the HXC, and connection of the cooling water and electrical power for the pumping skids.  For the MLC and its HXC, for the next several months there will be continuation of testing without beam, to further characterize microphonics and to test compensation by tuning adjustments with piezos.  This has involved moving the related pump skid into a new location and constructing a new pumping line, but this is ready for operation.  The exact final location of all pumping skids has not yet been decided upon, but the rearrangement of this part of the plumbing and utilities connection should be a straightforward task once the MLC has been moved into its final configuration for the ring.From the perspective of the cryogenics group, the relocation of the cryogenic transfer lines between the valve box and the HXC for the MLC is the primary specialized task remaining to be completed when the MLC gets moved to the final location for the CBETA ring.  The position of these components has already been decided, however, and the path for routing the cryogen transfer line from the valve box has also been allowed for.  The addition of 3 additional pump skids and the position change for the helium pumping lines will require some time to accomplish, but are much simpler to install than the vacuum insulated transfer lines.
\section{Heat Exchanger Cans in greater detail}

\subsection{Construction and Interfacing}

The HXC's for the ICM and MLC currently use a near-identical design, the original HXC for the ICM having been modified prior to the current installation to take advantage of some performance improvements introduced subsequent to the initial manufacture, during the course of many test experiments done in a smaller test cryostat, and then the production of the HXC for the MLC.  Each of the cans is about 1m in external diameter and 1.2m high, but situated on legs somewhat over a meter above floor level to provide a convenient connection into the cryomodules relative to the height of the beamline.  Each of these cans accepts through a vacuum insulated transfer line an inlet flow of LN$_2$ (at around 30psig and 77K) and LHe  (at about 3psig and 4.5K), and returns a stream of cold He vapor from the can through another tube in this same transfer line to a low-pressure input to the Wilson Laboratory helium refrigeration system.  It also accepts a stream of high-pressure helium gas (up to 20 bar at 300K), which is passed through a series of heat exchangers to provide coolant streams at various lower temperatures that are circulated through the associated cryomodule before being returned through the heat exchangers in the HXC and finally returned to the low-pressure input of the building helium compressors.  And finally, a portion of the 4.5K LHe that enters the HXC is fed through an additional heat exchanger at slightly above atmospheric pressure, subcooled to around 3K by returning low-pressure gas coming back from the cryomodule, and dropped to around 2K by a Joule-Thompson expansion valve on delivery to the cryomodule at a pressure of around 20 mbar (MLC) or 40 mbar (ICM) maintained by a room- temperature pumping system fed by the return flow through the low temperature heat exchanger, the gas being returned to near room temperature by electric heating before leaving the HXC, in order to avoid condensation on the length of room-temperature pumping line.

\begin{figure}[htbp]
\centering
\includegraphics[width=0.9\textwidth]{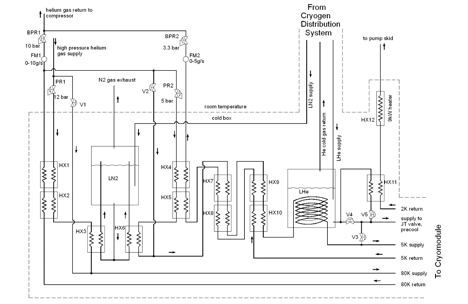}
\caption[]{Schematic view of the helium flow through the HXC for the MLC.}
\label{fig:helium_flow_schematic}
\end{figure}

High-pressure pure helium gas from the laboratory compressors enters the HXC through a pressure regulator PR1, typically set to about 12 bar, into the 80K system, whose output is maintained at typically 10 bar by BPR1.  The gas is pre-cooled by returning gas flow from the cryomodule through brazed-plate exchangers HX1 and HX2 before being finally cooled to 77K by heat exchange with LN$_2$ in HX3.  The flow of LN$_2$ is maintained by a thermosiphon from a LN$_2$ pot maintained at a constant fill level through a pressure-activated valve controlled from a level meter in the pot.  The flow rate of the gas through this part of the system is monitored by the mass-flow meter FM1, and may be remotely adjusted by a pressure-actuated valve situated in the cryomodule.  In practice we have chosen the set point of this valve manually rather than through a feedback system, although the latter plan would have been possible.
Part of the returning gas stream is then sent through pressure regulator PR2 at a pressure of around 5 bar back into the 5K cooling system of the HXC, while the remainder is returned to the low pressure inlet to the compressor system.  Similarly to the 80K cooling stream, this 5K coolant stream is first passed through a series of heat exchangers to pre-cool to 80K, then counter-cooled with returning cold gas through another series of 4 brazed-plate heat exchangers to near 5K, then finally to about 4.5K by passage through a spiral of copper tubing immersed in a 4.3K helium bath, before finally being sent on to the cryomodule.  As in the case of the 80K system, the flow rate is set by a pressure-actuated valve mounted on the cryomodule.  The outlet pressure from the HXC is maintained at about 3.3 bar by BPR2, and the exiting gas returned to the compressor system input.

Finally, cooling for the cavities is provided by a pumped liquid helium system.  Helium from the previously mentioned 4.3K helium bath is passed through a Hampson-style heat exchanger to a temperature around 3K, but still at near-atmospheric pressure, and fed through an actuated throttle valve in the cryomodule which performs a Joule-Thomson expansion to drop the pressure to about 20 mbar, with the throughput of the valve controlled by feedback to maintain a constant liquid helium level in the cryomodule.  The temperature of the helium is controlled by adjusting the speed of the Roots blowers which are the first stage of the pumping system with a properly tuned PID feedback system, using the pressure of the helium as a measure of the temperature.  It is desired to control the temperature of the pumped helium bath to the fraction of a mK level, not so much because the properties of the superconducting cavities depend that strongly on the temperature, but because variations of the pressure on the outer walls of the cavities, which are immersed in the helium bath, slightly affects the geometry of the cavities and hence the natural frequency, and the efficiency of coupling to the RF system.
The discussion of the HXC's up to this point has been discussing is steady-state operation.  During the process of cool-down from room temperature to the normal operating point, it is important to have very uniform cooling to avoid distortions and excessive stresses because of differential thermal contraction.  In practice, this means quite slow cool-down rates (on the order of 10K/hr, combined with a very uniform ramping of the temperature of the input fluids.  This is achieved by mixing the high-pressure gas flows after the cooling to 80K with a variable amount of room-temperature gas flow, starting with almost 100\% 300K gas initially, reducing finally to 100\% 80K gas.  This is done by manual control of valves V1 and V2.  Also, during the pre-cool operations, a considerable proportion of the gas which would normally all pass through the 5K system is diverted through V3 through the cavity helium system to pre-cool the cavities and the helium gas return pipe in the cryomodules.  This works out rather conveniently, because there is rather little thermal mass in the thermal anchors which are cooled by the 5K system, so cooling those takes very little of the mass flow.  Once the system in the cryomodules is down below 100K, there is very little further thermal expansion to be concerned with, so below this point the cooling may be done more rapidly, and with less concern about any thermal gradients which may temporarily occur.
The primary difference between this scheme for the MLC HXC and the one for the ICM is that in the latter case the flow control valves for all systems are located in the HXC rather than in the cryomodule.  Also, the temperature of the pumped helium in the ICM is 2.0K rather than 1.8K, since without energy recovery for this module, there are fewer gains for the RF system by operating the system at lower temperature, so the reduced cryogenic costs by operating as slightly higher temperature outweigh any added RF costs.

\subsection{Prior performance}
The testing which has already been carried out in testing both the ICM and the MLC has indicated that we should be in good shape as far as producing adequate high pressure flow for both the 5K and 80K systems.  The situation with the heat exchanger for the 2K system in the ICM has also been thoroughly tested.  The heat exchanger for the 1.8K system has only been checked at half capacity because of limits on availability of more pump skids.  Extrapolation from the half-flow performance leaves us confident that it will work fine when we get the additional pump skids.

\section{Cryomodules in greater detail}
\subsection{Construction and Interfacing}

\begin{figure}[htbp]
\centering
\includegraphics[width=0.9\textwidth]{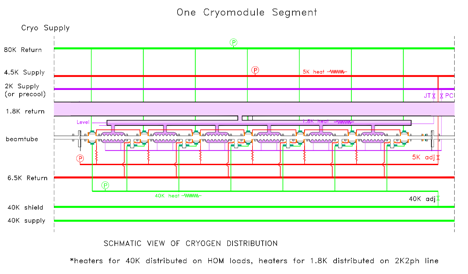}
\caption[]{Schematic view of the MLC as it would be placed in a string of many cryomodules.  In the case of the CBETA testing, there is just one cryomodule, with end caps on each of the main supply and return lines.}
\label{fig:MLC_schematic}
\end{figure}

The cartoon view of the MLC shown below in \Fig{fig:MLC_schematic} illustrates the main features of the cryogenic distribution within the cryomodule.  A more complete description may be found in the following reference \Ref{Eichhorn15_02}.  The outer vacuum jacket of the cryomodule is a steel cylinder approximately 1m in diameter and 10m long.  The series of six seven-cell niobium SRF cavities, each in its own surrounding helium chamber is supported from a 250mm titanium tube, the Helium Gas Return Pipe (HGRP) which doubles as a stable mechanical mount and part of the pumping system for maintaining the superfluid helium at 1.8K (approximately 20 mbar pressure). At one end of each cavity is an input coupler (IC) for introducing the 1.3GHz power into the cavity, and between each cavity and the next (and at each end of the string of cavities) is a higher-order mode absorber (HOM) to remove energy from unwanted higher-frequency RF fields which are generated by interaction between the fields in the cavities and the very short bunched structure of the electrons in the beam.  All the cryogenic components are enclosed in a cooled 80K aluminum shield to intercept the room-temperature thermal radiation load, and specific thermal intercepts are placed on all the support structures to remove most of incoming heat loads at the higher temperatures of 80K and 5K, rather than at the operational temperature of 1.8K.  All the cold structures are additionally wrapped in 10-30 layers of "superinsulation" in order to further reduce the heat transport by thermal radiation.  Because the cavity performance is adversely affected by the presence of static magnetic fields above the milligauss level, there are also magnetic shielding layers at low temperatures around each cavity, and at room temperature outside the 80K shield.  The HGRP is supported from the vacuum vessel by 3 cylindrical G10 posts, equipped with mechanisms for external positioning adjustment to allow fine control of the beamline location.  The central support post is fixed in location, while the end posts allow for longitudinal motion of the HGRP relative to the vacuum vessel caused by thermal contraction of about 20mm in the length of the HGRP upon cooling to operational temperature from room temperature.  Bellows sections incorporated into the HOM absorbers allow for the smaller changes in position relative to the HGRP experienced by the cavities.  Because of the magnitude of the heat loads being absorbed at both 5K and 80K under normal operating conditions, it was necessary to get the coolant fluids very near to the source of heat generation.  Cooling through copper braids as is commonly done in low duty cycle machines would in this machine have produced excessive temperature gradients.  Thus convective flow of helium through an array of parallel tubing flow channels was used to get the heat exchange area very near to the sources of heat production.  With multiple parallel flow paths, thermal runaway problems must always be considered, as if there is a greater heat load in one flow channel, the density of the coolant will decrease, and the viscosity decrease, thus diminishing the flow, in turn increasing the temperature rise.  The key to avoiding this is to insert a flow-limiting impedance in the supply side of the lines prior to the point where the heat load is introduced, establishing effectively a current source.  This has been implemented and tested in our MLC.  Also, the 80K heating in the HOM's is much greater than that in the IC's, so to simplify the plumbing the 80K cooling loop on each IC is placed in series and in front of the 80K cooling on the adjacent HOM,  This gives then a very small temperature rise in the IC, which is desirable in part because there is necessarily a stronger thermal connection between the 80K and 5K anchors in the IC's than in the HOM's.The input flows into the 5K and 80K manifolds is adjusted by remotely controlled electro-pneumatic valves, and may be adjusted to provide an adequate cooling for the current operating conditions of the machine, which may vary significantly with beam current.The cooling for the cavities to be maintained at 1.8K is supplied by an input stream from the associated HXC at roughly 3K and slightly above 1 bar absolute pressure, and restricted by a JT valve, remotely controlled, usually by automated PID feedback to maintain a constant level in the 100 mm diameter so-called "2K-2phase line" (2K2$\phi$) which is physically slightly above the string of helium vessels for each cavity, each connected to the line by a short connecting line of the same diameter.  The top of this 2K2$\phi$ line is attached to the HGRP again by a short connecting line, and the pressure of the gas is regulated by adjusting the blower speed of the room-temperature pumping system connected to the line.  Because the superfluid helium is an excellent thermal conductor, temperatures gradients between the evaporating surface and the cavities is no more than the mK range under normal operating conditions.From a cryogenic perspective, the ICM differs only in minor ways from the MLC.  It is about half the length, has five two-cell cavities rather than six seven-cell ones, has a different model of HOM from what is used in the MLC, and more significantly has much higher power dissipation in the IC's, because there is no energy recovery so vastly more power must be supplied from the RF system.  There are also two IC's per cavity, both to symmetrize the RF fields in this lower energy regime, and to cut in half the amount of power required to be handled by each IC.  Also, the four valves to regulate the flow of the cryogen streams into the cryomodule are in this case located in the associated HXC, rather than in the cryomodule itself.  The cryogenic loads for the two types of cryomodule are actually remarkably similar at the nominal 40mA operating conditions as may be seen from Table 1, the main difference being in more 5K cooling required for the input coupler operation (and less expensively, somewhat more 80K cooling also for the IC's).  Because the $Q$ of the cavities used in the ICM is about a factor of 4 lower than those in the ICM (for reasons still not completely understood), the performance depends much less critically on the cavity temperature -- there is essentially no performance improvement from operating at 1.8K instead of 2K.  Thus, it will be run at 2K, since this puts much less demands on the pumping system (the vapor pressure of the helium is twice as high).  We therefore have a lower capital cost (fewer pump skids), and somewhat lower operational costs.  If we were operating in an energy-recovery mode, the lower cavity $Q$ would be disastrous, because there would be much higher loading placed on the RF system, but in this case, almost all the power delivered by the rf system goes directly into the electron beam rather than internal losses in the cavities anyway.

\subsection{Prior performance}

At this point the ICM has seen several years of testing (the performance of the cryomodules from other than the cryogenic aspects may be seen in more detail in other sections of the design review document),  and the cryogenic system has been demonstrated to be adequate for essentially all aspects of operation which will be required for the CBETA project.  In particular, it has been above the full 6MeV level which will be utilized, and at or above the same average beam current.  This means that the capabilities of the 2K pumping system and heat exchanger have already been fully tested, as have the cooling of the 5K and 80K intercepts for the input couplers.  The 80K cooling power of the HOM loads has been less well investigated, however, because the HOM power is expected to vary as the square of the bunch charge and only linearly in duty cycle.  So in the CBETA operation where only one bunch in four is populated, but at four times the average charge, the average HOM power will be 4 times what has actually been tested under beam loading.  However, if this power goes according to theoretical expectations, we have tested with direct electrical heaters that it will also has the capacity to handle this heat load.Much of the required performance of the MLC cryogenic system has also been demonstrated, but budgetary limitations have prevented us from testing some of the performance parameters.  We do not yet have enough High Power Amplifiers (HPA's) to simultaneously test all cavities at full RF power, nor enough pump skids to provide the level of 1.8K helium throughput if we did.  However, all cavities have been individually tested at the needed RF power levels, and power demand levels and pumping characteristics have met design expectations.  We are quite confident that when the new HPA's and pump skids are purchased and installed, there should be no difficulties with the 1.8K pumping system.  We have not yet put a beam through the entire system, but have been able to test with electrical heating of resistive loads that we will be able to handle the requisite demands on the 5K and 80K cooling loops.

\section{Sensors and Controls for the cryogenics system}

Both cryomodules and both HXC's have a large number of installed thermal sensors installed (probably more than might be needed in a production model, but very useful for diagnostics in a prototyping phase).  In general, thermometers for 80K and above have been platinum resistance thermometers, and for thermometers which also need to cover the range below 80K, we have used Cernox thermometers.  These have been read out with CryoCon modules which can each handle 8 sensors with quite adequate readout speed (1 sec or less).  We have measured helium pressures on the pumped helium with capacitance manometers operated at room temperature, and we have used the output from these to regulate pumping speed, and hence control temperature.  Level measurement of LN$_2$ baths in the HXC's has been done with commercial capacitive level sensors.  Level measurement of the helium has been with commercial superconducting wire level monitors (because of the crucial nature of this output, and the great difficulty of making replacements, each cryomodule has 4 sensors installed, though we have not yet had failures -- there have been occasional failures in similar level sticks installed in many other cryostats in the overall facility over a period of many years).  For the high-pressure helium system we are using room-temperature mass-flow meters based on a thermal transport principle.  For measurement of mass flow on the 1.8K/2K pumping systems, we have been using a combination of similar mass-flow meters, or for higher precision on testing of cavity $Q$'s by thermal means, mechanical gas-flow meters on the output of the pumping skids.  There are some somewhat more specialized systems, built in-house for measuring transverse movements of the beamline/cavities upon cool-down with capacitive measurement of spacing from a stretched wire, piezoelectric systems for fine-tuning of cavity frequencies, low-temperature stepping motors for doing coarser control of cavity frequencies.  Vacuum measurements are made, depending on system and pressure range with cold cathode gauges, pirani or convection gauges, ion-pump currents.Generally, the outputs from all the sensors are fed locally to a PLC, which also controls many local control loops, and then are further integrated to a general EPICS control system which allows full remote control, readout, and archiving of signals.

\section{Safety Issues (that have already been considered for our prior test operations)}

As is general with cryogenic systems, the primary considerations for personnel protection are issues of direct contact with cold surfaces, avoidance of any possibility of asphyxiation from reduced oxygen because of displaced air by release of cryogenic gases, and assurance of pressure relief on all systems which might either be closed off, or have sudden pressure buildup because of sudden heat load.  Because both the ICM system and the MLC system have been the object of extended testing in Wilson Lab, extensive prior consideration has been given to these questions, and have been incorporated into Wilson Lab safety plans.  In general, it has been more straightforward than many of the laboratory cryogenic safety situations, because neither system has a very large volume of cryogenic fluids contained in it, and it is enclosed in a room with a very large volume (much easier to ensure safety against sudden cryogen release than some of the systems in the tunnel).  Only mild adjustments in the existing safety plans have been necessitated by moving cryostats into a new location.  One less major consideration, condensation from cryogenic transport lines, typically only occurring start-up operations, have also been addressed by catching basins and drain tubes in strategic areas to avoid risk of water dripping on electrical circuits, or providing slippery spots on floors where workers might be walking.

\section{Aspects still under development}

There are two potential areas that might impact readiness of the cryogenic system for CBETA operation.  One is that there was a small leak between the helium space and the insulation vacuum in the ICM during the last several years of operation.  This was small enough so that operation could be maintained (though with a somewhat higher static heat load) by continually pumping on the insulation vacuum space with a turbo pump.  Unfortunately the size of this leak was never documented at room temperature.  Between the last run of the ICM and the present time, the system was partially disassembled to improve the uniformity of the high-pressure flow distribution for 5K and 80K cooling.  A small leak was localized at this time, but in a place very hard to repair in situ, and it was decided that this was probably the same leak that has been there all along, so was not repaired.  We have added considerably greater pumping capacity, but until we cool the system back down, we will not be 100\% sure this is indeed the same, unchanged leak we saw before.  This will be verified in the first half June, but if the leak rate turns out to be too high, it may be necessary to again remove the ICM to another location and perform repairs, which would result in a minimum of a month's work for several people, or potentially several times that if the repair should require more extensive disassembly.The second aspect of things is that the tuning sensitivity of several of the cavities in the MLC to microphonics has been marginal.  There has not yet been time to thoroughly examine this issue.  While not enough to make the immediate 1 mA beam current impossible, it is probably necessary to either eliminate some of the sources of the microphonic noise, or else to use active piezoelectric control to compensate if we are to achieve the full 40 mA goal.  This should be done as soon as there is an opportunity (depends somewhat on both budget and personnel availability).

%
%
%
%
%





\ifdefined \buildingFullDocument

\renewcommand{\FiguresDirectory}{vacuum/figures}

\else
\newcommand{\FullDocumentRoot}{..}
\newcommand{\FiguresDirectory}{figures}

\begin{document}
\fi

\chapter{Vacuum System \Leader{Yulin}}\label{chapter:vacuum}

\section{Vacuum system layout and sections}

\begin{figure}[tb]
\centering
\includegraphics[width=0.9\textwidth]{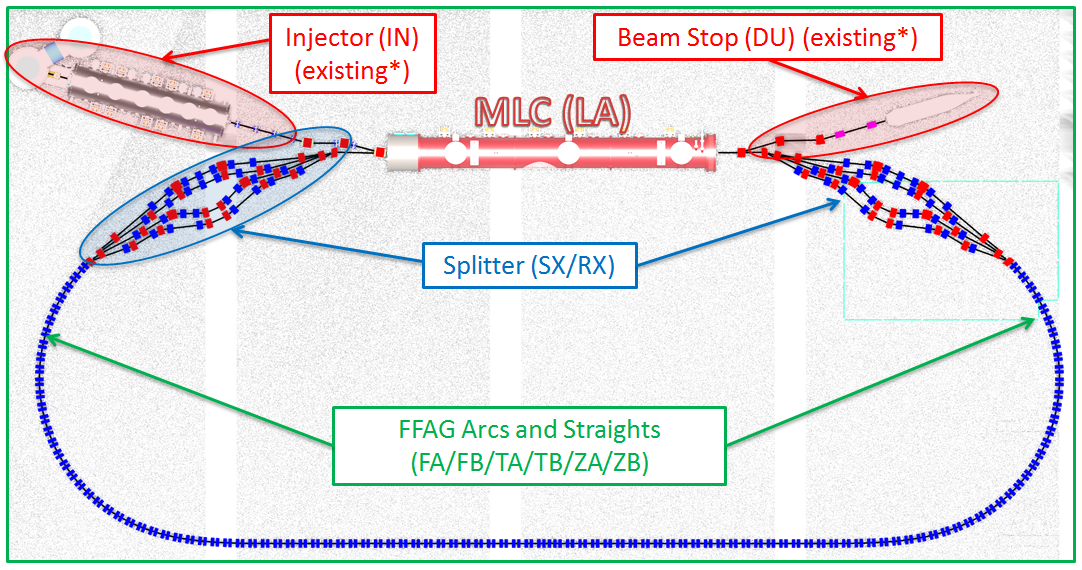}
\caption{CBETA Layout and vacuum sections.}
\label{fig:vacuum_layout}
\end{figure}

The vacuum system layout will conform to the accelerator lattice layout.  Accordingly, the vacuum system will consist of the following sections, as shown in \Fig{fig:vacuum_layout}:
\begin{itemize}
	\item	Injector and merger (IN section)
	\item	Main linac cryomodule (LA section)
	\item	Demerger and beam stop (DU section)
	\item	Splitters (SX and RX sections)
	\item	FFAG arcs (FA, TA, TB and FB sections) and straight (ZA and ZB sections)
\end{itemize} 
 
The injector (IN section) includes the electron gun and the  Injector Cryo-Module (ICM).  This section already exists from the Cornell Prototype ERL Injector project, and has been relocated and reused for the CBETA project. Full details are described in \Section{chapter:injector}.  The short merger beam line that connects the injector to the main linac (LA) is also described in that section.

The Main Linac Cryomodule (MLC) is a self-contained accelerator section (LA section) from a vacuum system point of view, and is described in \Section{chapter:linac}.  The interface between the MLC and the rest of the CBETA vacuum system will be described in \Section{sec:vacuum_instrumentation} below.

The electron beams of 4 different energies (exiting and entering the MLC) will be split into four separated beam pipes and then recombined into a single beam pipe for optics and timing reasons (in the SX and RX sections).  One of important features of these sections is to provide beam path length adjustments for each of four energy electron beams.

The FFAG arcs (including pure arcs FA and FB, transition arcs TA and TB)  and long straight (ZA and ZB sections) consist of more or less repetitive structures of hybrid iron magnets.  Thus units of repetitive simple beam pipes will be designed for the FFAG sections.  

The energy recovered electron beam (with lowest energy) is demerged into a high power beam stop.  The beam stop transport beamline (DU section) also exists from the Cornell Prototype ERL Injector project, and is to be relocated and re-used for the CBETA project.

\section{Vacuum system requirements and design considerations}

Vacuum beam pipes are part of beam transport system.  A list of vacuum requirements and design considerations is given below.
\begin{itemize}
	\item Produce adequate level of vacuum, through proper beam pipe material selection and preparation, and vacuum pumping.  The required level of vacuum will be determined by acceptable beam losses due to residual gas scattering, among other factors.
	\item Aluminum (6061 or 6063, in -T6 or -T4 temper) is preferred material for the beam pipes for its good electric conductivity (resistive-wall), no residual radioactivity (from beam losses) and low magnetization (from cold work and welding etc.)
	\item Provide sufficiently large beam apertures, while allowing adequate clearance to the magnets for their position adjustments.
	\item Periodically distributed bellows are integrated in the beam pipe design to facilitating modular installation of various sections.
	\item With high beam current and closely spaced electron bunches, design efforts will be made to keep low beam impedance, including smooth beam pipe inner profiles, RF shielded bellows and gate valves, gentle transitions between different beam pipe cross sections, etc.
	\item Beam pipes will host various beam instrumentation and diagnostics, such as BPMs and instrumentation ports (for beam viewers, etc.)  
\end{itemize}
 
\section{Vacuum system construction, installation, and operation} 
 
\subsection{FFAG Arcs and Straight}
 
\begin{figure}[tb]
\centering
\includegraphics[width=0.95\textwidth]{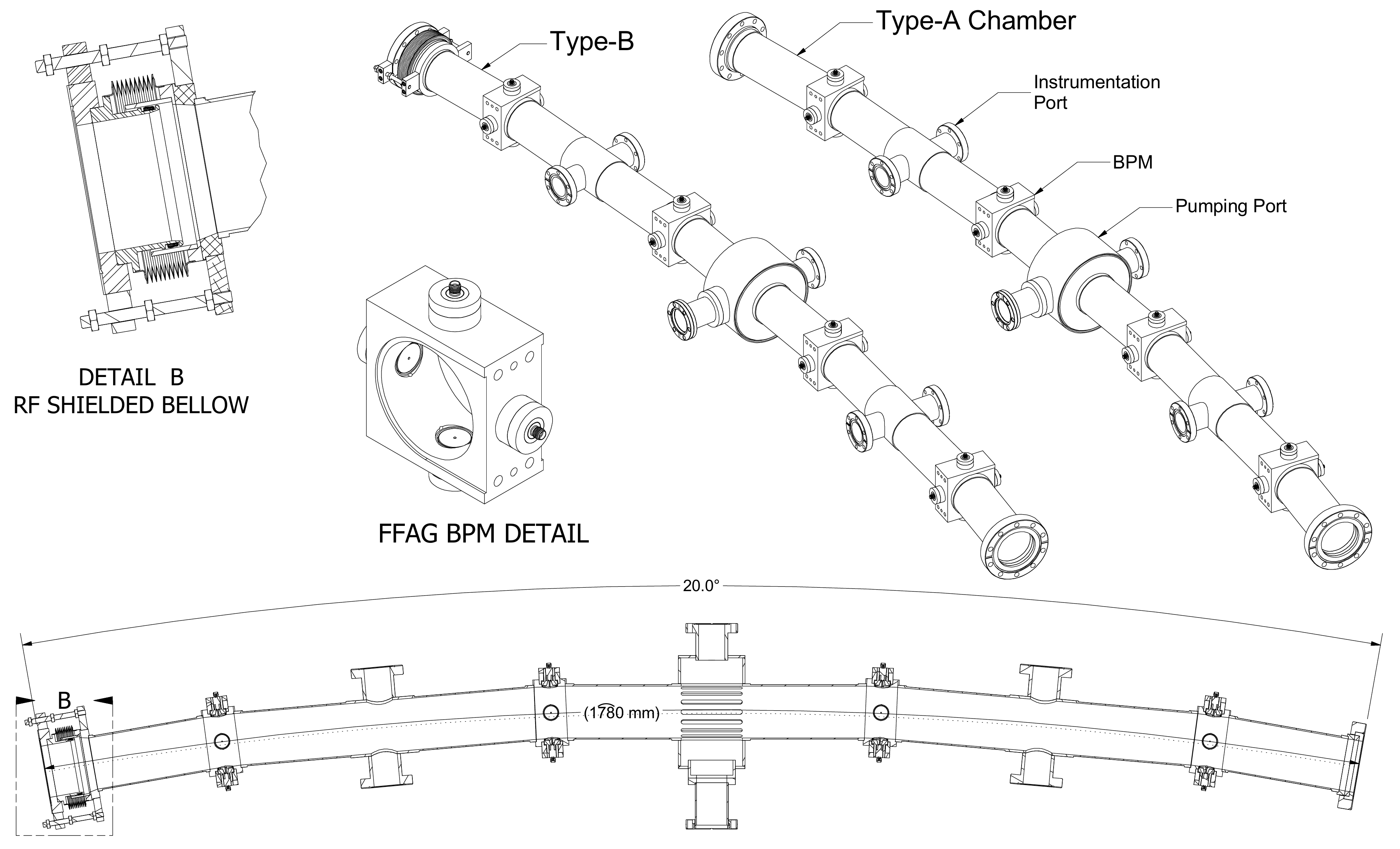}
\caption{A FFAG arc 4-cell beam pipe design.  The 1.78~m long beam pipe assembly houses 4 sets of 4-button BPMs on the 7~cm drifts between QF and QD magnets.  One pump port and two beam instrumentation ports are on the 12~cm middle drifts between FFAG cells.}
\label{fig:cell_vacuumchamber}
\end{figure}

\begin{figure}[tb]
\centering
\includegraphics[width=0.85\textwidth]{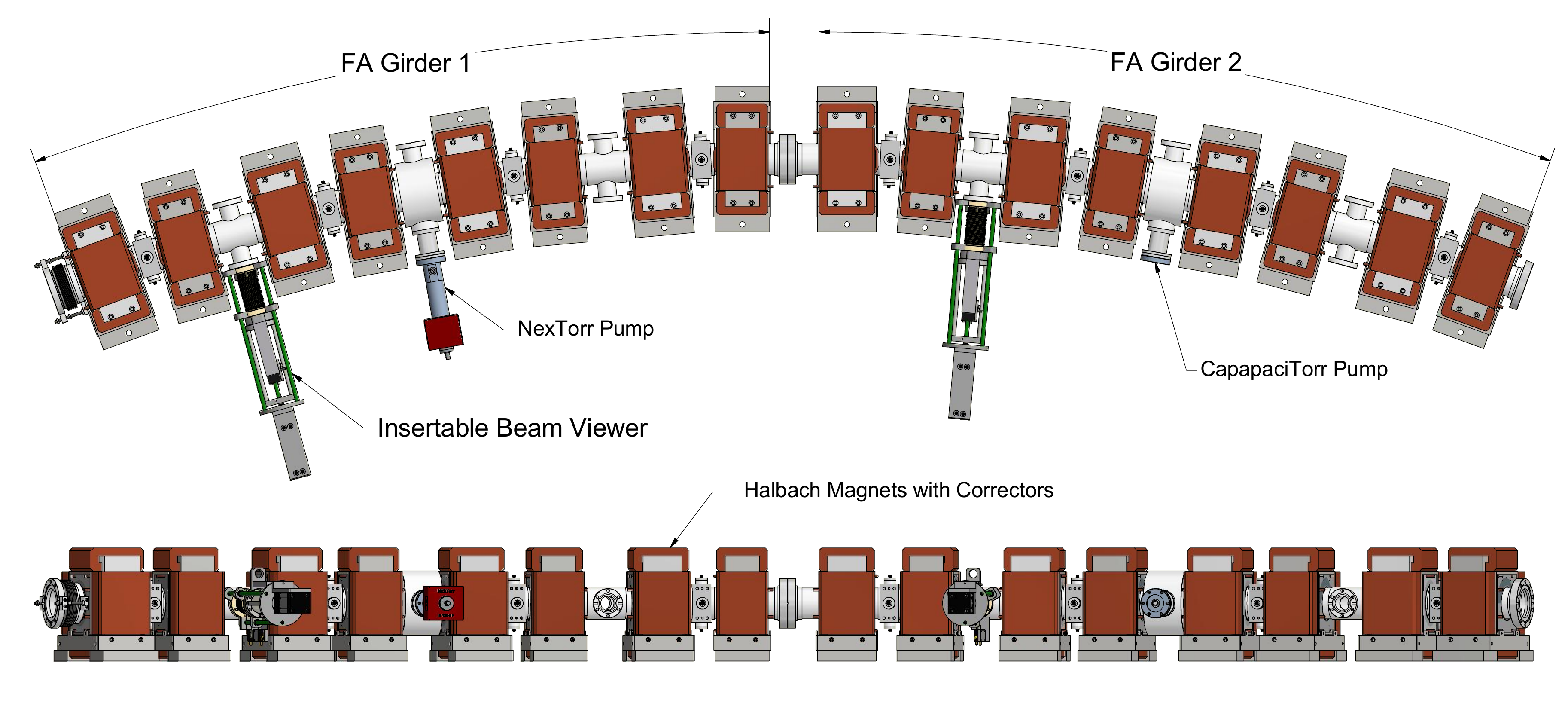}
\caption{Illustration of the 4-cell beam pipes in two conceptual arc FFAG girders, each with 4 FFAG Halbach magnet cells.}
\label{fig:cell_magnets}
\end{figure}

In the FFAG arcs and straight, permanent magnets are arranged in more or less periodic double-magnet cells.  The relatively short drifts between magnets are reserved as much as possible for vacuum pumping, beam instrumentation.  Therefore, it is efficient use of these drifts by constructing beam pipe assembly through multiple FFAG magnet cells, reducing the number of beam pipe flanges.  In the FFAG arcs (including FA, FB, TA and TB sections), a typical 4-cell FFAG arc beam pipe assembly is depicted in \Fig{fig:cell_vacuumchamber}.  As shown, there are two versions of 4-cell arc chambers, which are, a `Type-A' chamber without a bellow and a `Type-B' chamber with a RF-shielded bellow.  The FFAG beam pipes may be made of extruded 6061-T6 (or -T4) aluminum with cross section designed to meet both required beam apertures and magnet clearances, as shown in \Fig{fig:cell_magnets}. 

\begin{figure}[tb]
\centering
\includegraphics[width=0.95\textwidth]{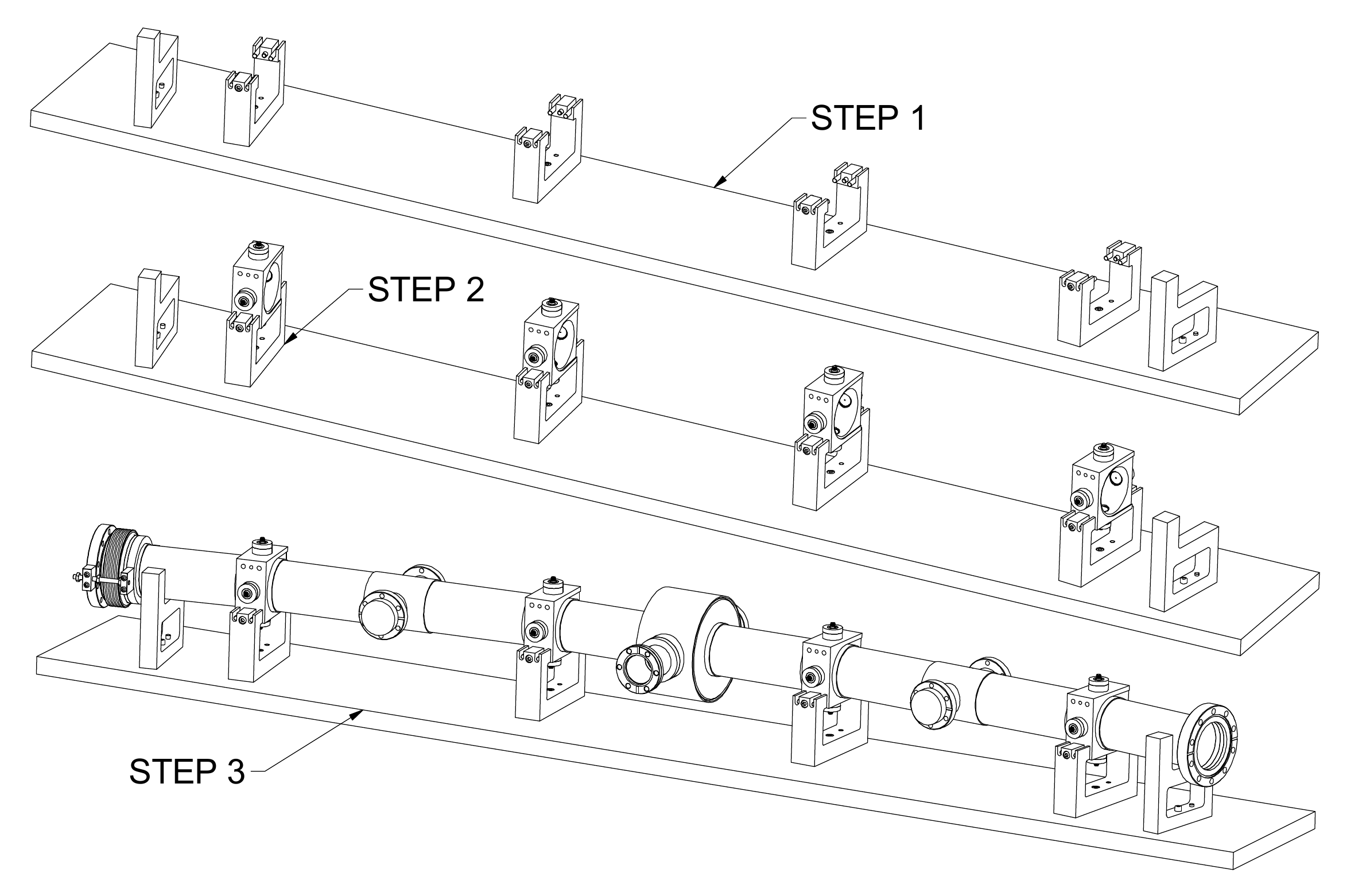}
\caption{FFAG 4-cell beampipe welding setup: Step 1 - BPM jig blocks are pinned to location on a base jig plates.  The pin-holes on the base plate are precised located for beampipes at various FFAG sections.  Step 2 - Pre-fabricated FFAG BPM assemblies are loaded onto the jig blocks; Step 3 - Beam pipes are inserted into the open slots on the BOM blocks.  The lenghs and end angles of the beam pipes are according to the lattice at various locations in the FFAG sections.}
\label{fig:beampipe_welding}
\end{figure}

In this arc chamber design scheme, identical 4-button BPM assemblies will be used for FA, TA, TB and FB sections.  Special assembling fixtures will be designed to set these BPM blocks to desired geometries at various location in the arc sections. Simple straight extrusions are to be machined to fit between these BPM blocks, and then welded (together with flanges and other functional ports) to complete beam pipe assemblies.  This scheme (dubbed as the block-and-straight scheme, as illustrated in \Fig{fig:beampipe_welding}) is to avoid complicated bending of extrusions, especially in the TA and TB sections where bending radii changes from cell to cell.  

In the FFAG straight (ZA and ZB sections), there is no horizontal offset between QF and QD magnets, and beam orbits of four energies are merged together.  However, beampipe construction similar to the arcs and transition arcs will be used for the straight section, as using the same BPMs and beampipes will yield significant savings both in engineering and fabrication costs.  
 
\subsection{Splitters}

\begin{figure}[tb]
\centering
\includegraphics[width=0.9\textwidth]{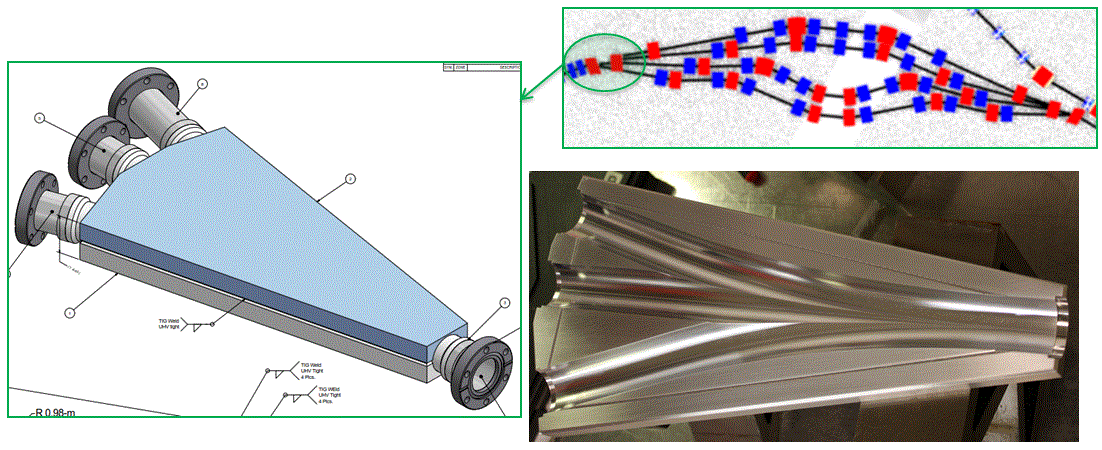}
\caption{Example design for beam splitting and combing chamber, by welding two machined aluminum halves with smooth beam path transitions.}
\label{fig:example_splitter}
\end{figure}

To keep low beam impedance, the beam splitting and combining vacuum chambers may be made of aluminum alloy (6061-T6) with smooth beam path transitions, as illustrated by an example vacuum chamber used for the Cornell Prototype Injector project, in \Fig{fig:example_splitter}.

The separated beam pipes will be designed to allow independent beam path length adjustment by using sets of 4 RF-shielded bellows in CBETA's final configuration.  However, during staged beam operations starting from single beam to multiple beams, simple beam pipes with low cost non-RF-shielded bellows may be used for various initial beam path configurations.

Beam collimators (or/and scrappers) may also installed in these single beam chambers, together with BPMs and other beam instruments.
 
\subsection{RF-shielded bellows}

\begin{figure}[tb]
\centering
\includegraphics[width=0.65\textwidth]{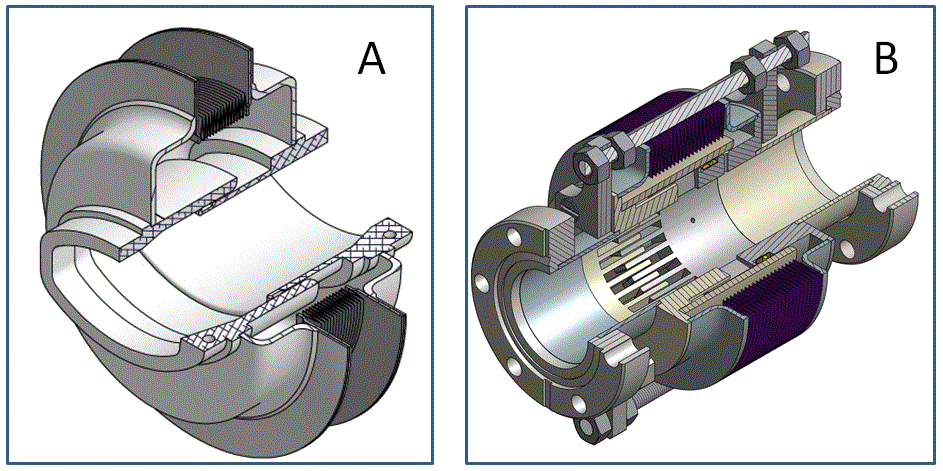}
\caption{RF shielded bellows used in CESR (A) and in Cornell Prototype injector (B).}
\label{fig:shielded_bellows}
\end{figure}

Bellows with RF shields are needed to provide adequate flexibility of the vacuum system in vacuum component installation and operations.  Two examples of RF shielded bellows have been successfully constructed and operated in the Cornell photoinjector and in the Cornell Electron storage Ring (CESR), as shown in \Fig{fig:shielded_bellows}.  An example of implementation of CESR-style sliding joint is illustrated in \Fig{fig:cell_vacuumchamber}.

\subsection{Ion clearing electrodes} 

\begin{figure}[tb]
\centering
\includegraphics[width=0.9\textwidth]{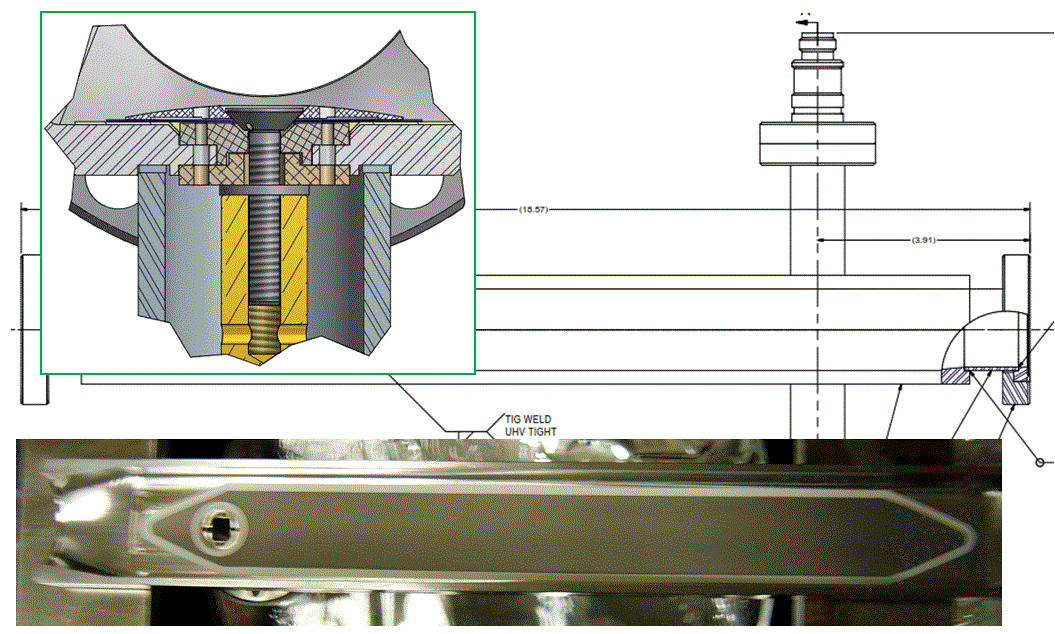}
\caption{Low impedance ion clearing electrode chamber tested in the Cornell prototype ERL injector.  The electrodes are made of a tungsten thin film (0.1~mm in thickness) on top of a thin (approx. 0.2~mm) alumina substrate.  Both alumina and tungsten thin films are deposited via thermal-spray technique.  A low profile electric connection is shown in the insert.}
\label{fig:clearing_electrode}
\end{figure}

Ion trapping may not be avoidable without active clearing method, due to the nature of in the final CBETA CW high beam current beam operations.  Low impedance clearing electrodes may be deployed at various locations to reduce ill-effect from the ion trapping.  Thin electrodes directly deposited onto the interior walls of beam pipes have been successfully implemented in CesrTA and Super KEKB.  A clearing electrode beam pipe of this style was made and tested in the Cornell prototype ERL injector (see \Fig{fig:clearing_electrode}).  No clearing electrode will be implemented in the initial CBETA.  The necessity and locations of these electrodes is being studied by simulations.  

\subsection{Vacuum system construction and installation}
	
	All vacuum beam pipes will be fabricated following stringent ultra-high vacuum (UHV) procedure and practice.  All beam pipe assemblies will be certified to be leak-free, and will be baked in vacuum up to 150~C.  Most of the beam pipes will be delivered to BNL to be assembled onto the girder units.  The baked beam pipes will be back-filled with chemically filtered nitrogen (with moisture and THC at ppb level) and sealed (with pinch-off fittings) for transportation and girder assembling.  The same nitrogen system must be used to purge the beam pipes whenever any flange is to be opened for connection, etc.  Operational experiences of Cornell prototype ERL injector, as well as CESR vacuum systems have demonstrated that in situ bake-out is not necessary with above beam pipe preparation and installation procedures. 

	 The injector, the ICM and the beam dump are to be installed and surveyed into locations.  These sections are equipped with RF-shielded gate valves.  

	The splitter and the FFAG sections will be installed in corresponding girder units.  The detailed installation sequence will be developed during the engineering design process to minimize air exposure of the vacuum beam pipes.  Further system and cost optimization will be carried out to decide if any additional RF-shielded gate valves are needed in these sections.  During the initial operations low cost UHV gate valves without RF-shield may also be used at locations where staged reconfigurations is expected, such as the splitter section. 
 
\subsection{Vacuum pumping}

\begin{figure}[tb]
\centering
\includegraphics[width=0.8\textwidth]{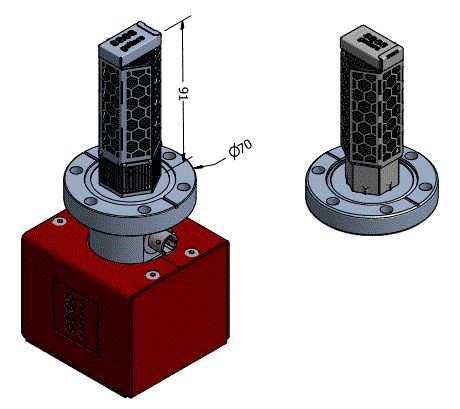}
\caption{Typical vacuum pumps for CBETA vacuum system, with a 200-l/s NexTorr (left) and a 200-l/s CapaciTorr (right).}
\label{fig:vacuum_pumps}
\end{figure}

	With very limited spaces between magnets, compact and high capacity non-evaporable getter (NEG) pumps will be used, such as CapaciTorr (sintered NEG modular pump) and NexTorr  (combination NEG and ion pump), see \Fig{fig:vacuum_pumps} for example pumps.  In the FFAG sections, modified NexTorr pumps with extension tubes are used, to ensure that the ion pumps are not interfered by strong magnetic field of the ion magnets, as shown in \Fig{fig:cell_magnets}.  The locations of pumps will optimized during CBETA engineering designs, with aid of vacuum simulations (see \Section{sec:vacuum_pumping_simulations}).  

\subsection{Vacuum instrumentation and operation}\label{sec:vacuum_instrumentation}

Ionization vacuum gauges and ion pumps will be used as primary vacuum signals for monitoring of the vacuum system performance.  Residual gas analyzers (RGAs) will also installed in strategic location for vacuum system diagnostics and in-situ trouble-shooting.   Vacuum system inter-lock based on combinations of ion gauges, ion pumps and low vacuum gauges (such as Pirani gauges) will be implemented to protect critical accelerator components, such as the DC photo-cathode electron gun, ICM and MLC, etc.

\begin{landscape}
\begin{figure}[tb]
\centering
\includegraphics[width=1.4\textwidth]{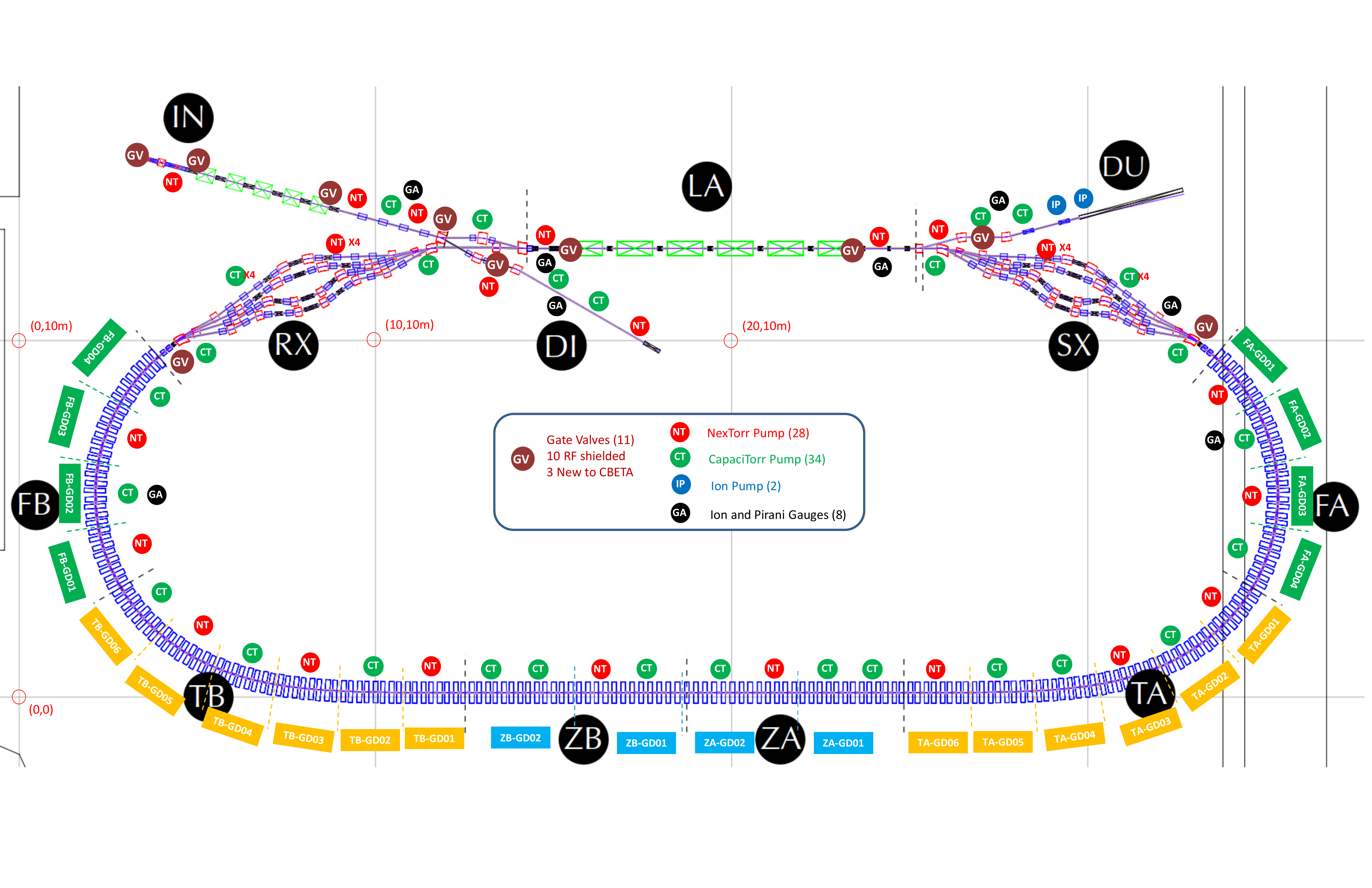}
\caption{A proposed distribution of CBETA vacuum pump, gauges, and gate valves.  Also shown are labels of FFAG girders, such as Girder 01 in FA section: FA-GD01. }
\label{fig:vac_instr}
\end{figure}
\end{landscape}

One proposed distribution of vacuum pumps and instrumentation is given in  \Fig{fig:vac_instr}.

\section{Vacuum pumping and performance simulations}\label{sec:vacuum_pumping_simulations}

With highest beam energy of 200~MeV, synchrotron radiation induced gas desorption from the beam pipe wall is negligible.  Therefore thermal outgassing is the only source of gas in the beam pipes.  With pre-installation bake-out of all beam pipes, and venting/purging with chemically filtered nitrogen, low thermal outgassing rate ($ < 1\times10^{-9}$ torr-l/s-cm$^2$) can be achieved within 24-hour of pump-down, and continuing decrease with time as $\dot{q} = q_i t^{-\alpha_i}$  (with $\alpha_i \approx 1$).

\begin{figure}[tb]
\centering
\subfloat{\includegraphics[width=0.45\textwidth]{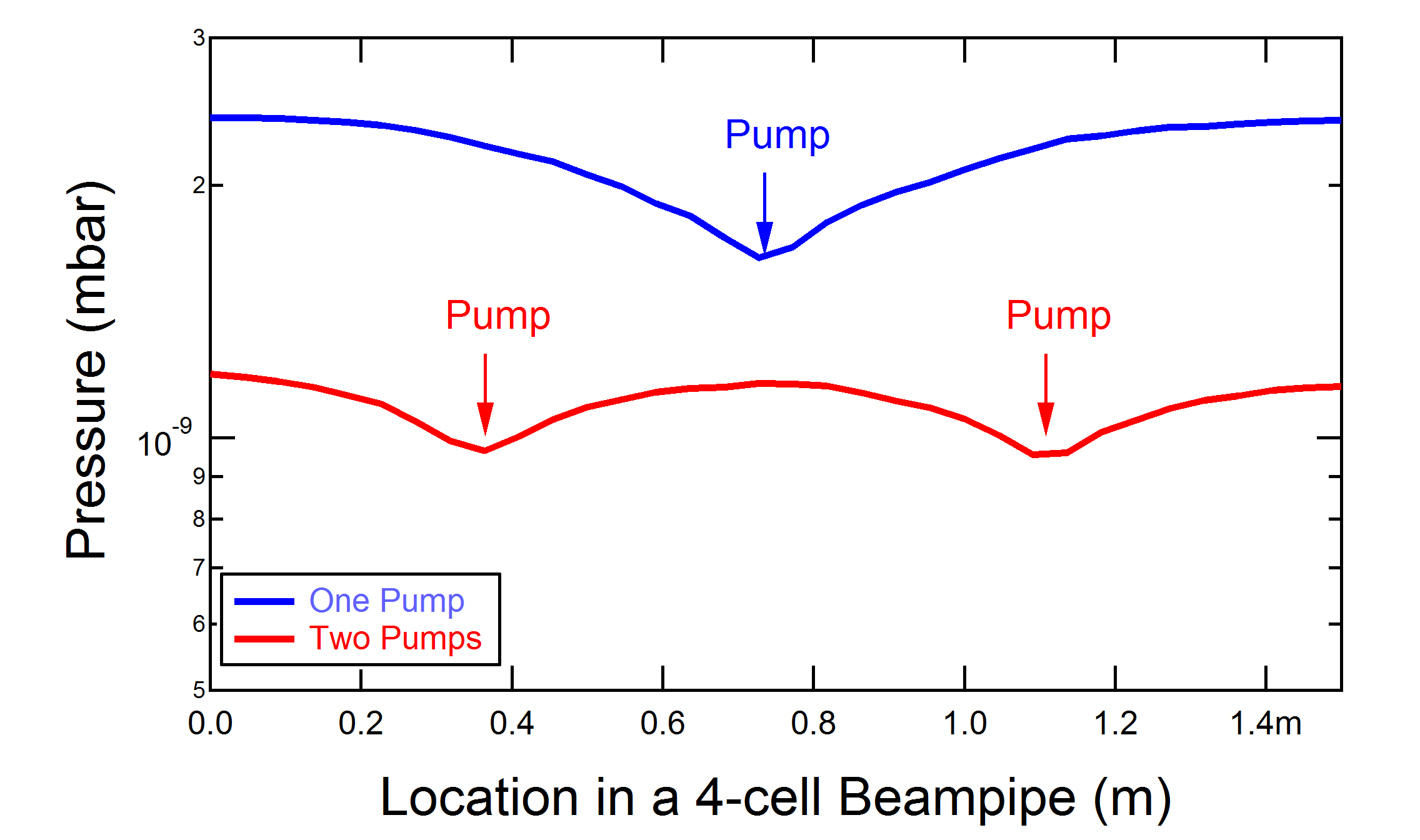}\label{fig:pumping_calcs1}}
\hspace{0.05\textwidth}
\subfloat{\includegraphics[width=0.45\textwidth]{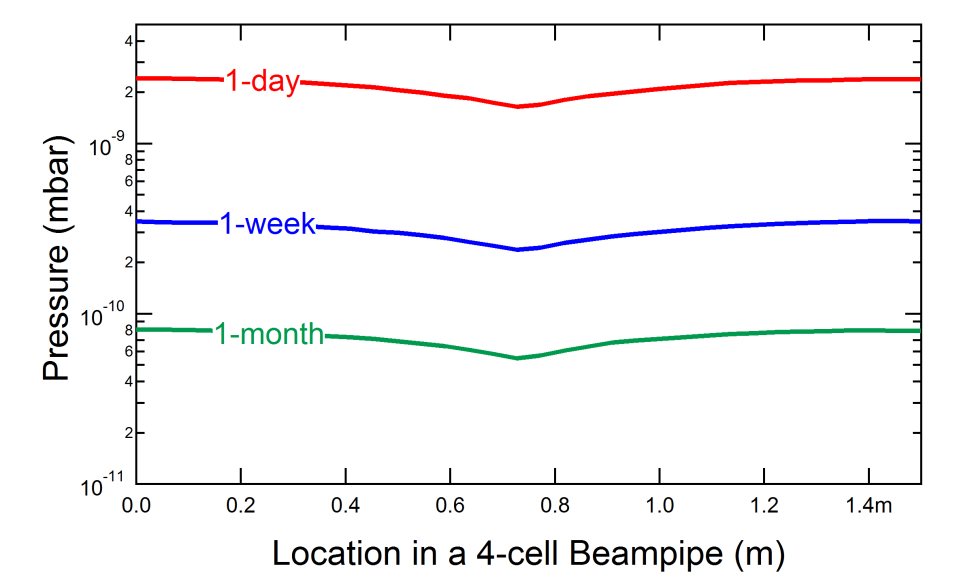}\label{fig:pumping_calcs2}}

\caption{a) Simulated pressure profiles in a 4-cell FFAG beam pipe, comparing one-pump and two-pump per beam pipe configurations and b) Simulated vacuum improvement with time for a 4-cell FFAG beam pipe in one-pump configuration.}

\end{figure}

Vacuum system performance design will be aided by 3D vacuum simulations, using a 3D tracking program, MolFlow+ \Ref{bib:molflow}.  As examples, \Fig{fig:pumping_calcs1} compares simulated pressure profiles with two different pumping configurations in a 4-cell FFAG beam pipe, and the results showed one pump per 4-cell beam pipe is sufficient.  \Figure{fig:pumping_calcs2} demonstrates the continuing improvement over time for a 4-cell FFAG beam pipe with one pump per cell.  Vacuum simulations will be carried out for all the CBETA sections as more engineering details of the beam pipes becoming available.

\ifdefined \buildingFullDocument

\renewcommand{\FiguresDirectory}{diagnostic_and_control/figures}

\else
\newcommand{\FullDocumentRoot}{..}
\newcommand{\FiguresDirectory}{figures}

\begin{document}
\fi

\chapter{Diagnostics and Control\Leader{Dobbins}\label{chapter:diagnostics_and_control}}

\section{Introduction}

The salient feature of CBETA for diagnostics is the presence of multiple beams which need to be individually diagnosed and controlled. The following sections describe the components of the beam diagnostics and control system with particular attention given to how these components will handle multiple beams.

\section{Beam position measurement system}

The transverse beam size in the CBETA FFAG arcs is measured in tens of microns. Lattice errors in the arcs that are a fraction of the beam size can lead to orbit oscillations which will degrade the beam quality. Somewhat larger lattice errors can lead to beam loss. Commissioning and operation of CBETA will require the ability to measure and correct orbits at the micron level. While these tolerances are tight, they are within the state of the art in presently operating accelerators. CBETA presents the additional requirement of measuring the orbit of multiple beams simultaneously.  It will also support multiple operating modes, as described in section \ref{sect:bunch-patterns}.

The proposed CBETA bunch patterns have in common the presence of a probe bunch separated in time from other bunches. The position of the probe bunch is observed for each energy pass at each BPM. We will be relying on the probe bunch to be an accurate representative of the other bunches in the loop. The BPM signal processing will be a time-domain based system (as opposed to an RF system) in order to distinguish closely spaced bunches. The minimum useful separation in time of the probe bunch from other bunches is determined by the capabilities of the BPM electronics as discussed below. The planned separation in time is 9.5 RF cycles at 1300 MHz, approximately 7.3 nS.

In the FFAG arc the BPMs must provide accurate position information for multiple orbits over a large horizontal aperture, approximately  +/- 21 mm from the nominal center. 

\subsection{BPM Pick-ups}

The BPM pick-ups will be 18 mm diameter buttons in a round beam-pipe of inner diameter 70 mm. Buttons will be rotated 45$^{\circ}$ from horizontal/vertical axes to avoid synchrotron radiation.

\subsection{BPM Electronics}

The electronic design concept is to use BNL designed V301 modules. Each V301 module has four 400 MSPS analog to digital converters (ADC).

\begin{figure}[tb]
\centering
\includegraphics[width=0.95\textwidth]{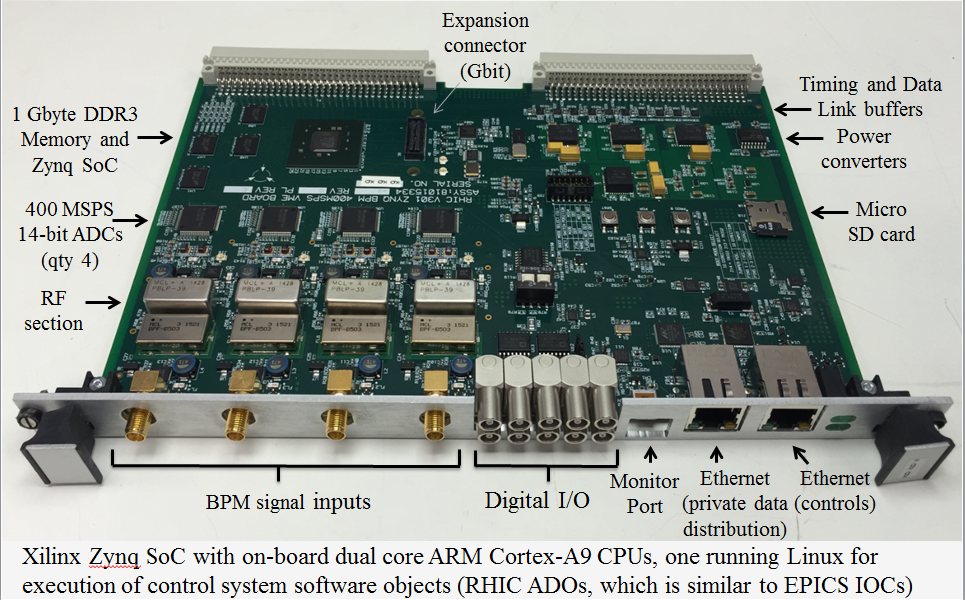}
\caption[]{BNL V301 BPM electronics module}
\label{fig:BNL_V301}
\end{figure}

In order to distinctly measure each beam, sufficient longitudinal separation between associated probe bunches is necessary.  Probe bunches will be separated by approximately 7.3 ns. The ADC clock will be synchronized to the beam and generate the ADC trigger on the peak of the bunch signal of interest.  A low pass filter will be used to broaden the pulse width, but any ringing of the signal must settle to zero before the next bunch arrives.  
    
Configuring the ADC timing to the peak of the selected bunch is not trivial.  A low jitter beam synchronous clock will be required, and continuous tracking of the peak will be needed. Timing for each accelerating/decelerating energy beam will be different and it may be possible to measure only a single accelerating or decelerating energy beam at any given time.  Selection of the specific beam for measurement and automatic coordination of timing changes will be required.  Automatic sequencing of beam selection may be possible.

Initial testing of the V301 module with beam at Cornell has demonstrated the viability of this approach. The V301 module has been successfully clocked from the existing timing system and used to acquire BPM signals from the beam in the CBETA injector.

\section{Bunch Arrival Monitors}

Operation of CBETA requires measurement and control of phase of the beam arrival at the linac relative to the RF.

In previous work frequency domain beam monitors have been developed for the injector which provide accurate beam phase information. These monitors achieved a phase accuracy of 0.1$^{\circ}$ under typical operation conditions. The monitors compare the 1.3 GHz frequency component of the beam signal to the RF master oscillator signal. Similar monitors can function in the presence of two beams by measuring both the 1.3 GHz and 2.6 GHz frequency components of the beam signal. Placing these monitors near the end of end of the splitters (where there are only two beams) just before the linac will provide the required bunch arrival time information.

The bunch arrival time is controlled via adjustment of the path length through the splitters.

\section{Beam Size}

View screens with resolution of about 30 um will be used to obtain the beam profile for low beam currents. While the beam size in the FFAG arcs is comparable to this the screens may still be useful for detecting conditions which cause the beam size to grow. In the splitters the horizontal beam size will be larger due to dispersions and view screens will then provide useful quantitative measurements.

Two different screen materials will be used to cover several orders of magnitude in beam current: BeO that is sensitive to sub-nano-Ampere average currents, and less sensitive CVD diamond screens that can take up to 1 $\mu$A of beam current. RF shielded assemblies will be employed throughout when the screens are retracted to minimize wakefields and and heating effects.

View screens are a destructive monitor, observation of the beam results in elimination of all subsequent passes. For this reason one may not observe decelerating beams in CBETA on view screens, i.e. a screen inserted to intercept a decelerating beam will first intercept the accelerating beam. Additionally, in the FFAG arcs the lowest and second energy beam have very little horizontal separation, hence the view screens will not practically allow observation of the second energy energy beam. Therefore the view screens in the FFAG  are intended for first pass commissioning while learning to use the BPMs. In the splitters the view screens will be useful for all accelerating beams.

\section{Beam Loss Monitors}

Beam losses in an accelerator with continuous injection can contain high power. A high energy, single point loss of a 20 um size electron beam of 40 mA is capable of burning through a 3 mm thick aluminum beam pipe in as as little as a few uS \Ref{Wesch}. Note that the energy deposited in the beam pipe does not strongly depend on the beam energy, high energy electrons will mostly pass through the beam pipe as "minimum ionizing particles" with an energy deposit per unit path length which rises only slightly with beam energy. The thermal relaxation time for aluminum is approximately 65 uS.  For beam loss events with duration less than this thermal conduction does not play a significant role. For such events beam loss can be described in terms of beam current lost times the duration. The short duration, single point loss damage threshold is approximately $10^{-7}$ A*s.

For lower levels of beam loss thermal conduction becomes significant. Thermal simulations using Ansys indicate that a continuous single point loss of 1 $\mu$A will produce a maximum temperature in the beam pipe below 150 degrees C, a value considered safe for mechanical integrity.

The radiation associated with partial and distributed beam losses can also make operation of the accelerator problematic, limiting energy recovery and presenting a radiation hazard to nearby equipment. Thus beam loss monitoring is required for both machine protection and informing operations. A beam loss monitor system should be capable of detecting beam loss approaching the damage threshold with a sub uS response time and should also provide information at lower loss levels on a longer time scale. 

For beam loss monitors to be useful they should:
\begin{enumerate}
  \item Provide comprehensive coverage, i.e. beam loss should be detected wherever it occurs.
  \item There should be at least a crude relation between detected radiation and beam loss.
\end{enumerate}

The use of scintillating optical fiber is being evaluated for detecting radiation associated with beam loss in CBETA.  It provides a cost effective way of providing comprehensive coverage. We have used Geant4 software to simulate the radiation pattern produced by beam loss in the FFAG sections of CBETA. In particular we have investigated how the detected radiation depends on the loss point, varying the azimuthal angle, the angle of incidence relative to the beam pipe, and the z-position with respect to the magnet period. Saint-Gobain scintillating fiber BCF-60 has a published sensitivity of approximately 7000 photons per MeV of absorbed radiation. A small fraction, perhaps 2 percent, of these photons can be collected at the end of the fiber. Note that the useful fiber length is limited by the self-absorption attenuation length of 3.5 meters. These numbers combined with the results of the simulation show that such a fiber running parallel to the beam line just outside the FFAG magnets could be an effective beam loss monitor. Beam test are planned to verify this performance. The attenuation length of scintillating plastic fiber becomes shorter when exposed to radiation, doses measured in tens of kRad being significant. There are locations in CBETA where the accumulated radiation dose could affect the performance of such fibers. As the fibers are a small part of the system cost one could envision replacing damaged fibers on a limited basis. Another alternative to explore is using optical fibers collecting Cerenkov light as radiation monitors. They offer the potential for much greater radiation hardness though the signal is orders of magnitude lower. Beam tests of such fibers are also planned.

\begin{figure}[tb]
\centering
\includegraphics[width=0.95\textwidth]{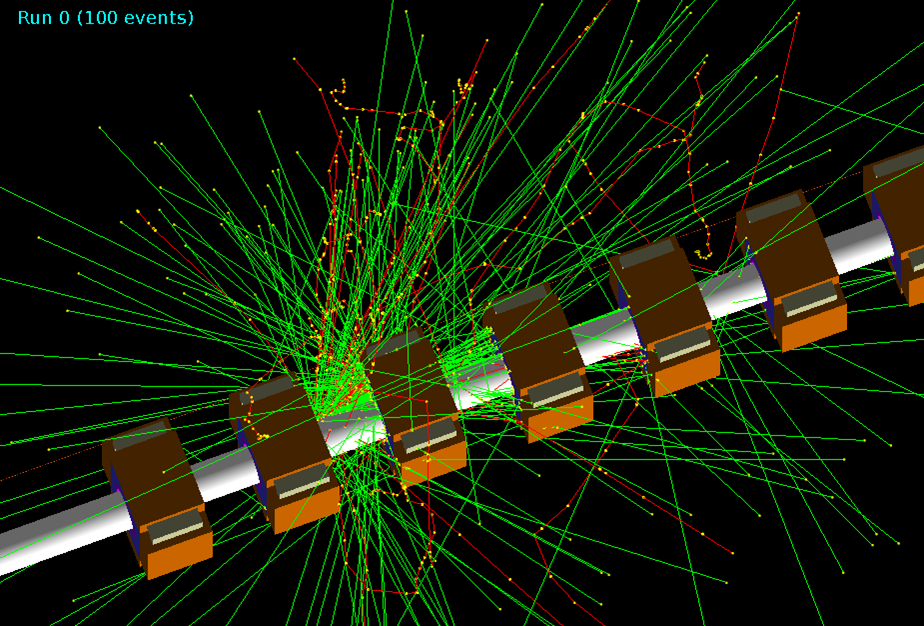}
\caption[]{Geant4 simulation of beam loss in section of Halbach magnets}
\label{fig:Geant4 Simulation of Beam Loss}
\end{figure}

In the injector and splitter sections of CBETA strong magnets of wide field range provide many possible modes of beam loss. Loss simulations using Geant4 will also be performed for these areas. It is expected that a combination of distributed scintillating fiber and single point scintillation detectors will provide effective beam loss monitoring for these sections.

Accumulated dose monitoring of radiation due to beam loss will also be implemented for equipment and permanent magnets as necessary. Relatively inexpensive monitors similar to film badges are available which are capable of measuring kRad doses.

Beam current measurements comparing the current out of the injector to the current delivered to the dump can provide a measure of the total beam loss around the ring. The usefulness of such a measurement depends on:
\begin{enumerate}
  \item The accuracy with which one can cross calibrate multiple beam current monitors
  \item The dependence of the measured current on the transverse beam position
  \item The response time of the monitor
\end{enumerate}
We are investigating multiple options for implementing beam current monitors including resistive wall current monitors, beam current transformers, and beam position monitor based current measurement. A differential current measurement, injector current minus current to dump, might achieve an accuracy of a fraction of one percent. This would be valuable as a supplement to radiation detector based beam loss monitoring, especially at high beam currents.

\section{Machine Protection System}

The operation of CBETA presents various hazards to the accelerator equipment itself. The machine protection system for CBETA will be an extension of the existing injector machine protection system. Some of these hazards are specific to beam operations while others are not. Certain sub-systems such as cryogenics and RF include their own monitoring to insure safe operation independent of beam operations. These protections are implemented as a combination of dedicated hardware and programmable logics controllers, PLCs.

The additional hazards of beam operation are handled via a permit system for the laser which controls redundant shutters between the laser and the gun photo-cathode. An equipment ready chain which include inputs from:
\begin{enumerate}
  \item Gun (high voltage power, SF6)
  \item Vacuum System (pressures, gate valve states)
  \item Magnets and magnet power supplies
  \item RF system
  \item Cryogenics System
  \item Dump (temperature, cooling water)
\end{enumerate}
must be complete before the laser shutters can be opened. After the laser shutters are open the fast shut-down system system stands ready to close the shutters. The fast shut-down system provides a means of closing the laser shutter with sub uS response time. It has inputs from the beam loss monitor system as well other fast inputs reflecting, for example, the state of the gun high voltage and the accelerating cavity fields.

\section{Integrated Control System}

The Control System for CBETA will extend the existing EPICS based controls at Cornell. EPICS provides a suite of software tools for data acquisition and control, operator displays, alarm state notification, data archiving, and more. EPICS  is a demonstrated and scalable controls platform supported by an large active open community of programmers and users.







\ifdefined \buildingFullDocument

\renewcommand{\FiguresDirectory}{safety/figures}

\else
\newcommand{\FullDocumentRoot}{..}
\newcommand{\FiguresDirectory}{figures}

\begin{document}
\fi

\chapter{Personnel Safety\Leader{Heltsley}}\label{chapter:safety}

\section{General considerations}
 
   All CBETA-related activities will be organized and operated within CLASSE and
   its safety protocols. CLASSE holds no higher priorities than ensuring the
   health and safety of all. These values have been woven into the fabric of
   laboratory administration and operation. Synergistic relationships with
   CLASSE and Cornell University provide important policy guidance,
   institutional support, and oversight. The CLASSE approach to workplace safety
   is built around three overlapping commitments: to continuously provide a \textit{safe laboratory environment}, to engender an abiding \textit{culture of safety}
   in all personnel, and to address and anticipate safety challenges with \textit{proactive management}. 
   
   The first line of defense against potential hazards is a \textit{safe
   laboratory environment}. Exterior doors are locked outside of business hours;
   entry at off-hours is by keycard access or explicit permission of a staff
   member. Fire alarms are tied into the University centralized systems. State
   fire officials conduct inspections of the entire laboratory on an annual
   basis. Designated staff members are trained for specific roles in emergency
   situations. Only trained and/or licensed personnel operate industrial
   equipment, such as cranes, forklifts, and large vehicles. Machine tools are
   periodically inspected for correct operation and presence of appropriate
   guards. A spill control plan is in place for oil-filled transformers. An
   arc-flash hazard study of laboratory high-voltage AC distribution panels is
   complete and related needs implemented. A lock/tag/verify program is in place
   to cover work near equipment with remote power control, as is a policy
   governing hot work and welding. Personal protective equipment and safety
   training specific to their tasks is made available to workers who need
   it. Fume hoods for handling chemical samples are used, and detailed written
   safety procedures exist for hazardous tasks. Safety Data Sheets are stored in notebooks near where the hazardous substances are used. Cryogenic installations and transfer lines are protected with sufficient insulation, redundant overpressure protections, oxygen-deficiency measurements where useful, and a multitude of instrumentation monitoring details of operation.
   
   Radiation safety at CLASSE is regulated by Cornell's Environmental Health and
   Safety policies and authorized via a permit system, which are, in turn,
   governed by the New York State Department of Health (NY is an NRC-agreement
   state, and the regulations are embodied in 10 NYCRR Part
   16 \Ref{bib:NYCRR}). Cornell University is licensed by New York State to
   internally regulate radioactive material (RAM) and radiation-producing
   equipment (RPE), and has its own Radiation Safety Manual \Ref{bib:NYRadiationSafetyManual}, Radiation Safety Officer (RSO), and Radiation Safety Committee (RSC). 
   
   Mitigation of radiation hazards from RPE is
   dealt with via redundant engineering controls and administrative controls as
   well. Permanent shielding, generally consisting of concrete, lead, and/or
   iron, surrounds all RPE so as to restrict potential exposure to personnel
   in public areas to below 2 mrem in one hour or 100 mrem in one
   year. Locations just outside the shielding where radiation dose rates are
   expected to be below those listed above but which are considered potentially
   vulnerable to higher levels, are designated as \textit{controlled areas}, 
   in accordance with Cornell University policy. Access to controlled areas is
   restricted to authorized personnel wearing radiation badges or those
   accompanying a CLASSE host with a real-time-readout dosimeter. Entrances to
   controlled areas are clearly signed in accordance with regulations. \textit{Exclusion areas}, inside which personnel
   are not permitted be present during RPE operation, are protected by more
   sophisticated access controls: all entryways are equipped with interlocked
   gates and/or light beams that, if tripped during RPE operation, cut power to
   the RPE and cause audible and visible alarms. Radiation detectors monitor the
   radiation in controlled areas, and trip off the RPE if conservative
   thresholds are exceeded. Exclusion area interlocks cannot be set until a full
   in-person search has been conducted; the integrity of interlock  operation is
   verified by periodic operational tests of interlock components. Emergency shutoff controls are
   present inside exclusion areas for use in the unlikely case that someone
   remains inside when RPE operation is about to commence.
   
   CLASSE seeks to establish and maintain a \textit{culture of safety}, which entails much more than simple compliance with a set of rules. A culture of safety is embodied by: each of us taking responsibility for our own safety and that of people we work with, supervise, or host; safety being valued on par with scientific achievement and/or task completion; safety concerns always being taken seriously and promptly addressed; safety challenges being approached with intellectual rigor; new activities being planned from the start with safety in mind; new participants receiving relevant safety training immediately. Such practices are self-reinforcing, but can be undermined by even occasional lapses, so considerable vigilance on the part of supervisory personnel is required.
   
   \textit{Proactive management} ensures that: specific safety responsibilities of each staff member, student, user, or visitor are clearly delineated and communicated; appropriate training and resources are provided to those who need it; mechanisms are in place to maintain accountability and establish and publicize appropriate safety-related policies; compliance with relevant University and governmental safety and environmental regulations and ordinances is attained; and intra-university resources are leveraged when helpful. 
   
   CLASSE has an extensive online Safety
   Handbook \Ref{bib:CLASSESafetyHandbook}. A central safety document database
   has been implemented (using the CERN EDMS system) and is home to procedures,
   radiation permit applications, meeting minutes, internal incident reports,
   and more. Conversion to a University-wide Learning Management System (LMS)
   occurred in November 2016. The new LMS uses a Saba cloud-hosted and
   browser-based solution known as CULearn that manages courses, classes,
   learner transcripts, and web-based content and assessments, allowing learners
   and administrators alike access to what they need to achieve safety
   objectives. 
 
   Clear lines of accountability for performance related to safety have been shown to be crucial to superior safety achievement, especially in academic research settings. The CLASSE Safety Committee and CLASSE Safety Director, which set, communicate, and implement laboratory safety policy, speak and act with the imprimatur of the CLASSE Laboratory Director, who appoints both. Each staff member is accountable to a supervisor, and each student to an advisor.

\section{Accident Rate and Training Compliance}

\begin{figure}[tb]
\centering
\includegraphics[width=\textwidth]{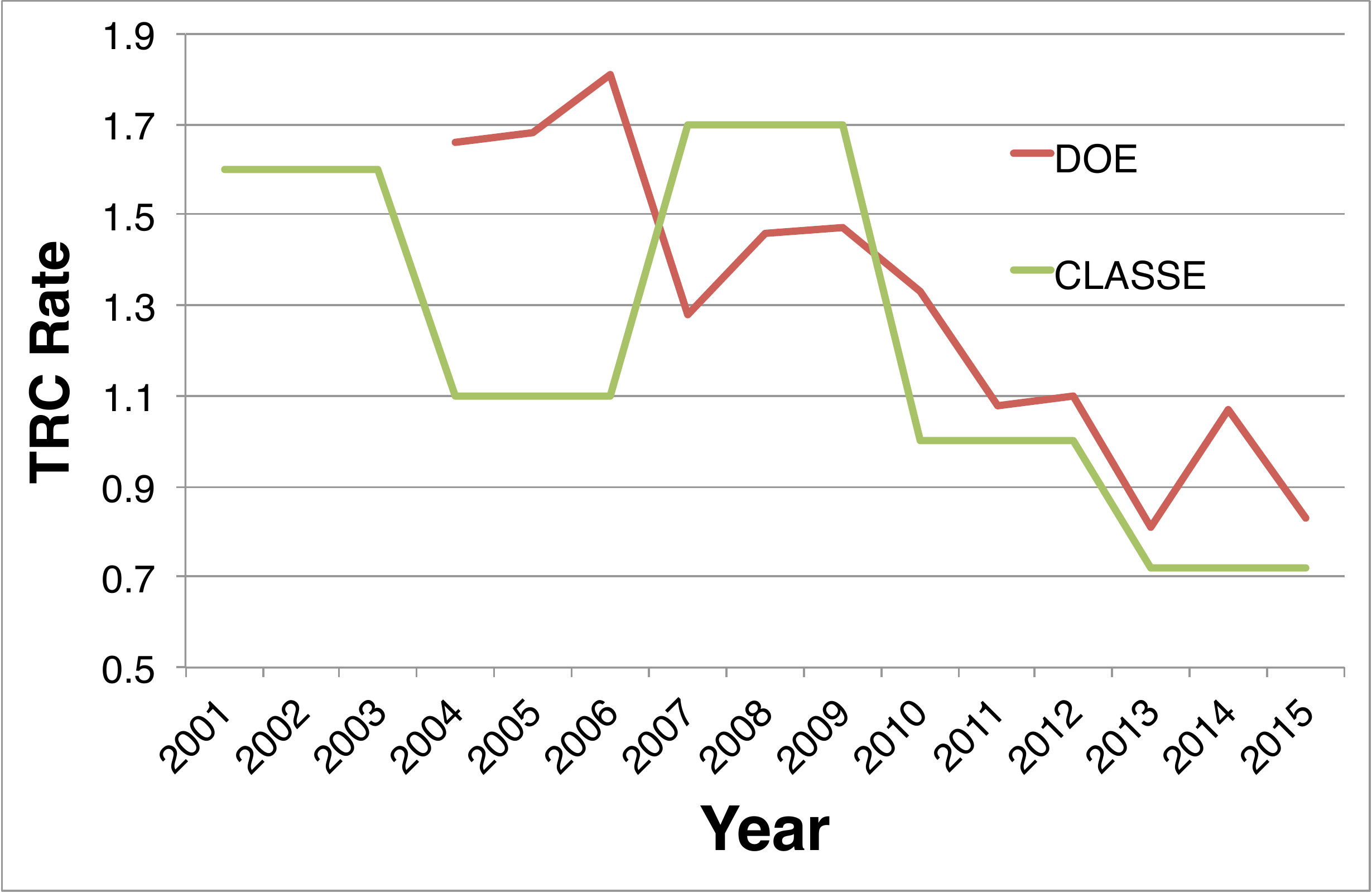}
\caption[TRCRate]{Accident rate at CLASSE and DOE-Lab-average vs. time. The TRC
   rate counts accidents that result in either lost time or treatment beyond
   first aid, or both, per 100 workers per year.}
\label{fig:TRCRate}
\end{figure}

   Cornell University tracks all accidents, injuries and exposures for the
   campus as a whole, and CLASSE tracks its approximately 300 workers
   separately. One metric that allows for comparison with other institutions is
   that of OSHA's \textit{Total Recordable Cases} (TRC), defined as the number of
   accidents, injuries, or exposures that results in treatment beyond first aid
   or lost work time, per 100 workers per year. Campus-wide the TRC rate has
   hovered just above 2 for many years, in line with other colleges and
   Universities. CLASSE has maintained, for its $\sim$300 personnel, a TRC rate below 2.0 for 15 years 
   (as long as it has been tracked), and below 1.0 since 2013, comparable to 
   the DOE-lab-average trend over 15 years, as seen in \Fig{fig:TRCRate}.

   CLASSE currently administers 36 internal safety trainings, each
   specific to a particular hazard at the Laboratory. Cornell University
   Environmental Health and Safety also provides a large assortment of general
   safety trainings, of which about two dozen are taken by CLASSE
   personnel. Most such trainings are recurring (meaning the training must be
   retaken periodically), with recurrence times ranging from one to five years
   (comporting with state and/or federal regulations when applicable). Many
   trainings are in-person instructor-led trainings (ILT) and many are web-based
   trainings (WBT). The new LMS has eased access to these trainings while
   minimizing the administrative burden. The learning plan of each worker is
   populated with courses appropriate to her duties: the average CLASSE worker
   has about eight such courses on their learning plan, with large variations
   from one person to the next, depending on tasks performed. In recent years we have
   tracked compliance on a person-trainings basis (treating each course assigned
   to any worker as the countable item), and on a people basis (treating each
   worker as 100\% compliant with all required trainings or not as the countable
   item). Since tracking these metrics began, compliance has improved
   near 98\% and 88\%, respectively.
   Due to recurring
   courses, individuals can cycle out of compliance when a course expires but
   regain it by refreshing the training, so the individuals out of compliance
   are a constantly churning group of different people temporarily in that
   state. New workers are not permitted to engage in hazardous activities until
   relevant training is completed.

\section{CBETA-specific issues}

\subsection{New collaborators}
   
   CLASSE already has a large variety of personnel using its facilities:
   technical staff, faculty, postdocs, undergraduate and graduate students,
   Cornell and non-Cornell CHESS users, and visiting scientists. New
   onsite CBETA collaborators will be subject to the same requirements as
   everyone else with regard to training and compliance. We have already
   provided for non-Cornellian access to buildings and to required trainings;
   training will be tracked in CULearn.

\subsection{Cryogenic safety}

  The CBETA cryogenic plant will be part of the larger CLASSE central
  facilities, which already satisfy CESR and existing ERL/MLC operations.
  Liquid helium and liquid nitrogen lines and volumes are designed to avoid
  trapped volumes and equipped with overpressure relief and backup burst disks,
  as elsewhere in the Laboratory. Particular attention will have to be paid to
  worst-credible-case accidents inside a shielding tunnel, should one be constructed.

\subsection{Electrical safety}

  The electrical plant and procedures is (as are all stored-energy locations) required to be compliant with the CLASSE
  Lock/Tag/Verify program, requiring individual-worker lockout of
  remotely-powered components prior to work or near on such components.

  All electrical lines are designed and built with best-practice criteria,
  including separation of high-voltage and low-voltage cable/conduit
  runs. Remotely powered equipment will be serviced with OSHA-compliant
  lock/tag/verify capabilities and procedures.

\subsection{Permanent Magnet Safety}

  Delivery of permanent magnets to Cornell must be preceeded by a
  written Safety Plan submitted to the CLASSE Safety Committee 
  describing hazards posed by the Halbach magnets,
  their safe handling, storage, assembly, and installation.

  Administrative measures related to safety in the vicinity of the CBETA
  permanent magnets will have to be established. At minimum will be adequate
  signage, clear notification and/or delineation of the 5~Gauss lines. 
  Iron-containing tools and equipment will be prohibited
  from the vicinity of the Halbach magnets.

\subsection{Equipment for personnel protection system}

At this time, we have specified eight pairs of gamma (Canberra Gamma GP 100) and
neutron (Canberra Neutron NP100H, with a helium gas detector) probes for
the CBETA personnel protection system, interfaced to the interlock system and analog diagnostics through four Canberra iR7040
intelligent ratemeters (newer than the Canberra ADM616 model in use elsewhere
in the laboratory). Each iR7040 can handle up to four total probes (either gamma
or neutron). There will likely be two entry/exit points to the exclusion area,
each equipped with gates and light-beam interlocks. All personnel safety wiring
is restricted to run through dedicated, labeled metal conduit.

\subsection{Radiation safety}
   The ERL linac, including its gun, accelerating components (ICM), and beam
   dump(s), as well as the CBETA SRF MLC have already been operating safely and
   compliantly in the L0E high-bay area at Wilson Laboratory for several months;
   aside from the MLC, those components operated similarly in the adjacent L0
   high-bay area for up to a decade prior. A Cornell University RPE permit is in force, and will be amended as upgrades toward CBETA progress. The current layout of shielding blocks, actively interlocked radiation monitors, area dosimeters, and entry interlocks will all be modified, in stages, as the project progresses.

  Going forward, personnel safety systems will be designed using similar criteria and equipment
  as for other facilities at CLASSE, meeting University and NY State standards.
  CLASSE has experience shielding electron beams at CESR, which have higher
  energy (5.3 GeV) and comparable currents ($\sim$100 mA). The CBETA location in
  L0E is well-shielded on the north, east, and half of the south sides by
  concrete walls and a high ground level outside. However, as L0E is adjacent to
  other laboratory activities, particular the loading area in the southwest
  corner and the routinely occupied CHESS user areas on the west end, shielding
  must guarantee safe radiation levels in those areas as well as keep locations
  outside the building below radiation levels for a public area.

  Required equipment are several gamma and neutron radiation probes attached to
  their interlocked readout electronics. These will be strategically placed
  outside the exclusion area at
  areas particularly prone to radiation leakage. Typically initial positions are
  selected for the probes but enough extra cable length left in place so that
  their exact locations may move by 10-15 feet. These will be interlocked so as
  to trip off the RPE at instantaneous levels of 2~mrem/h and monitored so as to
  keep average levels below 0.05~mrem/h.

  Baseline shielding is being designed
  with the largest credible beam loss scenario in mind. The simulation
  tool MCNP6 is being employed for studying details of this matter. 
  The laboratory standard for
  shielding blocks are iron-loaded
  concrete with dimensions of $2'\times4'\times4'$, which are generally stacked
  into walls. A shielding roof for at least portions of the CBETA ring is under consideration. The
  baseline plan is to enclose CBETA on the south and west sides with a shielding
  wall of at least two-foot thickness (iron-loaded concrete) that is at least
  eight feet tall. Additional shielding will be added to the plan as studies
  show it to be necessary; these may include covering the ring with a concrete
  plus steel roof, and/or a 12-foot shielding height on the southwest corner,
  and/or extra shielding around merger/splitter areas and/or the
  MLC. Additional localized radiation
sources may require lead and or concrete shielding to be put in place.

  Our experience with new RPE tells us that administrative limits on 
  CBETA operating parameters will evolve with shielding
  configuration so as to maintain safe and compliant radiation levels outside
  the shielding wall. The basic lesson is that not all radiation sources, or
  especially the total amount of beam losses, can be known beforehand in a frontier
  accelerator. CLASSE has for many years operated the ERL components 
  in sequential, separately approved \textit{stages}: administrative parameters for each stage are set by a
  \textit{Stage Review Committee} upon review of radiation surveys taken during
  low-current and/or after-hours operations. The current and energy operating envelopes
  are permitted to grow as smooth operations and adequate shielding are jointly
  improved.

\begin{figure}[tb]
\centering
\includegraphics[width=\textwidth]{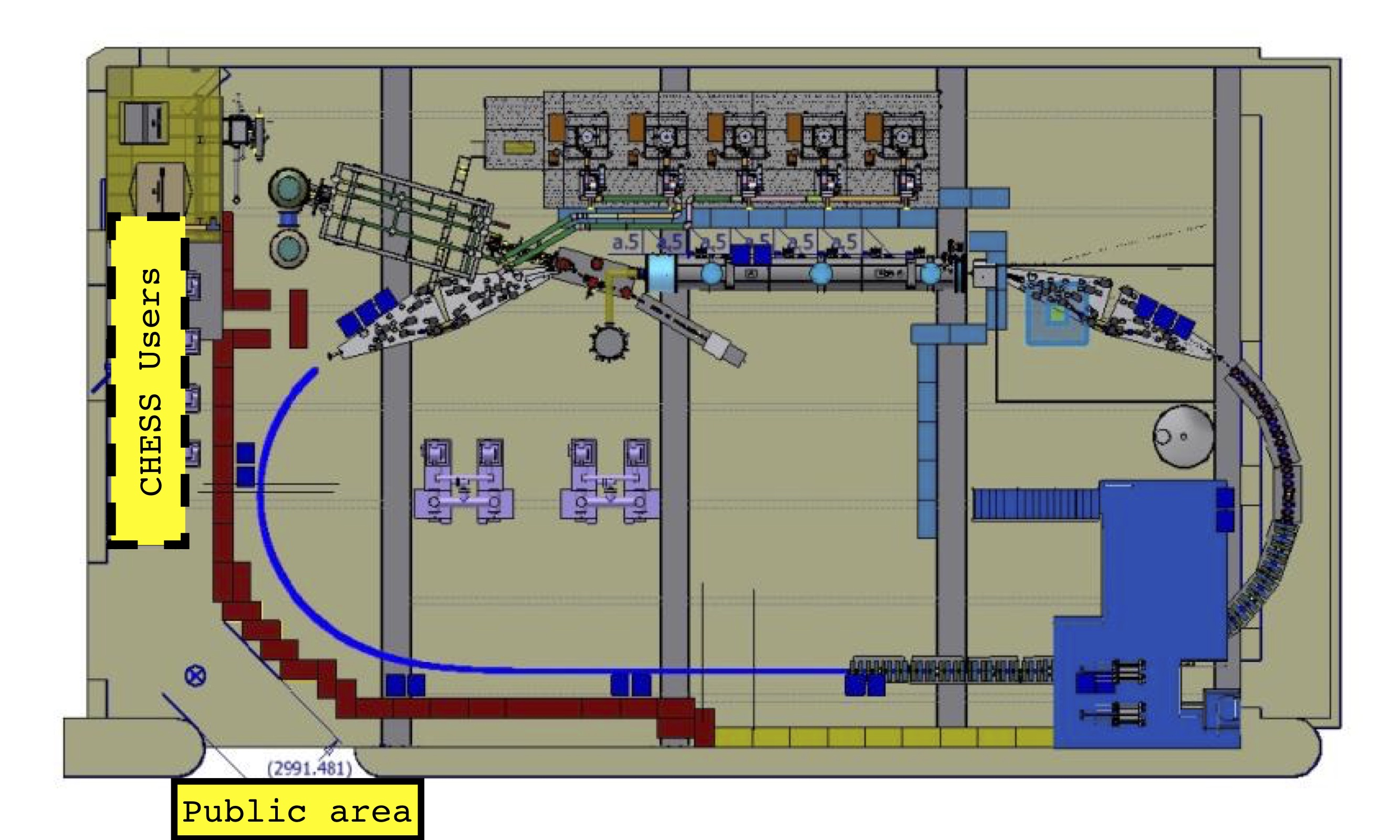}
\caption[Shielding]{A candidate layout for the baseline shielding configuration.}
\label{fig:Shielding}
\end{figure}

Procedures used for testing a new configuration are by now well-established. Generally
taking place after 4:00~pm or on the weekend, the
area is roped off and signed to restrict occupancy to those conducting the
test. The equipment is then brought to the new, as yet unapproved (for daytime operation) state gradually,
and usually at the smallest current or energy that can still yield reliable and
useful data on radiation outside the exclusion area. When stable conditions are
achieved, gamma and neutron radiation levels are manually measured with portable
meters at about ten
pre-agreed locations, plus any additional ones that may be of particular
interest for this particular test (prioritizing possible gaps between shielding
blocks or spots of potentially inadequate shielding). If there is time and radiation levels permit,
another current or energy point may be taken. This information is subsequently presented
by the RPE Permit Holder (presently Prof. Ivan Bazarov)
to the Stage Review Committee with a specific request for extending the envelope
of operations. The committee, composed of the Safety Director, Facility
Engineer, Radiation Safety Specialist, CESR Technical Director, and CHESS Safety
Officer, has in the past responded in several ways. It may recommend additional localized shielding and/or administrative or engineering limits. It
may restrict further operations in the new configuration to evenings or
weekends. It may
simply approve the proposed administrative envelope for routine daytime operations.

\subsection{Baseline shielding plan}

As detailed in the following section, the baseline shielding layout is
still under study. Figure~\Fig{fig:Shielding} shows a preliminate
candidate for such a layout for perimeter shielding.
As mentioned previously, the main
constraints are keeping the both public area at the southwest corner 
and the CHESS User area on the west side below an average dose rate of
50~$\mu$rem/h. 

Operational experience at Jefferson Lab suggests that a reasonable
assumption of continuous losses is $\sim$1~W/m for a 150~MeV, 8~mA
machine. Scaling this fractional beam power loss to the 42~MeV, 1~mA scenario results in 20~W/m
losses; for the full 150~MeV, 40~mA machine, 5~W/m losses would be expected.

The layout
shown should suffice for all operational modes until the final CBETA
milestone approaches. 
The preliminary simulations shown below suggest
that such operation would be safely shielded with the pictured
shielding for a 42~MeV beam with 20~W/m losses. 
Progressing to the full 150~MeV energy and 5~W/m losses would require 
another two feet of heavy concrete shielding along the
southwest corner and west side (each additional foot of thickness
provides a factor of $\sim$25 reduction in dose rate). 
Although space is tight, allocating space for
an additional 2' thickness of 
further shielding in the final CBETA stages is essential. The stage
review process will determine the timing of the shielding additions.

\subsection{Shielding simulations}
\begin{figure}[bt]
\centering
\includegraphics[width=0.8\textwidth]{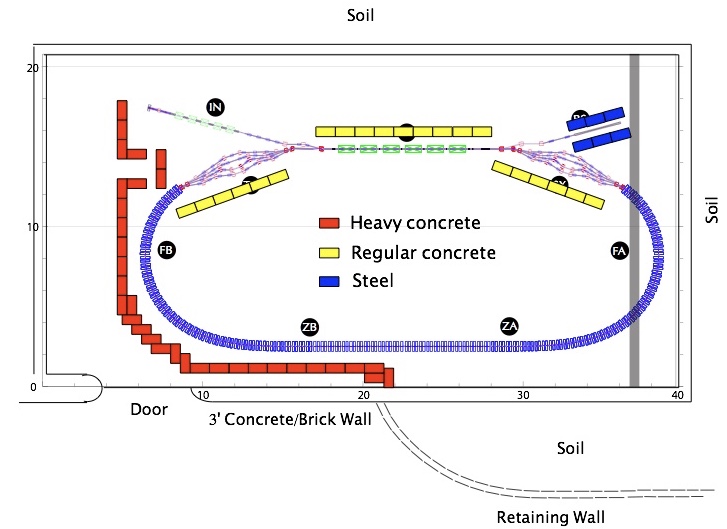}
\caption{Places in L0E that will require additional shielding. }
\label{fig:CBETA_shielding} 
\end{figure}

CBETA is located in L0E hall in Wilson Laboratory. L0E is approximately 124' by 72' with a 48' high ceiling. The roof consists of precast concrete T-sections that are 6' wide. The top is 10.5" thick, the T descender is 3'8" long and 8" thick. The L0E walls are shielded on the North and East by 2' thick concrete walls that are backed by soil. The South wall is 3' thick and partially backed by soil.\footnote{Architect drawings, 053-252-S01 to S03} (L0E originally served as the target room of the Cornell 12 GeV synchrotron, and is quite well shielded.) Figure \ref{fig:CBETA_shielding} shows locations in L0E that will require additional shielding.

In what follows, shielding calculations were carried out using the Los Alamos Monte Carlo code MCNP6~\Ref{MCNP6}.
 In CBETA, the injector  produces a 6 MeV, 40 mA electron beam that is accelerated by the MLC to 42, 78, 114 and 150 MeV and then decelerated four times back to 6 MeV and collected. The power in each of these beams is 0.24, 1.68, 3.12, 4.56 and 6.00 MW. 

A continuous source of radiation, always present during operation, is due to the 6 MeV, 40 mA beam collected by the electron beam dump. In addition to this source, it is possible to lose electrons during normal operation from the beam halo, the Touschek effect and electron-residual gas scattering. 

In addition to these, one has to consider unlikely, but possible scenarios of beam loss that include the following: a) any one of the beams can be fully lost at some point around the ERL, b) if for some reason any one of the beams loses say 1 MeV, then the energy cannot be recovered, and one has beams close to the full power that will be lost at some point and c) each of the six MLC cryomodules is fed 5 kW of rf power, and if half of the power is lost, then 15 kW of beam power will be lost at some point, or along the ERL, i.e. 0.357 mA at 42 MeV, 0.192 mA at 78 MeV, 0.132 mA at 114 MeV and 0.100 mA at 150 MeV.

\subsubsection{Electron beam dump}

\begin{figure}[tb]
\centering
\includegraphics[width=0.8\textwidth]{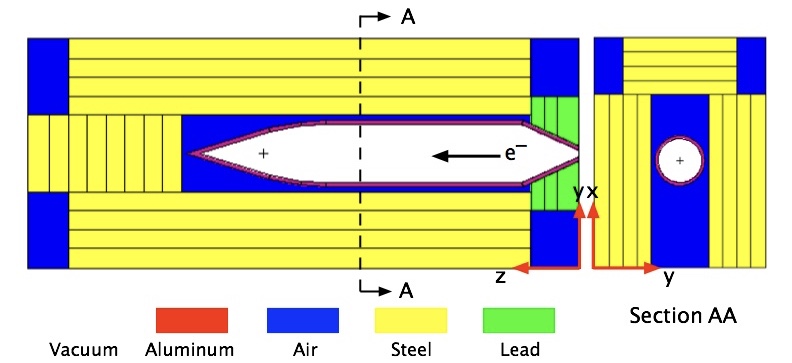}
\caption{MCNP6 input geometry for the CBETA electron beam dump.}
\label{fig:dump_input}
\end{figure}

\begin{figure}[tb]
\centering
\includegraphics[width=0.8\textwidth]{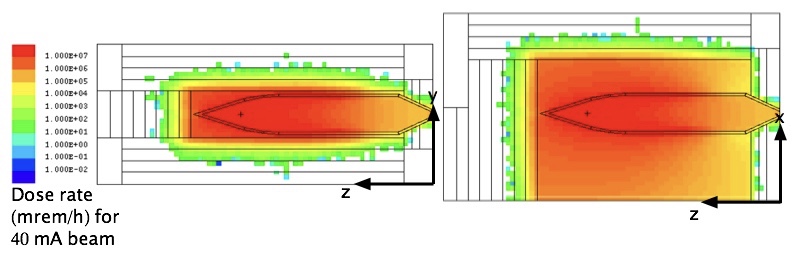}
\caption{Gamma dose rate contours in mrem/h or a 6 MeV, 40 mA electron beam.}
\label{fig:gamma_dump}
\end{figure}

Figure \ref{fig:dump_input} shows the MCNP6 input geometry of the electron beam dump. This geometry is identical to the beam dump shielding present until recently in L0E. The gamma dose rate contours in mrem/h for a 6 MeV, 40 mA electron beam collected by the dump are shown in figure \ref{fig:gamma_dump} 

The 2' thick steel shielding blocks are more than sufficient to shield against the gamma radiation. In this case, the neutron dose rate was not calculated as the 6~MeV electron beam is below the neutron production threshold in aluminum and steel. Due to the distant location of the beam dump and its shielding, the dump is more than adequately shielded.
 
\subsubsection {Continuous electron beam loses}
\begin{figure}[tb]
     \centering
     \subfloat[][Unshielded gamma dose rate at 1 m from CBETA vacuum chamber.]{\includegraphics[width=0.45\textwidth]{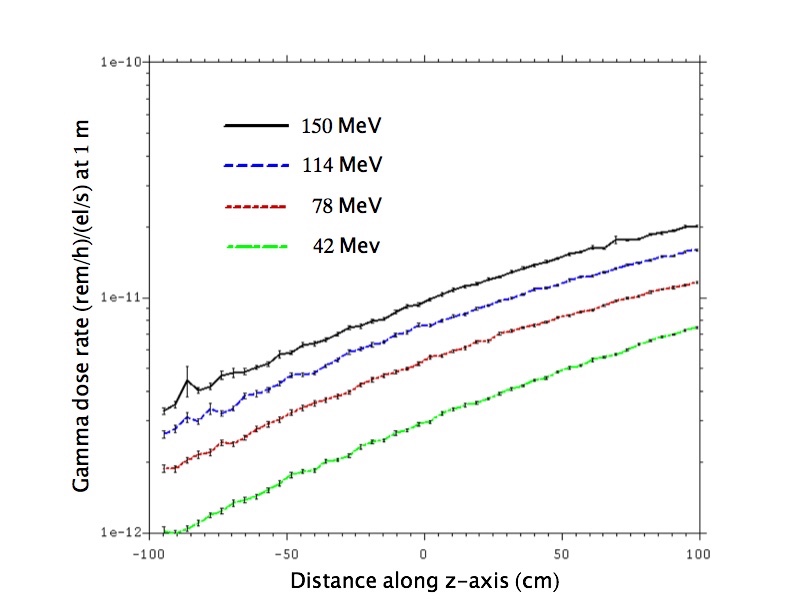}}
     \subfloat[][Unshielded neutron dose rate at 1 m from CBETA vacuum chamber.]{\includegraphics[width=0.45\textwidth]{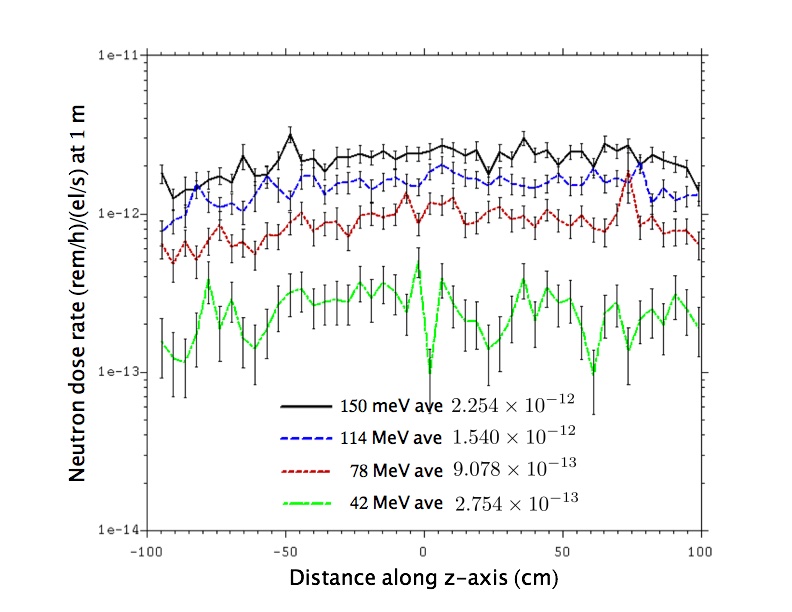}}
     \caption{}
     \label{fig:unshielded_doses}
\end{figure}

Before discussing these loses, it is instructive to consider the contribution of the various energy beams present in CBETA to the gamma and neutron dose rates. \Figure{fig:unshielded_doses} shows the gamma and neutron dose rates due to electrons continuously striking a 2~m long section of the 3~mm thick aluminum wall of the CBETA vacuum chamber at less than a mrad in angle. (To simplify the calculations, the electrons are assumed to originate continuously along and inside the aluminum vacuum chamber.)

\begin{figure}[h]
\centering
\includegraphics[width=0.8\textwidth]{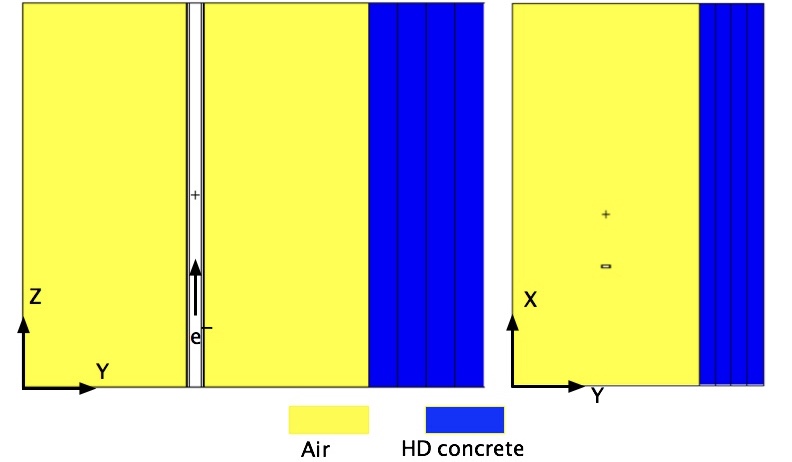}
\caption{Input geometry for a continuously lost electron beam. The HD concrete wall is located 90 cm from the CBETA vacuum chamber.}
\label{fig:MCNP6_geo}
\end{figure}

From the figure, one sees that the maximum dose rates occur, not surprisingly, for the 150~MeV electron beam. Accordingly, calculations will be restricted to this energy. \Figure{fig:MCNP6_geo} shows the input geometry for the following calculations.

\begin{figure}[h]
\centering
\includegraphics[width=0.8\textwidth]{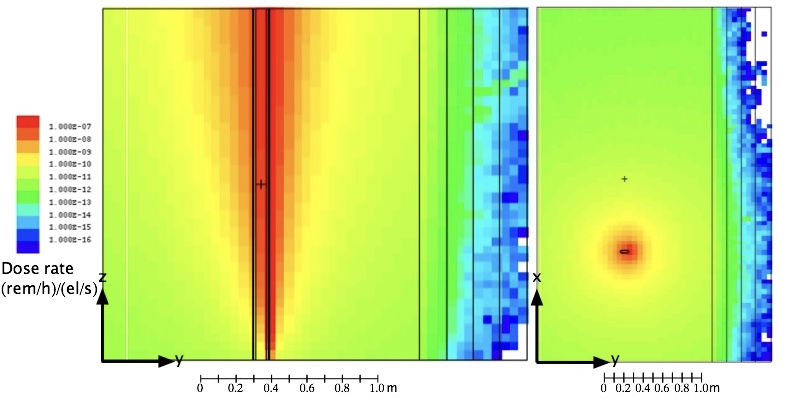}
\caption{Gamma dose rate contours in the plane of the electron beam.}
\label{fig:erl_gamma}
\end{figure}

\begin{figure}[h]
\centering
\includegraphics[width=0.8\textwidth]{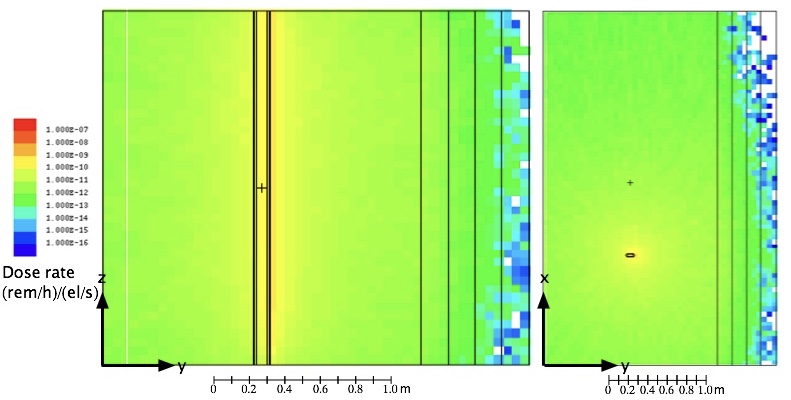}
\caption{Neutron dose rate contours in the plane of the electron beam.}
\label{fig:erl_neutron}
\end{figure}

The gamma and neutron dose rates in the plane of the electron beam are shown in figures \ref{fig:erl_gamma} and \ref{fig:erl_neutron}. Just outside the shielding wall, the dose rates have fairly large error bars. By treating each mesh point just inside of the outside wall in the calculation as a separate ``experiment'', average values and the relative error for the dose rates outside the shielding wall were obtained. The gamma dose rate is $4.263\times10^{-15}$ (rem/h)/(el/s) with a relative error of 0.100, and the neutron dose rate is $6.562\times10^{-14}$ (rem/h)/(el/s) with a relative error of 0.122. Thus the total dose is  $7.00\times10^{-14}$ (rem/h)/(el/s).

If the dose rate just outside the wall is to be 0.050~mrem/h, then the maximum number of electrons that can be lost is $7.2\times10^{8}$/s. Because the energy per 150~MeV electron is $2.4\times10^{-11}$ Joule, this would require a loss of the order or 0.02~W/m. Thus the requirement that the dose rate be kept at 0.05~mrem/h just outside the shielding wall may be unrealistic for a 2' thick shielding wall.

However, if the shield thickness is increased to three or even four feet, the situation improves greatly. If one fits the dose rate attenuation through the 2' thick heavy concrete to \Eq{eqn:dose_decay}, one gets that $\mu=0.1052\unit{cm^{-1}}$, and $D_{o}=1.1051\times10^{-7}$. (In terms of $\lambda$ in \unit{g/cm^2} this corresponds to 37.2\unit{g/cm^2} for a concrete density of 3.912\unit{g/cm^3}. This value is slightly lower than the 45\unit{g/cm^3} at a concrete density of 3.7\unit{g/cm^3} for giant resonance neutrons given in APS-LS-141 Revised.) 

\begin{figure}[h]
\centering
\includegraphics[width=0.8\textwidth]{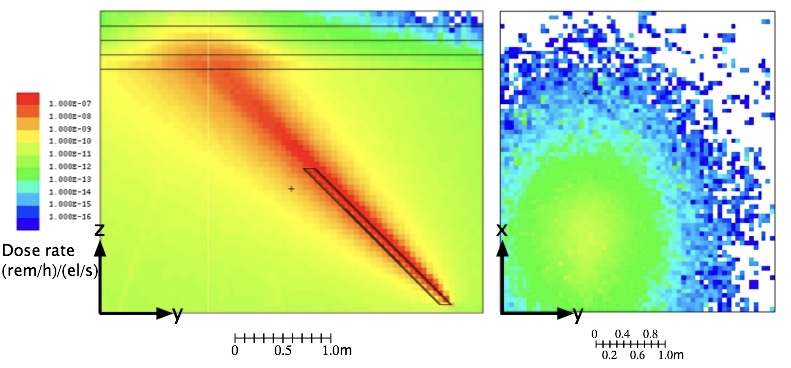}
\caption{Left Figure: Gamma dose rate contours in the plane of the electron beam. Right Figure: Gamma dose rate contours perpendicular to the electron beam outside the shielding wall.}
\label{fig:45deg_gam_hit}
\end{figure}

\begin{figure}[h]
\centering
\includegraphics[width=0.8\textwidth]{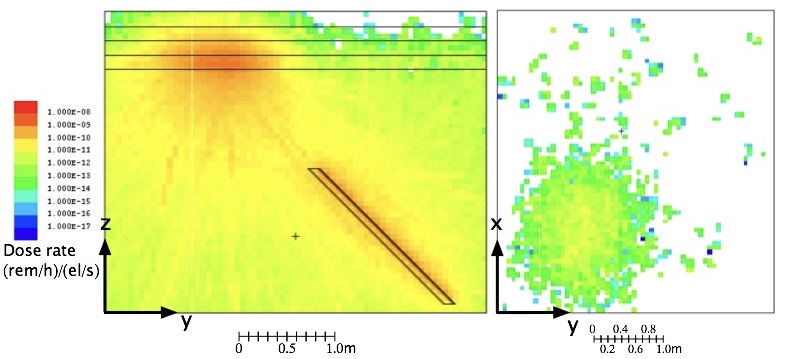}
\caption{Left Figure: Neutron dose rate contours in the plane of the electron beam. Right Figure: neutron dose rate contours perpendicular to the electron beam outside the shielding wall.}
\label{fig:45deg_neut_hit}
\end{figure}

Thus, using the above linear attenuation coefficient $\mu$, the neutron dose rate becomes $6.3\times10^{-16}$ (rem/h)/(el/s) if the shielding thickness is increased to 3' and $2.6\times10^{-17}$ (rem/h)/(el/s) at 4'. For a dose rate of 0.05~mrem/h just outside the wall, a 3' thick shield would allow a loss rate of 1~W/m, and a 4' thick shield, 23 W/m.
 \begin{equation}
 D(x)=D_{o}e^{-\mu x}
 \label{eqn:dose_decay}
 \end{equation}
\clearpage
It is of some interest to see what happens to the dose rates if the vacuum chamber is at $45^\circ$ with respect to the shielding wall. In this case the dose rate contours are shown in figures \ref{fig:45deg_gam_hit} and \ref{fig:45deg_neut_hit}. The maximum gamma dose in the center of the right handed figure in \Fig{fig:45deg_gam_hit} is $1.2\times10^{-11}$ (rem/h)/(el/s) and similarly, the maximum neutron dose rate in \Fig{fig:45deg_neut_hit} is $5.0\times10^{-11}$ (rem/h)/(el/s). Thus the total maximum dose rate is $6.2\times10^{-11}$ (rem/h)/(el/s). These are very high dose rates. It should be pointed out that in this case the beam line is 1.5~m from the wall, whereas in reality, the closest $45^{\circ}$ distance would be more like 3.5~m. 

\clearpage
From \Fig{fig:45deg_gam_hit} one sees that there is a significant flux of gammas that have been generated in the shielding wall. These gammas in turn produce neutrons, as can be seen from the left figure in \Fig{fig:45deg_neut_hit}. The situation will be even worse if the wall is perpendicular to the beam, as in case of the CBETA straight section along the door  wall shown in \Fig{fig:CBETA_shielding}. In this and the $45^\circ$ incidence case, the wall may have to be locally lined with a few inches of lead. This would greatly reduce gamma and neutron production in the shielding wall.

\subsubsection{Catastrophic loss of beam}
\begin{figure}[h]
\centering
\includegraphics[width=0.8\textwidth]{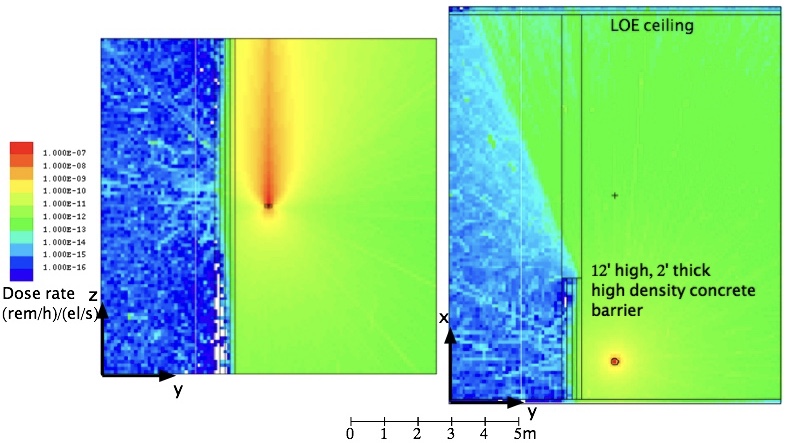}
\caption{}
\label{fig:hall_gamma}
\end{figure}

\begin{figure}[h]
\centering
\includegraphics[width=0.8\textwidth]{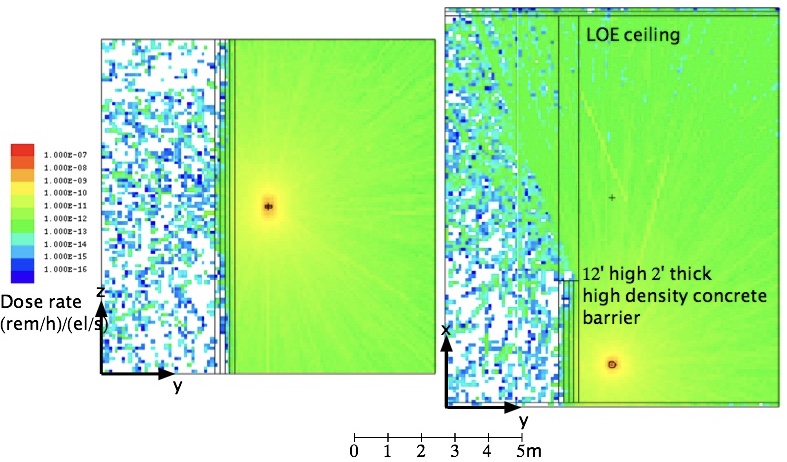}
\caption{}
\label{fig:hall_neut}
\end{figure}

The CSDA range of 150~MeV electrons in aluminum\footnote{\url{http://physics.nist.gov/cgi-bin/Star/e_table.pll}} is 39.72\unit{g/cm^2} This means that the electrons would travel about 15~cm in aluminum. However, a 2~mm diameter, 40~mA, 150~MeV electron beam has a power density of 191\unit{MW/cm^2}. Such a beam would penetrate the vacuum chamber and hit elsewhere. In order to get some idea of the dose rates generated by the full 150 MeV beam hitting something, the beam is assumed to hit a steel cylinder, 20 cm in diameter and 10 cm long. (The CSDA range in iron is 31.46\unit{g/cm^2}, or 4~cm.) Whether or not this is realistic may be questioned, however, it does give an idea of the dose rates produced. Figures~\ref{fig:hall_gamma} and \ref{fig:hall_neut} show the gamma and neutron dose rate contours due to a 150~MeV electron beam striking a steel target located 1 m from the 12' tall, 2' thick, heavy concrete shielding wall located in L0E. (The section of L0E hall modeled is $10\unit{m}\times10\unit{m}$ by 11.84~m high.)

\begin{figure}[h]
     \centering
     \subfloat[][Gamma dose rate outside shielding wall in LOE.]{\includegraphics[width=0.45\textwidth]{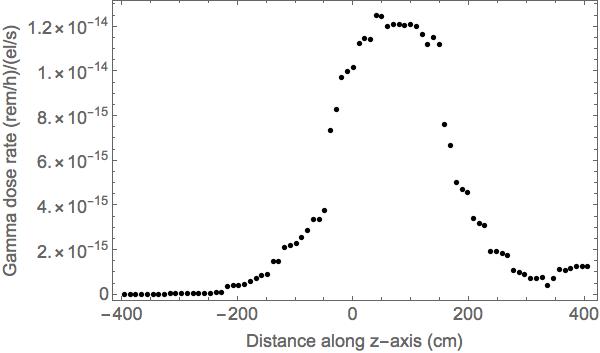}}
     \subfloat[][Neutron dose rate outside shielding wall in LOE.]{\includegraphics[width=0.45\textwidth]{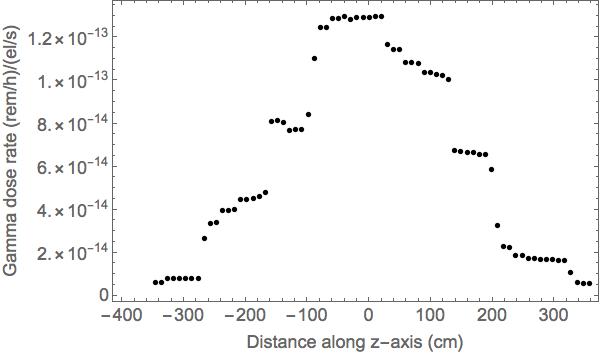}}
     \caption{}
     \label{fig:iron_dose_rates}
\end{figure}

The dose rates just outside the shielding wall have very high errors associated with them, however, by ``smoothing'' the data using the MovingAverage function in Mathematica, the dose rates shown in \Fig{fig:iron_dose_rates} were obtained. The gamma dose rate is $1.2\times10^{-14}$ (rem/h)/(el/s), and the neutron dose rate is $1.25\times10^{-13}$ (rem/h)/(el/s). The total dose rate is  $1.4\times10^{-13}$ (rem/h)/(el/s). Because 40~mA corresponds to $2.6\times10^{17}$ el/s, the dose rate is $3.5\times10^{4}$~rem/h. However, in one second, the dose is 9.7~rem. One would assume that in such case of catastrophic failure the photocathode laser could be shut off much faster than in one second.




\ifdefined \buildingFullDocument

\renewcommand{\FiguresDirectory}{commissioning/figures}

\else
\newcommand{\FullDocumentRoot}{..}
\newcommand{\FiguresDirectory}{figures}

\begin{document}
\fi

\chapter{Commissioning\Leader{Adam Bartnik}}\label{chapter:Commissioning}

\section{Concepts and Philosophy}

An ERL is a non-equilibrium system that lacks a closed orbit and may not possess global transverse or longitudinal stability. Dynamically it is more closely related to time-of-flight spectrometers and injector systems than the conventional linear and circular accelerators that it superficially resembles. ERLs therefore encounter numerous unique operational challenges \Ref{Douglas09_01, Douglas14_01}. Firstly, longitudinal motion dominates the dynamics: timing and energy control set the system architecture, and thus RF phase and gradient control must be assured, as must the lattice momentum compaction,  the correlation of time of flight with energy. Secondly, the non-equilibrium nature of an ERL means that stability is a significant challenge. Thirdly, halo effects dominate high power operation (much as they do in injector chains); losses can be performance limiting: activation, damage (burn-through), and background for potential users are all issues. Finally, as inherently multi-pass systems, ERLs must control multiple beams with different properties (energy, emittance, position, phase) during transport through and handling in common beamline channels. Successful machine operation thus requires a comprehensive strategy for machine commissioning, monitoring machine health, system stabilization, and machine protection.

ERL operation comprises a series of phases: commissioning, beam operation, and machine tuning/recovery. During each phase, system behavior falls into various classes that can be differentiated by the time scales on which they are manifest: ``DC'' conditions, those associated with the machine set point, ``drift'' effects, slow wandering of the set-point (due to, for example, thermal effects) degrading system output, and ``fast'' effects (at acoustical to RF time scales), resulting in beam instabilities. A fourth class, that of transient effects (for example, RF loading during beam on/off transitions and fast shut-down in the event of sudden beam loss for machine protection purposes), can occur throughout all operational cycles. 

Machine commissioning has combined goals of validating system design architecture and defining a recoverable system operating point. For an ERL, this requires demonstration of the control of phenomena of concern such as BBU and the micro-bunching instability ($\mu$BI), while generating settings for hardware components. 

Following pre-commissioning ``hot'' checkout of accelerator components and testing  of hardware subsystems, beam operations commence with threading of low power beam so as to establish a beam orbit and correct it to specified tolerances. At this stage, system performance is error dominated. Some errors (such as RF phases or magnet installation errors) can be readily detected and corrected; others are subliminal, below the resolution of diagnostics or individual measurements, and will accumulate. Thus, some corrections are local (eliminating the error) and others must be global (such as the compensation of cumulative errors).

This requires orbit correction systems based on beam position monitors and steerers (typically every quarter-betatron wavelength); unique to a multi-pass ERL with common transport of multiple beams in a single beam line is the requirement that the system correct perturbations locally so that the multiple passes respond identically and the orbits not diverge unacceptably from turn to turn. Similarly, a baseline for longitudinal beam control must be established, by synchronizing the beam to the RF using recirculator arcs as spectrometers for precision measurements of energy gain. Any path length adjustments needed to set RF phases and insure energy recovery per the design longitudinal match are thus determined.

 With a 6-D phase space reference orbit thus defined, the beam and lattice behavior is tuned and validated. Lattice performance is measured, tuned, and certified using differential orbit/lattice transfer function measurements; these, too, will require pass-to-pass discrimination for beams in common transport. Both transverse and longitudinal measurements (using phase transfer function diagnostics \Ref{Benson11_01}) are necessary for a full analysis of lattice behavior. Corrections must be applied to `rematch the beam to the lattice acceptance' and bring both transverse (betatron motion/focusing) and longitudinal (timing/momentum compaction) motion into compliance with design (or to establish an alternative working point). 
 
Certification of lattice performance allows analysis, tuning, and validation of beam parameters, and matching of the beam to the lattice. This requires measurements of both betatron (emittance, beam envelope functions) and longitudinal (bunch length/energy spread/emittance, phase/energy correlation) properties. If beam properties differ excessively from specification, matching of the beam to the lattice is performed using appropriate correction algorithms. As with orbit correction, perturbations will likely require local correction so as to avoid excessive pass-to-pass divergence of beam properties. Given a validated working point, beam power scaling is performed, with currents increased from tune-up levels to full power CW. 

\section{Goals and Overview}

The basic goal of commissioning is to demonstrate reproducible/recoverable machine operation with specific performance parameters. Given the precedence of longitudinal dynamics in an ERL, this requires, at the highest level, the proper adjustment of linac phases and gradients and transport system momentum compactions to confirm that they are in compliance with the design. Provision must therefore be made for measurement of time of flight and energy as a function of phase, with adequate resolution to set phases and calibrate RF gradients to the tolerances required for successful FFAG operation. This must be done on a pass-by-pass basis, initially setting RF parameters, and meeting subsequent phasing requirements by adjusting turn-to-turn path lengths. 
	This is an iterative process, in which an initial orbit is established (the beam is ``threaded''), and the RF is phased to the available resolution. Performance assessments (for both machine and beam) are then used to evaluate if iteration is required (to improve steering, phasing, and beam quality, correct machine errors, or improve address beam physics effects) or if the commissioning process moves forward. 
	Commissioning activities move forward until beam quality is sufficiently degraded that progress slows or stops, or losses exceed tolerances for safe system operation. Characterization and analysis procedures are then applied to determine if performance limitations are due to the linac, the transport system, or are rooted in beam dynamical effects. This will involve measurements of beam energy, timing (phase) and time of flight, beam properties, and lattice characterization (typically using differential orbit measurement).
	In the following sections, a more detailed list of tasks and goals will be presented. The commissioning is phased, and each section will explain exactly what part of the beamline has been built, which diagnostics are installed, and what the goals are for that particular phase. 
	
\section{Commissioning Flow}

There are five phases of commissioning, ending with 1-pass energy recovery at 1 mA average beam current, satisfying the Key Performance Parameter (KPP). Futher operation, including multiple-pass operation, will not be described here as it falls outside of the scope of this report. Each of the following five phases will be separately described in the following sections.

\begin{itemize}
	\item Gun-ICM-Dump Recommissioning
	\item MLC and Diagnostic Line
	\item Partial Arc Energy Scan
	\item Full Arc Energy Scan
    \item KPP: Single Pass Energy Recovery
\end{itemize}

\subsection{Gun-Linac-Dump Recommissioning}

\begin{figure}[htbp]
\centering
\fbox{\includegraphics[width=0.95\textwidth]{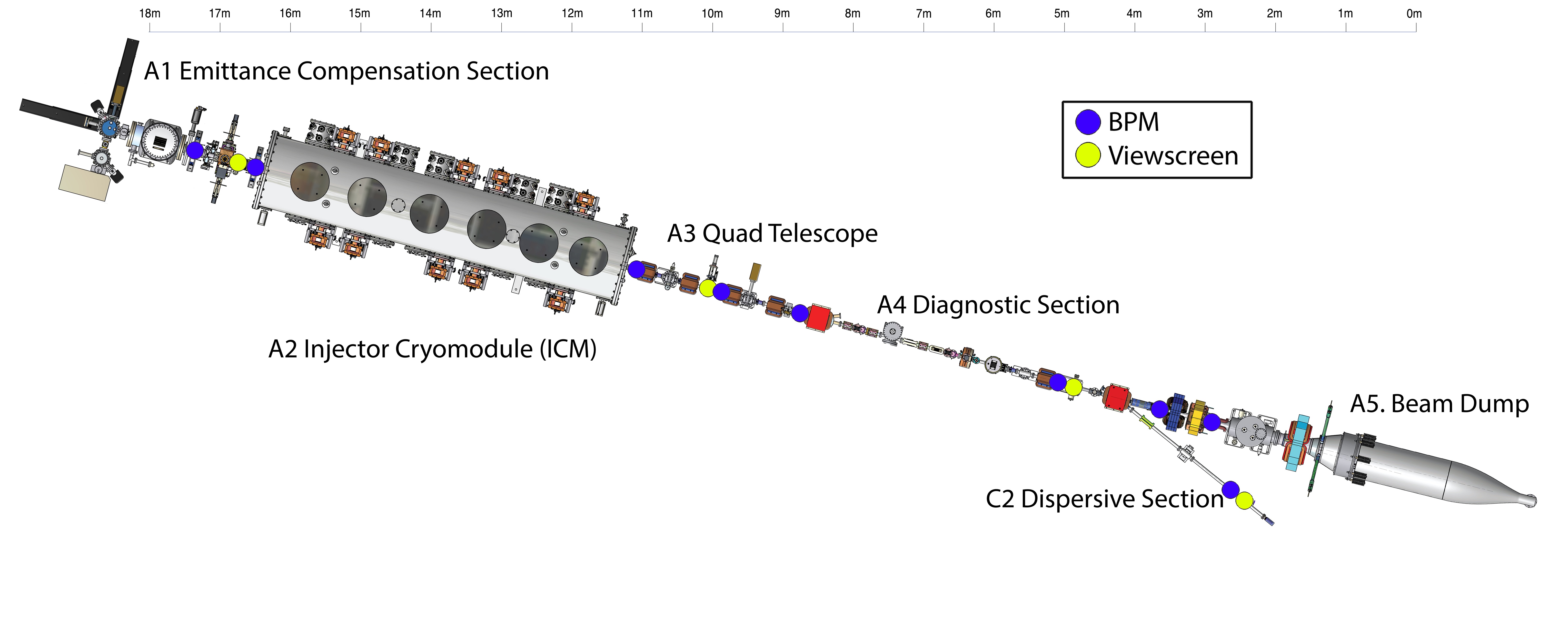}}
\caption[]{Beamline layout during gun-linac-dump recommissioning. BPMs and viewscreens are highlighted with blue and yellow circles, respectively.}
\label{fig:gunlinacrecommish}
\end{figure}

The first phase of commissioning will occur right after the gun, SRF injector cryomodule (ICM), and high power beam dump have been reinstalled in their new location. 
Because prior operation of the Cornell injector occurred at a different location, incompatible with the full CBETA ring design, we were forced to disassemble the injector and rebuild it at its present location. As a result, the first goal is to recommission typical operation of the injector. Thus, the primary goal of this phase is to check each machine subsystem with beam, discover what problems are present, and fix them. 

The injector has set records for both low emittance and high average current, and recovering machine settings for both of these is important, but somewhat incompatible with the short timescale of this phase. As a result, beam emittance will not be measured, and we will focus on recommissioning the high current ability of the injector. So, we will forgo any detailed bunch diagnostics, and instead delay that until it can be done in the diagnostic line test, covered in the next commissioning section.

One of the largest advantages to this commissioning phase is the lack of the Main Linac Cryomodule (MLC). High current operation through the MLC requires full energy recovery, and thus a completed return loop to the linac. So, as soon as the MLC is installed, all high current operation is halted until the entire ring is complete. So, this phase of the project is both the first chance to test the gun and ICM at high current in the new location, and also the last chance to test it without a completed ring.

To that end, we will temporarily install a straight transport line after the ICM, identical to our previously tested design in the injector's prior location, and fully install the high power beam dump along with its water systems and radiation shielding. Both of these will be removed after this test, to allow installation of the merger, diagnostic line, and MLC. Diagnostics during this phase, as shown in Fig. \ref{fig:gunlinacrecommish}, are limited to viewscreens and BPMs. All BPMs in this phase will have both position and phase information, using the same design as has been used in past injector operation. In addition, the high power beam stop also has two unique diagnostics, a quadrant current detector at its entrance, to detect small amounts of beam scraping, and an array of 8x10 thermocouples surrounding the dump, which give a rough position sensitivity to where and how the beam is hitting the dump. Both of these beam stop diagnostics will be tested during this phase. The rough procedure will be as follows. 

Beginning with the setup of the gun and cathode system, two types of cathodes are required for operation, a large area cathode for initial alignment, and an small area off-center cathode for high current. At least three of each type will be produced. Initial operation will use a large ($\approx 1$ cm) active area photocathode, both to allow the laser to be aligned within the gun, and to reduce beam asymmetries inherent in off-center operation. These asymmetries do not affect machine performance, and can be easily tuned away later, but can confuse the initial beam alignment procedure, so it's useful to avoid them whenever possible.

After this, we would normally perform a somewhat lengthy alignment procedure for our initial low energy optics. Good alignment is only truly important when low emittance is desired, and that is not yet a requirement. So, without using the translation and angle adjustment motors on the solenoid magnets, a low duty factor beam will be centered as best as possible though the solenoids, while ignoring alignment in the buncher. Alignment in the first SRF cavity will be quickly adjusted and verified by ensuring the beam position downstream is not sensitive to variations of $\pm 10^\circ$ in cavity phase. The alignment in the additional downstream SRF cavities will not be checked and is not required.

SRF cavities will be individually set to 1 MeV energy gain, and phased on crest. This is not the intended CBETA operating point, and does not minimize the beam emittance, but is sufficient to test basic operation of the ICM at high current. There is no deflector installed, nor emittance measurement equipment, so the properties of the resulting bunch will not be studied in detail. Instead, the beam will be guided to the dump, and optics tuned as needed to minimize loss. When this is finished, the large active area cathode will be swapped out to an off-center small-area photocathode, appropriate for high current. Still with low duty cycle, the laser will be re-guided to the new location of the active area, and the electron beam orbit will be adjusted back to the previously set reference orbit. 

Finally, the duty factor will be changed to $100\%$, and current will be increased by turning up the bunch charge. Of course, the properties of the bunch are changing as a function of charge, but since we have a short, straight beamline, the details of the bunch do not affect its ability to be transported successfully to the beam dump. The RF couplers into the SRF cavities will be set to their lowest current setting, since this requires the least amount of time for processing and setting up. With this setting, the   klystrons will probably be unable to support beam currents higher than 5 mA, since almost all power will be reflected. But, this is sufficient to test our ability to reach the beam current requirements of the KPP, and any significant problems in high current operation can be addressed in this early phase of commissioning.

\subsection{MLC and Diagnostic Line}

\begin{figure}[htbp]
\centering
\fbox{\includegraphics[width=0.95\textwidth]{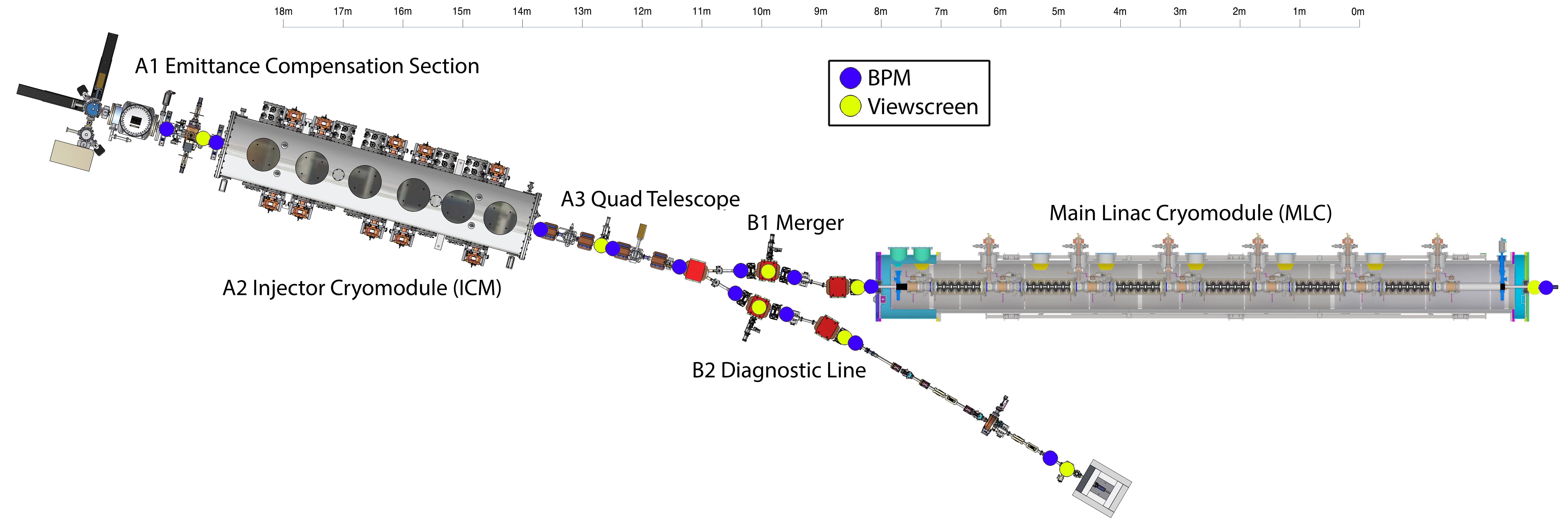}}
\caption[]{Beamline layout during MLC and diagnostic line commissioning. BPMs and viewscreens are highlighted with blue and yellow circles, respectively.}
\label{fig:mlccommish}
\end{figure}

The next major commissioning phase is the MLC and diagnostic line test. The goal of this test is to fully commission the injector to the intended CBETA operating point. This requires full characterization of the bunch at a variety of bunch charges in the diagnostic beamline, and demonstrated cavity-by-cavity acceleration in the MLC at the target cavity setpoint. At the end of the test, we should have a machine setpoint ready to be used for first pass operation of CBETA. 

The diagnostics for this phase (Fig. \ref{fig:mlccommish}) include the same position and phase BPMs used in the previous commissioning phase-- that is, the switch to the new BPM electronics design has not yet been performed. But, importantly, one prototype of this new design will be installed at the end of the diagnostic line, along with the pipe and BPM button dimensions appropriate for use in the FFAG arc. This phase will preset our first opportunity to test this design, and the location of this BPM right before a viewscreen will allow the design to be full calibrated, and its nonlinear position correction to be tested. 

All bunch characterization will be done in the diagnostic line (B2), which is designed to be a mirror image of the merger line (B1) to the MLC, so if the magnets are put at identical setpoints, then the bunch should have nearly identical properties as in the actual CBETA arc. We will use our Emittance Measurement System (EMS), described elsewhere, to characterize both horizontal and vertical 2D phase spaces of the beam, the longitudinal current profile of the bunch. This allows us to know both the emittance and the twiss parameters of the bunch as it would have on entering the MLC. All of those parameters are critical to match into the rest of the CBETA arc, and are very difficult to predict in this low energy part of the machine where space charge still dominates the dynamics. After acceleration to 42 MeV, space charge is greatly suppressed, so having the full phase space knowledge of the beam is perhaps most important at this point. Importantly, this will also be the last point in the machine where the bunch can be fully characterized in this way.

We will adhere to the rule that at least three usable cathodes should be available at all times. To begin, a large area photocathode will be reinstalled into the gun, and the laser re-centered onto the photocathode. Now that low emittance will be required, alignment in the early low-energy beamline is crucial. The first week of commissioning will be spent aligning the solenoids with their motors, and the buncher and first two SRF cavities using dipole corrector magnets. The alignment procedure exists and has been used successfully multiple times in the past and will not be detailed here. Verification of the alignment is performed by checking beam emittance at $\ll 1$ pC bunch charge, at a charge where there is no emittance degradation from space charge. The emittance is measured in the diagnostic line, so this will be a good first test of the EMS. Once we demonstrate that this cathode emittance is preserved through the merger, it will be time to tune up the beam at non-trivial bunch charge.

Using the machine settings from simulation as a starting point, we will tune up the machine initially at 25 pC bunch charge. This is sufficient bunch charge to reach the Key Performance Parameter (KPP) of 1 mA with a bunch rate of $1300/31 = 41.9$ MHz, which is the maximum rate in the eRHIC bunch pattern mode of operation. Beam duty factor during this phase will be kept $\lesssim 0.1\%$, to minimize radiation, and indeed will be kept this low throughout all commissioning phases. The primary goal of the tune up is to reach a normalized emittance below $1 \mu m$, as measured in the diagnostic line, while also matching desired beam twiss parameters needed to match optics in the CBETA lattice. The tuning of the machine will be primarily manual tuning, but guided by simulated predictions around the ideal setpoint in simulation. The capability to predict machine response already exists in our simulation and has been used similarly in the past.

Once the setpoint at 25 pC is determined, we will explore methods to maintain this twiss match as a function of bunch charge. Simulations predict that at moderate bunch charges, below $\approx 75$ pC, good twiss match can be maintained by adjusting only the laser spot size, buncher voltage, and first SRF cavity phase as a function of bunch charge, and we will attempt to verify this with the beam. If true, this would present a remarkably simple way to tune the machine for different bunch charges, and for high current operation, it might allow current to be ramped up by increasing bunch charge. Whether this simple method works or not, low-emittance and twiss-matching machine setpoints will be determined at a variety of bunch charges, including at least 0.7, 3, and 25 pC, allowing flexibility during future commissioning phases. 0.7 pC is sufficient to meet the KPP 1 mA current with a 1.3 GHz laser, while 3 pC is a rough minimum charge for reliable BPM measurements.

Using the 3 pC setpoint, the beam will be steered towards the Main Linac Cryomodule (MLC). Initially, the MLC will be left unpowered, and the beam steered through to the downstream BPM and viewscreen, which are the only diagnostics after the MLC. Additional diagnostics were prohibited during this phase, due to space conflicts with existing infrastructure. If a good twiss-match was achieved, as stated above, the beam should be able to be guided through the MLC without any loss, even without powering any MLC cavities.

After this, each MLC cavity will be individually powered, one-by-one, to the desired CBETA operating point. In simulation, the cavity phase can be adjusted by at least $\pm 60^\circ$ from on-crest without significant defocusing of the beam. We will phase each cavity on-crest by adjusting the cavity phase and monitoring the arrival time at the downstream BPM. With 0.1 $\mu A$ of beam current, our BPMs are capable of measuring arrival time with an uncertainty of $0.3^\circ$ of RF phase. With $\pm 60^\circ$ of RF phase adjustment in the cavity, we expect from 3 to 10 degrees of arrival phase change, depending on the distance of the cavity to the BPM (i.e. depending on which cavity we are phasing), which should be easily measurable. This should enable us to phase the cavity on-crest within $\approx 5^\circ$, which is sufficient for this phase of commissioning. Later, when the splitter is installed, further refinement of the phase can be performed by using the splitter as a spectrometer.

\subsection{Partial Arc Energy Scan}

\begin{figure}[htbp]
\centering
\fbox{\includegraphics[width=0.95\textwidth]{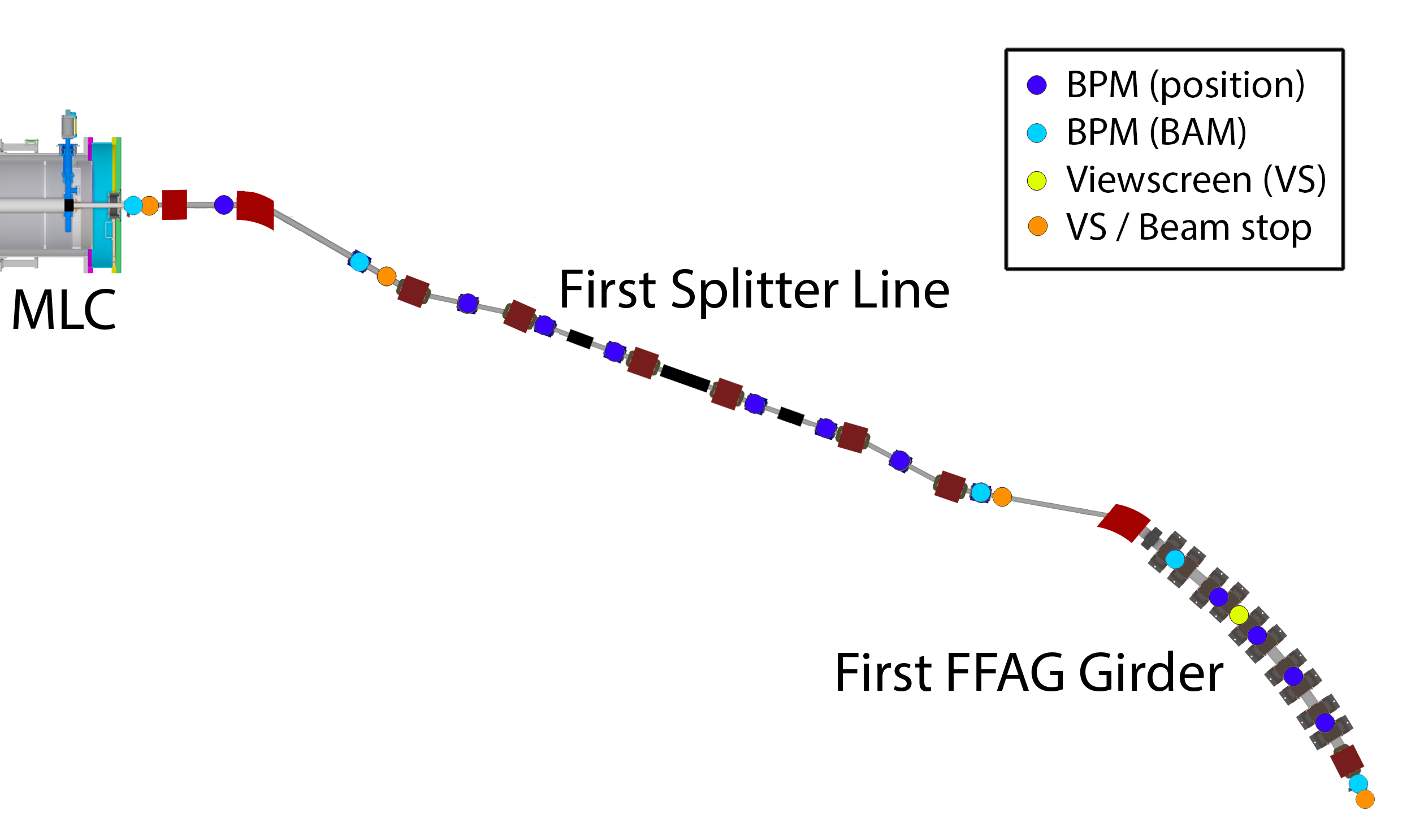}}
\caption[]{Beamline layout during partial arc commissioning. The ``flat pass'' layout of the first splitter line is shown, allowing energy recovery after a single pass.}
\label{fig:partialarccommish}
\end{figure}

Prior to the beginning of this commissioning phase, the first arm of the splitter and the first FFAG magnet girder will have been installed. The layout of the beamline after the MLC is shown in Fig. \ref{fig:partialarccommish}, and the location of diagnostics therein. The layout of the pre-MLC beamline is identical to that in the previous commissioning phase. The BPMs in the section will now be using the new electronics hardware and button design, previously tested during the MLC test. That means that there is now two possible types of BPMs, which are labeled as BPM (position) and BPM (BAM) in the figure. The former type is capable of only measuring the beam position, but can separately get the position of potentially 7 different beams at 4 energies,  during later operation with multiple beam passes. During this phase, of course, there is only one beam present at all of these BPMs, so their operation will be much simpler. The second type, referred to as Beam Arrival Monitors (BAM), is capable of seeing up to two temporally separated beams and gives both position and phase information. Thus, during later multi-pass operation, these will still give full position and phase information of the accelerating and decelerating beams in each spreader line. Importantly, in the spreader, each BAM is positioned right before a combination viewscreen and insertable beam stop, which will allow the beam to be safely parked and the current increased as needed to increase BPM position and phase resolution. 

The goals for this phase are to thread a beam though this FFAG girder and verify the expected optics through that section as a function of beam energy delivered by the MLC. This energy will include the target 42 MeV of the first pass, and be increased as high as the MLC can be operated. Among the important properties to verify in the splitter are that the dispersion is properly zeroed on exit, and that both the value of the momentum compaction (change of arrival time with energy) and its tunability is as predicted in the model. The orbit through the single FFAG girder is likely to be dominated by injector orbit error, rather than magnet position or strength errors, due to the short length of the section. But, it will still be valuable to check the periodicity of the orbit, and compare to model, even within only the first four cells.

The injector will be set up identically to how it was in the previous commissioning phase. Importantly, we will now have all high power amplifiers for the MLC installed, and can power all cavities simultaneously for the first time. So, using the previously determined 3 pC machine setpoint, we will finalize the previous cavity phasing, now that the dispersive splitter section can be used as a spectrometer. Beginning with 42 MeV total beam energy, the beam will then be threaded through the splitter, ending initially in the insertable beam stop at the end of the splitter. With the beam parked in the beam dump, the current will be increased enough to provide acceptable BPM resolution. Difference orbits, difference arrival time, and dispersion measurements will be performed for each of the splitter magnets and each BPM.

Barring any large unexpected differences with simulation, the beam will then be threaded through the end of the splitter and through the first FFAG girder, terminating at the beam stop after the girder. During that threading, we will keep the FFAG corrector magnets unpowered, as they should not be needed unless the FFAG magnets have major installation problems. But, after threading, similar difference measurements will now be performed to verify the full set of installed optics, including the FFAG correctors. 

At this point, the MLC energy gain will be slowly increased while ramping up the splitter magnet currents. The dependence of the orbit in the FFAG section will be monitored, and in this way the energy acceptance of the FFAG will be determined. If large systematic deviations from the expected orbit are observed, the FFAG magnets will now be moved, using predictions from simulation. In the process of correction, the origin of any systematic problems will need to investigated and understood, in order to guide further FFAG magnet installation. 

Finally, the energy will be brought back to 42 MeV, and the operation of the path length adjustment in the 1st splitter will be investigated. Importantly, we will need to determine how to keep the orbit stable during path length adjustment, so that we can later use that adjustment to fine tune energy recovery, without disturbing the orbit through the FFAG arc.

\subsection{Full Arc Energy Scan}

\begin{figure}[htbp]
\centering
\fbox{\includegraphics[width=0.95\textwidth]{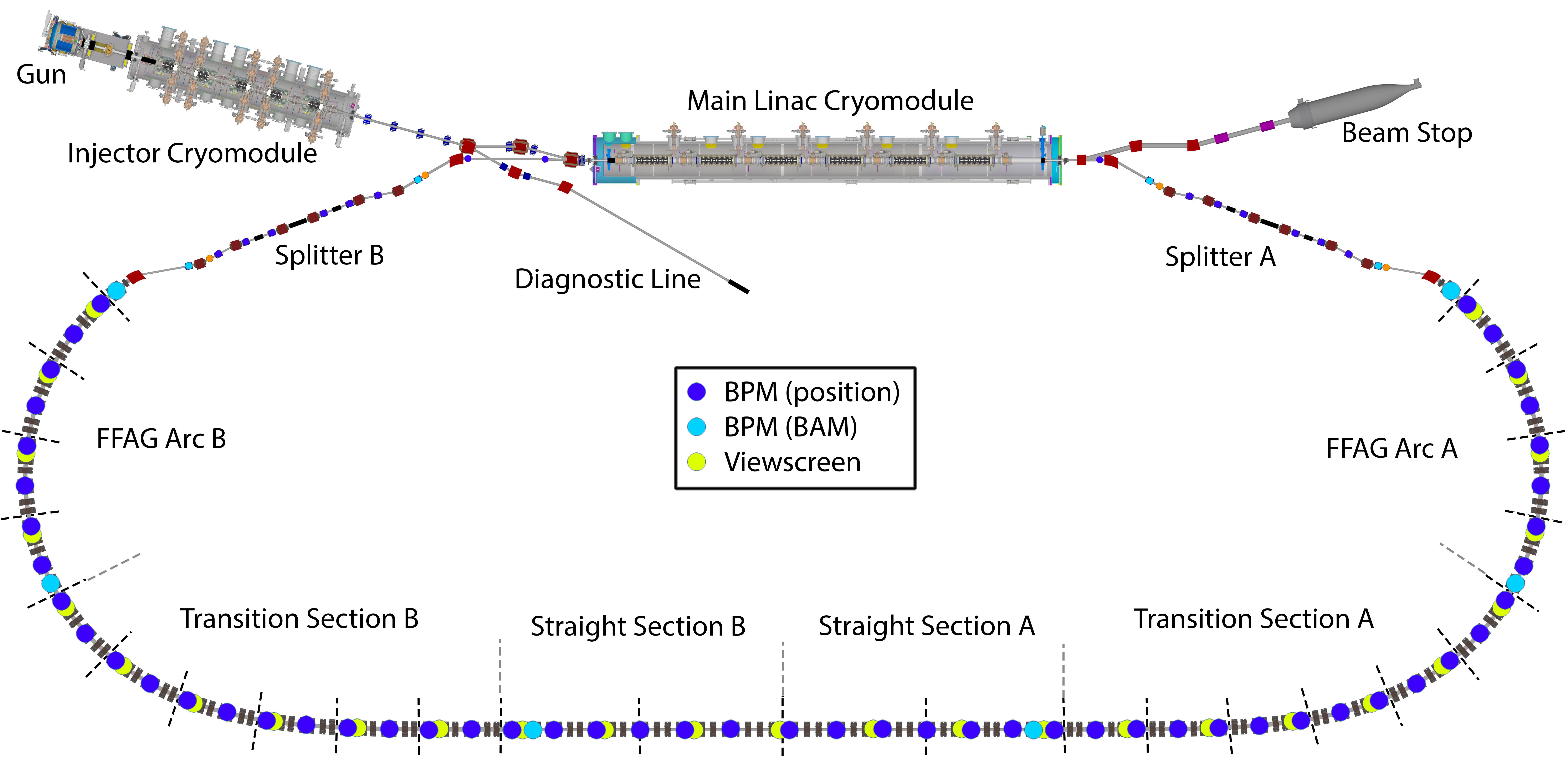}}
\caption[]{Beamline layout during full arc commissioning. The ``flat pass'' layout of the first splitter lines is shown, allowing energy recovery after a single pass.}
\label{fig:fullarccommish}
\end{figure}

At this point, the entire FFAG arc will have been completed, and the first pass splitter line on both sides of the arc completed and connected to the MLC (Fig. \ref{fig:fullarccommish}). The goals of this phase are identical to the partial arc energy scan, except now including the entire FFAG arc. So, the effect of FFAG magnet position errors and injection errors from the splitter will now be greatly magnified. So, the goals will be the same, but the needed precision in tuning will be much higher.

Diagnostics in the FFAG will generally follow the guidelines of one viewscreen per FFAG girder, and one BPM per cell. Importantly, only every other BPM will be wired to readout electronics, as shown in Fig. \ref{fig:fullarccommish}, with the possibility of swapping electronics or adding additional electronics at a later time. All of those BPMs will be position-only, capable of seeing potentially all 7 beams when in multi-pass configuration. A few of the BPMs in the unwired girder position will be used as Beam Arrival Monitors, as in Fig. \ref{fig:fullarccommish}, placed roughly at the beginning of each FFAG section. Though only useful in the initial single-pass mode of operation, these BPMs will still be a valuable way to measure arrival phase separately at each of these parts of the arc. Diagnostic details of the second splitter, though not shown here, are identical to the details of the first splitter as shown in Fig. \ref{fig:partialarccommish}.

Beginning with the 42 MeV, 3 pC machine setting determined in the last phase, the beam will be progressively threaded through the rest of the FFAG arc. This will likely be a long, iterative process. In each section of the FFAG we will have the ability to compare to an online model of the ring, able to predict any needed corrector settings. In addition, the arc sections have the benefit that each cell should have the same orbit, and this periodicity will give us a simple way to check the beam's behavior. Similarly, in the straight section the orbit is centered in the pipe, which is somewhat simpler conceptually, and perhaps easier in practice to tune. Also by virtue of the centered orbits, if the beam has the correct orbit, then the corrector magnets will have no effect on the orbit, which gives us an additional indirect orbit check at each magnet. The transition section orbit is likely to be the most challenging, since the orbit is neither periodic nor centered, so we will have to rely on comparisons to the model. Throughout this initial threading in the FFAG, it is possible that only viewscreen and rough BPM measurements will be possible until the beam has been fully threaded through the arc, as the max operational current may be limited to prevent accidentally dosing and demagnetizing the permanent magnets. Beam loss monitors around the ring may prove to be a useful, though indirect, steering tool as the initial goal might be to simply minimize loss. 

After successfully getting the beam through the FFAG sections, and into the second splitter, the beam will be safely parked in the splitter's first beam stop, in order to allow the current to be increased enough for higher precision BPM measurements. At this point, the orbit will be fine-tuned using the FFAG corrector magnets, using the predicted magnet responses from simulation. The strengths needed to correct the orbit will give us information about any systematic problems with our FFAG magnet installation, and depending on their severity, we will again evaluate the need to adjust the FFAG magnet positions. Once a satisfactory orbit has been achieved, difference orbits, dispersion, and arrival time measurements will be taken using each of the FFAG correctors, at each of the available BPMs, and compared to model.

Finally, provided that the machine appears to be in its intended operating point, and the optics are verified to needed accuracy, we will once again slowly increase the energy of the MLC, while the splitter magnet strengths are similarly ramped up. The dependence of the orbit through the FFAG will be checked against model in the same manner as in the partial arc test. This will be the final test of the acceptance of the FFAG before future multiple-pass operation.

\subsection{KPP: Single Pass Energy Recovery}

This phase of commissioning occurs immediately after the previous phase, and no changes to the beamline or diagnostic layout will have been performed. At this point, the primary remaining difficulty is achieving energy recovery. The orbit through the second splitter will be threaded and then verified in the same manner as the first splitter, and then the beam will be guided into the MLC. If there is sufficient energy recovery for the beam to enter the dump beamline, then the path length for best energy recovery can be fine-tuned using the path length adjustment in both splitters. If either the adjustment range of the splitters is insufficient, or the path length is so inaccurate that the beam does not even enter the dump beamline, then more drastic changes will need to be made to the splitter beamlines, to add or subtract additional path length. 

After the beam is successfully threaded into the dump, the next challenge is to raise the current. There are two strategies to do this: increase the number of bunches at a fixed charge, or increase the bunch charge at full duty factor. The first option requires the laser to achieve a large rejection ratio of laser pulses, so that the rejected part of the beam does not add up to a significant current. The rejected beam will be at a much smaller bunch charge than the desired part of the beam, and therefore be badly twiss-matched to the FFAG lattice and likely be lost. This beam loss can be a potentially limiting source of radiation, and might make this approach impractical. However, the second option will require us to find a path in optics settings from zero to full bunch charge that preserves the lattice matching. If too many machine settings need to be varied, or if impractical changes to the initial laser properties are required, then this may also prove to be difficult. Both will have to be tested and evaluated.

Traditionally, the most difficult aspect of increasing the current is reducing the radiation from beam loss of the halo. Up until this phase of commissioning, we will not have had any direct way to measure the presence and effect of halo. The location of the beam loss will be known only indirectly by beam loss monitors, and simulation will have to be used as a guide for optics changes to reduce it. In past operation of our injector, problems with halo only became limiting when the beam current exceeded roughly 10 mA, and as such we may discover that the KPP current of 1 mA is below the threshold where it becomes important. But, due to the factor of 10 increase both in energy and length of the accelerator, compared to our previous experience, we cannot completely rely on that as a guide. 

If it appears that using the eRHIC bunch pattern to reach the KPP current goal (i.e. 25 pC at 41.9 MHz) causes too many problems with halo, we can switch to a higher repetition rate laser and proportionally smaller bunch charge. We have an existing 1.3 GHz laser oscillator, which would offer the smallest possible bunch charge at the KPP 1 mA beam current. Using the previously determined injector setting at 0.7 pC, we can still attempt to turn up the current to 1 mA. At such a small bunch charge, and at such a large bunch rate, the precision of the BPMs will be greatly reduced. So, in this mode of operation, if the optics need to be greatly adjusted, the laser would need to be temporarily switched back to the eRHIC rate in order to have accurate BPM readings. We have had two laser oscillators running simultaneously in the past, with the ability to quickly switch repetition rate, so this capability is possible.

In the end, regardless of the method used to reach 1 mA beam current, we will still attempt to raise the current as high as possible. That may require using an entirely new operating bunch charge, or potentially a similar switch to a high repetition rate laser to keep using the same bunch charge. Exploring the maximum usable current will be the final goal of this phase of operation.

\ifdefined \buildingFullDocument

\renewcommand{\FiguresDirectory}{survey/figures}

\else
\newcommand{\FullDocumentRoot}{..}
\newcommand{\FiguresDirectory}{figures}

\begin{document}
\fi

\chapter{Survey and Alignment\Leader{Frank Karl and Dejan Trbojevic}}\label{chapter:survey}

\section{Survey Procedure}
 
A procedure is described and methods which yield toward alignment of
positions of all CBETA components within their tolerances. Major principles
governing survey and alignment measurement space are briefly shown and their
relationship to a lattice coordinate system. Then a discussion of the activities
involved in a step-by-step sequence from initial layout to final alignment is
described. In the vertical plane, the curvature of the earth does not need to be
considered as the dimensions of the whole CBETA project are to small. The
leveling is done with respect to gravity. Larger ERLs as in \Ref{Cornell-ERL-PDDR} would have to consider curvature effects.

\subsection{Network Design Philosophy}

\begin{figure}[htbp]
\centering
\subfloat[]{\includegraphics[width=0.5\textwidth]{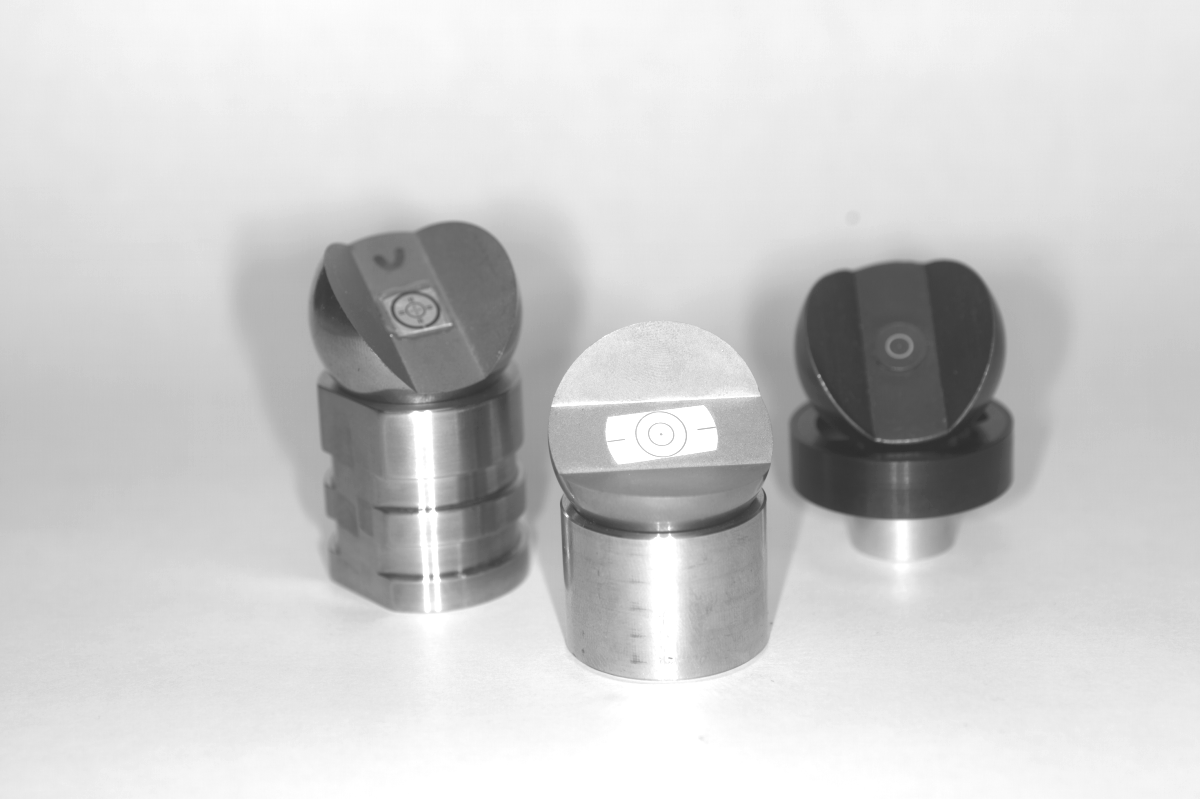}\label{fig:theodolite_target}}
\subfloat[]{\includegraphics[width=0.5\textwidth]{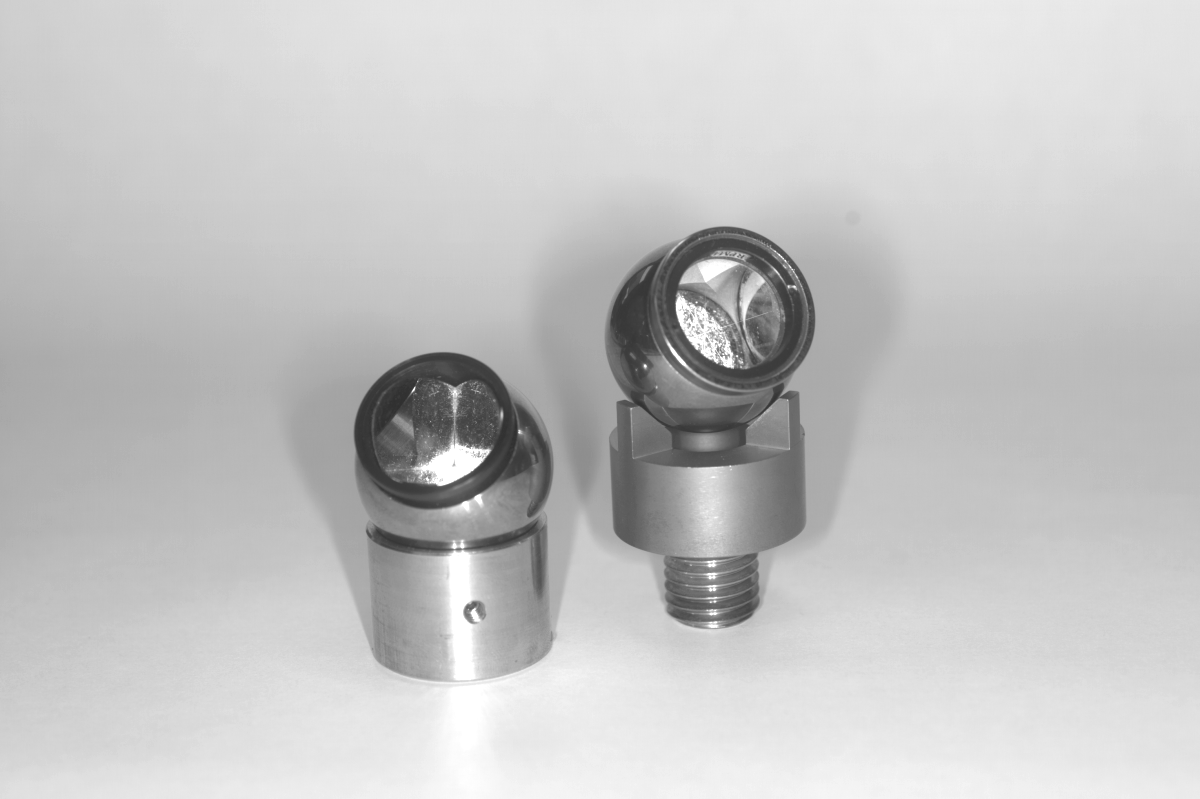}\label{fig:reflectors}}
\caption[]{Left: Sphere mounted theodolite target. 
Right: Sphere mounted glass and air reflectors.}
\label{fig:surveytargets}
\end{figure}

\begin{figure}[tb]
\centering
\includegraphics[width=\textwidth]{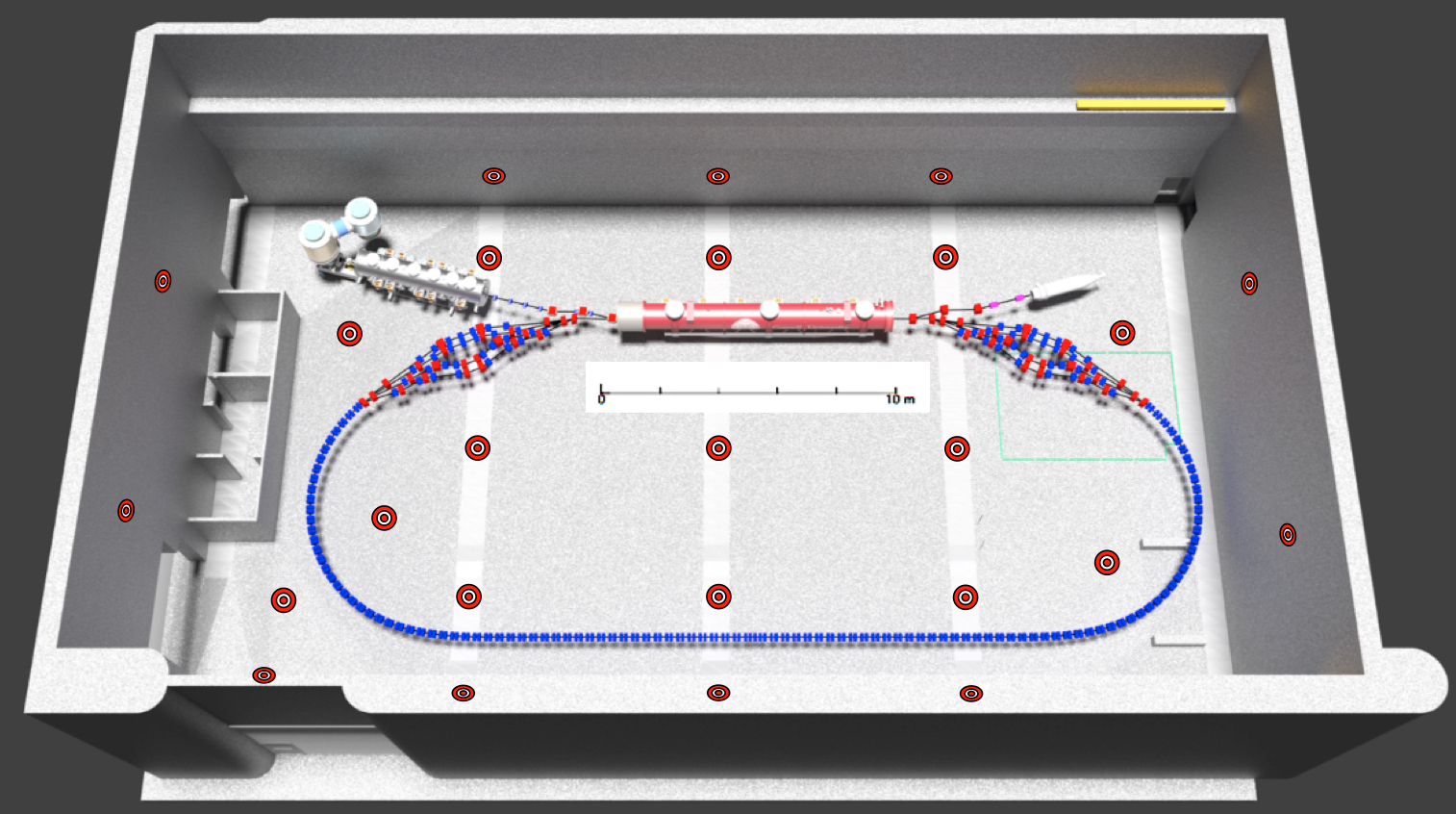}
\caption[network_layout]{CBETA survey network layout.}
\label{fig:network_layout}
\end{figure}

A 3D ``free stationing'' is chosen, rather than setting up the instrument over
a known point. The instrument's position is flexible and chosen only following
considerations of geometry, line of sight and convenience. To determine the
instrument position, at least three points, whose coordinates are already known or
are part of a network solution, need to be included in the measurements. The
approach does not require forced centered instrument set-ups, thus eliminating the
need for set-up hardware and their systematic error contribution. Removable
heavy-duty metal tripods, translation stages, CERN sockets and optical plummets are
not needed (see \Figure{fig:surveytargets}). The geometry should also permit observing each
target point from at least three different stations. The following sketch (\Figure{fig:network_layout})
shows a typical section of the layout. A triplet of monuments is always placed in
the tunnel cross section containing a corrector magnet. One monument will be
placed close to the corrector magnet on the floor, the second one mounted at
instrument height to the aisle wall, and the third monument mounted to the back
wall also at instrument height.

\subsection{Lattice Coordinate System}

The LCLS lattice is designed in a right handed beam following coordinate system,
where the positive y-axis is perpendicular to the design plane, the z-axis is pointing in the
beam direction and perpendicular to the y-axis, and the x-axis is perpendicular to both the
y and z-axes.

\subsection{Tolerance Lists}

The relative positioning tolerances $\sigma_x$, $\sigma_y$, $\sigma_z$ of the girders, BDs, QFs are listed
in \Table{tab:position_tolerance} below.

\begin{table}[htbp]
\caption[]{CBETA positioning tolerances.}
\begin{tabular*}{\columnwidth}{@{\extracolsep{\fill}}lcccc}
\toprule
 & $\sigma_x$ & $\sigma_y$ & $\sigma_r$ & $\sigma_{x/z}$ \\
 & [$\mu$m] & [$\mu$m] & [mrad] & [$\mu$m/m] \\
\midrule
Straightness of girders & 15 & 30 & 1 & n/a \\
Relative alignment between girders & 50 & 50 & 1 & n/a \\
Global straightness of CBETA arcs & 100 & 500 & 2 & 100/10 \\
Linac straightness & n/a & n/a & n/a & 150/15 \\
Quadrupole \textit{ab initio} & 50 & 50 & n/a & n/a \\
\bottomrule
\end{tabular*}
\label{tab:position_tolerance}
\end{table}

\subsection{Relationship between Coordinate Systems}

The relationship between the surveying and lattice coordinate systems is given by
the building design and machine layout parameters. The result is a transformation matrix
(rotations and translations).

\section{Fiducializing the CBETA Magnets}

\begin{figure}[tb]
\centering
\includegraphics[width=\textwidth]{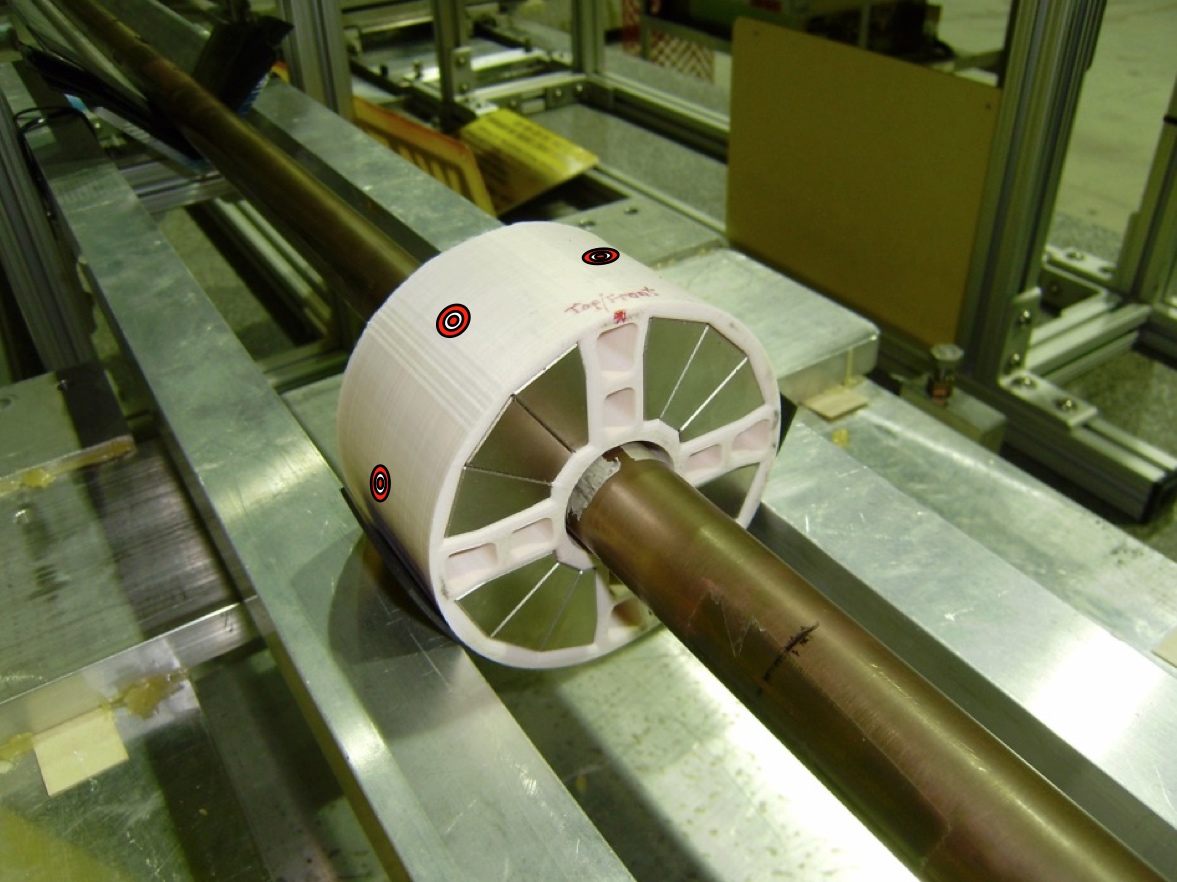}
\caption[network_layout]{Harmonic coil measurements and magnet fiducialization.}
\label{fig:fiducials}
\end{figure}

The correct fiducialization of magnets is as important as their correct alignment since an
error in either task will affect the particles' trajectory and cannot be distinguished from
each other. Fiducialization can be accomplished either through opto-mechanical and
opto-electrical measurements or by using fixtures, which refer to a magnet's reference
features. The opto-electrical measurements are the ``Harmonic coil'' measurement
preformed together with the survey measurements of the magnetic field center, magnet
fiducials with respect to the harmonic coil center as shown in \Figure{fig:fiducials}.

\section{CBETA Absolute Positioning}

Common to all parts of the machine, free-stationed laser trackers, oriented to at
least four neighboring points, are used for the absolute positioning measurements. The
tracking capabilities of these instruments will significantly facilitate the control of any
alignment operation (moving components into position).

\subsection{Pre-alignment of Girder Supports and Magnet Movers}

The girders will be
supported by adjustment systems sitting directly on top of concrete piers. The adjustment
systems are based on the Bowden camshaft design. Two individually controlled camshaft
pairs and two single camshafts provide five degrees of freedom per girder. The camshaft
design doesn't compromise the rigidity of the supports and, consequently, doesn't show a
resonance in an undesirable frequency range. This mover system comes in two horizontal
slices. The bottom piece consists of a mounting plate, which holds the shafts. The top part
is integrated into the girder by mounting the kinematic cams to the girder. The girder is
held onto the shafts by gravity.

To accommodate easy installation, the bottom parts of the movers, set to mid
range, have to be aligned relative to each other. The required relative position tolerance
of these, however, is fairly loose, since the two axes cams are only paired with a single
axis cam. On the other hand, to retain as much magnet mover range as possible, the
bottom part of the magnet movers should be within 0.5mm of their nominal positions.

To facilitate placing a pedestal such that its top is within 0.5mm of its nominal
position, a widely used method can be used. Here, the base plate of the bottom part of a
mover is mounted to the pier by four standoff screws, which are grouted/epoxied into the
concrete. The vertical/horizontal pre-alignment of the base plate is accomplished by the
following sequence of steps: After the four bolts are epoxied into the concrete, a nut with
a washer on top is screwed onto each bolt. These nuts are set to their nominal heights by
a simple level operation. Next the base plate is set on the nuts, and a set of washers and
nuts is then screwed on the bolts to fasten it down. However, the top nuts remain only
hand tight at this point. Next, the elevation and tilts of the base plate are set by adjusting
the position of the lower nuts, and subsequently checked with a level with respect to local
benchmarks. Then a total station with a ``free station Bundle'' software package is used to
determine the horizontal offset and to simultaneously double-check the vertical offset of
the base plate from its nominal position. Finally, the base plate is moved into horizontal
alignment using a clamp-on adjustment fixture (push - push screw arrangement), and the
nuts are tightened to the prescribed torque. To vibrationally stiffen the set-up, the space
between the pier and the base plate should be filled with non-shrinking grout after the
alignment has been confirmed.

\subsection{Quality Control Survey}

Once the above step is completed, the mover positions
will be mapped. If the positional residuals exceed the tolerance, a second iteration can be
``jump started'' by using the quality control map to quantify the position corrections,
which need to be applied. Should a second iteration be necessitated, a new quality control
survey is required after completion of the alignment process.







\end{document}